\pdfoutput=1
\documentclass[a4paper,11pt,fleqn]{book}
\usepackage[T1]{fontenc}
\usepackage[utf8]{inputenc}
\usepackage[french,english]{babel}

\usepackage{fourier} 
\usepackage{setspace} 

\usepackage{graphicx,xcolor}
\graphicspath{{images/}}

\usepackage{subfig}
\usepackage{booktabs}
\usepackage{lipsum}
\usepackage{microtype}
\usepackage{url}
\usepackage[final]{pdfpages}

\usepackage{fancyhdr}

\fancypagestyle{plain}{
	\fancyhf{}

	\fancyfoot[EL,OR]{\thepage}}
\fancypagestyle{addpagenumbersforpdfimports}{
	\fancyhead{}
	
	\fancyfoot{}
	\fancyfoot[RO,LE]{\thepage}
}

\usepackage{listings}
\makeatletter
\def\cleardoublepage{\clearpage\if@twoside \ifodd\c@page\else
    \hbox{}
    \thispagestyle{empty}
    \newpage
    \if@twocolumn\hbox{}\newpage\fi\fi\fi}
\makeatother \clearpage{\pagestyle{plain}\cleardoublepage}

\usepackage{color}
\usepackage{tikz}
\usepackage[explicit]{titlesec}
\newcommand*\chapterlabel{}
\titleformat{\chapter}[display]  
	{\normalfont\bfseries\Huge} 
	{\gdef\chapterlabel{\thechapter\ }}     
 	{0pt} 
 	  {\begin{tikzpicture}[remember picture,overlay]
    \node[yshift=-8cm] at (current page.north west)
      {\begin{tikzpicture}[remember picture, overlay]
        \draw[fill=black] (0,0) rectangle(35.5mm,15mm);
        \node[anchor=north east,yshift=-7.2cm,xshift=34mm,minimum height=30mm,inner sep=0mm] at (current page.north west)
        {\parbox[top][30mm][t]{15mm}{\raggedleft $\phantom{\textrm{l}}$\color{white}\chapterlabel}};  
        \node[anchor=north west,yshift=-7.2cm,xshift=37mm,text width=\textwidth,minimum height=30mm,inner sep=0mm] at (current page.north west)
              {\parbox[top][30mm][t]{\textwidth}{\color{black}#1}};
       \end{tikzpicture}
      };
   \end{tikzpicture}
   \gdef\chapterlabel{}
  } 

\titlespacing*{\chapter}{0pt}{50pt}{30pt}
\titlespacing*{\section}{0pt}{13.2pt}{*0}  
\titlespacing*{\subsection}{0pt}{13.2pt}{*0}
\titlespacing*{\subsubsection}{0pt}{13.2pt}{*0}

\newcounter{myparts}
\newcommand*\partlabel{}
\titleformat{\part}[display]  
	{\normalfont\bfseries\Huge} 
	{\gdef\partlabel{\thepart\ }}     
 	{0pt} 
 	  {\setlength{\unitlength}{20mm}
	  \addtocounter{myparts}{1}
	  \begin{tikzpicture}[remember picture,overlay]
    \node[anchor=north west,xshift=-65mm,yshift=-6.9cm-\value{myparts}*20mm] at (current page.north east) 
      {\begin{tikzpicture}[remember picture, overlay]
        \draw[fill=black] (0,0) rectangle(62mm,20mm);   
        \node[anchor=north west,yshift=-6.1cm-\value{myparts}*20mm,xshift=-60.5mm,minimum height=30mm,inner sep=0mm] at (current page.north east)
        {\parbox[top][30mm][t]{55mm}{\raggedright \color{white}Part \partlabel $\phantom{\textrm{l}}$}};  
        \node[anchor=north east,yshift=-6.1cm-\value{myparts}*20mm,xshift=-63.5mm,text width=\textwidth,minimum height=30mm,inner sep=0mm] at (current page.north east)
              {\parbox[top][30mm][t]{\textwidth}{\raggedleft \color{black}#1}};
       \end{tikzpicture}
      };
   \end{tikzpicture}
   \gdef\partlabel{}
  } 

\usepackage{amsmath}
\makeatletter
\def\resetMathstrut@{%
  \setbox\z@\hbox{%
    \mathchardef\@tempa\mathcode`\(\relax
      \def\@tempb##1"##2##3{\the\textfont"##3\char"}%
      \expandafter\@tempb\meaning\@tempa \relax
  }%
  \ht\Mathstrutbox@1.2\ht\z@ \dp\Mathstrutbox@1.2\dp\z@
}
\makeatother

\usepackage{cite}
\usepackage{array}
\usepackage{url}
\usepackage[utf8]{inputenc}  
\usepackage{mathtools}
\usepackage{amsfonts}
\usepackage{amssymb}
\usepackage{algpseudocode} 
\usepackage{marginnote}
\usepackage{rotating}
\usepackage{float}

\DeclareUnicodeCharacter{00A0}{~}

\def \({\left(}
\def \){\right)}
\def \[{\left[}
\def \]{\right]}
\newcommand{\tbf}[1]{{\textbf{#1}}}
\newcommand{\bsy}[1]{{\boldsymbol{#1}}}
\newcommand{\txt}[1]{\text{#1}}
\newcommand{\defeq}{\vcentcolon=}
\newcommand{\eqdef}{=\vcentcolon}
\newcommand{\cv}[1]{\underline{#1}}

\newcommand{\bq}{{\textbf {q}}}
\newcommand{\bm}{{\textbf {m}}}
\newcommand{\br}{{\textbf {r}}}
\newcommand{\bF}{{\textbf {F}}}
\newcommand{\bff}{{\textbf {f}}}
\newcommand{\bG}{{\textbf {G}}}

\newcommand{\bh}{{\textbf {h}}}
\newcommand{\bw}{{\textbf {w}}}
\newcommand{\bv}{{\textbf {v}}}
\newcommand{\bA}{{\textbf {A}}}
\newcommand{\bB}{{\textbf {B}}}
\newcommand{\bhx}{\hat {\textbf {x}}}
\newcommand{\bX}{{\textbf {X}}}
\newcommand{\bx}{{\textbf {x}}}

\newcommand{\by}{{\textbf {y}}}
\newcommand{\bz}{{\textbf {z}}}

\newcommand{\bs}{{\textbf {s}}}

\newcommand{\bR}{{\textbf {R}}}
\newcommand{\beps}{{\boldsymbol {\epsilon}}}
\newcommand{\bxigma}{{\boldsymbol{\Sigma}}}
\newcommand{\bxi}{{\boldsymbol{\xi}}}
\newcommand{\bzero}{{\textbf{0}}}
\newcommand{\ba}{{\textbf {a}}}

\newcommand{\bxii}{{\boldsymbol{\xi}}}
\newcommand{\bre}{{\textbf {e}}}
\newcommand{\bytilde}{{\tilde{\textbf {y}}}}

\newcommand{\be}{\begin{equation}}
\newcommand{\ee}{\end{equation}}
\newcommand{\bea}{\begin{eqnarray}}
\newcommand{\eea}{\end{eqnarray}}
\newcommand{\mE}{\mathbb{E}}
\begin{document}
\frontmatter
\begin{titlepage}
\begin{center}
\sffamily

\null\vspace{2cm}
{\LARGE STATISTICAL PHYSICS AND\\[3pt]APPROXIMATE MESSAGE PASSING ALGORITHMS\\[3pt]FOR SPARSE LINEAR ESTIMATION PROBLEMS\\[12pt]IN SIGNAL PROCESSING AND CODING THEORY}
    
\vfill

\begin{tabular} {cc}
\parbox{0.3\textwidth}{\includegraphics[width=2.65cm]{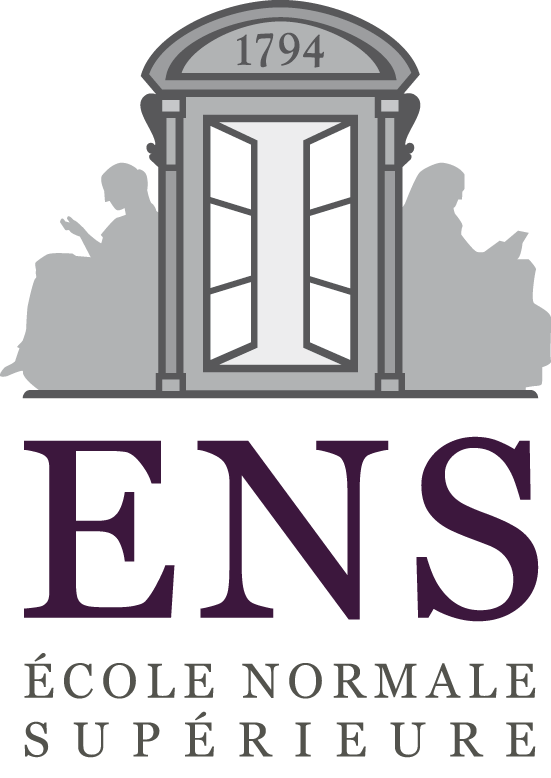}\\[6pt]
					   \includegraphics[width=2.65cm]{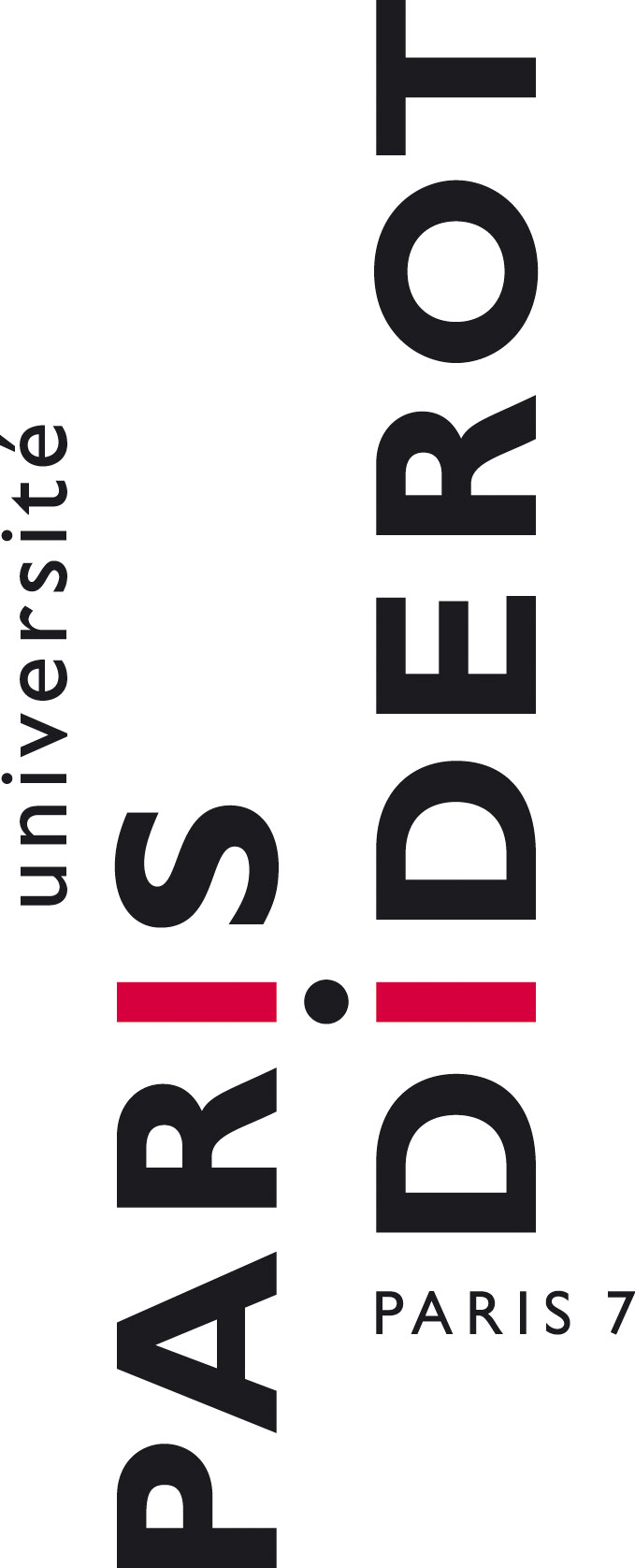}}
&
\parbox{0.7\textwidth}{%
	Thèse n. \\
	présentée le 18 septembre 2015\\
	à l'\'Ecole Normale Supérieure de Paris.\\
	Travail effectué au laboratoire de physique statistique\\
	de l'\'Ecole Normale Supérieure de Paris\\
	sous la direction de Prof. Florent KRZAKALA\\
	et au sein de l'école doctorale physique en île de France\\[6pt]	
	Université Paris Diderot (Paris 7) Sorbonne Paris cité\\
	pour l'obtention du grade de Docteur ès Sciences\\
	spécialité physique théorique, par\\[6pt]
	\null \hspace{3em} Jean BARBIER\\[6pt]
\small
acceptée sur la proposition du jury:\\[6pt]
    Prof. Laurent DAUDET, examinateur\\
    Prof. Silvio FRANZ, examinateur\\
    Prof. Florent KRZAKALA, directeur\\
    Prof. Marc LELARGE, examinateur\\
    Prof. Nicolas MACRIS, rapporteur\\
    Prof. Marc M\'EZARD, examinateur\\ 
    Prof. Federico RICCI-TERSENGHI, examinateur\\
    Prof. David SAAD, rapporteur\\[6pt]               
Paris, \'Ecole Normale Supérieure, 2015.}
\end{tabular}
\end{center}
\vspace{2cm}
\end{titlepage}
\cleardoublepage
\thispagestyle{empty}

\vspace*{3cm}

\begin{raggedleft}
    	La science c'est plutôt cool quand même...\\    		
\end{raggedleft}

\vspace{4cm}

\begin{center}
    A mes parents qui m'ont tout donné et mes soeurs que j'aime par dessus tout.\\
    Merci de m'avoir supporté jusqu'ici...
\end{center}
\cleardoublepage
\chapter*{Remerciements}
\markboth{Remerciements}{Remerciements}
\addcontentsline{toc}{chapter}{Remerciements}

Je tiens avant tout à remercier mes parents qui m’ont toujours laissé totalement libre de mes choix et m’ont tout donné, qui m’ont soutenu pendant cette longue période souvent difficile, parfois très difficile et toujours merveilleuse qu’ont été mes études. Merci à mes soeurs, Louise et Virginie. Merci à toute ma famille, présente dans les joies et difficultés.

Je remercie mon directeur Florent, le maitre Jedi, un ami. Merci de m’avoir fait confiance, de m’avoir enseigné par la pratique le vrai sens critique, de m’avoir présenté tant de personnes incroyables, de m’avoir offert ces expériences enrichissantes en école et ailleurs, d’avoir été insupportable quand il le fallait vraiment et d’avoir toujours su trouver l’équilibre entre pression et liberté, travail et humour, entre aide et indépendance. Je n’aurais réellement pas pu vivre une meilleure expérience pour mon doctorat.

Merci à Eric Tramel et Francesco Caltagirone pour leur sympathie et leurs explications.

Je tiens à remercier ceux qui ont rendu ma thèse encore plus agréable par leur amitié, qui ont transformé tous mes repas (et soirées) en moments toujours plus marrants, qui m’ont presentés leurs amis... Merci Thim, Thomas, le petit Christophe, Alaa, Alice, Antoine, Sophie, Ralph et merci à tous les autres aussi.

Je voudrais également remercier ceux qui m'ont permis d'en arriver là. En particulier Riccardo Zecchina qui m'a fait découvrir les domaines de l'inférence et de la belle physique statistique du désordre, Alessandro Pelizzola qui a tout fait pour rendre mon année à Turin si enrichissante et simple, Silvio Franz et Emmanuel Trizac pour m'avoir fait confiance en m'envoyant en Italie, merci à Marc Mézard pour son influence directe ou indirecte dans tous les travaux auxquels j'ai pu m'intéresser pendant ces trois années et demie, sur tous les papiers que j'ai pu lire et pour m'avoir montré ce que c'est que de vraiment savoir skier... Merci à Lenka Zdeborova pour ces collaborations fructueuses, Laurent Daudet pour m'avoir aidé à prendre des décisions importantes. 

Merci aussi à Rüdiger Urbanke et Nicolas Macris pour leur accueil à Lausanne. J'attends avec impatience les années à venir... Merci aux autres membres de mon jury de prendre le temps pour ma soutenance et avoir la patience de lire ma thèse: David Saad, Federico Ricci-Tersenghi et Marc Lelarge.

Je n’oublie pas les membres de mon premier lieu de travail, le laboratoire de physico-chimie théorique de l’École Supérieure de physique et chimie de Paris, en particulier Élie Raphaël et Thomas Salez pour leur sympathie perpétuelle, ainsi que Justine et Antoine.

Je dois beaucoup à mes colocataires qui m’auront supporté malgré mes crises de nerfs suite à trop de message-passing et avec qui j’aurais tellement rigolé: Charlotte, PH et Manon, vous êtes les meilleurs.

Merci Auré d'avoir rendu mes études encore plus intéressantes, du début à la fin. 

Je n'oublie pas Brian et tous les serveurs de Chez Léa sans qui cette période de rédaction aurait été très différente...

Je tiens également à remercier la DGA de m'avoir financé.

La liste pourrait encore continuer longtemps ayant rencontré tellement de gens intéressants et sympathiques durant ces années, merci à vous tous..

\bigskip
 
\noindent\textit{Paris, 20 Mai 2015}
\hfill J.~B.
%

\chapter*{Abstract}
\addcontentsline{toc}{chapter}{Abstract (English/Français)} 
This thesis is interested in the application of statistical physics methods and inference to signal processing and coding theory, more precisely, to sparse linear estimation problems.
\vskip0.2cm

The main tools are essentially the graphical models and the approximate message-passing algorithm together with the cavity method (referred as the state evolution analysis in the signal processing context) for its theoretical analysis. We will also use the replica method of statistical physics of disordered systems which allows to associate to the studied problems a cost function referred as the potential of free entropy in physics. It allows to predict the different phases of typical complexity of the problem as a function of external parameters such as the noise level or the number of measurements one has about the signal: the inference can be typically easy, hard or impossible. We will see that the hard phase corresponds to a regime of coexistence of the actual solution together with another unwanted solution of the message passing equations. In this phase, it represents a metastable state which is not the true equilibrium solution. This phenomenon can be linked to supercooled water blocked in the liquid state below its freezing critical temperature.
\vskip0.2cm

Thanks to this understanding of blocking phenomenon of the algorithm, we will use a method that allows to overcome the metastability mimicing the strategy adopted by nature itself for supercooled water: the nucleation and spatial coupling. In supercooled water, a weak localized perturbation is enough to create a crystal nucleus that will propagate in all the medium thanks to the physical couplings between closeby atoms. The same process will help the algorithm to find the signal, thanks to the introduction of a nucleus containing local information about the signal. It will then spread
as a "reconstruction wave" similar to the crystal in the water.
\vskip0.2cm

After an introduction to statistical inference and sparse linear estimation, we will introduce the necessary tools. Then we will move to applications of these notions. They will be divided into two parts.
\vskip0.2cm

The signal processing part will focus essentially on the compressed sensing problem
where we seek to infer a sparse signal from a small number of linear projections of it that can be noisy. We will study in details the influence of structured operators instead of purely random ones used originally in compressed sensing. These allow a substantial gain in computational complexity and necessary memory allocation, which are necessary conditions in order to work with very large signals. We will see that the combined use of such operators with spatial coupling allows the implementation of an highly optimized algorithm able to reach near to optimal performances. We will also study the algorithm behavior for reconstruction of approximately sparse signals, a fundamental question for the application of compressed sensing to real life problems. A direct application will be studied via the reconstruction of images measured by fluorescence microscopy. The reconstruction of "natural" images will be considered as well. 
\vskip0.2cm

In coding theory, we will look at the message-passing decoding performances for two distincts real noisy channel models. A first scheme where the signal to infer will be the noise itself will be presented. The second one, the sparse superposition codes for the additive white Gaussian noise channel is the first example of error correction scheme directly interpreted as a structured compressed sensing problem. Here we will apply all the tools developed in this thesis for finally obtaining a very promising decoder that allows to decode at very high transmission rates, very close of the fundamental channel limit.

\vskip0.5cm
\textbf{Keywords:} Statistical physics, disordered systems, mean field theory, signal processing, Bayesian inference, statistical learning, coding theory, linear estimation, sparsity, approximate sparsity, compressed sensing, spatial coupling, Gaussian channel, error correcting codes, sparse superposition codes, approximate message passing algorithm, cavity method, state evolution analysis, replica method.
%
%
%
\begin{otherlanguage}{french}
\cleardoublepage
\chapter*{Résumé}
Cette thèse s'intéresse à l'application de méthodes de physique statistique des systèmes désordonnés ainsi que de l'inférence à des problèmes issus du traitement du signal et de la théorie du codage, plus précisément, aux problèmes parcimonieux d'estimation linéaire.
\vskip0.2cm

Les outils utilisés sont essentiellement les modèles graphiques et l'algorithme approximé de passage de messages ainsi que la méthode de la cavité (appelée analyse de l'évolution d'état dans le contexte du traitement de signal) pour son analyse théorique. Nous aurons également recours à la méthode des répliques de la physique des systèmes désordonnées qui permet d'associer aux problèmes rencontrés une fonction de coût appelé potentiel ou entropie libre en physique. Celle-ci permettra de prédire les différentes phases de complexité typique du problème, en fonction de paramètres externes tels que le niveau de bruit ou le nombre de mesures liées au signal auquel l'on a accès: l'inférence pourra être ainsi typiquement simple, possible mais difficile et enfin impossible. Nous verrons que la phase difficile correspond à un régime où coexistent la solution recherchée ainsi qu'une autre solution des équations de passage de messages. Dans cette phase, celle-ci est un état métastable et ne représente donc pas l'équilibre thermodynamique. Ce phénomène peut-être rapproché de la surfusion de l'eau, bloquée dans l'état liquide à une température où elle devrait être solide pour être à l'équilibre. 
\vskip0.2cm

Via cette compréhension du phénomène de blocage de l'algorithme, nous utiliserons une méthode permettant de franchir l'état métastable en imitant la stratégie adoptée par la nature pour la surfusion: la nucléation et le couplage spatial. Dans de l'eau en état métastable liquide, il suffit d'une légère perturbation localisée pour que se créer un noyau de cristal qui va rapidement se propager dans tout le système de proche en proche grâce aux couplages physiques entre atomes. Le même procédé sera utilisé pour aider l'algorithme à retrouver le signal, et ce grâce à l'introduction d'un noyau contenant de l'information locale sur le signal. Celui-ci se propagera ensuite via une "onde de reconstruction" similaire à la propagation de proche en proche du cristal dans l'eau.
\vskip0.2cm

Après une introduction à l'inférence statistique et aux problèmes d'estimation linéaires, on introduira les outils nécessaires. Seront ensuite présentées des applications de ces notions. Celles-ci seront divisées en deux parties.
\vskip0.2cm

La partie traitement du signal se concentre essentiellement sur le problème de l'acquisition comprimée où l'on cherche à inférer un signal parcimonieux dont on connaît un nombre restreint de projections linéaires qui peuvent être bruitées. Est étudiée en profondeur l'influence de l'utilisation d'opérateurs structurés à la place des matrices aléatoires utilisées originellement en acquisition comprimée. Ceux-ci permettent un gain substantiel en temps de traitement et en allocation de mémoire, conditions nécessaires pour le traitement algorithmique de très grands signaux. Nous verrons que l'utilisation combinée de tels opérateurs avec la méthode du couplage spatial permet d'obtenir un algorithme de reconstruction extrêmement optimisé et s'approchant des performances optimales. Nous étudierons également le comportement de l'algorithme confronté à des signaux seulement approximativement parcimonieux, question fondamentale pour l'application concrète de l'acquisition comprimée sur des signaux physiques réels. Une application directe sera étudiée au travers de la reconstruction d'images mesurées par microscopie à fluorescence. La reconstruction d'images dites "naturelles" sera également étudiée. 
\vskip0.2cm

En théorie du codage, seront étudiées les performances du décodeur basé sur le passage de message pour deux modèles distincts de canaux continus. Nous étudierons un schéma où le signal inféré sera en fait le bruit que l'on pourra ainsi soustraire au signal reçu. Le second, les codes de superposition parcimonieuse pour le canal additif Gaussien est le premier exemple de schéma de codes correcteurs d'erreurs pouvant être directement interprété comme un problème d'acquisition comprimée structuré. Dans ce schéma, nous appliquerons une grande partie des techniques étudiée dans cette thèse pour finalement obtenir un décodeur ayant des résultats très prometteurs à des taux d'information transmise extrêmement proches de la limite théorique de transmission du canal.

\vskip0.5cm
\textbf{Mots clefs:} Physique statistique, systèmes  désordonnés, théorie du champ moyen, traitement du signal, inférence Bayesienne, apprentissage statistique, théorie du codage, estimation linéaire, parcimonie, parcimonie approximative, acquisition comprimée, couplage spatial, canal Gaussien, codes correcteurs d'erreurs, codes de superposition parcimonieuse, algorithme approximé de passage de messages, méthode de la cavité, analyse d'évolution des états, méthode des répliques.
\end{otherlanguage}

%
\tableofcontents
\addcontentsline{toc}{chapter}{List of figures}
\makeatletter
\renewcommand*\l@figure{\@dottedtocline{1}{1.5em}{3em}}
\let\l@table\l@figure
\makeatother
\listoffigures
\markboth{List of symbols and acronyms}{List of symbols and acronyms}
\addcontentsline{toc}{chapter}{List of symbols and acronyms}

\chapter*{List of symbols and acronyms}

\begin{flalign*}
a \ \ \ \ &: \ \ \ \ \txt{a generic quantity, usually a scalar if not precised further.}&\\
\tbf a \ \ \ \ &:\ \ \ \ \txt{a vector.} \\
\tbf A\ \ \ \ &:\ \ \ \ \txt{a matrix.} \\
a_{i} = (\tbf a)_i \ \ \ \ &:\ \ \ \ \txt{the $i^{th}$ component of $\tbf{a}$. The two notations are equivalent.}\\
A_{ij}\ \ \ \ &:\ \ \ \ \txt{the element at the $i^{th}$ line and $j^{th}$ column of}\ \tbf{A}. \\
\tbf A_{i,\bullet}\ \ \ \ &:\ \ \ \ \txt{the $i^{th}$ line of $\tbf A$.} \\
\tbf A_{\bullet,i}\ \ \ \ &:\ \ \ \ \txt{the $i^{th}$ column of $\tbf A$.} \\
\bs \ \ \ \ &:\ \ \ \ \txt{generally the signal or message to infer.} \\
\bx \ \ \ \ &:\ \ \ \ \txt{an intermediate variable for estimating $\bs$.} \\
\hat \bx\ \ \ \ &:\ \ \ \ \txt{the final estimate of the signal $\bs$.} \\
\by\ \ \ \ &:\ \ \ \ \txt{the measurement vector or codeword.} \\
\bF\ \ \ \ &:\ \ \ \ \txt{the measurement or coding operator.} \\
\hat a\ \ \ \ &:\ \ \ \ \txt{the estimate of the quantity $a$.} \\
K\ \ \ \ &:\ \ \ \ \txt{number of non zero components in the sparse signal} \ \bs.\\
N\ \ \ \ &:\ \ \ \ \txt{number of scalar components of the signal or message to reconstruct} \ \bs.\\
L\ \ \ \ &:\ \ \ \ \txt{number of vector components of the signal or message to reconstruct} \ \bs.\\
M\ \ \ \ &:\ \ \ \ \txt{number of scalar components of the measure or codeword} \ \by. \\
\defeq\ \ \ \ &:\ \ \ \ \txt{equal by definition.} \\
\rho \defeq K/N \ \ \ \ &:\ \ \ \ \txt{density of non zero components of} \ \bs. \\
\alpha \defeq M/N \ \ \ \ &:\ \ \ \ \txt{the measurement rate.}
\end{flalign*}
\begin{flalign*}
a | b \ \ \ \ &:\ \ \ \ \txt{$a$ "given" (or such that) $b$.} \\
\mathcal{N}(u|\mu,\sigma^2)\ \ \ \ &:\ \ \ \ \txt{a Gaussian distribution for the random variable $u$}\\ 
&\ \ \ \ \ \ \ \txt{with mean $\mu$ and variance $\sigma^2$.}&\\
\delta(x)\ \ \ \ &:\ \ \ \ \txt{the delta Dirac function (which is formally a distribution)} \\ 
&\ \ \ \ \ \ \ \txt{that is a probability density giving a infinite weight to the single value $x=0$.} \\
\delta_{ij}\ \ \ \ &:\ \ \ \ \txt{the kronecker symbol, which is one if $i=j$, $0$ else.}\\
\partial i \defeq \{j:(ij)\in E\} \ \ \ \ &:\ \ \ \ \txt{the set of neighbors nodes to node $i$ in the}\\ 
&\ \ \ \ \ \ \ \txt{graphical representation of the problem ($E$ is the set of edges).}\\
\partial i\backslash j \ \ \ \ &:\ \ \ \ \txt{the set of neighbors nodes to node $i$ except the node $j$}\\ 
&\ \ \ \ \ \ \ \txt{in the graphical representation of the problem.}\\
u \sim P (u | \bsy{\theta})\ \ \ \ &:\ \ \ \ \txt{$u$ is a random variable with distribution $P(u | \bsy{\theta})$}\\ 
&\ \ \ \ \ \ \ \txt{that depends on a vector of parameters} \ \bsy{\theta}. \\
\mE_x(y(x))=\mE_P(y(x))\ \ \ \ &:\ \ \ \ \txt{the average of the function $y(x)$ with respect to the random}\\
&\ \ \ \ \ \ \ \txt{variable $x\sim P(x)$. The two notations are used equivalently}\\
&\ \ \ \ \ \ \ \txt{when there are no possible ambiguities.}\\
<\bx> \defeq 1/N\sum_i^N x_i\ \ \ \ &:\ \ \ \ \txt{the empirical average of the vector $\bx$ (here there are $N$ components).}\\
\big\{x_i | f(x_i)\big\}_{g(i)}^b\ \ \ \ &:\ \ \ \ \txt{the ensemble made of $\{x_i\}$ that verify the conditions}\\
&\ \ \ \ \ \ \ \txt{$f(x_i) \ \forall \ i\in \{1,...,b | g(i) \ \txt{is true}\}$.}\\
&\ \ \ \ \ \ \ \txt{As for any operations such as sums, products, etc,}\\
&\ \ \ \ \ \ \ \txt{if the lower bound of the index is not explicited, it means it starts from $1$.}\\
&\ \ \ \ \ \ \ \txt{For example, $\{a_i|a_i>0\}_i^N=\{a_i|a_i>0\}_{i=1}^N$, $\sum_{i}^N x_i=\sum_{i=1}^N x_i$, etc.}\\
\big[x_i | f(x_i)\big]_{g(i)}^b\ \ \ \ &:\ \ \ \ \txt{the concatenation of $\big\{x_i | f(x_i)\big\}_{g(i)}^b$ to form a vector}\\
&\ \ \ \ \ \ \ \txt{of cardinality that depends on $g$ and $b$.}\\
[a, b]\ \ \ \ &:\ \ \ \ \txt{the simple concatenation of $a$ and $b$ to form a vector.}\\
\mathbb{I}(E)\ \ \ \ &:\ \ \ \ \txt{the indicator function, which is $1$ if the condition $E$ is true, $0$ else.}\\
\Delta\ \ \ \ &:\ \ \ \ \txt{the variance of the i.i.d Gaussian measurement noise.} \\
\bsy \xi \defeq \big[\xi_\mu \sim \mathcal{N}(\xi_\mu|0,\Delta)\big]_\mu^M\ \ \ \ &:\ \ \ \ \txt{the i.i.d Gaussian measurement noise vector.} \\
||\tbf y||_p \defeq 1/M \bigg(\sum_i^M |y_i|^p\bigg)^{1/p}\ \ \ \ &:\ \ \ \ \txt{the rescaled $\ell_p$ norm of $\tbf y$, which is here of size} \ M.\\
\rm{snr} \defeq ||\tbf y||^2_2 / \Delta \ \ \ \ &:\ \ \ \ \txt{the signal to noise ratio, where $y$ is the codeword.}& \\
z \in O(u)\ \ \ \ &:\ \ \ \ z \ \txt{is a quantity of the same order as} \ u \ \txt{i.e.}\ z = Cu \\ & \ \ \ \ \ \ \ \txt{where $C$ is a constant that does not scale with the problem size $N$.}
\end{flalign*}
\begin{flalign*}
z \in o(u)\ \ \ \ &:\ \ \ \ z \ \txt{is a quantity at least an order smaller than} \ u \ \txt{i.e.} \lim \frac{z}{u} = 0 \\
&\ \ \ \ \ \ \ \txt{in a proper limit that depends on the context.}\\
a \approx b \ \ \ \ &:\ \ \ \ \txt{$a$ and $b$ are equal up to a negligible difference $\in o(a)$.} \\
\bx_{\backslash i}\defeq \big[x_j\big]_{j\neq i}^N\ \ \ \ &:\ \ \ \ \txt{the vector made of the $N-1$ components}\\
&\ \ \ \ \ \ \  \txt{of $\bx$ that are not the $i^{th}$ one.}\\
\bx_{a \backslash i}\defeq \big[x_j\in \partial a\big]_{j\neq i}\ \ \ \ &:\ \ \ \ \txt{the vector of components of $\bx$ that are neighbors}\\
&\ \ \ \ \ \ \  \txt{of the factor $a$, except the $i^{th}$ one.}\\
d\bx \defeq \prod_i^N dx_i\ \ \ \ &:\ \ \ \ \txt{integration over all the components of the size $N$ vector $\bx$.}& \\
\mathcal{D}\bx \defeq \prod_i^N \mathcal{D}x_i = \prod_i^N dx_i \mathcal{N}(x_i|0,1)\ \ \ \ &:\ \ \ \ \txt{integration over all the components of the size $N$ vector $\bx$}\\
&\ \ \ \ \ \ \  \txt{with unit centered Gaussian measure.} \\
\bX^{\intercal} \ \ \ \ &:\ \ \ \ \txt{the transpose of a vector or matrix.}\\
\bx^{a}\defeq \big[x_i^a\big]_i^N \ \ \ \ &:\ \ \ \ \txt{the component wise power operation for a vector (or matrix).}\\
\tbf X\tbf Y\defeq \tbf Z \ \txt{where $Z_{ij}=X_{ij}Y_{ij}$} \ \ \ \ &:\ \ \ \ \txt{the component wise product between matrices}\\
&\ \ \ \ \ \ \ \txt{or vectors of same size.}\\
\bx^{\intercal} \by = \bx \by^{\intercal} \defeq \sum_i^N x_iy_i \ \ \ \ &:\ \ \ \ \txt{the scalar product between two vectors of size $N$.}\\
&\ \ \ \ \ \ \ \txt{The two notations are equivalent as we always assume the}\\
&\ \ \ \ \ \ \ \txt{vectors to have proper dimensions to apply the product.}\\
\bF \bx \defeq \big[\sum_i^N F_{\mu i} x_i\big]_\mu^M\ \ \ \ &:\ \ \ \ \txt{the matrix product between $\bF$ of size $M\times N$ and $\bx$ of size $N$.}\\
(\bF \bx)_\mu \defeq\sum_i^N F_{\mu i} x_i\ \ \ \ &:\ \ \ \ \txt{the $\mu^{th}$ component of the vector $\bF \bx$.}\\
\txt{Var}_P(u)\defeq \mathbb{E}_P(u^2)-\mathbb{E}_P(u)^2\ \ \ \ &:\ \ \ \ \txt{the variance of the random variable $u$ with distribution $P$.}\\
\txt{inv}\(\bA\)\ \ \ \ &:\ \ \ \ \txt{the inverse of the matrix $A$.}\\
\partial_x \ \ \ &:\ \ \ \ \txt{the partial derivative with respect to $x$.}\\
|\bx| \ \ \ &:\ \ \ \ \txt{the number of components of $\bx$ or the cardinality of an ensemble.}\\
a\propto b \ \ \ &:\ \ \ \ \txt{$a$ is proportional to $b$, i.e. they are equal up to a constant.}
\end{flalign*}
\newpage
\begin{flalign*}
\txt{CS}\ \ \ \ &:\ \ \ \ \txt{compressed sensing.}&\\
\txt{MSE}\ \ \ \ &:\ \ \ \ \txt{mean squarre error.}\\
\txt{MMSE}\ \ \ \ &:\ \ \ \ \txt{minimum mean squarre error.}\\
\txt{MAP}\ \ \ \ &:\ \ \ \ \txt{maximum à posteriori.}\\
\txt{i.e.}\ \ \ \ &:\ \ \ \ \txt{id est.}\\
N\txt{-d}\ \ \ \ &:\ \ \ \ N\txt{-dimensional.}\\
\txt{BP}\ \ \ \ &:\ \ \ \ \txt{belief propagation.}\\
\txt{AMP}\ \ \ \ &:\ \ \ \ \txt{approximate message passing.}\\
\txt{SE}\ \ \ \ &:\ \ \ \ \txt{state evolution analysis.}\\
\txt{i.i.d}\ \ \ \ &:\ \ \ \ \txt{independent and identically distributed.}\\
\txt{AWGN}\ \ \ \ &:\ \ \ \ \txt{additive white Gaussian noise.} \\
\txt{EM}\ \ \ \ &:\ \ \ \ \txt{expectation maximization.}
\end{flalign*}
\setlength{\parskip}{1em}
%
%
\mainmatter
\setcounter{page}{0}
\part[Main contributions and structure of the thesis]{Main contributions and\\structure of the thesis}
\chapter{Main contributions}
My work has been concentrated around two main axis: i) signal processing through compressed sensing and its application in image reconstructions and ii) coding theory over real channels and its links to compressed sensing. I will present here my main contributions in these fields, dividing my work into practical achievements through algorithms design and the theoretical and asymptotic studies. In addition, I've worked on a combinatorial optimization problem, namely the independent set problem, in order to get familiar with the cavity method and the diverse phase transitions that occur in such problems. I will start by briefly present this piece of work that I won't detail in this thesis. This choice has been made for sake of coherence of the thesis: all the problems I've worked on, except this one, belong to the class of sparse linear estimation problems and a common methodology is used, based on the approximate message-passing algorithm and the state evolution and replica analyzes for the asymptotic studies. 
\section{Combinatorial optimization}
\subsection{Study of the independent set problem, or hard core model on random regular graphs by the one step replica symmetry breaking cavity method}
In this work \cite{barbier2013hard}, we have studied the NP-hard independent set problem on random regular graphs, the dual of the vertex cover problem better known as the hard-core model in the physics literature. This model is of great interest as it can be seen as a lattice version of the hard spheres, a fundamental model in physics. The aim of this theoretical work was to reconciliate the two extreme regimes corresponding to the high and low connectivities of the graph. Both were known for a long time but each with a totally different behavior. While in the low connectivity regime, the problem displays a continuous full replica symmetry breaking transition as the density of particles increases in the graph, it was proven in the mathematical literature that in the high connectivity limit, the opposite phenomenon happens: the space of solution breaks discontinuously into exponentially many well separated components, a behavior typically found in glassy systems, at a density which is the half of the maximum one. 

The main result obtained through the cavity method is the obtention of the full phase diagram of the problem for all connectivities. The computation of the different phase transitions in the problem by population dynamics in the 1RSB framework shows that the change in behavior between a continuous full RSB regime and the appearance of the discontinuous 1RSB transition happens at connectivity $K=16$. It appears that between $16 \le K < 20$, despite the existence of a stable 1RSB glassy phase, the continous transition remains if the density of particles is too high until for $K \ge 20$, the 1RSB phase becomes stable for all densities until the maximum one. This shows that this model is the simplest mean field model of the glass and jamming transitions, and can be used to get insights on more complex models such as the hard spheres in high dimensions. In addition, the asymptotic analysis in the cavity framework is in perfect agreement with the rigorous results at high connectivity, which supports further the validity of the cavity method in such problems despite it is not yet rigorously established.
\section{Signal processing and compressed sensing}
\subsection{Generic expectation maximization approximate message-passing solver for compressed sensing}
\textbf{Practical achievements :} I've implemented a modular AMP solver for compressed sensing in MATLAB, that includes a lot of different possible priors for the signal model. In addition, most of the free parameters in these priors and the noise variance can be learned efficiently through expectation maximisation. All the algorithms can be found at \url{https://github.com/jeanbarbier/BPCS_common}.
\subsection{Approximate message-passing for approximate sparsity in compressed sensing}
\textbf{Practical achievements :} My first work \cite{BarbierKrzakalaAllerton2012} during this thesis was focused on the study of the AMP performances and behavior when dealing with signals that are only approximately sparse, sometimes referred as compressible. We implemented a specifically designed prior for approximate sparsity, and the expectation maximisation learning of all the parameters of this prior.

\textbf{Theoretical results :} We performed the static and dynamical asymptotic analyzes thanks to the replica and state evolution techniques respectively. We extracted how the AMP performances change as a function of the variance of the small components part of the signal, a kind of effective noise, and what are the best possible results from the Bayesian point of view. A first order phase transition blocking the AMP solver under some measurement ratio appears, but we have shown how the spatial coupling strategy can restore the optimality of the AMP solver and until which level of effective noise it makes sense to use this strategy. 
\subsection{Influence of structured operators with approximate message-passing for the compressed sensing of real and complex signals}
\textbf{Practical achievements :} A large amount of work has been put into the combination of this AMP solver with structured operators, based of fast Hadamard and Fourier transforms \cite{barbierSchulkeKrzakala}. Furthermore, I've developed a set of routines for applying the spatial coupling strategy in combination with these structured operators. The result is a very fast AMP solver yet optimal from the information theoretic point of view. It is able to deal with very large signals as the use of such operators allows to side step the memory issues that quickly arise working with the large matrices that one must store if not structured.

\textbf{Theoretical results :} Side to side with the developement of the AMP solver combined with full or spatially-coupled structured operators, we have studied how the use of such operators influence the performances of AMP in noiseless compressed sensing of real or complex signals. The point is that AMP together with the state evolution analysis has originally been derived for i.i.d matrices but we have numerically shown that despite the state evolution does not describe properly the AMP dynamic with structured operators, it remains an accurate predictive tool for its final performances as the reconstruction quality is the same than with i.i.d matrices. Furthermore, it appeared that structured operators improves the rate of convergence of AMP. In addition, the study of the spatial coupling strategy have shown that it performs very well with such operators as well and allows to make optimal inference as long as the signal density is not too large.
\subsection{"Total variation" like reconstruction of natural images by approximate message-passing}
\textbf{Practical achievements :} In order to reconstruct "natural images" (i.e. that are sparse in the discrete gradient space) in the compressive regime, I've worked on an AMP implementation mimicing the total-variation optimization algorithms. The result is an algorithm able to compete with the best optimization solvers in terms of reconstruction results, but with fewer parameters to tune as most of them can be learned efficiently.
\subsection{Compressive fluorescence microscopy images reconstruction with approximate message-passing}
\textbf{Practical achievements :} Still in the field of compressive imaging, I've developed an AMP implementation for reconstruction of images measured by compressive fluorescence microscopy, based on an approximate sparsity prior. The images here are highly sparse in the direct pixel space. The AMP overcomes the $\ell_1$ optimization solvers in terms of reconstruction quality, speed and minimum undersampling ratio to get good results. Furthermore, all the free parameters of the model can be learned efficiently.
\section{Coding theory for real channels}
\subsection{Error correction of real signals corrupted by an approximately sparse Gaussian noise}
\textbf{Practical achievements :} My first introduction to the field of coding theory is through a work \cite{barbierRobustErrorCorrection13} that naturally followed the study of approximate sparsity in compressed sensing. In this work, the aim is error correction over a channel that adds to a real transmitted message an approximately sparse Gaussian noise with some large components, the others being a smaller amplitude background noise. The aim is the reconstruction of the noise in order to cancel it at the end. We naturally used our previously developed AMP solver for approximately sparse signals to design an efficient decoder. In addition, the use of spatial coupling in this context allowed the decoder to perform at high rate, well above the results obtained with previously developed convex optimization based solvers.

\textbf{Theoretical results :} Based on the state evolution analysis for compressed sensing of approximately sparse signals, we predicted the asymptotic performances of our AMP based decoder, which shows that it is robust to the background noise, in the sense that the reconstruction error grows continuously with the variance of the small components of the noise.
\subsection{Study of the approximate message-passing decoder for sparse superposition codes over the additive white Gaussian noise channel}
\textbf{Practical achievements :} We presented the first decoder based on AMP for the sparse superposition codes, a capacity achieving error correction scheme over the AWGN channel. 
We exposed the first close connection between the compressed sensing theory and error correcting codes, as the sparse superposition codes decoding can directly be interpreted as a compressed sensing problem for signals with structured sparsity, or equivalently of signals with vectorial components instead of scalar ones, for which compressed sensing has originally been developed.

Our first paper \cite{barbier2014replica} on this topic studied the decoder when the coding matrix is i.i.d Gaussian and the power allocation of the transmitted message is constant. Despite the presence of a first order phase transition blocking the AMP decoder well before the capacity, the numerical results have shown that the perfomances are overcoming a previous decoder based on soft thresholding methods, the adaptative successive decoder. This decoder exhibits very poor results with respect to AMP with full coding matrices for any reasonnable codeword sizes, despite being asymptotically capacity achieving. 

In a second more in-depth study of our decoder \cite{DBLP:journals/corr/BarbierK15}, we included both non constant power allocation of the signal and the spatial coupling strategy to our scheme. Numerical studies suggested that despite improvements thanks to the power allocation, a well designed spatially-coupled coding matrix allows for better results both in terms of the rate of transmission and in robustness to noise. Numerical tests also show that the combined use of power allocation and spatial coupling lower the efficiency of the scheme with respect to power allocation or spatial coupling used alone.
In addition, we tested our structured Hadamard-based spatially-coupled operators that allow to perform Bayesian optimal decoding. The finite size effects in this setting have been quantified and show that this strategy is very efficient and allows to decode perfectly at very high rates, even for small sizes of the codeword. 

\textbf{Theoretical results :} Relying on this connection with compressed sensing, we derived the state evolution analysis of the decoder in the most general setting. It allows the prediction of the asymptotic dynamical behavior of the decoder for power allocated signals, encoded with or without spatially-coupled i.i.d operators. Based on our work on structured operators, we conjectured that the final perfomances of the decoder can be accurately predicted by this analysis, despite small descrepancies during the dynamic. Again, it appeared that structured operators allows for faster convergence. 

In addition, we performed the heuristic replica analysis of the coding scheme in order to compute the performances of the minimum mean square error estimator. This analysis is coherent with the previous rigorous results on the scheme as it shows that this Bayesian optimal estimator reaches asymptotically the capacity of the channel. The results suggest that the Bayes optimal estimator converge to the capacity with a rate following a power law as a function of the section size B, a fundamental parameter of the scheme. Both the derived state evolution recursions and replica potential are actually quite general, and can be applied for the prediction of the AMP behavior on any problem dealing with group sparsity, where the groups of variables are not overlaping each other.
\chapter[Structure of the thesis and important remarks]{Structure of the thesis\\and important remarks}
This thesis is decomposed into three main parts. I wrote the first one, "Fundamental concepts and tools" keeping constantly the following question in mind: 

{\it What would have been really useful for me to know in order to gain a lot of time starting my PhD three years ago ?} 

I have thus tried to make a (very subjective) introduction to what I consider as fundamental methods and ideas useful to a starting PhD student with a statistical physics background as me and who wants to work in the fascinating field of statistical inference and graphical models. I assume very few (if not at all) knowledge about the general theory of statistical inference but also that the reader have some notions in statistical physics of disordered systems and spin glasses, as I will make efforts to establish links with this fundamental field of research. My goal is not to explain the physics of disordered systems, as many great books already exist and can be found in the references, but more to see how it can be of great interest in the apparently unrelated field of sparse linear estimation, the main subject of this thesis.

I want to emphasize an important point, not only true for this first part but for all this manuscript:

{\it This work does not aim at mathematical rigor.}

It is worth to mention it as the kind of problems treated in this thesis are classicaly studied by people of the computer science and signal processing, information and coding theory or applied mathematics communities who are used to more, or even perfectly rigorous treatments. But I am (hopefully at this time...) a statistical physicist, and the tools developed in this field, at least part of them, are not yet proven rigorously. This remark leads to another important one:

{\it Despite the use of not (yet) rigorous methods, the theoretical results presented in this manuscript are conjectured to be exact.}

It must be understood that the rigor in the treatment is not a necessary condition for the results to be exact. The statistical physics methods used in this thesis, mainly the replica method used for asymptotic studies, are 
not rigorous but there exist an incredibly large amount of work and models where it has been proven to be exact, even sometimes rigorous, especially in the field of combinatorial optimization problems. Furthemore, even when not proven rigorously, numerical studies are always supporting the results of these methods. In opposite the cavity method, referred as the state evolution in the signal processing literature (and in this thesis as well) {\it is} rigorous for the prediction of the AMP behavior for sparse linear estimation problems (except for the vectorial components cases, but yet conjectured exact in this case).

The next two parts expose most of the original results and applications of my thesis, in addition to the asymptotic results of this first part. All the tools presented in the first part will be applied here.

The second part "Signal processing" is mostly related to compressed sensing. Chapter six presents a study of the influence of approximate sparsity in compressed sensing solved thanks to AMP. The seventh chapter studies how the use of structured operators such as Hadamard and Fourier ones change the AMP behavior in compressed sensing of real and complex signals. Furthermore, we will see that these can be combined with the spatial coupling strategy to perform optimal inference both from the theoretical and algorithmic point of views. Chapter eight is devoted to my work on compressive imaging, where the AMP algorithm is applied to the reconstruction of two different kind of images: i) the so called "natural" images, that have a compressible discrete gradient and ii) sparse images in the pixel basis, which have been obtained by fluorescence microscopy technic.

The last part "Coding theory" contains all my results on how the AMP algorithm can be used as a very efficient decoder for real noisy channels. In the chapter nine, the superposition codes for the additive white Gaussian noise channel are studied in depth. These represent the first direct link between error correction and compressed sensing. In this chapter, all the presented analytical tools are used to perform the asymptotic study of the decoder and the algorithmic tools as well: the spatial coupling and Hadamard-based structured operators are combined with AMP to get a capacity achieving decoder with very good finite size performances as shown by numerical studies. The last chapter introduce a different real channel model which adds gross Gaussian distributed errors to the signal in addition to a small Gaussian background noise. The algorithm developed in the chapter about approximate sparsity will be combined to spatial coupling to perform error correction at high rates.
\part{Fundamental concepts and tools}
\chapter[Statistical inference and linear estimation problems for the physicist layman]{Statistical inference and\\linear estimation problems\\for the physicist layman}
\vspace{1cm}
This first chapter which is voluntarily not technical at all is a general introduction to the main questions and techniques related to statistical inference and which are relevant to the present thesis. It can be skept by the reader familiar with statistical inference and compressed sensing as it does not include any original results. It is oriented towards a statistical physics student interested in working on inference related problems. Effort will be made in order to draw connections with physics, especially the statistical physics of disordered systems. It is thus assumed that the reader posseses a basic knowledge of this field.

First, I will define the general problem of statistical inference (or statistical estimation) and give the main distinctions among inference problems. I will also explain the difference between a direct and an inverse problem and show why in the context of statistical inference, only inverse problems really matter as opposed to statistical physics which has been created to deal with direct problems and later on extended to inverse ones. Canonical examples, yet very important both from the applicative and theoretical point of views will be presented. I will also discuss the notion of bias-variance tradeoff, which will help us to understand the fundamental limitations of statistical inference.

I will then focus on the model which is at the core of the present thesis, namely linear sparse estimation problems and compressed sensing, with a particular emphasis on the applications of compressed sensing in modern technologies.

I will introduce some basic notions of complexity theory, discussing the important tradeoff between statistical and computationnal efficiency of an algorithm. This question is essential in the modern context of "Big data" generating technologies where the sets of data produced become so large that solving the desired problem is not enough anymore: it must be done in a fast way as well.

I will then present two distinct methodologies to deal with sparse linear estimation. First I will very briefly present the convex optimization approach to compressed sensing which has been used and studied since the appearance of the field in 2006, and which remains an important tool for nowadays applications. I will try to give some insights behind the principle of $\ell_1$ norm optimization for inducing sparsity in the solution.

After a short introduction to some useful concepts of information theory, especially the notion of entropy and mutual information between random variables, I will move on to the main methodology underlying the techniques used in this thesis, namely the theory of Bayesian inference. The main principles will be exposed and then the modelisation of the sparse linear estimation problem thanks to these tools is discussed. I will underline the flexibility and the advantages of the method compared to an optimization approach. I will discuss the notion of estimator and give insights about why the minimum mean square error estimator is the appropriate choice in the continuous framework.

Finally I discuss the coding theory for the additive white Gaussian noise channel in the probabilistic Bayesian setting. The problem of communication through a noisy channel and the notion of capacity will be presented. Then we end up discussing the linear coding strategy and give a geometrical interpretation of the decoding problem.
\section[What is statistical inference ?]{What is statistical inference ?}
Before to enter the details of the problems studied in the present thesis, we present very briefly what are {\it statistical inference} problems, also referred as {\it inverse problems}, {\it estimation problems} or {\it learning} depending on the community. Great introductions can be found such as \cite{mackay2003information,james2013introduction,DBLP:journals/corr/TramelKGM14,wasserman2010statistics}.
\subsection{General formulation of a statistical inference problem}
Inference refers to the process of drawings conclusions about some system or phenomenon in a rational way from observations related to it, and a possible a priori knowledge about it. We are here interested in statistical inference, that will be mainly applied to signal processing problems. Assume you have access to some data $y$, that have been generated through some process $f$ related to some system properties of interest represented by the so-called {\it signal} $s$. The general relation linking these objects is given by:
\begin{align}
	y(\bsy{\theta}_{s},\bsy{\theta}_{f},\bsy{\theta}_{out}) &= P_{out}\(f\(s(\bsy{\theta}_{s}) | \bsy{\theta}_{f}\) | \bsy{\theta}_{out}\) \label{eqIntro:inferenceComplicated}	
\end{align}
$y$ is also referred as the observations, responses of the system or measurements in the present thesis, $s$ as the input, the predictors or signal in the present signal processing context. These two quantities can be a scalar, vectors, matrices, sets of labels, etc. $f$ models the {\it deterministic} part of the data generating process and can be any function, whereas $P_{out}$ is a {\it stochastic} function linking the processed signal to the actual observations, used to model some kind of {\it noise}. We will refer to it as the {\it channel}, this terminology coming from the communication theory, where the noise models how the non ideal transmitting channel alters what is travelling through it. In full generality, noise just means an incontrolable and undesired source of randomness that alters the observations of the system. In some cases, the noise can be correlated in some way to the signal $s$ or process $f$ such as in blind calibration \cite{DBLP:journals/corr/SchulkeCZ14}, but in all the remaining, we will always consider the noise to be additive and uncorrelated with them. $\bsy{\theta}_{s}$ are parameters of the signal such as some of its statistical properties like mean, variance, etc that can be known a priori or not. $\bsy{\theta}_{f}$ are those of the process $f$, for example the number of Fourier coefficients taken in a partial discrete Fourier transform and finally $\bsy{\theta}_{out}$ are parameters of the channel, usually the statistical properties of the noise.

The problem is to estimate the signal $s$ from the knowledge of the observations $y$, the process $f$ and the channel model $P_{out}$. The signal can be thought as a fixed realization of a random process, for example a message some emitter has sent you or some image.
The channel $\bsy{\theta}_{out}$, signal $\bsy{\theta}_{s}$ and processing function $\bsy{\theta}_{f}$ parameters can be unknown too and it is a part of the task to learn them as well. This can be done efficiently by statistical procedures such as {\it expectation maximization} based methods that will be detailed in sec.~\ref{sec:EMlearning}.

For sake of readibility, we drop these parameters dependencies and re-write the general statistical inference model as:
\begin{align}
	y = P_{out}\(f(s)\)
	\label{eqIntro:inference}	
\end{align}
and the dependency on all the parameters of the problem $\bsy \theta \defeq[\bsy{\theta}_{s}, \bsy{\theta}_{f}, \bsy{\theta}_{out}]$ is now always implicit. 

An important remark is that in the present thesis, we mostly consider the signal $s$ to be the unknown and $f$ to be known as this interpretation is more relevant in the problems studied here. But actually, it would be perfectly equivalent to change their respective roles and let the process $f$ becoming the unknown and $s$ to be some known inputs. Depending on the problem it can be more natural to think of $f$ as the unknow process that gave rise to the observations from controled inputs. It is really a matter of tastes. For example, in the classical reference \cite{james2013introduction}, $f$ is most of the time the object of interest that we want to infer, but in \cite{DBLP:journals/corr/TramelKGM14}, the authors adopted the same convention as in the present thesis of always considering the signal $s$ as the unknown. In examples where the unknown will be way more easily interpreted as $f$, it will be explicited, but in the rest $s$ is always the infered quantity. Let us now define more precisely the different kind of statistical inference problems and give some vocabulary.
\subsection{Inverse versus direct problems}
The statistical inference problem (\ref{eqIntro:inference}) is by nature an {\it inverse problem}, in the sense that it consists in estimating properties of the system {\it from} some noisy observations about it, as opposed to the {\it direct problem} which is to obtain these observations. Getting observations about a complex system is usually quite easy compared to the associated inverse problem.

Let us give some examples to emphasize this disymmetry in complexity between direct and inverse problems. It is easy to gather data about the past stock prices which are observations correlated to many parameters of the market and to the behavior of plenty of buyers and sellers with their own strategies, but it is highly difficult to infer from these the future prices, that must be in some way correlated to previous ones. It is nowadays quite easy to measure time series of the activity of many neurons in parallel, but the inverse problem consisting in infering the network of connections between the neurons from which result these activities is very hard \cite{sakellariou2013inverse}. In an epidemic spreading of some disease, we can partially know at some time $t$ who are contaminated or not and have some idea of the network of connections between people, from which we would like to infer back the source of the disease: the patient(s) zero \cite{guggiola2015minimal}. The same question can be asked for the identifications of the source of an internet virus, where it is even more easy to get the network of connections between computers. These are highly non trivial inverse dynamical problems. 

Statistical physics arised at the beginning of the $19^{th}$ to deal with direct problems. The aim was to link the microscopic properties of the system to its macroscopic ones, impossible to derive directly from the quantum mechanics, so the knowledge about the fundamental interactions between the atoms to the physical observables and order parameters like temperature, pressure, average magnetization, etc. But as we will see, the methodology of statistical physics and especially its tools to compute thermodynamical averages over some disorder is really useful in the signal processing and inverse problems context, where the atoms are replaced by the signal components, the interactions by the constraints extracted from the observations that must verify these variables and the order parameter or observable that we would like to predict is the typical error we will make in the inference of the signal. Fig.~\ref{figChIntro:statPhysVSinference} is a table summarizing the connections between quantitities and notions of statistical physics and those of inference and signal processing (defined in this chapter).
\begin{figure}[tb]
	\begin{center}
		\begin{tabular}{l|l}
 		 \hline
 		 \tbf{Statistical physics} & \tbf{Inference}\\
 		 \hline
  		Hamiltonian & Cost function \\
  		\hline
  		Particules, atoms, spins & Signal components \\
  		\hline
  		Microstates & All the possible measured signals \\
  		\hline  	
  		Macrostate & The final signal estimate $\bhx$, or estimator\\
  		\hline  	
  		Physical phases: liquid, solid, gas, glass, etc & Computational phases: Easy, hard,\\ &impossible inference \\
  		\hline
  		Boltzmann distribution & Posterior distribution \\
  		\hline
  		Partition function $Z(\by,\bF,\bsy\theta)$ & Absolute probability of the measure $P(\by | \bsy \theta, \bF)$ \\
  		\hline
  		External field & Prior distribution \\  		
  		\hline
  		External parameters: temperature, volume, & Noise variance $\Delta$, \\chemical potential, etc & signal to noise ratio ${\rm snr}$, \\ & measurement rate $\alpha$, signal densisty $\rho$, etc \\
  		\hline
  		Order parameter: average magnetization, & Mean square error $MSE$, \\
  		correlation functions, Edwards-Anderson & bit error rate, etc \\
  		order parameter for spin glasses, etc \\ 		
  		\hline
  		Quenched disorder: spin interactions, & Observations, sensing or coding matrix \\impurities in the medium, etc & and noise realizations \\
  		\hline
  		Free energy/entropy & Potential function \\
  		\hline
		\end{tabular}
	\end{center}
	\caption[Relations between the statistical physics quantities and vocabulary with the inference and signal processing one]{Relations between the statistical physics quantities and vocabulary with the inference and signal processing one, focused on the quantites useful in the present thesis, mainly related to compressed sensing and error correcting codes.}
	\label{figChIntro:statPhysVSinference}
\end{figure}
\subsection{Estimation versus prediction}
Inference can be important for two main reasons. In one hand, one could aim at accurately estimating the signal that gave rise to the observations. If the signal models some system, inference really is about {\it understanding} it. For example in seismology, the signal $s$ of interest could be the $3$-d density field of the floor in some area. Perturbations by located explosions could be performed and the vertical displacement $y$ of the floor in some places could be measured. The relation between the signal and the measures $f$, even non trivial is a priori obtainable (at least approximately) from the physics of waves propagations in complex media and the locations of the explosions. The noise here comes from the approximations in the modelisation of $f$ and the partial measurements.

In another hand, one aim could be to perform {\it predictions}. In this setting, it is easier to think as the signal $s$ to be known and it is the process $f$ which becomes the unknown object of interest. The goal is to get an estimate of it $\hat f$ which is able to accurately output responses to new, yet unobserved signals. This is for example the case for economical purposes. A trader is not really interested in understanding the complex relations defining the market, so to estimate accurately $f$ but more to be able to predict future prices $y$ from the knowledge of previous ones $s$ thanks to an estimator of the market process $\hat f$ that have a good predictive potential, despite it can have few common features with the true market behavior $f$, that can be way too complex to infer anyway from few partial observations.

All the problems that will be studied in depth in this thesis are estimation ones: we will always infer a signal that will be processed through some known transform $f(\bs)=\bF\bs$, a matrix product.
\subsection{Supervised versus unsupervised inference}
All the previously discused examples and the model (\ref{eqIntro:inference}) belong to the class of {\it supervised } learning: problems where both observations $y$ and the process $f$ are known. It is thus a {\it fitting} problem and we can interpret the observations $y$ associated with $f$ as a training data set, that allows to "teach" the inference algorithm to perform its task in a supervised way. Again, the $s$ and $f$ roles can be switched without loss of generality. An example of a supervised problem is {\it classification} where one seeks for an algorithm able to class data in groups. For example if one want to design an algorithm able to distinguish between pictures of boys and girls, the algorithm would be fed with a large amount of pictures, each with its corresponding label: boy or girl. Then, the algorithm $\hat f$ is expected to perfom the task well if fed with new unlabeled data, so to perform predictions.

In constrast, in an {\it unsupervised} inference task, one has only access to pure data $y$ without any associated training signal $s$ (as in the classification task) or process $f$ (as in the seismology example). So fitting loses its sense. The aim is more to find patterns in the data that can be interpreted a posteriori. The paradigmatic problem in this class is {\it clustering} where you have access to data you wish to class into a relatively small number of groups compared to the number of data points \cite{DBLP:journals/corr/KrzakalaMMNSZZ13,2014arXiv1406.1880S,james2013introduction}. For example, this is fundamental in recommander systems and collaborative filtering techniques that aim at clustering a set of buyers into groups, each representing a quite different consumer profile with different buying habits. Then, with some data about a new buyer (what he bought, when, the fequency, etc) he can be labeled with one of these typical profiles extracted from the clustering analysis, and thus the large amount of information gathered from the other consumers associated to this group can be used to predict products that will match this particular buyer need with high probability. Another classical problem falling in this class it the {\it community detection} where the data is a set of relations (a graph of connections) between unlabeled variables and one aims at labeling these points \cite{Fortunato201075,2014arXiv1409.2290D,moore2011nature,DBLP:journals/corr/KrzakalaMMNSZZ13}. This is performed everyday by Facebook which want to infer your potential friends (so labeled as Friend of Mr. You, in contrast with not friend of Mr. You) from the knowledge of the friendship connections among all their users. 

The assumption behind this kind of techniques is that despite each data points are different, in reality a small number of parameters (the labels) is enough to describe the entire data set accurately: this is called a {\it dimensionality reduction} assumption and stands at the roots of many modern techniques dealing with very large data sets. 
\subsection{Parametric versus non parametric inference}
\label{sec:paramVSnonParam}
Another important distinction is between {\it parametric} and {\it non parametric} problems. In the non parametric setting \cite{Wasserman:2006:NS:1202956,james2013introduction}, no or very few prior knowledge is assumed about the object to infer, so no prior model is assumed to perform the inference. For example in prediction, the estimate $\hat f$ of $f$ is chosen among {\it all} possible functions that are "smooth" enough, with only parameter being the level of smoothness $\lambda$. This is a very large space, denoted by $\mathcal{S}(\lambda)$. Non parametric fitting is generally performed by applying a regularizing kernel to the observations. The obtained processing function $\hat f$ is thus "just" a smoother version of the observations and it thus does not require any particular a priori structure or shape for $f$. 

On the opposite, parametric inference \cite{wasserman2010statistics,james2013introduction} makes strong assumptions about the structure of the infered object. For example, in the linear regression problem, ones assumes that $\by = f(\bX)=\tbf{f}^{\intercal}\tbf X$, where $\tbf f$ is a vector of coefficients linking the vector of observations to the matrix $\tbf X$, where each column is an input. The inference task is thus here to estimate these coefficients. Thus $f$ is now chosen among a way more restricted space which is here $\mathbb{R}^N$, instead of $\mathcal{S}(\lambda)$. One could even more constrain the functionnal space of the model, requiring for example that only a small fraction of the coefficients of $\tbf f$ are different from zero, i.e. that $\tbf f$ is {\it sparse}. In this example, it is clear that calling $f$ the process or the signal is irrelevant.

A natural question that arise from this discussion is: {\it Why would anyone constrain the functionnal space in which the processing function is chosen ?} that can be rephrased as: {\it Why would anyone prefere parametric to non parametric inference ?} Of course more degrees of freedom in the choice of $f$ or $s$ allows a priori for a better fit of the data but a more flexible model lower its {\it interpretability} i.e. it makes difficult the task of interpreting the relations between the data and observations. In contrast, a quite constrained sparse linear model is directly interpretable: the observations depend (approximately at least) only on a small subset of the inputs components, identified by the non-zero coefficients of the estimate of $\tbf f$. This can be very useful in a medical application. For example, we could gather observations $[y_i]_1^N \in\{1,0\}^N$ telling if yes or not the patient $i$ has some given cancer. The lines of the input matrix $\bX$ could correspond to conditions possibly correlated to this particular cancer such as health features or habits: is the person smoking, obese, doing sport regularly, a man, an O blood type, etc : $x_{ki}\in \{1,0\}$ tells if yes or not the $i^{th}$ patient verifies the condition $k$. Then, if the inference outputs a sparse vector of coefficients $\hat{\tbf f}$ such that $\by = \hat{\tbf f}^{\intercal} \bX$, one gets deep insights about which features participate or not to this cancer (the features corresponding to the non zero coefficients of $\hat{\tbf f}$), and with which weight (the amplitudes of the non zero coefficients of $\hat{\tbf f}$): here model interpretability is absolutely essential to identify the true causes of the cancer. But going back to the trader example that want to make accurate predictions about future stock prices, he does not not really care about the interpretability, only good predictions matter. In this case, non parametric inference can be more appropriate.

Another disadvantage of non-parametric inference is its high potential of {\it overfitting} the noise. It means that it is a difficult task to estimate the proper smoothing parameter $\lambda$ of the observations: a too smooth model will have a very poor predictive potential whereas a too rough model will fit the noise in the data instead of the data itself which again will generate a bad model. This is probably one of the most fundamental problem in any statistical learning problem, reffered as the {\it bias-variance tradeoff} \cite{james2013introduction}.
\subsection{The bias-variance tradeoff: what a good estimator is ?}
\label{sec:biasVarTradeoff}
A fundamental question in any statistical estimation problem is: {\it Can we quantify the error we will make using a given estimator for the quantity we wish to infer ?} This is in general very difficult to answer in a practical setting, actually most of the theoretical part of this thesis will be dedicated to this specific question. In spite of that, it appears that in such problems, we can always differentiate three distincts sources of error, namely the {\it bias}, the {\it variance} and the {\it irreducible error}. The problem is of course to quantify them. Let us demonstrate what they are and how they can be interpreted, so that we can assert what are the best results we can hope to obtain. We first restrict the fully general model (\ref{eqIntro:inference}) to additive noise channels which is of interest in the present thesis and quite a general model in the continuous framework: $P_{out}(\tilde \by) = \tilde \by + \bsy \xi$. We assume first that we want to infer the processing function. The appropriate object to quantify this estimation error is called the {\it prediction risk} :
\begin{align}
	R_p&\defeq\mathbb{E}_{\by}\(||\by-\hat{\bsy{y}}||_2^2\)\label{eq:predictionRisk}\\
	&=<\mathbb{E}_{y_\mu}\((y_\mu-\hat y_\mu)^2\)>
\end{align}
where we used that all the measurements are independent. It is the average mean square error between the observation $\by(\bs,\bsy\xi|f)$ given by (\ref{eqIntro:inference}) and the prediction $\hat{\by} \defeq \hat f(\bs|\by)$. The average is performed over the problem realization $\by$, i.e. over the noise $\bsy\xi$ and the input data $\bs$. We could consider also the case where $\bs$ is fixed, it would not change the analysis, just at the end the averages with respect to $\bs$ in the various sources of error that we will identify would disappear. $f$ is of course independent of $\bs$ and $\bsy\xi$. Let us derive the equations for only one component, the final result being the average over all the components $\mu \in \{1,\ldots,M\}$. We denote $f\defeq \(f(\bs)\)_\mu$ and $\hat f\defeq\(\hat f(\bs|\by)\)_\mu$ and skeep the $\mu$ index for sake of readibility:
\begin{align}	
	\mathbb{E}_{y}\((y - \hat y)^2\)&= \mathbb{E}_{\xi,\bs}\((f + \xi -\hat f)^2 \)\\
	&=\mathbb{E}_{\xi,\bs}\(\xi^2 + f^2 + \hat{f}^2 - 2f\hat f + 2\xi(f - \hat f)\)\\
	&=\Delta + \mathbb{E}_{\bs}\(f^2-2f\mathbb{E}_\xi(\hat f) + \mathbb{E}_{\xi}(\hat f^2) \)\\
	&=\Delta+ \mathbb{E}_{\bs}\(\mathbb{E}_\xi(\hat f^2) -\mathbb{E}_{\xi}(\hat f)^2 + f^2-2f\mathbb{E}_\xi(\hat f) +\mathbb{E}_{\xi}(\hat f)^2 \)\\
	&= \Delta + \txt{Var}(\hat f) + \mathbb{E}_{\bs}\((f - \mathbb{E}_{\xi}(\hat f))^2\)\\
	&\eqdef R_{p,\mu}\\
	\Rightarrow R_p &= \frac{1}{M} \sum_\mu^M R_{p,\mu}
	\label{eqChIntro:biasVarTrade}
\end{align}
where we have used the fact that the noise has zero mean and variance $\Delta$ and that the input $\bs$ is independent of the noise $\bsy\xi$ so $\mathbb{E}_{\xi}(f)=f$. Three quantities with transparent interpretation appeared:
\begin{itemize}
\item The {\it variance} $<\txt{Var}(\hat f)> = <\mathbb{E}_{\bs}\(\mathbb{E}_\xi(\hat f^2) -\mathbb{E}_{\xi}(\hat f)^2\)>$ which quantifies the average (over the signal realization) sensitivity of the estimator to fluctuations in the observations due to the noise. An high variance estimator would change a lot as a function of small perturbations in the observations and is thus not robust.
\item The squared {\it bias} $<\mathbb{E}_{\bs}\((f - \mathbb{E}_{\xi}(\hat f))^2\)>$ which represents the systematic error induced by the estimator if its average with respect to the noise differs from the true processing function $f$, thus a constant shift between the estimator and the observations.
\item The {\it irreducible error} $\Delta$ which is the error induced by the precense of the noise (the noise is i.i.d thus $\Delta_\mu=\Delta$). It is called irreducible as it is purely random and inherently present in the observations, and thus cannot be canceled in any manner and should not be fitted.
\end{itemize}
\begin{figure}[t!]
	\centering
	\includegraphics[width=.6\textwidth]{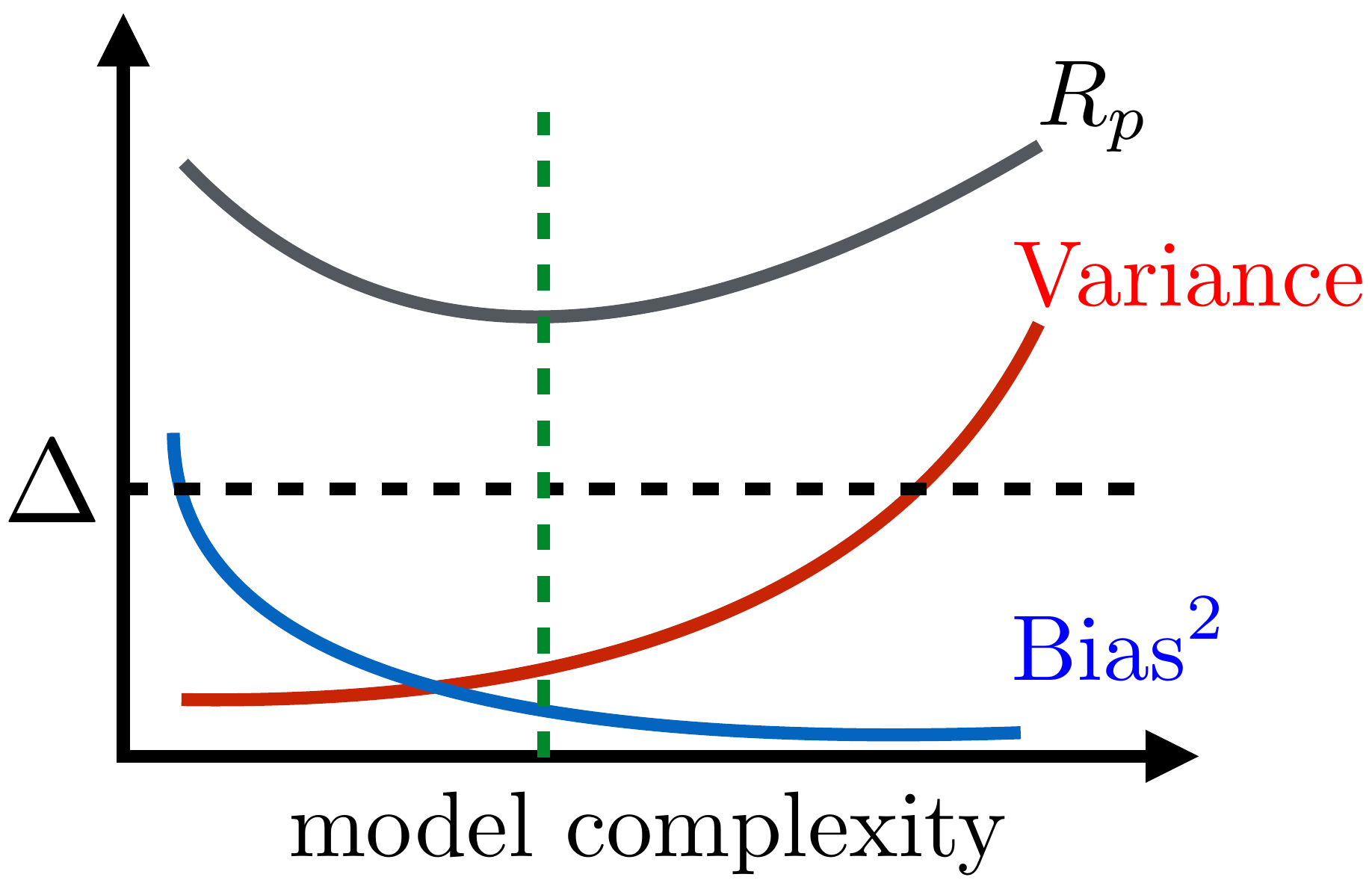}
	\caption[The bias-variance tradeoff]{Graphical representation of the bias-variance tradeoff. The horizontal axis is the model complexity, the black dashed curve is the irreducible error $\Delta$ and the grey curve, the prediction risk which is the sum of $\Delta$, the red variance curve and the blue squared bias curve. The sum of the squared bias and variance terms is the reducible error, that can be asymptotically canceled if we have access to a very large number of data points or if one has directly access to the data generating model. The optimal estimator is given by the model with complexity corresponding to the minimum of $R_p$, represented by the green dashed line.}
	\label{figChIntro:biasVarTrad}
\end{figure}

The sum of the variance and squared bias terms is the {\it reducible error} as it is the term that can be lowered by adjusting the estimator. This quantity is a function of the {\it model complexity}, i.e. of the number of degrees of freedom of $\hat f$ or its "roughness" in the non parametric case. As the model complexity increases, the bias decreases monotically as the observations are fitted more accurately but the pay-off is that the variance monotically increases until a point where it actually overcomes the gain in error due to the bias decrease. The optimal complexity of the estimator is the one that enables the estimator to fit enough the observations such that the bias is low but not too much such that the variance is not too high. This is summarized by the Fig.~\ref{figChIntro:biasVarTrad}. The green line which represents the optimal complexity (i.e. which minimizes the prediction risk) separates an underfitting regime on its left from the overfitting regime. The associated optimal estimator is denoted as $\hat f_{opt}$. The prediction risk cannot fall under the irreducible error due to the noise, the black dashed line on the plot. The reducible error is the gap between $\Delta$ and the prediction risk at the optimal complexity on the plot Fig.~\ref{figChIntro:biasVarTrad}.

What about the case of interest in the present thesis, the inference of the signal? The error estimate of interest in this case is the {\it risk} associated to the mean square error. Here one must be very careful: depending on the authors and especially on the adopted point of view (frequentist or Bayesian statistics), the risk can be defined in different ways. Here we place ourselves in the Bayesian framework and assume that we have access to data $\by$ but not to the true signal $\bs$. We represent the signal by an auxilliary variable $\bx$ to which we associate a posterior distribution $P(\bx|\by)$, see sec.~\ref{sec:BayesianInferenceForCS} for details. In this framework, the definition of the risk $R(\bhx|\by)$ of an estimator $\bhx$ is the average of its mean square error $MSE$ with respect to $\bx$ weighted by its posterior:
\begin{align}
	MSE(\bhx, \bs) &\defeq ||\bhx -\bs||_2^2 = \frac{1}{N}\sum_i^N(\hat x_i-s_i)^2= <(\bhx-\bs)^2>\label{eqIntro:MSE}\\
	R(\bhx|\by)&\defeq \mathbb{E}_{\bx|\by}\{MSE(\bhx, \bx)\} = \frac{1}{N}\sum_i^N\int dx_i P(x_i|\by) (\hat x_i-x_i)^2\label{eqIntro:RISK}
\end{align}
The true $MSE$ (\ref{eqIntro:MSE}) cannot be computed as $\bs$ in unknown but $R(\bhx|\by)$ can be if we are able to estimate $P(\bx|\by)$. $P(x_i|\by)$ is the marginal posterior of $x_i$. This risk, which is the one we refer to in this thesis, is linked to the so-called {\it Bayesian risk} $R_B(\bhx)$ as:
\begin{align}
	R_B(\bhx) = \int d\by P(\by) R(\bhx|\by) \label{eq:BayesRisk}
\end{align}
where we average over the data for which it is supposed that we have a prior distribution $P(\by)$. But in this thesis we always consider the data to be fixed. See \cite{wasserman2010statistics,DBLP:journals/corr/TramelKGM14} for frequentist definitions of the risk. 

In the special case of a linear orthogonal $f$ and defining the inverse of $f$ as $g\defeq \txt{inv}(f)$ one can write: 
\begin{equation}
	s_i = g(y_i-\xi_i) = g(y_i) + \bar\xi_i 
\end{equation}
which as the same form as (\ref{eqIntro:inference}) and where $\bar \xi_i$ is a new effective noise. In properly rescaled problems, $\bar\xi_i$ has also a variance $\in O(\Delta)$. Thus all the previous discussion and demontration remains valid (considering only the average over the noise) and we obtain that the prediction risk and the risk are equal up to a multiplicative factor $R_p=cR$ \cite{DBLP:journals/corr/TramelKGM14}, and thus the three sources of error remain the same with identical interpretations. This is a priori not justified when $f$ is not invertible as in the present thesis, where highly underdetermined linear systems will be studied but nevertheless, the three sources of errors actually remain the same. See \cite{james2013introduction} for a very nice introduction about statistical learning and the different sources of error in inference. \cite{2015arXiv150503941M} recently studied the influence of the reducible error in the prior mismatch case in the Bayesian framework, see sec.~\ref{sec:BayesianInference}.
\subsection{Another source of error: the finite size effects}
\label{sec:finiteSizeErrors}
There are two ways to cancel the reducible error: to have access to an infinite number of observations generated from the same system or having directly access to the data generating model. For example if one wants to infer the signal $\bs$, $P_{out}$, $f$, and all the parameters of the problem $\bsy \theta$ including those of the signal $\bsy \theta_s$ in (\ref{eqIntro:inference}) must be known. But as the data is finite $N<\infty$, even when the model is perfectly known and thus the reducible error is inexistant, there can remain another source of error related to the algorithm that performs inference: the finite size effects that induce a {\it finite size error} $\epsilon(N)$, where $\lim_{N\to\infty}\epsilon(N)= 0$. In the present thesis, the inference algorithm that we will use is the approximate message-passing algorithm derived and discussed in sec.~\ref{sec:AMP}. It is based on "law of large numbers-like" arguments and is only rigorous in the limit $N\to\infty$. Thus when using it on finite size systems, the algorithm becomes an approximation and this can induce finite size errors.

The optimal estimator (i.e. which has no reducible error) thus verifies:
\begin{equation}
	\hat x_{i,opt} = s_i + \epsilon(N) + \bar \xi_i
	\label{eqChIntro:xopt}
\end{equation}
where $\bar\xi_i $ is an effective error with variance $\in O(\Delta)$ as well when the system is properly scaled. This discussion shows that to get a good estimator, one must reduce as much as possible the reducible error and also study carefully the finite size effects associated to the inference algorithm used to get the estimate.
\subsection{Some important parametric supervised inference problems}
All the problems treated in the present thesis belong to the class of parametric supervised problems of the form (\ref{eqIntro:inference}). A non exhaustive list of such problems could include:

\tbf{Denoising} : $f(\bs) = \bs$, $P_{out}(\ | \bsy \theta_{out})$\\ 	 
This is the simplest (in terms of definition) parametric statistical inference problem where one aims at reconstructing a corrupted signal by a noisy channel $P_{out}(\ | \bsy \theta_{out})$, such as an AWGN channel of particular interest in this thesis: $P_{out}(\tilde{\tbf y}| \Delta) = \tilde{\tbf y} + \bsy \xi$ where $\Delta$ is the variance of the AWGN $\bsy \xi$ with i.i.d components of zero mean. 

\tbf{AWGN corrupted linear estimation problems} : $f(\bs|\bF) = \bF\bs$, $P_{out}(\tilde \by | \Delta) = \tilde \by + \bsy \xi$ \\	
This model, amongst which belongs AWGN compressed sensing and linear
error correcting codes over the AWGN channel, is at the core of this thesis. This is also a parametric problem as there are few (possibly unknown) free parameters, here the noise variance and some others, parametrizing the prior knowledge about $\bs$, see the section on Bayesian inference sec.~\ref{sec:BayesianInference} for more details about the notion of prior. Denoising over an AWGN channel can be seen as a particular instance of this problem where $\bF$ is the identity matrix.

\tbf{Binary symmetric or erasure channel models} : $f(\bs|\tbf H) = \txt{mod}(\tbf H\bs,2)$, $P_{out}(\tilde \by| \epsilon) = \tbf z$\\
where $\txt{mod}(\ ,2)$ is the modulo $2$ component-wise operation and $z_i = \tilde y_i$ with probability $(1-\epsilon)$, $z_i =\star\ \txt{or} \ -\tilde y_i$ with probability $\epsilon$ for the binary erasure channel or the binary symmetric channel respectively. Here $\star$ means lack of information. These are classical models in communication theory.

\tbf{Matrix completion} : $f(\tbf S) = \tbf S$, $P_{out}(\tilde{\tbf Y}| \epsilon) = \tbf Z$ \\	
where $Z_{ij} = \tilde Y_{ij}$ with probability $(1-\epsilon)$, $Z_{ij} =\star$ with probability $\epsilon$, usually close to one. This problem is fundamental in the field of recommander systems and collaborative filtering, where prior knowledge about the matrix $\tbf S$ can be that it is low rank and one seeks for the missing entries. Despite its usefulness in common fields, this problem is different from the unsupervised tasks of clustering or community detection.

\tbf{Classification} : $f(\tbf z|\tbf \bs, \mathcal{C}) = \[c_{z_i}| c_{z_i} \in \mathcal{C}\]_i^N$, $P_{out}(\tilde{\tbf y}) = \tilde{\tbf y}$ \\	
The classification problem is a canonical problem of parametric supervised learning. The classifier $f(z_i|\tbf \bs, \mathcal{C})$ ouputs the class $c_{z_i}\in\mathcal{C}$ to which belongs the item $z_i$. Here the vector to infer $\bs$ is actually a set of parameters allowing the classifier to perform its task properly. To do so, one must have a set on inputs objects $\tbf z^{train}$ and their associated classes $\{c_{z_i^{train}}\}_i^N$ to teach the classifier how to distinguish between the classes, i.e. learn $\bs$ which basically draws plans between the different classes of $\mathcal{C}$ in the items space. This problem stands at the roots of modern image recognition and deep neural networks theory.

\tbf{The inverse Ising problem} : $f(\bs|\bh,\tbf J) = \{m_i,\{c_{ij}\}_{j\neq i}\}_i^N$, $P_{out}(\tilde{\tbf y}|\epsilon,\Delta) = \tilde{\tbf z}$ \\
where $\tilde z_i=\tilde y_i + \xi$ with probability $1-\epsilon$, $\tilde z_i=\star$ with probability $\epsilon$ and $\xi\sim\mathcal{N}(\xi|0,\Delta)$. Here, one has access to partial noisy observations of the means $\{m_i\}_i^N$ and two point correlations $\{\{c_{ij}\}_{j\neq i}\}_i^N$ of the variables $\{s_i\}_i^N$ (for example measured experimentally). The aim is reconstructing the pairwise network of interactions $\{\{J_{ij}\}_{j\neq i}\}_i^N$ between these and the external fields $\{h_i\}_i^N$, such that the averages and correlations of the variables computed with respect to the Ising measure $P(\bs|\bh,\tbf J)\propto\exp\(-\sum_{i,j\neq i}^{N,N} J_{ij}s_is_j-\sum_{i}^{N} h_is_i\)$ matche the observed ones. This is a problem of great interest especially in phylogenetics and neurosciences. It is useful in any network reconstruction problems where one does not have access to higher than second order statistics about the variables that form the networks which is generally the case due to the restricted size of the samples in biology. This Ising form of the measure is derived by maximum entropy criterion as we shown in sec.~\ref{sec:maxEntropyCrit}.

Let us now focus on the specific problem of interest in this thesis, namely sparse linear estimation with i.i.d AWGN corruption, but it must be understood that the general methodology discussed hereafter (particularly Bayesian inference and message-passing algorithms) could be applied in most of these problems as they are special instances of the general model (\ref{eqIntro:inference}). For example, some references sharing the same methodology as the one developed in this work could include \cite{DBLP:journals/corr/TramelDK15} for classification, \cite{2011PhRvL.107f5701D} for clustering (interesting links between message-passing and spectral methods can be found \cite{DBLP:journals/corr/KrzakalaMMNSZZ13,2014arXiv1406.1880S}) or \cite{richardson2008modern,mezard2009information} for all the details about inference in communications over binary channels. See \cite{sakellariou2013inverse,2010PhDT474S} and the references therein for applications of the inverse Ising problem, including in biology and \cite{2012JSMTE..08..015R} for a review of efficient algorithms to deal with this problem.
\section[Linear estimation problems and compressed sensing]{Linear estimation problems and compressed sensing}
In this thesis, the studied problems belong to the class of noisy linear estimation problems under AWGN corruption which general form can be written as:
\begin{align}
  \by &= \bF\bs +\bsy \xi \\
  \Leftrightarrow y_\mu &= \sum_i^N F_{\mu i} s_i + \xi_\mu = (\bF\bs)_\mu + \xi_\mu \ \forall \mu \in \{1,\ldots,M\}
  \label{eqIntro:AWGNCS}
\end{align}
where $\xi_\mu \sim \mathcal{N}(\xi_\mu|0,\Delta) \ \forall \mu \in\{1,\ldots,M\}$ and we place ourselves in the general continuous framework $\bF \in \mathbb{R}^{MN}$, $\bs \in \mathbb{R}^{N}$. Let us start by underlying some fundamental differences between interpreting (\ref{eqIntro:AWGNCS}) as a linear {\it fitting} model, i.e. we fit some data as a linear combination of some basis $\bF$ vectors, the approximate coefficients being the signal we want to infer or as the true {\it generating} model of the data, i.e. we know that the data has been generated through (\ref{eqIntro:AWGNCS}) and we look for the unknown signal $\bs$. Despite some common features in terms of algorithmic techniques that can be used for estimation in both cases, there is a deep difference in the behavior of the problem as in the latter case, there exist a particular state, the true solution $\bs$, which can have a larger statistical weight than any other approximate solutions. In statistical physics, the construction of a problem from a given solution such as in inference defines the so called {\it planted ensemble}, see \cite{mezard2009information,KrzakalaZdeborova09} for the statistical physics of this ensemble.
\subsection{Approximate fitting versus inference}
For this discussion, we place ourselves in the noiseless setting $\Delta=0$. Algebra tells us that the system (\ref{eqIntro:AWGNCS}) can be solved exactly if the number of measurements is {\it at least equal} to the number of variables $M\ge N$ and are linearly independent one of the other. This can be done by simple matrix inversion, selecting a subset of $N$ lines of the original matrix and the associated measures and to inverse the resulting system. But this is true only if the observations {\it were really generated} through a linear model such as (\ref{eqIntro:AWGNCS}), which implies that there indeed exists a solution to the system. But one could just have access to data without really knowing its generating process and want to approximately fit these data as a linear combination of some basis functions or atoms (the columns of $\bF$), that form the so-called dictionnary in this context. 

Let us consider (\ref{eqIntro:AWGNCS}) as an approximation model. If the number of observations is $M<N$, the system is said {\it underdetermined}: there are too many possible solutions, usually an infinite number. If in contrast, $M>N$ the system is {\it overdetermined} and there could be no solution verifying all the observations exactly. To deal with these situations, one can use the {\it least square estimate} $\ \bhx_{LS}$. It consists in finding the linear combination of the basis functions that minimize the empirical prediction risk (\ref{eq:predictionRisk}):
\begin{align}
	\bhx_{LS} &= \underset{\bx}{\txt{argmin}}\ ||\by - \bF\bx||_2^2 \\
	&= \txt{inv}\(\bF^{\intercal}\bF\)\bF^{\intercal} \by \\
	&\defeq \bF^{*}\by
\end{align}
$\bF^*\defeq\txt{inv}\(\bF^{\intercal}\bF\)\bF^{\intercal}$ is the so-called pseuso-inverse of $\bF$. This solution has some caveats such as the fact that it won't produce any sparsity in the solution which can make the interpretation of the solution quite complicated as previously discussed in sec.~\ref{sec:paramVSnonParam}. More advanced methods such as the sparsity inducing LASSO which will be presented in sec.~\ref{sec:convexOptimization}, solved by linear programming techniques, can be used to improve the interpretability by sparsifying the solution. If an approximately sparse solution is found, it means that the data can be thought essentially as linearly depending on a small subset of features, the basis vectors associated to the estimated non zero components of the fit $\bhx$, selected among an original larger set of possibilities: this is called model selection. An ensemble of advanced techniques exist to deal with fitting linear models, details can be found in \cite{wasserman2010statistics,james2013introduction}.

But what happens if we know that the data has actually been generated by the linear model (\ref{eqIntro:AWGNCS})? The notion of overdetermined system loses its senses. For any $M\ge N$, finding the solution is trivial because it {\it exists}. Inference for sparse linear models thus deals with situations where $M<N$, or even $M\ll N$ but we know that the linear system have a solution for sure as the data was produced in this way, which makes all the difference. One could think of the least square estimate as a strategy, but in an inference problem, the aim is not to minimize the prediction risk but the mean square error (\ref{eqIntro:MSE}) and a minimum prediction risk solution can (and {\it will} in most of the cases) have a very high $MSE$.

So how to do? Still, albebra requires as many constraints as variables to infer over. Hopefully, additional input information about the solution can counterbalance the missing constraints: sparsity is our savior.
\subsection{Sparsity and compressed sensing}
A new paradigm in signal processing is the notion of {\it sparsity} and {\it compressibility} : a signal is said to be $K$-sparse if there exist a basis $\bsy \Psi$ in which the representation of the signal in it has only $K$ components that are non zero, that we call its support. A $K$-compressible signal is a signal that is "well" approximated by a $K$-sparse one. More precisely, if the signal is approximated by keeping only its $K$ components with largest amplitude in an appropriate basis, then the $\ell_p$ norm of the difference between this approximation and the signal in this basis decays as a power law:
\begin{equation}
  ||\bs_K - \bs||_p < C K^{-u}
  \label{eq_compressible}
\end{equation}
for some constants $C$ and $u>0$ where $\bs_K$ denotes the best $K$-sparse approximation of the compressible signal $\bs$. It basically means that the amplitude of the sorted components of $\bs$ decays at least as a power law, see \cite{modelBasedCSBaraniuk2008} for more details on this notion. In this thesis, I will sometimes use the terminology of sparse signals even for compressible ones.

Compressed sensing, introduced 10 years ago in a series of papers by Donoho, Candès, Tao and Romberg \cite{Candes:2005um,CandesTao:06,CandesRombergTao06,Donoho:06} is the theory and ensemble of techniques behind the measurement protocol and reconstruction process of sparse and compressible signals from few (noisy or not) linear observations. Mathematically speaking, it is the field of research focused on solving a priori undetermined systems of linear equations, using sparsity assumptions about the solution. A very nice and simple introduction to compressed sensing (seen from the convex optimization point of view) can be found in \cite{4472240}, see \cite{KrzakalaPRX2012,KrzakalaMezard12} for the probabilistic point of view as adopted in this thesis.

Another fundamental aspect of compressed sensing, complementary to sparsity, is the notion of {\it coherence}. In order to be able to infer back the sparse signal $\bs$ from a model like (\ref{eqIntro:AWGNCS}), the measurement (or "sensing") matrix $\bF$ must be as incoherent as possible with the sparsifying basis $\bsy \Psi$ of $\bs$. It means that each basis vector of $\bF$ must be as orthogonal as possible to {\it all} the basis vectors of $\bsy \Psi$ at the same time or equivalently any basis vector of $\bF$ {\it cannot} be expressed nor well approximated by a sparse linear combination of the basis vectors of $\bsy \Psi$. In this way, all measurements $y_\mu$ will contain an approximately equal amount of information about all the components of $\bs$ expressed in $\bsy \Psi$. Intuitively, the $O(K)$ measurements select a set of possible solutions to the linear system (\ref{eqIntro:AWGNCS}) and the sparsity assumption select among these the sparsest one which can be unique in the noiseless setting. As this solution has only $K$ non zero values, the $O(K)$ measurements are enough to fix their amplitudes. The coherence between two matrices $\tbf A$ and $\tbf B$ is formally defined as:
\begin{equation}
	\mu(\tbf A, \tbf B) = \sqrt{N} \max_{1\le k,l\le N}|(\tbf A_{\bullet,k})^{\intercal} \tbf B_{\bullet,l}|
\end{equation}
which is thus a direct measure of the correlation between the matrices.

Constructing a maximally incoherent sensing matrix for a given sparsifying basis $\bsy \Psi$ is a computationally very hard problem and cannot be solved in general. But here the intuition suggesting that a purely random sensing matrix (that we will always take i.i.d Gaussian) must be highly incoherent with any $\bsy \Psi$ with high probability (i.e. tends to one as the matrices size diverge) is valid. Indeed, drawing a random i.i.d Gaussian matrix will give rise to basis vectors uncorrelated with the $\bsy \Psi$ ones, exactly as a white noise has a flat spectrum in any basies. This is one among many others advantages of working in high dimensions. It is quite instinctive to see that in a very high dimensional space parametrized by some basis, if you draw a random vector, it has very low probability to be close to aligned to one of the vector basis as there are so many available directions. Working with random Gaussian i.i.d matrices has another great advantage: it allows to perform analytical predictions in the large size limit $N\to \infty$ using techniques mainly based on "law of large numbers like" arguments, standing at the roots of the state evolution analysis, see sec.~\ref{sec:stateEvolutionGeneric}.

Reconstructing $\bs$ from the knowledge of the measurement matrix $\bF$ and few AWGN corrupted observations $\by$ given by (\ref{eqIntro:AWGNCS}) is thus a compressed sensing problem as long as $\bs$ is sparse and $\bF$ is "random enough" (this notion will be studied in great details in the chapter about structured operators in compressed sensing, chap.~\ref{chap:structuredOperators}). The knowledge of the sparse nature of the signal allows one to solve the reconstruction problem in an efficient manner. 

What does an efficient manner actually means? First, compressed sensing theory shows that for a $K$-sparse signal of size $N$, its reconstruction can be performed from a number of linear observations that grows with $K$ instead of $N$. So for very sparse signals with a density $\rho\defeq K/N \ll 1$, the theoretically required number of measurements can be very low. This is litterally a revolution in signal processing as it overcomes the Shannon-Nyquist theorem that states that if some physical signal's highest represented frequency is $f$ (the so-called Nyquist rate), i.e. its Fourier coefficients are all zeros for any frequency higher than $f$, then at least $2f$ discrete samples of this signal are required for perfect reconstruction. Basically, it means that {\it without any prior knowledge about a signal} of size $N$ in its discrete representation, $O(N)$ measurements are required to estimate it, and this independently of its true informational content, usually carried by a small support of size $K\ll N$.

Assume that you want to measure a pure sinusoidal signal with very high frequency oscillations. The Shannon-Nyquist theorem a priori constrain you to perform many measurements of this signal. But if in addition you have now the prior knowledge that this signal is a pure sinusoide, you can use this supplementary information on the sparsity of this signal in the Fourier space to infer it from far fewer measurements. This is what compressed sensing is all about: performing the very least number of operations for estimating a signal. 

The terminology compressed sensing comes from this paradigmatic change. In usual sensing/compression strategies, one performs many measurements, where again "many" is fixed by the Shannon-Nyquist theorem. Once the signal has been estimated, one can try {\it a posteriori} to find a sparsifying basis for it and thus locate its zero or small components: this is performed thanks to compression algorithms such as JPEG2000 for images or Fourier analysis for sounds. At the end one stores only the informative support of the signal. This two-steps procedure is quite inefficient: most of the $O(N)$ performed measures contain highly redundant information as the zeros of the signal do not contribute to them, and thus they correlate only the $K$ informative components. This is why {\it afterwards} compression can be done, thanks to this inherent redundancy contained in the observations obtained by usual sampling techniques applied to sparse signals. 

Compressed sensing is an "all in one" procedure that optimally uses all the knowledge one have about the measured signal, at least in the Bayesian framework (convex optimization procedures discussed in sec.~\ref{sec:convexOptimization} usually only assume sparsity as opposed to Bayesian inference sec.~\ref{sec:BayesianInference} that allows to integrate more information in the model). By carefully designing the measurement protocol, one can maximally reduce the redundancy inside each sample which drastically lower the required number of them. This allows to directly reconstruct the most compressed form of the signal in a fixed sparsifying basis, chosen thanks to the a priori knowledge about the signal: {\it The signal I want to measure is a natural image, so I know that it is a priori sparse in the wavelet basis. I will thus try to directly estimate the few important coefficients of the signal in this basis. My signal is a sound, so it should be sparse in the Fourier basis, etc.}.
\subsection{Why is compressed sensing so useful?}
\label{sec:usefulnessCS}
In many applications, measurements can be costly in energy, time and money and reducing the required number of such samples can have a great impact. But one could ask that despite compressed sensing is a beautiful mathematical theory that should allow for great improvements in signal processing, is it actually relevant for real life applications? The anwser is yes. The notion of sparsity or compressibility of a signal is not only a great theoretical property but also an actual feature of virtually all signals in the nature. The following hypothesis appears to be almost always validated: if a signal carries some information, it must have some kind of structure in an appropriate representation, i.e. it is not pure noise, the only signal for which it exists no sparsifying basis. We will focus in this manuscript on some important applications such as image reconstructions (see chap.~\ref{chap:images}) and error correcting codes for communications (see part.~\ref{part:coding}) but we can name many others. A non hexaustive list of relevant examples could include:
\begin{itemize}
\item Medical imagery such as in magnetic resonance imagery \cite{Lustig07compressedsensing} where the measurements are very long and costly. Sparsity of the image in the wavelet or discrete cosine basis can be exploited to lower the number of required measured Fourier coefficients to reconstruct good quality images.
\item Deep space imagery and radio interferometry \cite{Wiaux21052009}. Probing the deep space is highly costly as massive telescopes are required, preferably all over the world to perform independent measurements. But many features of interest are highly sparse in appropriate basies such as intensity fields of compact astrophysical objects or the imprint of cosmic strings in the temperature field of the cosmic microwave background radiation.
\item New optical devices such as one pixel cameras \cite{duarte2008single}. Imagine that some highly accurate new sensor is really costly. Thanks to compressed sensing, a unique sensor can be used to measure images. 
\item Still in the field of optics, people are nowadays trying to use the randomness of physical media such as layers of white paint as the compressive imaging device \cite{2014NatSR...4E5552L}. Scattering media are thus promising candidates for designing efficient and compact compressive imager. In parallel, algorithms for the estimation of the sensing matrix generated by natural scattering media are developed \cite{2015arXiv150203324D}. 
\item Compressed sensing is also applied in acoustics, for example in problems of vibrating source localization \cite{2012ASAJ..132.1521C} or sampling of the $3$-d acoustic field and the room impulse responses \cite{MignotDaudet11}, that characterize the reverberation properties of the room. In both applications, the number of microphones required can be greatly reduced using compressed sensing techniques because, for example, the room impulse responses can be considered sparse in the time domain. This could be applied in virtual reality, video games and electronic music, where the use of a space-varying reverberation extracted from real environments improve the impression of immersion.
\item Compressed sensing has a great potential in group testing with applications in genetic screening, compressed genotyping or optimal blood testing \cite{DBLP:journals/corr/abs-1302-0189}. Imagine you have many samples of blood and you know that few of them are infected. How to optimally mix samples to reduce the number of required tests to find the infected samples? Compressed sensing theory answers this question.
\item The applicability of compressed sensing theory is nowadays also extended to problems with high computational cost. It can help to greatly lower the complexity of some matrix reconstruction problems such as the computation of sparse Hessian matrices, useful in density functional theory \cite{doi:10.1021/oc5000404}. 
\end{itemize}
The list could go on for a while: radar detection, efficient measurements and reconstructions of complex wave functions, data compression, efficient analog-to-information conversion, etc. A massive list of applications can be found online: \url{http://dsp.rice.edu/cs}, \url{http://nuit-blanche.blogspot.fr}.

Summing up, compressed sensing theory answers the questions: how one should design an optimal measurement process for a signal for which we have some information, such as sparsity. And how to reconstruct it efficiently from this few optimal measures. Compressed sensing will ouput the sparsest solution to a problem, and is in this sense a kind of modern Occam's razor: it finds the "minimal solution" to the problem and this from the minimal amount (or close to) of information theoretically required to solve the problem.

Let us now focus on the second fundamental efficiency aspect of compressed sensing. Now it is well defined, one could argue: {\it All that is really nice, at least on the paper, but can the estimation problem be actually solved in an amount of time which is not of the order of the age of the Universe?} In other words, are there efficient algorithms that can solve a given compressed sensing problem in a fast way? Fortunatly yes. An extensive subfield of research in compressed sensing focuses on developing such computationally efficient and yet accurate solvers, like in the present thesis. We will expose the two actual main ways of solving a compressed sensing problem, but before that we present some basic notions of complexity theory.
\section{The tradeoff between statistical and computationnal efficiency}
\label{sec:tradeoffStatisticalComputationalEfficiency}
With the explosion of the size of the data sets generated in modern scientific, medical, social and economical applications, a major point to consider in todays inference techniques is the 
tradeoff between {\it statistical} and {\it computationnal} efficiency. A statistically efficient algorithm is able to infer the desired quantity accurately while remaining robust in spite of the precense of noise. This can be quantified by error estimators such as the $MSE$ (\ref{eqIntro:MSE}). A computationnally efficient algorithm has a low complexity, i.e. it requires a number of fundamental operations to perform its task that scales "nicely" with the size of the input and output data. Lets precise this point by introducing the very basics of complexity theory, without the goal to be complete nor rigorous.
\subsection{A quick detour in complexity theory and worst case analysis : P $\neq$ NP ?}
Complexity theory aims at grouping into complexity classes the set of all problems that can be solved by computers. It exists a full zoology of such complexity classes. Let us formalize a bit this idea. Let us denote a generic problem by $\Psi$. For example, considering the core of this thesis, the AWGN corrupted linear estimation problem, $\Psi$ would be given by $\Psi=$ "find $\bs$ from the generic model (\ref{eqIntro:AWGNCS})". $\psi$ denotes a given {\it realization} or {\it instance} of $\Psi$ which would be in the present case a particular realization of the random noise vector $\bsy \xi$, of the sensing matrix $\bF$ and of the measured signal $\bs$ in (\ref{eqIntro:AWGNCS}).

Complexity theory is based on the notion of {\it worst case analysis} : a problem $\Psi$ belongs to a given complexity class if its "{\it most difficult instance}" $\tilde \psi$ belongs to this class. The most difficult instance is basically the one that requires the largest number of operations to be solved among all the instances $\{\psi\}$ of $\Psi$. The main complexity classes of interest for us are the so-called polynomial time $P$ and non-deterministic polynomial time $NP$ classes. In order to define them properly, we would need to introduce concepts such as Turing machines which are out of the scope of the present thesis and can be found in any text book on complexity theory or computer science such as the very nice books \cite{moore2011nature,papadimitriou1994computational}. Let us define them in a more handwavy way.

We define a problem $\Psi$ to belong to the $P$ class if there exist an algorithm able to solve {\it any} of its instances $\{\psi\}$ performing a number of fundamental operations (such as additions, multiplications, etc) that scales as a polynomial in the size of the problem $O(N^k)$ where $N$ is the number of variables of the solution we are looking for. We would say that a problem in $P$ is "easy" i.e. can be solved efficiently. This idea of simplicity of problems in $P$ is referred as the Cobham's thesis. Such problems include testing whether a number is prime, calculating the greatest common divisor or finding a maximum matching in a graph. In contrast, problems in $NP$ are usually referred as "hard" problems. A problem is said to belong to the $NP$ class if no algorithm is known {\it yet} to solve all of its instances (inluding the hardest one) in a number of operations bounded by a polynomial in the size of the solution, but if one is given an a priori solution to any instance, it can be checked efficiently if this is actually a solution as declared or not. The complexity of $NP$ problems usually scale as an exponential in the problem size $O(e^N)$ which becomes {\it very} quickly intractable by brute force combinatorial methods.

From the moment an algorithm can provably solve efficiently a problem, it is known to be in $P$ but it is really difficult to assert that a problem is {\it not} in $P$, or is in $NP$ as it would require to prove that no efficient algorithm exist for the hardest instances of this problem. This emphasize the most fundamental question of complexity theory and computer science: Is $P\neq NP$? No one knows a proof or disproof of this assertion despite that most scientists think nowadays that there actually exists a fundamental difference between these two classes, i.e. there are problems that {\it are} really difficult, and will remain difficult in the future.
\subsection{Complexity of sparse linear estimation and notion of typical complexity}
All these are very general considerations, but what about the core problem of this thesis (\ref{eqIntro:AWGNCS})? Unfortunately, this problem is thought to be $NP$ which makes it a priori intractable. This is also what makes this problem interesting from a theoretical point of view, in addition of its great applicability as we have seen in sec.~\ref{sec:usefulnessCS}. Fortunatly there exist efficient ways to solve such problems, and it is the subject of this thesis. As we said, complexity classes are based on the notion of worst cases, but what about {\it typical} cases? A typical case is an instance $\bar \psi$ that you would pick by selecting one at random in the set of all possible instances $\{\psi\}$ of the problem $\Psi$ when $N$ is large. It represents a kind of "average case". In many inference or combinatorial optimization problems, hard instances can be quite paradoxally very difficult to generate, most of the instances being easy: the hardest instances define the problem $\Psi$ to be formally in $NP$ but the typical instances place $\Psi$ inside $P$ in an effective way. And hopefully, many problems have lower typical complexity in proper regimes than what their worst case analyzes tell us. Further details about the typical complexity of the problem will be presented when discussing the phase diagram of the problem in sec.~\ref{sec:typicalComplexity}.
\section{The convex optimization approach}
\label{sec:convexOptimization}
How to solve a compressed sensing problem efficiently? The first methods were based on convex optimization methods.
Let us review the very basics of this appoach that is ${\it not}$ the kind of methods used and studied in this thesis but quickly presented here for sake of completeness. Very nice and complete reviews can be found such as \cite{Boyd:2004:CO:993483,van2009convex,journals/corr/abs-1007-3753,james2013introduction}.
\subsection{The LASSO regression for sparse linear estimation}
The goal is thus to find the sparsest solution among a huge set of admissible solutions to (\ref{eqIntro:AWGNCS}). Mathematically one must be able to find the (hopefully) unique solution $\hat \bx_0$ to (\ref{eqIntro:AWGNCS}) that has the smallest $\ell_0$ norm which is defined as the number on non zero components of a vector. Unfortunately, this problem is $NP$ and non convex: there exist no convex function such that its minimum is provably given by $\hat \bx_0$ for any measurement rate $\alpha \defeq M/N> \rho \defeq K/N$, the information theoretical bound. To face this, the problem is relaxed by replacing the constraint of $\ell_0$ minimization to an $\ell_1$ minimization, which makes the problem convex. This problem is referred as the LASSO (Least Absolute Shrinkage and Selection Operator) regression:
\begin{align}
  &\hat \bx_1 = \underset{\bx}{\txt{argmin}} \ ||\bx||_1 \ \txt{such that}\ \by=\bF\bx
  \label{eqIntro:l1problem}
\end{align}
which becomes in the noisy setting
\begin{align}
  &\hat \bx_1 = \underset{\bx}{\txt{argmin}} \ ||\bx||_1  \ \txt{such that}\ ||\by - \bF\bx||^2_2 < \epsilon \label{eqIntro:l1problem_relax1}\\
  \Leftrightarrow &\hat \bx_1 = \underset{\bx}{\txt{argmin}} \ ||\by - \bF\bx||^2_2 + \lambda ||\bx||_1 \label{eqIntro:l1problem_relax2}
\end{align}
where the second form is totally equivalent to the first one for a properly selected slack parameter $\lambda$, which is here to balance the relative weight of the sparsity constraint with the observations. $\epsilon$ is some small threshold relaxing the hard contraint of perfectly verifying the linear constraints, which takes into account the precense of noise. It appears that for a measurement rate $\alpha > \alpha_{DT}(\rho) > \rho$, the solution to the LASSO problem is provably equal to the one of the intractable $\ell_0$ optimization problem: $\bhx_1 = \bhx_0$. $\alpha_{DT}(\rho)$ is called the Donoho-Tanner transition \cite{2009RSPTA.367.4273D} and defines the best performances one can reach asymptotically with convex optimization based methods, see sec.~\ref{sec:typicalPhaseTransitions}: polynomial-time optimization solvers or greedy algorithms can solve the problem from $O(K \log(N/K))$ measurements, not $O(K)$. Unfortunately this transition is far from the optimality $\alpha = \rho$ and this is why we need different methods to improve the results such as Bayesian inference (sec.~\ref{sec:BayesianInference}) combined with spatial coupling (sec.~\ref{sec:spatialCoupling}) which is asymptotically optimal. If one knows some more information about the solution than the simple sparsity, more advanced models such as group sparsity, tree structures, etc can be constructed and solved by convex optimization \cite{modelBasedCSBaraniuk2008}. This can reduce the number of required measurements but in the general setting of purely sparse signals, convex optimization based approaches are {\it not} optimal from an information theoretical point of view.
\subsection{Why is the $\ell_1$ norm a good norm ?}
Why is the $\ell_1$ norm the appropriate sparsity inducing one? Smaller $p$-norm are non convex but nothing prevents from taking a higher order norm such as the $\ell_2$ one. In this case the convex problem to solve is:
\begin{align}
	\hat \bx_2 = \underset{\bx}{\txt{argmin}} \ ||\by - \bF\bx||^2_2 + \lambda ||\bx||_2^2 \label{eqIntro:l2problem}
\end{align}
Solving (\ref{eqIntro:l2problem}) is refered as the {\it ridge regression} problem \cite{james2013introduction}. Let us try to understand why is the $\ell_1$ norm minimization the proper choice, i.e. why LASSO overcomes ridge regression in solving sparse linear estimation problems. Let us define $\bs = [1, \epsilon]$.We assume $\epsilon>0$ without lose of generality. Its $\ell_1$ and $\ell_2$ norms are:
\begin{align}
	||\bs||_1 &= 1+\epsilon \\
	||\bs||_2^2 &= 1+\epsilon^2
\end{align}
What happens to these if we reduce one of its component by a small positive quantitiy $\delta$ such that $\delta < \epsilon \ll 1$.
\begin{align}
	||\bs - [0, \delta]||_1 &= 1+\epsilon-\delta = ||\bs||_1-\delta \\
	||\bs - [\delta,0]||_1 &= 1+\epsilon-\delta = ||\bs||_1-\delta\\
	||\bs - [0, \delta]||_2^2 &= 1+\epsilon^2-2\delta\epsilon + \delta^2 = ||\bs||_2^2-2\delta\epsilon+ \delta^2 \\
	||\bs - [\delta,0]||_2^2 &= 1+\epsilon^2-2\delta + \delta^2= ||\bs||_2^2-2\delta+ \delta^2
\end{align}
So we now understand that reducing small components almost does not affect the $\ell_2$ norm that tries in opposite to reduce larger ones and spread the power and thus would prevent sparsity whereas the $\ell_1$ norm is affected in the same manner as small or large components are reduced, allowing to cancel easier components that should be. So it is more that higher order norms prevent sparsity, $\ell_1$ optimization does not.
\begin{figure}
	\begin{center}
		\includegraphics[width=.8\textwidth]{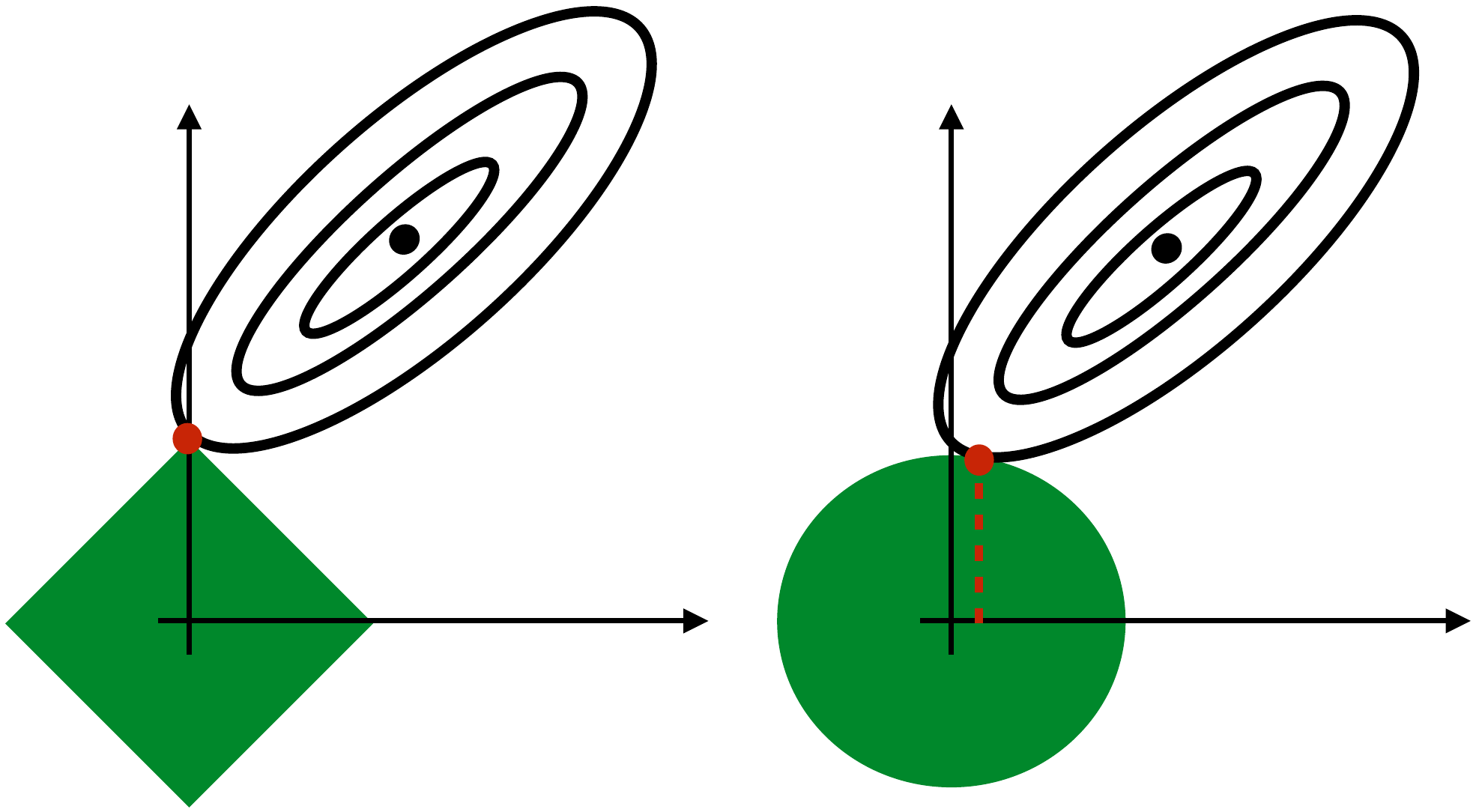}
	\end{center}
	\caption[Geometrical interpretation of $\ell_1$ and $\ell_2$ norms]{Geometrical interpretation of why the $\ell_1$ norm is the appropriate sparsity inducing one. The ellipses are iso-MSE lines. The green parts represents subspace with bounded $\ell_1$ and $\ell_2$ norm on the left and right plots respectively. The red point is the intersection between the further allowed iso-MSE line and the region bounding the norm of the vectors inside. We see that the intersection point which is the final estimate of the procedure cancels a components in the $\ell_1$ optimization case whereas it would not choosing the $\ell_2$ norm.}
	\label{figChIntro:geomtricalNorm}
\end{figure}

It can be understood in a more geometrical way as well. The optimization problem (\ref{eqIntro:l1problem_relax1}) has two parts. The first one enforces the linear observations to be fulfilled up to some error $\epsilon$. This selects a subspace with smooth bounds (due to the $\ell_2$ norm) such as the ellipses represented on Fig.~\ref{figChIntro:geomtricalNorm} which extent is fixed by the relaxation parameter $\epsilon$. Then one has to find the vector with smallest norm intersecting this region (the red point on the figure): it is the estimate $\bhx$. Vectors with a bounded $\ell_2$ norm belong to a ball (the disk on the figure), whereas with an $\ell_1$ norm they belong to a polytop with sharp corners on the axis of the frame. So with high probability, the intersection point between the two regions will be on a axis in the $\ell_1$ norm case and thus some components will be put to zero, whereas no components will be canceled in the $\ell_2$ norm case due to its smooth nature. This phenomenon is even more pronounced in higher dimensions.
\subsection{Advantages and disadvantages of convex optimization}
\label{sec:advantagesDisadvantagesConvex}
In the present thesis, the methods that we will use are based on the Bayesian inference (sec.~\ref{sec:BayesianInference}) due to its improved performances and phase transitions, see sec.~\ref{sec:typicalPhaseTransitions}. Nevertheless convex optimization methods are still used by many people and the field of research still very active. This is because despite not being optimal, convex optimization approaches have many advantages. The first one is the fact that optimization of the form (\ref{eqIntro:l1problem_relax1}) can be easily converted into linear programs and there exist a massive set of very efficient techniques and solvers combined with a well known theory of linear programming \cite{Boyd:2004:CO:993483,decodingByLinearProgCandesTao}. Furthermore, by the very definition of what a convex problem is, the problem to solve (\ref{eqIntro:l1problem_relax1}) always have a unique well defined solution. Modern solvers such as $\ell_1$ magic \cite{Candes06robustuncertainty} or 
NESTA \cite{journals/siamis/BeckerBC11} can solve compressed sensing instances in a quite fast way with convergence guarantees. 

Another strong advantage of these methods are their robustness. Robust in the sense that a given solver can be used for a very large class of convex problems. "Details" of the instance such as the sensing matrix realization or the structure of the signal that can be more complicated than just sparse are mostly irrelevant: any convex optimization solver will do the job and output a solution, despite being usually not the best one in the sense that more advanced solvers, such as based on Bayesian inference, could have found a lower MSE solution for the same problem if properly used. The pay-off is that Bayesian inference requires the computation of problem dependent quantities, but it is what makes it more powerful as well: a Bayesian solver is specifically designed for a problem and thus performs generally better than a more general convex optimization solver.
\section{The basics of information theory}
\label{sec:basicsInfoTheory}
Before to present the methodology of Bayesian inference, which is a statistical method, we present some fundamental concepts of information theory. The aim here is not completeness but really focus on the notions relevant for the present thesis and that are deeply connected to the statistical physics concepts. Very nice and complete books can be found as \cite{mackay2003information,richardson2008modern} for a computer science and communications point of view and \cite{NishimoriBook,mezard2009information} for an emphasize on the links with statistical physics. In this section, we use capital letters for the random variables, small letters for their realizations, or events.
\subsection{Incertitude and information: the entropy}
The fundamental object to quantify the information carried by some random variable $X\sim P_X(x|\bsy \theta)$, interpreted as a message sent to a receiver in the field of communication theory, is its {\it Shannon entropy} \cite{shannon48} or just entropy defined as:
\begin{equation}
	H(X|\bsy \theta) \defeq \mathbb{E}_{X}\(\log_2\(1/P_X(x|\bsy\theta)\) \) = -\int dx P_X(x|\bsy\theta) \log_2\(P_X(x|\bsy\theta) \)
	\label{eqChIntro:entropy}
\end{equation}
The object $\log_2\(1/P(x|\bsy\theta)\)$ can be interpreted as the {\it suprise} of the event $x$: the less probable $x$ is, the larger the surprise of observing it is. The informative content of $X$ is its average surprise, the entropy. It can be interpreted the other way around: the entropy measures how much incertitude (i.e. lack of information) we have about $X$ before its realization $x$ is observed, thus how much information we gain once observed. We thus speak of information, uncertainty or incertitude in the same manner, defined as the entropy. Information is more appropriate when $X$ has actually been observed, and incertitude when it has not yet. We could say that there is a fundamental conservation law linking information to incertitude: the information gained in observing some random variable realization compensates exactly the incertitude we had about it before the observation, measured by the entropy. We can make a parallel with an isolated mechanical system. Its total energy is conserved, and its dynamic is the result of the conversion of potential energy into kinetic one. The incertitude can be interpreted as the potential energy and the information as the kinetic one. 

A deterministic event has no entropy: we know everything about it, and thus gain nothing when observed. The other extreme case is the equidistributed random variable: we dont have any clue about what we will observe so our uncertainty about it reaches its maximum and thus we gain a maximum information observing its occurence. This is formalized by the second equality that verifies entropy in the next properties.

Why is this logarithm? It can be justified by rational arguments. Actually, the entropy is the {\it only} function verifying a number of necessary conditions for a coherent definition of what information is, including the previous remarks. We assume that $X$ have $n$ possible outcomes (we skeep the possible dependence on parameters):
\begin{align}
	&H(X) \ge 0 \ \txt{with equality only if X is deterministic.}\\
	&H(X) \le \log_2(n) = H(U), \ U \sim 1/n, \ \txt{the constant distribution over $n$ events.}\\
	&H(X,Y)=H(Y,X)\\
	&H(X,Y)=H(X)+H(Y|X)=H(Y)+H(X|Y) \label{eq:entropyProp1}\\
	&H(X,Y)\le H(X)+H(Y) \ \txt{with equality only if $P_{XY}(x,y)=P_X(x)P_Y(y)$} \label{eqChIntro:additH}\\
	&H(Z|X,Y)\le H(Z|X) \ \txt{with equality only if $P_{ZY}(z,y)=P_Z(z)P_Y(y)$}
\end{align}
The logarithm is in base $2$ because the natural unit of information in communication is the bit as messages are usually coded in binary form. Thus an equiprobable binary random variable carries 1 bit of information by definition.
This convention can also be justified interpreting the entropy as the number of dichotomic operations to perform (or minimum number of necessary yes/no questions to ask) to find the answer to a problem where all the answers are equiprobable, so in the worst case in a sense. 

The first equality tells that there is no such thing as negative information: we cannot lose information from any new observation of a random variable realization, at worst we gain nothing in the case of a deterministic event. Let us demonstrate the fourth one for the example:
\begin{align}
	&H(X,Y)= -\int dxdyP_{XY}(x,y)\log_2\(P_{XY}(x,y)\)\\
	&= -\int dxdyP_{XY}(x,y)\log_2\(P_{X|Y}(x|y))\)-\int dxdyP_{XY}(x,y)\log_2\(P_{Y}(y)\)\\
	&=\underbrace{-\int dyP_{Y}(y)\int dxP_{X|Y}(x|y)\log_2\(P_{X|Y}(x|y))\)}_{\defeq H(X|Y)}-\int dyP_{Y}(y)\log_2\(P_{Y}(y)\)\\
	&=H(X|Y) + H(Y)
\end{align}
where we skept the possible dependency in parameters.
This equality tells that the total information or incertitude carried by a couple of random variables $(X,Y)$ can be decomposed as the entropy of $Y$ plus the conditional entropy $H(X|Y)$: the remaining incertitude about $X$ once $Y$ has been observed, or equivalently the additional information that the $X$ observation would bring that has not already been obtained through the $Y$ observation alone. The role of $X$ and $Y$ can be switched from the third equality. The fifth equality tells that the incertitude about a couple of random variable is maximum when they are totally independent, so knowing one's realization does not help to infer anything about the other one. Finally the last equality, that seems natural now we understand what information and incertitude means, tells that knowledge about more random variables realizations can only lower the incertitude about another one, and at worst does not bring any new information when they are independent. Equivalently it means that the information brought by observing a random variable realization having already observed two other ones cannot be larger than the information brought by observing it having already observed just one other random variable realization.

Links with the Boltzmann entropy $S = k_B \log\(W\)$ of statistical physics can be established ($W$ is here the number of microstates of the system). From the information theoretical point of view, it can be interpreted as the number of bits (up to a multiplicative constant as the natural logarithm is used in physics) required to encode all the accessible microstates of the physical system with constant energy, i.e. in the microcanonical ensemble. The Boltzman constant is just here to fulfill the dimensionality requirement that the Boltzmann entropy times the temperature has the dimension of an energy. The entropy of a physical system can also be seen as its associated incertitude as it quantifies the information one would gain by measuring precisely its microstate, which is impossible in practice.
\subsection{The mutual information}
\label{sec:mutualInfo}
Now we have formalized the measure of information or incertitude carried by random variables, a natural question is the definition of a measure $I(X,Y)$ of the degree of correlation between random variables, i.e. how much information observing one of them brings about the other one: the {\it mutual information}. One could think that the previous definition of conditionnal entropy would do the job, but it is not symmetric with respect to $X$ and $Y$.

Assume you first observe $Y$ (the output of a noisy channel for example), then what is the remaining incertitude $H(X|Y)$ on $X$ (the sent codeword)? It is the total incertitude about $X$, $H(X)$ minus the information gained about $X$ (or equivalently minus the incertitude lost about $X$) from the observation of $Y$'s realization, $I(X,Y)$. Thus:
\begin{align}
	H(X|Y) &= H(X) - I(X,Y)\\
	\Rightarrow I(X,Y) &= H(X) - H(X|Y)\\
	&=  H(X) + H(Y) - H(X,Y) \label{eq:mutualInfo_sumEnt}\\
	&=H(Y) - H(Y|X)\label{eq:mutualInfo_sumEnt_2}
\end{align}
where the two last equalities were obtained using the property (\ref{eq:entropyProp1}) of the entropy. We obtain a symmetric measure of the information contained in each variable about the other one, or equivalently how much the observation of one of the two variable reduce the incertitude on the other one. The equality (\ref{eq:mutualInfo_sumEnt}) can be interpreted rewriting it as $H(X,Y) = H(X) + H(Y) - I(X,Y)$: the total information carried by the couple $(X,Y)$ is the sum of the individual informations minus the information counted twice, the mutual information, due to the correlations between the two variables. The mutual information is null when the two variables are independent, so observing one does not bring any information about the second one. Thus mutual information can be seen as a measure of how much the correlated couple of variables with probability measure $P_{XY}(x,y)$ deviates from independent ones with measure $P_X(x)P_Y(y)$. Is there a way to formalize this notion? The answer is given by the Kullback-Leibler divergence.
\subsection{The Kullback-Leibler divergence}
The appropriate object for estimating "distances" between distributions $P(\bx)$ and $Q(\bx)$ is the Kullback-Leibler divergence. It measures how well the distribution $Q$ describes the probabilistic structure of $P$. It is defined as:
\begin{equation}
	KL(P||Q) \defeq \mathbb{E}_{P}\(\log_2\(\frac{P(\bx)}{Q(\bx)}\)\)=\int d\bx P(\bx) \log_2\(\frac{P(\bx)}{Q(\bx)}\)
	\label{eq:KL_generic}
\end{equation}
This is not really a distance as it is not symmetric nor verify the triangular inequality but still verifies the properties we are interested in: $KL(P||Q) \ge 0$ with equality only if $P=Q$. Now we can estimate how much $P_{XY}$ differs from a factorizable measure $P_XP_Y$ like if $(X,Y)$ were independent:
\begin{align}
	&KL(P_{XY}||P_XP_Y) = \int dxdy P_{XY}(x,y) \log_2\(\frac{P_{XY}(x,y)}{P_{X}(x)P_{Y}(y)}\)\\
	&=\int dxdy P_{XY}(x,y) \log_2\(\frac{P_{X|Y}(x|y)}{P_{X}(x)}\)\label{eq:KL_mutualInfo} \\
	&=\int dxdy P_{X|Y}(x|y)P_{Y}(y) \log_2\(P_{X|Y}(x|y)\)-\int dxdy P_{XY}(x,y) \log_2\(P_{X}(x)\) \\
	&=\int dy P_{Y}(y)\int dx P_{X|Y}(x|y)\log_2\(P_{X|Y}(x|y)\)-\int dx P_{X}(x) \log_2\(P_{X}(x)\)\\
	&= -H(X|Y) + H(X)\\
	&= I(X,Y)
\end{align}
where we have used the marginalization property $\int dy P(x,y) = P(x)$. We find back the mutual information, the measure of how much random variables are correlated, i.e. how much their joint distribution is "far" from the factorized form: the greater the Kullback-Leibler divergence is between $P_{XY}$ and $P_XP_Y$, the more correlated are $X$ and $Y$ and thus their mutual information is larger.
\section{The Bayesian inference approach}
\label{sec:BayesianInference}
Bayesian inference stands at the roots of today's most sophisticated inference algorithms and signal processing techniques for matrix reconstruction problems such as compressed sensing and dictionary learning, error correcting codes, artificial intelligence, statistical arbitrage and classification, large scale analysis of economical market or astrophysical data, bio-infomatics and phylogenetics algorithms, decision making in automated systems such as planes, pattern recognition, optimal control theory, decision helping for judges in courtroom or physicians with automated diagnostics. It is even used in philosophy and social sciences. The list could go on. 

The strength of Bayesian inference resides in its very definition of being a general method for combining in a mathematical model all the observed data and the a priori information one have about the studied phenomenon or system. Let us formalize the basic principles of Bayesian inference, that are useful in the present context. We thus focus on the sparse linear estimation problem (\ref{eqIntro:AWGNCS}) but the method presented here is quite general. Many very nice references can be found for more details \cite{mezard2009information,mackay2003information}. In particular, \cite{wasserman2010statistics} discusses the advantages and weaknesses of Bayesian inference with respect to the frequentist statistical methods.
\subsection{The method applied to compressed sensing}
\label{sec:BayesianInferenceForCS}
Again, the problem is estimating $\bs$ as accurately as possible from the knowledge of the finite data $\by$ generated from the linear relation (\ref{eqIntro:AWGNCS}) where the sensing matrix $\bF$ is known too. Estimating the matrix as well can be of interest such as in matrix factorization or blind calibration problems \cite{GribonvalChardon11,KrzakalaMezard13b,KabashimaKMSZ14,DBLP:journals/corr/SchulkeCKZ13,DBLP:journals/corr/SchulkeCZ14,DBLP:journals/corr/abs-1301-5898,DBLP:journals/corr/BilenPGD13} but is out of the scope of the present thesis. This is done by optimizing some deterministic cost function (\ref{eqIntro:l1problem_relax1}) in the convex optimization approach, but in the Bayesian setting, we use a {\it probabilistic} point of view. To do so we define an intermediate vector $\bx$ to represent the signal, with its associated probability distribution $P(\bx | \by, \bsy \theta)$ given the observed data and some parameters linked to the prior knowledge about $\bs$ (the dependence on $\bF$ is implicit). From the simple yet very powerful Bayes formula which form the core of the Bayesian methodology, this {\it posterior} distribution is obtained as the product of the {\it prior distribution} $P_0(\bx | \bsy \theta)$ and the so called {\it likelihood} $P(\by | \bx, \bsy \theta)$:
\begin{align}
	P(\bx | \by, \bsy \theta) = \frac{P_0(\bx | \bsy \theta)P(\by | \bx, \bsy \theta)}{\int d\bx P_0(\bx | \bsy \theta)P(\by | \bx, \bsy \theta)} = \frac{P_0(\bx | \bsy \theta)P(\by | \bx, \bsy \theta)}{P(\by|\bsy \theta)}  \label{eqChIntro:BayesFormula}
\end{align}
The likelihood is the probability of the observed data given that the input would have been $\bx$ and the model parameters $\bsy \theta$. It is obtained from the generating model knowledge: it enforces the signal estimate to verify the system (\ref{eqIntro:AWGNCS}), i.e. to give back the actual observations. In the AWGN case, the proper form is naturally given by a product of Gaussian densities: the measures are independent of each other from the random design of the sensing matrix and the authorized fluctuations of the estimated observations $\tilde{\tbf y}\defeq \bF\bx$ around the actual ones $\by$ are Gaussian distributed due to the Gaussian nature of the noise:
\begin{align}
	P(\by | \bx, \Delta) &= \prod_\mu^M \mathcal{N}(y_\mu | (\bF \bx)_\mu, \Delta ) \\
	&= \frac{1}{(2\pi\Delta)^{M/2}} \exp\(-\frac{M}{2\Delta}||\by -\bF\bx||_2^2\)
	\label{eqIntro:likelihood}
\end{align}
where we remind that $||\tbf y||_p \defeq 1/M \bigg(\sum_i^M |y_i|^p\bigg)^{1/p}$ is the rescaled $\ell_p$ norm. The prior allows to include assumed knowledge about the solution $\bs$. From now on, the parameters $\bsy \theta$ are considered fixed but we will see in sec.~\ref{sec:EMlearning} how these can be learned efficiently in the Bayesian framework if unknown. So if one assumes sparsity about the signal and the fact that each of its entries have been generated independently from the same distribution, a proper factorizable prior would be of the form:
\begin{equation}
	 P_0(\bx | \bsy\theta) = \prod_i^N \[(1 - \rho) \delta(x_i) + \rho \phi(x_i | \tilde {\bsy \theta})\]
	 \label{eqIntro:priorDef}
\end{equation}
where the parameters $\bsy \theta = [\rho,  \tilde {\bsy \theta}]$ are the probability $\rho$ for a component to be part of the support (sometimes referred as the density of the signal) and the parameters $\tilde {\bsy \theta}$ that parametrize the distribution $\phi(\ | \tilde {\bsy \theta})$ associated to the support components. This distribution can be shaped as desired. This is in part thanks to the flexibility of the prior that Bayesian inference overcomes convex optimization procedures, if used properly. This distribution is defined independently of the observations: using the data to estimate the prior would be a mistake as the information contained in the likelihood and the prior would be redundant.

Finally $P(\by|\bsy \theta)=Z(\by,\bsy \theta)$ is the unknown probability of the data independently of the signal (at fixed parameters $\bsy \theta$), which can be interpreted as a partition function. $P(\bx | \by, \bsy \theta)$ is called the posterior because it is defined afterwards the data $\by$ has been obtained.

Convex $\ell_1$ optimization procedures generally just "know" about sparsity of the solution, whereas way more information about the solution structure can be included in the Bayesian setting through the prior, such as how sparse the signal is (thanks to $\rho$) or what is the precise distribution of the support components. We could even design a prior enforcing hard constraints, i.e. that strongly correlates the components of $\bx$, such as in the superposition codes studied later in this thesis sec.~\ref{sec:superCodes} or component-wise priors, assuming that all the signal components could have been generated from different distributions. This is for example studied in \cite{DBLP:journals/corr/TramelDK15} but is out of the scope of this thesis, where we always consider the signal components to be i.i.d. This additionnal information (if matching well the true features of the signal) allows inference from less data than convex optimization requires. As we will see, Bayesian inference actually allows asymptotically to reach optimality in two distinct senses: $i)$ It allows to solve the inference problem (\ref{eqIntro:AWGNCS}) from the {\it lowest possible} sampling rate $\alpha = \rho$ corresponding to as many samples as support components. This will be possible thanks to message-passing sec.~\ref{sec:AMP} combined with {\it spatial coupling} sec.~\ref{sec:spatialCoupling}, a technique intensively used in this thesis. $ii)$ It can find the solution that has the lowest $MSE$ among all approximate solutions, i.e. the minimum mean square error $MMSE$ estimator. This is possible only if the prior is the true distribution that has generated the random signal $\bs$: we call this the {\it prior matching} or {\it Nishimori} condition where the second denomination comes from the statistical physics vocabulary. In this case, the estimator is said to be {\it Bayes optimal}: the reducible error is canceled and it remains only the irreducible and finite size errors, see sec.~\ref{sec:biasVarTradeoff}.
\subsection{Different estimators for minimizing different risks: Bayesian decision theory}
\label{sec:estimators}
Let us assume that we are able in some way to obtain the true posterior distribution (\ref{eqChIntro:BayesFormula}), which is actually a $NP$ problem (see sec.~\ref{sec:tradeoffStatisticalComputationalEfficiency}). Actually this thesis is mainly about the approximate message-passing algorithm derived in sec.~\ref{sec:AMP} which is able to efficiently solve it.

The question now is how can we actually use the posterior to perform inference and estimate $\bs$? The answer resides in the {\it Bayesian decision theory}. Very nice courses on the subject are \cite{FigueiredoCOurse,KrzakalaCOurse,mackay2003information}. From this posterior, three different decisions seem naturals, each minimizing a different risk definition. We already defined the risk (\ref{eqIntro:RISK}) in sec.~\ref{sec:biasVarTradeoff} but its definition can be actually extended. We remain in the Bayesian framework considering the data $\by$ fixed as in sec.~\ref{sec:biasVarTradeoff}. In full generality, the risk associated to a {\it loss} $E(\bhx,\bx)$ (i.e. an error estimate) is its average with respect to the posterior distribution of the signal at fixed data:
\begin{align}
	R(\bhx|\by) \defeq \int d\bx P(\bx|\by) E(\bhx,\bx)
	\label{eq:riskGeneric}
\end{align}
It depends on the estimator $\bhx$ and the data. The auxilliary vector $\bx$ for which we have the posterior $P(\bx|\by)$ (\ref{eqChIntro:BayesFormula}) represents the signal $\bs$ that we don't know. The posterior can depend on parameters $\bsy \theta$, but we drop this dependency for simplicity. (\ref{eqIntro:RISK}) is thus (\ref{eq:riskGeneric}) where the loss $E(\bhx,\bx)$ is taken to be the $MSE(\bhx,\bx)$ (\ref{eqIntro:MSE}). The associated Bayes risk is (\ref{eq:BayesRisk}).

A Bayesian decision is just an estimator that minimizes some risk. An important remark is that from (\ref{eq:BayesRisk}), it is easy to see that if $\bhx^*$ is an estimator minimizing the Bayes risk (\ref{eq:BayesRisk}) then:
\begin{align}
	\partial_{\bhx} R(\bhx)\big|_{\bhx^*} &= 0\\
	&= \(\partial_{\bhx} \int d\by P(\by)R(\bhx|\by)\)\big|_{\bhx^*} \\
	&= \int d\by P(\by)\partial_{\bhx} R(\bhx|\by)\big|_{\bhx^*} \\
	\Rightarrow \partial_{\bhx} R(\bhx|\by)\big|_{\bhx^*}&= 0\\
	\Rightarrow \bhx^* &= \underset{\bhx}{\txt{argmin}} ~R(\bhx) = \underset{\bhx}{\txt{argmin}} ~R(\bhx|\by )
\end{align}
Thus minimizing the risk or the Bayes risk to take a decision by defining an estimator $\bhx^*$ is actually perfectly equivalent.
\subsubsection{The MAP estimator}
One could think about taking the mode of the posterior, i.e. the signal that maximizes it. This is referred as the {\it maximum-a-posteriori} $MAP$ estimator:
\begin{align}
	\bhx_{MAP} = \underset{\bx}{\txt{argmax}}\ P(\bx|\by) \label{eq:MAPestim}
\end{align}
The $MAP$ estimator is appropriate in cases where the information resides in the overall state of the full vector, i.e. each individual or subset of components does not bring any information, only the full vector has a meaning. The risk minimized by the $MAP$ estimator is associated to the following loss: 
\begin{align}
	E_{MAP}(\bhx,\bx) = 1 - \delta_{\bhx,\bx}
\end{align}
Indeed:
\begin{align}
	&\underset{\bhx}{\txt{argmin}}~\int d\bx P(\bx|\by)(1 - \delta_{\bhx,\bx})\\
	=&\underset{\bhx}{\txt{argmin}}~\(1 - \int d\bx P(\bx|\by)\delta_{\bhx,\bx}\)\\
	=&\underset{\bhx}{\txt{argmin}}\(1 - P(\bhx|\by)\)\\
	=&\underset{\bhx}{\txt{argmax}}~P(\bhx|\by)
\end{align}
we find back the $MAP$ estimator (\ref{eq:MAPestim}).
\subsubsection{The MARG estimator}
A related estimator is the minimal error assignments $MARG$ estimator:
\begin{align}
	\bhx_{MARG} = \big[\underset{x_i}{\txt{argmax}}\ P(x_i|\by)\big]_{i}^N
\end{align}
which is the component-wise $MAP$ estimator, i.e. it the concatenation of the $MAP$ estimates of the marginals $P(x_i|\by)$ defined as:
\begin{align}
	P(x_i|\by) \defeq \int d\bx_{\backslash i}P(\bx|\by) 
	\label{eqChIntro:marginal}
\end{align}
We have that $\bhx_{MAP}=\bhx_{MARG}$ if the posterior is factorizable over the estimate components, i.e. $P(\bx|\by)=\prod_{i}^NP(x_i|\by)$. The $MARG$ estimator minimizes the risk associated to the number of incorrectly infered components:
\begin{equation}
	E_{MARG}(\bhx,\bx) = \sum_i^N (1 - \delta_{x_i,\hat x_i} )
\end{equation}
This is well suited when the components are i.i.d discrete like in the binary channel models, that are classical noise models \cite{richardson2008modern,DBLP:journals/corr/abs-0704-2857} in communications. 
\subsubsection{The MMSE estimator}
In the general model (\ref{eqIntro:AWGNCS}), we are interested in continuous signal and sensing matrix elements and thus the notion of "exactness" of the solution is not really meaningful nor essential. Anyway, as the precision of a computer is finite, we cannot hope to infer exactly a real value as opposed to discrete ones. A better suited loss in this case is the $MSE$ (\ref{eqIntro:MSE}). The associated estimate is denoted as the minimum mean square error $MMSE$ estimator. The $MSE(\bhx,\bs)$ of the $MMSE$ estimate $\bhx$ can be interpreted as the empirical variance of a Gaussian distribution centered around the solution $\bs$ that would have been sampled to generate the i.i.d components of $\bhx$. It is a quite natural loss to use in the continuous framework and even more when the observations were corrupted by an AWGN such as in our case (\ref{eqIntro:AWGNCS}) because if we are performing inference under the prior matching condition, the $MSE(\bhx,\bs)$ becomes a measure correlated to the variance of the AWGN $\bsy \xi$. 

What is the expression of the $i^{th}$ component of the $MMSE$ estimator when we observed $\by$? As before, we minimize the risk (\ref{eqIntro:RISK}) that can be estimated from the knowledge of the posterior distribution. Differentiating it we obtain the estimator $\hat x_i(\by)$:
\begin{align}
	&\partial_{\hat x_i} \mE_{\bx|\by} <(\bhx - \bx)^2> = 2/N\  \mE_{\bx|\by}(\hat x_i - x_i) = 0 \\
	\Rightarrow &\hat x_i = \mE_{\bx|\by}(x_i ) = \mE_{x_i|\by}(x_i )
	\label{eqIntro:estimator}
\end{align}
where $\mE_{x_i|\by}$ denotes the average with respect to the posterior marginal distribution of $x_i$ given $\by$ (\ref{eqChIntro:marginal}). Thus in order to perfom $MMSE$ estimation, we need the true posterior marginals. If the estimated marginals are equal to the true ones, we say that the estimation is Bayes optimal, i.e. it is the true $MMSE$ and no solution can statistically make a better estimate given the data.
\begin{figure}[t!]
	\centering
	\includegraphics[width=.8\textwidth]{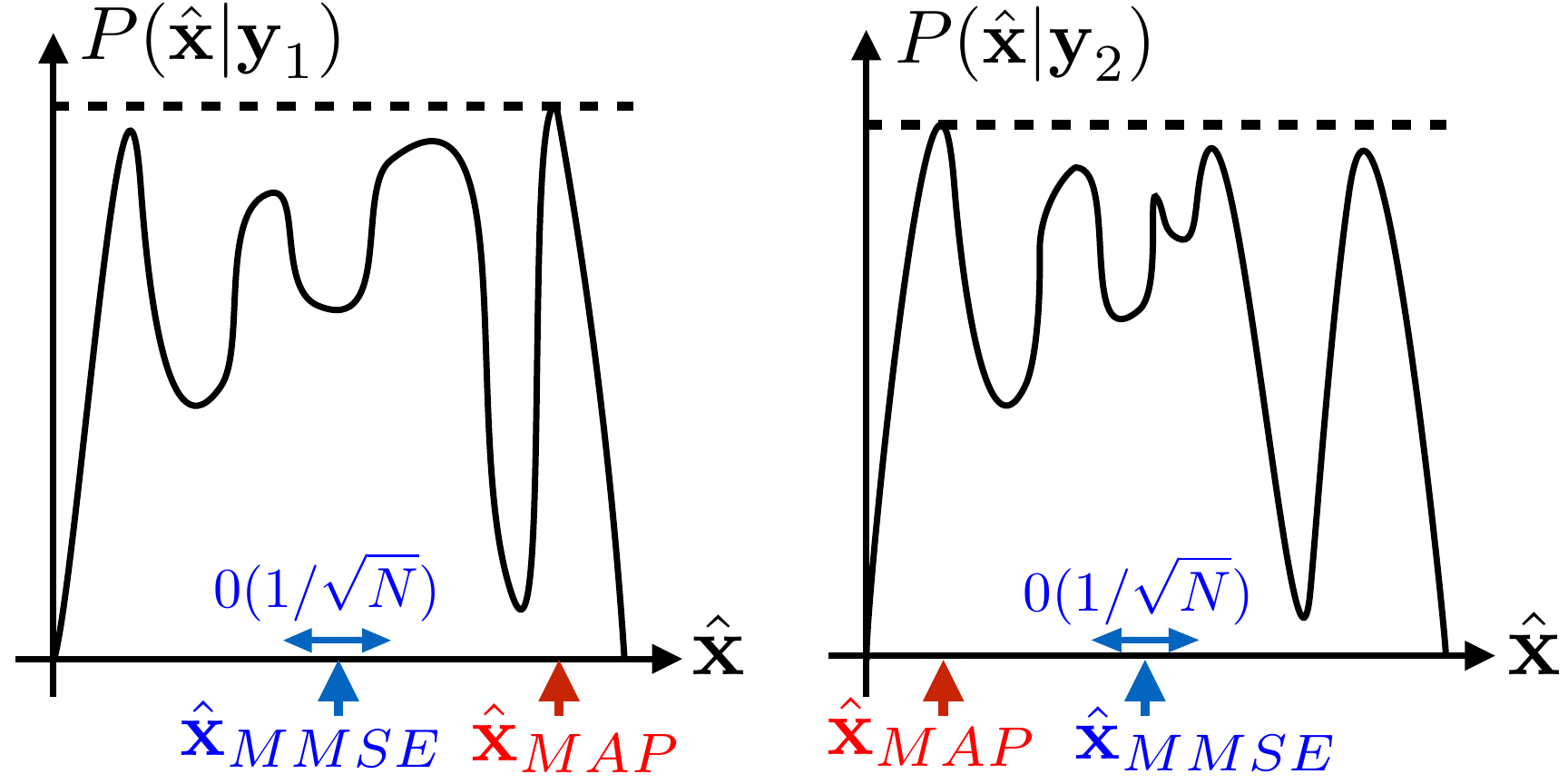}
	\caption[Minimum mean square error and maximum-a-posteriori estimators]{These plots present two complex posterior distributions corresponding to two different observation vectors $\by_1$ and $\by_2$ both related to the same signal $\bs$, the difference coming from the noise $\bsy\xi$ and measurement matrix $\bF$ realizations. Are also represented the minimum mean square error $MMSE$ and maximum-a-posteriori $MAP$ estimators in both cases. The $MAP$ estimator corresponds to the mode of the distribution, meanwhile the $MMSE$ is the typical value, i.e. averaged with respect to the posterior. We observe that due to the fluctuations in the problem realization, small changes in the shape of the posterior can induce large changes in the $MAP$ estimator meanwhile the fluctuations of the $MMSE$ one between different observation realizations are small $\in O(1/\sqrt{N})$.}
	\label{figChIntro:MMMSEVSMAP}
\end{figure}
\subsection{Why is the minimum mean square error estimator the more appropriate ? A physics point of view}
\begin{figure}[t!]
	\centering
	\includegraphics[width=.4\textwidth]{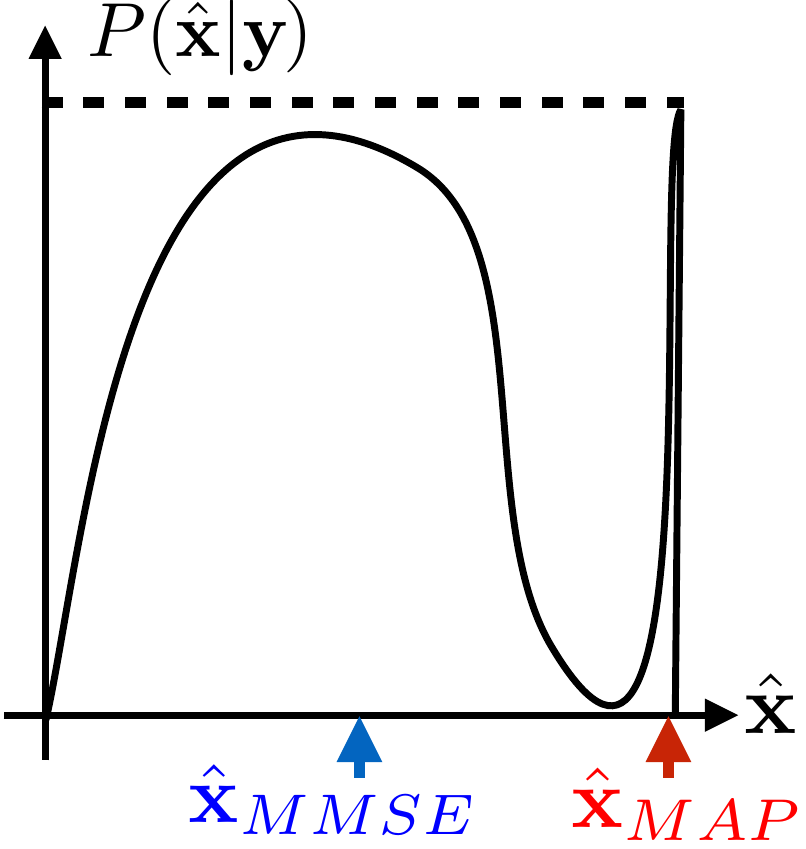}
	\caption[Posterior with a mode given by a very rare event]{This strange posterior illustrates that without taking the entropic constribution into account by selecting the $MAP$ estimator, we would miss the essential contribution of the solutions weighted by the big bump of the distribution. Instead, the $MMSE$ estimator properly weights each solution and gives the typical estimate.}
	\label{figChIntro:MMMSEVSMAP_2}
\end{figure}

Now we can give a more fundamental justification for preferring the $MMSE$ estimator to the $MAP$ or $MARG$ ones. The problems we are trying to solve are noisy, thus there exist an all set of possible solutions to (\ref{eqIntro:AWGNCS}), each weighted by its posterior probability. In such problems, the posterior can have a very complex shape which details depend on the observations $\by$. For example on Fig.~\ref{figChIntro:MMMSEVSMAP}, we show two different posterior distributions computed from two different measurement vectors $\by_1$ and $\by_2$ obtained from the {\it same} signal: the differences in the observations come from the noise and measurement matrix realizations. The point is that despite that these two measurements correspond to the same signal, their fluctuations modify the estimated posterior which mode can fluctuate a lot. It can be that there are many local maxima in the posterior with small differences but corresponding to totally different signal estimates and depending on the observations, the mode changes radically. In constrast, the $MMSE$ is robust to such fluctuations of the observations as it an averaged quantity, which thus cancels out the fluctuations in the thermodynamic limit. In statistical physics, we would say that the $MMSE$ estimate is {\it self-averaging}, like most of the thermodynamical quantities (average energy, average number of particules, total magnetization, etc). It means that its value in finite size problems converges to its asymptotic value as $N$ increases (which is the same as its average value with respect to the disorder, here the noise and matrix realizations) and its relative fluctuations around this mean are $\in O\(1/\sqrt{N}\)$. Another way of seeing it is that the $MAP$ estimator is a zero temperature quantity: it does not take into account the entropic contribution, i.e. there is no notion of average over the thermal fluctuations interpreted here as the various weighted solution estimates. Fig.~\ref{figChIntro:MMMSEVSMAP_2} is a strange posterior distribution with the mode being a rare event: sampling this distribution, we would obtain the $MAP$ estimate very rarely as opposed to many realizations that would be close to the $MMSE$ estimate as there are so many of them, despite being a bit less probable.

After this dicussion, we understand better why the physics community became interested in these topics: through the link with spin glass physics where the distributions are also very rough, like in constraint satisfaction problems as well \cite{zecchinaKsatScience,mezard2009information,barbier2013hard}. Actually, compressed sensing itself can also be seen as a finite temperature constraint satisfaction problem or a densely connected (i.e. with infinite range interactions) spin glass model of continuous spins in an external field, where the interactions defined by the measurement matrix and observations enforce the state of the spins to verify the linear system (\ref{eqIntro:AWGNCS}) (up to noise, interpreted as the temperature) and the external field would be the prior in the Bayesian setting. The problem of inferring the signal that generated the measurements (or "planted" solution in the physics language) by estimating the $MMSE$ solution is equivalent to sampling from the Boltzmann measure of the appropriate spin glass model given by the Hamiltonian (\ref{eqChIntro:hamiltonianLinearEstim}), whereas the maximum-a-posteriori estimate is given by its ground state (see \cite{KrzakalaPRX2012,KrzakalaMezard12} for a more detailed discussion of the links between compressed sensing and spin glass physics). 

Similar mappings can be established for many other computer science, inference and machine learning problems \cite{KabashimaKMSZ14,DBLP:journals/corr/KrzakalaMMNSZZ13,barbier2013hard,2001AIPC..553...89S,2002PhRvE..66d6120F} where the typical phenomenology of spin glasses is observed: phase transitions and dynamical slowing down of the reconstruction algorithms near the critical "temperature" (the critical measurement rate in compressed sensing), see sec.~\ref{sec:typicalPhaseTransitions}. Furthermore, message-passing algorithms such as belief propagation presented in sec.~\ref{sec:canonicalBP} can be interpreted in terms of the cavity method used on single instances (see sec.~\ref{sec:understandingBPwithCavityGraphs}), although the cavity method has been originally developed for computing thermodynamical quantities (i.e. averaged over the source of disorder) in spin glasses \cite{mezard2009information,MezardParisi87b}.

So one needs to compute the marginal posterior distributions (\ref{eqChIntro:marginal}) in an efficient manner to perform $MMSE$ estimation. This is in general a very difficult problem, as it is equivalent to compute the normalization $P(\by|\bsy \theta)$ of the posterior which plays the role of the partition function (\ref{eq1:fullZ}). Hopefully, a sub-field of research in inference is interested in the developement of efficient algorithms specifically designed for this task. Classical methods are based on monte carlo algorithms that directly sample the posterior for estimating it. But this distribution can have a very rough shape with many local minima like in Fig.~\ref{figChIntro:MMMSEVSMAP} which can block this kind of dynamical algorithms (it must be understood that Fig.~\ref{figChIntro:MMMSEVSMAP} is just a projection, in reality the distribution is defined over a very high dimensional space). Fortunatly it exists methods that are way more efficient such as the so-called message-passing algorithms that will be explained in great details and fully derived in this thesis.
\subsection{Solving convex optimization problems with Bayesian inference}
It is important to notice that the canonical convex optimization problems can be solved in the Bayesian framework as well. Looking at the LASSO regression (\ref{eqIntro:l1problem_relax2}) problem, we see that it is perfectly equivalent to find the mode, i.e. the $MAP$ estimate associated to the posterior distribution $P(\bx|\by,\Delta,\lambda) \propto P(\by|\bx,\Delta) P_0(\bx|\lambda)$:
\begin{align}
	\bhx_1 = \underset{\bx}{\txt{argmax}} ~P(\bx|\by,\Delta,\lambda) =\underset{\bx}{\txt{argmax}}~\exp\(-\frac{M}{2\Delta}||\by-\bF\bx||_2^2 - \lambda' N||\bx||_1\) \label{eq:lassoBayes}
\end{align}	
if we put $\Delta = M/2$ and $\lambda' = \lambda/N$, where we have used the likelihood (\ref{eqIntro:likelihood}) and the factorizable prior is $P_0(\bx) \propto \exp\(-\lambda' N ||\bx||_1\)$. This prior is refered as the double exponential or {\it Laplace} prior. The same is true for the ridge regression (\ref{eqIntro:l2problem}) using an i.i.d Gaussian prior $P_0(\bx) = \prod_i^N \mathcal{N}\(x_i|0,1/(2\lambda)\)$. In the case of this Gaussian prior, the posterior mode is also the posterior mean of $P(\bx|\by,\Delta,\lambda)$, but it is not the case with the Laplace prior. In this case the posterior mode solves the LASSO regression and leads to a sparse solution whereas the posterior mean of $P(\bx|\by,\Delta,\lambda)$ is not sparse \cite{james2013introduction}.
\section{Error correction over the additive white Gaussian noise channel}
We now present very briefly coding theory and error correcting codes. A very complete and comprehensive book on the subject is \cite{richardson2008modern}. Error correcting codes are of interest in the statistical physics community for a long time as the decoding problem can be interpreted as computing magnetizations of disordered spins systems \cite{2001PhyA..302...14S,SOURLAS_errorCorrection_book,0295-5075-45-1-097,PhysRevE.60.132,2006PhRvE..73b6122M,2001AIPC..553...89S,1999PhRvE..60.5352V}. Very nice and modern references emphasizing the numerous links between coding theory and diluted spin-glass models can be found in \cite{2002PhRvE..66d6120F,DBLP:journals/corr/abs-0704-2857}.

The aim of coding theory is to find the "best way" to encode a message such that once sent to some receiver through a noisy channel, it can be decoded, i.e. recovered despite of the errors induced by the channel. What does the best way means? Three main considerations must be taken into account for a coding/decoding scheme: $i)$ its robustness to the noise, $ii)$ its symbolic cost and $iii)$ the existence of an efficient decoder. A good scheme thus requires to be robust, in the sense that in regimes where the scheme should work (this notion of "favorable regime" will be discussed in details in sec.~\ref{sec:typicalComplexity}), its performances should decrease smoothly as the noise influence increases: a scheme which performances are highly dependent on the precise noise level or which correction does not improve as the noise lowers is not reliable. Then must be considered its symbolic cost, i.e. what is the true quantity of information that one can send reliably thanks to this scheme each time a symbol is sent through the noisy channel. This is important in practical situations as each sent symbol has a cost in energy and time. If one could send symbols without any cost, the error correction would be easy: just send many times the same message (this is called a repetition code), then the receiver just makes a majority choice by selecting for each symbol of the message the one that has been received the most or by averaging over all the received noisy realizations of the symbols in the continuous setting. Finally, even these two conditions are optimized, it is useless if the receiver has no way to decode the message, i.e. it must exist a decoder which performs quickly, is itself robust to noise and has good finite size properties, see sec.~\ref{sec:biasVarTradeoff}.
\subsection{The power constrained additive white Gaussian noise channel and its capacity}
\label{sec:capacityAWGN}
\begin{figure}[t!]
	\centering
	\includegraphics[width=1\textwidth]{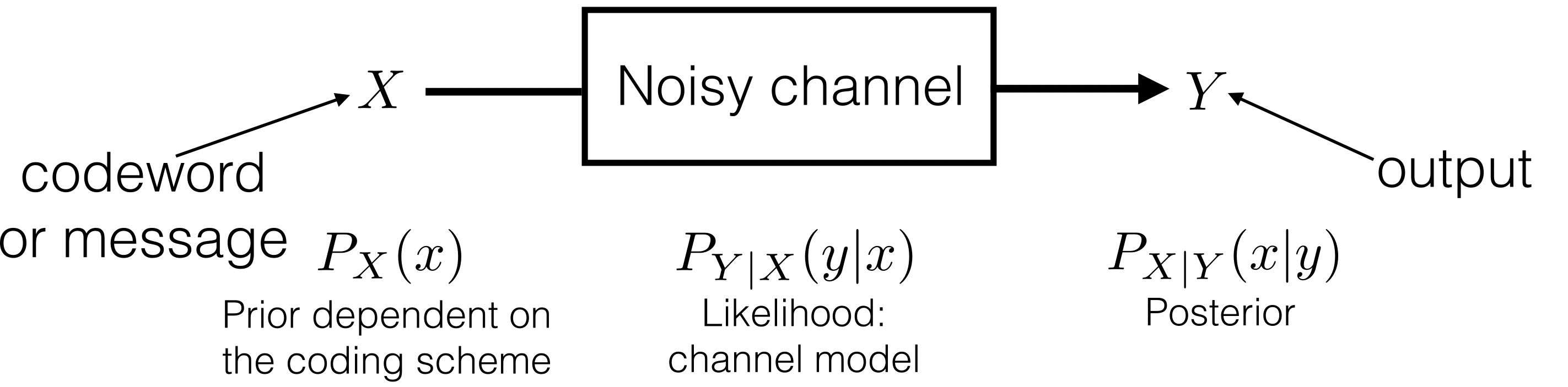}
	\caption[Generic noisy channel in the probabilistic framework]{Graphical representation of a generic noisy channel in the probabilistic framework: the codeword or input $x$ is designed through its prior $P_X$, the channel by the likelihood $P_{Y|X}$ of the output $y$ and the aim is to estimate the posterior distribution $P_{X|Y}$ of the input knowing the output in order to perform decoding through estimators.}
	\label{fig:noisyChannel}
\end{figure}
The communication channel model we are interested in is the i.i.d AWGN channel with zero mean and variance $\Delta$, a classical model in communication also extensively studied by the physics community, as in \cite{2007JPhA...4012259A,2000JPhA...33.1675K}. A generic noisy channel model in the probabilistic framework is represented on Fig.~\ref{fig:noisyChannel}. It is modeled in the Bayesian framework through the likelihood of an output $\by$ given the input $\tilde \by$ of the channel, which is in the i.i.d AWGN case:
\begin{equation}
	P(\by|\tilde\by, \Delta) = \prod_\mu^M \mathcal{N}(y_\mu | \tilde y_\mu, \Delta)
\end{equation}
A channel which likelihood is factorizable due to the independence assumption like here is referred as a memoryless channel \cite{richardson2008modern}: all output symbols (the components of $\by$) are corrupted independently one of the other. The input is called the {\it codeword}, and is the encoded version of the original message $\bs$ we want to send (the codeword can be the message itself, as in chap.~\ref{chap:robustErrorCorrection}). A natural question is wether there exist a maximum rate of information one can send reliably through this commmunication channel, and this independently of the coding/decoding scheme. This notion called the {\it capacity} $C$ has been formalized by Shannon in his celebrated paper \cite{shannon48} which started the field of communication and information theory. The noisy-channel coding theorem states that for any $\epsilon > 0$ and for any transmission rate $R<C$, there exist an encoding/decoding scheme transmitting data at rate $R$ which error probability is less than $\epsilon$, for a sufficiently large block length, the size of the codeword. Also, for any rate $R>C$, the probability of error at the receiver goes to one as the block length goes to infinity, for any coding/decoding scheme.

As discussed in sec.~\ref{sec:mutualInfo}, the mutual information between the noisy ouput $\by$ and the message $\bs$ represents the information gained about $\bs$ when we observed $\by$. As long as it is positive, it means that some information about the message is accesible through the observation of the channel output: a good coding/decoding scheme increases this mutual information maintaining the existence of an associated efficient decoder, able to maximally exploit it. A capacity achieving scheme allows to communicate asymptotically until the capacity of the channel.

Of course, we always consider that the coding scheme $\tilde \by=f(\bs)$ is a bijection and thus finding back the codeword $\tilde \by$ is equivalent to decode $\bs$ and vice-versa. The capacity is thus naturally defined as the maximum mutual information of the couple $(\bs,\by)$ or equivalently of the couple $(\tilde\by,\by)$ as $\tilde \by$ is a deterministic function of $\bs$. Maximum over what? As the likelihood is an inherent characteristic of the channel, the only degree of freedom is the codeword design $P_0(\tilde \by)$ that directly follows from the message design $P_0(\bs)$ and coding scheme $f$. Thus the capacity of a communication channel is:
\begin{equation}
	C\defeq\underset{P_{\tilde \by}}{\txt{max}} \ I(\tilde \by, \by) = \underset{\{P_{\bs},f\}}{\txt{max}} \ I(\tilde \by(\bs,f), \by)
	\label{eq:capacityGeneric}
\end{equation}
The second equality underlines that it is perfectly equivalent to consider directly the design of the codeword or the design of the message and the coding scheme, which is the usual way. But in chapter chap.~\ref{chap:robustErrorCorrection} the codeword is directly the signal, so the first equality is more appropriate. An important remark is that if one could send codewords with components of arbitrary amplitude through the channel, error correction would be useless as the relative noise (relative to the codeword amplitude) could be set to arbitrary small values. We thus always consider the codeword to be power constrained, i.e. we fix its power:
\begin{align}
	||\tilde \by||_2^2=\int d\tilde y P(\tilde y) \tilde y^2 = P
	\label{eq:powerFIxed}
\end{align}
In this way it can be compared to the noise variance to know its relative importance. The first equality in the power definition comes from the fact that the $\tilde \by$ are i.i.d. from the i.i.d assumption of the matrix elements in (\ref{eqIntro:AWGNCS}). A larger power requires more energy to input in the channel. The only relative parameter of interest is thus the so-called {\it signal to noise} ratio defined as ${\rm snr} \defeq P/\Delta$.
\subsubsection{Computation of the capacity of the i.i.d additive white Gaussian noise power constrained channel}
Let us compute the capacity of the power constrained AWGN channel. To do so, starting from the mutual information, we define a Lagragian to enforce the distribution of the codeword to be normalized and to have fixed power. We place ourselves in the scalar case, the codeword having i.i.d components, we just need to obtain the distribution of one component:
\begin{align}
	\mathcal{L}&=I(\tilde y, y)+\lambda\(\int d\tilde y P(\tilde y) \tilde y^2 - P\)+\gamma\(\int d\tilde y P(\tilde y) - 1\)\\
	&=\int d\tilde y dy P(\tilde y) P(y|\tilde y) \log_2\(\frac{P(y|\tilde y)}{P(y)}\) +\lambda\(\int d\tilde y P(\tilde y) \tilde y^2 - P\)+\gamma\(\int d\tilde y P(\tilde y) - 1\)
\end{align}
where we used the form (\ref{eq:KL_mutualInfo}) for the mutual information. Now we perform the functionnal derivative of the Lagrangian to find its optimum with respect to the codeword/input distribution, that we are looking for:
\begin{align}
	\frac{\delta \mathcal{L}}{\delta P(\tilde y^*)}& = \int d\tilde y \delta(\tilde y-\tilde y^*)\(\int dy P(y|\tilde y) \log_2\(\frac{P(y|\tilde y)}{P(y)}\) + \lambda \tilde y^2 + \gamma\)\nonumber \\
	&- \int d\tilde y dy P(\tilde y) P(y|\tilde y) \frac{P(y|\tilde y^*)}{P(y)} = 0 \ \forall \ \tilde y^*
	\label{eqChIntro:optLagragian_Py}
\end{align}
where the last term has been obtained using:
\begin{align}
	\frac{\delta P(y)}{\delta P(\tilde y^*)} &= \frac{\delta}{\delta P(\tilde y^*)} \int d\tilde y P(\tilde y)P(y|\tilde y) = \int d\tilde y \delta(\tilde y- \tilde y^*) P(y|\tilde y) = P(y|\tilde y^*)
	\label{ecChIntro:relationOnPy_1}
\end{align}
Now we notice that this last term of (\ref{eqChIntro:optLagragian_Py}) is equal to $-1$. Plugging the fact that the likelihood of the channel ouput is Gaussian in it and after integrating over $\tilde y$, (\ref{eqChIntro:optLagragian_Py}) simplifies to:
\begin{equation}
	\int dy\mathcal{N}(y|\tilde y^*, \Delta) \log_2\(P(y)\) = \lambda (\tilde y^*)^2 + \tilde \gamma
\end{equation}
where we used $\int dy\mathcal{N}(y|\tilde y^*, \Delta) \log_2\(\mathcal{N}(y|\tilde y^*, \Delta)\) = -1/2(1 + \log_2(2\pi\Delta))$ and we have put all the constants in $\tilde \gamma$. Using a Taylor expansion for $\log_2(P(y)) = a_0 + a_1 y + a_2 y^2 + a_3 y^3+ ...$, this last equality can be obtained only if the expansion is such that $a_i = 0 ~ \forall ~ i \neq \{0, 2\}$, thus the ouput distribution $P(y)$ is Gaussian with $0$ mean and a unknown variance $\sigma^2$ to find. We re-write it using the likelihod:
\begin{align}
	P(y) &= \mathcal{N}(y|0, \sigma^2) = \int d\tilde y P(y|\tilde y) P(\tilde y)= \int d\tilde y \mathcal{N}(y|\tilde y, \Delta) P(\tilde y)
\end{align}
The last equality can only be fulfilled if the codeword distribution $P(\tilde y) = \mathcal{N}(\tilde y|0, P)$ is a centered Gaussian, its variance being fixed by the power contraint (\ref{eq:powerFIxed}). It implies that the channel ouput variance is the sum of the power and noise variance $\sigma^2 = P+\Delta$ as the noise and inputs are independent. Now we know the best codeword distribution, we can compute the capacity from (\ref{eq:capacityGeneric}) and and (\ref{eq:mutualInfo_sumEnt_2}):
\begin{align}
	C &= - H(Y|\tilde Y) + H(Y)\\
	&= \int d\tilde y dy P(\tilde y) P(y|\tilde y) \log_2\(P(y|\tilde y)\) - \int dy P(y) \log_2\(P(y)\)\\
	&=\int d\tilde y dy\mathcal{N}(\tilde y|0,P) \mathcal{N}(y|\tilde y,\Delta)\log_2\(\mathcal{N}(y|\tilde y,\Delta)\) \nonumber\\
	&- \int dy \mathcal{N}(y|0,P+\Delta)\log_2\(\mathcal{N}(y|0,P+\Delta)\)\\
	&=-\frac{1}{2}\(1 + \log_2(2\pi\Delta)\) +\frac{1}{2}\(1 + \log_2(2\pi(P+\Delta))\) \\
	&=\frac{1}{2} \log_2\(1+{\rm snr}\) \label{eq:AWGN_C}
\end{align}
This is the maximum quantity of information in bits one can hope to transmit reliably per symbol sent through the i.i.d AWGN channel, and it increases with the ${\rm snr}$ as it should. One goal in communication theory over the AWGN channel is thus to find an encoder $f$ for the message $\bs$ such that the codeword is Gaussian distributed in order to get as close as possible to the capacity. Also one must derive an associated decoder to find back the message $\bs$ from the observation of the noisy observation $\by$ of the codeword. Such a strategy, the sparse superposition codes and the associated message-passing decoder will be studied in this thesis in great details, see sec.~\ref{sec:superCodes}. 
\subsection{Linear coding and the decoding problem}
\begin{figure}[t!]
	\centering
	\includegraphics[width=.9\textwidth]{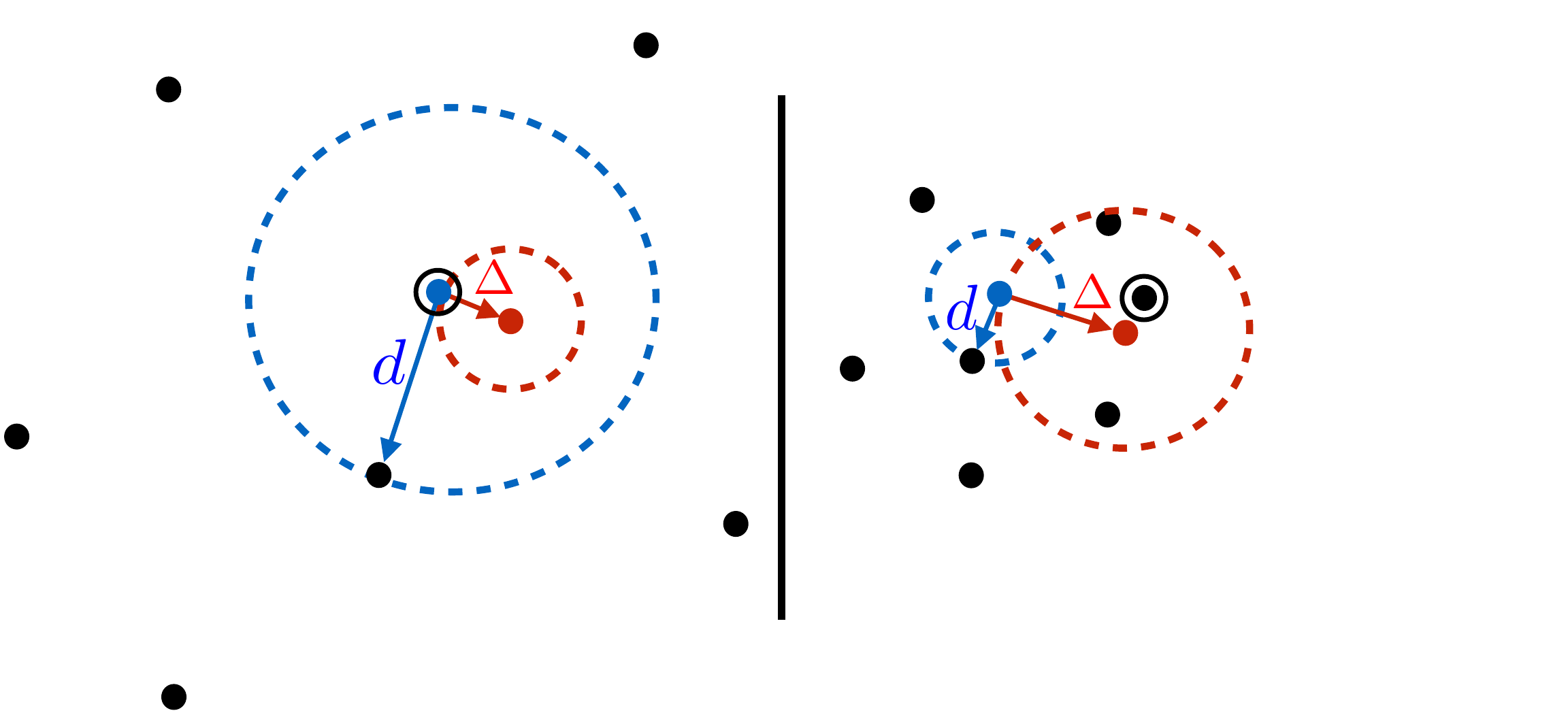}
	\caption[Reliable and unreliable codebooks]{Two different codes are represented, the codewords being the black dots. The code on the left is reliable as the minimal distance $d$ between two codewords is larger than (two times) the noise variance which is the typical distance between the vector that outputs the AWGN channel (the red dot) and the actual sent codeword, the blue one. Here, the decoder that will output the closest neighboring codeword of the channel output selects the encircled codeword which is the transmitted one. On the opposite, the right code is not reliable, as there are many codewords closest to the channel output than the transmitted codeword as $d<\Delta$. The decoder will output the encircled codeword which is a mistake.}
	\label{figChIntro:codebooks}
\end{figure}

Now we have presented the AWGN channel and quantified the maximum rate for reliable communication on this channel, the question is how to reach it? Many coding strategies are possible, but of particular interest in this thesis is linear coding, $\tilde \by = f(\bs) = \bF\bs$. The codeword is thus a linear combination of the basis $\bF$ elements, which vector of coefficients is the message. This scheme is of interest for diverse reasons. First, the encoding procedure is trivial, it requires only a matrix multiplication which is of complexity $O(N^2)$ in the general case, but using structured operators, it can be reduced to $O(L\log(N))$. Second, this coding strategy has good {\it minimal distance} $d$. This notion is fundamental and allows for a geometrical interpretation of the error correction problem. First we define a code $\mathcal{C}$ (also referred as a codebook) as the ensemble of allowed codewords by the coding scheme. The minimal distance of a code is the minimal distance between two codewords of the code. The distance is expressed in Hamming distance in the discrete case, i.e. the number of different components between codewords but in the present continuous case, an appropriate distance is the $\ell_2$ squared norm between two codewords: 
\begin{equation}
	d \defeq \underset{\tilde\by,\tilde\by'\in\mathcal{C}}{\txt{min}}||\tilde\by-\tilde\by'||_2^2
\end{equation}
Basically, a decoder will output the closest neighboring codeword of the noisy channel output $\by$ in the codebook: 
\begin{equation}
	\hat \by = \underset{\tilde\by\in \mathcal{C}}{\txt{argmin}}~||\tilde \by-\by||_2^2	
\end{equation}
Thus an error in decoding is highly probable if $d$ is small compared to the noise variance as the noisy channel will typically ouput a vector at a distance $\in O(\Delta)$ of the transmitted codeword and thus if $d$ is of the same order or smaller, there is no chance to distinguish between the good codeword and its closest neighbors, as represented on the right part of Fig.~\ref{figChIntro:codebooks}. In opposite, if $d>\Delta$ as on the left part of the figure, the code is reliable because the closest neighbor of the channel output is the transmitted codeword with high probability.
The distance can be related to the ${\rm snr}$ noticing that:
\begin{align}
	&d \le ||\tilde\by||_2^2 + ||\tilde\by'||_2^2 \le 2P\\
	\Rightarrow &d'\defeq \frac{d}{\Delta} \le 2~{\rm snr}
\end{align}
where $d'$ is a rescaled distance by the noise variance. Thus if the ${\rm snr}$ is too small, there is no way to have a minimal distance large enough to avoid wrong decoding. The quality of the error correction thus depends also strongly on the performances of the decoder and its ability to distinguish between codewords with small distance between them (but larger than $\Delta$). Hopefully there exist very efficient algorithms such as the approximate message-passing algorithm (see sec.~\ref{sec:AMP}) which relies on the linearity of the constraints, another reason for choosing linear coding.

A last remark follows from the geometrical understanding given by the Fig.~\ref{figChIntro:codebooks} and answers the question: {\it Why not to directly send the message? Why is encoding necessary?} The coding strategy is here to project the messages in a higher dimensional space, such that the distances between the codewords bijectively associated to the messages are larger than the initial distances between messages, and in this way, the decoding of the codewords is more robust to the noise influence.
%
%
%
%
%
%
%
%
%
%
%
%
%
%
%
%
%
%
%
%
%
%
%
%
%
%
%
%
%
%
%
%
%
%
%
%
%
%
%
%
%
%
%
%
%
%
%
%
%
%
%
%
%
%
%
%
%
%
%
%
%
%
%
%
%
%
%
%
%
%
%
%
%
%
%
%
%
%
%
%
%
%
%
%
%
%
%
%
%
%
%
%
%
%
%
%
%
%
%
%
%
%
%
%
%
%
%
%
%
%
%
%
%
%
%
%
%
%
%
%
%
%
\chapter{Mean field theory, graphical models and message-passing algorithms}
\label{chap:MF_graphs_MessPass}
This chapter is devoted to the mean field theory and message-passing algorithms for graphical models, the central algorithmic tools used in this thesis.

We start by introducing the mean field theory and the variationnal method used to compute approximations of the free energy. We will see that the free energy allows to recast inference problems as optimization ones.
In addition, algorithms can be derived as fixed point equations associated with these approximated free energies. We will see how this kind of approximation is naturally justified when finite statistics is known about a system thanks to the maximum entropy criterion. We will apply the methodology to compressed sensing after having defined the Hamiltonian of the problem, and we will derive the mean field algorithm for compressed sensing. We then introduce the notion of factor graphs, a nice tool to represent complex statistical dependencies among variables and very useful to understand how message-passing works.

Then we will push further the variational method using a more advanced approximation that takes into account dependencies among variables, namely the tree graph or Bethe approximation. We will start by presenting the belief propagation algorithm, a very powerful inference tool for solving the marginalization problem, among others. We will see how the belief propagation equations can be derived when the probability measure to sample is assumed to be of the Bethe form that will be presented. We will then show that the Bethe free energy and belief propagation fixed points are the same. The Bethe free energy will be written in terms of the quantities computed iteratively by belief propagation.

We will realize that this algorithm cannot be straightforwardly applied to problems defined over continuous variables on dense graphs, and we will thus derive an appropriate algorithm for this case, the main tool of this thesis: the approximate message-passing algorithm. The derivation will be performed in two different ways, both starting from the belief propagation equations. We will present in a simple fashion how this algorithm works and give the building blocks necessary to construct an approximate message-passing algorithm for a given problem. Finally we will derive the asymptotic limit of the Bethe free energy on dense graphs with linear constraints, which fixed points are the ones of the approximate message-passing and which can be expressed in terms of the quantities computed by the algorithm. This will be useful in order to derive learning equations for unknow parameters in the problem through the expectation maximization procedure. In this method, the paramaters to learn are updated in the direction that optimize this free energy, on in general, the cost function of the problem.
\section[Bayesian inference as an optimization problem and graphical models]{Bayesian inference as an optimization problem and graphical models}
We have presented in sec.~\ref{sec:basicsInfoTheory} the notion allowing for quantification of the infomation carried by a probability distribution or equivalently its uncertainty, the entropy, and defined the equivalent of a distance between distributions. Let us see how we can use these tools to perform inference by approximating the true posterior distribution (\ref{eqChIntro:BayesFormula}) which is most of the time very hard to compute exactly but yet required to perform minimum mean square error estimation, see sec.~\ref{sec:estimators}: it selects among all the possible $\theta$ values the most probable one.
\subsection{The variational method and Gibbs free energy}
\label{sec:variationalMethod}
When the posterior $P(\bx|\bsy\theta)$ (or any other distribution) is too complex to compute exactly, one need to approximate it in some way. To do so, we define an approximated distribution as $Q(\bx|\bsy \theta_Q)$ that can depend on some parameters $\bsy \theta_Q$. Now the question is, how to choose them? We can use the natural idea of minimizing the "distance" between our approximated distribution and the true posterior: we will thus optimize the Kullback-Leibler divergence (\ref{eq:KL_generic}) between the two. We want to compute $KL(Q||P)$. Why not $KL(P||Q)$ instead, as the Kullback-Leibler divergence is not symmetric? Despite we know the formal expression of the posterior, we could not compute the required averages with respect to it (or the variationnal method would be useless) as it is equivalent to compute the partition function. In opposite, we can compute averages with respect to $Q$ if it is simple enough. It is actually chosen in this purpose. Without loss of generality, we assume a Boltzmann form for the posterior $P(\bx|\bsy\theta) = \exp\(-E(\bx|\bsy\theta)\)/Z(\bsy\theta)$. Forgetting about the $\log(2)$ basis in (\ref{eq:KL_generic}) as it does not change the fixed points of the free energy, we obtain:
\begin{align}
	KL(Q||P) &= \int d\bx Q(\bx|\bsy \theta_Q)\log\(\frac{Q(\bx|\bsy \theta_Q)}{P(\bx|\bsy\theta)}\) \label{eqChIntro:KLvaria_base}\\
	&=\int d\bx Q(\bx|\bsy \theta_Q)E(\bx|\bsy\theta) +\int d\bx Q(\bx|\bsy \theta_Q)\log\(Q(\bx|\bsy \theta_Q)\)+ \log\(Z(\bsy\theta)\)\\
	&=\mathbb{E}_{Q|\bsy \theta_Q}\(E(\bx|\bsy\theta)\) - H(Q|\bsy \theta_Q) - F(\bsy\theta)	\label{eqChIntro:KLvaria}
\end{align}
where we recognized the entropy $H(Q|\bsy \theta_Q)$ (\ref{eqChIntro:entropy}) and the {\it Helmotz free energy} (or just free energy) at fixed parameters $\bsy \theta$, the true potential function of the problem:
\begin{align}
	F(\bsy\theta)&\defeq- \log\(Z(\bsy\theta)\)\\
	&= \mathbb{E}_{P|\bsy \theta}\(E(\bx|\bsy\theta)\) - H(P|\bsy \theta)
\end{align} 
The second equality is obtained by writing the entropy (\ref{eqChIntro:entropy}) of the posterior $P$ in its Boltzmann form. This free energy is not computable as its knowledge is equivalent to the computation of the true partition function $Z(\bsy\theta)$, or equivalently of the posterior. But this form suggests the definition of a {\it variational free energy} or {\it Gibbs free energy} associated to the approximate distribution $Q$:
\begin{equation}
	F_Q(\bsy \theta_Q)\defeq \mathbb{E}_{Q|\bsy \theta_Q}\(E(\bx|\bsy\theta)\) - H(Q|\bsy \theta_Q)
	\label{eqChIntro:GibbsF}
\end{equation} 
From this and (\ref{eqChIntro:KLvaria}) we obtain the following important equality that stands at the roots of the variational method:
\begin{equation}
	F_Q(\bsy \theta_Q)=F(\bsy\theta) + KL(Q||P)
	\label{eqChIntro:FQ_F_KL}
\end{equation}
and as $KL(Q||P)\ge 0$ with equality only if $Q=P$, we have that $F_Q(\bsy \theta_Q) \ge F(\bsy\theta)$. This validates a posteriori that the best parameters for $Q$ are given by those minimizing $KL(Q||P)$ as it correponds to the ones that minimize the variational free energy, lower bounded by the true one. The advantage with the variational formalism is that the Gibbs free energy (\ref{eqChIntro:GibbsF}) can be quite easy to compute for an appropriate $Q$ and the free parameters $\bsy \theta_Q$ optimal values are computed by optimizing it with respect to them.
\subsection{The mean field approximation}
The most natural approximation for $Q$ in the variational method is the so-called {\it mean field} approximation, where one assumes that $Q$ is factorizable over subsets of variables $\{\bx^a \defeq [x_1^a,\ldots,x_{n_a}^a]\}_a^G$ that can overlap, where $n_a$ is the number of variables in the subset $a$:
\begin{align}
	Q(\bx|\bsy \theta_Q) &= \frac{1}{Z(\bsy \theta_Q)}\prod_a^G \psi_a(\bx_a|\bsy \theta_Q) \label{eqChIntro:meanFieldQ}\\
	&= \frac{1}{Z(\bsy \theta_Q)} \exp\(-\sum_a^G E_a(\bx_a|\bsy \theta_Q)\) \\
	&= \frac{1}{Z(\bsy \theta_Q)} \exp\(-E(\bx|\bsy \theta_Q)\)
\end{align}
where $\psi_a(\bx_a|\bsy\theta_Q)$ is some function of the subset $\bx_a$. We call it the {\it compatibility function}, {\it constraint} or {\it factor} $a$. The second form is the associated Boltzmann form, where we define the "energy" assocated to the compatibility function $a$: 
\begin{align}
	&E_a\(\bx_a|\bsy \theta_Q)\) \defeq -\log\(\psi_a(\bx_a|\bsy \theta_Q)\) \label{eqChIntro:defEnergyGraph1}\\
	\Rightarrow & E\(\bx|\bsy \theta_Q)\) = \sum_a^G E_a\(\bx_a|\bsy \theta_Q)\) \label{eqChIntro:defEnergyGraph2}
\end{align}
$E\(\bx|\bsy \theta_Q)\)$ is the total energy of the system i.e. the Hamiltonian, referred as the {\it cost function} in statistical inference and computer science. The easiest mean field approximation to deal with, sometimes referred as the {\it naive mean field approximation}, corresponds to consider the subsets as being the individual variables, so to write $Q$ as a fully factorizable distribution over the signal components, considered as independent:
\begin{equation}
	Q(\bx|\bsy h) = \prod_i^N Q_i(x_i|h_i)
	\label{eqChIntro:NaiveMeanFieldQ}
\end{equation}
where $Q_i$ is the approximate marginal distribution of $x_i$ ($Q_i$ can be trivially computed by normalizing any factor $\psi_i$, in this way $Q$ is already normalized). The denomination of mean field approximation comes from the interpretation of the parameters $\{h_i\}_i^N$ as local fields felt by the variables that summarize the interactions with the other ones.
\subsection[Justification of mean field approximations by maximum entropy criterion]{Justification of mean field approximations by maximum entropy criterion}
\label{sec:maxEntropyCrit}
Assume that you have some partial knowledge about some complex system made of the interacting variables $\bx$, such as the first and second order statistics of $\bx$. What is the best mean field approximation $Q$ you can do of the true unknown distribution $P$ (here even its formal expression can be unknown)? A possible answer resides in the {\it maximum entropy criterion}. It is a kind of formalization of the Occam's razor: if one have some knowledge about a system, he should use a model that is in agreement with it, but does not assume anything additional. So in a sense, one should use the minimal model that fits the assumptions or the knowledge about the system. As we speak about statistical models represented by distributions, the natural object to quantify how much we constrain a distribution is through its entropy (\ref{eqChIntro:entropy}).

As we will derive, the Ising model is the mean field approximation of maximum entropy when only the first and second order statistics of the variables $\{x_i\}_i^N$ are known, at least empirically. To see that, we use the method of Lagrange multipliers and define the Lagrangian of a distribution $Q$ starting from its entropy and defining Lagrange multipliers $\{h_i\}_i^N$ that fix its marginals $\{m_i\defeq \int dx_i x_iQ_i(x_i)\}_i^N$ and $\{J_{ij}\}_{ij}^{N,N}$ for the second moments $\{\Sigma_{ij}\defeq \int dx_idx_j x_ix_jQ_{ij}(x_i,x_j) \}_{i,j}^{N,N}$ in addition of $\Gamma$ for its normalization:
\begin{align}
	\mathcal{L}(Q,\bh, \textbf J, \Gamma)&=H(Q) + \sum_i^N h_i\(\int dx_i x_iQ_i(x_i) - m_i\)\nonumber \\
	&+\sum_{i,j}^{N,N} J_{ij}\(\int dx_idx_j x_ix_jQ_{ij}(x_i,x_j) - \Sigma_{ij}\)+\Gamma\(\int d\bx Q(\bx)-1\) \label{eq:Lag_Ising}
\end{align}
Now to find the extremum of this object, we weakly perturbe the distribution $Q\to Q+\delta Q$ and compute the new Lagrangian which is:
\begin{align}
	&\mathcal{L}(Q+\delta Q,\bh, \textbf J,\Gamma)= \mathcal{L}(Q,\bh, \textbf J,\Gamma)\\
	&+\int d\bx \delta Q(\bx)\[-\log\(Q(\bx)\) -1 +\sum_i^N h_ix_i+\sum_{i,j}^{N,N} J_{ij}x_ix_j +\Gamma\] + O(\delta Q^2)
\end{align}
At the maximum, the first order must cancel out for any $\bx$, thus the integrand must always be zero, giving the shape of the distribution $Q$:
\begin{align}
	0&=-\log\(Q(\bx)\) -1 +\sum_i^N h_ix_i +\sum_{i,j}^{N,N} J_{ij}x_ix_j+\Gamma\\
	\Rightarrow  Q(\bx|\bh,\textbf J) &= \frac{1}{Z_Q(\bh,\textbf J)}\exp\({\sum_i^Nh_ix_i+\sum_{i,j}^{N,N} J_{ij}x_ix_j}\) \\
	&=\frac{1}{Z_Q(\bh,\textbf J)}\prod_i^N\psi_i(x_i|h_i)\prod_{i,j}^{N,N}\psi_{ij}(x_i,x_j|J_{ij})
	\label{eqChIntro:maxEntMeanField}
\end{align}
where we have put all the $\bx$ independent terms into the normalization constant $Z_Q(\bh,\textbf J)$. We find back a mean field approximation of the form (\ref{eqChIntro:meanFieldQ}), and thus understand now that it corresponds to the minimal model with fixed finite statistics. From this we interpret the lagrange multipliers $\{h_i\}_i^N$ as external fields and $\{J_{ij}\}_{i,j}^{N,N}$ as two points interactions between the variables. Taking into account higher order statistics would lead to more complex models, but in most of the practical situations, higher than second order statistics are difficult to extract from finite data because the number of samples required increases quickly with the order of the moment we want to compute. This follows from the fact that the larger the moment, the larger the amplitude of the fluctuations of its empirical estimate around its true value.

A remark is that interpreting the constraints we want to enforce for the distribution $Q$ through (\ref{eq:Lag_Ising}) as the average energy part in the variational free energy (\ref{eqChIntro:GibbsF}), the Lagrangian could be interpreted as a negative variational free energy. But the method is different in the sense that when we associate a Gibbs free energy $F_Q$ to a distribution, we {\it assume its form} that can depend on parameters and the optimization of $F_Q$ gives the best parameters that one should use in conjunction with this particular form of distribution. In the method of the maximum entropy, one {\it assumes constraints} that must verify the distribution, but {\it not its form}, and the optimization of the Lagrangian gives the form of the distribution one should use. Then, in order to find the best parameters introduced during the derivation to enforce the contraints, the Lagrange multipliers, one can a posteriori use the variational method. The two methods are thus complementary in a sense.
\subsection{The maximum entropy criterion finds the most probable model}
Let us give a more precise sense to the maximum entropy principle when used to find the value of a parameter of some variational distribution depending on it. Assume you have access to some data $\by$ that you assume generated by $P(\by|\theta)$ parametrized by a parameter $\theta$. The maximum entropy criterion states that its "best" value $\theta^*$ is the one maximizing the entropy of its posterior $P(\theta|\by)\propto P_0(\theta)P(\by|\theta)$:
\begin{align}
	\theta^* &= \underset{\theta}{\text{argmax}} \[ -\int d\by P(\theta|\by) \log_2\(P(\theta|\by)\)\]\\
	\Rightarrow 0 &=-\int d\by\frac{\partial P(\theta|\by)}{\partial \theta}\(\log_2\(P(\theta|\by)\) + 1\)\\
	\Rightarrow 0 &= \frac{\partial P(\theta|\by)}{\partial \theta}	
\end{align}
The last equality shows that the maximum entropy criterion for choosing $\theta$ is thus equivalent to take the maximum of the posterior of $\theta$, or just its likelihood $P(\by|\theta)$ if no prior is assumed. The maximum entropy criterion is thus perfectly equivalent to the maximum-a-posteriori $MAP$ principle, discussed in sec.~\ref{sec:estimators}.
\subsection{Factor graphs}
\label{sec:factorGraphs}
Let us see now how we can define graphical representations of complex functions such as posterior distributions in the Bayesian framework. The appropriate tool are the so called {\it factor graphs}. In full generality, they are used to represent the dependency structure among variables encoded through a factorizable function $Q(\bx)$, that can depends on many parameters, so of the form (\ref{eqChIntro:meanFieldQ}) (the partition function $Z$ can be forgotten for a generic function, but must be present if we want $Q$ to be a probability distribution). 

\begin{figure}[t!]
	\centering
	\includegraphics[width=.6\textwidth]{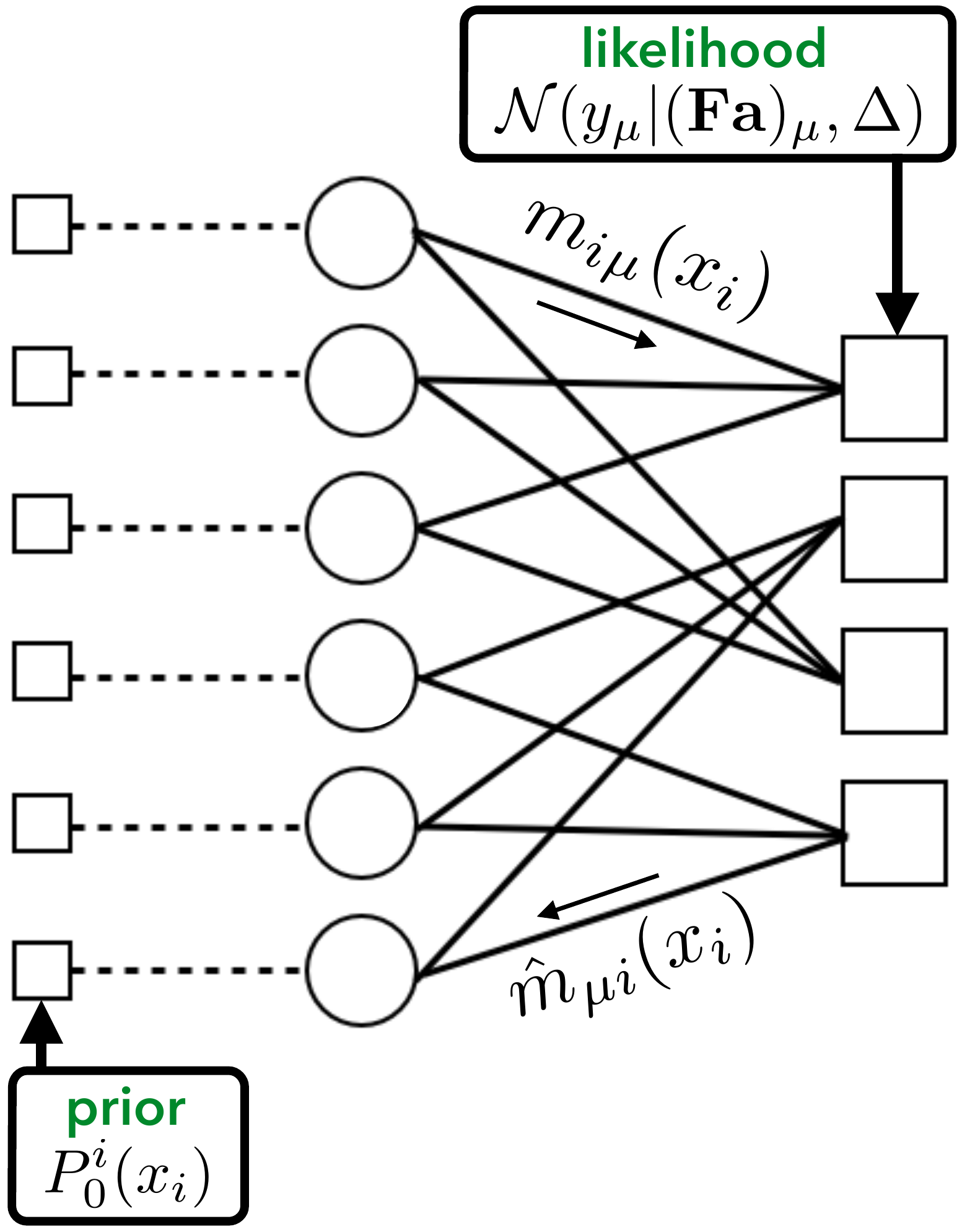}
	\caption[Factor graph of linear estimation problems under i.i.d AWGN corruption]{Factor graph associated with a linear estimation problem under i.i.d AWGN corruption. The squares are the factors or constraints. Here they are associated to the prior and likelihood terms. The factors are connected to the variables (circle nodes) they depend on in their functionnal representation. The node-to-factor $m_{i\mu}(x_i)$ and factor-to-node $\hat m_{\mu i}(x_i)$ cavity messages are represented. They should stand on the same edge as they share same indices but we put them on different ones for sake of readibility.}
	\label{figChIntro:factorGraphCS}
\end{figure}

A factor graph is a bipartite graph $G=\{\mathcal{V},\mathcal{F},\mathcal{E}\}$ where en edge $e\in \mathcal{E}$ is present between a variable node $v\in \mathcal{V}$ and a factor node $f\in \mathcal{F}$ only if the factor $\psi_f(\bx_{f\backslash v}, x_v)$ present in the factorized form of $Q$ depends on $x_v$. The circle nodes are associated to the variables while squares represent the factors. For example, if we want to associate a factor graph to the posterior distribution of the compressed sensing problem, we write it in its factorized form. Using (\ref{eqChIntro:BayesFormula}) combined with (\ref{eqIntro:likelihood}) and (\ref{eqIntro:priorDef}), we see that the posterior decomposes as a product of functions over the single components due to the prior part and over the ensemble of components due to the likelihood. The associated factor graph is given by Fig.~\ref{figChIntro:factorGraphCS}, and the factorizable structure of the posterior becomes clear. On the graph are representd objects that will be defined in sec.~\ref{sec:canonicalBP}, namely the cavity messages.
\subsection{The Hamiltonian of linear estimation problems and compressed sensing}
In the Bayesian framework, we can associate an Hamiltonian to the linear estimation problem (and thus to compressed sensing). We start by rewriting the posterior distribution in the Boltzmann form:
\begin{align}
	&P(\bx|\by,\bsy\theta) = \frac{1}{Z(\bsy \theta)}\exp\(-E(\bx|\by,\bsy\theta)\)\\ \Rightarrow &E(\bx|\by,\bsy\theta) = -\log\(P_0(\bx|\bsy\theta)\) -\log\(P(\by|\bx,\bsy\theta)\)
\end{align}
From (\ref{eqChIntro:BayesFormula}) combined with (\ref{eqIntro:likelihood}) and (\ref{eqIntro:priorDef}) we get the Hamiltonian, denoted by $E$ not to confuse it with the entropy $H$ (the notation generally used in physics for the Hamiltonian, the entropy symbol being $S$):
\begin{align}
	E(\bx|\by,\bsy\theta) &=-\sum_i^N \log\(P_0^i(x_i)\) + \frac{1}{2\Delta} \sum_\mu^M \(y_\mu-(\bF\bx)_\mu\)^2+ \frac{M}{2} \log\(2\pi\Delta\)\\
	&=-\sum_i^N \log[(1-\rho)\delta(x_i) + \rho \phi(x_i)] + \frac{1}{2\Delta} \sum_\mu^M \(y_\mu-(\bF\bx)_\mu\)^2 + \frac{M}{2} \log\(2\pi\Delta\)
	\label{eqChIntro:hamiltonianLinearEstim}
\end{align}
\subsection{Naive mean field estimation for compressed sensing}
\label{sec:naiveMFforCS}
Let us now derive the naive mean field solution, approximating the true posterior as fully factorizable (\ref{eqChIntro:NaiveMeanFieldQ}). In order to find what are the best parameters of the approximate distribution, we first need to write the variational free energy for the present problem. Using the variational free energy definition (\ref{eqChIntro:FQ_F_KL}) we obtain:
\begin{align}
	&F_Q(\bsy\theta_Q) = -\sum_i^N \mathbb{E}_{Q_i}\(\log(P_0^i(x_i))-\log(Q_i(x_i))\) \nonumber\\
	&+ \frac{1}{2\Delta}\sum_\mu^M \mathbb{E}_{Q}\(\(y_\mu-(\bF\bx)_\mu\)^2\)+ \frac{M}{2} \log\(2\pi\Delta\) \\
	&=\sum_i^N KL(Q_i||P_0^i) + \frac{M}{2} \log\(2\pi\Delta\) \\
	&+\frac{1}{2\Delta}\sum_\mu^M \Bigg(y_\mu^2 - 2y_\mu(\bF\ba)_\mu + \sum_{i,j\neq i}^{N,N}F_{\mu i}F_{\mu j} a_ia_j + \sum_i^NF_{\mu i}^2(v_i + a_i^2) \Bigg) \\
	&=\sum_i^N KL(Q_i||P_0^i) + \frac{M}{2} \log\(2\pi\Delta\)+ \frac{1}{2\Delta}\sum_\mu^M \([y_\mu - (\bF\ba)_\mu]^2 + (\bF^2\bv)_\mu \)
	\label{eq:naiveMF_F}
\end{align}	
where we have used the Kullback-Leibler divergence definition (\ref{eqChIntro:KLvaria_base}), the additivity property of the entropy for independent variables (\ref{eqChIntro:additH}) together with the definition of the marginal means and variances of $Q$:
\begin{align}
	a_i &\defeq \int dx_i x_i Q_i(x_i)\\
	v_i &\defeq \int dx_i x_i^2 Q_i(x_i) - a_i^2
\end{align}
Now that we have the naive mean field Gibbs free energy, we can figure out the expression of the marginals $\{Q_i(x_i)\}_i^N$. We perturbe one of the marginals $Q_i\to Q_i+\delta Q_i$, the associated perturbed distribution is denoted as $\tilde Q$. The perturbation term of the Gibbs free energy must cancel at the minimum for any $\bx$:
\begin{align}
	&F_Q(\bsy\theta_Q) - F_{\tilde Q}(\bsy\theta_Q) = 0 \nonumber\\
	\Rightarrow & \int dx_i \delta Q_i(x_i)\[ \frac{1}{2\Delta}\sum_\mu^M \left\{x_i2F_{\mu i} \(\sum_{j\neq i}^N F_{\mu j}a_j-y_\mu\) + x_i^2F_{\mu i}^2\right\} + 1 + \log\(\frac{Q_i(x_i)}{P_0^i(x_i)}\)\] = 0\nonumber\\
	\Rightarrow & Q_i(x_i|\bF,\ba) = \frac{1}{Z_i(\bF,\ba)}P_0^i(x_i)\exp\(\frac{x_i}{\Delta}\sum_\mu^M F_{\mu i}\(y_\mu - \sum_{j}^N F_{\mu j}a_j + F_{\mu i}a_i\) -\frac{x_i^2}{2\Delta}\sum_\mu^MF_{\mu i}^2\)\label{eqChIntro:nonSqFromMF}
\end{align}
Defining the following quantities:
\begin{align}
	\Sigma_i^2 &\defeq \frac{\Delta}{\sum_\mu^MF_{\mu i}^2}\\
	R_i &\defeq a_i + \frac{\Sigma_i^2}{\Delta} \sum_\mu^M F_{\mu i} \(y_\mu - (\bF\ba)_\mu\)
\end{align}
we obtain after simplification of (\ref{eqChIntro:nonSqFromMF}) the following form of the marginal distributions for compressed sensing under the naive mean field approximation: a product of the prior and a Gaussian mean field that summarize the influence of all the other variables on the $i^{th}$ one:
\begin{equation}
	Q_i(x_i|R_i,\Sigma_i^2) = \frac{1}{Z_i(R_i,\Sigma_i^2)}P_0^i(x_i) \exp\(-\frac{(x_i - R_i)^2}{2\Sigma_i^2}\)
\end{equation}
From this we can compute the Kullback-Leibler divergence (\ref{eq:KL_generic}) appearing in the previous mean field free energy (\ref{eq:naiveMF_F}):
\begin{align}
	\sum_i^N KL(Q_i||P_0^i) &= \sum_i^N \int dx_i Q_i(x_i) \log\(\frac{\exp\(-\frac{(x_i - R_i)^2}{2\Sigma_i^2}\)}{Z_i}\)\\
	&=\sum_i^N \int dx_i Q_i(x_i) \[-\frac{(x_i - R_i)^2}{2\Sigma_i^2}-\log\(Z_i\)\]\\
	&=-\sum_i^N \(\log\(Z_i\) + \frac{v_i + (a_i - R_i)^2}{2\Sigma_i^2}\)
\end{align}
Adding a time index to these equations gives us the mean field algorithm for compressed sensing Fig.~\ref{figChIntro:algoMF}, where we use the non linear thresholding functions:
\begin{align}
	f_a(\Sigma^2_i,R_i) &\defeq \frac{1}{Z(R_i,\Sigma^2_i)} \int dx_i ~x_i~ P_0^i(x_i) \exp\(-\frac{(x_i - R_i)^2}{2\Sigma_i^2}\)\\
	f_c(\Sigma^2_i,R_i) &\defeq \frac{1}{Z(R_i,\Sigma^2_i)} \int dx_i~ x_i^2~ P_0^i(x_i) \exp\(-\frac{(x_i - R_i)^2}{2\Sigma^2_i}\) - f_a(\Sigma^2_i,R_i)^2
\end{align}
with normalization
\begin{align}
	Z(R_i,\Sigma^2_i) \defeq \int dx_i P_0^i(x_i) \exp\(-\frac{(x_i - R_i)^2}{2\Sigma^2_i}\)	
\end{align}
This algorithm directly computes the estimates $\ba$ of the signal components.
The only non linear part in the mean field algorithm is the component-wise computations of $\ba$ and $\bv$ through these thresholding functions, the rest is linear and can be written in a efficient way with matrix operations which implies a parallel updating scheme of the estimates. It is possible to think about a randomized updating scheme as well which can sometimes help the convergence in message-passing algorithms as discussed in \cite{DBLP:journals/corr/ManoelKTZ14,DBLP:journals/corr/CaltagironeKZ14} but the payoff is a slowing down of the algorithm as matrix operations are greatly optimized. In all this thesis, we will always consider a parallel updates scheme, but it must be kept in mind that other strategies are possible with their own paybacks and advantages.
\begin{figure}[!t] 
	\centering
	\begin{algorithmic}[1]
	\State $t\gets 0$
	\State $\delta \gets \epsilon + 1$
	\State $\Sigma_i^2 \gets \frac{\Delta}{\sum_\mu^MF_{\mu i}^2} \ \forall \ i$ 
	\While{$t<t_{max} \ \textbf
	{and } \delta_{max}>\epsilon$}
	\State $R^{t+1}_i \gets  a_i^t + \frac{\Sigma_i^2}{\Delta} \sum_\mu^M F_{\mu i} \(y_\mu - (\bF\ba^t)_\mu\)$	
	\State $a^{t+1}_i \gets f_{a}\left((\Sigma_i)^2,R_i^{t+1}\right)$
	\State $t \gets t+1$
	\State $\delta \gets ||\ba^t - \ba^{t-1}||_2^2$
	\EndWhile
	\State \textbf{return} $\ba^t$
	\end{algorithmic}
	\caption[The mean field algorithm for compressed sensing]{The mean field (or iterative thresholding) algorithm for compressed sensing. $\epsilon$ is the accuracy for convergence and $t_{max}$ the maximum number of iterations. A suitable initialization for the quantities is ($a_i^{t=0}=\mathbb{E}_{P_0}(x_i)$). Once the algorithm has converged, i.e.  the quantities do not change anymore from iteration to iteration, the estimate of the $i^{th}$ signal component is $a^{t}_i$. The nonlinear thresholding function $f_{a}$ take into account the prior distribution $P_0(\bx)$.} \label{figChIntro:algoMF}
\end{figure}
The study of the perfomances of this algorithm written in the present form and the comparisons with the approximate message-passing algorithm can be found in \cite{DBLP:journals/corr/KrzakalaMTZ14} and it appears that despite good results in compressed sensing, the approximate message-passing algorithm that will be derived in sec.~\ref{sec:AMP} has a greater potential. The mean field approximation though can be asymptotically exact in models where the variance of the mean field felt by the variables goes to zero in the thermodynamic limit, for example in spin models such as the Ising fully connected ferromagnet \cite{Opper_and_Saad:2001}. It also worth noticing that this mean field algorithm Fig.~\ref{figChIntro:algoMF} is exactly equivalent to the iterative thresholding algorithm \cite{DBLP:journals/corr/abs-0805-0510,DBLP:journals/corr/KrzakalaMTZ14} if $\bF$ is properly rescaled such that $\sum_\mu F_{\mu i}^2=1\ \forall\ i$ because defining the residual $\tbf z^t = \by - \bF \ba^t$, the iterations become: 
\begin{align}
	\tbf z^t &= \by - \bF \ba^t\\
	\ba^{t+1}&=\eta_\Delta(\bF^\intercal\tbf z^t + \ba^t)
\end{align}
where $\eta_\Delta(x) = f_a(\Delta, x)$, which is the classical form of the iterative thresholding algorithm.
\section{Belief propagation and cavities}
Let us now present a more advanced mean field algorithm, the so called belief propagation algorithm BP, which allows to reach way better performances than the naive mean field approximation in many problems. The assumption behind it is that the factor graph associated to the distribution we want to sample has a tree structure: a tree is a graph such that there exist a {\it unique} path between two variables in the graph. In this case BP is exact \cite{mezard2009information}, as we will show in sec.~\ref{sec:BP_BetheF}. But more generally, BP is justified (but not strictly exact) when the distribution to sample is a Bethe measure as we will see in sec.~\ref{sec:understandingBPwithCavityGraphs}. From now on, we use $m_i(x_i)$ for the BP estimates of the true marginals $P_i(x_i)$.
\subsection{The canonical belief propagation equations}
\label{sec:canonicalBP}
The canonical BP equations, that allow to estimate the marginals $\{P_i(x_i)\}_i^N$ of a factorized distribution of the form (\ref{eqChIntro:meanFieldQ}) with an associated tree-like factor graph, i.e. a graph such that locally its structure is a tree despite the existence of "long" (of extensive size) loops in the graph, are given by:
\begin{align}
	\hat m_{a i}^{t}(x_i) &= \frac{1}{\hat z_{ai}^t} \int d\bx_{a\backslash i}~\psi_a(\bx_{a\backslash i}, x_i) \prod_{j\in \partial a \backslash i} m_{ja}^{t}(x_j)\label{ecChIntro:BP_f2n_mess}\\
	m_{ia}^{t+1}(x_i) &= \frac{1}{z_{ia}^{t+1}} \prod_{b\in\partial i \backslash a} \hat m_{b i}^{t}(x_i)\label{ecChIntro:BP_n2f_mess}
\end{align}
These quantities allow for the estimation of the marginals at time step $t$ through:
\begin{align}	
	m_i^{t}(x_i) &= \frac{1}{z_{i}^t}\prod_{a\in\partial i} \hat m_{ai}^t(x_i)\label{ecChIntro:BP_marg} \\
	&= \frac{1}{z_{ia}'^t}\hat m_{ai}^t(x_i)m_{ia}^{t+1}(x_i) \  \ \txt{for any $a\in \partial i$} \\
	m_a^{t}(\bx_a) &= \frac{\psi_a(\bx_a)}{z_a^t} \prod_{i \in \partial a} m_{ia}^{t}\label{ecChIntro:BP_marg_2}
\end{align}
A graphical representation of the BP equations is depicted on Fig.~\ref{figChIntro:BP}. The distributions (\ref{ecChIntro:BP_f2n_mess}) and (\ref{ecChIntro:BP_n2f_mess}) that are iteratively computed are the so called {\it cavity messages} (or simply messages), this vocabulary coming from the cavity method of statistical physics \cite{MezardParisi87b,mezard2009information}. This is because BP can be thought as the replica symmetric cavity equations on a single graph, or equivalently the replica symmetric cavity equations are the BP equations on an infinitely large graph or averaged over an infinite number of finite random graphs, each representing a random instance of the problem of interest. This last remark is true only if the problem is replica symmetric, i.e. the number of fixed points of the BP equations on an instance of the problem is sub-exponential in $N$. In the present thesis, the BP fixed point is unique when starting from a random initial condition of the messages. This does not imply the uniqueness of the BP fixed point, there can be another one but that require the messages to be initialized "close" to it in order to see the messages converge to this other fixed point as we will see in sec.~\ref{sec:typicalPhaseTransitions}: there is a regime where the message-passing always converges to a wrong solution despite the existence of another fixed point corresponding to the true solution of the problem. Another replica symmetric example with multiple fixed points is the Ising ferromagnet case where there exist two fixed points below the critical temperature corresponding to positive and negative average magnetizations. In the replica symmetric case, the messages (or any other thermodynamical quantity) are self-averaging: their fixed points values do not depend on the graph details (the measurement matrix $\bF$ realization in linear estimation) if large enough, the fluctuations being $\in O(1\sqrt{N})$. This is why the $MMSE$ estimator discussed in sec.~\ref{sec:estimators} is self averaging as well: the approximate message-passing algorithm (see sec.~\ref{sec:AMP}) used to compute it is directly derived from BP and applied to inference problems, which are replica symmetric under proper conditions (at least under the prior matching condition \cite{KabashimaKMSZ14}, see sec.~\ref{sec:BayesianInference}).
\subsection{Understanding belief propagation in terms of cavity graphs}
\label{sec:understandingBPwithCavityGraphs}
\begin{figure}[t!]
	\centering
	\hspace{-1cm}
	\includegraphics[width=1\textwidth]{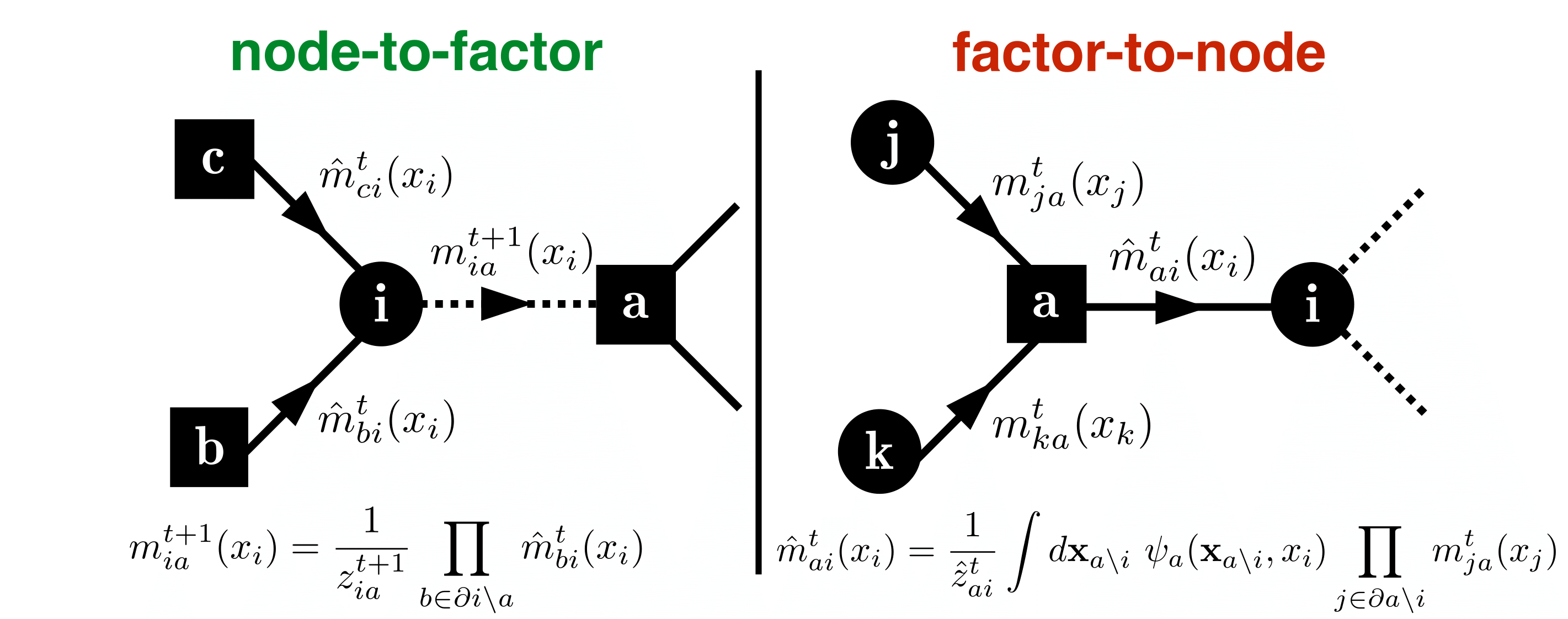}
	\caption[The belief propagation equations in terms of cavity graphs]{Graphical representation of the belief propagation equations, with the associated equations below. The dashed lines are edges present in the original factor graph which are considered not present in the cavity graph in which the cavity message is computed. On the left is the node-to-factor cavity message, the probability distribution of the $x_i$ variable in a cavity graph where the edge $(i,a)$ has been removed, i.e. a graph where $x_i$ is connected to all its neighbor factors except the factor $a$. On the right is the factor-to-node cavity message, the probability distribution of the $x_i$ variable in a cavity graph where all the edges connected to $x_i$ are removed except $(a,i)$, i.e. a graph where $x_i$ is only connected the factor $a$, not its other neighbors in the original factor graph.}
	\label{figChIntro:BP}
\end{figure}
The factor-to-node cavity message (\ref{ecChIntro:BP_f2n_mess}) is interpreted as the (approximate) probability distribution of $x_i$ in a modified graphical model, refered as a {\it cavity graph}, where $x_i$ is only connected to $a$, not anymore to its other factor neighbors in the original graph, see right part on Fig.~\ref{figChIntro:BP}. This distribution carries the information on how strongly the factor $a$ depends on the variable $x_i$, so its influence on $x_i$.
The node-to-factor cavity message (\ref{ecChIntro:BP_n2f_mess}) is the probability of $x_i$ in another cavity graph where $x_i$ is connected to all its neighbors except $a$, the edge between them being removed, see left part on Fig.~\ref{figChIntro:BP}. This message carries the information on the influence of all the neighbors factors on $x_i$ except $a$.

With this interpretation, the BP equations can be understood easily. As the graph is assumed to be a tree, the set of factor-to-node messages coming on $x_i$ are conditionally independent, thus they just multiply. The node-to-factor message $m_{ia}(x_i)$ of $x_i$ with the edge $(i,a)$ removed is thus naturally given as the product of all the factor-to-node messages $\{\hat m_{bi}(x_i)\}_{b\in \partial i \backslash a}$ except the one coming from $a$ (\ref{ecChIntro:BP_n2f_mess}). Now the factor-to-node message $\hat m_{ai}(x_i)$: we consider the cavity graph where $x_i$ is only connected to the factor $a$, which is equivalent to assume $m_{ia}(x_i)=C$ in this cavity graph, i.e. the distribution of $x_i$ in this cavity graph where we additionally removed the edge $(i,a)$ is uniform ($x_i$ does not feel any constraint). Now, the joint distribution of all the neighbors of the factor $a$ is (up to a normalization) the product of their cavity distributions $\prod_{j\in \partial a}m_{ja}(x_j)\propto\prod_{j\in \partial a\backslash i}m_{ja}(x_j)$, i.e. their joint distribution in a graph where the factor $a$ is not here, that we multiply by the compatibility function $\psi_{a}(\bx_a)$ to include back its influence. Thus the cavity message $\hat m_{ai}(x_i)$ is the marginalization of this joint distribution with respect to the neighbors other than $x_i$ that we must normalize, from which we get the equation (\ref{ecChIntro:BP_f2n_mess}). Finally the marginal (\ref{ecChIntro:BP_marg}) is given as the product of all the individual influences of the factors neighbors to $x_i$.
\subsection{Derivation of belief propagation from cavities and the assumption of Bethe measure}
\begin{figure}[t!]
	\centering
	\hspace{-1cm}
	\includegraphics[width=1\textwidth]{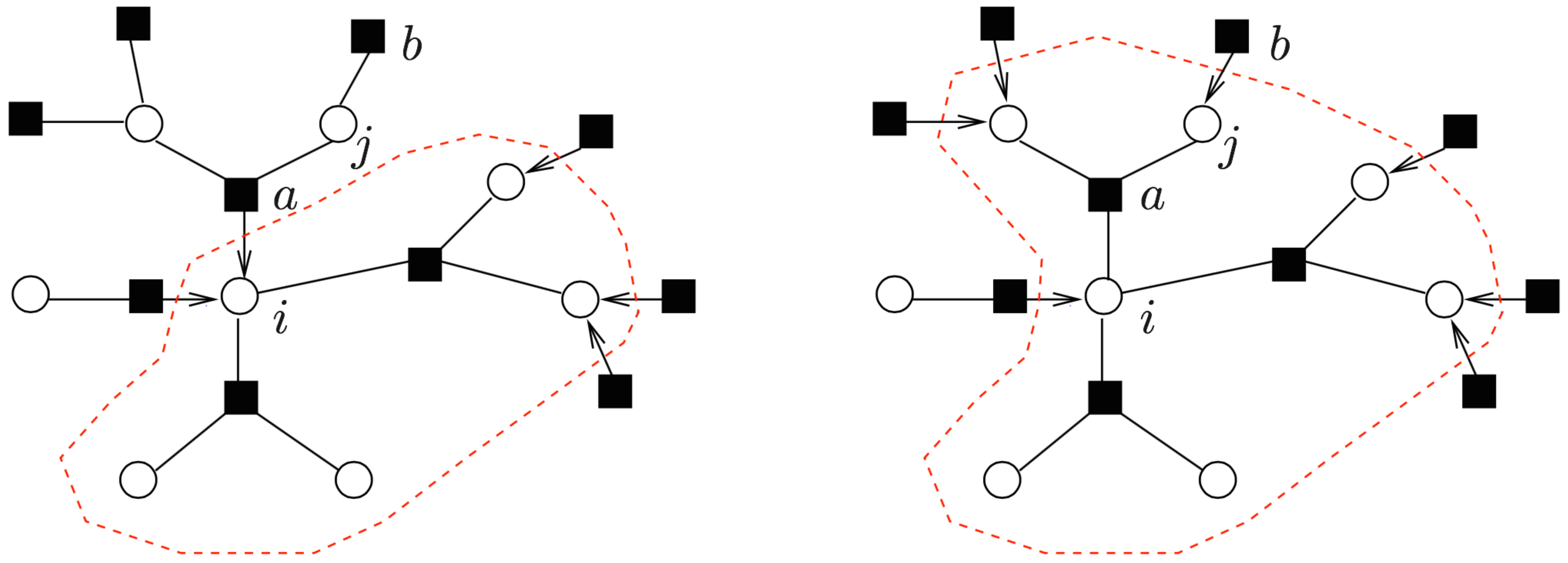}
	\caption[A cavity graph extension]{Figure taken from \cite{mezard2009information}. Two examples of cavity graphs. A cavity graph must include all the variable nodes neighbors to any factor in it. The cavity on the left is extended to include one more factor, and thus the two additional variable nodes $i$ and $j$ as well for it to remain a cavity. The consistency constraints of the Bethe measure for these two cavities imply the belief propagation equations for the messages sent from the boundary of the first cavity.}
	\label{figChIntro:cavity}
\end{figure}

We gave here arguments justifying a posteriori the BP equations thanks to cavity graphs, but is there a way to directly derive the BP equations starting from cavity graphs? The answer is given by a very insightful exercise (exercise 19.1) extracted from \cite{mezard2009information}. Let us assume we have a distribution $P(\bx)$ associated to a factor graph $G=(\mathcal{V},\mathcal{F},\mathcal{E})$ (see sec.~\ref{sec:factorGraphs} for the definition of factor graphs). A {\it cavity} $C=(\mathcal{V}_C,\mathcal{F}_C,\mathcal{E}_C)$ of the graph $G$ is a sub-graph of $G$ such that if any factor $a$ is included in $C$, then all its neigboring variable nodes $\partial a$ are in $C$ as well:
\begin{align}
	a \in \mathcal{F}_C \Rightarrow \partial a \in \mathcal{V}_C
\end{align}
The boundary $\partial C$ of the cavity is the set of edges $\{(a,i)\}_{a\not \in \mathcal{F}_C, i\in \mathcal{V}_C}$ connecting variable nodes inside the cavity to factors outside of it, see Fig.~\ref{figChIntro:cavity}. The probability distribution $P(\bx)$ is a {\it Bethe measure} if there exist a set of cavity messages $\{\hat m_{ai}(x_i)\}$ such that if $P(\bx)$ is restricted to any non extensive cavity $C$, $P(\bx_{C})$ can be expressed (up to a small error that goes to zero in the thermodynamic limit) as a local bulk term and a boundary contribution obtained from the messages:
\begin{align}
	P(\bx_{C}) \approx \frac{1}{z_C}\prod_{a\in \mathcal{F}_C}\psi_a(\bx_a)\prod_{(a,i) \in \partial C} \hat m_{ai}(x_i)
	\label{eq:BetheMeasureCavity}
\end{align}
Now let us assume that we have defined some cavity $C$, for example the left one on Fig.~\ref{figChIntro:cavity} and decide to extend it including the factor $a$ in it, and thus its two other neighboring variable nodes that were not in $C$ as well in order to maintain the cavity definition. We obtain the new cavity $\tilde C =(\mathcal{V}_{\tilde C} =\mathcal{V}_{C} \cup \partial a,\mathcal{F}_{\tilde C} = \mathcal{F}_{C}\cup a,\mathcal{E}_{\tilde C} = \mathcal{E}_{C} \cup \{(a,i)\}_{i \in \partial a})$, the right one on Fig.~\ref{figChIntro:cavity}. As $P(\bx)$ is assumed to be a Bethe measure, the distribution of this new cavity can be expressed as the previous cavity distribution (\ref{eq:BetheMeasureCavity}) times the new factor included and the new boundary contribution. But one must be careful to divide the result by the boundary contribution $\hat m_{ai}(x_i)$ in $P(\bx_{C})$ that is now included in the bulk and thus overcounted:
\begin{align}
	\tilde P(\bx_{\tilde C}) = \tilde P(\bx_{C}, \bx_{a\backslash i}) \approx \frac{1}{z_{\tilde C}} P(\bx_{C}) \frac{\psi_a(\bx_{a\backslash i},x_i)}{\hat m_{ai}(x_i)} \prod_{j\in \partial a \backslash i} \prod_{b\in \partial j \backslash a} \hat m_{bj}(x_j)
\end{align}
For these distributions to be coherent, the marginalization constraint must be verified:
\begin{align}
	P(\bx_{C}) &=\int \tilde P(\bx_{C}, \bx_{a\backslash i}) d\bx_{a\backslash i} \\
	&= \frac{1}{z_{\tilde C}}P(\bx_{C}) \frac{1}{\hat m_{ai}(x_i)}\int d\bx_{a\backslash i}\psi_a(\bx_{a\backslash i},x_i) \prod_{j\in \partial a \backslash i} \prod_{b\in \partial j \backslash a} \hat m_{bj}(x_j)\\
	\Rightarrow \hat m_{ai}(x_i) &=\frac{1}{z_{\tilde C}} \int d\bx_{a\backslash i} \psi_a(\bx_{a\backslash i},x_i) \prod_{j\in \partial a \backslash i} \prod_{b\in \partial j \backslash a} \hat m_{bj}(x_j)\\
	&= \frac{1}{\hat z_{ai}}\int d\bx_{a\backslash i} \psi_a(\bx_{a\backslash i},x_i) \prod_{j\in \partial a \backslash i} m_{ja}(x_j)\\
	m_{ja}(x_j) &= \frac{1}{z_{ja}}\prod_{b\in \partial j \backslash a} \hat m_{bj}(x_j)
\end{align}
where we have been careful to always normalize distributions. We find back the belief propagation equations (\ref{ecChIntro:BP_f2n_mess}), (\ref{ecChIntro:BP_n2f_mess}) at their fixed point, which are thus implied by the assumption that the distribution is a Bethe measure.
\subsection{When does belief propagation work}
\label{sec:BPworks}
Despite that BP is exact only on trees, it can be a very good approximation for any tree-like graph or more generaly as long as the distribution to sample is a Bethe measure as shown in the previous section. When used on such graph, the algorithm is referred as loopy-BP and if the algorithm converges, the BP marginals (\ref{ecChIntro:BP_marg}) are estimates of the true ones under the tree approximation. These loops can a priori induce correlations not taken into account by the BP equations but if they are "long enough", the induced correlations decrease so fast to zero that BP becomes quickly almost exact \cite{krzakalaGibbsStates06,mezard2009information,krzakalaGibbsStates06}. When does this situation occur? Hopefully, such locally tree-like graphs appear naturally in many applications: combinatorial optimization problems, modern coding theory, neural networks, artificial intelligence, etc. The tree-like property of the factor graphs in all these fields is a consequence of a common feature: these are all sparse random graphs, i.e. these are randomly generated graphs with a fixed average connectivity which does not scale with $N$, the number of random variables in the problem. It can be shown \cite{mezard2009information,journals/corr/abs-0806-4112} that sparse random graphs have loops which size typically grows as $O(\log(N))$, thus the correlations in such graphs decay fast enough for BP to converge and accuractly approximate the marginal distributions.

Of course, these generic considerations are not always true. In many graphical models, typically in combinatorial optimization problems, this assumption of small correlations can break down in certain parameters regimes despite the sparsity of the graph and BP cannot sample anymore the marginals as the space of solution splits into an exponential (in $N$) number of disconnected clusters of solutions. This scenario is referred as replica symmetry breaking \cite{krzakalaGibbsStates06,mezard2009information,journals/corr/abs-0806-4112,MezardParisi87b,barbier2013hard} but is out of the scope of this thesis: inference problems always have a solution by definition, and this prevents the replica symmetry breaking phenomenon to occur, at least in the case of the prior matching condition when the true generating model of the signal is know \cite{KabashimaKMSZ14}. All the theoretical analisies in this thesis will assume this condition, see sec.~\ref{sec:AMP} and sec.~\ref{sec:replica}.

There exist different alternatives to include part of the correlations induced by the loops in the graph, not taken into accout by BP in its canonical form. A very general and popular one is refered as generalized belief propagation algorithms \cite{Yedidia:2005:CFA:2263425.2271720}, another technique is the loop corrected belief propagation \cite{2015arXiv150503504Z}. But in problems with a glassy phenomenology where anyway the system is deep in its replica symmetry broken phase where its measure really splits into exponentially many ones, this phenomenon {\it must} be taken into account and more advanced algorithms such as survey propagation should be considered \cite{zecchinaKsatScience,MezardParisiZecchina02,DBLP:journals/corr/cs-CC-0212002,2002cond.mat.12451B,journals/corr/abs-0806-4112,barbier2013hard}.
\subsection{Belief propagation and the Bethe free energy}
\label{sec:BP_BetheF}
We previously assumed that the BP algorithm, that can be thought as a dynamical process on a graph was exact on trees, i.e. its fixed point marginals were the exact ones $m_i(x_i)=\int d\bx_{\backslash i} P(\bx)$. Let us prove that it is actually the case. In the case of a tree factor graph, the probability distribution of the variables can be written exactly as a factorized distribution over the single variable marginals $\{P_i(x_i)\}_i^N$ and the "factor marginals" $\{P_a(\bx_a)\}_a^G$:
\begin{equation}
	P(\bx) = \frac{1}{Z}\prod_a^G \psi_a(\bx_a) = \prod_a^G P_a(\bx_a) \prod_i^N P_i(x_i)^{1-c_i}\label{eqChIntro:treeDistri}
\end{equation}
where $c_i$ denotes the connectivity of the node $i$, the number of factors to which the variable $x_i$ is connected to. The inductive proof can be found in \cite{mezard2009information}. From this we can compute the associated Gibbs free energy using (\ref{eqChIntro:GibbsF}). We skeep the possible dependencies in parameters. The average energy part is found using (\ref{eqChIntro:defEnergyGraph1}), (\ref{eqChIntro:defEnergyGraph2}):
\begin{align}
	\mathbb{E}_P\(\sum_a^G E_a(\bx_a)\) = -\sum_a^G \mathbb{E}_{P_a}\(\log\(\psi(\bx_a)\)\)
\end{align}
Then the entropy part using (\ref{eqChIntro:entropy}) with (\ref{eqChIntro:treeDistri}):
\begin{equation}
	H(P) = -\sum_a^G \mathbb{E}_{P_a}\(\log\(P_a(\bx_a)\)\) - \sum_i^N(1-c_i) \mathbb{E}_{P_i}\(\log\(P_i(x_i)\)\)
\end{equation}
Thus the Gibbs free energy for a tree (which is also its true Helmotz free energy as (\ref{eqChIntro:treeDistri}) is exact for a tree) is given by:
\begin{align}
	F_{Bethe}(\{P_a,\psi_a\}_a^G,\{P_i\}_i^N) = &-\sum_a^G \int d\bx_a P_a(\bx_a)\log\(\frac{\psi_a(\bx_a)}{P_a(\bx_a)}\) \nonumber\\
	&- \sum_i^N(c_i-1) \int dx_iP_i(x_i)\log\(P_i(x_i)\) \label{eq:BetheF_formal}
\end{align}
This free energy is also referred as the Bethe free energy. Let us compute the marginals, used in the parametrization (\ref{eqChIntro:treeDistri}). As discussed in sec.~\ref{sec:variationalMethod}, we thus minimize the Bethe free energy to find their expressions, but we also need to be careful to enforce their normalization and the marginalization conditions to get coherent definitions of probability distributions. These two conditions together imply the normalization of the full distribution, we dont need a additional Lagrange multiplier to enforce it. We thus create a Lagrangian from the Bethe free energy:
\begin{align}
	&\mathcal{L}_{Bethe}(\{P_a,\psi_a\}_a^G,\{P_i\}_i^N) = F_{Bethe}(\{P_a,\psi_a\}_a^G,\{P_i\}_i^N) + \sum_i^N \gamma_i \(\int dx_i P_i(x_i) - 1\) \nonumber \\
	&+ \sum_a^G \sum_{i\in \partial a}\int dx_i \lambda_{ai}(x_i)\(\int d\bx_{a\backslash i} P_a(\bx_a) - P_i(x_i) \)
\end{align}
Now we optimize it with respect to the one point marginal by functionnal derivation:
\begin{align}
	&\frac{\delta \mathcal{L}_{Bethe}}{\delta P_i(x_i^*)} = -(c_i - 1)\[\log\(P_i(x_i^*)\) + 1\] + \gamma_i - \sum_{a\in\partial i}\lambda_{ai}(x_i^*) = 0\\
	&\Rightarrow P_i(x_i^*) = \frac{1}{z_i}\exp\(-\frac{1}{c_i-1}\sum_{a\in\partial i}\lambda_{ai}(x_i^*) \) \label{ecChIntro:P_i_BP_Lag}
\end{align}
where we put all the constants in the normalization. And now the factor marginals:
\begin{align}
	&\frac{\delta \mathcal{L}_{Bethe}}{\delta P_a(\bx_a^*)} = \log\(\frac{P_a(\bx_a^*)}{\psi_a(\bx_a^*)}\) +1 + \sum_{i\in\partial a}\lambda_{ai}(x_i^*) = 0\\
	&\Rightarrow P_a(\bx_a^*) = \frac{\psi_a(\bx_a^*)}{z_a'}\exp\(-\sum_{i\in\partial a}\lambda_{ai}(x_i^*) \) \label{ecChIntro:P_a_BP_Lag}
\end{align}
Now using the following reparametrization in (\ref{ecChIntro:P_i_BP_Lag}) we get:
\begin{align}
	&\lambda_{ai}(x_i^*) = -\sum_{b\in\partial i\backslash a}\log\(\hat m_{bi}(x_i^*)\)\\
	\Rightarrow &\sum_{a\in\partial i}\lambda_{ai}(x_i^*) = -(c_i-1) \sum_{a\in\partial i}\log\(\hat m_{ai}(x_i^*)\)	\\
	\Rightarrow &P_i(x_i) = \frac{1}{z_i} \prod_{a\in \partial i} \hat m_{ai}(x_i) \label{eq:Pi_cav}\\
	&= m_i(x_i)
\end{align}
where we recognized the BP marginal expression (\ref{ecChIntro:BP_marg}). And now applying the same for the factor marginals (\ref{ecChIntro:P_a_BP_Lag}):
\begin{align}
	\Rightarrow &P_a(\bx_a) = \frac{\psi_a(\bx_a)}{z_a'}\prod_{i\in \partial a} \prod_{b\in\partial i\backslash a} \hat m_{bi}(x_i)\\
	&=\frac{\psi_a(\bx_a)}{z_a}\prod_{i\in \partial a} m_{ia}(x_i) \label{eq:marginal_cavityMess} \\
	&= m_a(\bx_a)
\end{align}
where we have used (\ref{ecChIntro:BP_n2f_mess}) and (\ref{ecChIntro:BP_marg_2}) {\it at the fixed point} of BP, i.e. droping the time index. We thus realize that the fixed point marginals $\{P_i\}_i^N$ and $\{P_a\}_a^G$ of the Bethe free energy (\ref{ecChIntro:P_i_BP_Lag}) gives back the BP marginals $\{m_i\}_i^N$ and $\{m_a\}_a^G$. A very nice review on graphical models and the links between belief propagation and the Bethe free energy is \cite{yedidia2003understanding}.

Belief propagation can be generalized to optimize more complex variational free energies, that take into acount more complex probabilistic models. Some classical mean field models include the Kikuchi and junction tree approximations. These message-passing algorithms are refered as generalized belief propagation algorithms \cite{Yedidia:2005:CFA:2263425.2271720}. In this framework, the Bethe free energy is a particular choice of parametrization (\ref{eqChIntro:treeDistri}) for the distribution of the model. But in the present thesis, dense graphical models with linear constraints are studied, and in this case the Bethe free energy is asymptotically exact as we will see in sec.~\ref{sec:AMP}.
\subsection{The Bethe free energy in terms of cavity messages}
The Bethe free energy (\ref{eq:BetheF_formal}) is a formal expression useful to show the equivalence with the BP fixed points as done in the previous section, but is not very practical from the computational point of view. We now derive another equivalent expression of it, expressed in terms of the cavity messages at their fixed point. This will be useful when deriving expectation maximization learning equations for example. We start from the expression (\ref{eq:BetheF_formal}). Using the expression of the marginals as a function of the cavity messages (\ref{eq:Pi_cav}), (\ref{eq:marginal_cavityMess}) the free energy becomes:
\begin{align}
	&F_{Bethe} = -\sum_a^G \int d\bx_a P_a(\bx_a) \log\(\frac{z_a}{\prod_{i \in \partial a} m_{ia}(x_i)}\) \nonumber \\
	&- \sum_i^N (c_i-1)\int dx_i P_i(x_i)\log\(\frac{\prod_{a \in \partial i} \hat m_{ai}(x_i)}{z_i}\) \\
	&=\underbrace{-\sum_a^G \log\(z_a\) -\sum_i^N\log\(z_i\)}_{\defeq \tilde F} + \sum_i^N c_i \log\(z_i\) \nonumber \\
	&+ \sum_{a}^{G} \sum_{i \in \partial a} \int d\bx_a P_a(\bx_a) \log\(m_{ia}(x_i)\) - \sum_i^N(c_i - 1) \int dx_i P_i(x_i) \log\(\prod_{a \in \partial i} \hat m_{ai}(x_i)\)
\end{align}
Now we use the following identity that is a direct consequence of the definition of the cavity messages:
\begin{align}
	\prod_{b \in \partial i} \hat m_{bi}(x_i) = m_{ia}(x_i) \hat m_{ai}(x_i) z_{ia} \ \ \txt{for any \ $a \in \partial i$}
\end{align}
where $z_{ia}$ is the normalization of $m_{ia}(x_i)$. Now using the fact that $\sum_{i}^N c_i f_i = \sum_{i}^{N}\sum_{a \in \partial i} f_i$ for a generic objecf $f_i$ that depends on the variable index (the connectivity can be replaced by an additional sum over the neighbors factor indices) and using the marginalization property of $P_a(\bx_a)$, we deduce:
\begin{align}
	&F_{Bethe} = \tilde F + \sum_{a}^{G}\sum_{i \in \partial a} \int P_i(x_i) \log\(m_{ia}(x_i)\) +\sum_{i}^{N}\sum_{a \in \partial i} \log\(z_i\) \nonumber\\
	&- \sum_{i}^{N}\sum_{a \in \partial i} \[\int dx_i P_i(x_i) \(\log\(m_{ia}(x_i)\) + \log\(\hat m_{ai}(x_i)\) \) + \log\(z_{ia}\) \] \nonumber \\
	&+ \sum_i^N \[\int dx_i P_i(x_i) \(\log\(m_{ia}(x_i)\) + \log\(\hat m_{ai}(x_i)\) \) + \log\(z_{ia}\)\]
\end{align}
Now we use that the sums $\sum_{a}^{G}\sum_{i \in \partial a} f_{ia} = \sum_{i}^{N}\sum_{a \in \partial i} f_{ia}$ are equal ($f_{ia}$ is any function depending on the variables and factors indices) and the identity:
\begin{align}
	P_i(x_i) &= \frac{z_{ia}}{z_i} m_{ia}(x_i) \hat m_{a i}(x_i)\\
	\Rightarrow z_i &= z_{ia} \int dx_i m_{ia}(x_i) \hat m_{ai}(x_i)
\end{align}
We thus obtain:
\begin{align}
	&F_{Bethe} = \tilde F +\sum_{i}^{N}\sum_{a \in \partial i} \[\log\(z_{ia}\)+ \log\( \int dx_i m_{ia}(x_i) \hat m_{ai}(x_i)\) \] \nonumber\\
	&- \sum_{i}^{N}\sum_{a \in \partial i} \[\int dx_i P_i(x_i) \log\(\hat m_{ai}(x_i)\) + \log\(z_{ia}\) \] \nonumber \\
	&+ \sum_i^N \[\int dx_i P_i(x_i) \(\log\(m_{ia}(x_i)\) + \log\(\hat m_{ai}(x_i)\) \) + \log\(z_{ia}\)\]\\
	&=\tilde F + \sum_{i}^{N}\sum_{a \in \partial i} \log\( \int dx_i m_{ia}(x_i) \hat m_{ai}(x_i)\) - \sum_{i}^{N}\sum_{a \in \partial i} \int dx_i P_i(x_i) \log\(\hat m_{ai}(x_i)\) \nonumber\\
	&+ \sum_i^N \[\int dx_i P_i(x_i) \(\log\(\frac{1}{z_{ia}}\prod_{b\in\partial i \backslash a} \hat m_{bi}(x_i)\) + \log\(\hat m_{ai}(x_i)\) \) + \log\(z_{ia}\)\]\\
	&=\tilde F + \sum_{i}^{N}\sum_{a \in \partial i} \log\( \int dx_i m_{ia}(x_i) \hat m_{ai}(x_i)\)
\end{align}
So the final expression of the Bethe free energy in terms of the cavity messages fixed point is:
\begin{align}
	F_{Bethe} &= -\sum_a^G \log\(z_a\) -\sum_i^N \log\(z_i\) + \sum_{i}^{N}\sum_{a \in \partial i} \log\(\tilde z_{ia} \) \label{eq:BetheF_cavMess} \\
	z_a &= \int d\bx_a \psi_a(\bx_a) \prod_{i\in\partial a}m_{ia}(x_i) \label{eq:BetheF_za} \\
	z_i &= \int dx_i \prod_{a\in\partial i} \hat m_{ai}(x_i) \label{eq:BetheF_zi} \\
	\tilde z_{ia} &\defeq \int dx_i m_{ia}(x_i) \hat m_{ai}(x_i) \label{eq:BetheF_zai}
\end{align}
This expression is only true at the fixed points of the messages, but at any time step $t$ of the algorithm, an approximated free energy can be computed plugging the messages at this time in this expression. This formula can be understood in the following way: the total free energy on a tree graphical model is the sum of the contributions of each factors and their associated neighborhood (edges and variables), of the individual variable and their adjacent edges contributions but as each edges has been overcounted, we remove each edge contribution once.
\subsection{Derivation of belief propagation from the Bethe free energy}
\label{sec:BPfromBethe}
This form of the Bethe free energy is more practical than the (\ref{eq:BetheF_formal}) because the belief propagation equations can be trivially derived as fixed point equations for this potential. Let us show it for sake of completeness. Starting from (\ref{eq:BetheF_cavMess}) and performing the functional derivative with respect to the cavity messages, we obtain at the fixed point:
\begin{align}
	\frac{\delta F_{Bethe}}{\delta \hat m_{ai}(x_i^*)} &= -\frac{1}{z_i} \int dx_i \delta(x_i-x_i^*) \prod_{b\in \partial i \backslash a} \hat m_{bi}(x_i) \nonumber\\
	&+ \frac{1}{\tilde z_{ia}} \int dx_i \delta(x_i -x_i^*)m_{ia}(x_i) = 0\\
	\Rightarrow m_{ia}(x_i^*) &= \frac{1}{z_{ia}} \prod_{b\in \partial i \backslash a} \hat m_{bi}(x_i^*)\\
	z_{ia} &= \frac{z_i}{\tilde z_{ia}} 
\end{align}
In the similar way:
\begin{align}
	\frac{\delta F_{Bethe}}{\delta m_{ia}(x_i^*)} &= -\frac{1}{z_a} \int dx_id\bx_{a\backslash i} \psi_a(x_i,\bx_{a\backslash i})\delta(x_i-x_i^*) \prod_{j\in \partial a \backslash i} m_{ja}(x_j)\nonumber \\
	&+ \frac{1}{\tilde z_{ia}} \int dx_i \delta(x_i -x_i^*)\hat m_{ai}(x_i) = 0\\
	\Rightarrow \hat m_{ai}(x_i^*) &= \frac{1}{\hat z_{ai}} \int d\bx_{a\backslash i}\psi_a(x_i^*,\bx_{a\backslash i}) \prod_{j\in \partial a \backslash i} m_{ja}(x_j)\\
	\hat z_{ai} &= \frac{z_a}{\tilde z_{ia}} 
\end{align}
which are exactly the BP equations (\ref{ecChIntro:BP_f2n_mess}), (\ref{ecChIntro:BP_n2f_mess}) at their fixed point.
\section{The approximate message-passing algorithm}
\label{sec:AMP}
If the only available information about the signal is the matrix $\bF$
and the vector of measurements $\by$ in (\ref{eqIntro:AWGNCS}), then the
information-theoretically best possible estimate of each signal
component is computed as a weighted average over all solutions of the
linear system, where the weight of each solution is given by the prior. Of course, the undetermined linear system (\ref{eqIntro:AWGNCS}) has exponentially many (in $N$) solutions and hence computing exactly the above weighted average is in general intractable.
The corresponding expectation to perform inference can be, however, approximated
efficiently via the approximate message-passing algorithm
\cite{Rangan10b,DonohoMaleki10,KrzakalaPRX2012,KrzakalaMezard12} that we will present now. But before, let us expose why belief propagation is not the right tool to use in the present context.

\begin{figure}[th!]
\centering
\includegraphics[width=10cm]{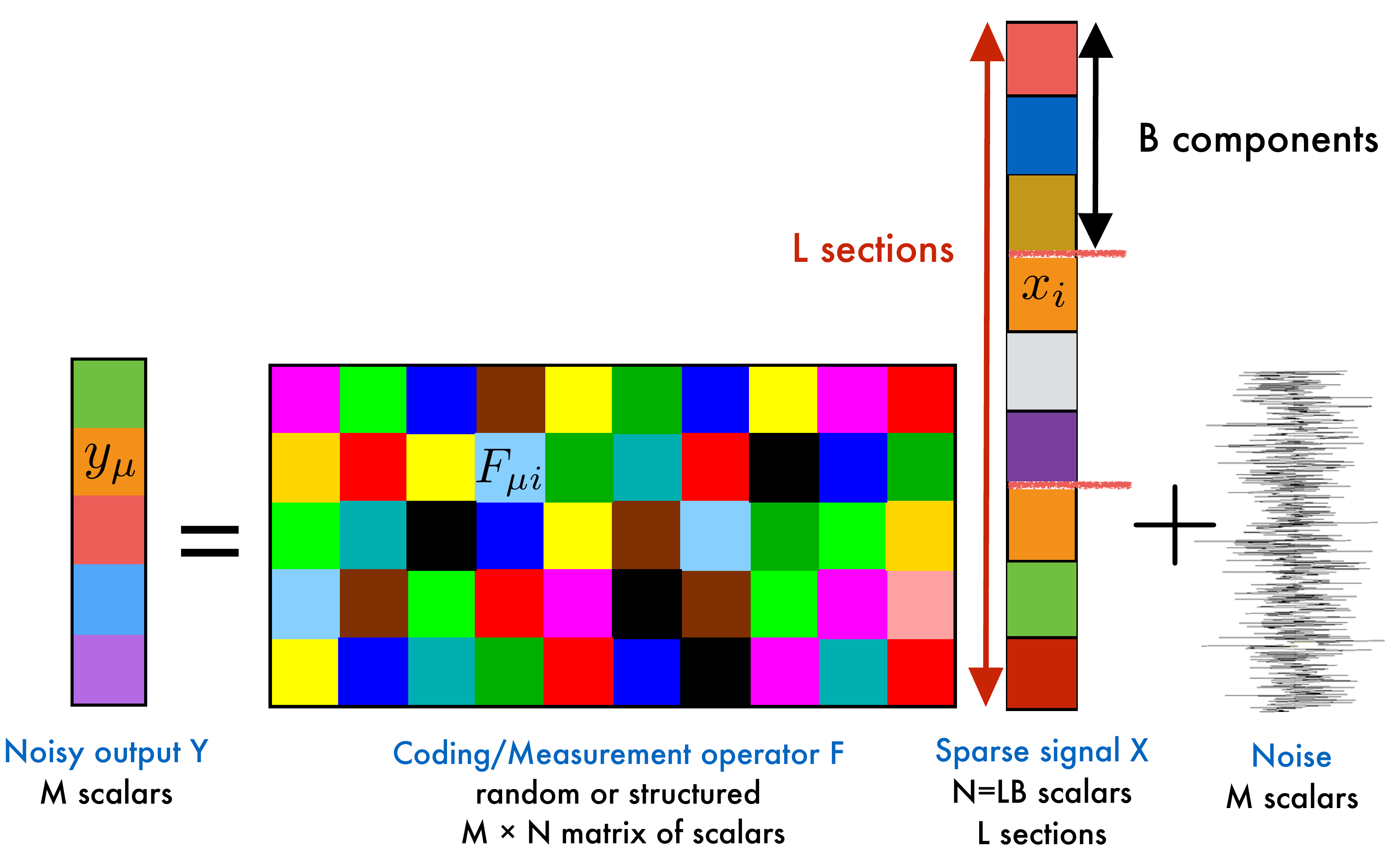}
\includegraphics[width=10cm]{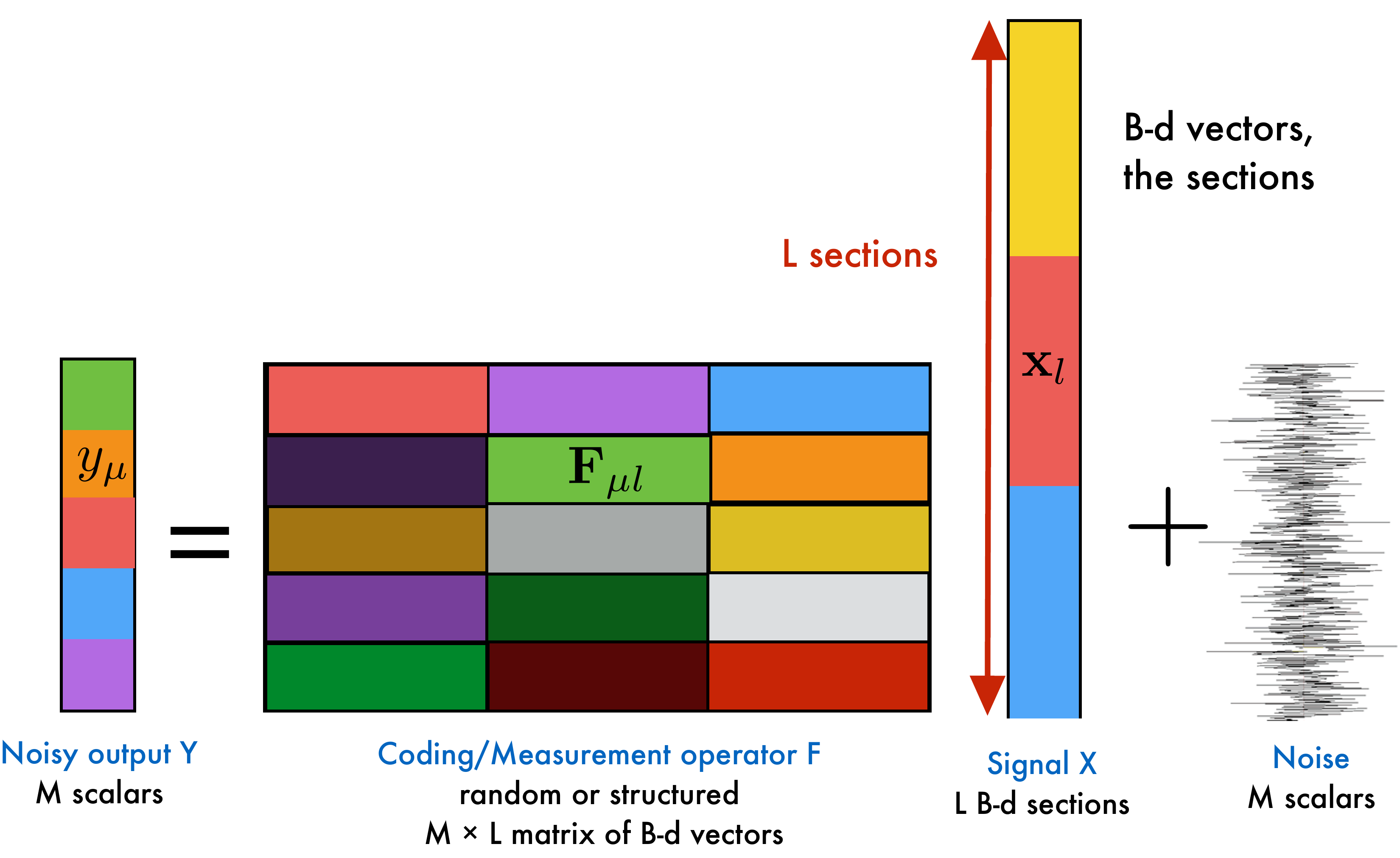}
\caption[Equivalence between the compressed sensing of prior-correlated scalar signals and i.i.d vectorial components signals]{\textbf{Up} : Representation of the linear estimation
problem over the i.i.d AWGN channel in terms of a signal and matrix with scalar components. The prior on this signal is factorizable over non overlaping groups of $B$ components, denoted as the sections. \textbf{Down} : Reinterpreting the same
problem in terms of $B$-d variables. Now, the matrix elements are concatenated to form $B$-d vectors that are applied
(using the usual scalar product for vectors) on the associated $B$-d
vectors representing the new components of the signal, the sections. In this new
setting, all the signal sections are uncorrelated by the prior.}
\label{figChIntro:1dOp}
\end{figure}

In order to be as general as possible, we now consider that the components of the signal we want to infer are $B$-dimensional vectors $\bx_l = [x_i]_{i\in l}$ where $l$ denotes both the vector variable index and the set of indices $\{i\in l\}$ of the scalar components of $\bx$ concatenated to form the new vector variable $\bx_l$. These new variables are called {\it sections}. We define $\bF_{\mu l}\defeq [F_{\mu i}]_{i \in l}$ as a vector of elements of the matrix $\bF$ that act on the section $\bx_l$ (see Fig.~\ref{figChIntro:1dOp}). Working with vectors is useful as we will work in this setting for the sparse superposition codes sec.~\ref{sec:superCodes} and the scalar equations can be recovered taking $B=1$ in the final equations. All the previous derivations (the BP algorithm, the Bethe free energy, etc.) would have been the same with vector variables so the obtained results remain valid. We will denote by $L$ the number of sections $\{\bx_l\}_l^L$, in order to keep the notation $N=LB$ for the number of $1$-d components of the signal. We take a generic factorizable prior over the sections. Fig.~\ref{figChIntro:1dOp} shows how a linear estimation problem with a scalar components signal which prior constrain groups of $B$ non overlaping components can be interpreted as an equivalent problem where now the components are $B$-d sections. The scalar matrix elements are concatenated in $B$-d vectors as well, and are applied to the vectorial signal components using the usual scalar product between vectors. This construction changes nothing to the scalar measurements, nor to the fact that the noise is i.i.d applied on the $1$-d components of the signal.
\subsection{Why is the canonical belief propagation not an option for dense linear systems over reals?}
Belief propagation is a very powerful inference algorithm but it has caveats. We will write the BP equations for the linear estimation problem and understand why the equations are intractable. 

The Hamiltonian we consider is thus a direct extension of (\ref{eqChIntro:hamiltonianLinearEstim}) to the vectorial case and we consider that the measurement matrix is full:
\begin{align}
	E(\bx) =-\sum_l^L \log\(P_0^l(\bx_l)\) + \frac{1}{2\Delta} \sum_\mu^M \(y_\mu-(\bF\bx)_\mu\)^2
\end{align}
A remark is that when there are factors $\{\phi(x_i)\}$ depending on single variables, the node-to-factor BP message (\ref{ecChIntro:BP_n2f_mess}) and the marginal (\ref{ecChIntro:BP_marg}) can be rewritten as:
\begin{align}
	m_{ia}^{t+1}(x_i) &= \frac{1}{z_{ia}^{t+1}} \phi(x_i)\prod_{b\in\partial i \backslash a} \hat m_{b i}^{t}(x_i)\label{ecChIntro:BP_n2f_mess_2}\\
	m_i^{t}(x_i) &= \frac{1}{z_{i}^t}\phi(x_i)\prod_{a\in\partial i} \hat m_{ai}^t(x_i)\\
	&= \frac{1}{z_{ia}'^t}\phi(x_i)\hat m_{ai}^t(x_i)m_{ia}^{t+1}(x_i) \ \ \txt{for any $a \in \partial i$}
\end{align}
which is a perfectly equivalent form as these single variable factors can be integrated in the set of previous factors $\{\psi_a\}$ considering that they do not receive any node-to-factor messages (i.e. the product of messages in (\ref{ecChIntro:BP_f2n_mess}) is $1$ and there is no marginalization to perform as they are only connected to one variable). This form will be more practical to use. Now using that the constraints are given by the likelihood of the observations (\ref{eqIntro:likelihood}) and that the prior is factorizable over the signal sections, we get the BP equations for vectorial linear estimation:
\begin{align}
	\hat m_{\mu l}^{t}(\bx_l) &= \frac{1}{\hat z_{\mu l}^t} \int \bigg[\prod_{k \neq l}^{L-1} d \bx_k m_{k \mu}^{t}(\bx_k)\bigg] e^{-\frac{{\rm{snr}}}{2} \left(\sum_{k\neq l}^{L-1} \bF_{\mu k}^{\intercal}  \bx_k + \bF_{\mu l}^{\intercal}  \bx_l - y_\mu \right)^2 } \label{eqChIntro:bp1} \\
	m_{l\mu}^{t+1}( \bx_l) &= \frac{1}{z_{l\mu}^{t+1}} P_0^l( \bx_l) \prod_{\gamma \neq \mu}^{M-1} \hat m_{\gamma l}^{t}( \bx_l) \label{eq1:bp2}\\
	m_{l}^{t}( \bx_l) &= \frac{1}{z_{l}^{t}} P_0^l( \bx_l) \prod_{\gamma}^{M} \hat m_{\gamma l}^{t}( \bx_l)\label{eq1:bp3}
\end{align}
Working with the noise variance or the ${\rm snr} = 1/\Delta$ is the same as we fix the power to be 1. What are the problems with these equations? There are essentially three:

$\bullet$ The factor graph associated to the linear estimation problem problem Fig.~\ref{figChIntro:factorGraphCS} (where the $1$-d $x_i$ variables are replace by the $B$-d $\bx_l$ ones) is densely connected. The likelihood constraints enforce all the variable nodes to be connected to all the likelihood factor nodes when the measurement matrix is full. This implies that the number of messages to store in the memory and exchange at each time step is $2ML \in O(L^2)$ (2 per edges) which is way too many and scales badly with the problem size.

$\bullet$ Furthermore, these messages are probability distributions over real variables. Such objects are really difficult to store on a computer as in general they have no analytical form that could be decomposed as simple functions. It would require to store a discretized version of each message, a histogram which discretization step is small enough to have an high numerical accuracy. This is impossible as the number of message is large and anyway, it would lower greatly the efficiency of the algorithm.

$\bullet$ As we are working in the continuous framework $\bx \in \mathbb{R}^{BL}$, $\bF \in \mathbb{R}^{M,BL}$ the computations of the factor-to-node cavity messages (\ref{eqChIntro:bp1}) require very high dimensional integrals to be performed ($(L-1)B$ integrals in the full matrix case) which are non analytic and thus would have to be computed numerically.

We understand that BP in this form is not an option. BP is useful when the factor graph is sparse and when the variables have few discrete states. In this case the integrals become sums over the states of few neighbors which is tractable. Furthermore, there are not too many messages to store due to the graph sparsity (this number scales as the number of variables) and each message can be easily stored as a small vector giving the probability of each discrete state.

Belief propagation based reconstruction algorithms were introduced in
compressed sensing by \cite{BaronSarvotham10}. The authors used sparse measurement matrices to reduce the number of messages and make the graph locally tree-like and then treated the BP messages as probabilities over
real numbers, that were represented by a histogram, one of the three major problems of BP discussed before, and that will face the approximate message-passing algorithm.
\subsection{Why message-passing works on dense graphs?}
\label{sec:whyMessagePassingWorksOnDenseGraphs}
Let us assume that we have an infinitely powerful computer with infinite memory, such that all the previous problems are not of concern anymore. Is the BP algorithm a good inference algorithm anyway for such linear estimation problems, where the factor graph is dense? We explained in sec.~\ref{sec:BP_BetheF} that BP finds the fixed point marginals of the Bethe free energy which is exact for trees, and suggests that BP is an accurate approximation in the case of sparse graphs because of their locally tree-like structure, see sec.~\ref{sec:BPworks}. But here it is not the case at all. It is even the opposite extreme case: the graph is {\it full} of loops. But actually, such very dense graphs share with tree-like ones the important common feature that makes BP the algorithm of choice for inference: the correlations between variables are very weak. Let us detail a bit more this notion.

These systems (tree-like and dense graphical models) are equivalent to infinite dimensional systems. This can be seen from the following fact: starting from any node in the system and moving with no coming back (without passing two times by the same edge), it is impossible to return at the starting point. It is trivial for a tree as it is the very definition of what a tree graph is and thus it becomes true with high probability for infinitely large tree-like graphs. On densely connected graphs, it becomes true with high probability as well as the graph size increases, because the number of paths becomes so large that taking one that luckily comes back to the initial point becomes infinitesimaly probable, which is not the case on a $2$-d graph like a grid for example.

This is why message-passing works on such graphs: the independence between neighboring variables assumed for computing the cavity messages is asymptotically valid, as the only possible paths that could correlate these variables in the cavity graphs have lenghts that diverge in the random sparse graphs case, and the variables are anyway almost independent in the densely connected case as each variable is connected to all other ones making the influence of each single one asymptotically null.
\subsection{Derivation of the approximate message-passing algorithm from belief propagation}
\label{sec:classicalDerivationAMP}
In order to get an algorithm capable of dealing with continuous variables and dense graphs with linear constraints, the approximate message-passing algorithm, we will start from BP and then perform two principal steps: $i)$ In order to face the problem of storing distributions over real variables, we will parametrize the cavity messages thanks to their first and second moments, i.e. project them on Gaussians. This step is exact in the large signal limit $L\to\infty$ as we will see, and is validated by "law of large numbers like" arguments as the number of incoming message on each factor is very large. $ii)$ We will expand the cavity quantities that depend on factors and variables indices around marginal quantities that depend only on the variables indices. The correction around these, the so called {\it Onsager reaction term} in statistical physics will be essential for the algorithm performance, and makes all the difference with the naive mean field approximation, see sec.~\ref{sec:naiveMFforCS}. The resulting algorithm, referred as the Thouless-Palmer-Anderson TAP equations in statistical physics, derived to deal with spin glasses \cite{ThoulessAnderson77,Opper_and_Saad:2001}, will be obtained for linear estimation: it is the approximate message-passing algorithm. Before the apparition of AMP, message-passing algorithms for dense graphs were already studied \cite{DBLP:journals/corr/abs-cs-0503070} but the strategy adopted here is different.
\subsubsection{Gaussian parametrization}
The cavity messages in BP (\ref{eqChIntro:bp1}), (\ref{eq1:bp2}) can be represented only by their mean and variance as done by \cite{GuoWang06,Rangan10}. Let us see how to derive iterative equations on these moments. 
For the rest of this derivation, we skeep the time index for sake of readibility, these will be added back at the end and justified in sec.~\ref{sec:alternativeAMPderivation}.

In order to fix the power to $1$ in (\ref{eqIntro:AWGNCS}), we use the following scaling for the matrix elements: $F_{\mu i} \in O(1/\sqrt{L}) \ll 1$, as we assume in the derivation that $L\gg 1$. Using this scaling, we will Taylor expand in the matrix elements the exponential appearing in the factor-to-node cavity message (\ref{eqChIntro:bp1}). But before that, after developing the square in this exponential, we need to apply the Hubbard-Stratanovitch transform to $w(\bx_{\backslash l})\defeq \sum_{k\neq l}^{L-1} \bF_{\mu k}^{\intercal}\bx_k$ to simplify the resulting expression, the aim being to linearize all the $\bx_l$ independent terms in the exponential so that the integrals become independent:
\begin{align}
e^{-\frac{w^2{\rm{snr}}}{2}} &= \sqrt{\frac{{\rm{snr}}}{2\pi}} \int d\lambda e^{-\frac{\lambda^2{\rm{snr}}}{2} + i{\rm{snr}}\lambda w}\\
\Rightarrow \hat m_{\mu l}(\bx_l) &= \frac{\sqrt{{\rm{snr}}}}{\sqrt{2\pi}\hat z_{\mu l}} e^{-\frac{{\rm{snr}}}{2}\left(\bF_{\mu l}^{\intercal} \bx_l - y_\mu\right)^2}\nonumber \\
&\int d\lambda e^{-\frac{\lambda^2{\rm{snr}}}{2}} \prod_{k\neq l}^{L-1} \underbrace{\left[\int d \bx_k m_{k\mu}( \bx_k) e^{{\rm{snr}}\bF_{\mu k}^{\intercal}  \bx_k\left(y_\mu - \bF_{\mu l}^{\intercal}  \bx_l + i\lambda\right)}\right]}_{\defeq u_k} \label{eq1:bp3_0}
\end{align}
To define the approximate messages Gaussian parametrization, we need the following vectorial objects:
\begin{align}
&\ba_{\Box} \defeq \int  \bx\ \! m_{\Box}(\bx) \ \! d\bx \label{eq:generic_a}\\
&\bv_{\Box} \defeq \int \bx^2 m_{\Box}(\bx) \ \! d\bx - \ba_{\Box}^2 \label{eq:generic_v}
\end{align}
where the square $\bullet^2$ is an elementwise operation as the inverse operation $\bullet^{-1}$ used later on. Expanding in $F_{\mu k}$ the $u_k$ appearing in (\ref{eq1:bp3_0}) and using the two previous definitions, the integral $u_k$ can be written as:
\begin{align}
u_k &\approx \int d \bx_k m_{k\mu}( \bx_k) \(1 + {\rm{snr}}\bF_{\mu k}^{\intercal}  \bx_k\left(y_\mu - \bF_{\mu l}^{\intercal}  \bx_l + i\lambda\right) + \frac{1}{2}\left[{\rm{snr}}\bF_{\mu k}^{\intercal}  \bx_k\left(y_\mu - \bF_{\mu l}^{\intercal}  \bx_l + i\lambda\right)\right]^2\)\nonumber \\
&=\(1 + {\rm{snr}}\bF_{\mu k}^{\intercal}  \ba_{k\mu}\left(y_\mu - \bF_{\mu l}^{\intercal}  \bx_l + i\lambda\right) + \frac{1}{2}(\bv_{k\mu} + \ba_{k\mu}^2)\left[{\rm{snr}}\bF_{\mu k}^{\intercal} \left(y_\mu - \bF_{\mu l}^{\intercal}  \bx_l + i\lambda\right)\right]^2\)\nonumber\\
&\approx \(1 + {\rm{snr}}\bF_{\mu k}^{\intercal}  \ba_{k\mu}\left(y_\mu - \bF_{\mu l}^{\intercal}  \bx_l + i\lambda\right) + \frac{\ba_{k\mu}^2}{2}\left[{\rm{snr}}\bF_{\mu k}^{\intercal} \left(y_\mu - \bF_{\mu l}^{\intercal}  \bx_l + i\lambda\right)\right]^2\)\nonumber \\
&\ \ \ \ \(1 + \frac{\bv_{k\mu}}{2} \left[{\rm{snr}}\bF_{\mu k}^{\intercal} \left(y_\mu - \bF_{\mu l}^{\intercal}  \bx_l + i\lambda\right)\right]^2\)\nonumber\\
&\approx e^{\ba_{k\mu}^{\intercal} \bF_{\mu k} {\rm{snr}} \left(y_\mu - \bF_{\mu l}^{\intercal}  \bx_l + i\lambda\right) + \frac{ {\rm{snr}}^2}{2}\bv_{k\mu}^{\intercal} \bF_{\mu k}^2 \left(y_\mu - \bF_{\mu l}^{\intercal}  \bx_l + i\lambda\right)^2} \label{eq:uExp}
\end{align}
where we kept only the terms up to $O(1/L)$. This allows us to write the cavity factor-to-node message as:
\begin{align}
\Rightarrow &\hat m_{\mu l}( \bx_l) = \frac{\sqrt{{\rm{snr}}}}{\sqrt{2\pi}\hat z_{\mu l}} e^{-\frac{{\rm{snr}}}{2}\left(\bF_{\mu l}^{\intercal} \bx_l - y_\mu\right)^2}\nonumber\\ 
&\int d\lambda e^{-\frac{{\rm{snr}}\lambda^2}{2}}  \prod_{k\neq l}^{L-1} \left[e^{\ba_{k\mu}^{\intercal} \bF_{\mu k} {\rm{snr}} \left(y_\mu - \bF_{\mu l}^{\intercal}  \bx_l + i\lambda\right) + \frac{{\rm{snr}}^2}{2}\bv_{k\mu}^{\intercal} \bF_{\mu k}^2 \left(y_\mu - \bF_{\mu l}^{\intercal} \bx_l + i\lambda\right)^2}\right]
\end{align}
The Gaussian integral over $\lambda$ can now be performed easily, and putting all the $\bx_l$ independent terms in the normalization constant $\hat z_{\mu l}$ we obtain:
\begin{align}
\hat m_{\mu l}( \bx_l) &= \frac{1}{\hat z_{\mu l}} e^{-\frac{1}{2} \bA_{\mu l}^{\intercal}  \bx_l^2  +\bB_{\mu l}^{\intercal}  \bx_l} \label{eq1:hat_mul_cav_notZ} \\ 
\hat z_{\mu l}&=\prod_{i\in l}^{B} \sqrt{\frac{2\pi}{A_{\mu i}}} e^{\frac{B_{\mu i}^2}{2A_{\mu i}}} \label{eq1:hat_mul_cav}\\
\bA_{\mu l} &\defeq \frac{\bF_{\mu l}^2}{1/{\rm{snr}} + \sum_{k\neq l}^{L-1}\bv_{k \mu}^{\intercal} \bF_{\mu k}^2} \label{eq1:bA} \\
\bB_{\mu l} &\defeq \frac{\bF_{\mu l}\(y_\mu - \sum_{k\neq l}^{L-1} \bF_{\mu k}^{\intercal} \ba_{k \mu} \) }{1/{\rm{snr}} + \sum_{k\neq l}^{L-1} \bv_{k \mu}^{\intercal} \bF_{\mu k}^2} \label{eq1:bB} 
\end{align}
We deduce the node-to-factor cavity message expression from (\ref{ecChIntro:BP_n2f_mess_2}):
\begin{align}
m_{l\mu}( \bx_l) &= \frac{1}{z_{l\mu}} P_0^l( \bx_l) e^{-\frac{1}{2} ( \bx_l^2)^{\intercal} \sum_{\gamma \neq \mu}^{M-1}\bA_{\gamma l}  +  \bx_l^{\intercal} \sum_{\gamma \neq \mu}^{M-1}\bB_{\gamma l} } \label{eq1:hat_lmu_cav_notZ}\\ 
z_{l\mu} &= \int d \bx_l P_0^l( \bx_l) e^{-\frac{1}{2} ( \bx_l^2)^{\intercal} \sum_{\gamma \neq \mu}^{M-1}\bA_{\gamma l}  +  \bx_l^{\intercal} \sum_{\gamma \neq \mu}^{M-1}\bB_{\gamma l} } \label{eq1:hat_lmu_cav_Z}
\end{align}
where sums of the form $\sum_{\gamma}\boldsymbol{\Gamma}_{\gamma l} \defeq \[\sum_{\gamma }\Gamma_{\gamma i}\]_{i\in l}$ are vectors of size $B$. We now have projected the set of cavity messages onto Gaussian distributions, fully parametrized by their first and second moments.

We define $l_i$ as the $B$-d section index (or the set of indices, depending on the context) to which the $i^{th}$ $1$-d signal component belongs to. We can now define a probability measure over the section $l$: $m_B((\bsy \Sigma_{l})^2,\bR_{l},\bx_l)$ and the corresponding $1$-d components marginals, the marginals of the $1$-d variables in the section: $\{m_i((\bsy \Sigma_{l_i})^2,\bR_{l_i}, x_i)\}_{i\in l}$. We also define the associated vector of averages $\bff_{a_l}$ and variances $\bff_{c_l}$ over these marginals:
\begin{align}
&m_B((\bsy \Sigma_{l})^2,\bR_{l},\bx_l) \defeq \frac{1}{z((\bsy \Sigma_{l})^2,\bR_{l})} P_0^l( \bx_{l}) e^{-([ \bx_{l}-\bR_{l}]^2)^{\intercal}(2\bsy \Sigma_{l}^2)^{-1}}\label{eq1:marginal_B} \\
&m_i((\bsy \Sigma_{l_i})^2,\bR_{l_i}, x_i) \defeq \int d \bx_{l_i\backslash i} m_B((\bsy \Sigma_{l_i})^2,\bR_{l_i},\bx_{l_i} ) \label{eq1:marginal}\\
&z((\bsy \Sigma_{l})^2,\bR_{l}) = z((\bsy \Sigma_{l_i})^2,\bR_{l_i}) = \int d \bx_l P_0^l( \bx_{l}) e^{-([ \bx_{l}-\bR_{l}]^2)^{\intercal}(2\bsy \Sigma_{l}^2)^{-1}}  \label{eq1:AMPzi}\\
&f_{a_i}((\bsy \Sigma_{l_i})^2,\bR_{l_i}) \defeq \int d x_i m_i((\bsy \Sigma_{l_i})^2,\bR_{l_i}, x_i)  x_i \label{eq1:fai}\\
&f_{c_i}((\bsy \Sigma_{l_i})^2,\bR_{l_i}) \defeq \int d x_i m_i((\bsy \Sigma_{l_i})^2,\bR_{l_i}, x_i)  x_i^2 - f_{a_i}((\bsy \Sigma_{l_i})^2,\bR_{l_i})^2 \label{eq1:fci}\\
&\bff_{a_l}((\bsy \Sigma_{l})^2,\bR_{l}) \defeq \left[f_{a_i}((\bsy \Sigma_{l_i})^2,\bR_{l_i}) \right]_{i\in l}\label{eq1:fa}\\
&\bff_{c_l}((\bsy \Sigma_{l})^2,\bR_{l}) \defeq \left[f_{c_i}((\bsy \Sigma_{l_i})^2,\bR_{l_i}) \right]_{i\in l}\label{eq1:fc}
\end{align}
Using (\ref{eq1:hat_lmu_cav_notZ}) together with these definitions and (\ref{eq:generic_a}), (\ref{eq:generic_v}) we get the second order BP iterations:
\begin{align}
\ba_{l\mu} &= \bff_{a_l}\left(\frac{1}{\sum_{\gamma\neq\mu}^{M-1}\bA_{\gamma {l}}}, \frac{\sum_{\gamma\neq\mu}^{M-1} \bB_{\gamma {l}}}{\sum_{\gamma\neq\mu}^{M-1} \bA_{\gamma {l}}}\right) \label{eq1:relaxedCav}\\ 
\bv_{l\mu} &= \bff_{c_l}\left(\frac{1}{\sum_{\gamma\neq\mu}^{M-1} \bA_{\gamma {l}}}, \frac{\sum_{\gamma\neq\mu}^{M-1} \bB_{\gamma {l}}}{\sum_{\gamma\neq\mu}^{M-1} \bA_{\gamma {l}}}\right)\label{eq1:relaxedCav_v}\\
\ba_l &= \bff_{a_l}\left(\frac{1}{\sum_{\mu}^{M}\bA_{\mu {l}}}, \frac{\sum_\mu^{M} \bB_{\mu {l}}}{\sum_\mu^{M} \bA_{\mu {l}}}\right)\label{eq1:relaxedMarg} \\ 
\bv_l &= \bff_{c_l}\left(\frac{1}{\sum_{\mu}^{M} \bA_{\mu {l}}}, \frac{\sum_\mu^{M} \bB_{\mu {l}}}{\sum_\mu^{M} \bA_{\mu {l}}}\right)
\end{align}
where the two last equations are the marginal mean (\ref{ecChIntro:BP_marg}) and associated variance, that takes into account all factors. At this stage, after indexing with the time, the algorithm defined by the set of equations (\ref{eq1:bA}), (\ref{eq1:bB}) and (\ref{eq1:relaxedCav}), (\ref{eq1:relaxedCav_v}) together with the definitions (\ref{eq1:marginal}), (\ref{eq1:fa}) and (\ref{eq1:fc}) is usually referred as relaxed-BP \cite{GuoWang06,rangan2010estimation,KrzakalaPRX2012,KrzakalaMezard12}, which is exact for linear estimation as the number of sections $L\to \infty$. After convergence, the final estimates are obtained through (\ref{eq1:relaxedMarg}). This first step thus solves the problem of storing the messages, as now each message is parametrized by just two numbers, its mean $\ba_{l\mu}$ and variance $\bv_{l\mu}$.
\subsubsection{Reduction of the number of messages: the TAP equations}
We can simplify further the equations, going from an algorithm where $2ML$ messages are exchanged at each time step to one with only $M+L$ messages per time step \cite{DonohoMaleki09}. The following expansion, the Thouless-Anderson-Palmer approximation in statistical physics of spin glasses \cite{ThoulessAnderson77} is exact in the large signal size limit, as the previous Gaussian parametrization. It starts by noticing that in the $L\to \infty$ limit (and thus the number M of factors diverges as well such that the measurement rate is constant), the cavity quantities (\ref{eq1:relaxedCav}), (\ref{eq1:relaxedCav_v}), (\ref{eq1:bA}) and (\ref{eq1:bB}) become almost independent of the $\mu$ index (which is equivalent to say that each factor's influence becomes infinitely weak as there are so many). We can thus re-write these objects as marginal quantities (that depend on single variable indices) keeping the proper first order correction in $F_{\mu i}$, the Onsager reaction term, essential for the efficiency of the AMP algorithm. 

We first define new useful quantities (again, all the operations such as $1/\bullet$ or the dot product $\tbf u \tbf v$ applied to vectors are elementwise, as opposed to $\tbf u^{\intercal} \tbf v$ which is the usual scalar product between vectors):
\begin{align}
w_\mu &\defeq \sum_{k}^L \bF_{\mu k}^{\intercal} \ba_{k\mu}  \label{eqChIntro:w}\\ 
\Theta_\mu &\defeq \sum_{k}^L (\bF_{\mu k}^2)^{\intercal} \bv_{k\mu} \label{eqChIntro:Theta}\\
(\bsy \Sigma_k)^2 &\defeq \frac{1}{\sum_{\mu}^M \bA_{\mu k}}\label{eqChIntro:defS2cav} \\ 
\bR_k &\defeq \frac{\sum_{\mu}^M \bB_{\mu k}}{\sum_{\mu}^M \bA_{\mu k}} \label{eqChIntro:defRcav} \\
(\bsy \Sigma_{k\mu})^2 &\defeq \frac{1}{\sum_{\gamma\not =\mu}^{M-1} \bA_{\gamma k}} \label{eqChIntro:defS2cav_cav}\\ 
\bR_{k\mu} &\defeq \frac{\sum_{\gamma\not =\mu}^{M-1} \bB_{\gamma k}}{\sum_{\gamma\not =\mu}^{M-1} \bA_{\gamma k}} \label{eqChIntro:defRcav_cav}
\end{align}
We always remain in the limit $B\in O(1)$ for the derivation, so if $B$ terms $\in O(1/\sqrt{L})$ are summed, the result is still $\in O(1/\sqrt{L})$.
We now expand the cavity quantities (\ref{eq1:relaxedCav}), (\ref{eq1:relaxedCav_v}) of the relaxed-BP algorithm considering the $\mu$'s factor influence weak. Let us start by the cavity averages:
\begin{align}
\ba_{l\mu} &= \bff_{a_l}((\bsy \Sigma_{l\mu})^2,\bR_{l\mu})\\
&\approx \bff_{a_l}((\bsy \Sigma_{l})^2,\bR_{l}) + ((\bsy \Sigma_{l\mu})^2 - (\bsy \Sigma_{l})^2) \nabla_{(\bsy \Sigma_{l})^2} \bff_{a_l}((\bsy \Sigma_{l})^2,\bR_{l}) \nonumber \\
&~ ~ ~ ~ + (\bR_{l\mu} - \bR_{l}) \nabla_{\bR_{l}} \bff_{a_l} ((\bsy \Sigma_{l})^2,\bR_{l})  \\
&=\ba_l + \left[\frac{\bA_{\mu l}}{(\sum_\gamma^M \bA_{\gamma l})(\sum_{\gamma}^M \bA_{\gamma l} - \bA_{\mu l})}\right] \nabla_{(\bsy \Sigma_{l})^2} \bff_{a_l}((\bsy \Sigma_{l})^2,\bR_{l}) \nonumber \\
&~ ~ ~ ~ + \left[\frac{\bB_{\mu l}(\sum_{\gamma}^M \bA_{\gamma l}) - \bA_{\mu l}(\sum_{\gamma}^M \bB_{\gamma l})}{(\sum_{\gamma}^M \bA_{\gamma l})(\sum_{\gamma}^M \bA_{\gamma l} - \bA_{\mu l})}\right] \nabla_{\bR_{l}} \bff_{a_l}((\bsy \Sigma_{l})^2,\bR_{l})+ o(1/\sqrt{L})
\end{align}
where we have used (\ref{eq1:relaxedMarg}), (\ref{eqChIntro:defS2cav}), (\ref{eqChIntro:defRcav}), (\ref{eqChIntro:defS2cav_cav}) and (\ref{eqChIntro:defRcav_cav}). The gradient operator outputs a vector, for example: $\nabla_{\bR_{l}} \bff_{a_l}((\bsy \Sigma_{l})^2,\bR_{l})\defeq \[\partial_{R_{i}} \bff_{a_l}((\bsy \Sigma_{l})^2,\bR_{l})\]_{i\in l}$. Now we use the fact that the $\bA_{\gamma l} \in O(1/L)$ is a strictly positive term and $\bB_{\gamma l} \in O(1/\sqrt{L})$ can be of both signs thus
$\sum_\gamma \bA_{\gamma l}$ and $\sum_\gamma \bB_{\gamma l}$ are both $\in O(1)$. Furthermore, using $(\bsy \Sigma_{l\mu})^2 = (\bsy \Sigma_{l})^2 + O(1/L)$ we obtain the first order corrections to $\ba_l$ and $\bv_l$ (following the same computation):
\begin{align}
&\ba_{l\mu} = \ba_l - \underbrace{(\bsy \Sigma_{l})^2\bB_{\mu l} \nabla_{\bR_{l}} \bff_{a_l}((\bsy \Sigma_{l})^2,\bR_{l})}_{\defeq \bsy \epsilon_{\ba_{l \mu}} \in O(1/\sqrt{L})}+ o(1/\sqrt{L}) \label{eq1:appCorrections_a}\\ 
&\bv_{l\mu} = \bv_l - \underbrace{(\bsy \Sigma_{l})^2\bB_{\mu l} \nabla_{\bR_{l}} \bff_{c_l}((\bsy \Sigma_{l})^2,\bR_{l})}_{\defeq \bsy \epsilon_{\bv_{l \mu}} \in O(1/\sqrt{L})}+ o(1/\sqrt{L}) \label{eq1:appCorrections}
\end{align}
$\beps_{\ba_{l\mu}/\bv_{l \mu}}\defeq \[\epsilon_{a_{i\mu}/v_{i\mu}}\]_{i\in l}$ is the (positive or negative) vector of corrections $\in O(1/\sqrt{L})$ linking $\ba_{l\mu}/\bv_{l\mu}$ to $\ba_l/\bv_l$. We need to express all the cavity quantities appearing in these corrections in terms of marginal quantities. In order to do so, we start by expanding (\ref{eqChIntro:defS2cav}) and (\ref{eqChIntro:defRcav}) in the $\bF$ elements and keeping only the $O(1)$ terms, we get:
\begin{align}
(\bsy \Sigma_{l})^2 &=\left[\sum_{\mu}^M \frac{\bF_{\mu {l}}^2}{1/{\rm{snr}} + \Theta_{\mu} - \bv_{{l}\mu}^{\intercal} \bF_{\mu {l}}^2 } \right]^{-1}\nonumber\\
&\approx \left[\sum_{\mu}^M \frac{\bF_{\mu {l}}^2}{1/{\rm{snr}} + \Theta_{\mu} } \right]^{-1}\\
\bR_{l} &= (\bsy \Sigma_{l})^2 \left[\sum_{\mu}^M \frac{\bF_{\mu {l}} (y_\mu - w_\mu + \bF_{\mu {l}}\ba_{{l}\mu}^{\intercal})}{1/{\rm{snr}} + \Theta_{\mu} - \bv_{{l}\mu}^{\intercal} \bF_{\mu {l}}^2 } \right]\nonumber \\
&\approx (\bsy \Sigma_{l})^2 \left[\sum_{\mu}^M \frac{\bF_{\mu {l}} \left(y_\mu - w_\mu\right)}{1/{\rm{snr}} + \Theta_{\mu}} + \underbrace{\sum_{\mu}^M \bF_{\mu {l}}\frac{(\bF_{\mu {l}})^{\intercal} \ba_{l}}{1/{\rm{snr}} + \Theta_{\mu}}}_{\bsy \Sigma_{l}
^{-2} \ba_{l} + O(1/\sqrt{L})} - \underbrace{\sum_{\mu}^M \bF_{\mu {l}}\frac{(\bF_{\mu {l}})^{\intercal} \beps_{\ba_{l \mu}}}{1/{\rm{snr}} + \Theta_{\mu}}}_{\in O(1/L)}\right]\nonumber \\
&\approx \ba_{l} + (\bsy \Sigma_{l})^2 \sum_{\mu}^M \frac{\bF_{\mu l}(y_\mu - w_\mu)}{1/{\rm{snr}} + \Theta_\mu}
\end{align}
These depend on the previously defined quantities (\ref{eqChIntro:w}) and (\ref{eqChIntro:Theta}) that we thus need to expand keeping only the terms in $O(1)$ as well to get marginal quantities:
\begin{align}
\Theta_{\mu} &= \sum_{k}^L (\bF_{\mu k}^2)^{\intercal}\bv_k - \underbrace{\sum_{k}^L (\bF_{\mu k}^2)^{\intercal}\beps_{\bv_{k\mu}}}_{\in O(1/L)} \nonumber \\
&\approx \sum_{k}^L (\bF_{\mu k}^2)^{\intercal}\bv_k \label{eq:appTheta}\\
w_\mu &=\sum_{k}^L \bF_{\mu k}^{\intercal}\ba_k - \sum_{k}^L \bF_{\mu k}^{\intercal}\beps_{\ba_{k\mu}} \nonumber\\
&\approx \sum_{k}^L \bF_{\mu k}^{\intercal}\ba_k - \sum_{k}^L \bF_{\mu k}^{\intercal} \[\frac{\bF_{\mu k}\(y_\mu - \sum_{k'}^{L} \bF_{\mu k'}^{\intercal} \ba_{k' \mu} + \bF_{\mu k}^{\intercal} \ba_{k \mu} \) }{1/{\rm{snr}} + \sum_{k'}^{L} \bv_{k' \mu}^{\intercal} \bF_{\mu k'}^2 - \bv_{k \mu}^{\intercal} \bF_{\mu k}^2} \underbrace{(\bsy \Sigma_{k})^2 \nabla_{\bR_{k}} \bff_{a_k}((\bsy \Sigma_{k})^2,\bR_{k})}_{=\bv_k}\]\nonumber\\
&\approx \sum_{k}^L \bF_{\mu k}^{\intercal}\ba_k - \frac{y_\mu - w_\mu}{1/{\rm{snr}} + \Theta_\mu} \sum_{k}^L (\bF_{\mu k}^2)^{\intercal} \bv_k \label{eq1:appW}
\end{align}
The last equality is obtained using (\ref{eq1:appCorrections_a}), (\ref{eq1:bB}), neglecting $o(1)$ terms and noticing that:
\begin{align}
\partial_{R_{i} }f_{a_i}((\bsy \Sigma_{l_i})^2,\bR_{l_i}) &= \partial_{R_{i} } \(\frac{\int d\bx P_0(\bx) x_i \exp\((\bx - \bR_{l_i})^{\intercal} (2(\bsy \Sigma_{l_i})^2)^{-1} \)}{\int d\bx P_0(\bx) \exp\((\bx - \bR_{l_i})^{\intercal} (2(\bsy \Sigma_{l_i})^2)^{-1} \)}\)\\
&=\frac{1}{(\Sigma_{i})^2} \(\mathbb{E}_{\bx|\by}(x_i^2) -R~\mathbb{E}_{\bx|\by}(x_i) - \mathbb{E}_{\bx|\by}(x_i)(\mathbb{E}_{\bx|\by}(x_i) - R) \)\\
&=\frac{1}{(\Sigma_{i})^2} \(\mathbb{E}_{\bx|\by}(x_i^2) - \mathbb{E}_{\bx|\by}(x_i)^2\)\\
&= \frac{1}{(\Sigma_{i})^2}f_{c_i}((\bsy \Sigma_{l_i})^2,\bR_{l_i})\\
\Rightarrow f_{c_i}((\bsy \Sigma_{l_i})^2,\bR_{l_i})&=v_i=(\Sigma_{i})^2 \partial_{R_{i} }f_{a_i}((\bsy \Sigma_{l_i})^2,\bR_{l_i}) \label{eq1:propertyfafc_1}\\
\Rightarrow \bff_{c_l}((\bsy \Sigma_{l})^2,\bR_{l})&=\bv_l = (\bsy \Sigma_{l})^2 \nabla_{\bR_{l} }\bff_{a_l}((\bsy \Sigma_{l})^2,\bR_{l}) \label{eq1:propertyfafc}
\end{align}
Adding back the time indices and rewriting the previous set of equations in terms of their single components, we get the AMP algorithm, see Fig.~\ref{algoCh1:AMP}. Here $\mathbb{E}_{\bx|\by}$ is the posterior average estimation by the AMP algorithm. The only problem-dependent objects in the AMP are the denoising functions $\{f_{a_i}, f_{c_i}\}$ that depend on the assumed prior, see the next section. The updating schedule of Fig.~\ref{algoCh1:AMP} is what we would have obtained keeping the time index when starting the derivation from the usual parallel BP updates (\ref{eqChIntro:bp1}), (\ref{eq1:bp2}) as we did. On Fig.~\ref{algoCh1:AMP}, we added a damping scheme controled by the parameter $\alpha$: at $\alpha=0$, we recover the derived AMP algorithm but for finite values $0<\alpha<1$, the algorithm can converge easier in situations where it experiences some troubles such as numerical oscillations. This scheme is validated empirically, but other ones could be tried. It must be understood that when $\alpha\neq 0$, the algorithm does not follow anymore the state evolution asymptotic predictions that we will derive in sec.~\ref{sec:stateEvolutionGeneric}. In all the theoretical analyzes of this thesis, we thus always consider $\alpha=0$ but in practical situations, a small $\alpha\approx 0.1$ can help.

The AMP algorithm has been generalized to any noise model in \cite{DonohoMaleki10,Rangan10b} and called GAMP for "generalized AMP". The algorithm presented here is equivalent to GAMP for the i.i.d AWGN channel. It is important to note that the name "approximate" message-passing is a little misleading since as we have shown through the derivation, for dense i.i.d random measurement matrices the AMP is asymptotically equivalent to BP, i.e. all the leading terms in $N$ are included in AMP.
\begin{figure}[!t]
\begin{algorithmic}[1]
\State $t\gets 0$
\State $\delta \gets \epsilon + 1$
\While{$t<t_{max} \ \textbf
{and} \ \delta>\epsilon$} 
\State $\tilde\Theta^{t+1}_\mu \gets \sum_{i}^{N}F_{\mu i}^2v_i^{t}$
\State $w^{t+1}_\mu \gets \alpha w^{t}_\mu + (1-\alpha)\(\sum_{i}^N F_{\mu i}a_i^{t} - \tilde\Theta^{t+1}_\mu\frac{y_\mu-w^{t}_\mu}{{1/{\rm{snr}}} + \Theta^{t}_\mu}\)$
\State $\Theta^{t+1}_\mu \gets \alpha \Theta^{t}_\mu + (1-\alpha)\tilde\Theta^{t+1}_\mu$
\State $\Sigma^{t+1}_i \gets \left[\sum_{\mu}^{M}\frac{F_{\mu i}^2}{{1/{\rm{snr}}} + \Theta_\mu^{t+1}}\right]^{-1/2}$
\State $R^{t+1}_i \gets a^{t}_i + (\Sigma^{t+1}_i)^2 \sum_{\mu}^{M} F_{\mu i}\frac{y_\mu - w^{t+1}_\mu}{{1/{\rm{snr}}} + \Theta^{t+1}_\mu}$
\State $v^{t+1}_i \gets f_{c_i}\left((\bsy \Sigma_{l_i}^{t+1})^2,\bR_{l_i}^{t+1}\right)$
\State $a^{t+1}_i \gets f_{a_i}\left((\bsy \Sigma_{l_i}^{t+1})^2,\bR_{l_i}^{t+1}\right)$
\State $t \gets t+1$
\State $\delta \gets ||\ba^{t+1} - \ba^{t}||_2^2$
\EndWhile
\State \textbf{return} $\ba^t$
\end{algorithmic}
\caption[Generic approximate message-passing algorithm with damping]{The approximate message-passing algorithm with damping controled by $0\le\alpha<1$. $l_i$ is the index of the section to which the $i^{th}$ $1$-d variable belongs to. In the scalar components case $B=1$, the sections are just the components. $\epsilon$ is the accuracy for convergence and $t_{max}$ the maximum number of iterations. A suitable initialization for the quantities is ($a_i^{t=0}=\mathbb{E}_{P_0}(x_i)$, $v_i^{t=0}=\txt{Var}_{P_0}(x_i)$, $w_\mu^{t=0}=y_\mu$). Once the algorithm has converged, i.e. the quantities do not change anymore from iteration to iteration, the estimate of the $l^{th}$ signal section is $\textbf{a}_l^{t}$. At $\alpha=0$, the usual approximate message-passing algorithm is recovered. $\{\tilde \Theta_\mu\}_\mu^M$ are auxilliary variables for the damping scheme.}   
\label{algoCh1:AMP}
\end{figure}
\subsection{Alternative "simpler" derivation of the approximate message-passing algorithm}
\label{sec:alternativeAMPderivation}
The previous derivation did not assume anything and proves the asymptotic equivalence between AMP and BP but is a bit long. We present here an alternative derivation of the AMP algorithm, which will use directly the assumption that the node-to-factor messages can be represented as Gaussians. This derivation is very close to the one of the non-parametric BP algorithm \cite{sudderth2010nonparametric} in the special case where only one Gaussian in used in the messages parametrization. Like the previous derivation, it starts from the factor-to-node cavity message (\ref{eqChIntro:bp1}). We first define $\bsy \Gamma_{\mu l}\defeq \sum_{k \neq l}^{L-1} \bF_{\mu k}^{\intercal} \bx_k$, the variable that appears in the exponential in (\ref{eqChIntro:bp1}). Now we assume that the cavity node-to-factor messages are Gaussians. In a sense we place ourselves directly at the step (\ref{eq1:hat_lmu_cav_notZ}) of the previous derivation:
\begin{equation}
	m_{k \mu}^t(\bx_k) = \mathcal{N}\(\bx_k \big| \ba_{k\mu}^t, \bv_{k\mu}^t\)
\end{equation}
As discussed in sec.~\ref{sec:understandingBPwithCavityGraphs}, when computing the factor-to-node messages we assume that the node-to-factor messages arriving on the factor of interest are conditionally independent (due to the tree approximation behind BP). This implies that $\bsy\Gamma_{\mu l}^t \sim \mathcal{N}\(\bsy\Gamma_{\mu l}^t \big| \bar{\bsy\Gamma}_{\mu l}^t, \bv_{\bsy\Gamma_{\mu l}}^t\)$ in (\ref{eqChIntro:bp1}) is also Gaussian distributed with moments given by:
\begin{align}
	\bar{\bsy\Gamma}_{\mu l}^t &= \sum_{k\neq l} \bF_{\mu k}^{\intercal} \ba_{k\mu}^t\\
	\bv_{\bsy\Gamma_{\mu l}}^t &= \sum_{k\neq l} (\bF_{\mu k}^2)^{\intercal} \bv_{k\mu}^t
\end{align}
We thus obtain the factor-to-node message $\hat m_{\mu l}^{t+1}(\bx_l)$ expression from (\ref{eqChIntro:bp1}):
\begin{align}
	\hat m^{t}_{\mu l}( \bx_l) &= \frac{1}{\hat z_{\mu l}{t}} \int \mathcal{N}\(\bsy\Gamma_{\mu l} \big| y_\mu - \bF_{\mu l}^{\intercal} \bx_l, 1/{\rm snr} \) \mathcal{N}\(\bsy\Gamma_{\mu l} \big| \bar{\bsy\Gamma}_{\mu l}^t, \bv_{\bsy\Gamma_{\mu l}}^t\) d\bsy\Gamma_{\mu l}\\
	&=\mathcal{N}\(\bx_l \big| \ba_{\mu l}^{t}, \bv_{\mu l}^{t}\)
\end{align}
where the first Gaussian is the likelihood part. The moments are given by:
\begin{align}	
	\ba_{\mu l}^{t} &= \frac{y_\mu - \sum_{k\neq l} \bF_{\mu k}^{\intercal} \ba_{k\mu}^t}{\bF_{\mu l}} = \frac{\bB_{\mu l}^t}{\bA_{\mu l}^t}\\
	\bv_{\mu l}^{t} &= \frac{1/{\rm snr}+ \sum_{k\neq l} (\bF_{\mu k}^2)^{\intercal} \bv_{k\mu}^t}{\bF_{\mu l}^2} = \frac{1}{\bA_{\mu l}^t}	
\end{align}
where we recognized the expressions (\ref{eq1:bA}), (\ref{eq1:bB}) obtained in the previous derivation, thus it is coherent. Now in order to compute the node-to-factor message from (\ref{eq1:bp2}), we need the fact that a product of $K$ Gaussians distributions over the same variable with respective means and variances $\{r_k, v_k\}_k^K$ gives a new Gaussian distribution with moments $\{r,v\}$ given by:
\begin{align}
	v &= \(\sum_k^K v_k^{-1}\)^{-1}\\
	r &= v\sum_k^K r_k v_k^{-1}
\end{align}
From this combined with (\ref{eq1:bp2}), we close the equations on the cavity node-to-factor means and variances:
\begin{align}
	m_{l \mu}^{t+1}(\bx_l) &= \frac{1}{z_{l\mu}^{t+1}} P_0^l(\bx_l) \prod_{\nu\neq\mu}\hat m_{\nu l}^t(\bx_l)\\
	&= \frac{1}{z_{l\mu}^{t+1}}P_0^l(\bx_l) \mathcal{N}(\bx_l|\tilde \ba_{l\mu}^{t+1},\tilde\bv_{l\mu}^{t+1}) \label{eq:lastStep_alternativeAMPderivation}\\
	\tilde\bv_{l\mu}^{t+1} &= \left[\sum_{\nu \neq \mu}\(\frac{\bF_{\nu l}^2 }{1/{\rm snr} + \sum_{k\neq l} (\bF_{\nu k}^2)^{\intercal} \bv_{k\nu}^t} \) \right]^{-1} = \frac{1}{\sum_{\nu \neq \mu} \bA_{\nu l}^t}\\
	\tilde\ba_{l\mu}^{t+1} &= \bv_{l\mu}^{t+1} \sum_{\nu \neq \mu} \frac{\bF_{\nu l} \(y_\mu - \sum_{k\neq l} \bF_{\nu k}^{\intercal} \ba_{k\nu}^t \)}{1/{\rm snr} + \sum_{k\neq l} (\bF_{\nu k}^2)^{\intercal} \bv_{k\nu}^t } = \frac{\sum_{\nu \neq \mu} \bB_{\nu l}^t}{\sum_{\nu \neq \mu} \bA_{\nu l}^t}
\end{align}
using again (\ref{eq1:bA}), (\ref{eq1:bB}). The last step is to compute the mean $\ba_{l\mu}^{t+1}$ and variance $\bv_{l\mu}^{t+1}$ of the cavity message (\ref{eq:lastStep_alternativeAMPderivation}): it gives exactly the relaxed-BP equations (\ref{eq1:relaxedCav}), (\ref{eq1:relaxedCav_v}) and thus the AMP algorithm after the TAP step like in the previous section.
\subsection{Understanding the approximate message-passing algorithm}
\begin{figure}[t!]
	\centering
	\includegraphics[width=1\textwidth]{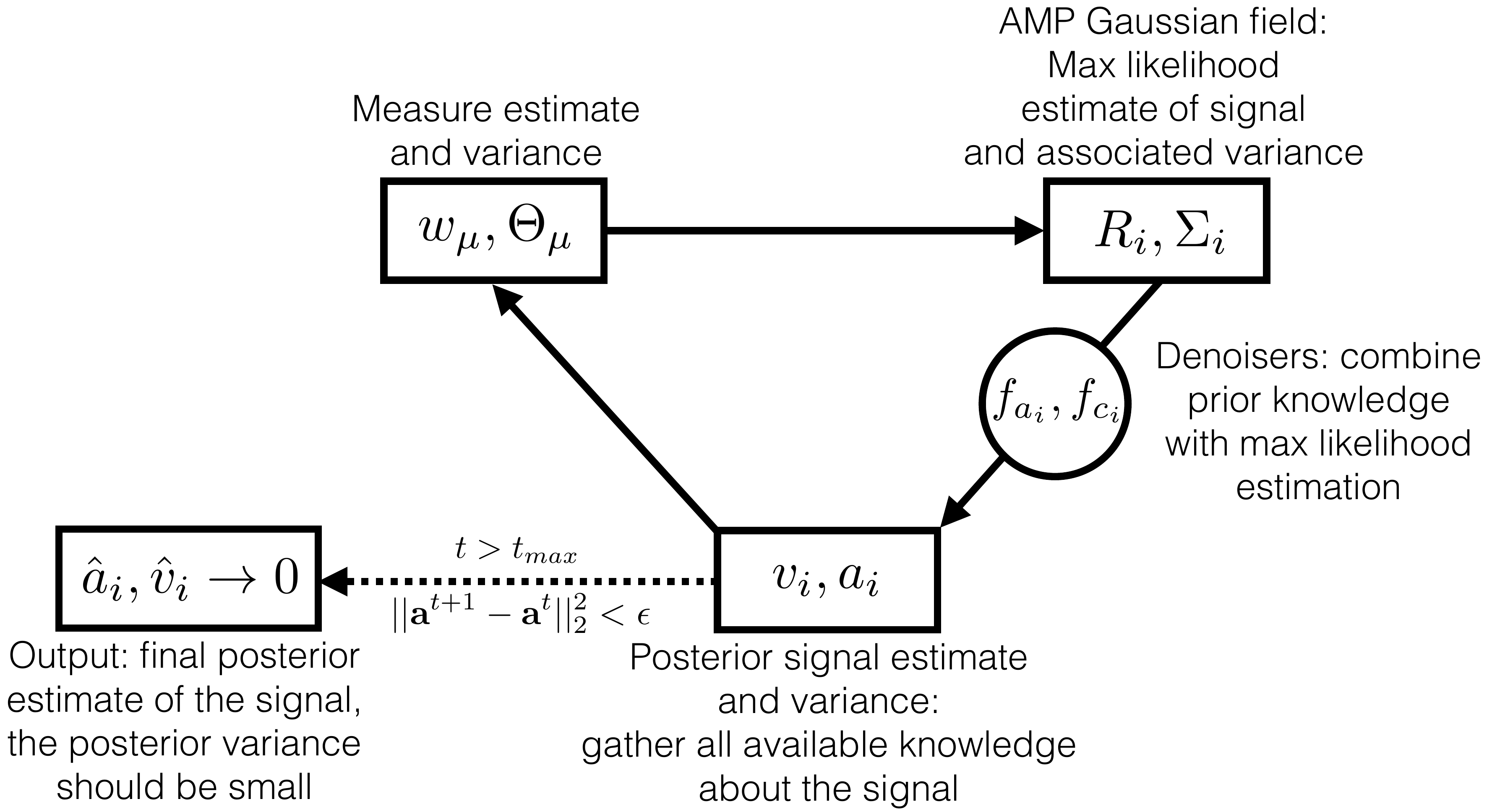}
	\caption[Graphical representation of the approximate message-passing algorithm]{Graphical representation of the approximate message-passing algorithm. It iteratively repeats three main steps until convergence: $i)$ computation of an estimate of the measurement ans its variance, $ii)$ computation of the maximum likelihood estimate of the signal and its variance, the AMP fields and then $iii)$ combine the previous maximum likelihood estimate with the prior through the denoisers to get the posterior estimate and variance.}
	\label{figChIntro:AMP}
\end{figure}
Let us explain what AMP is doing. It is an iterative algorithm that iteratively repeat three basic steps, each time computing an estimate of some quantity and its associated variance. This is represented on Fig.~\ref{figChIntro:AMP}:

\textbf{First step :} From the actual knowledge of the posterior signal estimate and variance, AMP computes the estimate of the measurement vector and its variance. This estimate $\bw^{t+1}$ (see Fig.~\ref{algoCh1:AMP}) is a sum of two terms: the first one is the measure one would get if the true signal was given by the actual posterior estimate $\ba^t$, the second one is a "gradient like" term that tends to {\it increase} the difference between the measure and its new estimate. This increase is proportional to the measurement variance and the difference. It can seem a priori strange to amplify this difference, but it is actually understandable from the use of this estimate in the second step. 

\textbf{Second step :} AMP computes what we call the AMP fields: $\bR^{t+1}$ and associated variances $(\bsy \Sigma^{t+1})^2$, the means and variances of a Gaussian mean field on each variable that summarizes the influence of all the likelihood factors, i.e. this Gaussian field tends to maximize the likelihood of the signal by enforcing its estimate to match the measurements. $\bR^{t+1}$ (see Fig.~\ref{algoCh1:AMP}) is the sum of the previous posterior estimate, the best estimate at the previous time step plus another  "gradient like" term which is proportional to the AMP field variance and the most recent difference between the measure estimate and the true measurement vector: now we understand that the amplification of this difference in the previous step actually leads to a stronger shift of the AMP fields $\bR^{t+1}$ in the proper direction to reduce the difference with the measurement vector, thus to increase the likelihood.

\textbf{Third step :} The last step takes as input the new AMP fields and combine them with the prior distribution of the signal to get the new posterior estimate that gather all the actual information about the signal. This is performed thanks to the denoiser $f_a$: this function averages over all the possible signal estimates properly weighted by their actual posterior distribution, given by the product of the AMP Gaussian field and the prior. $f_c$ computes the associated posterior variance that should converge to $0$ as the algorithm converges to its fixed point, hopefully given by the true posterior $MMSE$ estimate (under the prior matching condition, and above the BP transition if no spatial coupling is used, see sec.~\ref{sec:typicalPhaseTransitions} and sec.~\ref{sec:spatialCoupling}).

These steps are repeated until convergence or the maximum number of iterations is reached, and the estimator is the last posterior estimate of the signal. 
\subsection{How to quickly cook an approximate message-passing algorithm for your linear estimation problem}
\label{sec:cookAMP}
In the generic AMP algorithm, the only problem dependent part are the denoising functions $\{f_{a_i}, f_{c_i}\}$, but once adapted the algorithm Fig.~\ref{algoCh1:AMP} can be applied to a large class of linear estimation problems. We give here "blocks" for constructing such denoising functions. For a factorizable prior $P_0(\bx)=\prod_i^N P_0^i(x_i)$, we need the posterior partition function $z(\Sigma^2, R)$, the first $u(\Sigma^2, R)$ and second $v(\Sigma^2, R)$ non normalized moments. We consider that the prior $P_0^i(x)$ is a linear combination of different distributions $p_u(x)$:
\begin{align}
	P_0^i(x) &\propto \sum_u w_u p_u(x)
\end{align}
where $w_u$ is the weight of the distribution $p_u$ in the prior. These weights dont need to be normalized in the present construction. From this we can construct the posterior normalization:
\begin{align}	
	z(\Sigma^2, R) &\defeq \int dx~P_0^i(x)~\mathcal{N}(x|R, \Sigma^2) \nonumber\\
	&~=\sum_u w_u\int dx~p_u(x)~\mathcal{N}(x|R, \Sigma^2) \nonumber\\	
	&~=\sum_u w_u z_u(\Sigma^2,R)
\end{align}
In the same way we construct the first and second non normalized moments:
\begin{align}
	\gamma(\Sigma^2, R) &\defeq \int dx~P_0^i(x)~\mathcal{N}(x|R, \Sigma^2)x \nonumber\\
	&~=\sum_u w_u\int dx~p_u(x)~\mathcal{N}(x|R, \Sigma^2)x \nonumber\\
	&~=\sum_u w_u \gamma_u(\Sigma^2,R)\\
	\tau(\Sigma^2, R) &\defeq \int dx~P_0^i(x)~\mathcal{N}(x|R, \Sigma^2)x^2 \nonumber\\
	&~=\sum_u w_u\int dx~p_u(x)~\mathcal{N}(x|R, \Sigma^2)x^2 \nonumber\\
	&~=\sum_u w_u \tau_u(\Sigma^2,R)
\end{align}
From this contruction, we define the denoisers as:
\begin{align}
	f_{a_i}(\Sigma^2, R) &\defeq \frac{\gamma(\Sigma^2, R)}{z(\Sigma^2, R)}\\	
	f_{c_i}(\Sigma^2, R) &\defeq \frac{\tau(\Sigma^2, R)}{z(\Sigma^2, R)} - f_a(\Sigma^2, R)^2
\end{align}
see the Tab.~\ref{tab:prior} for possible triplets $(z_u, \gamma_u, \tau_u)$ to construct denoisers.
Numerically, it is careful to take for the variance denoiser $f_{c_i}(\Sigma^2, R)=\txt{max}\(f_{c_i}(\Sigma^2, R), \epsilon\)$ where $\epsilon$ is a very small constant, like $10^{-20}$. This avoids possible negative variances that can appear at the very beginning of the convergence.
\begin{table}[t!]		
	\centering
	\begin{tabular}{c|c}
	\hline
	\hline
	\hline
	\textbf{prior term $p_u(x_i)$} & $z_u(\Sigma^2, R)$ \\
	\hline
	\textbf{Dirac} & \\
	$\delta(x_i-m)$ & 
	$\mathcal{N}\(m|R,\Sigma^2\)$ \\
	\hline
	\textbf{Gauss} & \\
	$\mathcal{N}(x_i|m, v)$ & 
	$\mathcal{N}\(m|R,\Sigma^2+v\)$ \\
	\hline
	\textbf{Exponential} & \\
	$\lambda e^{-\lambda x_i} \mathbb{I}(x_i>0)$ & 
	$\frac{\lambda}{2} \exp\(\frac{\lambda}{2} (\lambda \Sigma^2-2 R)\)\txt{erfc}\left[\frac{\lambda \Sigma^2 - R}{\sqrt{2\Sigma^2}}\right]$ \\
	\hline
	\textbf{Laplace} & \\
	$\frac{\beta}{2} e^{-\beta |x_i|}$ & 
	$\frac{\beta e^{\beta^2 \Sigma^2 / 2}}{4}\(e^{-\beta R} \txt{erfc}\(\frac{\beta \Sigma^2-R}{\sqrt{2\Sigma^2}}\) + e^{\beta R} \txt{erfc}\(\frac{\beta \Sigma^2+R}{\sqrt{2\Sigma^2}}\) \)$ \\
	\hline
	\hline
	\hline	
	\textbf{prior term $p_u(x_i)$} & $\gamma_u(\Sigma^2, R)$ \\
	\hline
	\textbf{Dirac} & \\
	$\delta(x_i-m)$ & 
	$ m\mathcal{N}\(m|R,\Sigma^2\)$ \\ 
	\hline
	\textbf{Gauss} & \\
	$\mathcal{N}(x_i|m, v)$ & 
	$ \mathcal{N}\(m|R,\Sigma^2+v\) \frac{m \Sigma^2 + Rv}{\Sigma^2 + v}$ \\ 
	\hline
	\textbf{Exponential} & \\
	$\lambda e^{-\lambda x_i} \mathbb{I}(x_i>0)$ & 
	$\frac{\lambda e^{-R^2/(2\Sigma^2)}}{2\sqrt{\pi}} \(\sqrt{2\Sigma^2} + \sqrt{\pi}(R-\lambda\Sigma^2) e^{(R-\lambda \Sigma^2)^2/(2\Sigma^2)} \txt{erfc}\left[\frac{\lambda\Sigma^2-R}{\sqrt{2\Sigma^2}} \right] \)$ \\
	\hline
	\textbf{Laplace} & $\frac{\beta e^{\beta^2 \Sigma^2/2}}{4}\bigg(e^{-\beta R} (R-\beta\Sigma^2 ) \txt{erfc}\(\frac{\beta \Sigma^2-R}{\sqrt{2\Sigma^2}}\)$ \\
	$\frac{\beta}{2} e^{-\beta |x_i|}$ & 
	$+ e^{\beta R}(R+\beta\Sigma^2 ) \txt{erfc}\(\frac{\beta \Sigma^2+R}{\sqrt{2\Sigma^2}}\)\bigg)$ \\
	\hline
	\hline
	\hline	
	\textbf{prior term $p_u(x_i)$} & $\tau_u(\Sigma^2, R)$ \\
	\hline
	\textbf{Dirac} & \\
	$\delta(x_i-m)$ & 
	$m^2 \mathcal{N}\(m|R,\Sigma^2\)$ \\
	\hline
	\textbf{Gauss} & \\
	$\mathcal{N}(x_i|m, v)$ &
	$\mathcal{N}\(m|R,\Sigma^2+v\) \frac{m^2 \Sigma^4 + \Sigma^2 (2 m R + \Sigma^2) v + (R^2 + \Sigma^2) v^2}{(\Sigma^2 + v)^2}$\\
	\hline
	\textbf{Exponential} & $\frac{\lambda e^{-R^2/(2\Sigma^2)}}{2\sqrt{2\pi}} \big\{ 2 \Sigma(R-\lambda\Sigma^2) $ \\
	$\lambda e^{-\lambda x_i} \mathbb{I}(x_i>0)$ & 	
	$+ \sqrt{2\pi}\((R-\lambda\Sigma^2)^2+\Sigma^2\) e^{(R-\lambda \Sigma^2)^2/(2\Sigma^2)} \txt{erfc}\left[\frac{\lambda\Sigma^2-R}{\sqrt{2\Sigma^2}} \right] \big\}$\\
	\hline
	\textbf{Laplace} & $ \frac{\beta e^{\beta^2\Sigma^2/2 } }{4}\bigg(-4\beta\Sigma^3e^{-R^2/(2\Sigma^2) } + e^{-\beta R} (\Sigma^2 + (\beta\Sigma^2-R)^2) \txt{erfc}\(\frac{\beta\Sigma^2-R}{\sqrt{2\Sigma^2}}\)$ \\
	$\frac{\beta}{2} e^{-\beta |x_i|}$ & 
	$+e^{\beta R} (\Sigma^2 + (\beta\Sigma^2+R)^2) \txt{erfc}\(\frac{\beta\Sigma^2+R}{\sqrt{2\Sigma^2}}\)\bigg)$ \\
	\hline
	\hline
	\hline	
	\end{tabular}
	\caption{Examples of functions for the prior construction}
	\label{tab:prior}
\end{table}
\subsection{The Bethe free energy for large dense graphs with i.i.d additive white Gaussian noise}
\label{sec:BetheFforAMP}
For completeness, we now show how to derive an expression of the Bethe free energy that depends on the quantities appearing in the AMP algorithm. The resulting expression is only valid at the fixed point of the algorithm and is true asymptotically on dense graphs as the derivation uses the same assumptions that for AMP, see sec.~\ref{sec:classicalDerivationAMP}. We consider that the measurement matrix is homogeneous.
\subsubsection{The factors contributions}
We start from the free energy expression (\ref{eq:BetheF_cavMess}). Let us start by computing the term (\ref{eq:BetheF_za}). Using the likelihood of the compressed sensing with AWGN (\ref{eqIntro:likelihood}), we get:
\begin{align}
	z_\mu &= \int d\bx\prod_{i}^N m_{i\mu}(x_i) \frac{1}{\sqrt{{2\pi\Delta}}}\exp\(-\frac{(y_\mu - \sum_i^N F_{\mu i} x_i)^2}{2\Delta} \) \label{eq:zmu0}
\end{align}
which is almost (up to one additional variable) the (scalar) normalization of (\ref{eqChIntro:bp1}) at the fixed point. Thus the "clean" derivation is exactly the same as in sec.~\ref{sec:classicalDerivationAMP}: $i)$ apply the Stratanovitch transformation to linearize the squared sum that appears in the exponent. In this way there are no crossed terms between variables and the integrals can be perfomed independently. $ii)$ Expand the exponential up to second order using the fact the the $\bF$ elements are small, $iii)$ integrate the independent integrals and use the same trick as (\ref{eq:uExp}) to get a similar expression, except that the signal dependent term inside the parenthesis are not present anymore as all variables have been integrated and finally $iv)$ perform the Gaussian integral over $\lambda$, the auxiliary parameter introduced for the Stratanovitch transform and simplify the result.

Now the fastest way is to remind ourselves that the Bethe approximation is equivalent to assume the tree property of the graph and thus as discussed in sec.~\ref{sec:understandingBPwithCavityGraphs}, when a cavity is dug in the graph by removing a factor, the neighboring variables (i.e. all of them in the present dense case) are considered independent. Thus the central limit theorem implies that $\gamma \defeq \sum_i^N F_{\mu i} x_i$ appearing in the exponent in (\ref{eq:zmu0}) is a Gaussian random variable with mean $w_\mu\defeq\sum_i^N F_{\mu i}a_{i\mu}$ and variance $\Theta_\mu\defeq\sum_i^NF_{\mu i}^2v_{i\mu}$, the same as (\ref{eqChIntro:w}), (\ref{eqChIntro:Theta}) in the scalar case where $a_{i\mu}$ and $v_{i\mu}$ are the cavity mean and variance associated to $m_{i\mu}(x_i)$ and computed respectively through (\ref{eq1:relaxedCav}) and (\ref{eq1:relaxedCav_v}) in the AMP framework. Finally the result is:
\begin{align}
	z_\mu &= \int d\gamma \mathcal{N}(\gamma|y_\mu,\Delta)\mathcal{N}(\gamma|w_\mu,\Theta_\mu)\\
	&=\mathcal{N}\(w_\mu|y_\mu, \Delta+\Theta_\mu \)\label{eq:zmu}
\end{align}
\subsubsection{The nodes and edges contributions}
We now compute (\ref{eq:BetheF_zi}). From (\ref{eq1:hat_mul_cav_notZ}) that we got in the AMP derivation and as each variable is connected to each likelihood factor in the homogeneous measurement matrix case that we consider here, we have:
\begin{align}
	z_i & = \[\prod_{\mu}^M \hat z_{\mu i}\]^{-1}\int dx_i P_0(x_i)\exp\(-\frac{x_i^2}{2}\sum_\mu^M A_{\mu i} + x_i\sum_\mu^M B_{\mu i} \) \\
	&=\[\prod_{\mu}^M \hat z_{\mu i}\]^{-1}\tilde z_i \label{eq:lastzi}
\end{align}
where the prior factor must not be forgotten in (\ref{eq:BetheF_zi}) and we define:
\begin{align}
	\tilde z_i \defeq \int dx_i P_0(x_i)\exp\(-\frac{x_i^2}{2\Sigma_i^2} + x_i\frac{R_i}{\Sigma_i^2} \)
\end{align}
where we have used the definition of the AMP fields (\ref{eqChIntro:defRcav}), (\ref{eqChIntro:defS2cav}). Now we notice that using (\ref{eq:BetheF_zai}) with (\ref{ecChIntro:BP_n2f_mess}), we get at the fixed point that:
\begin{align}
	\tilde z_{i\mu} &= \frac{\tilde z_i}{z_{i\mu}\hat z_{\mu i}}	
\end{align}
It allows with (\ref{eq:lastzi}) to re-write the nodes and edges contributions to the free energy (\ref{eq:BetheF_cavMess}) as:
\begin{align}
	&-\sum_i^N \log\(z_i\) + \sum_{i}^N\sum_\mu^M\log\(\tilde z_{i\mu}\) \\
	=&-\sum_i^N \(\log\(\tilde z_i\) - \sum_\mu^M \log\(\hat z_{\mu i}\) -\sum_\mu^M\(\log\(\frac{\tilde z_i}{z_{i\mu}}\) - \log\(\hat z_{\mu i}\) \) \)\\
	=&-\sum_i^N \( \log\(\tilde z_i\) + \sum_\mu^M \log\(\frac{z_{i\mu}}{\tilde z_i} \) \) \label{eq:lastSecondContributionBetheF}
\end{align} 
It is easy to verify from (\ref{eqChIntro:defRcav}), (\ref{eqChIntro:defS2cav}), (\ref{eqChIntro:defRcav_cav}), (\ref{eqChIntro:defS2cav_cav}) that at first order:
\begin{align}
	\frac{1}{\Sigma_{i\mu}^2} &\defeq \sum_{\gamma\neq\mu}^M A_{\gamma i} \\
	&\approx \frac{1}{\Sigma_{i}^2}\(1 - \Sigma_{i}^2 A_{\mu i}\)\\
	\Rightarrow R_{i\mu} &\defeq \frac{\sum_{\gamma\neq\mu}^M B_{\gamma i}}{\sum_{\gamma\neq\mu}^M A_{\gamma i}}\\
	&=R_i \(1 + \Sigma_i^2 A_{\mu i}\) - \Sigma_i^2B_{\mu i}	
\end{align}
which implies also at first order:
\begin{align}
	\frac{R_{i\mu}}{\Sigma_{i\mu}^2}\approx \frac{R_i}{\Sigma_i^2} - B_{\mu i}
\end{align}
reminding that $A_{\mu i} \in O(1/N)$ and $B_{\mu i} \in O(1/\sqrt{N})$. From these results, we can simplify the node-to-factor partition function (\ref{eq1:hat_lmu_cav_Z}):
\begin{align}
	z_{i\mu} &= \int dx_i P_0(x_i)\exp\(-\frac{x_i^2}{2\Sigma_{i\mu}^2} + x_i\frac{R_{i\mu}}{\Sigma_{i\mu}^2} \)\\
	&\approx \int dx_i P_0(x_i)\exp\(-\frac{x_i^2}{2\Sigma_{i}^2} + x_i\frac{R_{i}}{\Sigma_{i}^2} \) \(1 - x_iB_{\mu i} + \frac{x_i^2}{2} \(B_{\mu i}^2 + A_{\mu i}\)\)\\
	&=\tilde z_i + \int dx_i P_0(x_i)\exp\(-\frac{x_i^2}{2\Sigma_{i}^2} + x_i\frac{R_{i}}{\Sigma_{i}^2} \) \(- x_iB_{\mu i} + \frac{x_i^2}{2} \(B_{\mu i}^2 + A_{\mu i}\)\)
\end{align}
The last equality allows to simplify the second sum appearing in (\ref{eq:lastSecondContributionBetheF}):
\begin{align}
	\sum_\mu^M \log\(\frac{z_{i\mu}}{\tilde z_i} \) &= \sum_\mu^M\log\(1 + \frac{1}{\tilde z_i} \int dx_i P_0(x_i)e^{-\frac{x_i^2}{2\Sigma_{i}^2} + x_i\frac{R_{i}}{\Sigma_{i}^2}} \(- x_iB_{\mu i} + \frac{x_i^2}{2} \(B_{\mu i}^2 + A_{\mu i}\)\) \)\\
	&\approx \frac{1}{\tilde z_i}\int dx_i P_0(x_i)e^{-\frac{x_i^2}{2\Sigma_{i}^2} + x_i\frac{R_{i}}{\Sigma_{i}^2}} \sum_\mu^M \(- x_iB_{\mu i} + \frac{x_i^2}{2} \(B_{\mu i}^2 + A_{\mu i}\)\)\nonumber \\
	&-\frac{1}{2} \sum_\mu^M B_{\mu i}^2 \Bigg(\underbrace{\frac{1}{\tilde z_i}\int dx_i P_0(x_i)e^{-\frac{x_i^2}{2\Sigma_{i}^2} + x_i\frac{R_{i}}{\Sigma_{i}^2}} x_i}_{= a_i}\Bigg)^2\\
	&\approx\frac{1}{\tilde z_i}\int dx_i P_0(x_i)e^{-\frac{x_i^2}{2\Sigma_{i}^2} + x_i\frac{R_{i}}{\Sigma_{i}^2}} \(- x_i\frac{R_i}{\Sigma_i^2} + \frac{x_i^2}{2} \(\sum_\mu^M \frac{F_{\mu i}^2 (y_\mu - w_\mu)^2}{(\Delta + \Theta_\mu)^2} + \frac{1}{\Sigma^2_i}\)\)\nonumber \\
	&-\frac{a_i^2}{2} \sum_\mu^M \frac{F_{\mu i}^2 (y_\mu - w_\mu)^2}{(\Delta + \Theta_\mu)^2}
\end{align}
\begin{align}
	&=- a_i\frac{R_i}{\Sigma_i^2} + \frac{v_i+a_i^2}{2} \(\sum_\mu^M \frac{F_{\mu i}^2 (y_\mu - w_\mu)^2}{(\Delta + \Theta_\mu)^2}+ \frac{1}{\Sigma^2_i}\)-\frac{a_i^2}{2} \sum_\mu^M \frac{F_{\mu i}^2 (y_\mu - w_\mu)^2}{(\Delta + \Theta_\mu)^2}\\
	&=- a_i\frac{R_i}{\Sigma_i^2} + \frac{v_i+a_i^2}{2\Sigma_i^2} + \frac{v_i}{2}\(\sum_\mu^M \frac{F_{\mu i}^2 (y_\mu - w_\mu)^2}{(\Delta + \Theta_\mu)^2}\)
\end{align}
where we have used (\ref{eqChIntro:Theta}), (\ref{eqChIntro:w}) inside (\ref{eq1:bB}) which gives at first order: 
\begin{align}
	\sum_\mu^M B_{\mu i}^2 \approx \sum_\mu^M \frac{F_{\mu i}^2 (y_\mu - w_\mu)^2}{(\Delta + \Theta_\mu)^2}
\end{align}
and where we have recognized the expressions of the marginal posterior mean $a_i$ and variance $v_i$ ((\ref{eq:generic_a}), (\ref{eq:generic_v}), (\ref{eq1:marginal_B})). Finally, as we want to express everything in terms of quantities computed by the AMP algorithm, we notice that we can re-write $\tilde z_i$ in terms of the AMP variable partition function (\ref{eq1:AMPzi}) that we call here $z_i^{AMP} = \tilde z_i \exp\(-R_i^2/(2\Sigma_i^2)\)$. Finally the free energy reads:
\begin{align}
	F&\approx -\sum_\mu^M \log\(z_\mu\) - \sum_i^N \[\log\(z_i^{AMP}\) + \frac{(R_i-a_i)^2 + v_i}{2\Sigma_i^2} + \frac{v_i}{2} \sum_\mu^M \frac{F_{\mu i}^2 (y_\mu - w_\mu)^2}{(\Delta + \Theta_\mu)^2}\]\\
	&\approx -\sum_\mu^M \log\(z_\mu\) - \sum_i^N \[\log\(z_i^{AMP}\) + \frac{(R_i-a_i)^2 + v_i}{2\Sigma_i^2} \]- \sum_\mu^M \Theta_\mu\frac{(y_\mu - w_\mu)^2}{2(\Delta + \Theta_\mu)^2}
\end{align}
using (\ref{eq:appTheta}) for the second equality. All the approximations are again exact in the large dense graph limit. Now we notice from Fig.~\ref{algoCh1:AMP} that the fixed point conditions are:
\begin{align}
	\Theta_\mu\frac{y_\mu - w_\mu}{\Delta+\Theta_\mu} &= \sum_i^N F_{\mu i} a_i - w_\mu \label{eq:wmuFixedPoint1}\\
	\Theta_\mu &= \sum_i^N F_{\mu i}^2 v_i \label{eq:fixedPointTheta}\\
	a_i &= f_{a_i}(\Sigma_i^2, R_i) \label{eq:constai}\\
	v_i &= f_{c_i}(\Sigma_i^2, R_i) \label{eq:constvi}
\end{align}
Using (\ref{eq:wmuFixedPoint1}), we get the final expression of the Bethe free energy in terms of the AMP quantities {\it at their fixed point}:
\begin{align}
	F&\approx -\sum_\mu^M \[\log\(z_\mu\) + \frac{\(\sum_i^N F_{\mu i} a_i - w_\mu\)^2}{2\Theta_\mu}\] - \sum_i^N \underbrace{\[\log\(z_i^{AMP}\) + \frac{(R_i-a_i)^2 + v_i}{2\Sigma_i^2} \]}_{=-KL\(P_i||P_0\)}\label{eq:BetheF_forAMP_0}\\
	&= \frac{M}{2}\log\(2\pi\Delta\)+\sum_\mu^M \[\frac{\(y_\mu - w_\mu\)^2}{2\(\Delta + \Theta_\mu \)} + \frac{1}{2} \log\(1+\frac{\Theta_\mu}{\Delta}\)-\frac{\(\sum_i^N F_{\mu i} a_i - w_\mu\)^2}{2\Theta_\mu}\]\nonumber\\
	& + \sum_i^N KL\(P_i||P_0\) \label{eq:BetheF_forAMP}
\end{align}
where we have replaced (\ref{eq:zmu}) in the last equality and:
\begin{align}
	P_i(x_i|\Sigma_i^2,R_i) \defeq \frac{1}{\tilde z_i} P_0(x_i) \exp\(-(R_i - x_i)^2/(2\Sigma_i^2)\)
\end{align}
is as usual the AMP posterior measure (\ref{eq1:marginal_B}) of the variable $x_i$ and we use the Kullback-Leibler divergence (\ref{eq:KL_generic}). Now let us simplify even further this expression by re-writing (\ref{eq:wmuFixedPoint1}) which is true at the fixed point as:
\begin{align}
	w_\mu = \frac 1 \Delta \(\tilde a_\mu (\Delta+ \Theta_\mu) -\Theta_\mu y_\mu\) \label{eq:wmuFixedPoint2}
\end{align}
where we define $\tilde a_\mu \defeq \sum_i^N F_{\mu i} a_i$. We want to simplify (\ref{eq:BetheF_forAMP}) by working out the term:
\begin{align}
	&\frac{\(y_\mu - w_\mu\)^2}{2\(\Delta + \Theta_\mu \)} -\frac{\(\tilde a_\mu - w_\mu\)^2}{2\Theta_\mu}\\
	=& \frac{y_\mu^2}{2(\Delta + \Theta_\mu)} - \frac{\tilde a_\mu^2}{2\Theta_\mu} + w_\mu \(\frac{\tilde a_\mu}{\Theta_\mu} - \frac{y_\mu}{\Delta + \Theta_\mu}\) - \frac{w_\mu^2\Delta}{2\Theta_\mu(\Delta  + \Theta_\mu)} \label{eq:tocombine}
\end{align}
Now replacing $w_\mu$ by (\ref{eq:wmuFixedPoint2}), we get after careful simplifications:
\begin{align}
	 w_\mu \(\frac{\tilde a_\mu}{\Theta_\mu} - \frac{y_\mu}{\Delta + \Theta_\mu}\) &= \frac{1}{\Delta}\(\tilde a_\mu^2 \frac{\Delta + \Theta_\mu}{\Theta_\mu} - 2y_\mu \tilde a_\mu + \frac{y_\mu^2\Theta_\mu}{\Delta + \Theta_\mu}\)\\
	- \frac{w_\mu^2\Delta}{2\Theta_\mu(\Delta  + \Theta_\mu)} &= -\frac{1}{2\Delta}\(\tilde a_\mu^2 \frac{\Delta+\Theta_\mu}{\Theta_\mu} - 2y_\mu \tilde a_\mu + \frac{y_\mu^2\Theta_\mu}{\Delta+\Theta_\mu} \)
\end{align}
Combining everything in (\ref{eq:tocombine}), we get after simplification:
\begin{align}
	\frac{\(y_\mu - w_\mu\)^2}{2\(\Delta + \Theta_\mu \)} -\frac{\(\tilde a_\mu - w_\mu\)^2}{2\Theta_\mu} = \frac{1}{2\Delta}\(y_\mu - \tilde a_\mu\)^2
\end{align}
which allows to simplify the Bethe free energy (\ref{eq:BetheF_forAMP}) using the fixed point condition (\ref{eq:fixedPointTheta}):
\begin{align}
	F(\{\Sigma_i^2, R_i\}_i^N) =&\frac{1}{2} \sum_{\mu}^{M}\[\frac{(y_\mu - \sum_i^N F_{\mu i} a_i)^2}{\Delta} + \log\(1+\frac{ \sum_i^N F_{\mu i}^2 v_i}{\Delta}\)\] \nonumber\\
	&+\frac{M}{2}\log\(2\pi\Delta\)+ \sum_i^N KL(P_i||P_0)
	\label{eq:BetheF_forAMP_last}
\end{align}
which is only true at the fixed point of the algorithm, i.e. when all the constraints (\ref{eq:wmuFixedPoint1}), (\ref{eq:fixedPointTheta}), (\ref{eq:constai}), (\ref{eq:constvi}) are verified.

This expression of the free energy is only valid in the thermodynamic limit as we used the AMP approximations (and thus becomes exact in the infinite dense graph limit). Its expression remains the same in the vector variables case. It is the same as (24) in \cite{DBLP:journals/corr/KrzakalaMTZ14} where it has been first derived and is formally valid only at a fixed point of the algorithm, which is a minimum of this expression. It is important to understand that during the dynamic of the algorithm, if we plug at each step the AMP quantities Fig.~\ref{algoCh1:AMP} in this free energy, it does not necessarily decrease monotonously in time but at a fixed point, it is guaranteed that the algorithm reached a minimum of (\ref{eq:BetheF_forAMP_last}): it is a variational expression. The reasons behind why the Bethe free energy is variational when imposing the constraints  (\ref{eq:wmuFixedPoint1}), (\ref{eq:fixedPointTheta}), (\ref{eq:constai}), (\ref{eq:constvi}) are discussed in \cite{DBLP:journals/corr/KrzakalaMTZ14}. This form or any of the two previous ones (\ref{eq:BetheF_forAMP_0}), (\ref{eq:BetheF_forAMP}) can thus be optimized to derive learning equations, see next section. This free energy is also at the core of the methods used in \cite{DBLP:journals/corr/VilaSRKZ14} to solve convergence issues by using an adaptative damping in AMP. 
\subsection{Learning of the model parameters by expectation maximization}
\label{sec:EMlearning}
In the estimation task, the prior model parameters $\bsy \theta$ in (\ref{eqIntro:priorDef}) or the noise variance can be unknown as well. In this case one needs to learn them in some way. This can be done in the Bayesian framework similarly to the classical expectation maximisation method. We consider that $\theta$ is the scalar parameter that we want to learn. It starts from the Bayes formula:
\begin{align}
	P(\theta|\by) &= \frac{P(\by|\theta)P(\theta)}{P(\by)}\propto Z(\theta)P(\theta)
\end{align}
where we used that the partition function $Z(\theta)\propto P(\by|\theta)$ from (\ref{eqChIntro:BayesFormula}). Thus if no prior is assumed about $\theta$, then maximizing $Z$ (\ref{eq1:fullZ}) or equivalently minimizing the Bethe free energy, considering the other parameters fixed, gives the most probable value of $\theta$. So the method is simple: $i)$ take your favorite form of the Bethe free energy $F$, $ii)$ solve $\partial_\theta F(\theta|\bsy\Gamma) = 0$ from which you extract a fixed point equation verified by $\theta$ that has been isolated: $\theta = f(\theta|\bsy\Gamma^t)$ where $\bsy\Gamma$ is the set of all the other quantities on which depend $F$ and $f$ is a given function, $iii)$ finally add the time: $\theta^{t+1} = f(\theta^t|\bsy\Gamma^t)$ to get an iterative learning equation.

As the Bethe free energy expressions given previously are variational when imposing all the fixed point conditions (\ref{eq:wmuFixedPoint2}), (\ref{eq:fixedPointTheta}), (\ref{eq:constai}) and (\ref{eq:constvi}), we can use them in the expectation maximization procedure to fix the new values of the learned parameters. It must be understood that depending on the used form of the Bethe free energy (that are all equivalent), and even for a given fixed form, there exist usually many different ways to write fixed point equations verified by $\theta$. Thus a learning equation must be tried empirically to assess its efficiency even if all fixed point equations are a priori equivalent.

Let us apply the method to the Gaussian noise variance to obtain its generic learning equation, that does not depend on the prior model but only on the AMP fields. As we will see, different learnings can be derived. For example starting from the Bethe free energy (\ref{eq:BetheF_forAMP_0}), the noise variance will appear only in the factor term given by (\ref{eq:zmu}). A possible fixed point equation \cite{KrzakalaMezard12} that arise naturally is in this case:
\begin{align}
	\Delta^{t+1} = \[\sum_\mu^M \frac{(y_\mu -w_\mu^t)^2}{\(1 + \frac{\Theta_\mu^t}{\Delta^t}\)^2}\] \[\sum_\mu^M \(1 + \frac{\Theta_\mu^t}{\Delta^t}\)^{-1}\]^{-1}
\end{align}
where the AMP quantities iterations are given on Fig.~\ref{algoCh1:AMP}. But if instead we were starting from the perfectly equivalent expression (\ref{eq:BetheF_forAMP}), the most straightforward learning would be:
\begin{align}
	\Delta^{t+1} =\[\frac{1}{M}\sum_\mu^M \(\frac{(y_\mu -w_\mu^t)^2}{\(\Delta^t + \Theta_\mu^t\)^2} + \frac{\Theta_\mu^t}{\Delta^t(\Delta^t+\Theta_\mu^t)} \) \]^{-1}
\end{align}
Many other forms could be derived as for any learned parameter. The first learning equation will be essentially used in this thesis despite the second is valid as well. It can be easily checked that both expressions are the same at the fixed point as they should: any of the two equations can be derived from (\ref{eq:BetheF_forAMP_0}), (\ref{eq:BetheF_forAMP}) or (\ref{eq:BetheF_forAMP_last}) after simplifications.
\subsection{Dealing with non zero mean measurement matrices}
Despite the AMP derivation (see sec.~\ref{sec:classicalDerivationAMP}) does {\it not} rely on the fact that the measurement matrix has zero mean (as opposed to the state evolution, see sec.~\ref{sec:stateEvolutionGeneric}), it appears that AMP experiences strong convergence issues when the matrix has a finite mean. Recent works have shed light on solving this issue \cite{DBLP:journals/corr/VilaSRKZ14,DBLP:journals/corr/ManoelKTZ14,DBLP:journals/corr/GuoX15}, but these advanced methods are not used in the present thesis. In this work, we used a more classical trick to deal with this problem which starts from (\ref{eqIntro:AWGNCS}) and noticing that:
\begin{align}
\frac{1}{M}\sum_\mu^M y_\mu &= \frac{1}{M}\sum_\mu^M \(\sum_i^N F_{\mu i} s_i +\xi_\mu\) \\
\Rightarrow <\by> &= \sum_i^N \frac{1}{M}\(\sum_\mu^M F_{\mu i}\) s_i + <\bsy \xi>\\
&= \sum_i^N <\bF_{\bullet, i}> s_i
\end{align}
where we have used that the noise has zero mean (if it not the case, its mean can be included in the next rescaling as well). Finally if we call $\tilde \by \defeq \by - <\by>$ and $\tilde \bF_{\bullet, i} \defeq \bF_{\bullet, i} - <\bF_{\bullet, i}>$, we obtain the rescaled system where now the measurement matrix has zero mean:
\begin{align}
	\tilde \by = \tilde \bF \bs + \bsy \xi
\end{align}
that can now be solved without convergence issues. So now we always consider the measurement matrix to have zero mean, as anyway this trick can be used if it is not the case.
%
\chapter[Phase transitions, asymptotic analyzes and spatial coupling]{Phase transitions, asymptotic analyzes and spatial coupling}
\label{chap:phaseTrans_asymptAnalysies_spC}
In this chapter we preset the tools allowing for the asymptotic analyzes of the linear estimation problems and of the approximate message-passing algorithm. We start by discussing how the statistical physics notion of disorder is related to noisy linear estimation problems and give a first example of phase diagram in compressed sensing. Furthermore we introduce the notion of typical complexity phase transitions in sparse linear estimation problems, which connects even more this discipline to statistical physics. It appears that three different complexity regimes are present in linear estimation. Their definitions and implications will be discussed.

Then we will present the replica method of statistical physics of disordered system and use it to compute the potential of the linear estimation problem (\ref{eqIntro:AWGNCS}) in the general case of a signal with vector components. This potential which is the Bethe free entropy contains the information about the typical complexity of the inference problem as a function of the external parameters, the measurement rate and the noise variance and thus allows to obtain the phase diagram of the problem.

We will derive the state evolution recursions that allow to predict the asymptotic dynamical reconstruction perfomances of the approximate message-passing algorithm,  still in the vectorial case. This will be done starting directly from the approximate message-passing algorithm and then starting from the cavity equations as it is usually done. After a discussion on the Bayesian optimality under the prior matching condition, the link between the state evolution and replica analyzes will be underlined: we will show that the fixed points of the state evolution equations give back the optima of the replica potential, i.e. that the two analyzes despite being totally different in their derivations contain the same information on the static properties of the problem.

Finally, we introduce a central notion in this thesis that allows to perfom optimal inference from the information theoretical point of view: the spatial coupling technique. We will discuss its relation to physical concepts such as the nucleation theory and the notion of metastability. Then the approximate message-passing in a form more adapted to use in combination with spatial coupling and the state evolution will be derived. We end up showing how to go from the form of the algorithm presented here to the well known Montanari's notations.
%
\section{Disorder and typical complexity phase transitions in linear estimation}
\label{sec:typicalComplexity}
Statistical physics of disordered systems has been specifically created to deal with complex systems that are drawn from some stochastic process. For example, in a disordered spin model such as the fully connected Sherrington-Kirckpatrick spin glass, the interactions amplitudes between spins are random (in this case we speak about interaction disorder), or in a combinatorial optimization problem such as the independent set problem \cite{barbier2013hard}, the disorder is the graph instance itself (we speak about structural disorder). The specifically designed replica method \cite{MezardParisi87b,mezard2009information} allows to perform the necessary thermodynamic averages over these new sources of randomness that are distincts from the pure thermal agitation, i.e. the entropic contribution. It allows to compute the free entropy (i.e. minus the free energy) of the system, its potential function which should not depend on the specific disorder instance in the thermodynamic limit due to the self-averageness of the thermodynamic quantities.

The sources of randomness in the linear estimation problem (\ref{eqIntro:AWGNCS}) are coming from the measurement noise (that plays the role of a temperature), the random measurement matrix and signal realizations (which induce an interaction disorder). So the replica method is the method of choice and can be "straightforwardly" used in the present context to compute the potential and extract from it the phase diagram of the problem. This potential is minus the Bethe free energy (\ref{eq:BetheF_formal}), (\ref{eq:BetheF_forAMP}) averaged over these disorder sources as in (\ref{eq:BetheF_formal}),(\ref{eq:BetheF_forAMP}) all the quantities are dependent on the specific problem instance. Equivalently it is its thermodynamic limit by the self-averageness property. The Bethe free energy in its form (\ref{eq:BetheF_formal}), (\ref{eq:BetheF_cavMess}) or especially (\ref{eq:BetheF_forAMP}) is adapted for single instances of the problem, such as when deriving learning equations, see sec.~\ref{sec:EMlearning}.

Before to present in details the replica computation of the potential in linear estimation problems, we present now the typical phase diagram for sparse linear estimation and discuss the nature of the different phase transitions and typical complexity phases that exist.
\subsection{Typical phase diagram in sparse linear estimation and the nature of the phase transitions}
\label{sec:typicalPhaseTransitions}
The replica computation of the next section allows to compute the Bethe free entropy $\Phi(E)$ of the problem, where $E$ is the $MSE$ (\ref{eqIntro:MSE}) of the reconstruction. This function contains the information on the different phase transitions that happen in the problem as the control parameters, the signal density of non zero (or "large" ones in the approximates sparsity setting, see chap.~\ref{chap:appSparsity}) components $\rho$ and the Gaussian noise variance $\Delta$ are tuned. These transitions separate three distinct phases of typical complexity: Fig.~\ref{eqChIntro:phaseDiagIntro} is the typical phase diagram in the $(\alpha,\rho)$ plane that we will encounter in this thesis. The details of the problem to which it corresponds will be exposed later on but it it not the point here, it is more to understand the general scenario that happens in linear estimation.

Glassy systems share common features with the present problem (\ref{eqIntro:AWGNCS}), such as the fact that in the out of equilibrium glass phase, local algorithms are inefficient for sampling the solution space exactly like in the hard phase of the present problem defined hereafter. But it is important to underline that the physics behind these two phases is different. The present inference problem is replica symmetric under the prior matching condition \cite{KabashimaKMSZ14} and thus {\it does not} formally present the typical glass phenomenology like the splitting of the solution space into exponentially many disconnected clusters, i.e. the replica symmetry breaking. Let us now present in details the different phases of the problem.
\subsubsection{The "very" easy and easy phases} 
\begin{figure}[!ht]
\centering
\includegraphics[width=.8\textwidth]{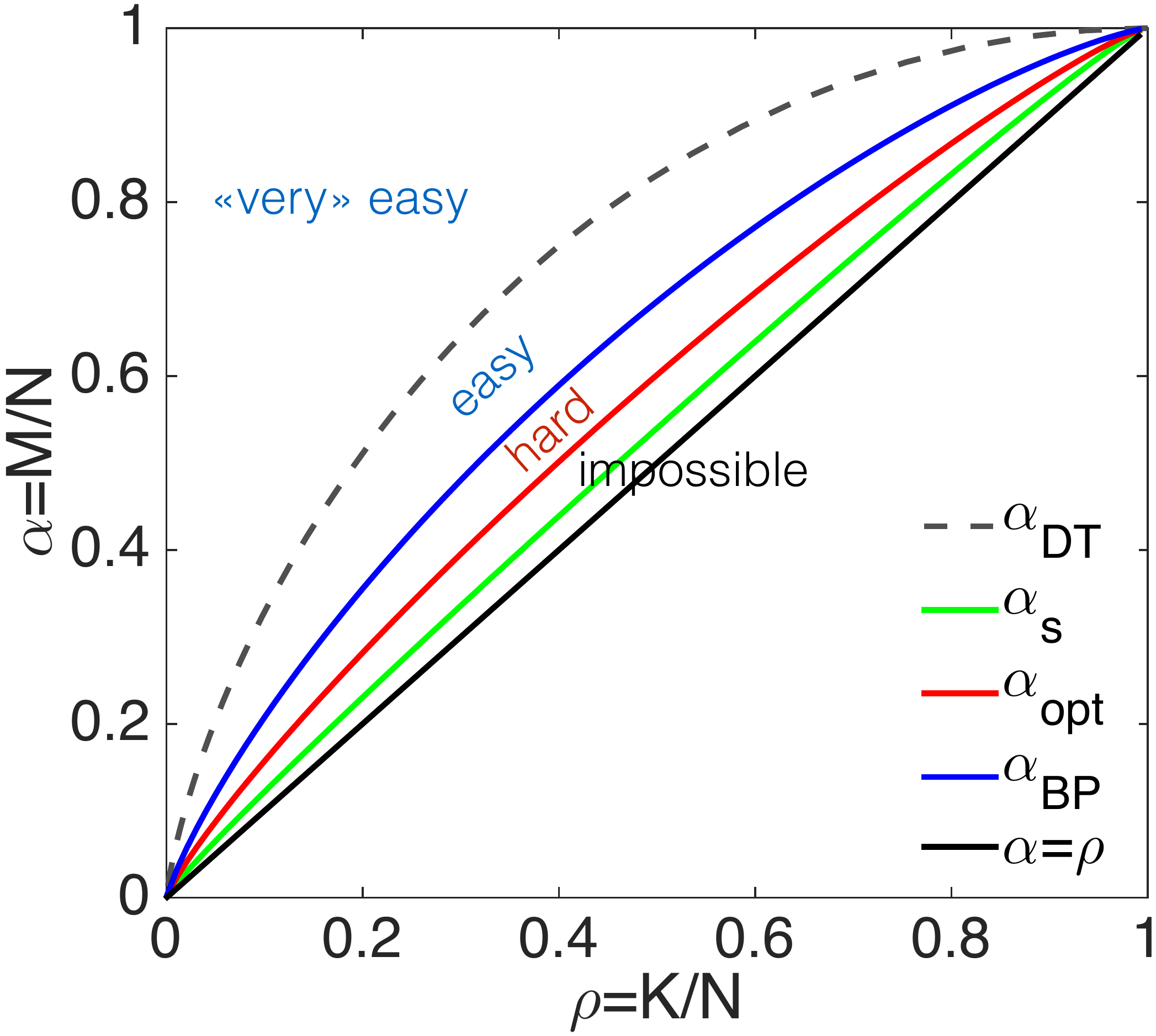}
\caption[Typical phase diagram in sparse linear estimation under sparsity assumption]{Typical phase diagram of linear estimation problems under sparsity assumption in the plane (density of the signal $\rho$, measurement rate $\alpha$). Are plotted the different phase transitions that appear in the problem and the typical complexity phases.
The Donoho-Tanner line $\alpha_{DT}(\rho)$ separates the "very" easy phase where $\ell_1$ optimization solvers are efficient from the easy phase, where the potential has a unique "low" $MSE$ maximum and thus message-passing is Bayes optimal (under the prior matching condition) with good reconstruction results whereas convex optimization is not. The first order BP phase transition $\alpha_{BP}(\rho)$ is the largest $\alpha$ for which the potential function $\Phi(E)$ has two coexisting maxima. Below this line in the hard phase, message-passing is blocked in a metastable state which is not the $MMSE$ estimate and reconstruction fails, at least without spatial coupling. This hard phase is the gap we want to close thanks to spatial coupling, see sec.~\ref{sec:spatialCoupling}. The optimal transition $\alpha_{opt}(\rho)$ is the $\alpha$ for which the two coexisting maxima of the potential have the same height, i.e. the smallest $\alpha$ at fixed $\rho$ (or highest $\rho$ at fixed $\alpha$) for which it is theoretically possible to find the signal. In the impossible phase, no algorithm can solve the estimation problem. Finally the static transition $\alpha_{s}(\rho)$ is defined as the smallest $\alpha$ for which the potential function $\Phi(E)$ has two local maxima. Below this line, all information about the signal is lost. The solid black curve correponds to the $\alpha = \rho$ line and is the fundamental limit of reconstruction (i.e. the optimal transition) in the noiseless limit.}
\label{eqChIntro:phaseDiagIntro}
\end{figure}
\begin{figure}[!t]
\includegraphics[width=.8\textwidth]{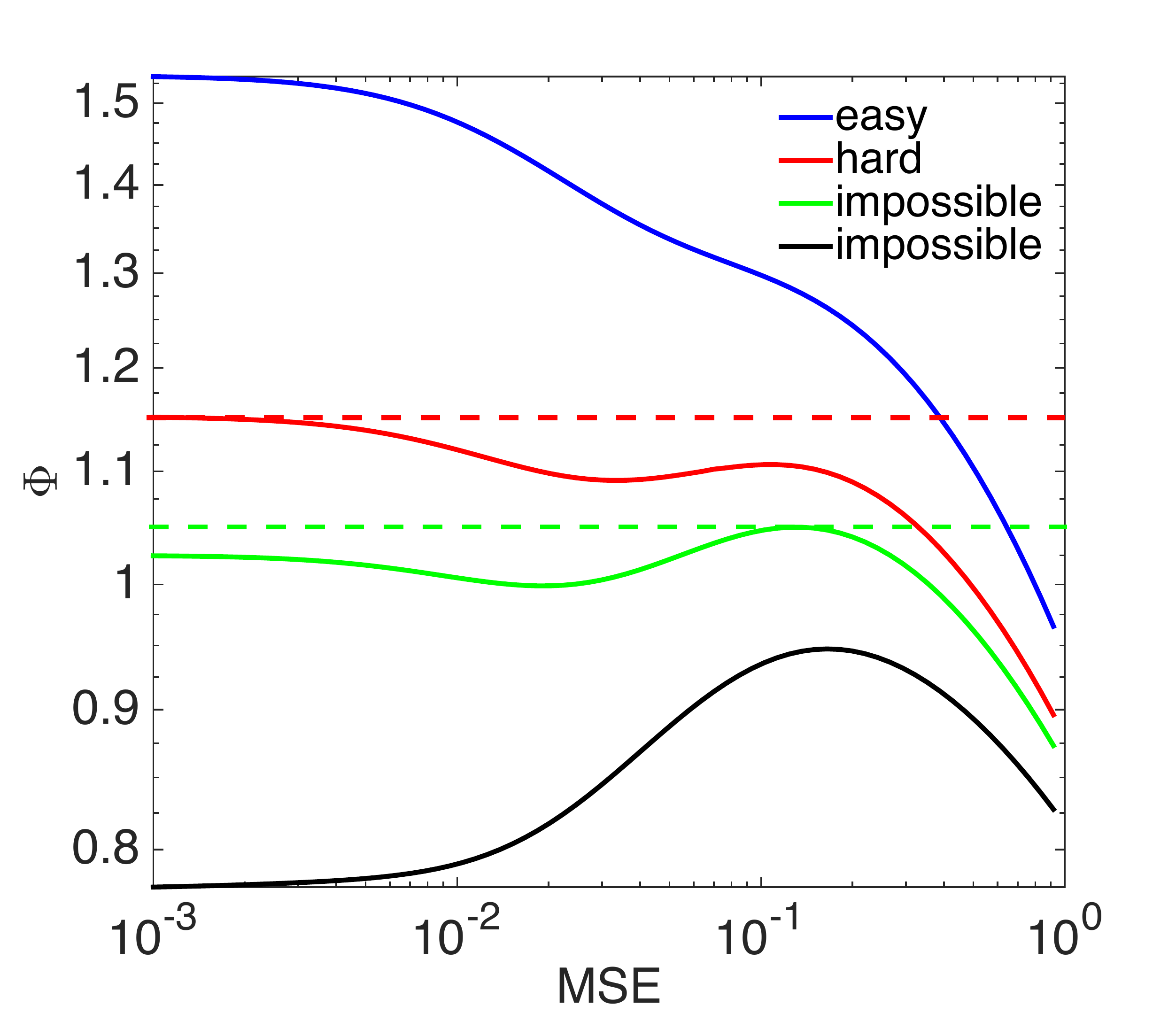}
\caption[Bethe free entropy shape in the different typical complexity phases]{The Bethe free entropy function $\Phi(E)$ for a noisy compressed sensing problem in the different typical complexity phases. In the easy phase, the $MMSE$ solution is unique, and the gradient ascent starting from high $MSE$ performed by the message-passing will find it. In the hard phase, the $MMSE$ solution is still the global maximum but it coexists with a local high $MSE$ fixed point blocking the message-passing (without spatial coupling sec.~\ref{sec:spatialCoupling}). In the impossible phase the equilibrium becomes the wrong solution and then the $MMSE$ metastable fixed point totally disappears below the static transition: no information about the solution is present anymore in the posterior distribution (\ref{eqChIntro:BayesFormula}).}
\label{eqChIntro:replicaPotentialIntro_phases}
\end{figure}
The first {\it "very" easy phase} corresponds to the region above the Donoho-Tanner transition, the grey dashed line on Fig.~\ref{eqChIntro:phaseDiagIntro}. In this region $\ell_1$ optimization solvers (see sec.~\ref{sec:convexOptimization}) are efficient for reconstructing the signal, but if sparsity only is known about the signal, they are not anymore in the {\it easy phase} which corresponds to the region between the Donoho-Tanner and BP transitions. In this region, the potential has a unique maximum corresponding to the $MMSE$ estimate at a "low" $MSE$, see the blue line on Fig.~\ref{eqChIntro:replicaPotentialIntro_phases}. By low, we mean compared to the second maximum of the potential appearing below the BP transition explained after. 

In a pure compressed sensing problem where the only knowledge about the signal is that it is sparse, the gap between the Donoho-Tanner and BP transitions is due to the fact that the minimum $\ell_1$ solution of the LASSO regression (\ref{eqIntro:l1problem_relax1}) does not match the minimizer of the $\ell_0$ equivalent of (\ref{eqIntro:l1problem_relax1}), i.e. the sparsest solution of the linear system, which is the true solution of the problem. If more knowledge is known about the signal, $\ell_1$ optimization solvers are anyway not Bayes optimal as they do not take it into account, it only seeks for the solution with minimum $\ell_1$ norm, as opposed to AMP.

In the easy phase, local algorithms such as message-passing or monte-carlo based methods are able to efficiently sample the posterior distribution which is Bayes optimal under the prior matching condition. The "easyness" of this region can be understood thanks to the Bethe free entropy. As we have shown in sec.~\ref{sec:BPfromBethe} the message-passing algorithm fixed points correspond to the optima of this potential (on a single instance). In sec.~\ref{subsec:repIsSE} we will show that it remains the case in the thermodynamic limit. Thus we can interpret the algorithm dynamics as a sort of gradient ascent of this potential starting from a high $MSE$ random initialization, until a fixed point is reached and the algorithm converges. Thus when the maximum is unique, AMP will find it and is thus Bayes optimal as it gives the true $MMSE$ estimate. This interpretation of gradient ascent is not rigorously correct in the sense that nothing proves that the Bethe free energy (\ref{eq:BetheF_forAMP}) is strictly decreased (or the Bethe free entropy strictly increased) at each time step by the message-passing but as BP (and thus AMP which is just its limit on dense graphs) is derived as fixed point equations of the potential (see sec.~\ref{sec:BPfromBethe}), if the AMP converges then the free entropy {\it must} have been increased until a maxima (local or global) during the dynamics of the message-passing. Further details about the Bethe free energy and its optimization during the message-passing dynamics can be found in \cite{DBLP:journals/corr/KrzakalaMTZ14}.
\subsubsection{The hard phase and the BP transition} 
\begin{figure}[!t]
\includegraphics[width=.9\textwidth]{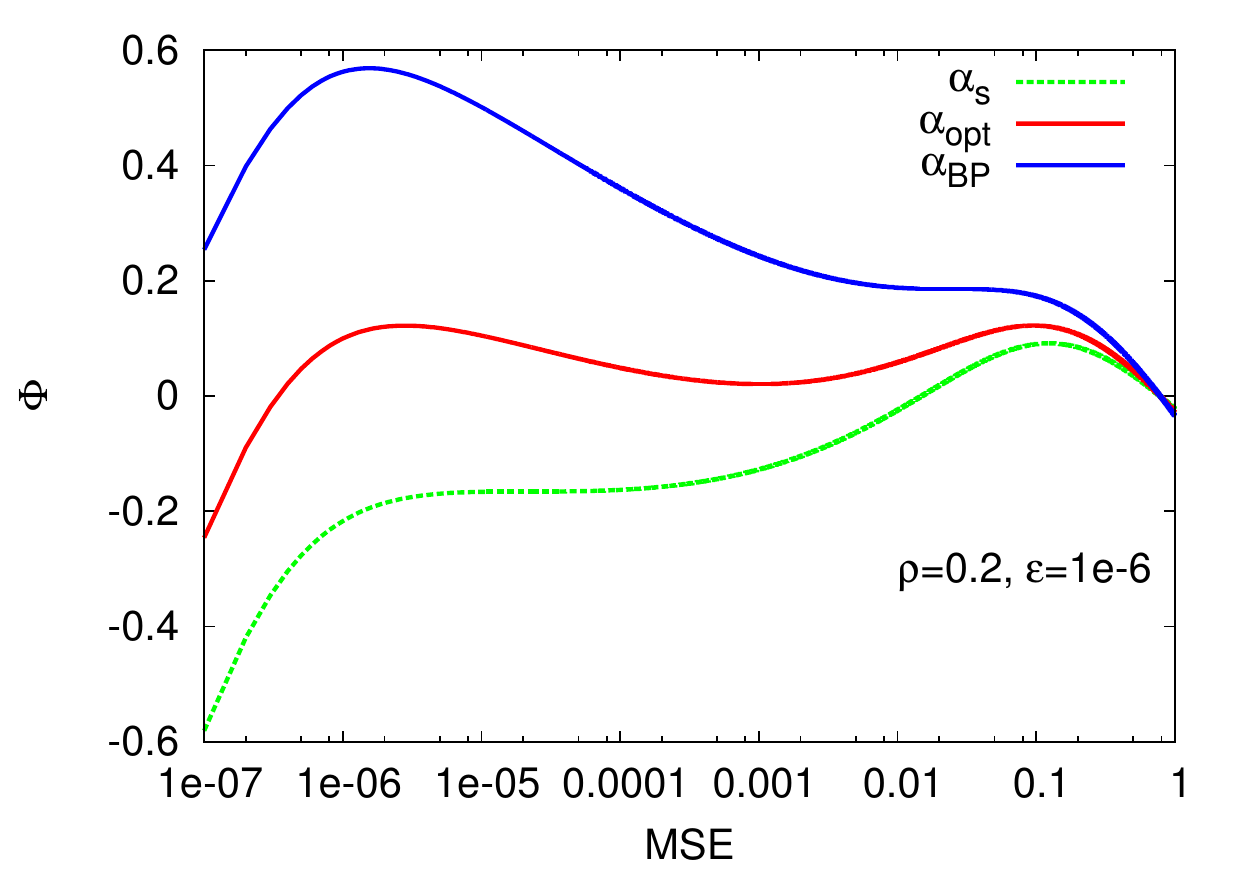}
\caption[Bethe free entropy at the different transitions]{The Bethe free entropy $\Phi(E)$ for signals of density $\rho=0.2$, with an effective noise of variance $\Delta \in O(10^{-6})$. The three lines depict the potential for three different measurement rates corresponding to the critical values: $\alpha_{BP}=0.3559$, $\alpha_{opt}=0.2817$, $\alpha_s=0.2305$. The two local maxima exists for $\alpha \in (\alpha_s,\alpha_{BP})$, at $\alpha_{opt}$ the low $MSE$ maxima becomes the global one. This scenario is typical in linear estimation problems.}
\label{eqChIntro:replicaPotentialIntro_trans}
\end{figure}
The second phase, right below the blue line and above the red one on the phase diagram Fig.~\ref{eqChIntro:phaseDiagIntro} is called the {\it hard phase}. In this region of parameters, the problem is theoretically solvable but local algorithms cannot anymore sample the posterior. It correponds to the regime where the potential has now two distincts maxima, see the red line in Fig.~\ref{eqChIntro:replicaPotentialIntro_phases}: the one at low $MSE$ corresponding to the Bayes optimal $MMSE$ estimate is present and still corresponds to the global maximum but it coexists with another spurious local maximum corresponding to a high $MSE$ solution of the message-passing equations. This solution is refered as the {\it metastable state} as it not the true equilibrium given by the free entropy global maximum. So the message-passing climbing the potential starting from high $MSE$ will reach this fixed point before being able to see the $MMSE$ one, and thus reconstruction will fail. With a monte carlo or any other local algorithm (which is of course initialized randomly, thus in the basin of attraction of the metastable wrong solution), only an exponentially long sampling of the solution space, passing through exponentially rare configurations with a low free entropy would allow the algorithm to ultimately reach the true equilibrium, at least without spatial coupling (sec.~\ref{sec:spatialCoupling}). This blocking of local algorithms is observed also in glassy phases of combinatorial optimization problems, but in this case the free entropy is way more rough than Fig.~\ref{eqChIntro:replicaPotentialIntro_phases} due to the spontaneous replica symmetry breaking.

The transition between the easy and hard phases is called the {\it BP transition} or {\it spinodal transition} as it corresponds to the separation between a region where message-passing based algorithms are Bayes optimal from one where it fails. Exactly at the transition point, the Bethe free entropy has typically the shape given by the blue line on Fig.~\ref{eqChIntro:replicaPotentialIntro_trans} with a plateau appearing at high $MSE$: $\alpha_{BP}(\rho)$ is defined as the largest $\alpha(\rho)$ for which the potential has two local maxima. This transition does {\it not} affect the Bayes optimal estimator, it is an algorithmic phase transition of the first order type: the order parameter which is the asymptotic reconstruction $MSE$ we would get by message-passing (without spatial coupling) jumps {\it discontinuously} from a low (in the noisy setting) or zero (in the noiseless setting) value when situated in the easy phase to a high value in the hard phase, the jump happening exactly on the transition line. But we will see in the next that when the noise is very large, this transition can become continuous (there is no more sharp transition), and only one maximum will exist at any measurement rate that smoothly moves to a lower $MSE$ solution as the measurement rate increases.

This plateau of the free entropy explains the phenomenon of {\it critical slowing down} close to the BP transition, a phenomenon typical from first order transitions: interpreting again the message-passing as a gradient ascent of the potential, as the paramaters of the problem get closer to the BP transition this plateau with very low gradient reduces greatly the convergence rate of the algorithm until it becomes infinitely long in the large limit exactly at the transition and then it fails. The same behavior happens in many other problems such as random K-SAT \cite{MezardParisiZecchina02} and is inherent to local algorithms facing first order transitions. The same phenomenon of appearance of a metastable state preventing the equilibrium to be reached occurs in the supercooled state, where a liquid is blocked in the liquid state despite the temperature is below its critical temperature of solidification.
\subsubsection{The impossible phase and the optimal transition} 
The third phase is all the region below the red line on Fig.~\ref{eqChIntro:phaseDiagIntro}. Is it the {\it impossible phase}. Here, no algorithm is able to find the solution as the potential is dominated by a high $MSE$ solution and thus even the Bayes optimal inference would fail, see the green and black curves on Fig.~\ref{eqChIntro:replicaPotentialIntro_phases}. 

The first order transition between the hard and impossible phases, called the {\it optimal transition} $\alpha_{opt}(\rho)$ happens when the two coexisting free entropy maxima have exactly the same height, i.e. at the exact parameters values where the $MMSE$ solution (the global maximum) jumps from a low $MSE$ to an high one, see the red curve on Fig.~\ref{eqChIntro:replicaPotentialIntro_trans}.

From an algorithmic point of view and running purely local algorithms (i.e. without considering spatial coupling), this transition will not be detected as the algorithm was failing before it in the hard phase, and continues to do so after in the impossible phase.
\subsubsection{The static transition} 
Despite the fact that in all this impossible phase the algorithm behaves the same, from the potential (and thus thermodynamical) point of view, another transition occurs denoted as the {\it static} transition: $\alpha_{s}(\rho)$ is defined as the smallest $\alpha$ for which the potential function $\Phi(E)$ has two local maxima. It separates the impossible phase into two distinct phases, "one more impossible than the other". In the impossible phase before that the static transition happens, the potential is dominated by the high $MSE$ solution (by definition of the impossible phase) but it remains a local maximum of the potential corresponding to the $MMSE$ estimate, which is now the metastable state. This maximum could be reached initializing the message-passing equations close enough to the solution, inside the basin of attraction of this state. But anyway, the true equilibrium correponds to a failure of the reconstruction. The static transition happens when this low error metastable state disappears, see the green curve on Fig.~\ref{eqChIntro:replicaPotentialIntro_trans}, and only remains the high $MSE$ state, see the black curve on Fig.~\ref{eqChIntro:replicaPotentialIntro_phases}. Below the static transition, even if we would initialize the reconstruction algorithm on the solution itself, it would converge to the high $MSE$ state: it does not remain any information about the signal in the posterior distribution (\ref{eqChIntro:BayesFormula}) because there are too few measurements for the signal density, there are too noisy or both at the same time.
\subsubsection{Sum up of the different transitions and algorithmic implications}
To summarize, the AMP algorithm exhibits a double phase transition. It is asymptotically equivalent to the optimal Bayes inference at large $\alpha_{BP}<\alpha<1$, where it matches the optimal reconstruction with a small value of the $MSE$. At
low values of $\alpha<\alpha_{opt}$ the AMP is also
asymptotically equivalent to the optimal Bayes inference as the potential has only one maximum, but in this low-sampling-rate region the optimal result leads to a
large $MSE$. In the intermediate region $\alpha_{opt}<
\alpha<\alpha_{BP}$, AMP leads to large $MSE$, but the optimal
Bayes inference leads to low $MSE$. This is the region where one needs to improve on AMP, using for example the spatial coupling technique discussed in sec.~\ref{sec:spatialCoupling}.

It must be understood that all these considerations and phase transitions except the Donoho-Tanner one are properties of the problem itself, not of the algorithm used to reconstruct the signal. This potential, the Bethe free entropy, is conjectured exact in the thermodynamic limit for such dense graphs. But it does not mean that the easy phase is easy for all reconstruction algorithms, it means that mean field methods are appropriate and are able to sample properly the posterior until the BP transition under the prior matching condition. But as we have seen, convex $\ell_1$ optimization based solvers cannot reconstruct averywhere in the easy phase: they experience another phase transition preventing the solver to reconstruct well before the BP transition.
%
\section{The replica method for linear estimation over the additive white Gaussian noise channel}
\label{sec:replicaAnalyisGeneric}
%
The replica method leads to asymptotically exact evaluation of the logarithm of the
partition function $Z$ (\ref{eq1:fullZ}), which is here the Bethe free entropy. In general, if the partition function can be evaluated precisely then the marginal means with respect to the true posterior and the associated $MSE$ of the optimal inference can be computed. More generally, it is a method for averaging logarithms of complex functions depending on some disorder. It does not always lead to a Bethe form of the potential but it appears that in cases where the factor graph corresponding to the model of interest is not a tree nor dense, the replica method is far too complex to be applied. It is thus usual to confound the potential extracted from the replica analysis and the Bethe free entropy as it can be applied only in cases where the Bethe free entropy is the true potential. This is the case in our problem (\ref{eqIntro:AWGNCS}), the Bethe free entropy (\ref{eq:BetheF_formal}), (\ref{eq:BetheF_forAMP}) is actually exact because the problem is dense and thus its average matches the potential extracted from the replica analysis. 

The optimal reconstruction in compressed sensing, information that we will extract from the replica potential, was studied extensively in \cite{WuVerdu11}. The replica method was used in compressed sensing similarly as in the present thesis in \cite{RanganFletcherGoyal09,GuoBaron09} for example. Let us now derive the potential of the problem (\ref{eqIntro:AWGNCS}). All the computations are made considering that the signal is made of $B$-d sections (see Fig.~\ref{figChIntro:1dOp}) and that the prior is factorizable over them, with the same prior $P_0$ independently of the section. Furthermore we assume the prior matching condition which as already explained implies the replica symmetry in inference problems.
\subsection{Replica trick and replicated partition function}
We start from the definition of the free entropy potential at fixed section size $B$:
\begin{align}
	\Phi_B &\defeq \lim_{L\to \infty} \frac{1}{L}\mathbb{E}_{\bF,\bxi,\bs}\{\log\(Z(\bF,\bxi,\bs)\)\}\\
	Z(\bF,\bxi,\bs) &= \int \bigg[\prod_{l}^L d \bx_l P_0(\bx_l)\bigg] \prod_{\mu}^M \sqrt{\frac{{\rm{snr}}}{2\pi}} e^{-\frac{{\rm{snr}}}{2}\(\sum_{l}^L \bF_{\mu l}^{\intercal}(\bs_l-\bx_l) + \xi_\mu\)^2} \label{eq1:fullZ}
\end{align}
where we have used (\ref{eqIntro:AWGNCS}) to replace $\by$ and $Z$ is the normalization constant of the full posterior distribution (\ref{eqChIntro:BayesFormula}), i.e. the partition function, a random variable of the disorder. It can be transformed using the so called replica trick, a trivial mathematical identity:
\begin{equation}
\Phi_B  = \lim_{L\to \infty}\lim_{n\to0}\frac{1}{L} \frac{\mathbb{E}_{\bF,\bxi,\bs}\{Z^n\}-1}{n}\label{eq1:replicaTrick}\\
\end{equation}
where $\mathbb{E}_{\bF,\bxi,\bs}$ is the average over all the sources of disorder in (\ref{eqIntro:AWGNCS}). So the problem of computing the free energy is converted into computing the $n^{th}$ moment of the partition function. $Z^n$ is the so-called replicated partition function as it can be interpreted as the partition function of $n$ independent systems drawn from the same distribution, refered as the replicas. The index associated with the replicas is $a \in \{1,..,n\}$:
\begin{align}
Z^n &~= \left({\rm{snr}/2\pi} \right)^{\frac{Mn}{2}} \int \bigg[\prod_{l,a}^{L,n} d\bx_l^a P_0(\bx_l^a) \bigg]\prod_{\mu}^M e^{-\frac{{\rm{snr}}}{2}\sum_{a}^n \(\sum_{l}^L \bF_{\mu l}^{\intercal}(\bs_l - \bx_l^a) + \xi_\mu \)^2} \label{eq:repZ0}
\end{align}
The average replicated partition function can be rearranged as:
\begin{align}
\mathbb{E}_{\bF,\bxi,\bs}\{Z^n\} &~= \left({\rm{snr}/2\pi} \right)^{\frac{Mn}{2}} \mathbb{E}_\bs \left\{\int \bigg[\prod_{l,a}^{L,n} d\bx_l^a P_0(\bx_l^a) \bigg]\prod_{\mu}^M X_\mu\right\}
\label{eq:repZ1}
\end{align}
where we have defined:
\begin{align}
	X_\mu &\defeq \mathbb{E}_{\bF,\bxi}\left\{e^{-\frac{{\rm{snr}}}{2}\sum_{a}^n \(\sum_{l}^L \bF_{\mu l}^{\intercal}(\bs_l - \bx_l^a) + \xi_\mu \)^2}\right\}\\
	&~= \mathbb{E}_{\bF,\bxi}\left\{e^{-\frac{{\rm{snr}}}{2}\sum_{a}^n (v_\mu^a)^2}\right\}\label{eq1:Xmu}\\
	v_\mu^a &\defeq \sum_{l}^L\bF_{\mu l}^{\intercal} (\bs_l - \bx_l^a) + \xi_\mu
\end{align}
In order to compute $X_\mu$ (\ref{eq1:Xmu}), we can apply the central limit theorem to the quantity $v_\mu^a$ which is a sum of independent terms as the measurement matrix is i.i.d. We thus need its first two moments to define its associated Gaussian distribution. Using the fact that both the measurement matrix and the i.i.d Gaussian noise have zero mean we get:
\begin{align}
\mathbb{E}_{\bF,\bxi} \{v_\mu^a\} &= 0 \\
\mathbb{E}_{\bF,\bxi} \{(v_\mu^a)^2 \} &= \mathbb{E}_{\bF,\bxi} \bigg\{\sum_{l,k}^{L,L} [\bF_{\mu l}^{\intercal} (\bs_l - \bx^a_l)]^{\intercal} \bF_{\mu k}^{\intercal} (\bs_k - \bx^a_k) + 2\xi_\mu \sum_{l}^L \bF_{\mu l}^{\intercal} (\bs_l - \bx^a_l) + \xi_\mu^2\bigg\}\nonumber\\
&= \sum_{l,k}^{L,L} \left[(\bs_l - \bx^a_l)^{\intercal} \mathbb{E}_{\bF}\left\{\bF_{\mu l} \bF_{\mu k}^{\intercal}\right\} (\bs_k - \bx^a_k) \right] + {1/{\rm{snr}}}
\end{align}
Using the fact that each element of the matrix is i.i.d with variance $1/L$, we find that only the diagonal elements of the matrix $\mathbb{E}_{\bF}\left\{\bF_{\mu l} \bF_{\mu k}^{\intercal}\right\}$ are non zero: 
\begin{align}
&\mathbb{E}_{\bF}\left\{\bF_{\mu l} \bF_{\mu k}^{\intercal}\right\} = \frac{\delta_{k,l}}{L} \boldsymbol{I}_B \\
\Rightarrow &\mathbb{E}_{\bF,\bxi} \{(v_\mu^a)^2\} = 1/L \sum_{l}^L(\bs_l - \bx^a_l)^{\intercal}(\bs_l - \bx^a_l) + {1/{\rm{snr}}} \label{eq1:vmu2}
\end{align}
where $\boldsymbol{I}_B$ is the identity matrix of dimension $B\times B$. Now we define new macroscopic order parameters which will be considered as the new degrees of freedom of the replicated system, instead of the individual replica states $\{\bx_a\}$. In this way, we average out the microscopic properties of the system (the randomness instance, i.e. the replica individual states) to get access to the macroscopic ones (the observables, such as the $MSE$), the goal of the replica methodology. This is referred as coarse-graining in physics:
\begin{align}
&m_a \defeq 1/L\sum_{l}^L (\bx_l^a)^{\intercal} \bs_l \label{eq1:repOrderParam_ma}\\ 
&Q_a \defeq 1/L\sum_{l}^L(\bx_l^a)^{\intercal} \bx_l^a \label{eq1:repOrderParam_Qa}\\ 
&q_{ab} \defeq 1/L \sum_{l}^L (\bx_l^a)^{\intercal} \bx_l^b\label{eq1:repOrderParam_qab}
\end{align}
$m_a$ is the overlap between the replica state $\hat\bx^a$ and the signal $\bs$, $Q_a$ is the power (or self-overlap) of the replica $a$ and $q_{ab}$ is the overlap between replicas $a$ and $b$. Rewriting the previous moment (\ref{eq1:vmu2}) in terms of these new quantities, we get:
\begin{align}
\mathbb{E}_{\bF,\bxi} \{(v_\mu^a)^2 \} &= <\bs^2> - 2m_a + Q_a + {1/{\rm{snr}}}
\label{eq:mom_v_1}
\end{align}
Exactly in the same way, we get the cross terms $\forall\ a\neq b$:
\begin{equation}
\mathbb{E}_{\bF,\bxi} \{v_\mu^a v_\mu^b\} = <\bs^2> - (m_a + m_b) + q_{ab} + {1/{\rm{snr}}}
\label{eq:mom_v_2}
\end{equation}
The dependence on the measurement index $\mu$ of $v_\mu^a$ (and thus $X_\mu$ as well (\ref{eq1:Xmu})) is lost due to the averaging. Thus we note $v_\mu^a=v^a$, $X_\mu = X$. We now apply the replica symmetric ansatz which is valid for inference problems (and more generally planted problems) on locally tree-like or highly dense graphs under the prior matching condition such as in the present case \cite{mezard2009information,KabashimaKMSZ14}. In the replica method, it is expressed by removing the replica indices in the macroscopic order parameters (thus the name of the ansatz):
\begin{align}
q_{ab}&=q \ \forall \ (a,b : a\neq b)\\ 
Q_a &= Q \ \forall \ a \\ 
m_a &= m \ \forall \ a
\end{align}
Now we have computed the necessary moments (\ref{eq:mom_v_1}), (\ref{eq:mom_v_2}) we can write the covariance matrix $\bG$ of $\{v^a\}$ under this ansatz which reads $\forall\ (a,b)$:
\begin{align}
G_{ab} &\defeq  \mathbb{E}_{\bF,\bxi} \{v^a v^b\} \\
&= <\bs^2> - 2m + {1/{\rm{snr}}} + q + (Q - q)\delta_{a,b} \\
\Rightarrow \bG &= \left(<\bs^2> - 2m + {1/{\rm{snr}}} + q\right)\boldsymbol{1}_n + (Q - q)\boldsymbol{I}_n
\end{align}
where $\boldsymbol{1}_n$ is a matrix full of ones of dimension $n\times n$. We thus have:
\begin{align}
X &= \mathbb{E}_{\bv} \{e^{-\frac{{\rm{snr}}}{2} \bv^{\intercal} \bv}\} \\ 
P(\bv) &= [(2\pi)^n \text{det}(\bG)]^{-1/2} e^{-\frac{1}{2}\bv^{\intercal} \bG^{-1} \bv}
\end{align}
The explicit computation of $X$ by Gaussian integral gives:
\begin{align}
X &= [(2\pi)^n \text{det}(\bG)]^{-1/2} \int d\bv e^{-\frac{1}{2}\bv^{\intercal} (\bG^{-1} + {\rm{snr}} \boldsymbol{I}_n) \bv} \\
&= \[\text{det}\left(\boldsymbol{I}_n + {\rm{snr}} \bG \right)\]^{-1/2}
\end{align}
The eigenvectors of $\bG$ are one eigenvector $[1]_{a}^n=[1,1,\ldots,1]$ with associated eigenvalue $Q - q + n \left(1 - 2m + {1/{\rm{snr}}} + q \right)$ and $n-1$ eigenvectors of the type $[0,..,0,-1,1,0,..,0]$ with the couples $[-1,1]$ shifting by one component from one eigenvector to the next. Their degenerated eigenvalue is $Q-q$. Therefore:
\begin{align}
\text{det}\left(\boldsymbol{I}_n + {\rm{snr}} \bG \right) = \frac{1 + {\rm{snr}} \left[Q-q + n \left(<\bs^2> - 2m + {1/{\rm{snr}}} + q \right) \right]}{\left[1 + {\rm{snr}} (Q-q)\right]^{1-n}}
\end{align}
from which we get:
\begin{equation}
\lim_{n\to0} X = e^{-\frac{n}{2} \left[ \frac{q - 2m + <\bs^2> + {1/{\rm{snr}}}}{Q - q + {1/{\rm{snr}}}}+ \log({1/{\rm{snr}}} + Q - q) - \log({1/{\rm{snr}}}) \right] }
\end{equation}
When computing (\ref{eq:repZ1}), we need to enforce the constraints that the new order parameters satisfy their definitions (\ref{eq1:repOrderParam_qab}). This is done by the usual trick of rewriting $1$ as the inverse Fourier transform of its Fourier transform and plugging this expression in the definition of the averaged replicated partition function that we are computing:
\begin{align}
1 &= \int \bigg[\prod_{a}^n dQ_ad\hat Q_a dm_a d\hat m_a\bigg] \bigg[\prod_{b,a<b}^{n,(n-1)/2} dq_{ab}d\hat q_{ab}\bigg] \nonumber\\
&\exp\bigg[-\sum_a^n \hat m_a(m_a L - \sum_{l}^L (\bx_l^a)^{\intercal} \bs_l) + \sum_a^n \hat Q_a (Q_aL/2 - 1/2 \sum_l^L(\bx_l^a)^{\intercal} \bx_l^a)\nonumber \\
&- \sum_{b,a<b}^{n,(n-1)/2} \hat q_{ab}(q_{ab}L - \sum_l^L (\bx_l^a)^{\intercal} \bx_l^b)\bigg] \label{eq1:1Fourier}
\end{align}
Plugging this into the average replicated partition function (\ref{eq:repZ1}) expression we get:
\begin{align}
&\mathbb{E}_{\bF,\bxi,\bs}\{Z^n\} = \left({\rm{snr}/2\pi} \right)^{\frac{Mn}{2}}\int \bigg[\prod_{a}^n dQ_a d\hat Q_a dm_a d\hat m_a \bigg]\bigg[\prod_{b,a<b}^{n,(n-1)/2} dq_{ab}d\hat q_{ab}\bigg] \label{eq1:avZ} \\
&\exp\left[L \left( {1\over2} \sum_a^n\hat Q_a Q_a - {1\over 2} \sum_{b,a\neq b}^{n,(n-1)}\hat q_{ab} q_{ab} - \sum_{a}^n \hat m_a m_a \right)\right] \bigg[\prod_{\mu}^M X \bigg]\nonumber \\
&\Bigg(\underbrace{\int_{\mathbb{R}^B} d\bs P_0(\bs) \int_{\mathbb{R}^{Bn}}\bigg[\prod_{a}^n d\bx^aP_0(\bx^a) \bigg] \exp\[-\frac{1}{2} \sum_{a}^n \hat Q_a (\bx^a)^{\intercal} \bx^a + \frac{1}{2} \sum_{b,b\neq a}^{n,(n-1)} \hat q_{ab} (\bx^a)^{\intercal} \bx^b  + \sum_{a}^n \hat m_a (\bx^a)^{\intercal} \bs\]}_{\defeq\Gamma}\Bigg)^L \nonumber
\end{align}
At this stage, it is worth noticing that thanks to the average over the disorder $(\bF,\bxi,\bs)$ and the introduction of the replica macroscopic orders parameters (\ref{eq1:repOrderParam_ma}), (\ref{eq1:repOrderParam_qab}), (\ref{eq1:repOrderParam_Qa}) we converted a sytem which partition function was (\ref{eq:repZ0}) i.e. made of $n$ i.i.d strongly disordered replicas each with their own interacting variables into an asymptotically (we averaged over the disorder and used the central limit theorem to get there) equivalent system made of $n$ interacting replicas but independent of the original sources of disorder. Furthermore, their variables are effectively independent one of the other inside the same replica but interact with their equivalent in the other replicas: for example the $\bx_l^a$ variable is effectively independent of $\{\bx_k^a\}_{k\neq l}^L$ but interacts with $\{\bx_l^b\}_{b \neq a}^n \ \forall b$ with effective interaction $\hat q_{ab}$, the dual variable of (\ref{eq1:repOrderParam_qab}), it interacts with itself through the self field $\hat Q_a$ (\ref{eq1:repOrderParam_Qa}) and with the signal through $\hat m_a$ (\ref{eq1:repOrderParam_ma}). The interactions between variables belonging to the same replica are "hidden" in the replica order parameters that are now variables which control the new coupling interactions between replicas $(\{\hat m_a, \hat q_{ab}, \hat Q_a\})$. 

These new effective interactions are way more easy to deal with as the dependence on the disorder (noise and measurement matrix realizations) has been averaged out when applying the central limit theorem to $v^a$, which naturally gave rise to the replica order parameters. Looking at $\Gamma$, we realize that there is a Gaussian coupling between the replicas. As usual in this situation (like is the AMP derivation sec.~\ref{sec:AMP}), we use the Stratanovitch transform to decouple them by linearizing the exponent, the payoff being an additional Gaussian integral to perform at the end. In $B$ dimensions the transform is given by:
\begin{align}
e^{\frac{\hat q}{2}\sum_{b,a\neq b}^{n,(n-1)} \bx_a^{\intercal} \bx_b} &= \prod_{i}^B e^{\frac{\hat q}{2} \sum_{b,a\neq b}^{n,(n-1)}  x_{a,i}  x_{b,i}} \\
&= \prod_i^B \int \mathcal{D}z_i \ \! e^{\sqrt{\hat q}\ \! z_i \sum_a^n  x_{a,i} - \frac{\hat q}{2} \sum_a^n { x_{a,i}}^2}\\
&= \int_{\mathbb{R}^B} \mathcal{D}\bz\ \! e^{\sqrt{\hat q}\ \!\bz^{\intercal}\sum_a^n \bx_a} e^{-\frac{\hat q}{2} \sum_a^n \bx_a^{\intercal} \bx_a}
\end{align}
where we remind that $\mathcal{D} \bz \defeq \prod_{i}^{B} \mathcal{D} z_i = \prod_{i}^{B} \mathcal{N}(z_i|0,1) dz_i$ is a unit centered $B$-d Gaussian measure and sums of the form $\sum_a^n \bx_a = [\sum_a^n  x_{a,i}]_{i}^B$ are vectors. Using the replica symmetric ansatz we obtain that: 
\begin{equation}
\Gamma = \int_{\mathbb{R}^B} d\bs \ \!\mathcal{D}\bz \ \!P_0(\bs) {\underbrace{\left[\int_{\mathbb{R}^B} d\bx P_0(\bx) e^{-\frac{1}{2} (\hat Q+\hat q) \bx^{\intercal}\bx + \hat m \bx^{\intercal} \bs + \bz^{\intercal} \bx \sqrt{\hat q} } \right]}_{\defeq f(\bz,\bs)}}^n\label{eq1:Y}
\end{equation}
In addition we have:
\begin{align}
&\Gamma \underset{n\to 0}{\approx}  \exp\(n \int_{\mathbb{R}^B} d \bs\mathcal{D}\bz P_0(\bs)  \log \(f(\bz,\bs)\)\)  \label{eq1:fz}
\end{align}
Combining (\ref{eq1:fz}) and (\ref{eq1:avZ}) under the replica symmetric ansatz, we get the expression of the averaged replicated partition function as $n\to 0$:
\begin{align}
&\mathbb{E}_{\bF,\bxi,\bx}\{Z^n\} \underset{n\to 0}{\approx} \int dQ d\hat Q dm d\hat m dq d\hat q e^{nL \tilde \Phi_B(m,\hat m,q,\hat q, Q, \hat Q)} \label{eq1:saddle}
\end{align}
where the replica potential, up to irrelevant constants independent on the order parameters is:
\begin{align}
\tilde \Phi_B(m,\hat m,q,\hat q, Q, \hat Q) &= \frac{1}{2} \left(\hat Q Q + \hat q q - 2 \hat m m \right) \nonumber\\
&- \frac{\alpha B}{2} \left(\frac{q - 2m + <\bs^2> + 1/{\rm{snr}}}{Q - q + 1/{\rm{snr}}} + \log(1/{\rm{snr}} + Q - q) \right)\nonumber \\
&+ \int_{\mathbb{R}^B} d\bs \mathcal{D}\bz P_0(\bs) \log\left( \int_{\mathbb{R}^B} d\bx P_0(\bx) e^{\hat m \bx^{\intercal} \bs + \sqrt{\hat q} \bz^{\intercal} \bx - \frac{1}{2}(\hat q + \hat Q)\bx^{\intercal}\bx} \right) \label{eq1:freeEnt}
\end{align}
where it must be kept in mind that the vectors in the last integral are one $B$-d section, not the overall vectors.
\subsection{Saddle point estimation}
For the replica trick (\ref{eq1:replicaTrick}) to be formally valid, the limit $n\to 0$ should be taken before the limit over $L$. But we need to estimate the integral (\ref{eq1:saddle}) by its saddle point as it is intractable otherwise. This can be justified only if the limit $L\to\infty$ is performed before the limit over $n$, but in the same time the expression (\ref{eq1:saddle}) has been obtained already considering $n$ very small.

We thus assume that $n$ is small enough for (\ref{eq1:saddle}) to be accurate but yet fixed. Then we assume that the limits commute and perform the saddle point estimation of the integral before to really let $n\to 0$. This is not rigorous, but heuristically verified in many different models, including inference problems. The saddle point estimate is performed by taking the optimum of $\tilde\Phi_B$, given by its fixed point value with respect to the different order parameters. The resulting potential actually corresponds to the desired Bethe free entropy as seen from (\ref{eq1:replicaTrick}):
\begin{equation}
\Phi_B\defeq \tilde\Phi_B(m^*,\hat m^*,q^*,\hat q^*, Q^*, \hat Q^*) 
\end{equation}
The optimization gives these fixed point values, denoted with stars:
\begin{align}
&\frac{\partial\tilde \Phi_B}{\partial m}=0 \Rightarrow \hat m^* = \frac{\alpha B}{Q^*-q^*+{1/{\rm{snr}}} } \label{eq:hatmstar_0}\\
&\frac{\partial\tilde \Phi_B}{\partial q}=0 \Rightarrow \hat q^* = \alpha B\frac{1/{\rm{snr}} + B<\bs^2> - 2m^* + q^*}{(Q^*-q^*+{1/{\rm{snr}}})^2}\label{eq:hatqstar_0}\\
&\frac{\partial\tilde \Phi_B}{\partial Q}=0 \Rightarrow \hat Q^* = \alpha B\frac{2m^* - B<\bs^2> - 2q^* + Q^*}{(Q^*-q^*+{1/{\rm{snr}}})^2}\label{eq:hatQstar_0}
\end{align}
to be plugged into the previous potential (\ref{fig_freeEnt}) to get its most general form. But as we assume the prior matching condition, further simplifications are possible.
\subsection{The prior matching condition}
\label{sec:priorMatchingCondInReplicaComp}
As shown after in sec.~\ref{sec:priorMatching} together with sec.~\ref{subsec:repIsSE}, the matching prior condition implies:
\begin{align}
q^* &= m^* \label{eq:qstar}\\ 
Q^* &= B<\bs^2> \label{eq:Qstar} \\ 
E &= <\bs^2> - \frac{m^*}{B} \label{eq1:nishi1}
\end{align}
which implies for their conjugate parameters (\ref{eq:hatmstar_0}), (\ref{eq:hatqstar_0}), (\ref{eq:hatQstar_0}):
\begin{align}
\hat q^* &= \hat m^*={\alpha B \over BE + 1/{\rm{snr}}} \label{eq1:hatmstar}\\ 
\hat Q^* &= 0
\end{align}
where $E\defeq<(\bs-\bx)^2>$ is the $MSE$, i.e. the observable of the system. We get the final expression of the Bethe free entropy:
\begin{align}
\Phi_B(E) = &-\frac{\alpha B}{2} \left(\log({1/{\rm{snr}}} + BE) + \frac{B<\bs^2> - BE}{{1/{\rm{snr}}} + BE} \right)\nonumber\\ 
&+ \int_{\mathbb{R}^B} d\bs P_0(\bs) \mathcal{D}\bz \log\left( \int_{\mathbb{R}^B} d\bx P_0(\bx) \exp\(\frac{\bs^{\intercal} \bx}{\Sigma(E)^2}  + \frac{\bz^{\intercal}\bx}{\Sigma(E)}  - \frac{\bx^{\intercal} \bx}{2\Sigma(E)^2} \)\right) \label{eq1:freeEnt2}
\end{align}
where:
\begin{align}
	\Sigma(E)^2\defeq 1/\hat m^* = \frac{{1/(B{\rm{snr}})}+ E}{\alpha} \label{eqChIntro:SIGMA2def}
\end{align}
This expression is general for linear estimation of fixed section size $B$ signals. It will be used in most of the theoretical studies in this thesis. The asymptotic variance (\ref{eqChIntro:SIGMA2def}) of the maximum likelihood estimate increases linearly both with the noise variance and the $MSE$. Furthermore, their common effect is enhanced as the measurement rate decreases.
%
%
\section{The cavity method for linear estimation: state evolution analysis}
\label{sec:stateEvolutionGeneric}
%
The state evolution analysis, referred as the cavity method in physics \cite{MezardParisi87b,mezard2009information} is a statistical analysis that allows to monitor the approximate message-passing algorithm dynamics and performance in the limit of reconstructing infinitely large signals, i.e. in the thermodynamic limit. We consider the case of i.i.d Gaussian matrices $\bF$ for which state evolution has been originally derived \cite{DonohoMaleki09} and then proved to be rigorous in large generality \cite{BayatiMontanari10} (see also \cite{GuoWang07,Rangan10b,KrzakalaMezard12}). As in the replica computation of the previous section, this assumption is essential in order to decouple the signal components in the analysis. Extension to more general ensembles such as row orthogonal matrices could be considered \cite{WenW14} but it is out of the scope of the present thesis. In addition, as we will show in chap.~\ref{chap:structuredOperators} in great details, the state evolution analysis derived in the i.i.d Gaussian matrices case is a good predictive tool of the reconstruction performances of the AMP algorithm even with structured operators, despite not predicting well the dynamics before convergence nor being rigorous. 

As before, we consider the general case of signals with $B$-d sections and assume the proper scaling to get a codeword with power $P=1$: $F_{ij} \sim \mathcal{N}\(F_{ij}|0,1/L\) \forall \ (i,j) \Rightarrow F_{ij} \in O(1/\sqrt{L})\ \forall \ (i,j)$. As in the replica analysis, we consider the case of a factorizable prior over the sections with the same prior for every sections. This drastically simplifies the analysis due to the induced symmetry between all the sections. We will look at the section index dependent prior case in the chapter about superposition codes in the sec.~\ref{sec:powA_SE}. Furthermore, we consider having perfect knowledge of the channel noise statistical properties as the prior which generated the signals as well so that we place ourselves under the prior matching case.
\subsection{Derivation starting from the approximate message-passing algorithm}
One can perform the analysis starting from the cavity quantities defined in the AMP derivation starting from BP, see sec.~\ref{sec:classicalDerivationAMP} as it is more classicaly done \cite{KrzakalaMezard12}. This will be done in the next section, but here we will follow another path starting from the AMP algorithm itself. We refer to Fig.~\ref{algoCh1:AMP} for the definitions of all the quantities that we will use in the derivation.

The aim is to evaluate the asymptotic AMP posterior estimate of a section at each time in order to compute the asymptotic mean square error $E^{t+1}(E^t)$ as a function of its value at time $t$. The posterior estimate is given by applying the denoising function $\tbf f_{a_l}$ to the variables $((\bsy \Sigma_l^{t+1})^2, \bR_l^{t+1})$, see Fig.~\ref{algoCh1:AMP}. We thus need to get an asymptotic estimate of $\bR_l^{t+1}$, the average of variable $\bx_l$ at time $t+1$ with respect to the likelihood. $(\bsy \Sigma_l^{t+1})^2$ is its associated variance. Injecting in $\bR_l^{t+1}$ the expression of $w_\mu^{t+1}$ as a function of the previous time quantities and using (\ref{eqIntro:AWGNCS}) to replace the measurement by its expression in terms of the signal, measurement matrix and noise (i.e. the disorder sources), we get the following expression for $\bR_l^{t+1}$:
\begin{align}
\bR_l^{t+1} &= \ba_l^t + (\bxigma^{t+1}_l)^2\sum_{\mu}^M \frac{\bF_{\mu l}}{1/{\rm{snr}}+ \Theta_\mu^{t+1}} \bigg[\sum_k^L \bF_{\mu k}^{\intercal}(\bs_k - \ba_k^t) + \xi_\mu + \underbrace{\Theta_\mu^{t+1} \frac{y_\mu - w_\mu^t}{1/{\rm{snr}} + \Theta_\mu^{t}}}_{\defeq \Lambda_\mu^t} \bigg] \label{eq:R_derivSE_full}
\\
&= \ba_l^t + (\bxigma^{t+1}_l)^2\sum_{\mu}^M \frac{\bF_{\mu l}}{1/{\rm{snr}}+ \Theta_\mu^{t+1}} \bigg[\bF_{\mu l}^{\intercal}(\bs_l - \ba_l^t) +\sum_{k\neq l}^L \bF_{\mu k}^{\intercal}(\bs_k - \ba_k^t) + \xi_\mu + \Lambda_\mu^t \bigg]
\end{align}
Now we use the fact that $\Theta_\mu^t$ is asymptotically independent of $\mu$ as we can replace the $F_{\mu i}^2$ elements by the matrix variance $1/L$:
\begin{align}
\Theta_\mu^t &\approx \Theta^t \defeq 1/L\sum_i^N v_i^t \\
\Rightarrow (\bxigma^{t+1}_l)^2 &\approx \frac{1/{\rm{snr}}+ \Theta^{t+1}}{B\alpha}\boldsymbol{1}_B
\end{align}
where we have used $M=LB\alpha$ and $\boldsymbol{1}_B$ is a vector of ones of size $B$. This simplifies the expression to:
\begin{align}
\bR_l^{t+1} &= \ba_l^t + \frac{1}{B\alpha} \sum_{\mu}^M \bF_{\mu l}\bigg[\sum_{k\neq l}^{L-1} \bF_{\mu k}^{\intercal}(\bs_k - \ba_k^t) + \xi_\mu +  \Lambda_\mu^t \bigg] + \frac{1}{B\alpha}\sum_\mu^M\bF_{\mu l} \bigg[\bF_{\mu l}^{\intercal}(\bs_l-\ba_l^t)\bigg]
\end{align}
Now we notice that we can simplify the last term in the previous equality:
\begin{align}
\sum_\mu^M\bF_{\mu l} \bigg[\bF_{\mu l}^{\intercal}(\bs_l-\ba_l^t)\bigg]&=\underbrace{\bigg[\sum_\mu^M F_{\mu i}^2(s_i-a_i^t) \bigg]_{i\in l}}_{=B\alpha(\bs_l - \ba_l^t)}+\bigg[\underbrace{\sum_\mu^M \sum_{j\in l:j\neq i}^{B-1} F_{\mu i}F_{\mu j}(s_j-a_j^t)}_{\in O(1/\sqrt{L})}\bigg]_{i\in l}
\label{eq_forgetSecPart}
\end{align}
We can neglect the second term ($B$ remains finite as $L$ diverges) as we will keep only $O(1)$ terms in the computation of the moments of the Gaussian fluctations of $\bR_l^{t+1}$ around $\bs_l$. This leads to the expression:
\begin{align}
\bR_l^{t+1} &\approx \bs_l + \frac{1}{B\alpha} \underbrace{\sum_{\mu}^M \bF_{\mu l} \bigg[\sum_{k\neq l}^{L-1} \bF_{\mu k}^{\intercal} (\bs_k - \ba_k^t) + \xi_\mu  + \Lambda_\mu^t  \bigg]}_{\defeq \br_l^{t+1}} \label{eq_rl}
\end{align}
Now we notice from (\ref{eq1:appW}) and the definition of $\Theta_\mu$ in the AMP algorithm Fig.~\ref{algoCh1:AMP} that: 
\begin{align}
&\Lambda^t_\mu \approx \sum_k^L \bF_{\mu k}^{\intercal}\beps_{a_{k\mu}}\label{eq_lambdaMu}
\end{align}
where $\beps_{a_{k\mu}} \in O(1/\sqrt{L})$ is given by (\ref{eq1:appCorrections_a}), (\ref{eq1:appW}). Using the independence assumption of the operator $\bF$ elements, we can apply the central limit theorem to $\br_l^{t+1}$ which is thus Gaussian distributed with moments that we compute now. We remind the reader that the noise has zero mean and we note that the $MSE(\ba,\bs)$ (\ref{eqIntro:MSE}) tends to its average over the disorder in the large size signals limit:
\begin{equation}
E^{t+1} = <\(\bs - \ba^{t+1}\)^2> \underset{L\to\infty}{\to} \frac{1}{B}\mathbb{E}_{\bF,\bxi,\bs}\left\{\(\bs_l - \ba_l^{t+1}\)^\intercal \(\bs_l - \ba_l^{t+1}\)\right\}
\label{eq:MSEtoRisk}
\end{equation}
where $\ba^{t+1} = \mathbb{E}^{t+1}_{\bx|\by}\{\bx\} = \[\tbf f_{a_l}\((\bsy\Sigma^{t+1}_l)^2,\bR_l^{t+1}\)\]_{l}^L$ is the AMP posterior estimate of the signal at time $t+1$ and we put the index $l$ to underline that we speak about a section, not the overall signal (which is the same for the $MSE$ in the thermodynamic limit). The matrix $\bF$ elements having $0$ mean, only the terms with even power of the matrix elements in the various sums that appear remain because of the average over the disorder that we will use so that, using (\ref{eq_lambdaMu}), (\ref{eq1:appW}) we obtain for its first moment:
\begin{align}
\mathbb{E}_{\bF,\bxi,\bs} \{\br_l^{t+1}\} &= \underbrace{\mathbb{E}_{\bF,\bxi,\bs} \left\{\sum_\mu^M \bF_{\mu l}\[\sum_{k\neq l}^{L-1} \bF_{\mu k}^{\intercal} (\bs_k - \ba_k^t) + \xi_{\mu}\]\right\} }_{=\bzero_B} + \mathbb{E}_{\bF,\bxi,\bs}\left\{\sum_{\mu}^M\bF_{\mu l} \sum_{k}^L \bF_{\mu k}^{\intercal} \beps_{a_{k \mu}}\right\} \label{eq_rFirstMom1}\\
&\approx \underbrace{\mathbb{E}_{\bF,\bxi,\bs} \left\{\sum_{\mu}^M (\bF_{\mu l}^3)^{\intercal} \bv_l\frac{y_\mu - w_\mu^t}{1/{\rm{snr}} + \Theta^t}\right\}}_{\in O(1/L)} \\
&= \bzero_B \label{eq_rFirstMom2}
\end{align}
Now the cross terms, with $l'\neq l$:
\begin{align}
\mathbb{E}_{\bF,\bxi,\bs}\{{\br_l^{t+1} \br_{l'}^{t+1}}\} &= \mathbb{E}_{\bF,\bxi,\bs}\Bigg\{\sum_{\mu,\nu}^{M,M} \bF_{\mu l} \bF_{\nu l'}\left[\sum_{k\neq l}^{L-1} \bF_{\mu k}^{\intercal} (\bs_k - \ba_k^t) + \xi_\mu  + \Lambda_\mu^t  \right]\nonumber\\
&\left[\sum_{k'\neq l'}^{L-1} \bF_{\nu k'}^{\intercal} (\bs_{k'} - \ba_{k'}^t) + \xi_\nu  + \Lambda_\nu^t  \right]\Bigg\} \label{eq_rSecMomCross1}\\
&=\mathbb{E}_{\bF,\bxi,\bs}\bigg\{\sum_{\mu}^M \bF_{\mu l} \bF_{\mu l'}\left[\bF_{\mu l'}^{\intercal} (\bs_{l'} - \ba_{l'}^t) + \xi_\mu  + \Lambda_\mu^t  \right]\\
&\left[\bF_{\mu l}^{\intercal} (\bs_{l} - \ba_{l}^t) + \xi_\mu  + \Lambda_\mu^t  \right] \bigg\} \label{eq_rSecMomCross2}\\
&=\underbrace{\mathbb{E}_{\bF,\bxi,\bs}\left\{\sum_{\mu}^M \bF_{\mu l}^2 \bF_{\mu l'}^2(\bs_{l'} - \ba_{l'}^t)(\bs_{l} - \ba_{l}^t) \right\}}_{\in O(1/L)} \nonumber \\
&\approx \bzero_B \label{eq_rSecMomCross3}
\end{align}
Here if the matrix elements were not i.i.d we would obtain non trivial crossed terms that would greatly complexify the analysis, as all the sections would become correlated.
Finally its diagonal second moment:
\begin{align}
\mathbb{E}_{\bF,\bxi,\bs}\{(\br_l^{t+1})^2\} &= \mathbb{E}_{\bF,\bxi,\bs}\Bigg\{\sum_{\mu,\nu}^{M,M} \bF_{\mu l}\bF_{\nu l} \left[\sum_{k\neq l}^{L-1} \bF_{\mu k}^{\intercal} (\bs_k - \ba_k^t) + \xi_\mu  + \Lambda_\mu^t  \right]\nonumber \\
&\left[\sum_{k'\neq l}^{L-1} \bF_{\nu k'}^{\intercal} (\bs_{k'} - \ba_{k'}^t) + \xi_\nu  + \Lambda_\nu^t  \right]\Bigg\} \label{eq_rVar1} \\
&=\mathbb{E}_{\bF,\bxi,\bs}\Bigg\{\sum_{\mu}^M \bF_{\mu l}^2 \left[\sum_{k\neq l}^{L-1} \bF_{\mu k}^{\intercal} (\bs_k - \ba_k^t) + \xi_\mu  + \Lambda_\mu^t  \right]\nonumber\\ 
&\left[\sum_{k'\neq l}^{L-1} \bF_{\mu k'}^{\intercal} (\bs_{k'} - \ba_{k'}^t) + \xi_\mu  + \Lambda_\mu^t  \right]\Bigg\}\label{eq_rVar2}\\
&=\mathbb{E}_{\bF,\bxi,\bs}\left\{\sum_{\mu}^M \bF_{\mu l}^2 \left[\sum_{k\neq l}^{L-1} \bF_{\mu k}^{\intercal} (\bs_k - \ba_k^t) \right] \left[\sum_{k'\neq l}^{L-1} \bF_{\mu k'}^{\intercal} (\bs_{k'} - \ba_{k'}^t) \right]\right\}\nonumber\\ 
&+ \frac{\alpha B}{{\rm{snr}}} + \underbrace{\mathbb{E}_{\bF,\bxi,\bs}\left\{\sum_\mu^M \bF_{\mu l}^2 (\Lambda_\mu^t)^2\right\}}_{\in O(L^{-3/2})} \nonumber\\
&+ 2\underbrace{\mathbb{E}_{\bF,\bxi,\bs} \left\{\sum_{\mu}^M \bF_{\mu l}^2 \Lambda_\mu^t \left[\sum_{k\neq l}^{L-1} \bF_{\mu k}^{\intercal} (\bs_k - \ba_k^t)\right] \right\}}_{=\boldsymbol{0}_B}\label{eq_rVar3}
\end{align}
\begin{align}
&\approx \mathbb{E}_{\bF,\bxi,\bs}\left\{\sum_{\mu}^M \bF_{\mu l}^2 \left[\sum_{k\neq l}^{L-1} (\bF_{\mu k}^2)^{\intercal} (\bs_k - \ba_k^t)^2 \right] \right\} + \frac{\alpha B}{{\rm{snr}}}\\
&\approx \frac{M}{L^2} \sum_{k}^L\mathbb{E}_{\bs}\{(\bs_k - \ba_k^t)^{\intercal}(\bs_k - \ba_k^t)\}+ \frac{\alpha B}{{\rm{snr}}} \nonumber \\
&=B\alpha \left({1/{\rm{snr}}} + B E^t\right)
\end{align}
where $\bzero_B$ is a vector of zeros of size $B$ and we have used (\ref{eq:MSEtoRisk}) with the $MSE$ definition (\ref{eqIntro:MSE}) and $N=LB$. Thanks to the independence assumptions and the performed averages, we get a variance that is independent of the component and that depends only on the global $MSE$. From all these computations, we can now write $\bR_l^t$ using (\ref{eq_rl}) as a Gaussian variable, dependent on the other components only through $E$:
\begin{align}
&\br_l^{t+1} \sim \mathcal{N}\(\br_l^{t+1}\bigg|\bsy 0_B,B\alpha \left({1/{\rm{snr}}} + B E^t\right)\bsy I_B \) \\
\Rightarrow &\bR_l^{t+1} \sim \mathcal{N}\(\bR_l^{t+1}\bigg|\bs_l,\frac{{1/({\rm{snr}}}B)+ E^t}{\alpha}\bsy I_B\)
\end{align}
where $\bsy I_B$ is the identity of size $B$. It remains to perform the average over the signal $\bs$ distribution. The equivalence between all the sections due to the prior which is the same for all of them implies that we can consider only the $MSE$ evolution in a single section instead of the overall one. Thus the state evolution in the matching prior case, with knowledge of the channel noise is:
\begin{align}
E^{t+1} &~= \frac{1}{B}\sum_{i\in l}^B\int_{\mathbb{R}^B} d\bs_l \ \!P_0(\bs_l) \int_{\mathbb{R}^B} \mathcal{D}\bz\ \! \bigg[f_{a_i}\left((\Sigma^{t+1})^2, \bR^{t+1}(\bz,\bs_l)\right) - s_i\bigg]^2 \label{eq_E1_1}\\
\Sigma^{t+1}(E^t) &\defeq \sqrt{\frac{1/({\rm{snr}}B)+ E^t}{\alpha}} \label{eqChIntro:defSigma2}\\
\bR^{t+1}(\bz,\bs_l) &\defeq \bs_l + \bz~\Sigma^{t+1}(E^t)\label{eqChIntro:defR}
\end{align}
where $\bz$ is a i.i.d unit centered Gaussian $B$-d vector. In sec.~\ref{sec:priorMatching} we will show that this recursion can be written in other equivalent ways under the prior matching condition. The definition (\ref{eqChIntro:defSigma2}) at its fixed point matches what we obtained in the replica computation of the Bethe free entropy (\ref{eqChIntro:SIGMA2def}).
%
%
%
\subsection{Alternative derivation starting from the cavity quantities}
We now re-derive the state evolution using the cavity quantities used to derive AMP, instead of starting from the final TAP equations, but the derivation sis very close. We start from the MSE (again, focusing on a unique section as they are all equivalent), which is the observable we want to predict:
\begin{align}
	E^t &= <(\tbf f_{a_l}((\bsy \Sigma_l^t)^2, \bR_l^t) - \bs_l)^\intercal (\tbf f_{a_l}((\bsy \Sigma_l^t)^2, \bR_l^t) - \bs_l) >\\
	&= \frac{1}{B}\mathbb{E}_{\bR_l^t}\left\{(\tbf f_{a_l}((\bsy\Sigma_l^t)^2, \bR_l^t) - \bs_l)^{\intercal}(\tbf f_{a_l}((\bsy\Sigma_l^t)^2, \bR_l^t) - \bs_l) \right\}
\end{align}
We thus need to derive the distribution of $\bR_l^t$, defined by (\ref{eqChIntro:defRcav}), $(\bsy\Sigma_l^t)^2$ being its variance. Plugging the definitions (\ref{eq1:bA}) and (\ref{eq1:bB}) into (\ref{eqChIntro:defRcav}) and using the fact that:
\begin{align}
	&\sum_{k \neq l} \bv_{k\mu} = \sum_{k \neq l} \bv_{k} + O(1)\\
	\Rightarrow &\sum_{k \neq l} (\bF_{\mu k}^2)^{\intercal} \bv_{k\mu} = 1/L\sum_{k \neq l} \bv_{k} + O(1/L)
\end{align}
coming from (\ref{eq1:appCorrections}) for the first equality, and the fact that the $\bF^2$ elements can be safely replaced by the matrix variance in the thermodynamic limit, we obtain after simple algebraic simplifications similar to the ones at the beginning of the previous section:
\begin{align}
	\bR_l^t &= \frac{\sum_\mu^M \bB_{\mu l}}{\sum_\mu^M \bA_{\mu l}}\\
	&= \bs_l + \frac{1}{B\alpha} \underbrace{\sum_\mu^M \bF_{\mu l} \(\xi_\mu + \sum_{j\neq l} \bF_{\mu j}^{\intercal} (\bs_j - \ba_{j \mu}^t ) \)}_{\defeq \tbf p_l^t}
\end{align}
where we used the definition of the measurement rate $\alpha \defeq M/BL$ and the relation between the measurement and the signal (\ref{eqIntro:AWGNCS}) to replace $y_\mu$. From the central limit theorem and the independence assumption of the $\bF$ elements, $\tbf p_l^t$ is a Gaussian random variable. Actually by identification with (\ref{eq_rl}), we recognize that $\tbf p_l^t=\br_l^t$ which moments have already been computed in the previous section and the final state evolution recursion (\ref{eq_E1_1}) is thus found back.
%
\subsection{The prior matching condition and Bayesian optimality}
\label{sec:priorMatching}
Let us us now discuss some implications of the prior matching condition, i.e. of the knowledge of the true generating model of the signal. We will demonstrate relations that are true when this assumption is verified thanks to the state evolution analysis and then discuss what it implies in terms of replicas. 
\subsubsection{The prior matching conditions from the state evolution}
We will now show that the prior matching condition asymptotically implies that at each time step during the reconstruction, the $MSE$ of the posterior estimate by AMP equals the posterior variance of this estimate. We start from the state evolution results for the asymptotic $MSE$ through time (\ref{eq_E1_1}), (\ref{eqChIntro:defSigma2}), (\ref{eqChIntro:defR}). The aim is to show that it is equal to the average posterior variance defined as:
\begin{align}
	V^{t+1} &\defeq \frac{1}{B}\sum_{i\in l}^B\int_{\mathbb{R}^B} d\bs_l \ \!P_0(\bs_l) \int_{\mathbb{R}^B} \mathcal{D}\bz\ \! f_{c_i}\left((\Sigma^{t+1})^2, \bR^{t+1}(\bz,\bs_l)\right) \label{eqChIntro:SE_V}\\
	&=\frac{1}{B}\[\mathbb{E}_{\by}\{\mathbb{E}_{\bx|\by}\{\bx_l^\intercal \bx_l\}\} - \mathbb{E}_{\by}\{\mathbb{E}_{\bx|\by}\{\bx_l\}^\intercal \mathbb{E}_{\bx|\by}\{\bx_l\}\} \]\label{eqChIntro:eqNishE1}
\end{align}
together with the pevious definitions (\ref{eqChIntro:defSigma2}), (\ref{eqChIntro:defR}), where $\mathbb{E}_{\bx|\by}\{\}$ is the asymptotic average with respect to the posterior estimate by AMP. Its dependence on the time dependent AMP fields $((\Sigma^{t+1})^2, \bR^{t+1})$ is implicit. Furthermore, we denote by $\mathbb{E}_\by\{\}\defeq \mathbb{E}_{\bF,\bxi,\bs}\{\}$ the disorder average, which is the integral over $P_0(\bs)$ and $\mathcal{D}\bz$ in (\ref{eqChIntro:SE_V}). Expanding the $MSE$ (\ref{eq_E1_1}) we get:
\begin{align}
	E^{t+1} &= \frac{1}{B}\sum_{i\in l}^B \bigg[\int_{\mathbb{R}^B} d\bs_l \ \!P_0(\bs_l) \int_{\mathbb{R}^B} \mathcal{D}\bz\ \! f_{a_i}\left((\Sigma^{t+1})^2, \bR^{t+1}(\bz,\bs_l)\right)^2 \nonumber\\
	&- 2\int_{\mathbb{R}^B} d\bs_l \ \!P_0(\bs_l) s_i \int_{\mathbb{R}^B} \mathcal{D}\bz f_{a_i}\left((\Sigma^{t+1})^2, \bR^{t+1}(\bz,\bs_l)\right) + \int_{\mathbb{R}^B} d\bs_l \ \!P_0(\bs_l) s_i^2\bigg] \label{eq:MSEexpandedSE}\\
	&= \frac{1}{B}\[\mathbb{E}_{\by}\{\mathbb{E}_{\bx|\by}\{\bx_l\}^\intercal\mathbb{E}_{\bx|\by}\{\bx_l\}\} - 2\mathbb{E}_{\by}\{\bs_l^\intercal\mathbb{E}_{\bx|\by}\{\bx_l\}\} + \mathbb{E}_{\bs}\{\bs_l^\intercal\bs_l\}\]
\end{align}
We start by proving the equality between the first and second term (up to the 2). From now on, we skeep the time index for sake of readibility. Using the definition of the denoiser $f_{a_i}$ (\ref{eq1:fai}), we get:
\begin{align}
	&\int d\bs_l \ \!P_0(\bs_l) \int \mathcal{D}\bz\ \! f_{a_i}\((\Sigma)^2, \bR(\bz,\bs_l)\)^2\nonumber\\
	&=\int d\bs_l \ \!P_0(\bs_l) \int \mathcal{D}\bz\ \!\frac{\left[\int d\bx_l \ \!P_0(\bx_l) \exp\(-\frac{\bx_l^{\intercal} \bx_l}{2 \Sigma^2} + \frac{\bx_l^{\intercal} (\bs_l + \bz \Sigma)}{\Sigma^2} \) x_i \right]^2}{\left[\int d\bx_l \ \!P_0(\bx_l) \exp\(-\frac{\bx_l^{\intercal} \bx_l}{2 \Sigma^2} + \frac{\bx_l^{\intercal} (\bs_l + \bz \Sigma)}{\Sigma^2} \) \right]^2}
\end{align}
where we have expanded the square in the exponent and simplified the $\bx_l$ independent term with its normalization one in the denoiser expression. Now we use the following change of variable: $\bz' \defeq (\bs_l+\bz \Sigma)$:
\begin{align}
	&=\int \mathcal{D} \bz'\int d\bs_l \ \!P_0(\bs_l)  \exp\(\frac{\bs_l^{\intercal}\bz'}{\Sigma} - \frac{\bs_l^\intercal \bs_l}{2\Sigma^2} \)\ \!\frac{\left[\int d\bx_l \ \!P_0(\bx_l) \exp\(-\frac{\bx_l^{\intercal} \bx_l}{2 \Sigma^2} + \frac{\bx_l^{\intercal} \bz'}{\Sigma} \) x_i \right]^2}{\left[\int d\bx_l \ \!P_0(\bx_l) \exp\(-\frac{\bx_l^{\intercal} \bx_l}{2 \Sigma^2} + \frac{\bx_l^{\intercal} \bz'}{\Sigma} \) \right]^2}\\
	&=\int \mathcal{D} \bz'\ \!\frac{\left[\int d\bx_l \ \!P_0(\bx_l) \exp\(-\frac{\bx_l^{\intercal} \bx_l}{2 \Sigma^2} + \frac{\bx_l^{\intercal} \bz'}{\Sigma} \) x_i \right]^2}{\left[\int d\bx_l \ \!P_0(\bx_l) \exp\(-\frac{\bx_l^{\intercal} \bx_l}{2 \Sigma^2} + \frac{\bx_l^{\intercal} \bz'}{\Sigma} \) \right]}
\end{align}
Now the second term of (\ref{eq:MSEexpandedSE}) using the same change of variable:
\begin{align}
	&\int \mathcal{D} \bz'\int d\bs_l \ \!P_0(\bs_l)s_i  \exp\(\frac{\bs_l^{\intercal}\bz'}{\Sigma} - \frac{\bs_l^\intercal \bs_l}{2\Sigma^2} \)\ \!\frac{\left[\int d\bx_l \ \!P_0(\bx_l) \exp\(-\frac{\bx_l^{\intercal} \bx_l}{2 \Sigma^2} + \frac{\bx_l^{\intercal} \bz'}{\Sigma} \) x_i \right]}{\left[\int d\bx_l \ \!P_0(\bx_l) \exp\(-\frac{\bx_l^{\intercal} \bx_l}{2 \Sigma^2} + \frac{\bx_l^{\intercal} \bz'}{\Sigma} \) \right]}\\
	&=\int \mathcal{D} \bz'\ \!\frac{\left[\int d\bx_l \ \!P_0(\bx_l) \exp\(-\frac{\bx_l^{\intercal} \bx_l}{2 \Sigma^2} + \frac{\bx_l^{\intercal} \bz'}{\Sigma} \) x_i \right]^2}{\left[\int d\bx_l \ \!P_0(\bx_l) \exp\(-\frac{\bx_l^{\intercal} \bx_l}{2 \Sigma^2} + \frac{\bx_l^{\intercal} \bz'}{\Sigma} \) \right]}
\end{align}
So the two terms are equal: 
\begin{align}
	\mathbb{E}_{\by}\{\mathbb{E}_{\bx|\by}\{\bx_l\}^\intercal\mathbb{E}_{\bx|\by}\{\bx_l\}\} = \mathbb{E}_{\by}\{\bs_l^\intercal\mathbb{E}_{\bx|\by}\{\bx_l\}\}
\end{align}
This allows to re-write the MSE as:
\begin{align}
	E^{t+1} &= \frac{1}{B}\sum_{i\in l}^B \bigg[\int_{\mathbb{R}^B} d\bs_l \ \!P_0(\bs_l) s_i^2 - \int_{\mathbb{R}^B} d\bs_l \ \!P_0(\bs_l) \int_{\mathbb{R}^B} \mathcal{D}\bz\ \! f_{a_i}\left((\Sigma^{t+1})^2, \bR^{t+1}(\bz,\bs_l)\right)^2 \bigg]\\
	&= \mathbb{E}_{\bs}\{\bs_l^\intercal\bs_l\} - \mathbb{E}_{\by}\{\mathbb{E}_{\bx|\by}\{\bx_l\}^\intercal\mathbb{E}_{\bx|\by}\{\bx_l\}\} \label{eqChIntro:eqNishE2}
\end{align}
Let us now show that under the prior matching condition, $\mathbb{E}_{\bs}\{s_i^2\} = \mathbb{E}_{\by}\{\mathbb{E}_{\bx|\by}\{x_i^2\}\}$ which will complete the proof. We do so by using again the same change of variable:
\begin{align}
	&\mathbb{E}_{\by}\{\mathbb{E}_{\bx|\by}\{x_i^2\}\}=\int d\bs_l \ \!P_0(\bs_l) \int \mathcal{D}\bz\ \!\frac{\left[\int d\bx_l \ \!P_0(\bx_l) \exp\(-\frac{\bx_l^{\intercal} \bx_l}{2 \Sigma^2} + \frac{\bx_l^{\intercal} (\bs_l + \bz \Sigma)}{\Sigma^2} \) x_i^2 \right]}{\left[\int d\bx_l \ \!P_0(\bx_l) \exp\(-\frac{\bx_l^{\intercal} \bx_l}{2 \Sigma^2} + \frac{\bx_l^{\intercal} (\bs_l + \bz \Sigma)}{\Sigma^2} \) \right]}\\	
	&=\int \mathcal{D} \bz'\int d\bs_l \ \!P_0(\bs_l) \exp\(\frac{\bs_l^{\intercal}\bz'}{\Sigma} - \frac{\bs_l^\intercal \bs_l}{2\Sigma^2} \)\ \!\frac{\left[\int d\bx_l \ \!P_0(\bx_l) \exp\(-\frac{\bx_l^{\intercal} \bx_l}{2 \Sigma^2} + \frac{\bx_l^{\intercal} \bz'}{\Sigma} \) x_i^2 \right]}{\left[\int d\bx_l \ \!P_0(\bx_l) \exp\(-\frac{\bx_l^{\intercal} \bx_l}{2 \Sigma^2} + \frac{\bx_l^{\intercal} \bz'}{\Sigma} \) \right]}\\
	&=\int \mathcal{D} \bz' \int d\bx_l \ \!P_0(\bx_l) \exp\(-\frac{\bx_l^{\intercal} \bx_l}{2 \Sigma^2} + \frac{\bx_l^{\intercal} \bz'}{\Sigma} \) x_i^2\\
	&=\int d\bx_lP_0(\bx_l) \exp\(-\frac{\bx_l^{\intercal} \bx_l}{2 \Sigma^2} \)x_i^2\int d\bz' \frac{1}{\sqrt{2\pi}} \ \! \exp\(\frac{\bx_l^{\intercal} \bz'}{\Sigma} - \frac{(\bz')^{\intercal} \bz'}{2} \)\\
	&= \int d\bx_lP_0(\bx_l)x_i^2\\
	&=\mathbb{E}_{\bs}\{s_i^2\}
\end{align}
where the last equality is true due to the matching prior condition. This with (\ref{eqChIntro:eqNishE1}) and (\ref{eqChIntro:eqNishE2}) implies that $E^t = V^t\  \forall\ t$. Of course this is true only if the two quantities are initialized with same value, but as long as they are, they remain equal at any time step: we say the the algorithm remains on the Nishimori line. We summarize the diverse relations that are true under the prior matching condition:
\begin{align}
	E^t &= V^t \ \forall\  t \label{eqChIntro:NISHI_1}\\
	\mathbb{E}_{\by}\{\mathbb{E}_{\bx|\by}\{\bx_l\}^\intercal\mathbb{E}_{\bx|\by}\{\bx_l\}\}&=\mathbb{E}_{\by}\{\bs_l^\intercal\mathbb{E}_{\bx|\by}\{\bx_l\}\}\label{eqChIntro:NISHI_2}\\
	\mathbb{E}_{\by}\{\mathbb{E}_{\bx|\by}\{\bx_l^\intercal\bx_l\}\}&=\mathbb{E}_{\bs}\{\bs_l^\intercal\bs_l\}\label{eqChIntro:NISHI_3}
\end{align}
To summarize, this implies that the three following forms of the state evolution are perfectly equivalent under the prior matching condition:
\begin{align}
	E^{t+1} = V^{t+1} &~= \frac{1}{B}\sum_{i\in l}^B\int_{\mathbb{R}^B} d\bs_l \ \!P_0(\bs_l) \int_{\mathbb{R}^B} \mathcal{D}\bz\ \! \bigg[f_{a_i}\left((\Sigma^{t+1})^2, \bR^{t+1}(\bz,\bs_l)\right) - s_i\bigg]^2 \label{eqChIntro:SE_form1}\\
	&= \frac{1}{B}\[\mathbb{E}_{\bs}\{\bs_l^\intercal \bs_l\} - \sum_{i\in l}^B\int_{\mathbb{R}^B} d\bs_l \  P_0(\bs_l)s_i\int_{\mathbb{R}^B}\mathcal{D}\bz\ f_{a_i}\left((\Sigma^{t+1})^2, \bR^{t+1}(\bz,\bs_l)\right) \] \label{eqChIntro:SE_form2}\\
	&= \frac{1}{B}\sum_{i\in l}^B\int_{\mathbb{R}^B} d\bs_l \ \!P_0(\bs_l) \int_{\mathbb{R}^B} \mathcal{D}\bz\ \! f_{c_i}\left((\Sigma^{t+1})^2, \bR^{t+1}(\bz,\bs_l)\right) \label{eqChIntro:SE_form3}
\end{align}
together with (\ref{eqChIntro:defSigma2}) and (\ref{eqChIntro:defR}). One form can be more practical to use depending on the denoising functions. The second form (\ref{eqChIntro:SE_form2}) can be computationally easier and faster to deal with but can be more dangerous to use than the two other forms of the state evolution because it can become negative if the difference is really small due to finite numerical precision. 
\subsubsection{The prior matching conditions in terms of replicas}
Let us discuss the physical meaning of the macroscopic order parameters introduced in the replica computation and make the connection between these and the equalities obtained thanks to the Nishimori condition (\ref{eqChIntro:NISHI_1}), (\ref{eqChIntro:NISHI_2}), (\ref{eqChIntro:NISHI_3}).

The replicas are interpreted as different possible solutions (microstates) to the noisy problem (\ref{eqIntro:AWGNCS}),  the fluctuations coming from the disorder. Using the replica symmetric ansatz is assuming that all the replicas belong to the same pure state, i.e. the space of solutions does not split into many disconnected components and thus each replicas have same statistical properties, given by the replica macroscopic order parameters. 

$\bullet$ The meaning of $m$ (\ref{eq1:repOrderParam_ma}) in the replica symmetric ansatz is the overlap between the signal and the signal estimate given by the average over all the replica states. We can thus identify $m$ with $m=\mathbb{E}_{\by}\{\bs_l^\intercal \mathbb{E}_{\bx|\by}\{\bx_l\}\}$.

$\bullet$ The order parameter $Q$ (\ref{eq1:repOrderParam_Qa}) is the average selfoverlap of the replicas and is thus naturally identified with $Q=\mathbb{E}_{\by}\{\mathbb{E}_{\bx|\by}\{\bx_l^\intercal \bx_l\}\} = \mathbb{E}_{\bs}\{\bs_l^\intercal \bs_l\}$.

$\bullet$ Finally, $q$ (\ref{eq1:repOrderParam_qab}) is the overlap between different replicas. In the phase where reconstruction is possible as enough information about the signal is contained in the measurement, the differences between the different replica states should be dominated by the noise, and the features present in all the replica states is the true information on the signal. So the overlap $q$ between infinitely large replicas cancels out the noise-induced fluctuations in average, and remains only the average over the replicas to the square, i.e. the squared signal estimate. Thus $q=\mathbb{E}_{\by}\{\mathbb{E}_{\bx|\by}\{\bx_l\}^\intercal \mathbb{E}_{\bx|\by}\{\bx_l\}\}=\mathbb{E}_{\by}\{\bs_l\intercal\mathbb{E}_{\bx|\by}\{\bx_l\}\}=m$.

The second equalities have been obtained thanks to the Nishimori conditions (\ref{eqChIntro:NISHI_1}), (\ref{eqChIntro:NISHI_2}) and (\ref{eqChIntro:NISHI_3}).
\section{The link between replica and state evolution analyzes: derivation of the state evolution from the average Bethe free entropy}
\label{subsec:repIsSE}
We now show that the fixed point conditions of the Bethe free entropy computed by the replica method are giving back the state evolution recursion of the $MSE$ at its fixed point. We restrict ourselves to the section independent prior case but the derivation for generic prior is done in a similar manner. 

As we place ourselves under the matching prior condition, the previous section sec.~\ref{sec:priorMatching} have shown that the conditions assumed in sec.~\ref{sec:priorMatchingCondInReplicaComp} are true. Thus starting from the replica potential expression (\ref{eq1:freeEnt}) at its fixed point with respect to all its parameters, we should be able to derive the state evolution, for example the form (\ref{eqChIntro:SE_form2}). By identification of (\ref{eqChIntro:SE_form2}) with (\ref{eq1:nishi1}), we should find that the fixed point of $m$ gives the integral part of (\ref{eqChIntro:SE_form2}). Let us prove it. The fixed point condition for $m$ around the optimal values of all the parameters of the Bethe free energy gives:
\begin{align}
&\frac{\partial\tilde \Phi_B}{\partial \hat m}\bigg|_{(\hat m^*, \hat q^*, \hat Q^*, m^*, q^*, Q^*)}=0 \\
&\Rightarrow m^*(E) = \int d\bs_l \mathcal{D}\bz P_0(\bs_l) \int d\bx_l P_0(\bx_l) {\bs_l^{\intercal}\bx_l\over Z(\bs_l, \bz, E)} \exp\left\{\frac{\bx_l^{\intercal} \bs_l}{\Sigma(E)^2} + \frac{\bz^{\intercal} \bx_l}{\Sigma(E)} - \frac{\bx_l^{\intercal} \bx_l}{2\Sigma(E)^2}\right\} \nonumber \\
&=\sum_{i\in l}^B \int d\bs_l \mathcal{D}\bz P_0(\bs_l) s_i \int d\bx_l P_0(\bx_l) x_i {1\over Z(\bs_l, \bz, E)} \exp\left\{\frac{\bx_l^{\intercal} \bs_l}{\Sigma(E)^2} + \frac{\bz^{\intercal} \bx_l}{\Sigma(E)} - \frac{\bx_l^{\intercal} \bx_l}{2\Sigma(E)^2}\right\} \label{eq:mstar_last}
\end{align}
using (\ref{eq1:freeEnt2}), (\ref{eqChIntro:SIGMA2def}) and where $Z(\bs_l, \bz, E)$ is a partition function:
\begin{equation}
Z(\bs_l, \bz, E) \defeq \int d\bx_l P_0(\bx_l) \exp\left\{\frac{\bx_l^{\intercal} \bs_l}{\Sigma(E)^2} + \frac{\bz^{\intercal} \bx_l}{\Sigma(E)} - \frac{\bx_l^{\intercal} \bx_l}{2\Sigma(E)^2}\right\}
\end{equation}
We recognize the expression of the denoiser $f_{a_i}$ (\ref{eq1:fai}) when we use the definitions (\ref{eqChIntro:defSigma2}) and (\ref{eqChIntro:defR}):
\begin{align}
	f_{a_i}(\Sigma^2, \bR) &= \frac{1}{\tilde Z(\Sigma^2, \bR)}\int d\bx_l x_i P_0(\bx_l) \exp\left\{-\frac{(\bx_l - \bR)^{\intercal}(\bx_l - \bR)}{2\Sigma^2}\right\} \\
	&= \frac{1}{\tilde Z(\bs_l, \bz, E)}\int d\bx_l x_i P_0(\bx_l) \exp\left\{-\frac{(\bx_l - \bs_l - \bz \Sigma)^{\intercal}(\bx_l - \bs_l - \bz \Sigma)}{2\Sigma^2}\right\}\\
	&=\frac{1}{ Z(\bs_l, \bz, E)}\int d\bx_l x_i P_0(\bx_l) \exp\left\{\frac{\bx_l^{\intercal} \bs_l}{\Sigma(E)^2} + \frac{\bz^{\intercal} \bx_l}{\Sigma(E)} - \frac{\bx_l^{\intercal} \bx_l}{2\Sigma(E)^2}\right\}	
\end{align}
From this together with (\ref{eq:mstar_last}) we get that at its fixed point, the replica order parameter $m$ verifies:
\begin{align}
	m^* &= \sum_{i\in l}^B \int_{\mathbb{R}^B} d\bs_l P_0(\bs_l)s_i\int_{\mathbb{R}^B}\mathcal{D}\bz  f_{a_i}(\Sigma(E)^2, \bR(\bz,\bs_l))
\end{align}
which is exactly the integral of (\ref{eqChIntro:SE_form2}). Thus using (\ref{eq1:nishi1}) we get:
\begin{align}
	E = <\bs^2> - \frac{1}{B}\sum_{i\in l}^B \int_{\mathbb{R}^B} d\bs_l P_0(\bs_l)s_i\int_{\mathbb{R}^B}\mathcal{D}\bz f_{a_i}(\Sigma(E)^2, \bR(\bz,\bs_l))
\end{align}
which is exactly the expression of the state evolution (\ref{eqChIntro:SE_form2}) as the empirical and true averages are equal in the thermodynamic limit. Thus the fixed point conditions verified at the optimum of the Bethe free entropy gives back the state evolution at its fixed point (when the time index is droped). 

From this analysis, we can now assert that using the state evolution analysis to find fixed points of the algorithm or extracting this information from the potential (\ref{eq1:freeEnt2}) are totally equivalent. In addition, this validates further the replica analysis as an exact procedure in the present context. The state evolution thus finds the optima of the potential $\Phi_B(E)$ and thus when it has two local maxima the fixed point of the state evolution depends on the initial condition of the recursion.
\subsection{Different forms of the Bethe free entropy for different fixed points algorithms}
It is worth noticing that the previous derivation is the asymptotic equivalent of the derivation in sec.~\ref{sec:BPfromBethe} of the belief propagation algorithm as fixed point equations extracted from the Bethe free energy on a single instance (\ref{eq:BetheF_cavMess}). Actually depending on the limit taken and the graph topology, we get different expressions of the Bethe free energy, and thus different fixed point equations and message-passings.

$\bullet$ When using the Bethe free entropy/energy parametrized by the cavity messages (\ref{eq:BetheF_cavMess}), justified on locally tree-like graphs, the fixed points equations directly give the canonical belief propagation for a single graph instance, see sec.~\ref{sec:BPfromBethe}.

$\bullet$ Now in the case of large densely connected graphs, the Bethe free entropy becomes (\ref{eq:BetheF_forAMP}) and deriving its fixed point equations would lead to the approximate message-passing algorithm, that has been instead directly derived from belief propagation in sec.~\ref{sec:AMP}. This free entropy and algorithm are dependent on the disorder instance.

$\bullet$ Finally, when the dense graph is really taken to be infinitely large (or equivalently averaged over the disorder by the self averageness property), the Bethe free entropy is the one extracted from the replica analysis (\ref{eq1:freeEnt2}) and the associated fixed point algorithm is the state evolution, the asymptotic AMP.

One could think also about a Bethe free entropy for tree-like graphs, but averaged over the disorder. In this case the fixed point equations predicting the BP algorithm behavior on infinite graphs is referred as density evolution, the equivalent of the state evolution for AMP.
\section{Spatial coupling for optimal inference in linear estimation}
\label{sec:spatialCoupling}
As already discussed, mean-field systems (systems for which mean-field techniques such as message-passing algorithms are appropriate) are of two types: the problems defined on sparse random graphs due to their tree-like property, such as the independent set \cite{barbier2013hard} or many other combinatorial optimization problems and problems defined on densely connected graphs such as in the present case (\ref{eqIntro:AWGNCS}). As discussed in sec.~\ref{sec:whyMessagePassingWorksOnDenseGraphs} these systems are equivalent to infinite dimensional homogeneous systems. Here nucleation (a local change of thermodynamical phase) cannot occur: an infinite number of dimensions implies that there is no notion of locality in these systems and thus no nucleus can spread as any apparition of a different phase nucleus inside another one has an infinite energy cost, as the "surface" of the nucleus is itself infinite.

As discussed in sec.~\ref{sec:typicalPhaseTransitions}, in inference the phases we are dealing with are computational: easy/hard/impossible inference phases. In order to have nucleation of an easy phase inside an hard one and to allow this nucleus to propagate inside the full system, one has to introduce a dimensionality, or structure in the problem. This is done by spatial coupling using a properly designed coding operator, see Fig.~\ref{fig_opSpCoupling}. It mimics the strategy employed by the nature. We use again the example of supercooled water which is blocked in the metastable liquid state by a first order transition despite it is below its critical temperature. If a nucleus of crystal appears somewhere in the fluid, the surface between the two phases has an energetic cost $C_1 2\pi R^2$ where $R$ is the radius of the nucleus that we consider spheric. But as the system is $3$-d, this cost remains always finite. Now if the nucleus is big enough, the reduction in energy by $C_2 4/3\pi R^3$ due to the bulk of the small crystal nucleus (the true equilibrium state at this temperature) counterbalances the surface energy term and the entropy loss $\Delta S <0$ due to the crystal: $\Delta F =F_{with ~nucleus} - F_{no ~nucleus}= C_1 2\pi R^2 - C_24/3\pi R^3 -T \Delta S< 0$ and the nucleus spreads in the overall system which finally reaches its true equilibrium macrostate. $C_1, C_2>0$ are constants that depend on the microscopic physical interactions between atoms.

Spatial coupling thanks to which the theoretical optimal thresholds can be saturated was developed in error-correcting codes
\cite{FelstromZigangirov99,KudekarRichardson10,KudekarRichardson12}
and has been extensively used in the compressed sensing setting as well \cite{KrzakalaPRX2012,KrzakalaMezard12,DonohoJavanmard11,DonohoMaleki10}. It rigorously allows to reach the information theoretical bound in LDPC codes \cite{KudekarRichardson10} and in compressed sensing in the random i.i.d measurement matrix case \cite{KudekarRichardson10,DonohoJavanmard11} as we will see. Rigorous proofs of this was worked out in
\cite{DonohoJavanmard11}. The robustness to measurement noise was also discussed in \cite{DonohoJavanmard11,KrzakalaMezard12}. In addition, spatial coupling is used as a proof technique for understanding properties of uncoupled ensembles \cite{DBLP:journals/corr/abs-1301-5676}. Furthermore, it is applicable in a very wide range of graphical models and allows to asymptotically solve constraint satisfaction problems until their satisfability threshold \cite{DBLP:journals/corr/abs-1105-0785,EPFL-REPORT-195693}, the last point in the phase diagram until which it is theoretically possible to find a solution to the problem, see sec.~\ref{sec:typicalPhaseTransitions}. The spatial coupling strategy is conjectured efficient to solve a problem only if a first order phase transition is present, which is the case in a very large class of interesting problems.
\subsection{The spatially-coupled operator and the reconstruction wave propagation}
%
\begin{figure}[!t]
\centering
\includegraphics[width=1\textwidth]{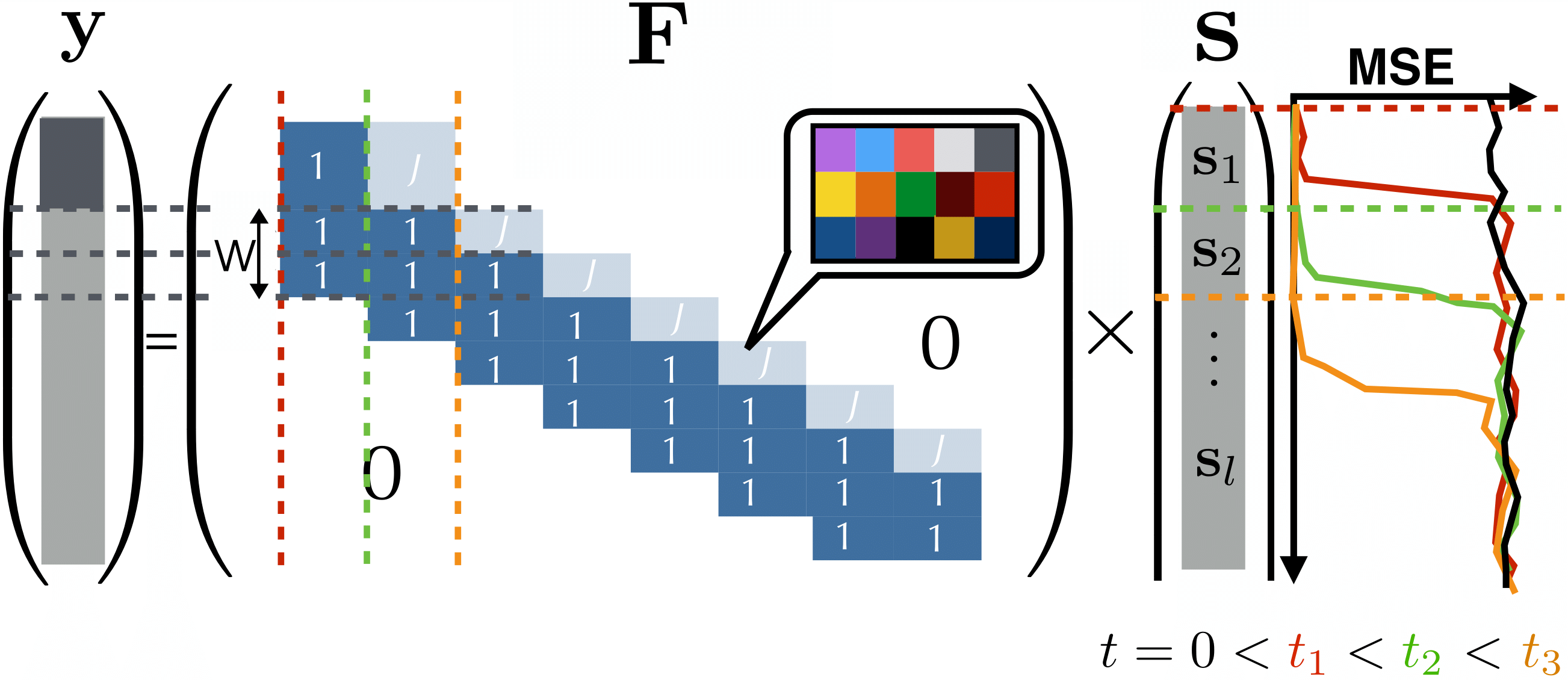}
\caption[Spatial coupling in sparse linear estimation and the reconstruction wave propagation]{Representation of the spatially-coupled sensing matrix used in this thesis. The operator is decomposed in
  $L_r\times L_c$ blocks, each being made of $N/L_c$ columns and
  $\alpha_{seed}N/L_c$ lines for the blocks of the first
  block-row, $\alpha_{rest}N/L_c$ lines for the following
  block-rows (these follow from the definition of $\alpha \defeq M/N$
  combined with (\ref{eq_alphaRest}) ), with
  $\alpha_{seed} > \alpha_{BP} \ge \alpha_{rest} > \alpha_{opt}$. The figure shows that each block is itself a small random operator with i.i.d elements. There is a number $w$ (the coupling
  window) of lower diagonal blocks with elements of variance $1$ as
  the diagonal blocks, the upper diagonal blocks have elements
  of variance $J<1$ where $\sqrt{J}$ is the coupling strength, all
  the other blocks contain only zeros. The matrix structure is thus encoded through a $L_r\times L_c$ matrix $\tbf J$ with element $J_{r,c}$ giving the variance $\in O(1)$ of the block at the $l^{th}$ block-row and $c^{th}$ block-column. The colored dotted lines help
  to visualize the block decomposition of the signal $\bs=[\bs_c]_c^{L_c}$ induced by the
  operator structure. Each block of the signal will be reconstructed
  at different times in the algorithm, as depicted by the right part of the plot which shows how the wave of reconstruction propagates though time, see the color code behind: at $t=0$, the $MSE$ is homogeneously high (black curve) and as time goes on, blocks are reconstructed from the seed until the end of the signal. The parameters that define the spatially-coupled operator ensemble are
   $(L_c,L_r,w,\sqrt{J},\alpha_{seed},\alpha_{rest})$. In the thesis, the matrix elements are rescaled by $1/\sqrt{L}$ such that the variances are rescaled by $1/L$ to enforce the measurements to be $\in O(1)$.}
\label{fig_opSpCoupling}
\end{figure}
Let us now describe how the spatial coupling is implemented in the context of sparse linear estimation through the measurement operator construction. The spatial structure decribed above is induced in the signal by the block structure of the measurement (or coding) matrix, see Fig.~\ref{fig_opSpCoupling}. Other designs are possible \cite{KrzakalaMezard12,DBLP:journals/corr/abs-1207-2853} but in the present thesis, spatially-coupled operators will always be of the form Fig.~\ref{fig_opSpCoupling} as they are empirically very efficient. Looking at Fig.~\ref{fig_opSpCoupling}, we clearly see that if we consider only the diagonal blocks on the matrix, from the measurements point of view the signal becomes the concatenation of independent sub-systems $\bs=[\bs_c]_c^{L_c}$ where $L_c$ is the number of such blocks. But the matrix has also blocks on the the upper and some lower diagonals as well: these couple the different sub-systems $\{\bs_c\}_c^{L_c}$. The blocks on the lower diagonals couple the signal block $\bs_c$ with the $w$ previous ones $\{\bs_{c-i},\}_i^w$, where $w$ is called the coupling window. They have elements with statistically the same amplitude as the diagonal blocks. The upper diagonal blocks weakly couple the block $\bs_c$ with the next one $\bs_{c+1}$. This forward coupling strength is tuned by the $J$ parameter, the variance of the elements of the random values inside the block. The matrix structure is thus fully encoded through a $L_r\times L_c$ matrix $\tbf J$ with element $J_{r,c}$ giving the (non rescaled by $1/L$) variance $\in O(1)$ of the block at the $l^{th}$ block-row, $c^{th}$ block-column of the matrix.

Let us assume that we want to solve a compressed sensing instance generated in the hard phase, where the equilibrium is given by a low $MSE$ estimate and reconstruction should succeed from a thermodynamic point of view. Unfortunately, the algorithm is stuck in the metastable high $MSE$ state by the first order BP transition. The nucleus of "crystal", i.e. of easy phase called the {\it seed}, is induced by the matrix first block-row: as seen on Fig.~\ref{fig_opSpCoupling}, these blocks are closer to square than the next block-rows, i.e. they have a higher effective measurement rate $\alpha_{seed} > \alpha_{rest}$, where $\alpha_{rest}$ is the effective measurement rate of the next block-rows which can be asymptotically as small as the Bayes optimal measurement rate $\alpha_{opt}$. As the forward coupling is relatively weak, the first measurements represented by the darker grey part of the measurement vector in Fig.~\ref{fig_opSpCoupling} contain essentially information about the components of the signal inside its first block $\bs_1$. As we enforce $\alpha_{seed} > \alpha_{BP}$, the "signal seed" $\bs_1$ is easily reconstructed by message-passing and then this new information which is coupled to the next signal block helps the reconstruction of this neighboring $\bs_2$ by increasing its effective measurement rate. This triggers a wave of reconstruction from the top of the signal where the seed is until its end. This is represented on Fig.~\ref{fig_opSpCoupling}: initially the $MSE$ is homogeneously high in all the signal (black curve on the $MSE$ plot on the right of the signal), and as the time steps increase, more and more blocks are reconstructed until the signal is fully reconstructed. \cite{caltagirone2014properties,journals/corr/CaltagironeFMZ13} present detailed studies of the spatial coupling in compressed sensing and the dynamical properties of the reconstruction wave front.

One has to be careful to ensure that these sub-systems $\{\bs_c\}_c^{L_c}$ remain large enough for the assumptions behind the approximate message-passing to be valid inside each of them: each $\bs_c =[\bs_l]_{l\in c}$ must itself be a mean-field system, which is true if $L\gg L_c$. Concurrently, however, the larger $L_c$, the better it is to get closer to the optimal treshold $\alpha_{opt}$. This is due to the fact that the overall measurement rate $\alpha$ is a weighted average of the effective measurement rate of the seed block $\alpha_{seed}$ and that of the remaining ones $\alpha_{rest}$:
\begin{equation}
\alpha_{rest}= \frac{\alpha L_c -\alpha_{seed}}{L_r - 1} \defeq \alpha\left(\frac{L_c - \beta_{seed}}{L_r - 1}\right)
\label{eq_alphaRest}
\end{equation}
In practice, $\alpha$ is fixed and $\alpha_{seed}\defeq \alpha\beta_{seed}$ as well as by fixing $\beta_{seed}$. $\alpha_{rest}$ is then deduced from (\ref{eq_alphaRest}). In the remaining of this thesis, we will define the spatially-coupled ensemble by $(L_c,L_r,w,\sqrt{J},\alpha, \beta_{seed})$ or $(L_c,L_r,w,\sqrt{J},\alpha_{seed},\alpha_{rest})$ equivalently. 
This relation (\ref{eq_alphaRest}) is equivalent to:
\begin{align}
	\alpha = \frac{(L_r-1) \alpha_{rest} + \alpha_{seed}}{L_c} \underset{L_c, L_r \to \infty}{\to} \alpha_{rest} > \alpha_{opt}\label{eq_alphaRest_2}
\end{align}
where $\alpha_{rest}$ can asymptotically be as small as $\alpha_{opt}$. Thus spatial coupling allows to {\it asymptotically} reach the optimal transition, the information theoretic limit of reconstruction. In this way, the gap between $\alpha_{opt}$ and $\alpha_{BP}$, the hard phase discussed in sec.~\ref{sec:typicalPhaseTransitions} is filled and AMP becomes asymptotically Bayes optimal for any $\alpha > \alpha_{opt}$.

In the case of vectorial components, one may be careful to induce the blocks such that the sections are not "cut" by the induced structure, but it appears empirically that it does not change anything, so we forget about this detail in the algorithm implementation.
\section{The spatially-coupled approximate message-passing algorithm}
We define $\textbf{e}_c$ with $c\in\{1,\ldots,L_c\}$, a vector of size $N/L_c$, as the $c^{th}$ block of $\textbf{e}$ (of size $N$) and $\textbf{f}_r$ with $r\in\{1,\ldots,L_r\}$, a vector of size $\alpha_rN/L_c$ as the $r^{th}$ block of $\textbf{f}$ (of size
$M$). For example, in Fig.~\ref{fig_opSpCoupling}, the signal
$\bs$ is decomposed as $\bs=[\bs_c]_c^{L_c}$. The notation $i\in c$ (resp. $\mu\in r$) means all the components of $\textbf{e}$ that are in $\textbf{e}_c$ (resp. all the components of $\textbf{f}$ that are in $\textbf{f}_r$). The algorithm (for scalar signals and matrices, see Fig.~\ref{algo_AMP_complex} for the complex case) requires four different operators performing the following operations:
\begin{align}
\tilde O_\mu(\textbf{e}_c) &\defeq  \sum_{i\in c}^{N/L_c} F_{\mu i}^{2} e_i \label{eq_fastOpDefs11}\\  
O_\mu(\textbf{e}_c) &\defeq  \sum_{i\in c}^{N/L_c} F_{\mu i} e_i \label{eq_fastOpDefs1} \\
\tilde O_i(\textbf{f}_r) &\defeq  \sum_{\mu\in {r}}^{\alpha_rN/L_c} F_{\mu i}^2 f_\mu \label{eq_fastOpDefs2}\\
O_i(\textbf{f}_r) &\defeq  \sum_{\mu\in {r}}^{\alpha_rN/L_c}F_{\mu i} f_\mu \label{eq_fastOpDefs22}
\end{align}
$\alpha_r$ is the measurement rate of all the blocks at the $r^{th}$ block-row, for example on Fig.~\ref{fig_opSpCoupling}, $\alpha_1 = \alpha_{seed}$ and $\alpha_{j} = \alpha_{rest} \ \forall \ j>1$.

This version of AMP is perfectly equivalent to Fig.~\ref{algoCh1:AMP} but underlines the spatially-coupled structure of the matrix. Furthermore, it will be useful when defining the structured operators making use of fast transforms such as Hadamard and Fourier operators, see chap.~\ref{chap:structuredOperators}. Actually, even in the case where no fast transforms are used, this form of AMP can be advantageous as the matrix is sparse, and thus it can avoid the useless time consuming products with the many zeros in the matrix.
\begin{figure}[!t]
\centering
\begin{algorithmic}[1]
\State $t\gets 0$
\State $\delta \gets \epsilon + 1$
\While{$t<t_{max} \ \textbf
{and} \ \delta>\epsilon$} 
\State $\Theta^{t+1}_\mu \gets \sum_{c}^{L_c}\tilde O_\mu(\textbf{v}_c^t)$
\State $w^{t+1}_\mu \gets \sum_{c}^{L_c}O_\mu(\textbf{a}_c^t) - \Theta^{t+1}_\mu\frac{y_\mu-w^t_\mu}{{1/{\rm{snr}}} + \Theta^t_\mu}$
\State $\Sigma^{t+1}_i \gets \left[\sum_{r}^{L_r}\tilde O_i\left([{1/{\rm{snr}}} + \boldsymbol{\Theta}_r^{t+1}]^{-1}\right)\right]^{-1/2}$
\State $R^{t+1}_i \gets a^{t}_i + (\Sigma^{t+1}_i)^2 \sum_{r}^{L_r} O_i\left(\frac{\textbf{y}_r - \textbf{w}^{t+1}_r}{{1/{\rm{snr}}} + \boldsymbol{\Theta}^{t+1}_r}\right)$
\State $v^{t+1}_i \gets f_{c_i}\left((\bsy \Sigma_{l_i}^{t+1})^2,\bR_{l_i}^{t+1}\right)$
\State $a^{t+1}_i \gets f_{a_i}\left((\bsy \Sigma_{l_i}^{t+1})^2,\bR_{l_i}^{t+1}\right)$
\State $t \gets t+1$
\State $\delta \gets ||\ba^{t+1} - \ba^{t}||_2^2$
\EndWhile
\State \textbf{return} $\ba^t$
\end{algorithmic}
\caption[The AMP algorithm written with operators]{The AMP algorithm written with operators. Here for example $\textbf{w}^{t+1}_r\defeq [w^{t+1}_\mu]_{\mu \in r}$ and $\ba_c^t\defeq [\ba_l^t]_{l\in c}$. This form underlines how AMP is operating when spatially-coupled operators are used instead of homogeneous matrices and takes advantage from this structure. $l_i$ is the index of the section to which the $i^{th}$ $1$-d variable belongs to. $\epsilon$ is the accuracy for convergence and $t_{max}$ the maximum number of iterations. A suitable initialization for the quantities is ($a_i^{t=0}=\mathbb{E}_{P_0}(x)$, $v_i^{t=0}=\txt{Var}_{P_0}(x)$, $w_\mu^{t=0}=y_\mu$). Once the algorithm has converged, i.e. the quantities do not change anymore from iteration to iteration, the estimate of the $l^{th}$ signal section is $\textbf{a}_l^{t}$. If needed, the damping scheme of Fig.~\ref{algoCh1:AMP} can be used.}  
\label{algoCh1:AMP_op}
\end{figure}
\subsection{Further simplifications for random matrices with zero mean and equivalence with Montanari's notations}
\label{sec:fullTap}
We can go further in the approximation of some quantities computed by the algorithm Fig.~\ref{algoCh1:AMP_op} by considering that the large signal limit allows to replace the elements $F_{\mu i}^2$ by the matrix variance $F_{\mu i}^2\approx J_{r_\mu,c_i}/L$. This allows to derive the so called full-TAP equations for AMP. This is justified by the fact that the average with respect to the matrix realization of all the quantities appearing in the algorithm depending on such squared elements (such as $\Theta_\mu$) are $\in O(1)$ whilst their variance $\in O(1/N)$, see \cite{KrzakalaMezard12}. Thus in the large signal limit, we can neglect their instance-dependent fluctuations using this simplification. Considering the most general version of AMP Fig.~\ref{algoCh1:AMP_op} where it is written in terms of the operators, the dependency in squared matrix elements is just in the $\tilde O_\mu$ (\ref{eq_fastOpDefs11}) and $\tilde O_i$ (\ref{eq_fastOpDefs2}) operators which can thus be approximated as:
\begin{align}
&\tilde O_\mu(\textbf{e}_c)\approx \tilde O_{r_\mu}(\textbf{e}_c) \defeq \frac{J_{r_\mu,c}}{L}\sum_{i\in {c}}^{N/L_c} e_i \label{eq:defTildeOmu}\\ 
&\tilde O_i(\textbf{f}_r) \approx \tilde O_{c_i}(\textbf{f}_r) \defeq \frac{J_{r,c_i} }{L}\sum_{\mu\in {r}}^{\alpha_rN/L_c} f_\mu
\end{align}
They now depend only on the block indices and thus also $\Theta_r$ and $\Sigma_c$ that are derived from them. From this, we can write a simplified form for the AMP algorithm Fig.~\ref{algo_AMP2}.
\subsubsection{Equivalence with Montanari's notations}
%
\begin{figure}[!t]
\centering
\begin{algorithmic}[1]
\State $t\gets 0$
\State $\delta \gets \epsilon + 1$
\While{$t<t_{max} \ \textbf
{and} \ \delta>\epsilon$} 
\State $\Theta^{t+1}_r \gets  \sum_{c}^{L_c}\tilde O_r(\textbf{v}_c^t)$
\State $w^{t+1}_\mu \gets \sum_{c}^{L_c}O_\mu(\textbf{a}_c^t) - \Theta^{t+1}_{r_\mu}\frac{y_\mu-w^t_\mu}{{1/{\rm{snr}}} + \Theta^t_{r_\mu}}$
\State $\Sigma^{t+1}_c \gets \left[\sum_{r}^{L_r}\tilde O_c\left([{1/{\rm{snr}}} + \Theta_r^{t+1}]^{-1}\right)\right]^{-1/2}$
\State $R^{t+1}_i \gets a^t_i + (\Sigma^{t+1}_{c_i})^2 \sum_{r}^{L_r} O_i\left(\frac{\textbf{y}_r - \textbf{w}^{t+1}_r}{{1/{\rm{snr}}} + \Theta^{t+1}_r}\right)$
\State $v^{t+1}_i \gets f_{c_i}\left((\Sigma^{t+1}_{c_i})^2,\bR_{l_i}^{t+1}\right)$
\State $a^{t+1}_i \gets f_{a_i}\left((\Sigma^{t+1}_{c_i})^2,\bR_{l_i}^{t+1}\right)$
\State $t \gets t+1$
\State $\delta \gets ||\ba^{t+1} - \ba^{t}||_2^2$
\EndWhile
\State \textbf{return} $\ba^t$
\end{algorithmic}
\caption[Simplified full-TAP version of the AMP algorithm with operators]{The simplified (with respect to Fig.~\ref{algoCh1:AMP_op}) full-TAP AMP algorithm, where we have approximated the squared elements of the matrix by the its variance. For example in Fig.~\ref{algoCh1:AMP_op}, $\bsy \Theta_r^{t+1}\defeq[\Theta_\mu^{t+1}]_{\mu\in r}$ was a vector of measure-index dependent components whereas now $\Theta_r^{t+1}$ is a scalar with same value $\forall \mu \in r$. Notations must not be confused: $l_i$ is the index of the unique section to which the $1$-d component $i$ belongs to, whereas $c_i$ is the index of the block to which the section $l_i$ belongs to, etc. If needed, the damping scheme of Fig.~\ref{algoCh1:AMP} can be used.}
\label{algo_AMP2}
\end{figure}
Now we show how to go from this simplified algorithm to the equivalent notation of Montanari \cite{montanari2012graphical} in the case of an homogeneous matrix. Starting from Fig.~\ref{algo_AMP2}, we plug the AMP field $R_i^{t+1}$ expression into the denoiser:
\begin{align}
a_{i}^{t+1} &= f_{a_i}\Bigg((\Sigma_{c_i}^{t+1})^2, \underbrace{\left[a^t_j + (\Sigma^{t+1}_{c_i})^2 \sum_{r}^{L_r} O_i\left(\frac{\bsy \tau_r^t}{{1/{\rm{snr}}} + \Theta^{t+1}_r}\right)\right]_{j\in l_i}}_{\defeq \bR_{l_i}^{t+1}}\Bigg)
\end{align}
where we define the residual $\bsy \tau_r^{t}\defeq \textbf{y}_r - \textbf{w}^{t+1}_r = \[\tau_\mu^{t} = y_\mu - w_\mu^{t+1}\]_{\mu\in r}$ and we have defined the blocks such that all the $1$-d components inside the same section are in the same block for the derivation. Now using the iteration of $\textbf{w}^{t+1}_r$ in Fig.~\ref{algoCh1:AMP_op} we get:
\begin{align}
\bsy \tau_r^t &= \textbf{y}_r - \left[\sum_{c}^{L_c}O_\mu(\textbf{a}_c^t) \right]_{\mu \in r} + {\Theta}^{t+1}_{r}\frac{\by_r-\textbf{w}^t_r}{{1/{\rm{snr}}} + {\Theta}^t_{r}} \\
&= \textbf{y}_r - \left[\sum_{c}^{L_c}O_\mu(\textbf{a}_c^t)  \right]_{\mu \in r} + \frac{\Theta^{t+1}_{r}\bsy \tau_r^{t-1}}{{1/{\rm{snr}}} + {\Theta}^t_{r}}
\end{align}
Using the definition of $\Theta^{t+1}_{r}$ from the algorithm Fig.~\ref{algo_AMP2} together with (\ref{eq:defTildeOmu}) we obtain:
\begin{align}
\Theta^{t+1}_{r} &= \frac{B}{L_c}\sum_c^{L_c} J_{r,c}<f_c^t>_c \\
&= \frac{B}{L_c}\sum_c^{L_c} J_{r,c} (\Sigma_c^{t})^2 <(f_a^t)'>_c \label{eq_defThetaAMP2}
\end{align}
where we have used the property (\ref{eq1:propertyfafc_1}) of the denoising function for the last equality and we define the shorthand notations $<>_c$ of the empirical average restricted to one block $c$ :
\begin{align}
<f_c^t>_c &\defeq \frac{L_c}{N}\sum_{i\in c}^{N/L_c} f_{c_i}\left((\Sigma_{c}^{t})^2,\bR_{l_i}^{t}\right) \\ 
<(f_a^t)'>_c &\defeq \frac{L_c}{N}\sum_{l\in c}^{L/L_c}\sum_{j}^{B} \frac{\partial f_{a_{l(j)}}\left(x,\textbf{y}\right)}{\partial y_j}\Bigg|_{(\Sigma_{c}^{t})^2,\bR_{l}^{t}}
\end{align}
where $l(j) \in \{1,\ldots,N\}$ is the index of the $j^{th}$ $1$-d variable belonging to the section $l$, where $j\in \{1,\ldots,B\}$. From here we can show that this form gives back the Montanari's one in the homogeneous operator case $(L_c=L_r=J_{r,c}=1)$. In this case, the quantities in the algorithm become:
\begin{align}
\Theta^{t+1} &= B (\Sigma^{t})^2 <(f_a^t)'> \label{eq:Thetatp1}\\
(\Sigma^{t+1})^2 &= \[\frac{1}{L} \sum_\mu^{\alpha LB} \frac{1}{1/{\rm snr} +\Theta^{t+1}}\]^{-1}\\
&= \frac{\Theta^{t+1} + 1/{\rm snr}}{B\alpha} \label{eq:Sigmatp1}\\
&=\frac{B (\Sigma^{t})^2 <(f_a^t)'> + 1/{\rm snr}}{B\alpha}
\end{align}
\begin{align}
a_{i}^{t+1} &= f_{a_i}\left((\Sigma^{t+1})^2, \left[a^t_j + (\Sigma^{t+1})^2 \sum_{\mu}^{M} \frac{F_{\mu j} \tau_\mu^t}{\Theta^{t+1} + 1/{\rm snr}} \right]_{j\in l_i} \right)\label{eq:ai_almostLast}\\
&=f_{a_i}\left((\Sigma^{t+1})^2, \left[a^t_j + \frac{1}{B\alpha} \sum_{\mu}^{M} F_{\mu j} \tau_\mu^t \right]_{j\in l_i} \right)\\
\tau_\mu^t &= y_\mu - \sum_{i}^N F_{\mu i}a_i^t + \tau_\mu^{t-1}\frac{\Theta^{t+1}}{{1/{\rm{snr}}} + \Theta^t} \\
&= y_\mu - \sum_{i}^N F_{\mu i}a_i^t + \tau_\mu^{t-1}\frac{B(\Sigma^t)^2 <(f_a^t)'>}{B\alpha (\Sigma^t)^2}\label{eq:taumu_almostLast} \\
&=y_\mu - \sum_{i}^N F_{\mu i}a_i^t + \tau_\mu^{t-1}\frac{<(f_a^t)'>}{\alpha}
\end{align}
where we used the definition of $\Sigma^{t+1}$ from Fig.~\ref{algo_AMP2} and (\ref{eq:Thetatp1}), (\ref{eq:Sigmatp1}) to simplify (\ref{eq:ai_almostLast}) and obtain (\ref{eq:taumu_almostLast}). The very last step is to rescale the coding matrix by dividing its elements by $B\alpha: \tilde \bF\defeq \bF / (B\alpha)$. The measure $\tilde \by$ is thus rescaled in the same way. We finally obtain the Montanari's form of AMP for homogeneous matrices and $B$-d vectorial components signals:
\begin{align}
\tilde \tau_\mu^t &= \tilde y_\mu - \sum_{i}^N \tilde F_{\mu i}a_i^t + \frac{\tilde \tau_\mu^{t-1}<(f_a^t)'>}{\alpha}\\
(\Sigma^{t+1})^2 &= \frac{(\Sigma^{t})^2 <(f_a^t)'> + 1/(B{\rm snr})}{\alpha}\\
a_{i}^{t+1} &= f_{a_i}\left((\Sigma^{t+1})^2, \left[a^t_j + \sum_{\mu}^{M} \tilde F_{\mu j} \tilde \tau_\mu^t\right]_{j\in l_i} \right)
\end{align}
where $\tilde{\bsy \tau}^t$ is the rescaled residual, and:
\begin{align}
<(f_a^t)'> &\defeq \frac{1}{N}\sum_{l}^{L}\sum_{j}^{B} \frac{\partial f_{a_{l(j)}}\left(x,\textbf{y}\right)}{\partial y_j}\Bigg|_{(\Sigma^{t})^2,\left[a^{t-1}_j + \sum_{\mu}^{M} \tilde F_{\mu j} \tilde \tau_\mu^{t-1} \right]_{j\in l_i}} \\
&= \frac{1}{N(\Sigma^{t})^2}\sum_{i}^{N} f_{c_i}\left((\Sigma^{t})^2,\left[a^{t-1}_j + \sum_{\mu}^{M} \tilde F_{\mu j} \tilde \tau_\mu^{t-1}\right]_{j\in l_i}\right)
\end{align} 
\section{State evolution analysis in the spatially-coupled measurement operator case}
\label{sec:spatiallyCoupledSE}
The derivation of the state evolution with spatial coupling is
very similar to the full operator case, see sec.~\ref{sec:stateEvolutionGeneric}. The difference is that now each block of the matrix Fig.~\ref{fig_opSpCoupling} can have a different variance, and thus one must be vigilant when performing the derivation. We give here the main steps, the details being similar to the full case. All the computations are done keeping in mind the limit $L\gg L_c, L_r$ for which AMP is valid with spatial coupling. We start from the algorithm Fig.~\ref{algoCh1:AMP_op} and the operators definitions (\ref{eq_fastOpDefs11}), (\ref{eq_fastOpDefs1}), (\ref{eq_fastOpDefs2}), (\ref{eq_fastOpDefs22}). As in sec.~\ref{sec:stateEvolutionGeneric}, we need to study the fluctuations of the AMP field, the random variable of the disorder that takes as input the denoisers:
\begin{align}
\bR_l^{t+1} &= \ba_l^t + (\bxigma^{t+1}_l)^2\sum_r^{L_r}\sum_{\mu\in r}^{\alpha_r N/L_c} \frac{\bF_{\mu l}}{1/{\rm{snr}}+ \Theta_\mu^{t+1}} \bigg[\sum_c^{L_c} \sum_{k\in c}^{L/L_c} \bF_{\mu k}^{\intercal}(\bs_k - \ba_k^t) + \xi_\mu + \Lambda_\mu^t \bigg]
\end{align}
where $\Lambda_\mu^t$ is defined in terms of the AMP quantities in (\ref{eq:R_derivSE_full}). The variance of the matrix elements depend only on the block to which they belong, thus in the thermodynamic limit when we replace the square matrix elements by their variance, it simplifies the variance of the measure estimate which end up depending only on the block line index:
\begin{align}
\Theta_\mu &= \sum_c^{L_c}\sum_{l\in c}^{L/L_c}(\bF_{\mu l}^2)^{\intercal}\bv_l \\&
\approx \frac{1}{L}\sum_c^{L_c} J_{r_\mu,c}\sum_{l\in c}^{L/L_c}\sum_{i\in l}^B v_i \\
&\eqdef \Theta_{r_\mu} \label{eq_thetar}\\
\Rightarrow \Lambda_{\mu}^t &=\Theta_{r_\mu}^{t+1}\frac{y_\mu - w_\mu^t}{1/{\rm{snr}} + \Theta_{r_\mu}^{t}}
\end{align}
where $J_{r,c} \in O(1)$ is the not yet rescaled by $1/L$ variance of the elements of the block of the spatially-coupled operator that is at the $r^{th}$ block-line, $c^{th}$ block-column, see Fig.~\ref{fig_opSpCoupling}. The notation $r_\mu$ ($c_l$) means the block index $r\in\{1,\ldots,L_r\}$ ($c\in\{1,\ldots,L_c\}$) to which the factor index $\mu$ (section index $l$) belongs to. The previous simplification implies from the definition of $(\bxigma_l^{t+1})^2$ in Fig.~\ref{algoCh1:AMP_op} that:
\begin{align}
	(\bxigma_l^{t+1})^2 &= \frac{L_c}{B} \left(\sum_r^{L_r} \frac{J_{r,c_l}\alpha_r}{1/{\rm{snr}} + \Theta_{r}^{t+1}}\right)^{-1}\boldsymbol{1}_B \label{eq_sigmalSq}\\
	&=(\Sigma_{c_l}^{t+1})^2 \boldsymbol{1}_B
\end{align}
which thus just depend on the block-column index $c_l$ to which the section $l$ belongs to. We deduce from (\ref{eq_sigmalSq}) the simplified expression of $\bR_l^{t+1}$:
\begin{align}
\bR_l^{t+1} &\approx \ba_l^t + (\Sigma^{t+1}_{c_l})^2 \sum_r^{L_r}\frac{1}{1/{\rm{snr}} + \Theta_{r}^{t+1}} \sum_{\mu\in r}^{\alpha_rN/L_c} \bF_{\mu l} \bigg[\sum_c^{L_c}\sum_{k\in c:k\neq l}^{L/L_c} \bF_{\mu k}^{\intercal} (\bs_k - \ba_k^t) + \xi_\mu  + \Lambda_{\mu}^t  \bigg] \nonumber \\ 
&+ (\Sigma^{t+1}_{c_l})^2 \underbrace{\sum_r^{L_r}\frac{1}{1/{\rm{snr}} + \Theta_{r}^{t+1}}\sum_{\mu\in r}^{\alpha_rN/L_c} \bF_{\mu l} \bigg[\bF_{\mu l}^{\intercal}(\bs_l-\ba_l^t)\bigg]}_{\defeq U}\\
&\approx \bs_l + (\Sigma^{t+1}_{c_l})^2 \sum_r^{L_r}\frac{1}{1/{\rm{snr}} + \Theta_{r}^{t+1}}\underbrace{\sum_{\mu\in r}^{\alpha_rN/L_c} \bF_{\mu l} \bigg[\sum_c^{L_c}\sum_{k\in c:k\neq l}^{L/L_c} \bF_{\mu k}^{\intercal} (\bs_k - \ba_k^t) + \xi_\mu  + \Lambda_{\mu}^t  \bigg]}_{\defeq \br_{r l}^{t+1}} \label{eq_Rltp1}
\end{align}
where we have used the same approximation as in (\ref{eq_forgetSecPart}) which implies:
\begin{align}
&U=(\Sigma_{c_l}^{t+1})^{-2} (\bx_l - \ba_l^t) + O(1/\sqrt{L})
\end{align}
Now we define:
\begin{equation}
\br_l^{t+1} \defeq (\Sigma^{t+1}_{c_l})^2 \sum_r^{L_r}\frac{\br_{rl}^{t+1}}{1/{\rm{snr}} + \Theta_{r}^{t+1}} \label{eq_rlSeeded}
\end{equation}
Using the independence assumption about the matrix elements we can now compute by central limit theorem the moments of the Gaussian distributed variables $\br_{r l}^{t+1}$ in order to deduce the moments of $\br_{l}^{t+1}$. As in sec.~\ref{sec:stateEvolutionGeneric}, we only keep the $O(1)$ terms. We can actually identify $\br_{r l}^{t+1}$ to $\br^{t+1}$ of (\ref{eq_rl}) and thus, the computations are exactly the same as in the full operator case up to the values of the variances that are different. Using again (\ref{eq1:appW}), (\ref{eq_lambdaMu}) remains valid, so that we get a similar result to (\ref{eq_rFirstMom1}) and (\ref{eq_rFirstMom2}):
\begin{align}
&\mathbb{E}_{\bF,\bxi,\bs} \{\br_{rl}^{t+1}\} \approx \bzero_B \\
\Rightarrow &\mathbb{E}_{\bF,\bxi,\bs} \{\br_{l}^{t+1}\} \approx \bzero_B 
\end{align}
Identifying in (\ref{eq_rSecMomCross1}) $\br_{rl}^{t+1}$ with $\br_l^{t+1}$ which is defined by (\ref{eq_rl}), the result (\ref{eq_rSecMomCross3}) implies that if $l'\neq l$ then $\forall \ r'$:
\begin{align}
	&\mathbb{E}_{\bF,\bxi,\bs} \{\br_{rl}^{t+1}\br_{r'l'}^{t+1}\} \approx \bsy 0_B  \label{eq:brl_always0}
\end{align}
Furthermore, in the case where $r'\neq r$, we have:
\begin{align}
	\mathbb{E}_{\bF,\bxi,\bs} \{\br_{rl}^{t+1}\br_{r'l}^{t+1}\} &=\mathbb{E}_{\bF,\bxi,\bs} \Bigg\{\sum_{\mu \in r}\sum_{\nu \in r'} \bF_{\mu l}\bF_{\nu l}\bigg[\sum_c^{L_c}\sum_{k\in c:k\neq l}^{L/L_c} \bF_{\mu k}^{\intercal} (\bs_k - \ba_k^t) + \xi_\mu  + \Lambda_{\mu}^t  \bigg]\nonumber\\
	&\bigg[\sum_{c'}^{L_c}\sum_{k'\in c':k'\neq l}^{L/L_c} \bF_{\nu k'}^{\intercal} (\bs_{k'} - \ba_{k'}^t) + \xi_\nu  + \Lambda_{\nu}^t  \bigg]\Bigg\}\\
	&= \bsy 0_B  \label{eq:brl_always0_1}
\end{align}
so $\mathbb{E}_{\bF,\bxi,\bs} \{\br_{rl}^{t+1}\br_{r'l'}^{t+1}\}$ can be different of zero only if $l=l'$ and $r=r'$. It implies that the cross terms cancel as well:
\begin{align}
\mathbb{E}_{\bF,\bxi,\bs}\{{\br_l^{t+1} \br_{l'}^{t+1}}\} &\approx \bzero_B
\end{align}
The only moment that changes with respect to the full matrix case is the variance term. Skipping some steps similar to (\ref{eq_rVar1}), (\ref{eq_rVar2}), we get:
\begin{align}
&\mathbb{E}_{\bF,\bxi,\bs}\{(\br_{rl}^{t+1})^2\} =\mathbb{E}_{\bF,\bxi,\bs}\left\{\sum_{\mu\in r}^{\alpha_rN/L_c} \bF_{\mu l}^2 \left[\sum_c^{L_c}\sum_{k\in c:k\neq l}^{L/L_c} \bF_{\mu k}^{\intercal} (\bs_k - \ba_k^t) \right] \left[\sum_{c'}^{L_c}\sum_{k'\in c':k'\neq l}^{L/L_c} \bF_{\mu k'}^{\intercal} (\bs_{k'} - \ba_{k'}^t) \right]\right\}\nonumber\\ 
&+ \mathbb{E}_{\bF,\bxi,\bs}\left\{\sum_{\mu\in r}^{\alpha_rN/L_c} \bF_{\mu l}^2 \xi_\mu^2\right\} + \underbrace{\mathbb{E}_{\bF,\bxi,\bs}\left\{\sum_{\mu\in r}^{\alpha_rN/L_c} \bF_{\mu l}^2 (\Lambda_\mu^t)^2\right\}}_{\in O(L^{-3/2})} \nonumber\\
&+ 2\underbrace{\mathbb{E}_{\bF,\bxi,\bs} \left\{\sum_{\mu\in r}^{\alpha_rN/L_c} \bF_{\mu l}^2 \Lambda_\mu^t \left[\sum_c^{L_c}\sum_{k\in c:k\neq l}^{L/L_c} \bF_{\mu k}^{\intercal} (\bs_k - \ba_k^t)\right] \right\}}_{=\boldsymbol{0}_B}\\
&\approx \mathbb{E}_{\bF,\bxi,\bs}\left\{\sum_{\mu\in r}^{\alpha_rN/L_c} \bF_{\mu l}^2 \left[\sum_c^{L_c}\sum_{k\in c:k\neq l}^{L/L_c} (\bF_{\mu k}^2)^{\intercal} (\bs_k - \ba_k^t)^2 \right] \right\} + \frac{\alpha_r B J_{r,c_l}}{{\rm{snr}}L_c}\\
&\approx \frac{\alpha_r B J_{r,c_l}}{{\rm{snr}}L_c} + \sum_{\mu\in r}^{\alpha_rN/L_c} \frac{J_{r,c_l}}{L}\Bigg[\sum_c^{L_c} \frac{J_{r,c_c}}{L} \underbrace{\sum_{k\in c:k\neq l}^{L/L_c} \mathbb{E}_{\bF,\bxi,\bs}\{(\bs_k - \ba_k^t)^{\intercal}(\bs_k - \ba_k^t)\}}_{\approx E_c^t N/L_c} \Bigg]\\ 
&= \frac{\alpha_r B J_{r,c_l}}{L_c} \left(\frac{1}{{\rm{snr}}} + \frac{B}{L_c}\sum_c^{L_c} J_{r,c_c} E_c^t\right) \label{eq_rl2last}
\end{align}
where $E_c^t$ is the $MSE$ at time $t$ of the block $c$ of the signal, see Fig.~\ref{fig_opSpCoupling}:
\begin{align}
	E_c^t\defeq\frac{L_c}{N}\sum_{k\in c}^{L/L_c} \mathbb{E}_{\bF,\bxi,\bs}\{(\bs_k - \ba_k^t)^{\intercal}(\bs_k - \ba_k^t)\}\label{eq_tildeEc}
\end{align}
The variance of $\br_{r}^{t+1}$ is deduced from (\ref{eq_rlSeeded}) using (\ref{eq:brl_always0_1}):
\begin{align}
\mathbb{E}_{\bF,\bxi,\bs}\left\{(\br_{l}^{t+1})^2\right\} &= (\Sigma_{c_l}^{t+1})^4\sum_{r,r'}^{L_r,L_r} \frac{\mathbb{E}_{\bF,\bxi,\bs}\left\{\br_{rl}^{t+1}\br_{r'l}^{t+1}\right\}}{(1/{\rm{snr}} + \Theta_{r}^{t+1})(1/{\rm{snr}} + \Theta_{r'}^{t+1})} \\
&= (\Sigma_{c_l}^{t+1})^4\sum_{r}^{L_r} \frac{\mathbb{E}_{\bF,\bxi,\bs}\left\{(\br_{rl}^{t+1})^2\right\}}{(1/{\rm{snr}} + \Theta_{r}^{t+1})^2} \label{eq_meanrl2}
\end{align}
We define the average variance of the posterior estimates inside the block $c$ as:
\begin{equation}
V_c^t\defeq {L_c \over N} \sum_{l\in c}^{L/L_c}\sum_{i \in l}^{B} v_i^t \label{eq_meanVc}
\end{equation}
The matching prior conditions (\ref{eqChIntro:NISHI_1}), (\ref{eqChIntro:NISHI_2}) and (\ref{eqChIntro:NISHI_3}) remain true "per block" as the derivation performed in sec.~\ref{sec:priorMatching} just assumed the definition of the denoisers and the prior matching condition: we would obtain the same replacing $\((\Sigma^{t+1})^2, \bR^{t+1}(\bz,\bs_l)\)$ by $\((\Sigma_c^{t+1})^2, \bR_c^{t+1}(\bz,\bs_l)\)$, the only block index dependent quantities. These conditions allow to greatly simplify the analysis. It implies an equality between the mean variance per block and the $MSE$ per block. It becomes $V_c^t=E_c^t \ \forall \ c \in \{1,\ldots,L_c\}$ at each time step if they are initialized with same value. From this and (\ref{eq_thetar}), we can re-write $\Theta_{r}$ as:
\begin{align}
\Theta_r^t &= {B \over L_c}\sum_c^{L_c} J_{r,c}V_c^t\\
&={B \over L_c}\sum_c^{L_c} J_{r,c} E_c^t
\end{align}
We plug this expression into (\ref{eq_rl2last}) and using (\ref{eq_meanrl2}), (\ref{eq_sigmalSq}) we get:
\begin{align}
\mathbb{E}_{\bF,\bxi,\bs}\left\{(\br_{l}^{t+1})^2\right\} &= (\Sigma_{c_l}^{t+1})^4 {B \over L_c}\sum_{r}^{L_r} \frac{\alpha_rJ_{r,c_l}(1/{\rm{snr}} + \Theta_{r}^{t+1})}{(1/{\rm{snr}} + \Theta_{r}^{t+1})^2} \\
&= (\Sigma_{c_l}^{t+1})^2 \boldsymbol{1}_B
\end{align}
So now we know the distribution of $\bR_l^{t+1}$ from (\ref{eq_Rltp1}):
\begin{align}
&\br_l^{t+1} \sim \mathcal{N}\(\br_l^{t+1}\big| \bsy 0_B,(\Sigma_{c_l}^{t+1})^2 \boldsymbol{I}_B\) \\
\Rightarrow &\bR_l^{t+1}\sim \mathcal{N}\(\bR_l^{t+1}\big| \bs_l,(\Sigma_{c_l}^{t+1})^2 \boldsymbol{I}_B\)
\end{align}
From the same arguments as in the full case derivation sec.~\ref{sec:stateEvolutionGeneric}, we finally obtain the following state evolution over the block mean square error $E_c^{t+1}$ inside the block $c$ in the $L\to\infty$ limit:
\begin{align}
E_c^{t+1} &= \frac{1}{B}\sum_{i\in l}^B\int_{\mathbb{R}^B} d\bs_l P_0(\bs_l) \int_{\mathbb{R}^B} \mathcal{D}\bz \bigg[f_{a_i}\left((\Sigma_c^{t+1})^2, \bR_c^{t+1}(\bz,\bs_l)\right) - s_i\bigg]^2 \label{eq_E1}\\
\Sigma_c^{t+1}\(\{E_{c'}^t\}_{c'}^{L_c}\) &= \left[B\sum_{r}^{L_r} \frac{\alpha_{r} J_{rc}}{{L_c/{\rm{snr}}} + B\sum_{c'}^{L_c} J_{rc'}E_{c'}^t}\right]^{-1/2} \label{eq_SEsigmaSeeded}\\
\bR_c^{t+1}(\bz,\bs_l) &\defeq \bs_l + \bz \Sigma_c^{t+1}\label{eqChIntro:defRseeded}
\end{align}
where $f_{a_i}$ is the denoiser (\ref{eq1:fai}). As explained previously, the prior matching conditions of sec.~\ref{sec:priorMatching} imply the same equalities {\it per block} as in the full case, so that like in sec.~\ref{sec:stateEvolutionGeneric} the two following forms of state evolution for spatially-coupled matrices are equivalent to (\ref{eq_E1}):
\begin{align}	
	E_c^{t+1} &~= \frac{1}{B}\[\mathbb{E}_{\bs_c}\{\bs_l^\intercal \bs_l\} - \sum_{i\in l}^B\int_{\mathbb{R}^B} d\bs_l P_0(\bs_l)\int_{\mathbb{R}^B}\mathcal{D}\bz f_{a_i}\left((\Sigma_c^{t+1})^2, \bR_c^{t+1}(\bz,\bs_l)\right)s_i\] \label{eqChIntro:SE_form2_seeded}\\
	&= {1 \over B} \sum_{i\in l}^B\int_{\mathbb{R}^B} d\bs_l P_0(\bs_l) \int_{\mathbb{R}^B} \mathcal{D}\bz f_{c_i}\left((\Sigma_c^{t+1})^2, \bR_c^{t+1}(\bz,\bs_l)\right) \label{eqChIntro:SE_form3_seeded}
\end{align}
together with (\ref{eq_SEsigmaSeeded}) and (\ref{eqChIntro:defRseeded}), where $\mathbb{E}_{\bs_c}\{\bs_l^\intercal \bs_l\} / B$ is the average of the squared signal's components of the block $c$, see Fig.~\ref{fig_opSpCoupling} (in all this thesis, the statistical properties of the signal are homogeneous). $f_{c_i}$ outputs the posterior variance and is given by (\ref{eq1:fci}).
%
%
%
%
%
%
%
%
%
%
%
%
%
%
%
%
%
%
%
%
%
%
%
%
%
%
%
%
%
%
%
%
%
%
%
%
%
%
%
%
%
%
%
%
%
%
%
%
%
%
%
%
%
%
%
%
%
%
%
%
%
%
%
%
%
%
%
%
%
%
%
%
%
%
%
%
%
%
%
%
%
%
%
%
%
%
%
%
%
%
%
%
%
%
%
%
%
%
%
%
%
%
%
%
%
%
%
%
%
%
%
%
%
%
%
%
%
%
%
%
%
%
%
%
%
%
%
%
%
%
%
%
%
%
%
%
%
%
%
%
%
%
%
%
%
%
%
%
%
%
%
%
\part{Signal processing}
\chapter{Compressed sensing of approximately sparse signals}
\label{chap:appSparsity}
Compressed sensing is designed to measure sparse signals directly in
a compressed form. However, most signals of interest are only approximately sparse, i.e. even though the signal contains only a small fraction of relevant large components, the other ones are not strictly equal to zero but are only close to it. In this
chapter we model the approximately sparse i.i.d signal with a sum of two Gaussian
distributions, one with large variance correponding to the informative support of the signal, the second for the small components (which act as an effective noise) with smaller variance, and we study its compressed sensing with dense random matrices. We use replica calculations to determine the mean square error of the Bayes optimal reconstruction
for such signals as a function of the variance of the small
components, the density of large components and the measurement rate. We then study the approximate message-passing algorithm for approximate sparsity and we quantify the region of parameters for which it achieves optimality (for large systems). Finally, we show that in the region where the AMP algorithm with the homogeneous measurement matrices is not optimal, the spatial coupling allows to restore optimality. Even though we
limit ourselves to this special class of signals and assume the prior matching condition, many qualitative features of the results stay true for other signal models and when the distribution of the signal-components is not
known as well \cite{VilaSchniter11}.

The $\ell_1$-minimization based algorithms \cite{CandesTao:06,Donoho:06} are widely used
for compressed sensing of approximately sparse signals. They are very general as discussed in sec.~\ref{sec:convexOptimization} and provide good performances in many situations. They, however, do not achieve optimal reconstruction even when the statistical properties of the signal are known, see sec.~\ref{sec:typicalPhaseTransitions}. 

As we shall see the AMP algorithm for homogeneous measurement matrices matches asymptotically the performance of the optimal reconstruction in a large part of the parameter space. However in some region of parameters that define the hard phase, it is suboptimal as the BP transition blocks the algorithm before the optimal one, see sec.~\ref{sec:typicalPhaseTransitions}. In compressed sensing the spatial coupling was first tested in \cite{KudekarPfister10} who did not observe any improvement for the present bi-Gaussian model for reasons that we will clarify later in this chapter. Basically, the spatial coupling provides improvements only if a first order phase transition is present, but for the variance of small components that was tested in \cite{KudekarPfister10} there is no such
transition: it appears only for slightly smaller values.
\section{The bi-Gaussian prior for approximate sparsity}
We study compressed sensing for approximately sparse signals: the
$N$-d scalar components signals (i.e. $L=N, B=1$ in the general equations for vector components signals presented in the previous theoretical chapters chap.~\ref{chap:MF_graphs_MessPass} and chap.~\ref{chap:phaseTrans_asymptAnalysies_spC}) that we consider have i.i.d components, $K$ of
these being drawn from a distribution $\phi(s)$ and the density of such components is $\rho=K/N$. The remaining $N-K$ components are Gaussian with zero mean and small variance $\epsilon$ :
\begin{align}
	P_0(\bs) &=\prod_{i=1}^N P_0(s_i)\\
	&= \prod_{i=1}^N [ \rho\phi(s_i) + (1-\rho) {\cal N}(s_i|0, \epsilon) ] \label{Px_appSparse}
\end{align}
Of course no real signal of interest is truly i.i.d. However, our analysis also applies to non i.i.d signals which empirical distribution of
components is converging to $P_0(s_i)$, this condition being sufficient \cite{DonohoJavanmard11}. We focus on the special case of a Gaussian $\phi(s_i) = {\cal N}(s_i| 0, \sigma^2=1)$ of zero mean and unit variance. Although the numerical results depend on the form of $\phi(s_i)$, the overall picture is robust with respect to this choice. We further assume the prior matching condition: the parameters of $P(\bs)$ are known and used in the algorithm. The bi-Gaussian model for approximately sparse signals (\ref{Px_appSparse}) was previously used in compressed sensing, for example in \cite{BaronSarvotham10,KudekarPfister10}. 

For simplicity we assume the measurements to be noiseless, the case of noisy measurements can be treated similarly by the AMP algorithm and learned by expectation maximization (see sec.~\ref{sec:EMlearning}) if unknown. We consider the measurement matrix $\bF$ having i.i.d components of zero mean and variance $1/N$. The measurements are obtained through (\ref{eqIntro:AWGNCS}). In the numerical experiments, the components of the matrix are Gaussian distributed, but the asymptotic analysis does not depend on the details of the components distribution, as long as they are i.i.d.  

The Bayes optimal estimation is intractable in the general case, but as discussed in sec.~\ref{sec:typicalPhaseTransitions} the AMP estimation is Bayes optimal under the matching prior condition before its spinodal transition, which blocks its convergence. We will use an asymptotic replica analysis of the Bayes optimal reconstruction, which allows to compute
the asymptotic $MSE$ as a function of the parameters of the signal distribution ($\rho, \epsilon$) and of the measurement rate $\alpha$. This allows to obtain the phase diagram of the problem.
\subsection{Learning of the prior model parameters}
If the prior parameters are unknown, they can be learned efficiently through expectation maximization described in sec.~\ref{sec:EMlearning}. Here we could start from the Bethe free energy (\ref{eq:BetheF_forAMP}) and derive fixed point equations but as the parameters have simple interpretations, we can derive learnings more easily. Actually, the easiest way is exactly as we will do later on in sec.~\ref{sec:learningMicroscopy} to which we refer: it just requires to compute the posterior probability estimates $\{P(x_i^t \in \mathcal{N})\}_i^N$ at time $t$ that the signal components have been generated by the Gaussian part of the prior:
\begin{align}
	P(x_i^t \in \mathcal{N}) = (1-\rho) \frac{\int dx_i\mathcal{N}(x_i|0,\epsilon)\mathcal{N}(x_i|R_i^t,(\Sigma_i^t)^2)}{\int dx_iP_0(x_i)\mathcal{N}(x_i|R_i^t,(\Sigma_i^t)^2)}
\end{align}
where $P_0(x_i)$ is given by (\ref{Px_appSparse}) and $(R_i^t,(\Sigma_i^t)^2)$ are the AMP fields at time $t$. Then from this, the parameters are easily derived, see sec.~\ref{sec:learningMicroscopy}. For example, the density of informative components (that have been generated by $\phi$ is (\ref{Px_appSparse})) is just (\ref{eq:rhoMicroLearn}). The other parameters are learned similarly.
\section{Reconstruction of approximately sparse signals with the approximate message-passing algorithm}
The generic AMP algorithm in its scalar form Fig.~\ref{algoCh1:AMP} is now studied. Only the denoisers $f_{a_i}$ and $f_{c_i}$ depend explicitely on the signal model $P_0(\bs)$ (\ref{Px_appSparse}). Referring to the sec.~\ref{sec:cookAMP} and using the table Tab.~\ref{tab:prior} for the prior construction, we get directly the denoisers for bi-Gaussian approximate sparsity:
\begin{align}
&f_{a}(\Sigma^2,R) = \frac{  \sum_{a=1}^2 w_a
e^{-\frac{R^2}{2(\Sigma^2+\sigma_a^2)}}
\frac{ R \sigma_a^2}{(\Sigma^2+\sigma_a^2)^{\frac{3}{2}}} }{   \sum_{a=1}^2 w_a \frac{1}{\sqrt{\Sigma^2+\sigma_a^2}}
e^{-\frac{R^2}{2(\Sigma^2+\sigma_a^2)}} } \label{eq:fa_appSparsity} \\
&f_{c}(\Sigma^2,R)=\frac{  \sum_{a=1}^2 w_a
e^{-\frac{R^2}{2(\Sigma^2+\sigma_a^2)}}
\frac{\sigma_a^2
  \Sigma^2 (\Sigma^2 + \sigma_a^2)+  R^2 \sigma_a^4 }{(\Sigma^2+\sigma_a^2)^{\frac{5}{2}}} }{    \sum_{a=1}^2 w_a \frac{1}{\sqrt{\Sigma^2+\sigma_a^2}}
e^{-\frac{R^2}{2(\Sigma^2+\sigma_a^2)}} }-f_{a}(\Sigma^2,R)^2 \label{eq:fc_appSparsity}
\end{align}
where we use the notation $f_a/f_c$ instead of $f_{a_i}/f_{c_i}$ of Fig.~\ref{algoCh1:AMP} in the scalar components signal case. For the approximately sparse signals that we consider here we have:
\begin{align}
	w_1&=\rho \\
	\sigma_1^2&=\sigma^2 = 1 \\
	w_2&=1-\rho\\ 
	\sigma_2^2&=\epsilon
\end{align}
\subsection{State evolution of the algorithm with homogeneous measurement matrices}
\label{evolution}
In the limit of large system sizes, i.e. when the parameters ($\rho,\epsilon,\alpha$) are fixed whereas $N\to \infty$, the evolution of the AMP algorithm is described exactly using the state evolution \cite{BayatiMontanari10} given by (\ref{eqChIntro:SE_form3}) (or any of the other two equivalent forms) under the prior matching condition, as it is the case here. When $B=1, L=N$ it becomes in the noiseless case ${\rm snr}\to\infty$:
\begin{align}
	E^{t+1} = \int ds P_0(s) \int {\cal D}z f_{c}\(\frac{E^t}{\alpha}, s+z\sqrt{\frac{E^t}{\alpha}}\) 
	\label{eq:SE_appSparse_0}
\end{align}
Now plugging the bi-Gaussian prior $P_0(s) = \sum_a^2 w_a \mathcal{N}(s|0,\sigma^2_a)$ in (\ref{eq:SE_appSparse_0}), using the fact that the sum of two independent Gaussian random variables is a new Gaussian random variable with mean and variance given by the sum of the means and variances of the original random variables plus the fact that:
\begin{align}	
	\int du  {\cal N}(u|m,v) f(u) = \int {\cal D}z f(\sqrt{v}z + m) \label{eq_NewGauss}
\end{align}
we directly obtain the final simplified state evolution recursion:
\begin{align}
    &E^{t+1} = \sum_{a=1}^{2} w_a  \int {\cal D}z f_c\left(\frac{E^t}{\alpha}, z\sqrt{\sigma_a^2  + \frac{E^t}{\alpha} }\right) \label{Et}   
\end{align}
where again ${\cal D} z = e^{-z^2/2}/\sqrt{2\pi} dz$ is a unit centered Gaussian measure. The initialization corresponding to the one for the algorithm is $E^{t=0}=\txt{Var}_{P_0}(x) = (1-\rho)\epsilon + \rho \sigma^2$.

\begin{figure}[!t]
\centering
\includegraphics[width=1\textwidth]{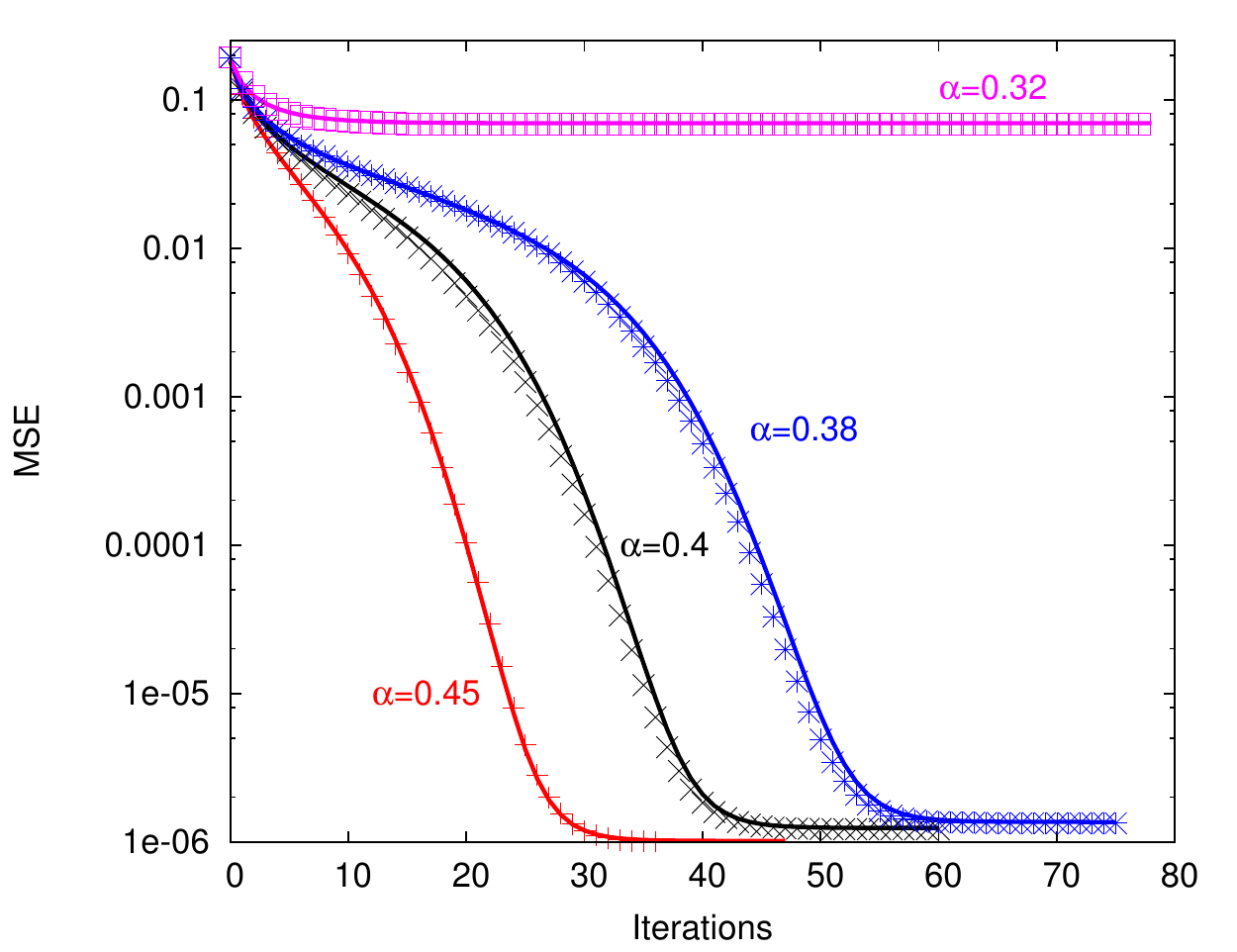}
\caption[State evolution for approximately sparse signals]{Time evolution of the $MSE$ the AMP algorithm achieves (crosses) compared to the asymptotic $N\to \infty$ evolution obtained from the state evolution (\ref{Et}) (full lines) for different measurement rates. Data are obtained for a signal with density of large component $\rho=0.2$ and variance of the small components $\epsilon=10^{-6}$. The algorithm was used for a signal of $N=3\cdot 10^4$ components.}
\label{fig_evolution}
\end{figure}

In Fig.~\ref{fig_evolution} we plot the analytical prediction for the
time evolution of the $MSE$ computed from the state evolution
(\ref{Et}), and we compare it to the one measured in one run of the
AMP algorithm for a system size $N=3\cdot 10^4$. The agreement for
such system size is already excellent. As we see, when the measurement rate is too low, the algorithm converges to an high $MSE$. Furthermore, we observe that when reconstruction succeeds, the $MSE$ falls to a value comparable to the small components variance, here $\epsilon = 10^{-6}$.
\subsection{Study of the optimal reconstruction limit by the replica method}
\label{replicas}
As discussed in sec.~\ref{sec:typicalPhaseTransitions}, for measurement rates below the BP transition and above the static transition $\alpha_{s}(\rho)<\alpha < \alpha_{BP}(\rho)$, the state evolution equation (\ref{Et}) has two different stable fixed points. In particular, if the iterations are initialized with $E\to 0$, one will reach a fixed point with much lower $MSE$ than initializing with large $E=1$. In fact, if $\alpha_{s}(\rho)\le\alpha_{opt}(\rho)<\alpha<\alpha_{BP}(\rho)$ the low error fixed point determines the $MSE$ that would be achieved by the exact Bayes optimal inference. Let us now compute the phase diagram of compressed sensing for bi-Gaussian approximately sparse signals from the Bethe free entropy using the replica method.
\begin{figure}[!t]
\hspace{-.3cm}
\includegraphics[width=.9\textwidth]{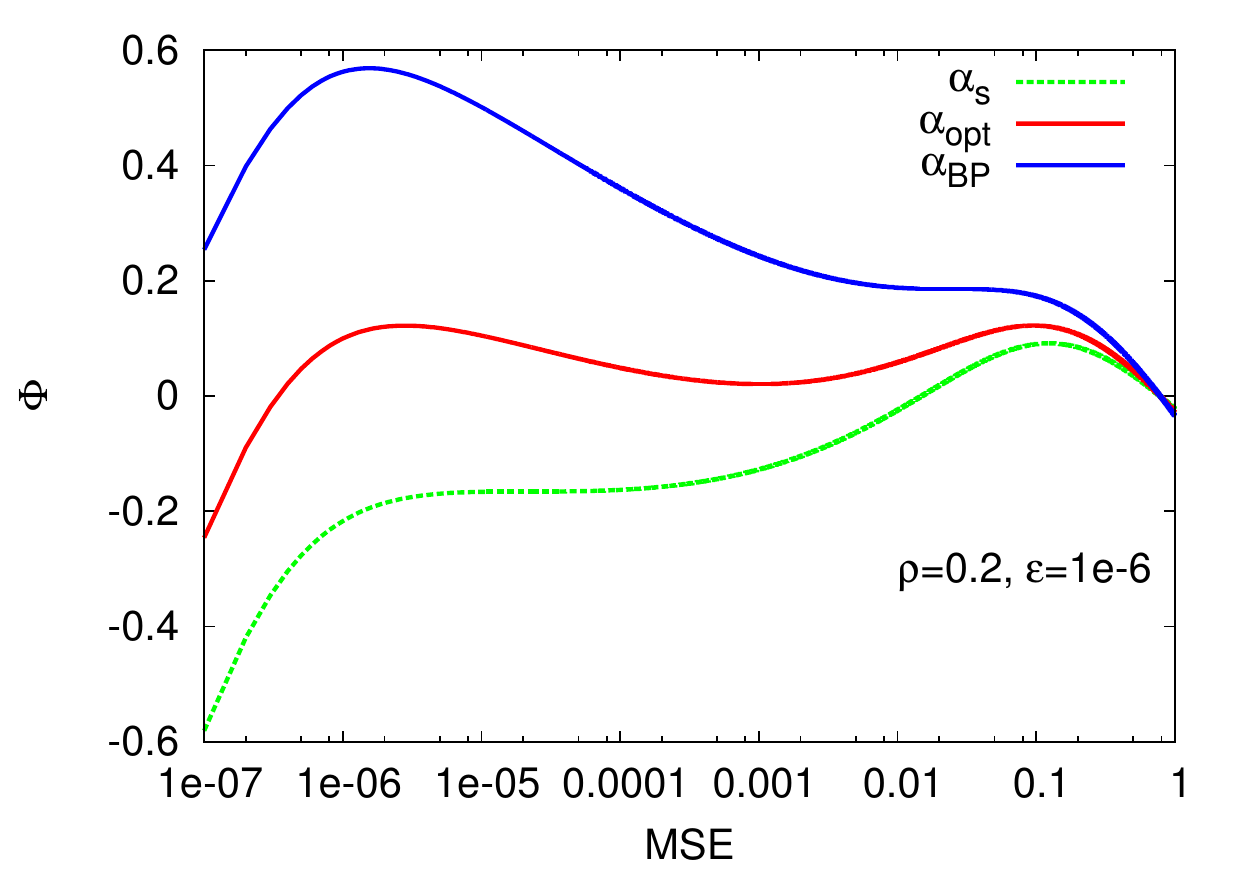}
\caption[Bethe free entropy for compressed sensing of approximately sparse signals at the phase transitions points]{The Bethe free entropy $\Phi(E)$ for compressed sensing of approximately sparse signals of density $\rho=0.2$, with variance of the small components $\epsilon= 10^{-6}$. The three lines depict the potential for three different measurement rates corresponding to the critical values: $\alpha_{BP}=0.3559$ below which AMP is not Bayes optimal anymore without spatial coupling, $\alpha_{opt}=0.2817$, $\alpha_s=0.2305$. The two local maxima exists for $\alpha \in [\alpha_s,\alpha_{BP}]$, and at $\alpha>\alpha_{opt}$ the low $MSE$ maxima is the global one, i.e. the $MMSE$ estimate. Below the static transition $\alpha<\alpha_s$, all information about the signal is lost and only remain the spurious solution at high $MSE$.}
\label{fig_potential}
\end{figure}
We start from the general potential valid under the prior matching condition (\ref{eq1:freeEnt2}), which becomes in the scalar components $B=1$ and noiseless ${\rm snr} \to \infty$ case:
\begin{align}
	\Phi(E) &= \Phi_{B=1}(E|{\rm snr} \to \infty) \nonumber \\
	&= -\frac{\alpha}{2}\(\log(E) + \frac{<\bs^2>}{E}\)+ \int P_0(s) {\cal D}z \log\(\int P_0(x) e^{\frac{sx}{\Sigma(E)^2} + \frac{zx}{\Sigma(E)} -\frac{x^2}{2\Sigma(E)^2}}\)
\end{align}
up to irrelevant constants that do not depend on the $MSE$. $\Sigma(E)=\sqrt{E/\alpha}$ is defined by (\ref{eqChIntro:SIGMA2def}). We define $I$ as the integral appearing in the previous expression. Now using the bi-Gaussian prior $P_0(s) = \sum_i^2 w_i \mathcal{N}(s|0,\sigma^2_i)$ we can compute $I$ by Gaussian integral: 
\begin{align}
	\int dx e^{-ax^2 + bx} = \sqrt{\frac{\pi}{a}}e^{\frac{b^2}{4a}}
\end{align}
to get:
\begin{align}
	I &= \sum_i^2 w_i \int ds {\cal D}z {\cal N}(s|0,\sigma_i^2) \log\(\sum_j^2  \frac{w_j}{\sqrt{2\pi \sigma_j^2}} \int dx e^{-\frac{x^2}{2}(1/\Sigma(E)^2 + 1/\sigma_j^2) + \frac{x}{\Sigma(E)} (s/\Sigma(E) + z)} \)\nonumber \\
	&=\sum_i^2 w_i \int ds {\cal D}z {\cal N}(s|0,\sigma_i^2) \log\(\sum_j^2  \frac{w_j}{\sqrt{\sigma_j^2/\Sigma(E)^2 + 1}}  e^{\frac{(s/\Sigma(E) +z)^2}{2(1 + \Sigma(E)^2/\sigma_j^2)} } \)
\end{align}
Now we use again the simple form of the moments of the sum of two independent Gaussian random variables together with (\ref{eq_NewGauss}) and defining $u\defeq s/\Sigma(E) + z$ we get:
\begin{align}
	I &= \sum_i^2 w_i \int du {\cal N}(u|0,\sigma_i^2/\Sigma(E)^2 + 1) \log\(\sum_j^2 \frac{w_j}{\sqrt{\sigma_j^2/\Sigma(E)^2 + 1}}  e^{\frac{u^2}{2(1 + \Sigma(E)^2/\sigma_j^2)} } \) \\
	&= \sum_i^2 w_i \int {\cal D}z \log\(\sum_j^2 \frac{w_j}{\sqrt{\sigma_j^2/\Sigma(E)^2 + 1}}  e^{\frac{z^2(1 + \sigma_i^2/\Sigma(E)^2)}{2(1 + \Sigma(E)^2/\sigma_j^2)} } \)
\end{align}
The final Bethe free entropy expression is thus:
\begin{align}
\Phi(E) = &- \frac{\alpha}{2} \left(\log(E) + \frac{w_1\sigma_1^2 + w_2\sigma_2^2}{E}\right) \nonumber\\
&+ \sum_i^2 w_i \int {\cal D}z \log\(\sum_j^2 \frac{w_j}{\sqrt{\sigma_j^2/\Sigma(E)^2 + 1}}  e^{\frac{z^2(1 + \sigma_i^2/\Sigma(E)^2)}{2(1 + \Sigma(E)^2/\sigma_j^2)} } \)
\end{align}
In Fig.~\ref{fig_potential} we plot the function $\Phi(E)$ for a signal of density $\rho=0.2$, variance of small components $\epsilon=10^{-6}$ and three different values of the measurement rate $\alpha$ corresponding to the critical values at which happen the different phase transitions. 
We will show next that at a fixed signal density $\rho$, for a variance of the small components lower than a critical value $\epsilon<\epsilon_c(\rho)$, the
optimal Bayes reconstruction has a transition at a critical value $\alpha=\alpha_{opt}(\rho)$ separating a regime with a small value (comparable to $\epsilon$) of the $MSE$ obtained at $\alpha>\alpha_{opt}(\rho)$ from a phase with a large value of the $MSE$ obtained at $\alpha<\alpha_{opt}(\rho)$. As discussed in sec.~\ref{sec:typicalPhaseTransitions}, this is a first order phase transition (as the BP transition) in the sense that the Bayes optimal $MSE$ jumps discontinuously at $\alpha=\alpha_{opt}(\rho)$.

In this intermediate hard region $\alpha_{opt}(\rho)<\alpha<\alpha_{BP}(\rho)$ the AMP performance can be improved with the use of spatially-coupled measurement matrices and with a proper choice of the parameters defining these matrices, one can approach the performance of the optimal Bayes inference in the large system size limit for any measurement rate. 

Finally for higher variance of the small components $\epsilon>\epsilon_c(\rho)$ there is no more phase transition for any $0<\alpha<1$. In this regime, AMP always achieves optimal Bayes inference and the $MSE$ that it obtains varies continuously from $0$ at $\alpha=1$ to
$O(1)$ values at low measurement rate $\alpha$.
\section{Phase diagrams for compressed sensing of approximately sparse signals}
\label{results}
\begin{figure}[!t]
\centering
\hspace{-1cm}
\includegraphics[width=.9\textwidth]{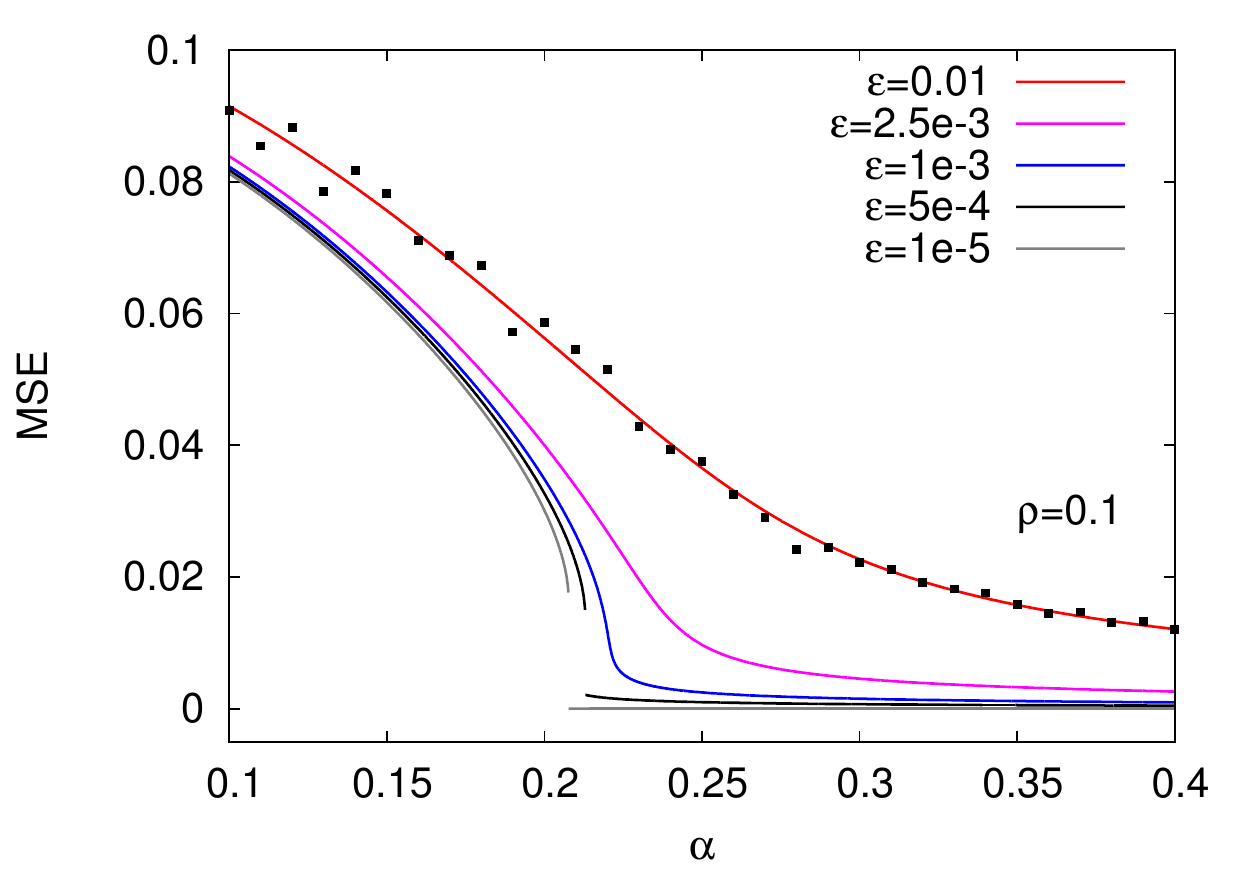}
\caption[From a first order to a continuous transition in compressed sensing with approximate sparsity]{The $MSE$ achieved by the AMP algorithm. The lines correspond to the evaluation of the $MSE$ from the state evolution (\ref{Et}), the data points to the $MSE$ achieved by the AMP algorithm on single instances with $N=3\cdot 10^4$. The data are for signals with density $\rho=0.1$ and several values of variance of the small components $\epsilon$ as a function of the measurement rate $\alpha$. The $MSE$ grows continuously as $\alpha$ decreases for $\epsilon>\epsilon_c(\rho=0.1)=0.00075$. For smaller values of the small components variance, a first order phase transition is present and the $MSE$ jumps discontinuously at $\alpha_{BP}(\rho=0.1,\epsilon)$.}
\label{fig_MSE1}
\end{figure}

In Fig.~\ref{fig_MSE1} we plot the $MSE$ to which the state evolution converges if initialized at large value of $MSE$ - such initialization corresponds to the iterations of AMP when the actual signal is not known. For $\epsilon=0.01$ we also compare explicitly to a run of AMP for a system size of $N=3\cdot 10^4$. Depending on the value of the density $\rho$ and variance $\epsilon$, two situations are possible: for relatively large $\epsilon$, as the measurement rate $\alpha$ decreases the final $MSE$ grows continuously from $E=0$ at $\alpha=1$ to $E=E^{t=0}$ at $\alpha=0$. For lower values of $\epsilon$ the $MSE$ achieved by AMP has a discontinuity at $\alpha_{BP}(\rho,\epsilon)$ at which the second maxima of $\Phi(E)$ appears. Note that the case of $\epsilon=0.01$ was tested in
\cite{BaronSarvotham10}, the case of $\epsilon=0.0025$ in \cite{KudekarPfister10}. This why the authors of \cite{KudekarPfister10} did not observe any improvement by spatial coupling: for these parameters $(\rho, \epsilon)$, there is no first order transition and thus spatial coupling is useless as AMP is anyway Bayes optimal at any measurement rate $\alpha$.
\begin{figure}[!t]
\includegraphics[width=.9\textwidth]{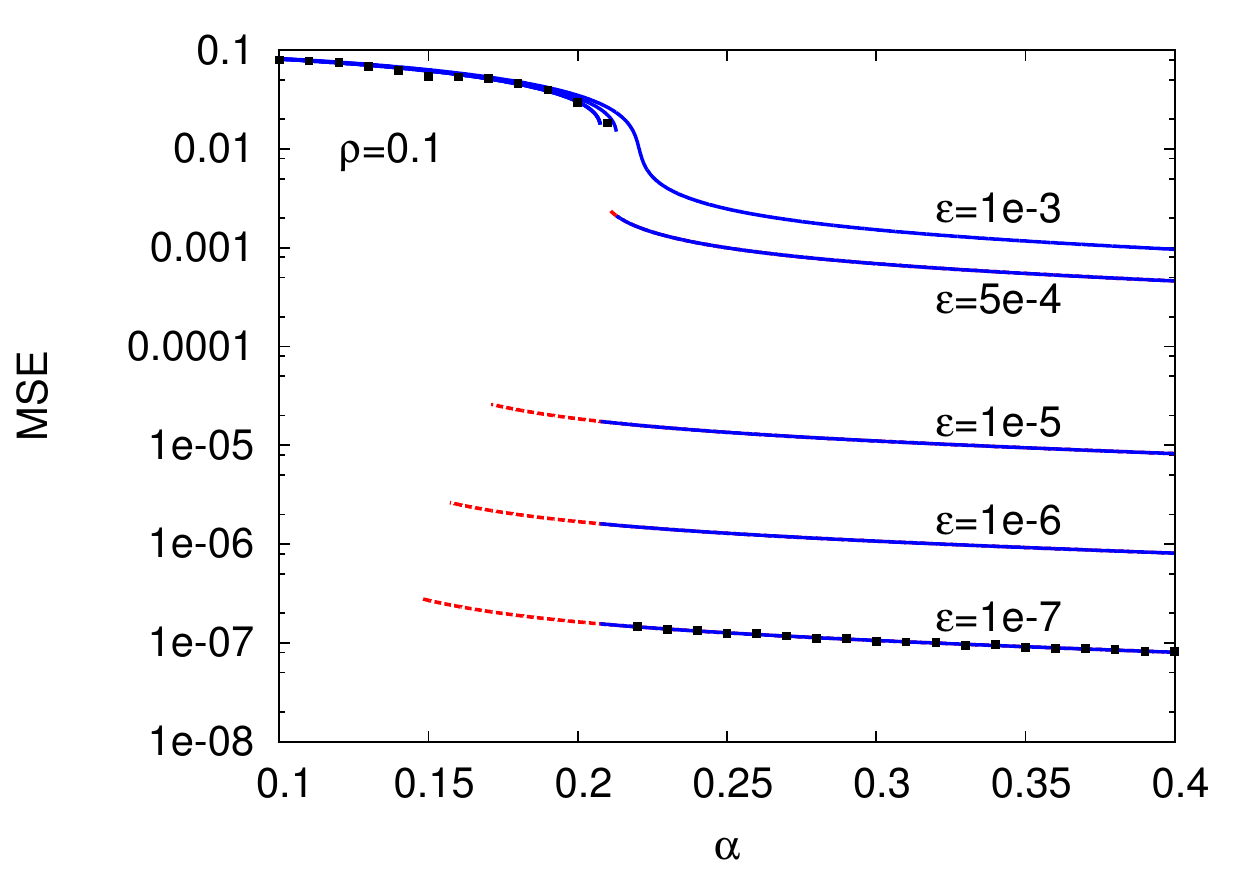}
\caption[Appearance of the BP transition as the small components variance is decreased and comparison of the AMP reconstruction performances with the Bayes optimal inference]{$MSE$ achieved asymptotically by the AMP (blue solid lines) compared to the $MSE$
achieved by the Bayes optimal inference (red dashed lines) as evaluated using the state evolution, initializing it from $E\to 0$ to get the Bayes optimal $MSE$ or $E=1$ for the AMP asymptotic $MSE$. The data points correspond to the $MSE$ achieved by the AMP algorithm for $N=3\cdot 10^4$. The optimal $MSE$ jumps at $\alpha_{opt}(\rho,\epsilon)$. Hence for $\epsilon<\epsilon_c(\rho=0.1)=0.00075$ there is a range of measurement rates $[\alpha_{opt}(\rho=0.1,\epsilon),\alpha_{BP}(\rho=0.1,\epsilon)]$ for which the AMP algorithm is asymptotically suboptimal. In this gap, spatial coupling can be used to restore the optimality of AMP.}
\label{fig_MSE2}
\end{figure}

In Fig.~\ref{fig_MSE2} we plot in solid blue line the $MSE$ to which the AMP asymptotically converges and compare it to the $MSE$ achieved by the optimal Bayes inference (in dashed red line), i.e. the $MSE$ corresponding to the global maximum of $\Phi(E)$. We see that, when the discontinuous transition point $\alpha_{BP}(\rho,\epsilon)$ exists, then in the region $\alpha_{opt}(\rho,\epsilon)<\alpha<\alpha_{BP}(\rho,\epsilon)$ AMP is suboptimal. We remind that in the limit $\epsilon \to 0$, exact reconstruction is possible for any $\alpha>\alpha_{opt}(\rho)=\rho$. We see that for $\alpha<\alpha_{opt}(\rho,\epsilon)$ and for $\alpha>\alpha_{BP}(\rho,\epsilon)$ the performance of AMP matches asymptotically the performance of the Bayes optimal inference. The two regions are, however, quite different as discussed in sec.~\ref{sec:typicalPhaseTransitions}. For $\alpha<\alpha_{opt}(\rho,\epsilon)$ the final $MSE$ is relatively large, whereas for $\alpha>\alpha_{BP}$ the final $MSE$ is of order $\epsilon$ and hence is this region the problem shows a very good stability towards approximate sparsity. 

\begin{figure}[!t]
\includegraphics[width=1\textwidth]{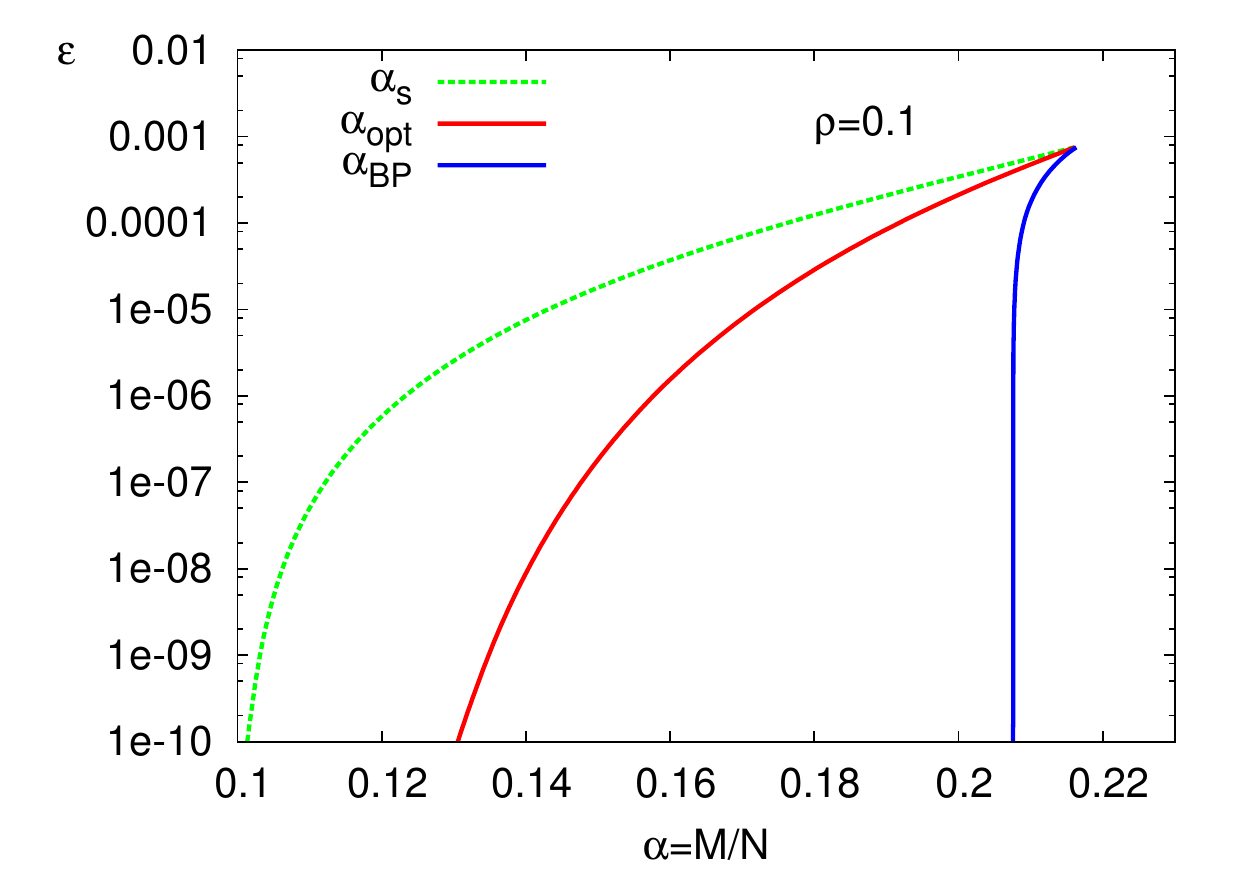}
\caption[Phase diagram for compressed sensing of approximately sparse signals in the $(\epsilon, \alpha)$ plane]{Phase diagram for compressed sensing of approximately sparse signals. The density of the large signal components is $\rho=0.1$, we are changing the measurement rate $\alpha$ and the variance of the small components $\epsilon$. The critical values of measurement rates $\alpha_{opt}(\epsilon,\alpha)$, $\alpha_{BP}(\epsilon,\alpha)$ and $\alpha_s(\epsilon,\alpha)$ are plotted. For homogeneous measurement matrices, AMP does not achieve optimal reconstruction in the area between $\alpha_{opt}(\epsilon,\alpha)$ (red curve) and $\alpha_{BP}(\epsilon,\alpha)$ (blue curve). For any measurement rate above the tri-critical point where the three transitions curves meet, there is no more phase transitions and AMP is always Bayes optimal for any $\epsilon$ and the $MSE$ becomes a continuous function of $\alpha$.}
\label{fig_phase_diagram}
\end{figure}

In Fig.~\ref{fig_phase_diagram} we summarize the critical values of
$\alpha_{BP}(\epsilon,\alpha)$ and $\alpha_{opt}(\epsilon,\alpha)$ for a signal of density
$\rho=0.1$ as a function of the variance of the small components and the measurement rate. Note that for $\epsilon>\epsilon_c(\rho=0.1)=0.00075$ there are no phase transitions anymore, hence for this large value of $\epsilon$, the AMP algorithm matches asymptotically the optimal Bayes inference at any $\alpha$. Note that in the limit of exactly sparse signal
$\epsilon\to 0$, the values $\alpha_{opt}(\rho)\to \rho$ and $\alpha_{s}(\rho)\to \rho$ whereas $\alpha_{BP}(\rho)\to 0.2076$, hence for $\alpha>0.2076$ the AMP algorithm is very robust
with respect to appearance of approximate sparsity since the
transition $\alpha_{BP}(\rho)$ has a very weak $\epsilon$-dependence, as
seen in Fig.~\ref{fig_phase_diagram}.
\begin{figure*}[t!]
\centering
\hspace{-1.5cm}
\includegraphics[width=.387\textwidth]{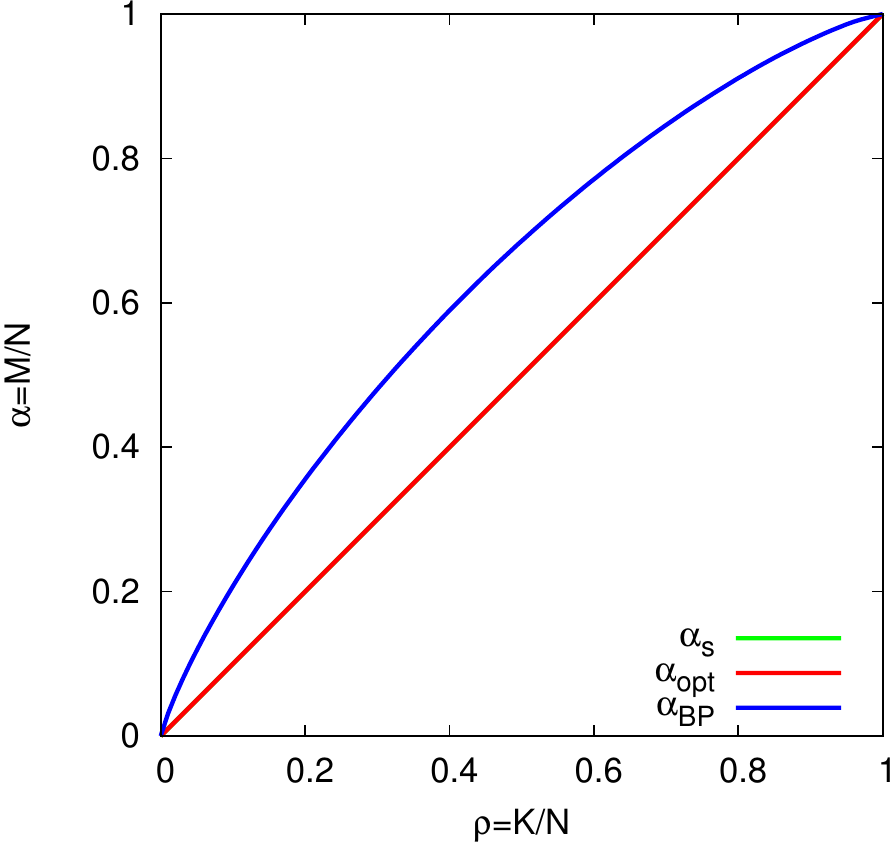}
\includegraphics[width=.35\textwidth, trim=25 0 0 0, clip=true]{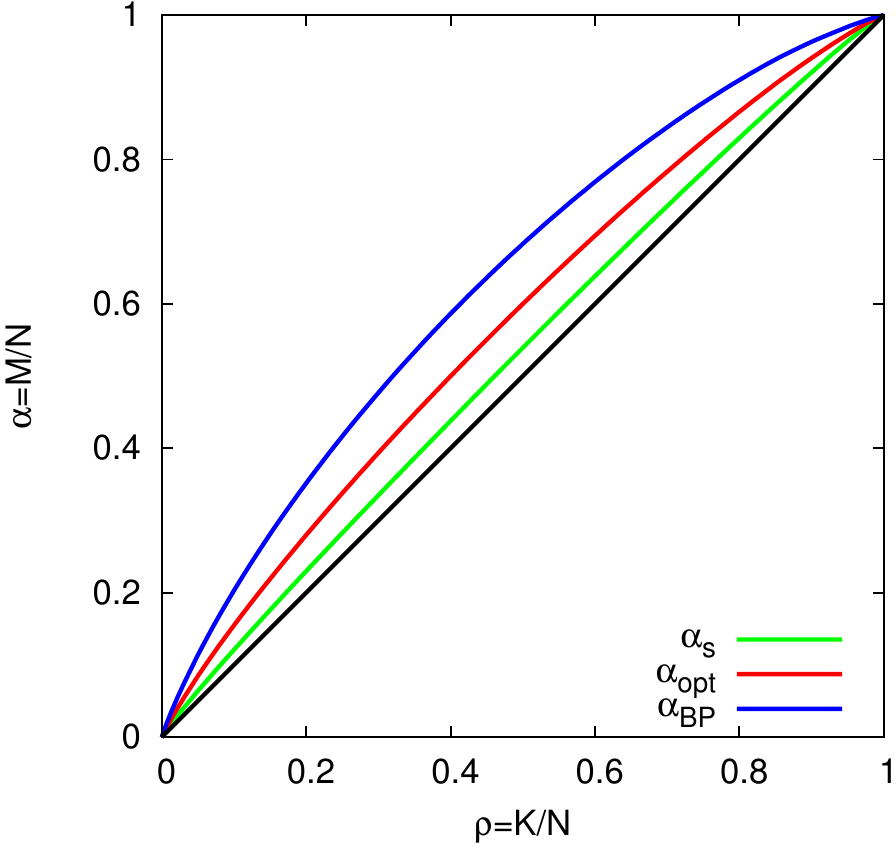}
\includegraphics[width=.35\textwidth, trim=25 0 0 0, clip=true]{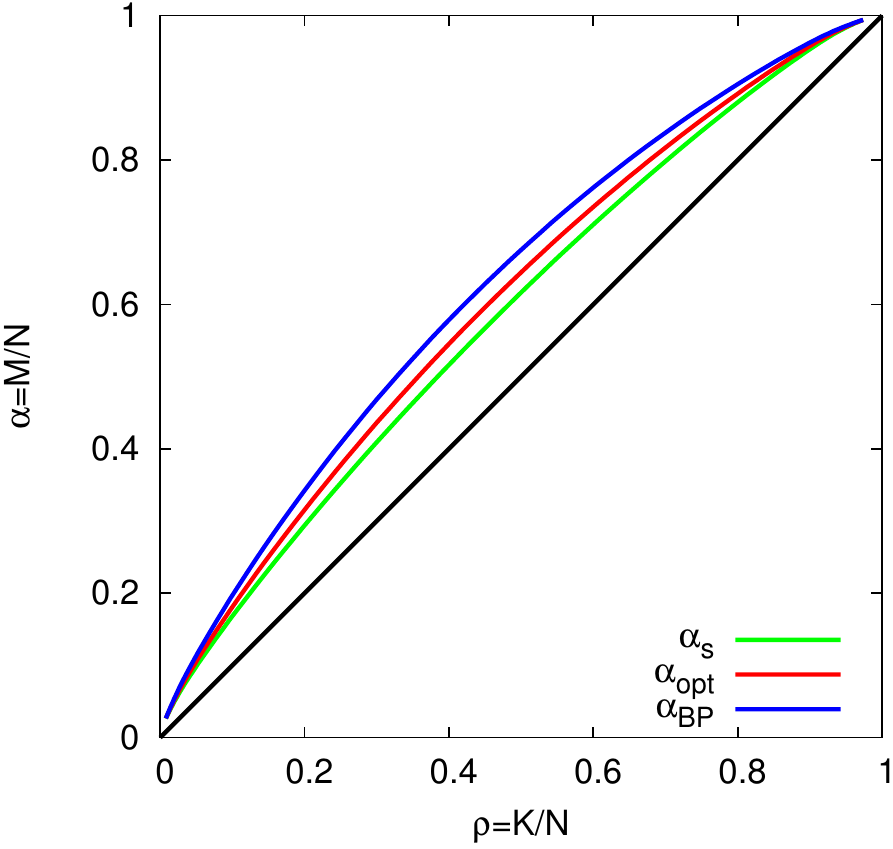}
\caption[Phase diagrams for compressed sensing of approximately sparse signals in the $(\alpha,\rho)$ plane for different small components variances]{Phase diagrams for compressed sensing of approximately sparse signals in the $(\alpha,\rho)$ plane with variance of small components $\epsilon=0$ (left), $\epsilon=10^{-6}$ (center) and $\epsilon=10^{-4}$ (right). As $\epsilon$ increases, the space for improvement of the AMP results by spatial coupling, situated between the optimal and BP transitions, shrinks until it will totally disappear.}
\label{fig_phase_general}
\end{figure*}

In Fig.~\ref{fig_phase_general} we plot the phase diagram at fixed
variance $\epsilon$ in the density $\rho$, measurement rate $\alpha$
plane. The only space for improvement is in the region $\alpha_{opt}(\rho,\alpha)<\alpha <\alpha_{BP}(\rho,\alpha)$, which shrinks as $\epsilon$ increases. In this region, AMP is not optimal because the potential $\Phi(E)$ has two maxima, and the iterations are blocked in the "wrong" metastable local maximum of the potential $\Phi(E)$ with the largest $E$.
\section{Reconstruction of approximately sparse signals with optimality
achieving matrices}
\label{seeding}
A first order phase transition that is causing a failure
(sub-optimality) of the AMP algorithm appears also in the case of
truly sparse signals \cite{KrzakalaPRX2012}, see sec.~\ref{sec:typicalPhaseTransitions}. In that case \cite{KrzakalaPRX2012} showed that with the so-called seeding (i.e. spatially-coupled) measurement matrices, the AMP algorithm is able to restore asymptotically optimal performance as discussed in sec.~\ref{sec:spatialCoupling}. This was proven
rigorously in \cite{DonohoJavanmard11}. Using arguments from the
theory of crystal nucleation, it was argued heuristically in
\cite{KrzakalaPRX2012} that spatial coupling provides improvement
whenever, but only if, a first order phase transition is present. 
Spatial coupling was first suggested for compressed sensing in
\cite{KudekarPfister10} where the authors tested cases without a first
order phase transition (see Fig.~\ref{fig_MSE1}), hence no improvement was observed. Here we show that for measurement rates $\alpha_{opt}(\rho,\epsilon)<\alpha <\alpha_{BP}(\rho,\epsilon)$, seeding matrices allow to restore optimality also for the inference of approximately sparse signals.
\subsection{Restoring optimality thanks to spatial coupling}
In order to restore the asymptotic optimality of AMP with approximately sparse signals, we use spatially-coupled measurement matrices of the form Fig.~\ref{fig_opSpCoupling}.
The state evolution for such block matrices have been derived in sec.~\ref{sec:spatiallyCoupledSE}. The general recursion is given by (\ref{eqChIntro:SE_form3_seeded}). As the derivation in the present setting is exactly the same as in the full measurement matrix case of sec.~\ref{evolution} up to the block index, we give here directly the spatially-coupled state evolution recursion for approximate sparsity:
\begin{align}
    E_c^{t+1} &= \sum_{a=1}^{2} w_a  \int {\cal D}z f_c\left((\Sigma_c^{t+1})^2, z\sqrt{\sigma_a^2  + (\Sigma_c^{t+1})^2 }\right) \label{Et_seeded_appSparsity}   \\
	\Sigma_c^{t+1}\(\{E_{c'}^t\}_{c'}^{L_c}\) &= \left[\sum_{r}^{L_r} \frac{\alpha_{r} J_{rc}}{\sum_{c'}^{L_c} J_{rc'}E_{c'}^t}\right]^{-1/2} \label{eq_SEsigmaSeeded_appSparsity}
\end{align}
where we have used (\ref{eq_SEsigmaSeeded}) in the noiseless case with $B=1$ and $\alpha_r$ is the measurement rate of all the blocks at the $r^{th}$ block-row, $J_{rc}$ the $O(1)$ variance of the $(r,c)$-block elements see Fig.~\ref{fig_opSpCoupling}. This kind of evolution belongs to the class for which threshold saturation (asymptotic achievement of performance matching the optimal
Bayes inference solver) was proven in \cite{YedlaJian12} (when $L_c\to \infty$,
$W\to \infty$ and $L_c/W \gg 1$).
This asymptotic guarantee is reassuring, but one must check if finite
$N$ corrections are gentle enough to be able to
perform compressed sensing close to $\alpha_{opt}(\rho,\epsilon)$ even for
practical system sizes, see sec.~\ref{sec:finiteSizeErrors}. We
hence devote the next section to numerical experiments showing that
the AMP algorithm is indeed able to reconstruct close to optimality with
spatially-coupled matrices.

\begin{figure}[!t]
\includegraphics[width=1\textwidth]{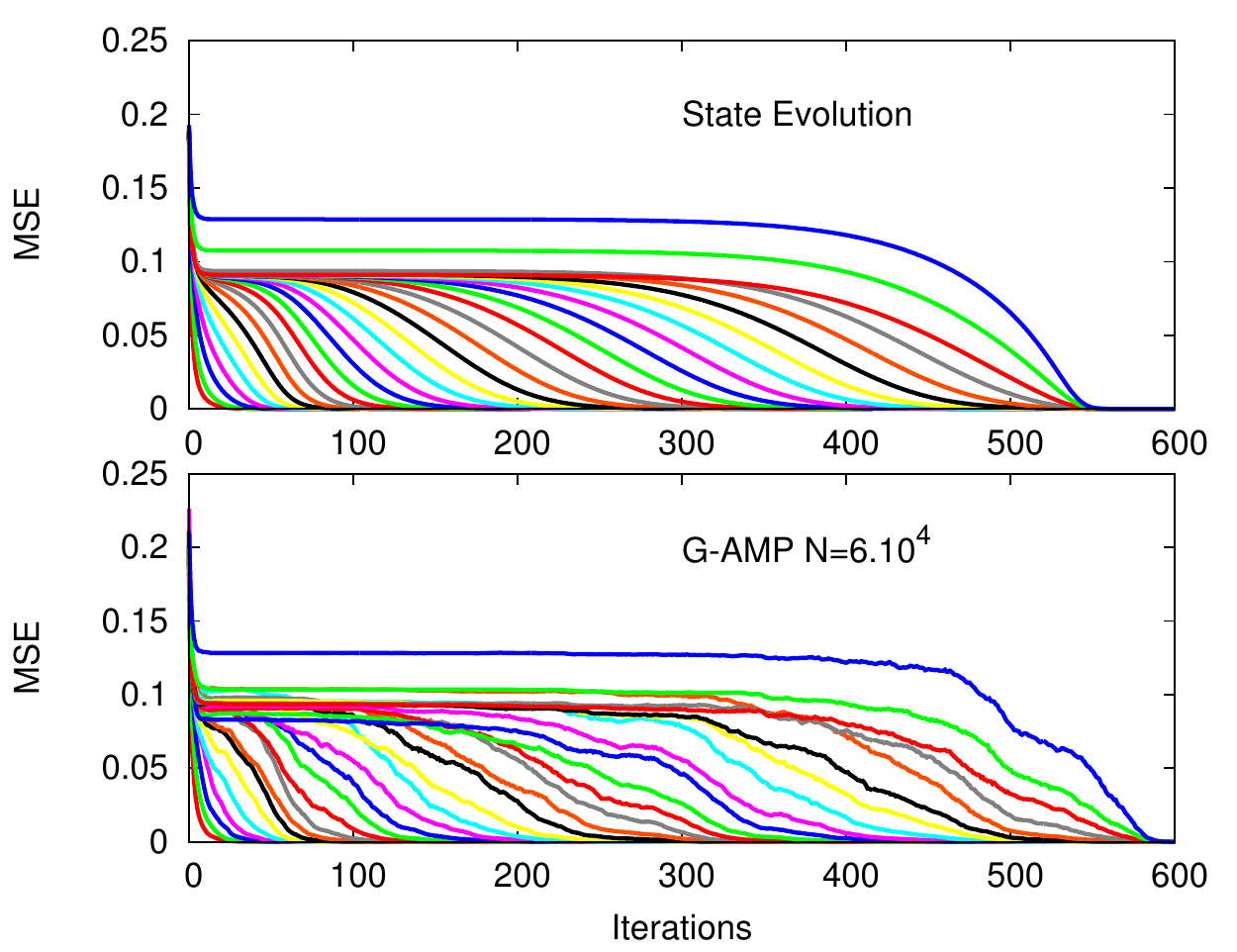}
\caption[State evolution for approximate sparsity with spatially-coupled matrices and comparison with the AMP results on finite size signals]{Evolution of the $MSE$ in reconstruction of an approximately sparse signal with density $\rho=0.2$, variance of small components $\epsilon=10^{-6}$ at measurement rate $\alpha=0.303$. The state evolution on the top is compared to the evolution of the algorithm for a signal size $N=6\cdot 10^4$ on the bottom. The measurement is performed using a seeding matrix with the following parameters: ($\alpha_{seed}=0.4$, $\alpha_{rest}=0.29$, $W=3$, $J=0.2$, $L_c =30$, $L_r =31$). Each colored curve correspond to a different signal block (see Fig.~\ref{fig_opSpCoupling}), and we clearly see the reconstruction wave propagating.}
\label{fig_wave}
\end{figure}

In Fig.~\ref{fig_wave} we show the spatially-coupled state evolution compared to the evolution of the AMP algorithm for system size
$N=6\cdot 10^4$. The signal was of density $\rho=0.2$ and
$\epsilon=10^{-6}$, the parameters of the measurement matrix are in
the second line of Tab.~\ref{param}, $L_c=30$ giving measurement rate
$\alpha=0.303$ which is deep in the region where AMP for homogeneous
measurement matrices is not Bayes optimal and gives large $MSE$ (for any $\alpha<\alpha_{BP}(\rho=0.2,\epsilon=10^{-6})=0.356$). We see finite size fluctuations, but the overall evolution corresponds well to the asymptotic curve, and we see the reconstruction wave propagation happening. 
\begin{figure}[!t]
\includegraphics[width=1\textwidth]{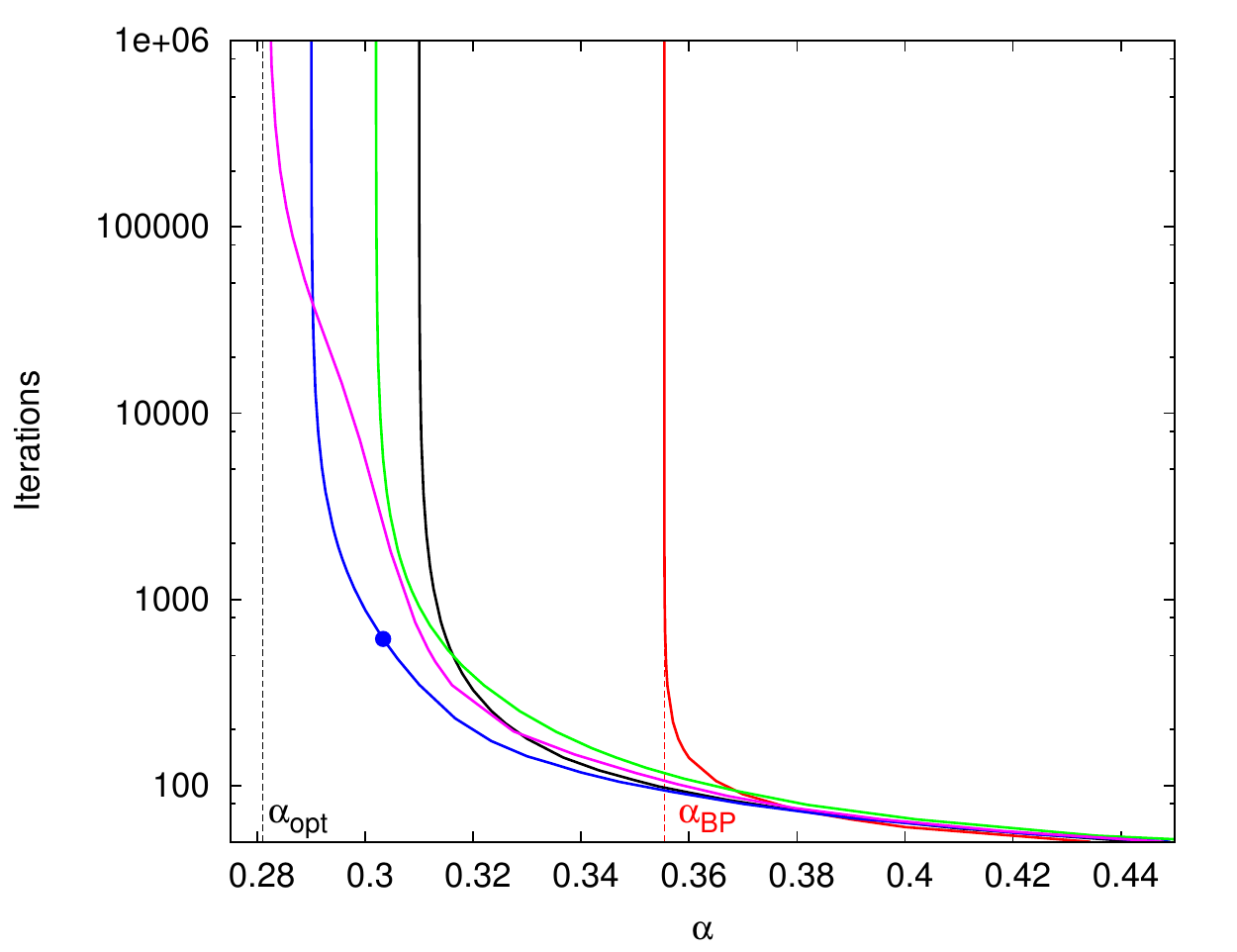}
\caption[Asymptotic convergence time required by AMP for reconstructing approximately sparse signals]{The convergence time of AMP for large system sizes estimated
  by the state evolution as a function of the measurement rate
  $\alpha$. Data are for signals with density $\rho=0.2$, variance of
  small components $\epsilon=10^{-6}$. The red line is obtained using
  an homogeneous measurement matrix, the vertical dashed line
  corresponds to the limit this approach can achieve $\alpha_{
    BP}(\epsilon=10^{-6},\rho=0.2)=0.3554$. All the other lines are obtained using a spatially-coupled matrix
  with parameters specified in Table~\ref{param} and varying $\alpha_{rest}$ by changing $L_c$ which are related by (\ref{eq_alphaRest}),
  the resulting measurement rate $\alpha$ is computed from
  (\ref{eq_alphaRest_2}). With these seeding matrices and using large
  $L_c$, reconstruction is possible at least down to $\alpha_{
    rest}=0.282$ which is very close to the measurement rate
  $\alpha_{opt}=0.2817$. The blue point corresponds to the
  evolution illustrated in Fig.~\ref{fig_wave}. The divergence of the convergence time of AMP approaching the phase transitions is the critical slowing down discussed in sec.~\ref{sec:typicalPhaseTransitions}, a typical behavior of local algorithms near first order phase transitions, like the present BP and optimal ones.}
\label{fig_time}
\end{figure}
\begin{table}[!ht]
\begin{center}
\begin{tabular}{|c|c|c|c|c|c|c||}
  \hline
  color & $\alpha_{seed}$ & $\alpha_{rest}$ & $J$ & $W$ & $L_r$ \\
  \hline
   purple & $0.4$ & $0.282$ & $0.3$ & $3$ & $L_c+2$ \\
   blue & $0.4$ & $0.290$ & $0.2$ & $3$ & $L_c+1$ \\
   green & $0.4$ & $0.302$ & $0.001$ & $2$ & $L_c+1$ \\
   black & $0.4$ & $0.310$ & $0.4$ & $3$ & $L_c+1$\\
\hline
\end{tabular}
\end{center}
\caption{Parameters of the seeding matrices used in Fig.~\ref{fig_time}. The $\alpha_{rest}$ is modified by changing $L_c$, the link between them being (\ref{eq_alphaRest}).\label{param}}
\end{table}

In Fig.~\ref{fig_time} we plot the asymptotic convergence time needed to achieve
reconstruction with $E \approx \epsilon$ for several sets of
parameters of the seeding matrices, see Tab.~\ref{param}. Each color corresponds to a different $L_c$, which changes the $\alpha_{rest}$ thanks to (\ref{eq_alphaRest}). With a proper choice of the
parameters, we see that we can reach an optimal reconstruction for
values of $\alpha$ extremely close to $\alpha_{opt}(\epsilon,\rho)$. Note, however,
that the number of iterations needed to converge diverges as $\alpha
\to \alpha_{opt}(\epsilon,\rho)$. This critical slowing down typical of first order phase transitions is discussed in sec.~\ref{sec:typicalPhaseTransitions}. This is very similar to what has been obtained in the
case of purely sparse signals in \cite{KrzakalaPRX2012,DonohoJavanmard11}.
\subsection{Finite size effects influence on spatial coupling performances}
It is important to point out that these theoretical analyzes
are valid for $N \to \infty$ only. Since we eventually work with finite
size signals, in practice, finite size effects slightly degrade this
asymptotic threshold saturation, see sec.~\ref{sec:finiteSizeErrors}. 
This is a well known effect in coding theory where a major question is
how to optimise finite-length codes (see for instance \cite{AmraouiMontanari2007,richardson2008modern}).

\begin{figure}[!ht]
\includegraphics[width=.9\textwidth]{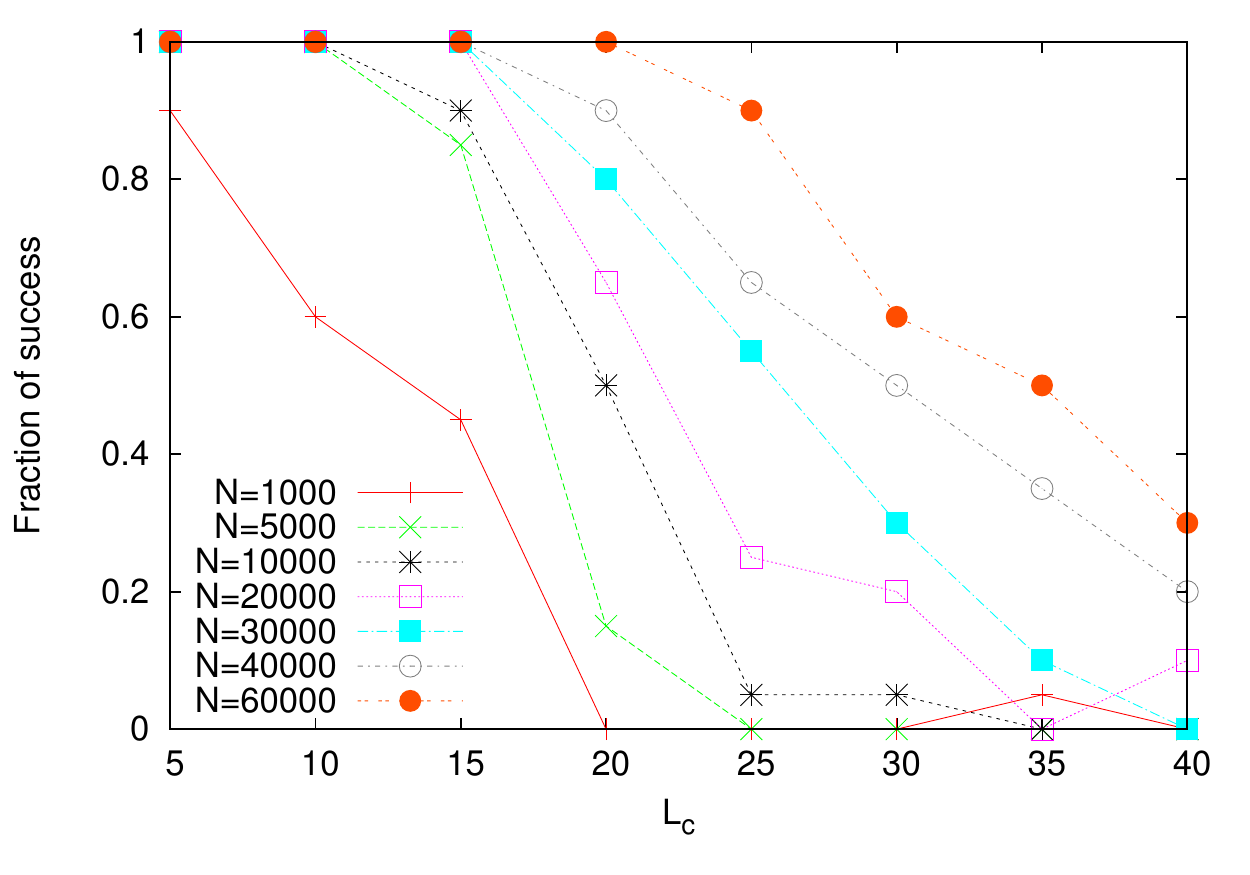}
\caption[Fraction of instances reconstructed with spatially-coupled matrices]{Fraction of instances (over $20$ attempts) that were solved
  by the algorithm in less than twice the number of iterations
  predicted by the density evolution for different system sizes, as a
  function of the number of blocks $L_c$. We used the parameters that
  lead to the blue curve in Fig.~\ref{fig_time} (i.e. second line of
  Table~\ref{param}). As $N \to \infty$, reconstruction is reached in
  all the instances, as predicted by the state evolution. For finite
  $N$, however, reconstruction is not reached when $L_c$ is too
  large. But in the same time, as $L_c$ increases we can theoretically decode closer to the optimal threshold as seen from (\ref{eq_alphaRest_2}). Thus there is a tradeoff between measurement rate decrease and probability of success in the reconstruction.}
\label{convergence}
\end{figure}

In Fig.~\ref{convergence} we plot the fraction of cases in which the
algorithm reached successful reconstruction for different system sizes
as a function of the number of blocks $L_c$. We see that for a given
size as the number of blocks is growing, i.e. as the size of one block
decreases, the performance deteriorates. As expected the situation
improves when the size increases. Analysies of the data presented in
Fig.~\ref{convergence} suggest that the size of one block that is
needed for good performance grows roughly linearly with the number of
blocks $L_c$. This suggests that the probability of failure to transmit the information to every new block is roughly inversely proportional to
the block size. The algorithm nevertheless reconstructs
signals at rates close to the optimal one for system sizes of practical interest. This figure emphasizes the tradeoff between measurement rate decrease and probability of success in the reconstruction because as $L_c$ increases, we can theoretically decode closer to the optimal threshold as seen from (\ref{eq_alphaRest_2}). But in the same time, it increases the finite size effects influence and thus lowers the probability of success.

\begin{figure}[!ht]
\hspace{.8cm}
\includegraphics[width=.8\textwidth]{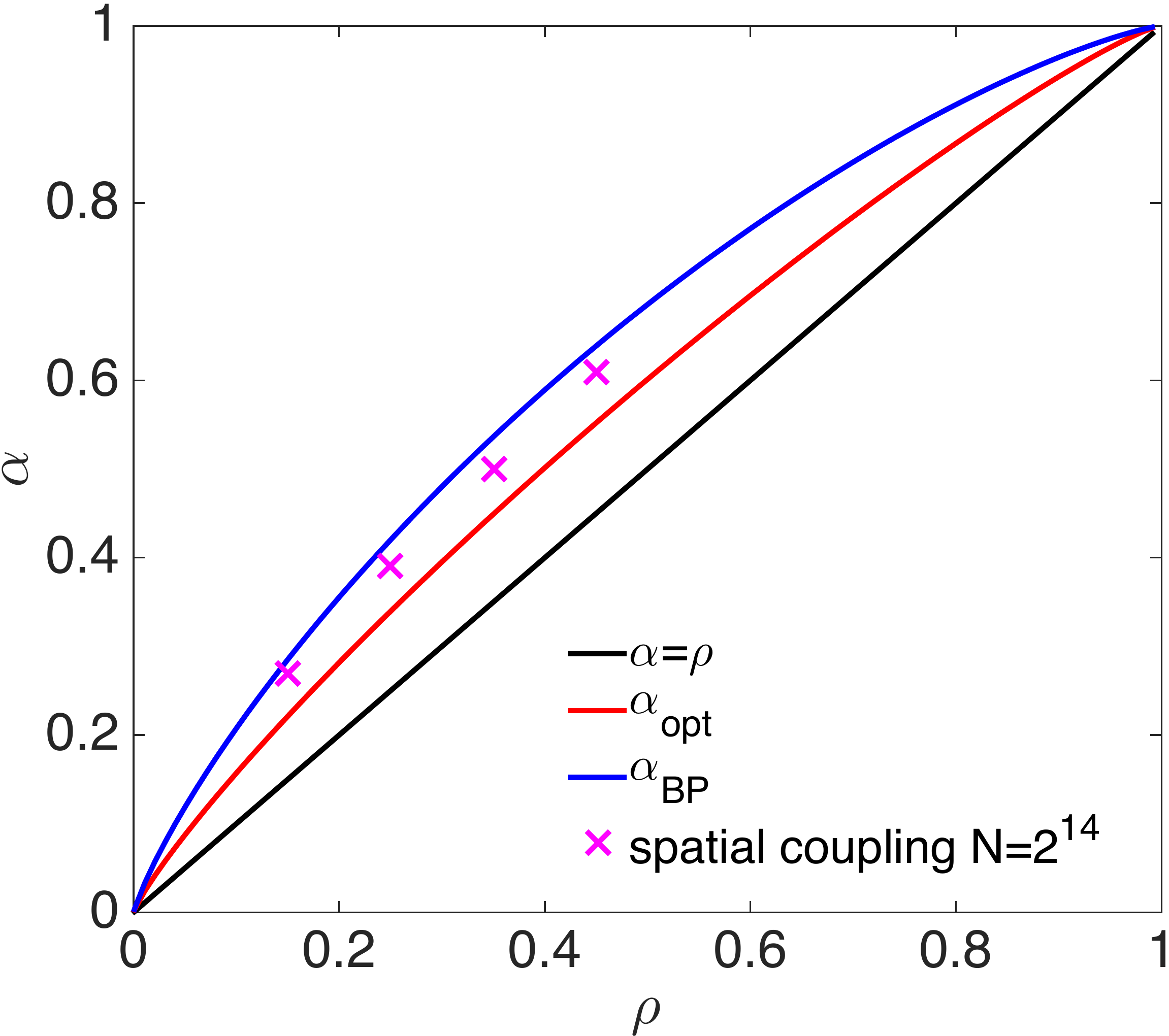}
\caption[Phase diagram with finite size instances solved by spatial coupling]{Phase diagram for a small variance $\epsilon=10^{-6}$ on the $(\alpha,\rho)$ plane, where we added crosses at parameters values where reconstruction have been successful thanks to spatial coupling for signals of size $N=2^{14}$. As the hard phase between the optimal and BP transitions decreases with $\rho$ and because the gap between the sucessful reconstruction line (the virtual line linking the pink crosses) and $\alpha_{opt}(\rho)$ is close to constant for fixed $N$, the gain in measurement rate obtained with spatial coupling decreases with $\rho$. The spatially-coupled random i.i.d Gaussian matrices were drawn from the ensemble $(L_c=32,L_r=33,w=2,\sqrt{J}=0.4,\alpha,\beta_{seed}=1.3)$.}
\label{fig_phaseDiag_appSparsSpC}
\end{figure}

Fig.~\ref{fig_phaseDiag_appSparsSpC} is the phase diagram for a variance of the small components $\epsilon=10^{-6}$ in the $(\alpha,\rho)$ plane and shows how with spatial coupling, we can reconstruct instances generated in the hard phase between the BP transition and the optimal one. We notice that the pink crosses corresponding to these finite size instances reconstructed by spatial coupling for fixed size $N=2^{14}$ are approximately at a constant distance of the optimal transition, and thus as the region allowing for improvement $[\alpha_{opt}(\epsilon=10^{-6},\rho),\alpha_{BP}(\epsilon=10^{-6},\rho)]$ gets smaller when $\rho$ decreases, the gain with respect to the BP transition decreases as well. But even for this small size, a non negligible gain is possible when $\rho$ is not too small.
%
%
%
%
%
%
%
%
%
\section{Some results on real images}
\begin{figure}[!h]
\centering
  \includegraphics[scale=0.4]{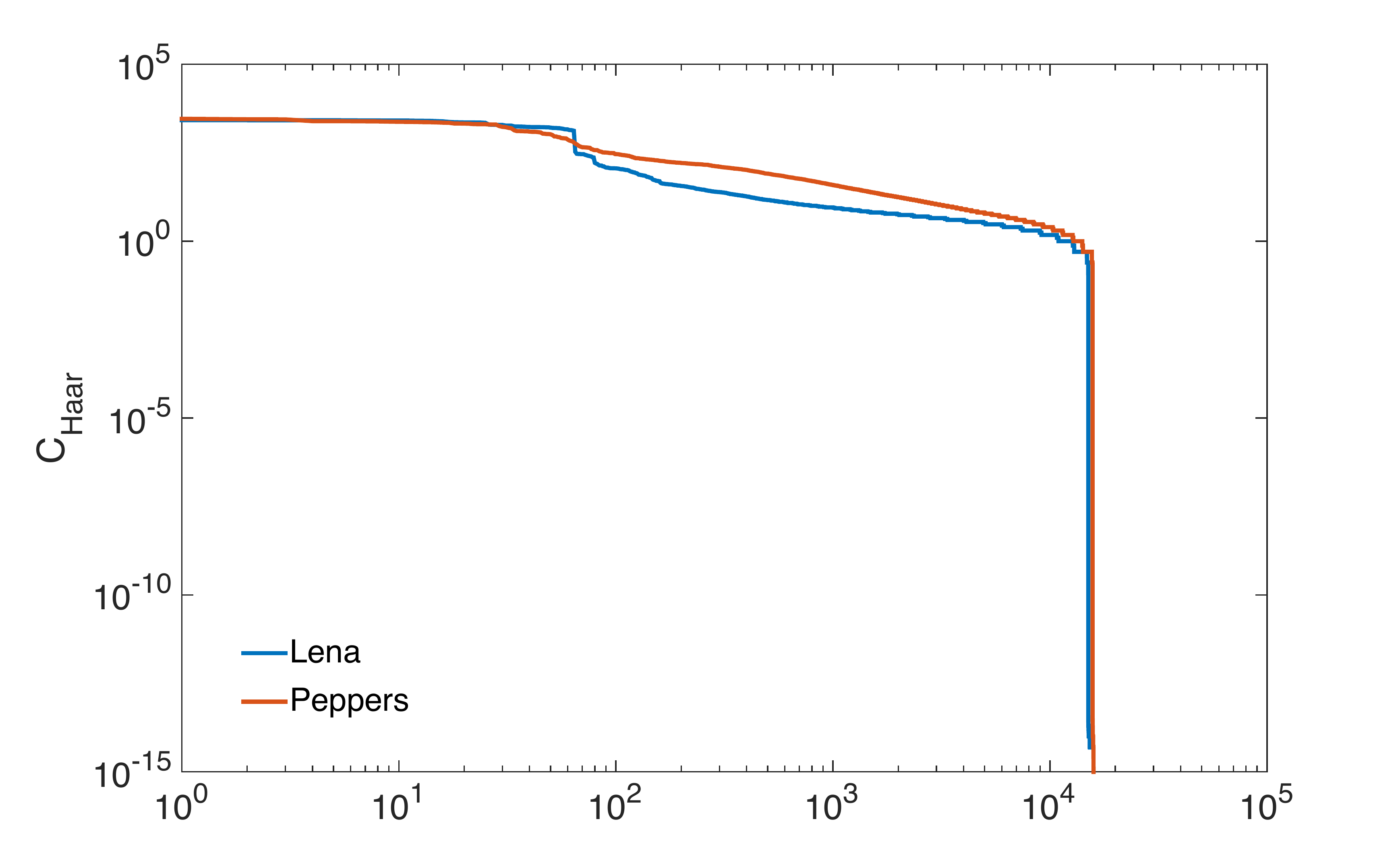}
  \caption[Sorted wavelet spectrum of the 4-steps Haar transformed Lena and peppers images]{Sorted wavelet spectrum of the 4-steps Haar transformed Lena and peppers images in double logarithmic plot. There are few high amplitude coefficients and a power law tail of smaller coefficients: this is typical of compressible signals.}
  \label{fig_spectrumImages}
\end{figure}
\begin{figure}[h!]
\centering
\includegraphics[scale=0.42]{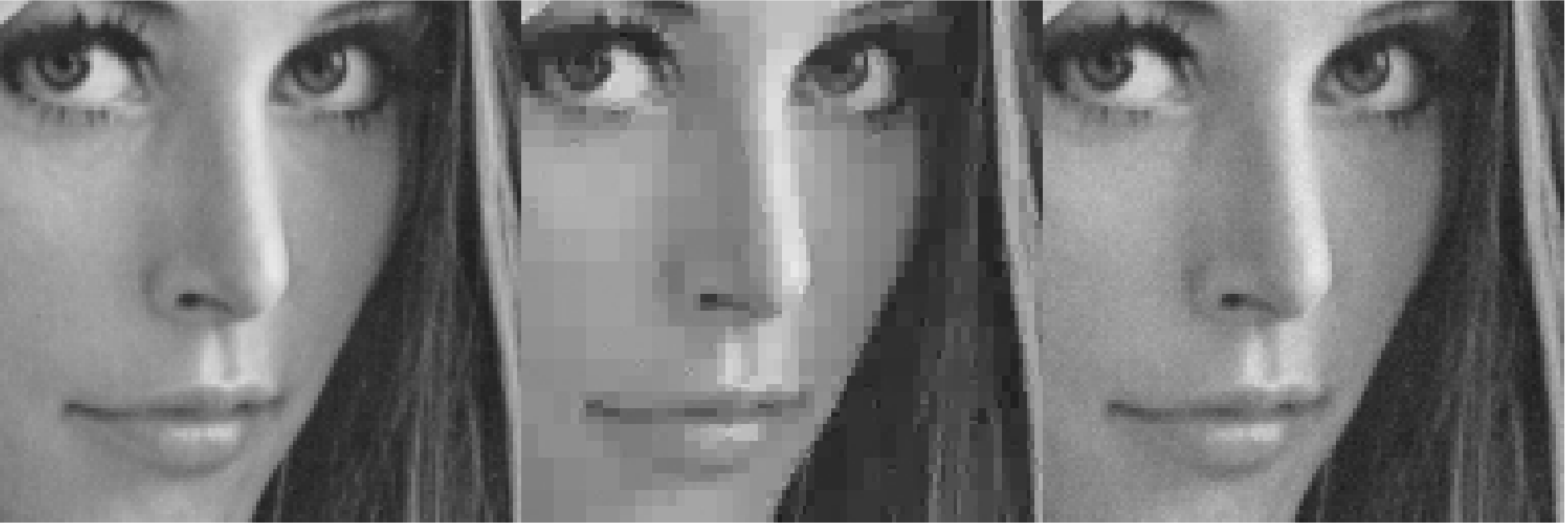}  
  \includegraphics[scale=1.47]{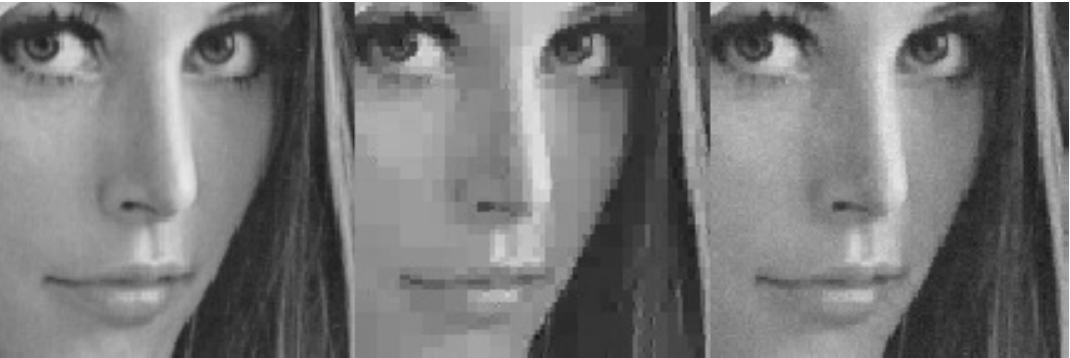}  
  \includegraphics[scale=0.42]{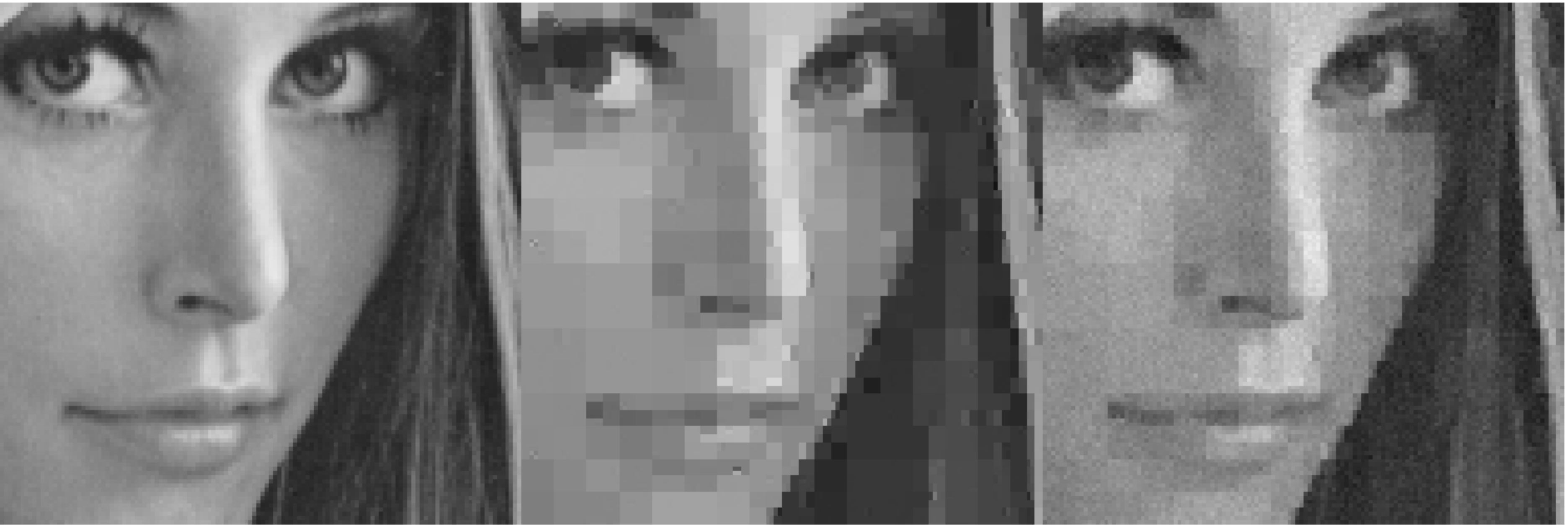}
  \caption[Reconstruction results of Lena using approximate sparsity in the Haar wavelet basis]{Comparisons between reconstruction results of Lena (left picture). Lena was first expressed in the 4-steps Haar wavelet basis, and the coefficients Fig.~\ref{fig_spectrumImages} were then reconstructed with AMP using a strict sparsity inducing prior (center) or the approximate sparsity prior (right) for different measurement rates. The measurement rates and final $MSE$ of the wavelet coefficient are: \tbf{Up}: $\alpha=0.8$, $MSE_{sparse} = 7.2 \times 10^{-4}$, $MSE_{app. sparse} = 2\times 10^{-4}$, \tbf{Middle}: $\alpha=0.65$, $MSE_{sparse} = 10^{-3}$, $MSE_{app. sparse} = 4.6\times 10^{-4}$ and \tbf{Down}: $\alpha=0.415$, $MSE_{sparse} = 1.8 \times 10^{-3}$, $MSE_{app. sparse} = 1.3 \times 10^{-3}$.}
  \label{fig:LenaAppSparsity}
\end{figure}
\begin{figure}[!h]
\centering
\includegraphics[scale=0.7]{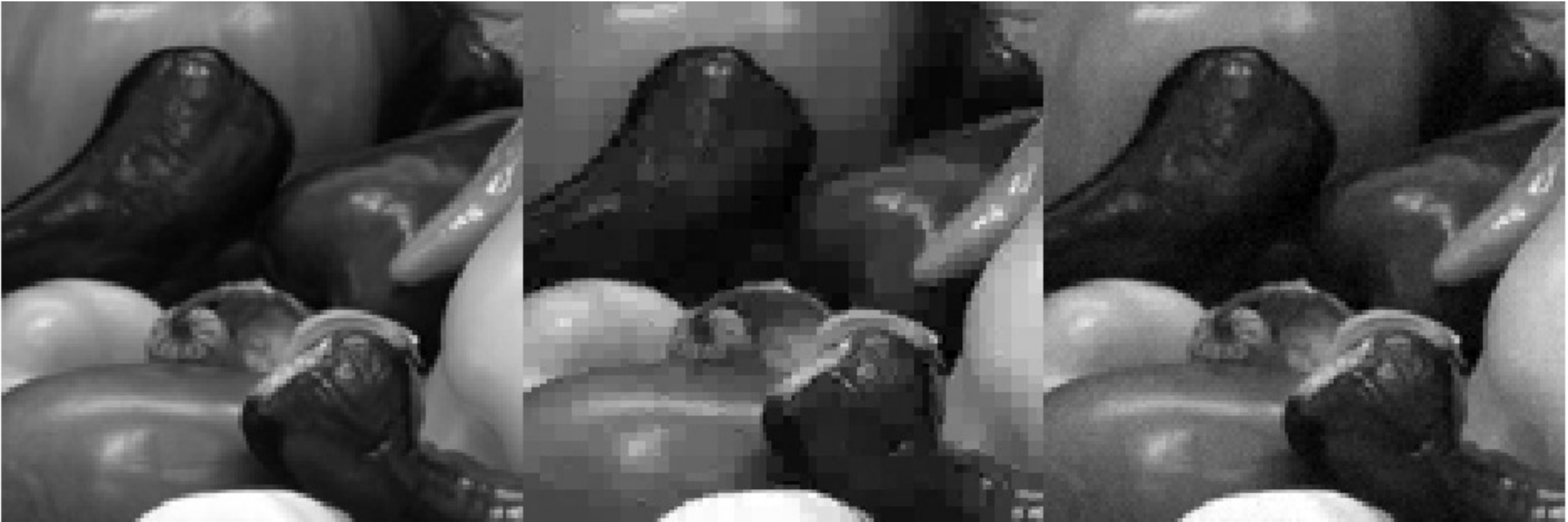}
\includegraphics[scale=0.7]{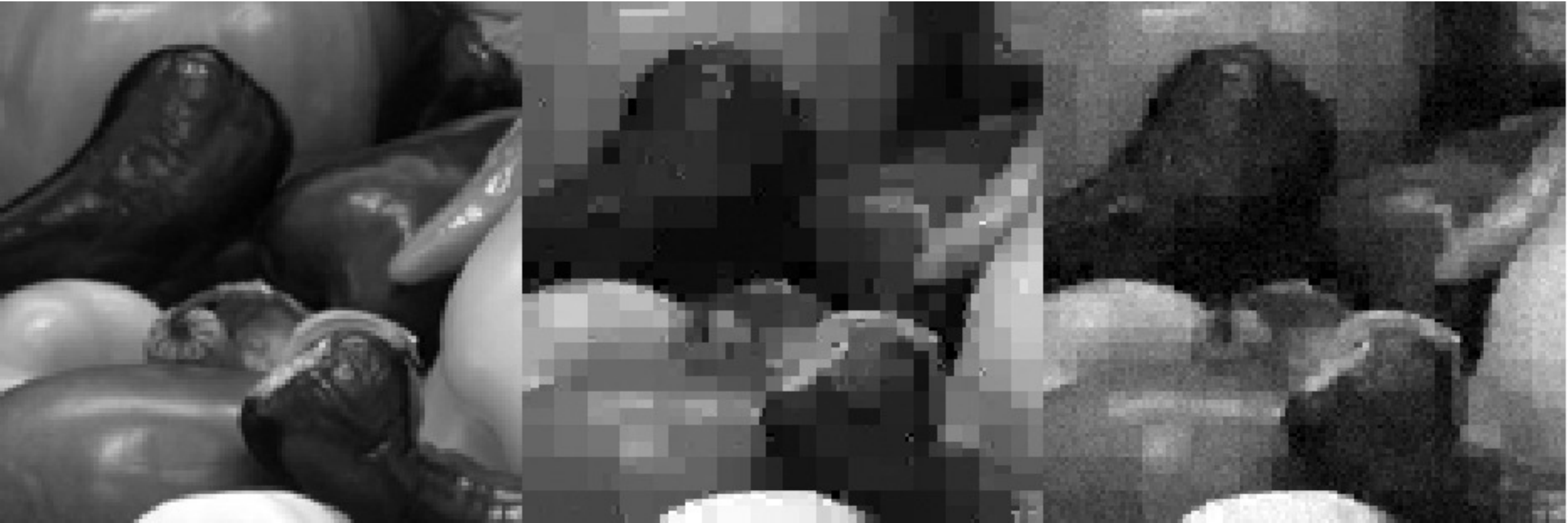}
  \caption[Reconstruction results of the peppers image using approximate sparsity in the Haar wavelet basis]{The same experiment as in Fig.~\ref{fig:LenaAppSparsity} but with the peppers image. The measurement rates and final $MSE$ of the wavelet coefficient are: \tbf{Up}: $\alpha=0.8$, $MSE_{sparse} = 6.6 \times 10^{-4}$, $MSE_{app. sparse} = 2.4 \times 10^{-4}$ and \tbf{Down}: $\alpha=0.4$, $MSE_{sparse} = 2.7 \times 10^{-3}$, $MSE_{app. sparse} = 2 \times 10^{-3}$.}
  \label{fig_recPeppersAppSpars}
\end{figure}
We now present some results on a potential application of the approximate sparsity prior: image reconstructions. The appropriate sparsifying basis for natural images is the wavelet basis, but the resulting signal is not truly sparse but compressible (\ref{eq_compressible}). This can be seen from Fig.~\ref{fig_spectrumImages} where we show the sorted coefficients of the 4-step Haar transformed Lena and peppers images, see Fig.~\ref{fig:LenaAppSparsity} and Fig.~\ref{fig_recPeppersAppSpars}. It appears that the energy is concentrated on few coefficients but the smaller ones follow a "power law-like" distribution. Compressed sensing is thus appropriate but perfect reconstruction is impossible in the compressed regime $\alpha<1$. We perform an experiment where the Lena and peppers images are 4-step Haar transformed, and the coefficients Fig.~\ref{fig_spectrumImages} are reconstructed for different measurement rates, with a purely sparsifying prior $P_0(\bx) = \prod_i^N[(1-\rho)\delta(x_i) + \rho \mathcal{N}(x_i|m,\sigma^2)]$ or with the approximate sparsity prior (\ref{Px_appSparse}). All the prior parameters (and the noise variance in the purely sparse case) are learned through expectation maximisation, see sec.~\ref{sec:EMlearning}. The results Fig.~\ref{fig:LenaAppSparsity} and Fig.~\ref{fig_recPeppersAppSpars}, both in terms of the $MSE$ of the reconstructed wavelets coefficients and in terms of the "by-eyes" quality of reconstruction are homogeneously better with the approximate sparsity prior at any measurement rate. In the case of Lena, it is even stronger as the by-eyes quality of reconstruction of Lena at $\alpha = 0.415$ with the approximate sparsity prior is better than the image reconstructed at $\alpha=0.65$ with the sparsity inducing prior and both the by-eyes quality and $MSE$ performances of reconstruction of Lena at $\alpha = 0.65$ with the approximate sparsity prior are better than the ones at $\alpha=0.8$ with the sparsity inducing prior. The conclusions are identical with the peppers image. So reconstructing with the approximate sparsity prior images expressed in the wavelet basis appears to be a good strategy. Looking at the pictures, we see that way more details are reconstructed with the approximate sparsity prior. This is due to the fact that these are contained in the high frequency coefficients in the wavelet basis which are hidden in the low energy tail of Fig.~\ref{fig_spectrumImages}. Because the approximate sparsity model considers this tail as being part of the signal rather than noise, it reconstructs part of these small coefficients, which induce this important gain in detailed imformation.

However, to become competitive with state-of-the-art algorithms \cite{DBLP:journals/corr/abs-1108-2632,DBLP:journals/corr/VilaSM15,DBLP:journals/tip/DongSLMH14,DBLP:journals/tsp/TanMB15,DBLP:journals/corr/MetzlerMB14} that can reconstruct wavelet coefficients for images, we also need to find better models for the signal coefficients, likely
including the fact that the approximately sparse components are highly structured for real images (wavelet coefficients are known to have a tree structure exploited by \cite{DBLP:journals/corr/abs-1108-2632,DBLP:journals/corr/VilaSM15} for example).
\section{Concluding remarks}
At this point we want to state that whereas all our results do
depend quantitatively on the statistical properties of the signal,
the qualitative features described here (for example the presence and the nature of the phase transitions) are valid for other signal models,
distinct from the bi-Gaussian case that we have studied here. This is even the case when the signal model does not match the statistical properties of the actual signal. This was illustrated for example for the noisy compressed sensing of truly sparse signal in \cite{KrzakalaMezard12}. In the same line,
we noticed and tested that if AMP corresponding to $\epsilon=0$ is runned for the approximately sparse signals, then the final $MSE$ is always larger than the one achieved by AMP with the right value of $\epsilon$, as seen on the images.

We studied the case of noiseless measurements, but the
measurement noise can be straightforwardly included into the analysis as in \cite{KrzakalaMezard12}. Again, the results would change
quantitatively, but not qualitatively. The point was really to understand the influence of the small components alone, as the influence of measurement noise was already extensively studied in compressed sensing \cite{KrzakalaMezard12}.

For small variance of the small components of the signal, the AMP
algorithm for homogeneous matrices does not reach optimal
reconstruction for measurement rates close to the theoretical limit
$\alpha_{opt}(\epsilon,\rho)$. The spatial coupling approach, resulting in the design of seeding matrices improves significantly the
performances. For diverging system sizes, optimality can be restored. We have shown that significant improvement is also reached for sizes of
practical interest. There are, however, non negligible finite size effects that should be studied in more details. The optimal design of the seeding matrix for finite system sizes (as studied for instance in
depth in the context of error correcting codes
\cite{AmraouiMontanari2007}) remains an important open
question.
\chapter{Approximate message-passing with spatially-coupled structured operators}
\label{chap:structuredOperators}
We now study the behavior of the approximate message-passing algorithm when the i.i.d matrices for which it has been specifically designed are replaced by structured operators, such as Fourier and Hadamard ones. The aim is in one hand to be able to tackle very large single instances of inference problems such as compressed sensing and in the other hand, to reach close to Bayes optimal reconstruction performances.

To work with large signals and matrices,
however, one needs fast and memory efficient solvers. Indeed, the mere
storage of the measurement matrix in memory can be problematic as soon as the signal size $N>O(10^4)$. A classical trick (see for instance
\cite{do2008fast}) is thus to replace the random sensing matrix with a
structured one, typically random modes of a fast transform such as Fourier-like matrices.
We will show empirically that after a proper randomization, the
structure of the operators does not significantly affect the
performances of the solver.

The use of fast transforms makes matrix
multiplications faster ($O(N\log N)$ instead of $O(N^2)$ operations), and
thus both speeds up the reconstruction algorithm and removes the need
to store the matrix in memory. This is also important for coding
applications where $O(N^2)$ operations can be burdensome for the
processor.

While using Fourier or Hadamard matrices has often been done with AMP (see
for example \cite{JavanmardMontanari12,kamilovsparse}), we provide here a
close examination of its performances with Fourier and Hadamard
operators for compressed sensing of complex and real sparse signals respectively. As
suggested by the heuristic replica analysis \cite{vehkapera2013analysis,wen2014analysis}, such matrices often lead to better performances than random ones. This will be confirmed through numerical investigation.

Furthermore, inspired by the Gabor construction of
\cite{JavanmardMontanari12} that allowed optimal sampling of a
random signal with sparse support in frequency domain, we
extend the construction of spatially-coupled matrices to a structured form
using fast Fourier/Hadamard operators, which allow to deal with large
signal sizes and up to the information theoretical limit. Given the lack of theoretical guaranties, we numerically
study this strategy on synthetic problems, and compare its performance
and behavior with those obtained with random i.i.d Gaussian matrices. The main
result is that after some randomization procedure, structured
operators appear to be nearly as efficient as random i.i.d
matrices. In fact, empirical performances are as good as those
reported in \cite{KrzakalaPRX2012,KrzakalaMezard12} despite the
drastic improvement in computational time and memory.
\section{Problem setting}
\label{sec:setting}
\begin{figure}
\centering
\includegraphics[width=.8\textwidth]{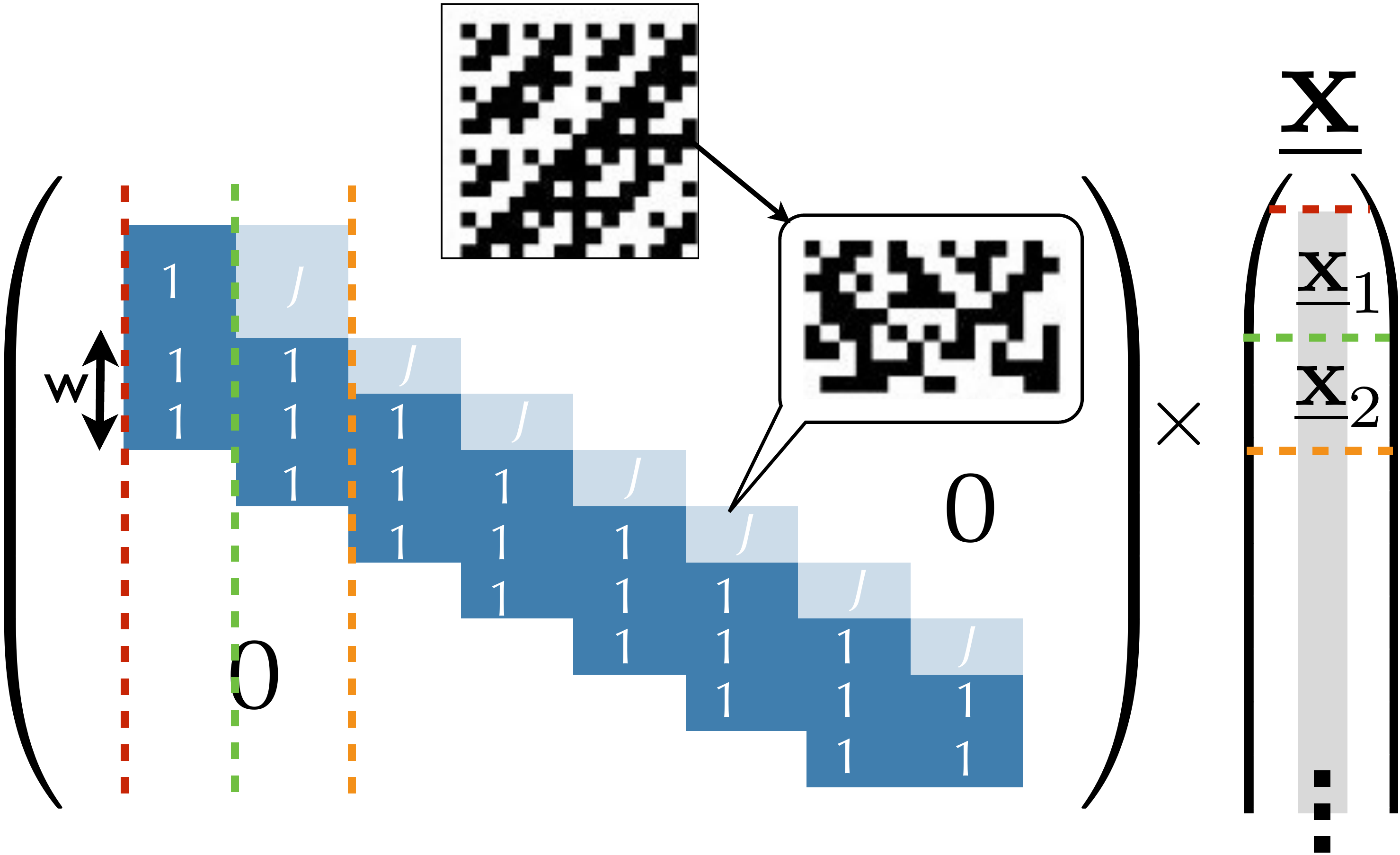}
\centering
\caption[spatially-coupled structured operator]{Representation of the spatially-coupled Hadamard sensing
  matrix used in our study, which structure is the same as Fig.~\ref{fig_opSpCoupling}. The operator is decomposed in $L_r\times
  L_c$ blocks, each being made of $N/L_c$ columns and $\alpha_{seed}N/L_c$ lines for the blocks of the first block-row,
  $\alpha_{rest}N/L_c$ lines for the following block-rows, with
  $\alpha_{seed}>\alpha_{BP}\ge\alpha_{rest}>\alpha_{opt}$. The figure shows how the
  lines of the original small Hadamard matrix (of size $N/L_c\times N/L_c$)
  are randomly selected, re-ordered and sign-flipped to form a given
  block of the final operator. There is a number $w$ (the coupling
  window) of lower diagonal blocks with elements $\in \{\pm 1\}$ as
  the diagonal blocks, the upper diagonal blocks have elements $\in
  \{\pm\sqrt{J}\}$ where $\sqrt{J}$ is the coupling strength, all the
  other blocks contain only zeros. The colored dotted lines help to
  visualize the block decomposition of the signal induced by the
  operator structure: each block of the signal will be reconstructed
  at different times in the algorithm (see Fig.~\ref{fig_expVsDE} main
  figure and Fig.~\ref{fig_opSpCoupling}). The procedure is exactly the same for constructing
  spatially-coupled Fourier operators, replacing the small Hadamard
  operator from which we construct the blocks by a small Fourier
  operator. The parameters that define the spatially-coupled operator
  ensemble are $(L_c,L_r,w,\sqrt{J},\alpha_{{{seed}}},\alpha_{{{rest}}})$.}
\label{fig_seededHadamard}
\end{figure}
In the following, complex variables will be underlined: $\cv{x}_j =
x_{j,1} + ix_{j,2} \in \mathbb{C}$. We will write $\cv{x} \sim \mathcal{C}\mathcal{N}(\cv{x}|\cv{\bar x},\sigma^2)$ if the real and
imaginary parts $x_1$ and $x_2$ of the random variable $\cv{x}$ are
independent and verify $x_1 \sim \mathcal{N}(x_1|\Re{\cv{\bar x}},\sigma^2)$
and $x_2 \sim \mathcal{N}(x_2|\Im{\cv{\bar x}},\sigma^2)$.

The generic problem we consider is noiseless complex compressed sensing. We consider that the signal components are scalars ($L=N$, $B=1$). We choose not to consider noise as the point here is really to study the influence of structure in the matrix. Noise or vectorial components can be trivially included in the model, and do not change the qualitative features presented here. This will be confirmed through the extensive use of the present structured Hadamard operator in the context of the superposition codes chap.~\ref{sec:superCodes} where vectorial components signals measured under high noise levels will be reconstructed.
In the noiseless complex case (\ref{eqIntro:AWGNCS}) becomes:
\begin{equation}
\label{eq_Y}
\cv{\by} = \cv{\bF} ~\cv{\bs}
\end{equation}  
We will use the following Gauss-Bernoulli distribution to
generate $\rho$-sparse complex random vectors:
\begin{equation}
\label{Px}
P(\cv{\bs}) =\prod_{j=1}^N \left[(1 - \rho) \delta(\cv{s}_j) + \rho \ \mathcal{C}\mathcal{N}(\cv{s}_j|\cv{\bar s},\sigma^2)\right] 
\end{equation}
Here we shall assume that the correct values for $\rho$, $\cv{\bar{s}}$, $\sigma^2$ as well as the empirical signal distribution (\ref{Px}) are known and thus we place ourselves under the prior matching condition in the theoretical analyzes. As discussed in sec.~\ref{sec:EMlearning}, these
parameters can be learned efficiently with an expectation maximization procedure if unknown.
\subsection{Spatially-coupled structured measurement operators}
As discussed in sec.~\ref{sec:spatialCoupling}, in order to asymptotically reach the optimal transition $\alpha_{opt}$, spatial coupling is the strategy of choice. The spatially-coupled structured operator is constructed as Fig.~\ref{fig_opSpCoupling} with the novelty that the blocks are not made of random i.i.d matrices anymore, but are replaced by sub-sampled fast operators, see Fig.\ref{fig_seededHadamard}: each of these blocks is constructed from the same original
operator of size $N/L_c\times N/L_c$ and the differences from one block to another comes from the selected modes, their permutation and signs that are randomly changed. In the case of an Hadamard construction, all the blocks are generated from the same original small Hadamard operator with the constraint that $N/L_c$ must be a power of two, intrinsic to the Hadamard construction.
\subsection{The approximate message-passing algorithm for complex signals}
In order to avoid confusions with the literature where variations of
AMP are already presented, we will refer in the present chapter to the Bayes-optimal AMP by
"BP" and "c-BP" for the real and complex case respectively,
and to the $\ell_1$-minimizing version by "LASSO" and "c-LASSO"
respectively. As the thresholding functions are applied
component-wise, the time-consuming part of the algorithm are the matrix
multiplications in the linear step. Here, we use Fourier and Hadamard
operators in order to reduce their complexity from $O(N^2)$ to $O(N \log N)$. The authors of
\cite{JavanmardMontanari12} have used a related, yet different way to
create spatially-coupled matrices using a set of Gabor transforms.

The generalization of the scalar algorithm Fig.~\ref{algoCh1:AMP_op} to the complex case is not sraightforward as in the derivation of sec.~\ref{sec:classicalDerivationAMP}, the real and imaginary parts of the different complex quantities appearing in the computations would couple through non diagonal covariance matrices. But it appears that many simplifications arise due to the independence assumption of the matrix elements, which make the final algorithm Fig.~\ref{algo_AMP_complex} look very similar to its scalar version. The derivation and study of this complex version can be found in \cite{maleki2012asymptotic,Schniter2012compressive,DBLP:journals/corr/abs-1108-0477,DBLP:journals/corr/SchniterR14}. The four different operators (\ref{eq_fastOpDefs11}), (\ref{eq_fastOpDefs1}), (\ref{eq_fastOpDefs2}), (\ref{eq_fastOpDefs22}) are respectively generalized to the complex case as:
\begin{align}
\tilde O_\mu(\textbf{e}_c) &\defeq \sum_{i\in c}^{N/L_c} |\cv{F}_{\mu i}|^{2} e_i\\
O_\mu(\textbf{e}_c) &\defeq \sum_{i\in c}^{N/L_c} \cv{F}_{\mu i} e_i \\
\tilde O_i(\textbf{f}_r) &\defeq \sum_{\mu\in r}^{\alpha_rN/L_c} |\cv{F}_{\mu i}|^2 f_\mu\\O_i^*(\textbf{f}_r) &\defeq \sum_{\mu\in r}^{\alpha_rN/L_c}\cv{F}_{\mu i}^* f_\mu
\end{align}
where $\cv{F}_{\mu i}^*$ is the complex conjugate of $\cv{F}_{\mu i}$. Because the value of $|\cv{F}_{\mu i}|^2$ is either $0$, $1$ or $J$ $\forall \ (\mu,i)$ depending on the block as we use Hadamard or Fourier operators (it can be read on Fig.~\ref{fig_seededHadamard}), all these operators 
do not require matrix multiplications as they are implemented as fast transforms ($O_\mu$ and $O_i^*$)  or simple sums ($\tilde O_\mu$ and $\tilde O_i$). It results in the updates for complex AMP \cite{Schniter2012compressive} with a generic operator, see Fig.~\ref{algo_AMP_complex}.
\begin{figure}[!t] 
\centering
\begin{algorithmic}[1]
\State $t\gets 0$
\State $\delta \gets \epsilon + 1$
\While{$t<t_{max} \ \textbf
{and } \delta>\epsilon$} 
\State $\Theta^{t+1}_\mu \gets \sum_{c}^{L_c}\tilde O_\mu(\textbf{v}_c^{t})$
\State $\underline{{w}}^{t+1}_\mu \gets \sum_{c}^{L_c}O_\mu(\underline{\textbf{a}}_c^{t}) - \Theta^{t+1}_\mu\frac{\underline{y}_\mu-\underline{{w}}^{t}_\mu}{\Delta + \Theta^{t}_\mu}$
\State $\Sigma^{t+1}_i \gets \left[\sum_{r}^{L_r}\tilde O_i\left([\Delta + \boldsymbol{\Theta}_r^{t+1}]^{-1}\right)\right]^{-1/2}$
\State $\underline{R}^{t+1}_i \gets \underline{a}^{t}_i + (\Sigma^{t+1}_i)^2 \sum_{r}^{L_r} O_i^*\left(\frac{\underline{\textbf{y}}_r - \underline{\textbf{w}}^{t+1}_r}{\Delta + \boldsymbol{\Theta}^{t+1}_r}\right)$
\State $v^{t+1}_i \gets f_{c}\left((\Sigma^{t+1}_i)^2,\underline{R}_i^{t+1}\right)$
\State $\underline{a}^{t+1}_i \gets \cv{f_{a}}\left((\Sigma^{t+1}_i)^2,\underline{R}_i^{t+1}\right)$
\State $t \gets t+1$
\State $\delta \gets <|\underline{\ba}^{t+1} - \underline{\ba}^{t}|^2>$
\EndWhile
\State \textbf{return} $\underline{\ba}^{t}$
\end{algorithmic}
\centering
\caption[Complex AMP algorithm with operators]{The complex AMP algorithm written with operators. Depending on whether it is used on a real or complex signal, with 
Bayes-optimal or sparsity-inducing thresholding functions $\cv{f_{a}}$ and $f_{c}$, we call it BP, c-BP, LASSO or c-LASSO.
$\epsilon$ is the accuracy for convergence and $t_{max}$ the maximum number of iterations.
A suitable initialization for the quantities is ($\cv{a}_i^{t=0}=\mathbb{E}_{P_0}(x) = 0$, $v_i^{t=0}=\txt{Var}_{P_0}(x)=\rho \sigma^2 $, $\cv{{w}}_\mu^{t=0}=\cv{y}_\mu$) where we have used the prior (\ref{Px}). Once the algorithm has converged, i.e. the quantities do not change anymore from iteration to iteration, the estimate of the $i^{th}$ signal component is $\cv{a}^{t}_i$.
The nonlinear thresholding functions $\cv{f_{a}}$ and $f_{c}$ take into account the prior distribution. In the case of compressed sensing, applying $\cv{f_{a}}$ to a $\underline{R}_i^{t+1}$ close to zero will give a result even closer to zero, while bigger inputs will be left nearly unchanged, thus favoring sparse solutions. If needed, the damping scheme of Fig.~\ref{algoCh1:AMP} can be used.}  
\label{algo_AMP_complex}
\end{figure}

Here, we give the functions
$\cv{f_{a}}$ and $f_{c}$ that are calculated by Gaussian integration from (\ref{Px}) and are thus Bayes-optimal, which is not the case for
LASSO and c-LASSO \cite{maleki2012asymptotic} as discussed in sec.~\ref{sec:typicalPhaseTransitions}.
For BP, they are trivially constructed from sec.~\ref{sec:cookAMP}. For c-BP, the signal is complex and drawn from the distribution (\ref{Px}), and the thresholding functions (which give posterior scalarwise estimates of the mean and variance) are given by:
\begin{align}
\cv{f_{a}}(\Sigma^2,\cv{R}) &= g\rho\chi^2\cv{M}/Z \\
f_{c}(\Sigma^2,\cv{R}) &= \left(g\rho\chi^2 \(|\cv{M}|^2 + 2 \chi^2\)/Z - |\cv{f_a}(\Sigma^2,\cv{R})|^2\right)/2
\end{align}
together with the following definitions:
\begin{align}
\cv{M} &\defeq (\sigma^2 \cv{R}+\Sigma^2 \cv{\bar x})/(\Sigma^2+\sigma^2)\\ 
\chi^2 &\defeq \Sigma^2\sigma^2/(\Sigma^2 + \sigma^2)\\
g &\defeq \exp\(-\frac{1}{2}\left(\frac{|\cv{\bar x}|^2}{\sigma^2} + \frac{|\cv{R}|^2}{\Sigma^2} - \frac{|\cv{M}|^2}{\chi^2}\right)\)\\ 
Z &\defeq \sigma^2(1-\rho)\exp\(-\frac{|\cv{R}|^2}{2\Sigma^2}\) + \rho\chi^2 g
\end{align}
where $\cv{R}$ and $\Sigma^2$ are the AMP fields and we have $\cv{\bar x} = \cv{\bar s}$. These functions are not identical to the ones for the real case since in the
prior distribution (\ref{Px}), the real and imaginary parts of the
signal are jointly sparse (i.e. have same support but independent values), which can be a good assumption, for instance in MRI. As in c-LASSO \cite{maleki2012asymptotic}, because the joint sparsity is more constrained and thus bring more information, it allows to lower the phase transition compared to when the real and imaginary parts of the signal are assumed to be fully independent.
\subsection{Randomization of the structured operators}
\label{sec:randomizationStructOp}
The implementation requires caution: the necessary "structure killing" randomization of the fast structured operator used to construct the blocks of the matrix Fig.~\ref{fig_seededHadamard} is obtained by applying a permutation of lines after the use of the fast operator.
For each block $(r,c)$, we choose a random subset of modes $\Omega^{r,c} = \{ \Omega^{r,c}_1 ,\ldots, \Omega^{r,c}_{N_r} \} \subset \{ 1,\ldots, N_c \}$.
The definition of 
$O_\mu(\textbf{e}_c)$ using a standard fast transform ${\rm FT}$ will be:
\begin{equation}
 O_\mu(\textbf{e}_c) \defeq   \({\rm FT}(\textbf{e}_c)\)_{\Omega^{r_{\mu},c}_{\mu - \mu_{r_{\mu}} + 1}}
\end{equation}
where $r_{\mu}$ is the index of the block-row that includes the index $\mu$, $\mu_{r_{\mu}}$ is the number of the first line of the block row $r_{\mu}$ and $(\tbf u)_{\mu}$ is the $\mu^{\rm th}$ component of $\tbf u$. For $O_i^*(\textbf{f}_r)$ instead:
\begin{equation}
 O_i^*(\textbf{f}_r)\defeq\({\rm FT}^{-1}( \tilde{\textbf{f}}_r )\)_{i - i_{c_i}+1}
\end{equation}
where $c_{i}$ is the index of the block-column that includes the index $i$, $i_{c_{i}}$ is the number of the first column of the block-column $c_{i}$,
${\rm FT}^{-1}$ is the standard fast inverse operator of ${\rm FT}$ and $\tilde{\textbf{f}}_r$ is defined in the following way:
\begin{equation}
  \forall \gamma \in \{ 1,\ldots,N_r \}, \quad \(\tilde{\textbf{f}}_r\)_{\Omega_{\gamma}^{r,c_i}} = \(\textbf{f}_r\)_{\gamma} \quad \txt{and} \quad  \forall k \notin \Omega^{r,c_i}, \, \(\tilde{\textbf{f}}_r\)_k = 0
\end{equation}
The $MSE$ achieved by the algorithm is: 
\begin{equation}
E^{t}\defeq||\cv{\ba}^{t} - \cv{\bs}||_2^2=<|\cv{\ba}^{t} - \cv{\bs}|^2>
\label{eq_MSE}
\end{equation}
and measures how well the signal is reconstructed.
\begin{figure}[!t]
\centering
\hspace{-.8cm}
\includegraphics[width=.9\textwidth]{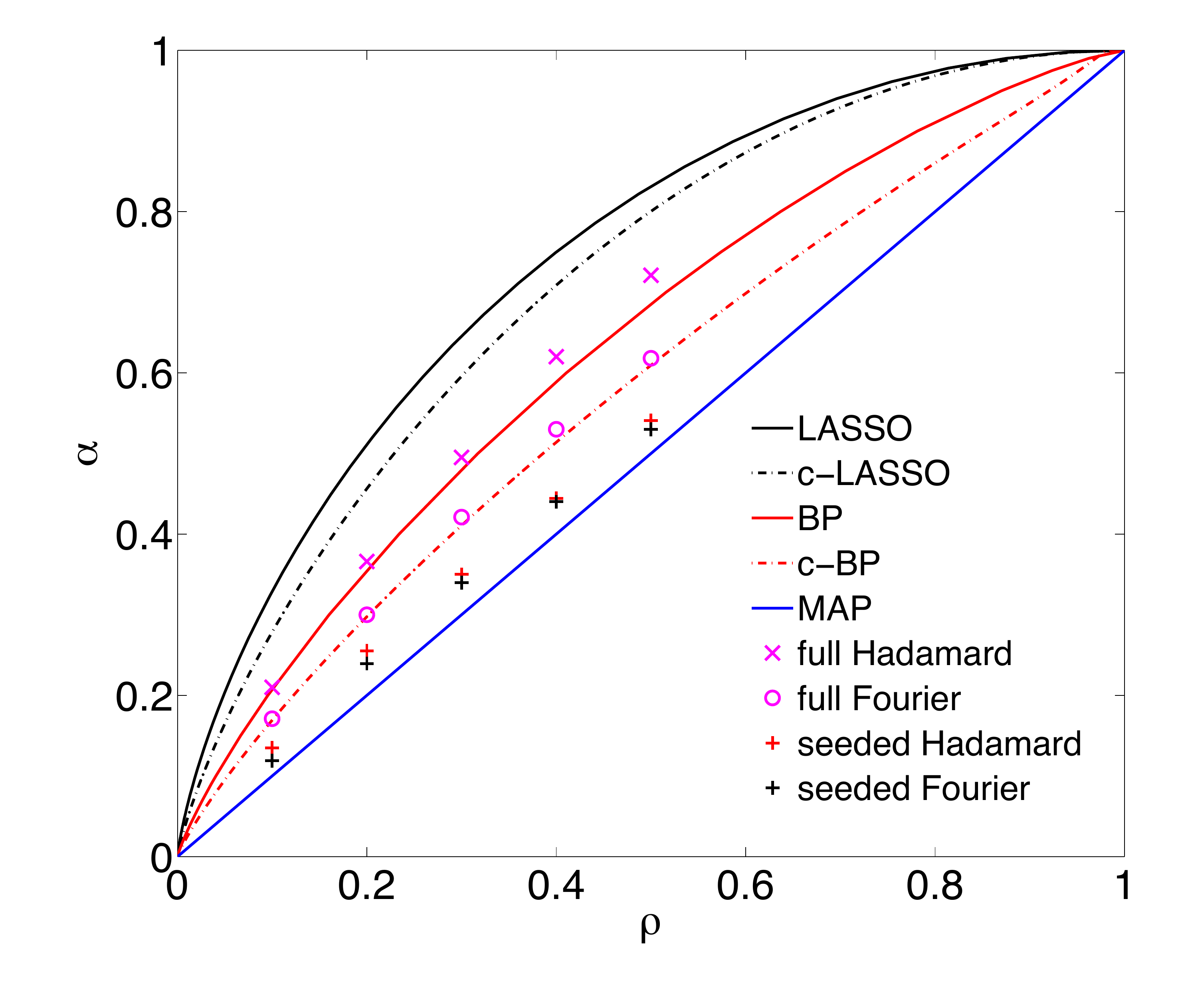}
\caption[Phase diagram of noiseless compressed sensing of real and complex signals, with instances solved with structured operators]{Phase diagram on the $(\alpha,\rho)$ plane for noiseless $\Delta=0$ compressed sensing. Lines are phase transitions predicted by the
state evolution analysis for i.i.d random Gaussian matrices, while markers are
points from experiments using structured operators with empirically optimized parameters. Good sets of parameters usually lie in the following sets: $(L_c\in\{8,16,32,64\}, L_r = L_c + \{1,2\}, w\in\{2,\ldots,5\}, \sqrt{J}\in[0.2,0.7],\beta_{{{seed}}}\in[1.2,2])$, see (\ref{eq_alphaRest}). As discussed in the previous chapter, with larger signals, higher values of $L_c$ are better as it allows to get closer to the optimal transition. Just as c-LASSO
allows to improve the usual LASSO phase transition when the complex
signal is sampled according to (\ref{Px}) (thanks to the joint
sparsity of the real and imaginary parts), c-BP improves the usual
BP transition. The line $\alpha=\rho$ is both the maximum-a-posteriori $MAP$ threshold for noiseless compressed sensing and the (asymptotic) optimal phase transition that can be reached with spatially-coupled
matrices. Pink experimental points correspond to perfectly reconstructed instances
using homogeneous Hadamard and Fourier operators (on the BP and c-BP
phase transition respectively), the black and red crosses using spatially-coupled ones (close to the $MAP$ threshold). Properly randomized structured operators appear to have similar performances as random measurement matrices.}

\label{fig_DE}
\end{figure}
\section{Results for noiseless compressed sensing}
\label{sec:results}
\begin{figure}
\centering
\includegraphics[width=1\textwidth]{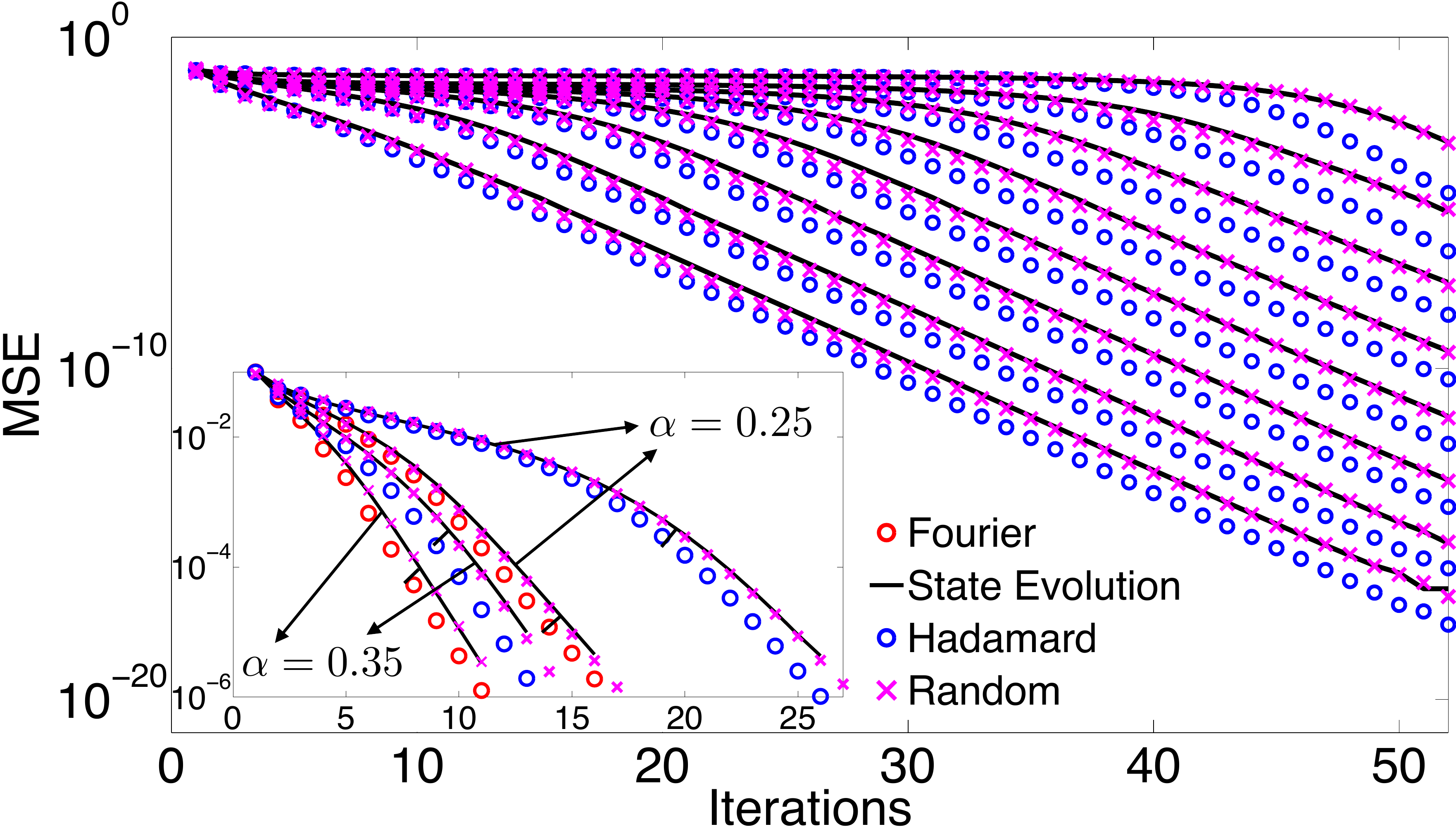}
\caption[State evolution for spatially-coupled operators and comparison with structured operators]{Comparison of the mean square error predicted by the state evolution (black curves) and the actual behavior of the algorithm for spatially-coupled matrices (main figure) and standard homogeneous
ones (inset), both with structured operators (circles) and random i.i.d
Gaussian matrices (purple crosses) in a noiseless compressed sensing setting. In both plots, the signal size is $N=2^{14}$ with the random i.i.d Gaussian matrices, and $N=2^{20}\approx 10^{6}$ 
with the operators and are generated
with ($\rho= 0.1$, $\cv{\bar x}= 0$, $\sigma^2=
1$). While experiments made with random i.i.d matrices fit very well the asymptotic predictions, those with
the structured operators are not described well by the state evolution, although
final performances are comparable. \textbf{Main:} For an Hadamard spatially-coupled matrix as in
Fig.~\ref{fig_seededHadamard} with ($L_c= 8$, $L_r = L_c+2$, $w = 1$, $\sqrt{J} = 0.1$, $\alpha = 0.22$, $\beta_{{{seed}}} = 1.36$). Each curve corresponds
to the $MSE$ tracked in a different block of the real reconstructed signal $\bs$ (see Fig.~\ref{fig_seededHadamard}). \textbf{Inset:} Reconstructions made
with structured homogeneous matrices at $\alpha = 0.35$ and $\alpha = 0.25$. The reconstruction
with the Fourier operator of a complex signal (instead of real with Hadamard) is faster thanks to the joint
sparsity assumption of (\ref{Px}). The arrows identify
the groups of curves corresponding to same measurement rate $\alpha$. Both in the Fourier and Hadamard cases, we observe that convergence is slightly faster than in the random i.i.d case as predicted by the state evolution.}
\label{fig_expVsDE}  
\end{figure}
When the sensing matrix is i.i.d random, or spatially-coupled with i.i.d random blocks, the evolution of $E^{t}$ in AMP is predicted in the large signal limit on
a rigorous basis by the state evolution \cite{BayatiMontanari10,DonohoJavanmard11,bayati2012universality}, see sec.~\ref{sec:stateEvolutionGeneric}. For c-BP with i.i.d Gaussian matrices, the derivation goes very much along the same lines and we shall report the results
briefly. The generalization to the complex case of the state evolution (\ref{eqChIntro:SE_form3}) for homogeneous matrices under the prior (\ref{Px}) is given by the following recursion:
\begin{equation}
E^{t+1} = \int \mathcal{D}\underline z \left[(1-\rho) f_{c} \left((\Sigma^{t+1})^2, \cv{R}_1^{t+1}(\underline z)\right)+\rho f_{c} \left((\Sigma^{t+1})^2, \cv{R}_2^{t+1}(\underline z) \right)\right]
\label{complexDE}
\end{equation}
together with:
\begin{align}
\underline z &\defeq z_1 + i z_2 \\ 
(\Sigma^{t+1})^2 &\defeq (\Delta+E^{t})/\alpha ,\nonumber\\
\cv{R}_u^{t+1}(\underline z) &\defeq \underline z\sqrt{\sigma^2\delta_{u,2} + (\Sigma^{t+1})^2}
\\ \mathcal{D}\underline z &\defeq dz_1dz_2 \frac{e^{-\frac{1}{2} (z_1^2 + z_2^2)}}{2 \pi}.
\end{align}
with $\Delta = 0$ in the noiseless case. (\ref{complexDE}) has been obtained exactly as in the previous chapter when we derived (\ref{Et}). Note that this state evolution equation is the same as given in \cite{maleki2012asymptotic}, despite slightly different update rules in the algorithm. 

For c-BP with spatially-coupled matrices with i.i.d Gaussian blocks, the expression involves the $MSE$ in each block $c'\in\{1,\ldots,L_c\}$, see sec.~\ref{sec:spatiallyCoupledSE}. The generalization of (\ref{eq_SEsigmaSeeded}) ,(\ref{eqChIntro:defRseeded}) and (\ref{eqChIntro:SE_form3_seeded}) to the complex case under the prior (\ref{Px}) is given by:
\begin{align}
\label{densEvo}
E_c^{t+1} &=\int \mathcal{D}\underline z \left[(1-\rho)f_c\((\Sigma^{t+1}_c)^2, \cv{R}_{c,1}^{t+1}(\underline z)\) +\rho f_c\((\Sigma^{t+1}_c)^2, \cv{R}_{c,2}^{t+1}(\underline z)\)\right]
\end{align}
where:
\begin{align}
\Sigma^{t+1}_c\(\{E_{c'}^t\}_{c'}^{L_c}\) &=\left[\sum_{r}^{L_r} \frac{ \alpha_{r} J_{rc}}{L_c\Delta + \sum_{c'}^{L_c} J_{rc'}E_{c'}^{t}}\right]^{-1/2} \\ 
\cv{R}_{c,u}^{t+1}(\underline z)&=\underline z \sqrt{\sigma^2 \delta_{u,2} + (\Sigma^{t+1}_c)^{2}} 
\end{align}
and $\alpha_{r}=\alpha_{rest} + (\alpha_{seed} - \alpha_{rest})\delta_{r,1}$, $J_{rc}$ is the variance of the elements belonging to the block at the $r^{th}$ block-row and $c^{th}$ block-column (1, $J$ or $0$ in Fig.~\ref{fig_seededHadamard}) and again $\Delta = 0$ in the noiseless case.
\begin{figure}[!t]
\centering
\includegraphics[width=1\textwidth]{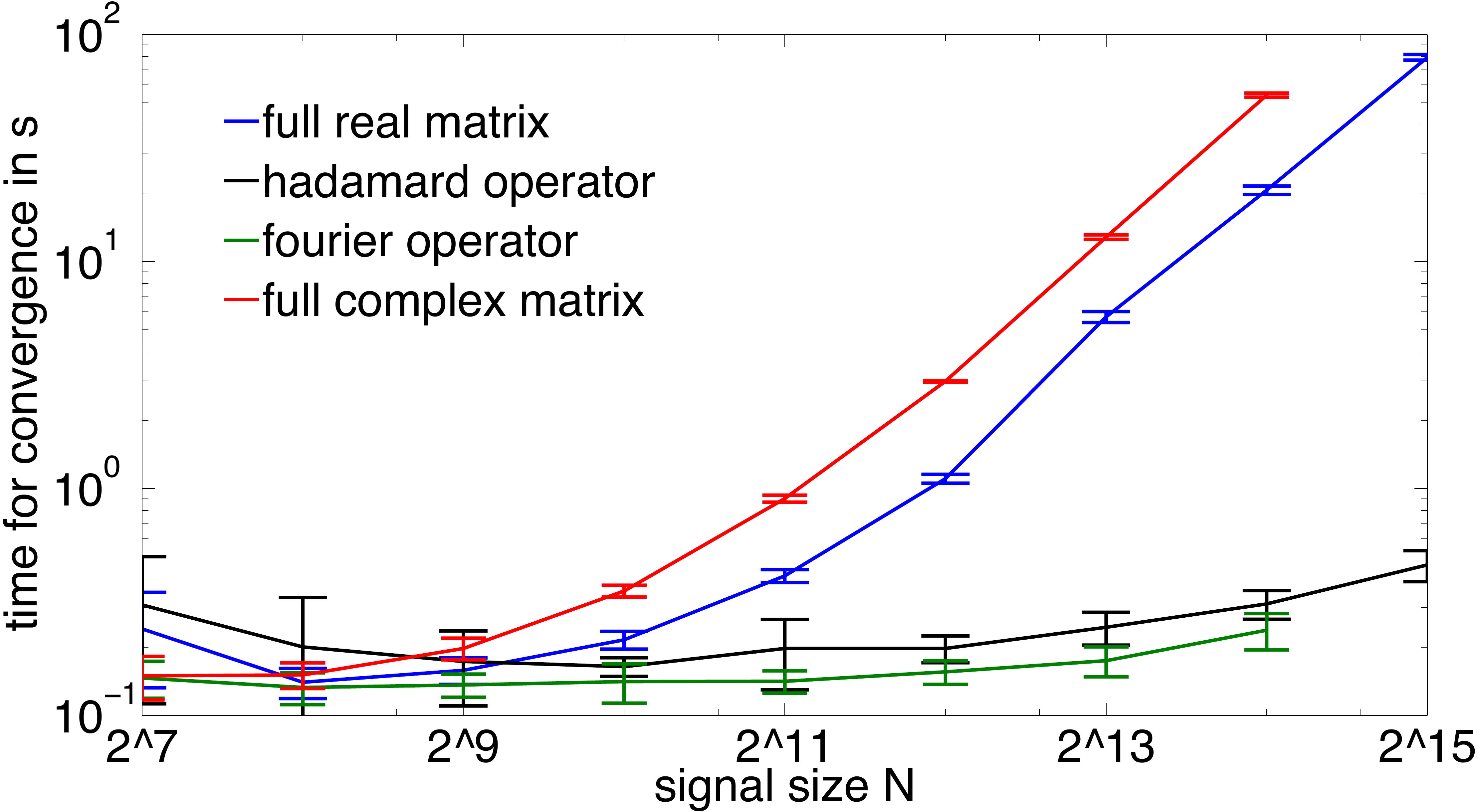}
\caption[Comparison of the running time with fast operators and random ones]{Time required for convergence (i.e. $MSE< 10^{-6}$) of the AMP algorithm in seconds as a function of the signal size, in the homogeneous matrix case for a typical compressed sensing problem.
The signal has distribution given by (\ref{Px}) and is real (complex) for the reconstruction with real (complex) matrices. 
The plot compares the speed of AMP with matrices (blue and red lines) to those of AMP using the structured operators (black and green lines). 
The points have been averaged over 10 random instances and the error bars represent the standard deviation with respect to these. 
The simulations have been performed on a personal laptop. As the signal size increases, the advantage of using operators becomes quickly obvious.}
\label{fig_timeComp}
\end{figure}

We now move to our main point. In the case of AMP with structured (Fourier or Hadamard) operators instead of i.i.d matrices,
the state evolution analysis cannot be made. Hence we experimentally compare the performances between
AMP with structured operators and i.i.d matrices. The comparison is shown in
Fig.~\ref{fig_expVsDE} which presents theoretical results from the state evolution and experimental ones obtained by
running AMP on finite size signals. On Fig.~\ref{fig_DE}, we
show the phase transition lines obtained by state evolution analysis in the $(\alpha,\rho)$
plane, and we added markers showing the position of instances actually recovered by the algorithm with spatially-coupled structured operators in the noiseless case $\Delta=0$. It appears that with structured operators, AMP is still able to decode really close to the optimal threshold as with random i.i.d matrices.
\subsection{Homogeneous structured operators}
Let us first concentrate on AMP with homogeneous (or full) structured operators. The
first observation is that the state evolution {\it does not} correctly describe the
evolution of the $MSE$ for AMP with full structured operators (inset Fig.~\ref{fig_expVsDE}).
It is perhaps not surprising, given that AMP has
been derived for i.i.d matrices. The difference is
small, but clear: $E^{t}$ decreases faster with structured operators than with i.i.d matrices.
 However, despite this slight difference in the dynamical behavior of the algorithm,
the phase transitions
and the final MSE performances for both approaches appear to be
extremely close. As seen in Fig.~\ref{fig_DE}, for small
$\rho$, we cannot
distinguish the actual phase transition with structured operators from the one predicted by state evolution. 
Thus, the state evolution analysis remains a good predictive tool of the AMP performances with structured operators.
\subsection{Spatially-coupled structured operators}
For spatially-coupled operators, the conclusions are similar (main plot on
Fig.~\ref{fig_expVsDE}). Again, $E_c^{t}$ in each of the blocks of the signal, induced by the spatially-coupled structure of the measurement matrix, decreases faster with structured operators than with random i.i.d matrices. 
But our empirical results are
consistent (see Fig.~\ref{fig_DE}) with the hypothesis that the
proposed scheme, using spatially-coupled Fourier/Hadamard operators, achieves correct
reconstruction as soon as $\alpha > \rho$ when $N$ is large. Indeed, we observe that the
gap to the $MAP$ threshold (the optimal threshold in the noiseless case) $\alpha_{opt}(\rho)=\rho$ decreases as the signal size increases upon
optimization of the spatially-coupled operator structure. The results in
Fig.~\ref{fig_DE} and Fig.~\ref{fig_expVsDE} are obtained with spatially-coupled
matrices of the ensemble: $(L_c=8, L_r = L_c + 1, w \in\{1,2\}, \sqrt{J} \in \[0.2, 0.5\], \beta_{seed} = [1.2,1.6])$. While these parameters do not
quite saturate the bound $\alpha=\rho$ (which is only possible for $L_c
\to \infty$
\cite{KudekarRichardson10,KrzakalaPRX2012,DonohoJavanmard11}, see sec.~\ref{eq_alphaRest_2}), they
do achieve near optimal performances.
This, as well as the substantial cut in running time (Fig.~\ref{fig_timeComp}) 
with respect to AMP with i.i.d matrices and the possibility to work with very large systems without saturating the memory strongly supports the advantages of the proposed implementation of AMP.
\section{Conclusion}
We have presented a large empirical study using structured Fourier and
Hadamard operators in sparse linear estimation. We have shown that combining these operators with a spatial coupling strategy allows to get very close to the information-theoretical limits. We have tested our algorithm for noiseless compressed
sensing of real and complex signals. The resulting algorithm is
more efficient than Fig.~\ref{algoCh1:AMP_op} both in terms of memory and running time. This allows us to deal with signal sizes as
high as $10^{6}$ and a measurement rate $\alpha \approx \rho$ on a personal laptop using MATLAB, and
achieve perfect reconstruction in about a minute. 
%
%
%
%
%
%
%
%
%
%
%
%
%
%
%
%
%
%
%
%
%
%
%
%
%
%
%
%
%
%
%
%
%
%
%
%
%
%
%
%
%
%
%
%
%
%
%
%
%
%
%
%
%
%
%
%
%
%
%
%
%
%
%
%
%
%
%
%
%
%
%
%
%
%
%
%
%
%
%
%
%
%
%
%
%
%
%
%
%
%
%
%
%
%
%
%
%
%
%
%
%
%
%
%
%
%
%
%
%
%
%
%
%
%
%
%
%
%
%
%
%
%
%
%
%
%
%
%
%
%
%
%
%
%
%
%
%
%
%
%
%
%
%
%
%
%
%
%
%
%
%
%
%
%
%
%
%
%
%
%
%
%
%
%
%
%
%
%
%
%
%
%
%
%
%
%
%
%
%
%
%
%
%
%
%
%
%
%
%
%
%
%
%
%
%
%
%
%
%
%
%
%
%
%
%
%
%
%
%
%
\chapter{Approximate message-passing for compressive imaging}
\label{chap:images}
In the present chapter, we focus on two distincts applications of the approximate message-passing algorithm to compressive imaging. The first one is the reconstruction of sub-sampled natural images which have a sparse discrete gradient, while the second one focuses on the reconstruction of point like objects, and thus directly sparse in the pixel domain, measured by fluorescence microscopy technique.
\section{Reconstruction of natural images in the compressive regime by "total-variation-minimization"-like approximate message-passing}
\label{sec:TV}
Total variation (TV) and the associated gradient-optimization-based algorithms have been the long-standing state-of-the-art approach to 
the reconstruction of images from compressive measurements. For signals
well-modeled by i.i.d priors, many recent works have presented optimal, or
near optimal reconstruction algorithms built instead from a statistical framework.
Here, we present a method for incorporating a TV-like structured prior 
in the image domain to allow for $MMSE$ image recovery using the approximate message-passing algorithm. 

In gradient-optimization-based methods for image reconstruction, instead of searching for a sparse signal in the pixel domain as in (\ref{sec:microscopySec}), we seek for a signal with a sparse discrete gradient. This kind of signals are supposed to represent well natural images which are piecewise constant or smooth. Specific algorithms to tackle image reconstruction in the compressive regime have been designed on the optimization side. For example the TV-AL3 \cite{Li2009} is very robust to noise and reduction of the measurement rate but is also fast. On the probabilistic side, the phase transitions for TV were investigated and a method, TV-AMP \cite{DJM2013}, was proposed. Furthermore, \cite{BP2013} proposed a structured prior in conjunction with the co-sparse model and developed the GrAMPA algorithm based on the approximate message-passing algorithm. Finally, some recent work have proposed a joint prior defined over nearest-neighbors for 
$1$-d signals in SS-AMP \cite{KJL2014}. Motivated by the very good performances of both \cite{KJL2014} and \cite{BP2013}, we develop here an approximate message-passing algorithm for $2$-d piecewise smooth signals reconstruction in the compressive regime.

The methodology presented here is closely related to the GrAMPA algorithm \cite{BP2013} as we will work also with the co-sparse model, defined in the next section, and AMP. The differences with \cite{BP2013} are essentially coming from the learning procedure of the noise that we use which allows for better reconstruction results, and the fact that we use a Gauss-Bernoulli prior for the dual variables (that represent the differences between neighboring pixels), whereas GrAMPA uses the SNIPE prior, the limiting distribution of the Gauss-Bernoulli one when the variance of the Gaussian part goes to infinity. As we will see, this prior does not improve on the Gauss-Bernoulli one, they give similar results. The improvement with respect to GrAMPA really comes from the noise learning which acts as a kind of annealing. 

We will show through intensive numerical experiments that when we push the TV-AL3 algorithm (considered as the state-of-the-art optimization algorithm) to its limits by optimizing all of its parameters, it gives almost exactly the same reconstruction results than our implementation which requires way less tuning. We will observe the same when using the SNIPE prior of GrAMPA in our implementation in conjunction with the noise learning. The results are so close that it suggests that we are reaching the limits of classical reconstruction methods based on first neighbors interactions.
\subsection{Proposed model}
\label{sec:algoTV}
We consider the model (\ref{eqIntro:AWGNCS}) where now $\bs$ is the reshaped image of size N (the original image being of size $\sqrt{N}\times\sqrt{N}$). In order to mimic the total-variation-minimization idea that enforces neighbors pixels to have identical or closeby values, we naturally take a Gauss-Bernoulli prior over the {\it differences} of the pixel values:
\begin{align}
&P_0(\bs|\sigma^2) \propto \prod_{(ij)\in E}\[(1-\rho)\delta(s_i-s_j) + \rho \mathcal{N}\(s_i-s_j|0,\sigma^2\)\]
\end{align}
where $E\defeq\{(ij):i \ \text{neighbor to} \ j \ \text{in the picture}\}$ is the set of pairs of pixels that are neighbors in the original image. This natural extension to the bi-dimensionnal case of \cite{KJL2014} suffers from convergence issues due to the short loops in the associated factor graph (a grid) that are created by these two-pixels dependent prior terms.
\begin{figure}[!t]
\centering
\includegraphics[width=0.3\textwidth]{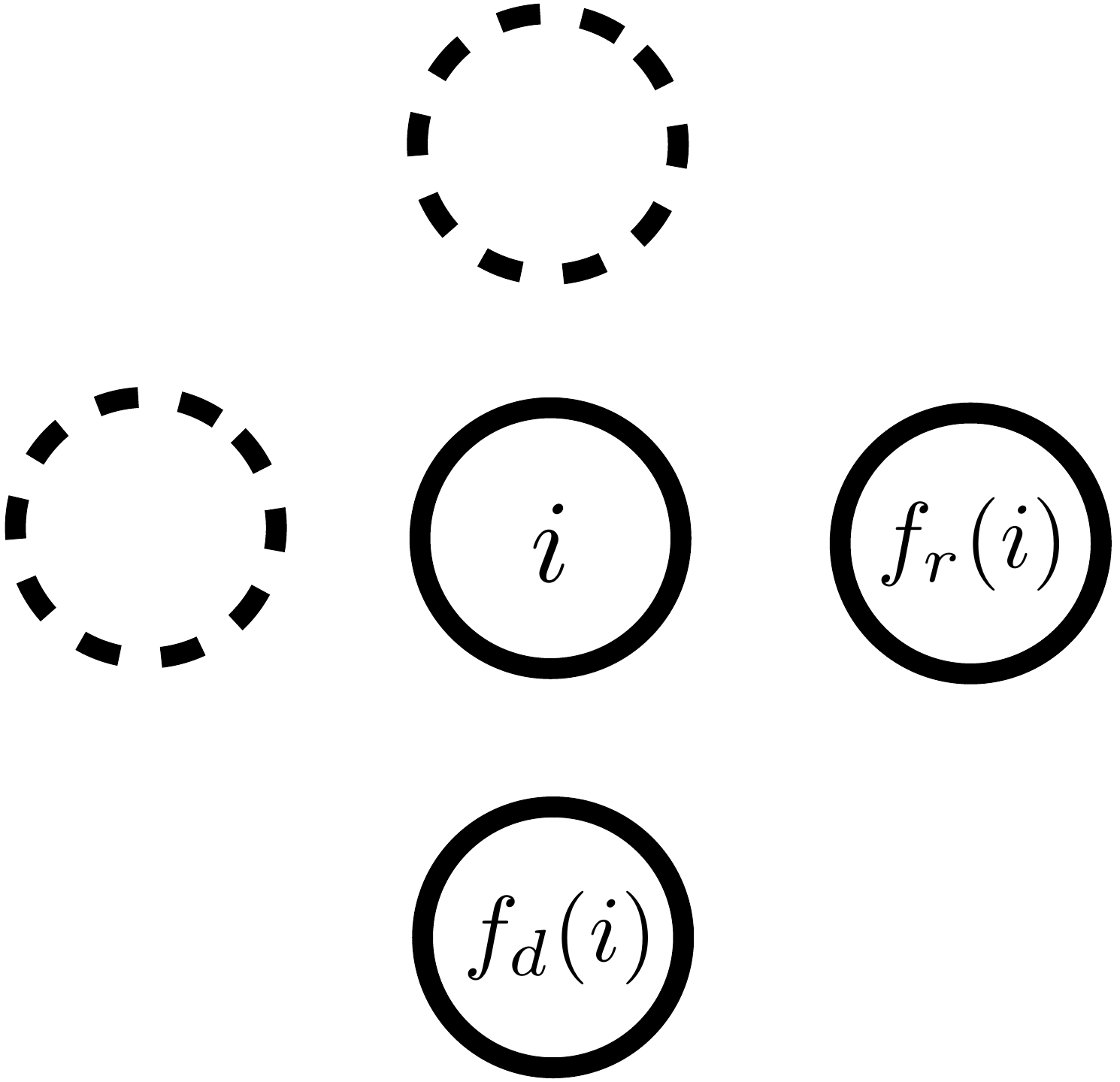}
\caption[The pixel $i$ with its four closest neighbors]{The pixel $i$ with its four closest neighbors, but when considering $i$ in the operator $\tilde{\tbf F}$ construction (\ref{eq:hatSyst}), only the interaction with its right neighbor $f_r(i)$ and down one $f_d(i)$ are taken into account not to consider twice each interactions. This choice is arbitrary and considering for example the left and up interactions would be the same, one just needs to design properly the mapping functions $f$ which encode the dependency structure between the image pixels and allow to design the difference matrix ${\tbf D}$ in (\ref{eq:hatSyst}).}
\label{fig:pixels}
\end{figure}
A natural idea to face this problem is to use new auxilliary variables. Defining $\{d_{ij}\defeq x_i-x_j : (ij)\in E \}$ as the differences variables, refered as {\it dual} variables, and $\bx \defeq [\bs, \tbf d]$ as the concatenation of the vectorized original image and the vector of dual variables, we now get back a factorized prior where the $s_i$'s are independent with uniform prior and the Gauss-Bernoulli prior is over the dual variables:
\begin{align}
P_0(\bx|\sigma^2, \mathcal{S}) &\propto \prod_{(ij)\in E}\[(1-\rho)\delta(d_{ij}) + \rho \mathcal{N}\(d_{ij}|0,\sigma^2\)\]\prod_{i=1}^{N}\mathcal{U}(s_i|\mathcal{S})
\label{eq:facPrior}
\end{align}
where $\mathcal{U}(s_i|\mathcal{S})$ is a uniform distribution over the set $\mathcal{S}$. From now on when the index $i$ of the $\bx$ component is such that if $i\le N$ it correponds to an image pixel variable, else it is a dual variable. Thus the prior can be re-written as:
\begin{align}
P_0(\bx|\sigma^2, \mathcal{S}) &\propto \prod_{i=1}^{N+|E|} P_0^i(x_i|\sigma^2, \mathcal{S})\\
&\propto \prod_{i=1}^{N+|E|}\left[\mathbb{I}(i\le N)\mathcal{U}(x_i|\mathcal{S}) + \mathbb{I}(i>N) \[(1-\rho)\delta(x_i) + \rho \mathcal{N}\(x_i|0,\sigma^2\)\] \right]\label{eq:facPrior_2}
\end{align}
and the linear constraints $\{d_{ij}-(s_i-s_j)=0\}_{(ij)\in E}$ are enforced through an extension of the original linear system (\ref{eqIntro:AWGNCS}) that becomes:
\begin{align}
\left( \begin{array}{c}
\by_{M,1}\\
\boldsymbol{0}_{|E|,1} \end{array}  \right)
&= \left( \begin{array}{cc}
\bF_{M,N} & \boldsymbol{0}_{M,|E|} \\
{\tbf D}_{|E|,N} & -\boldsymbol{I}_{|E|,|E|} \end{array}  \right)
\left( \begin{array}{c}
\bs_{N,1}\\
{\tbf d}_{|E|,1} \end{array}  \right)
+ \left( \begin{array}{c}
\bxii_{M,1}\\
\boldsymbol{0}_{|E|,1} \end{array}  \right) \nonumber \\
\Leftrightarrow \tilde \by &= \tilde \bF\bx + \tilde{\bsy \xi}
\label{eq:hatSyst}
\end{align}
where the dimension of each vector or matrix has been indicated to avoid confusions and where the tilde stands for these new extended objects. $\bF$ is the original operator, $\boldsymbol{0}_{a,b}$ is a matrix full of zeros of dimensions $a\times b$ and $\boldsymbol{I}_{a,a}$ is the identity matrix if size $a\times a$. ${\tbf D}$ is the "difference" matrix which is the concatenation of smaller matrices constructed as follows. In the present case we consider for each pixel its four closest neighbors (its left/right and up/down ones on the image). ${\tbf D}\defeq[{\tbf D}_r, {\tbf D}_d]^\intercal$ is the concatenation of ${\tbf D}_r$ which is made of zeros everywhere except on the diagonal where there are $1$'s, and $-1$'s on the $\{(i, f_r(i)):i\in \{1,\dots,N\}, f_r(i) \neq 0\}$ elements where $f_r(i)$ outputs the index (in the vectorized form of $\bx$) of the right neighbor of the $i^{th}$ pixel if it actually has a neighbor, $0$ else (it is the same for constructing ${\tbf D}_d$ thanks to the mapping $f_d(i)$), see Fig.~\ref{fig:pixels}. This new system to be solved is refered as the co-sparse model.
\subsection{The Hadamard operator and the denoisers}
Images are large signals so being able to deal with very large matrices with $O(N^2)$ elements is an issue in itself. We can thus use fast Hadamard-based operators of the form Fig.~\ref{fig_opSpCoupling}, see chap.~\ref{chap:structuredOperators} for a full study of their reconstruction abalities. Because of the fact that each Hadamard mode except the first one (which is a line of ones) have exactly zero mean, the system (\ref{eq:hatSyst}) would be invariant by a constant shift in the signal components without this mode. To break this symmetry we enforce the first mode to be present which fixes the mean of the signal, the other ones being selected totally randomly. This issue was never present in the other problems treated in this thesis because the prior was directly on the signal components whereas here, its on the differences (the prior is uniform over the pixels): the prior is invariant by a constant shift in the values of two neighbors and thus of the overall signal.

Once $\bF$ is designed (as well as $\bF^\intercal, \bF^2, (\bF^2)^\intercal$ required by AMP, see Fig.~\ref{algoCh1:AMP_op}), it is trivial to implement $\tilde{\tbf F}$ (and ${\tilde{\tbf F}}^\intercal, {\tilde{\tbf F}}^2, ({\tilde{\tbf F}}^2)^\intercal$). $\tilde{\tbf F}$ is also fast and does not generate memory issues because all its parts (except $\bF$ which is already a fast Hadamard-based operator) are highly sparse.

Now the co-sparse model (\ref{eq:hatSyst}) to solve is well defined with a factorized prior (\ref{eq:facPrior}), we can use AMP, Fig.~\ref{algoCh1:AMP_op}. 
Despite the extended measurement matrix $\tilde \bF$ is sparse and the derivation of AMP in sec.~\ref{sec:classicalDerivationAMP} is based on the high density of the matrix, nothing prevents us to use it anyway and as we will see, it gives very good results. From the definition of the denoising functions (\ref{eq1:fai}), (\ref{eq1:fci}) and using the prior (\ref{eq:facPrior_2}), we obtain:
\begin{align}
	f_{a_i}(\Sigma_i^2, R_i) &\defeq \int dx_i x_i P_0(x_i|\sigma^2,\mathcal{S})\mathcal{N}\(x_i|R_i, \Sigma_i^2\) \\
	&= R_i \ \mathbb{I}(i\le N) + \tilde f_{a_i}(\Sigma_i^2, R_i) \ \mathbb{I}(i>N) \\
	f_{c_i}(\Sigma_i^2, R_i) &\defeq \int dx_i x_i^2 P_0(x_i|\sigma^2,\mathcal{S})\mathcal{N}\(x_i|R_i,\Sigma_i^2\) - f_{a_i}(\Sigma_i^2, R_i)^2  \\
	&= \Sigma_i^2 \ \mathbb{I}(i\le N) + \tilde f_{c_i}(\Sigma_i^2, R_i) \ \mathbb{I}(i>N)
\end{align}
where the explicit form of $\tilde f_{a_i}$ and $\tilde f_{c_i}$ in this Gauss-Bernoulli case are obtained through the construction of sec.~\ref{sec:cookAMP}:
\begin{align}
	\tilde f_{a_i}(\Sigma^2, R) &= \frac{1}{z_i(\Sigma^2, R)}\frac{m \Sigma^2 + R\sigma^2}{\Sigma^2 + \sigma^2} \\
	\tilde f_{c_i}(\Sigma^2, R) &= \frac{1}{z_i(\Sigma^2, R)}\frac{m^2 \Sigma^4 + \Sigma^2 (2 m R + \Sigma^2) \sigma^2 + (R^2 + \Sigma^2) \sigma^4}{(\Sigma^2 + \sigma^2)^2} - \tilde f_{a_i}(\Sigma^2, R)^2\\
	z_i(\Sigma^2, R) &= \frac{\mathcal{N}\(m|R,\Sigma^2\)}{ \mathcal{N}\(m|R,\Sigma^2+\sigma^2\)} + 1
\end{align}
where we have used that the set $\mathcal{S}=\mathbb{R}$ and thus just replaced the uniform distribution in the prior by $1$. We could enforce positive values for the pixels, but it complicates the denoising functions expressions and seems to gives exactly the same final results anyway. 
\subsection{The learning equations}
The hyperparameters $\bsy \Delta\defeq[\Delta_s, \Delta_d]$ need to be learned, where the noise associated to the $\mu^{th}$ measurement is $\Delta_{\mu} \defeq \Delta_s \mathbb{I}(\mu\le M) + \Delta_d \mathbb{I}(\mu > M)$. $\Delta_s$ is the true noise variance associated to the measure, $\Delta_d$ is the artificial noise variance associated to the dual variables which measure the level of relaxation of the constraints that define them: at $\Delta_d=0$, the dual variables must exactly fulfill their definition, at finite value of $\Delta_d$, these linear constraints are relaxed. We decide to keep $\rho$ free being the control parameter of how smooth the final reconstructed picture is. Furthermore $\sigma^2$ can be learned but it appears empirically that fixing its value to $O(10^{-3})$ is sufficient and does not change the performances. For learning $\bsy\Delta$, we will use the expectation maximization learning of sec.~\ref{sec:EMlearning} optimizing the Bethe free energy. 
%
%
The learning of the noise is obtained by optimizing (\ref{eq:BetheF_forAMP_last}). For the moment we consider that there is a unique noise parameter $\Delta$. The only part $F_\Delta$ dependent on it in (\ref{eq:BetheF_forAMP_last}) can be rewritten as:
\begin{align}
	F_\Delta = \frac{1}{2} \sum_{\mu}^{M+|E|}\[\frac{(\tilde y_\mu - \sum_i^{N+|E|} \tilde F_{\mu i} a_i)^2}{\Delta} + \log\(\Delta+\sum_i^{N+|E|} \tilde F_{\mu i}^2 v_i\)\] 
\end{align}
where ($a_i,v_i$) are the posterior mean and variance of $x_i$. Then by optimizing it we obtain the fixed point equation for $\Delta$:
\begin{align}
&\frac{\partial F_\Delta}{\partial \Delta}=0\\
&\Leftrightarrow \sum_{\mu}^{M+|E|}\[\frac{1}{\Delta^2}\(\tilde y_\mu - \sum_{i}^{N+|E|} \tilde F_{\mu i}a_i\)^2 - \(\Delta + \sum_{i}^{N+|E|} \tilde F_{\mu i}^2 v_i\)^{-1}\] = 0 \label{eq:toOptNoiseImages}
\end{align}
A possible solution is to equate the two terms inside the sum for each component, which gives different solutions $\Delta_\mu$ for each $\mu$. We define the auxiliary functions:
\begin{align}
&\chi_\mu\defeq \(\tilde y_\mu - \sum_{i}^{N+|E|} \tilde F_{\mu i}a_i\)\\
&g_\mu\defeq \frac{1}{2}\(\chi_\mu^2 + \chi_\mu \sqrt{\chi_\mu^2+4\Theta_\mu}\) \label{eq:gmu}
\end{align}
reminding the definition of $\Theta_\mu=\sum_{i}^{N+|E|} \tilde F_{\mu i}^2 v_i$ (\ref{eq:fixedPointTheta}) at the fixed point. The solutions canceling each term inside (\ref{eq:toOptNoiseImages}) are given by the second order equations:
\begin{align}
	\Delta_\mu^2 - \chi_\mu^2 \Delta_\mu -\chi_\mu^2\Theta_\mu = 0
\end{align}
which exact solution is simply:
\begin{align}
	\Delta_\mu = g_\mu
\end{align}
Then we would average over these (the positive solutions are selected among the two possible ones) to get a single parameter $\Delta$. But now we remember that we need to consider two different noise parameters $\{\Delta_s,\Delta_d\}$: the noise learning is not the same for the pixels and the dual variables. To consider this, the average for the two different noise levels are performed over the proper sets of variables. We get the fixed point equations to which we add the time:
\begin{align}
\Delta_s^{t+1} &= \frac{1}{M}\sum_{\mu=1}^{M}g_\mu^t\\
\Delta_d^{t+1} &= \frac{1}{|E|}\sum_{\mu=M+1}^{M+|E|}g_\mu^t \label{eq:noiseLearn}
\end{align}
where $g_\mu^t$ is given by (\ref{eq:gmu}) with $\chi_\mu^t=\(\tilde y_\mu - \sum_{i}^{N+|E|} \tilde F_{\mu i}a_i^t\)$, $\Theta_\mu^t=\sum_{i}^{N+|E|} \tilde F_{\mu i}^2 v_i^t$ where ($a_i^t, v_i^t$) are the AMP posterior estimates at time $t$. In the implementation, these learnings are weakly damped.

It appears that this learning is essential for the perfomances of the algorithm. It is the main difference with the GrAMPA implementation \cite{BP2013}. Their prior model, refered as the SNIPE prior is different as well but as we will show in the experiments, it does not change the perfomances of AMP: the improvement in our implementation really comes from this noise learning.
%
%
\subsection{Numerical experiments}
\begin{figure}[ht!]
	\centering
		\includegraphics[width=0.35\textwidth]{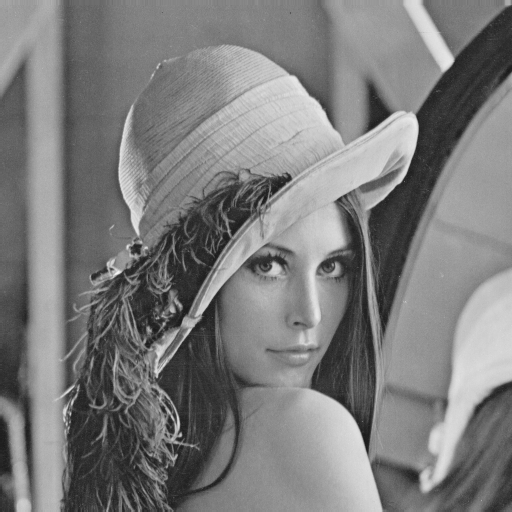}~
		\includegraphics[width=0.35\textwidth]{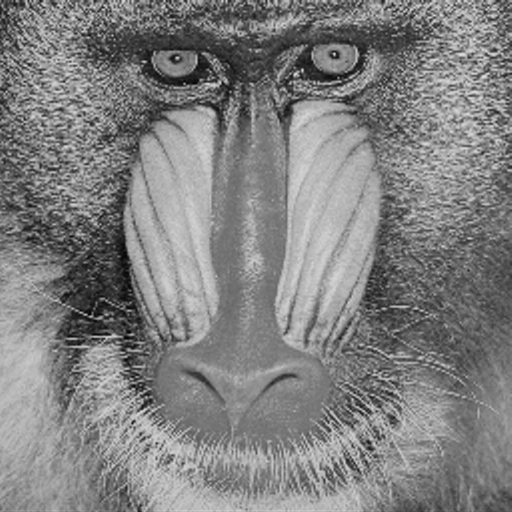}~
		\includegraphics[width=0.35\textwidth]{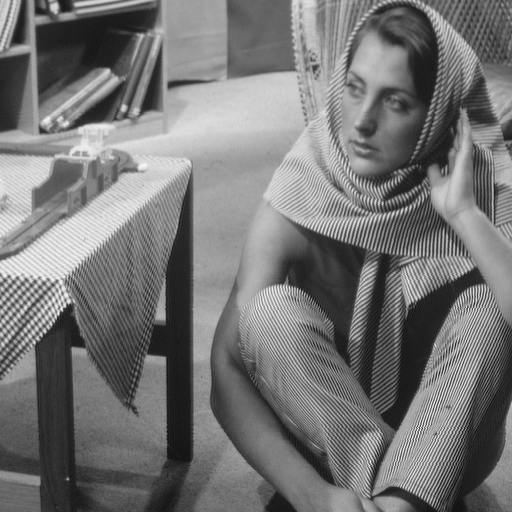}\\
		\includegraphics[width=0.35\textwidth]{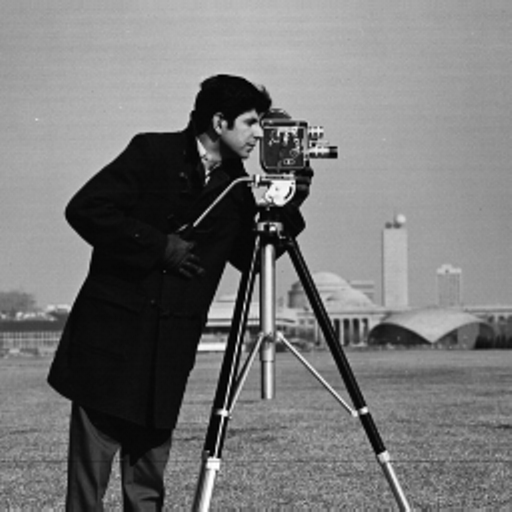}~
		\includegraphics[width=0.35\textwidth]{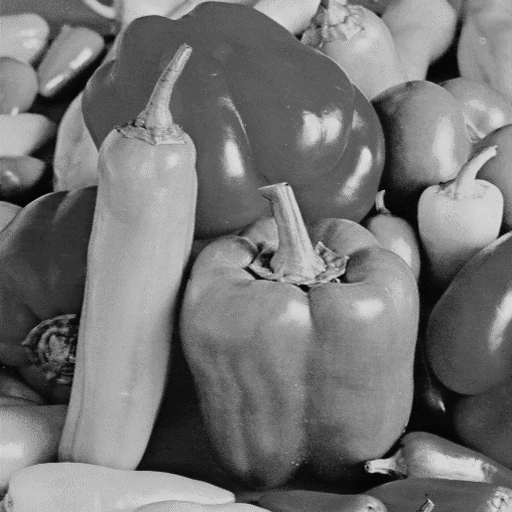}
		\caption[Images used for the TV-reconstruction comparisons]{Images used for the comparisons between TV-AL3 and our AMP implementation refered as DC-AMP for dual constraints AMP. Top, left to right:
		Lena, Baboon, Barbara. Bottom, left to right: Cameraman and Peppers \label{fig:imagesTests}}
\end{figure}

We now present results of a serie of intensive numerical experiments. We have selected classical test $512\times 512$ images, see Fig.~\ref{fig:imagesTests}. For each of them, we have compared the results obtained with the best TV-optimization-based algorithm, namely TV-AL3 \cite{Li2009} and our AMP implementation refered as DC-AMP for dual-constraints AMP. For each measurement rate, we have scanned different noise levels. For each point, $5$ instances with different measurement operators and noise realizations have been reconstructed by each algorithm and the result is the averaged normalize mean square error NSNR (the $MSE$ rescaled by the $\ell_2$ norm of the picture) in decibels. 
For TV-AL3 that depends on two slack parameters (see \cite{Li2009} for the details), we have optimized them for {\it each point} (for every images and for all measurement rates). The same has been done for the unique free parameter of DC-AMP, the smoothness parameter $\rho$. So these results represent the optimal reconstruction performances of both algorithms. Furthermore, the SNIPE prior have also been implemented in our code for comparison with the Gauss-Bernoulli one and its free $w$ parameter, which is the equivalent of $\rho$ for the Gauss-Bernoulli prior have also been optimized for each point, see \cite{BP2013} for details on this prior.

Looking at all the results and tables, it is striking how close are the performances. There are virtually no differences. This suggests strongly that there exist some bound for the performances of such methods based on gradient optimization. The two priors in AMP appear to give almost the same results as it can be sees from the figures at $\alpha=0.05$. This similarity remains true at higher measurement rates. When the GrAMPA implementation have been tested with the optimized value of $w$ as well, it appeared to always give worst results by at least one or two decibels than what presented here. This must be due to our noise learning, not present in their implementation. Furthermore, when we do not use this learning, we get similar results as GrAMPA with our implementation which is normal as the two algorithms are supposed to solve the same problem (\ref{eq:hatSyst}). In terms of running time, all algorithms are equivalent despite a small advantage for TV-AL3.

An important remark is that we have tried our implementation also with further interactions, including the $8$ first-neighbors in the prior model instead of $4$. It appears that it worsen the results by reconstructing too smooth solutions.
\subsection{Concluding remarks}
We have presented an AMP implementation for the reconstruction of natural images based on the co-sparse model, with a Gauss-Bernoulli prior on dual variables representing the differences between pixels. It appears that its optimal performances after optimization of its single free parameter are perfectly equivalent to the ones of the state-of-the-art TV-optimization algorithms which require more parameters to be tuned to get similar results. Furthermore, it seems that our results are weakly sensitive to the prior used with AMP and that Gauss-Bernoulli or its infinite variance limit give similar results. The two main points observed here are that $i)$ the proposed algorithm get better performances that those of the similar GrAMPA algorithm due to a noise learning that exactly minimizes the Bethe free energy at each step and which is essential to reach the best performances and $ii)$ it seems that we are reaching some bound on the perfomances of reconstruction algorithms for natural images in the compressive regime that are based on gradient-based models. 

It would be of great interest to study further the fundamental reasons behind this limit. Furthermore, when trying to learn the $\rho$ parameter, it seems to enter in conflict with the noise learning. This issue must be fixed to get a parameter-free algorithm with the best performances.
\clearpage
\subsubsection{\emph{Lena} Image}
\begin{figure}[H]
	\centering
		\includegraphics[width=0.35\textwidth, trim=0 17 0 0, clip=true]{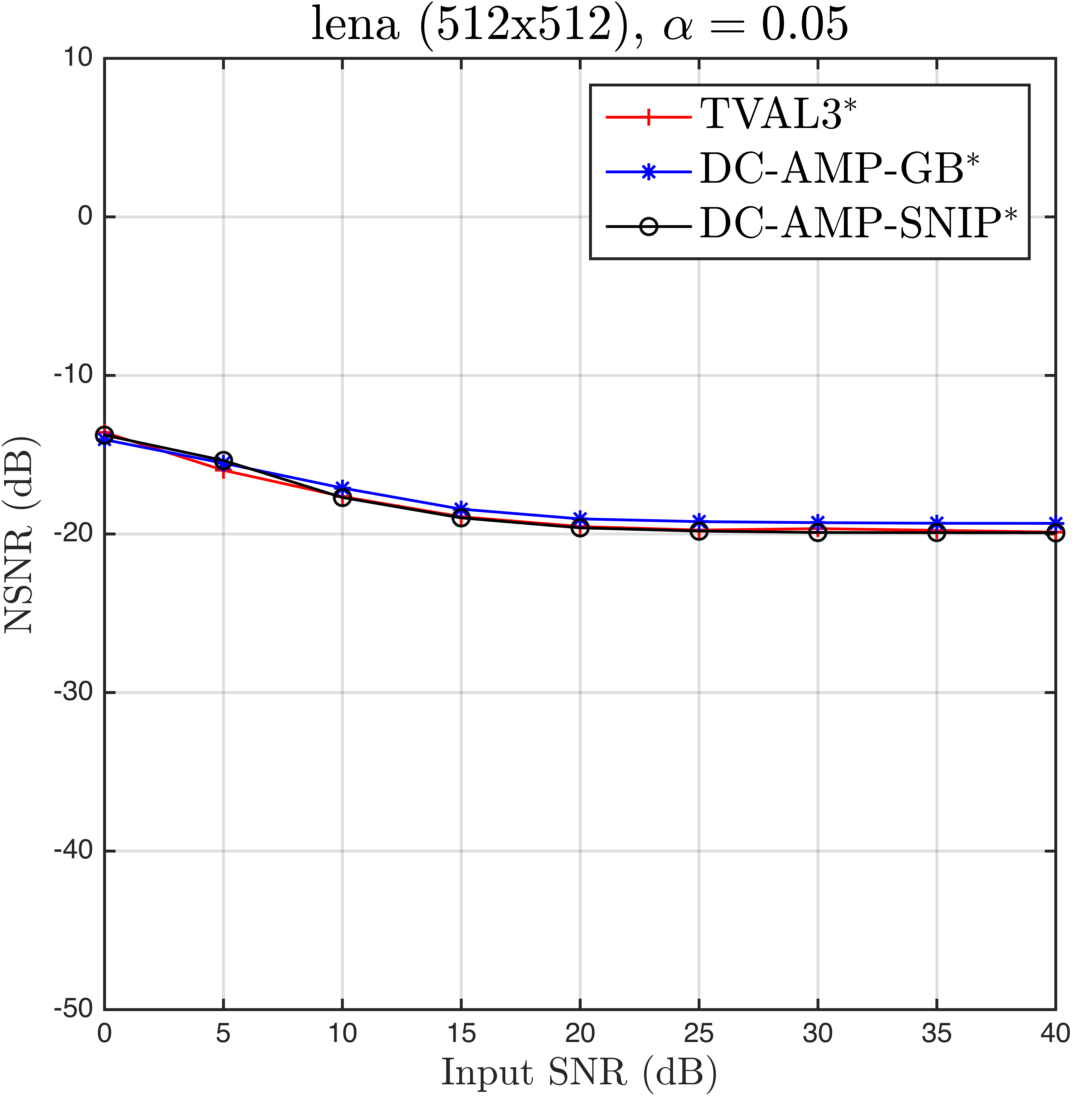}\quad
		\hspace{1.4cm}\includegraphics[width=0.45\textwidth, trim=35 30 45 0, clip=true]{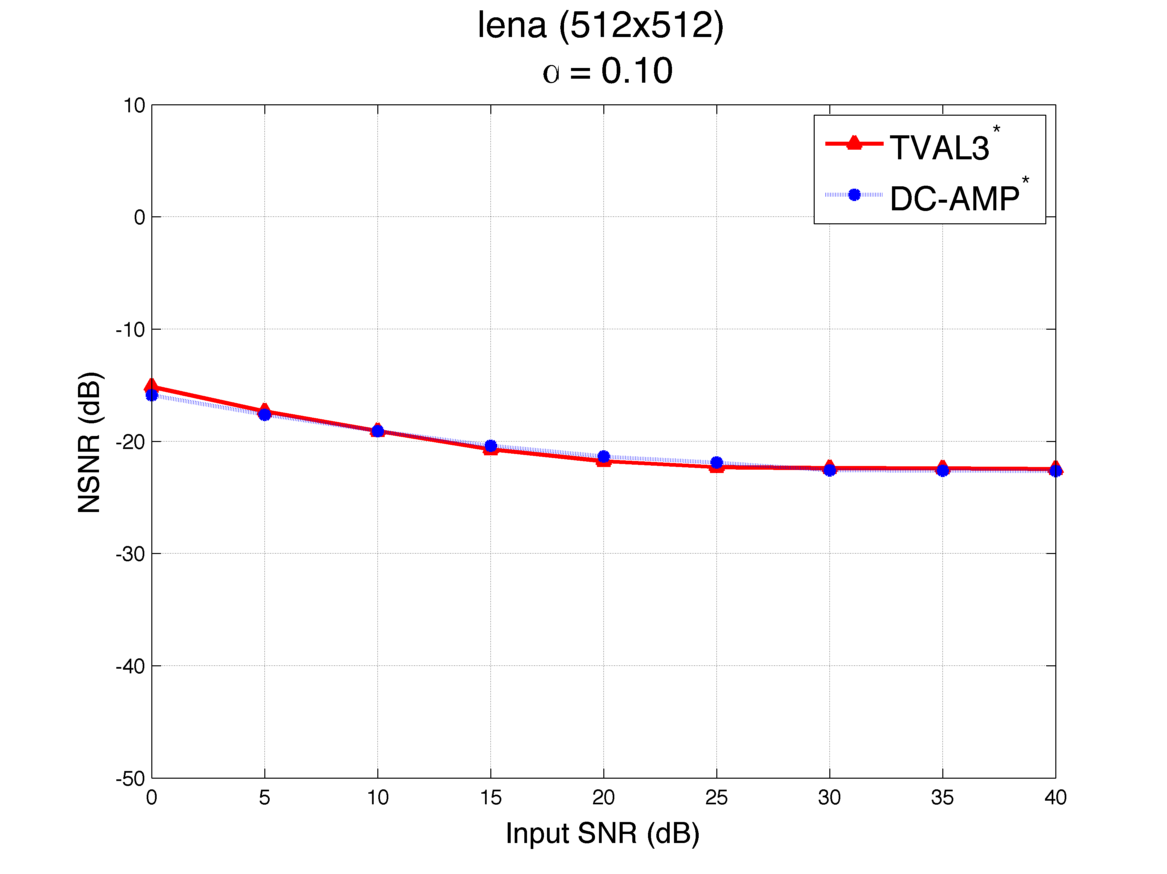}\\
		\includegraphics[width=0.45\textwidth, trim=35 30 45 25, clip=true]{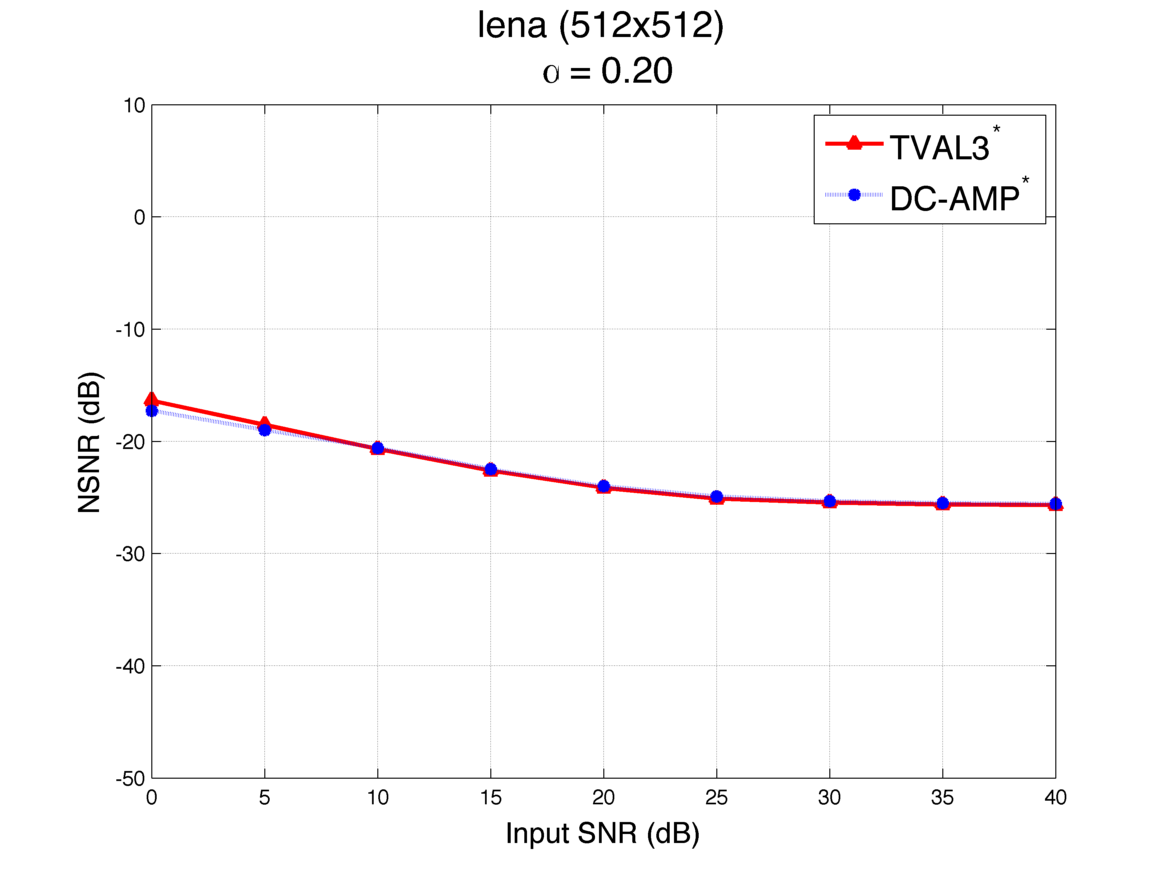}\quad
		\includegraphics[width=0.45\textwidth, trim=35 30 45 25, clip=true]{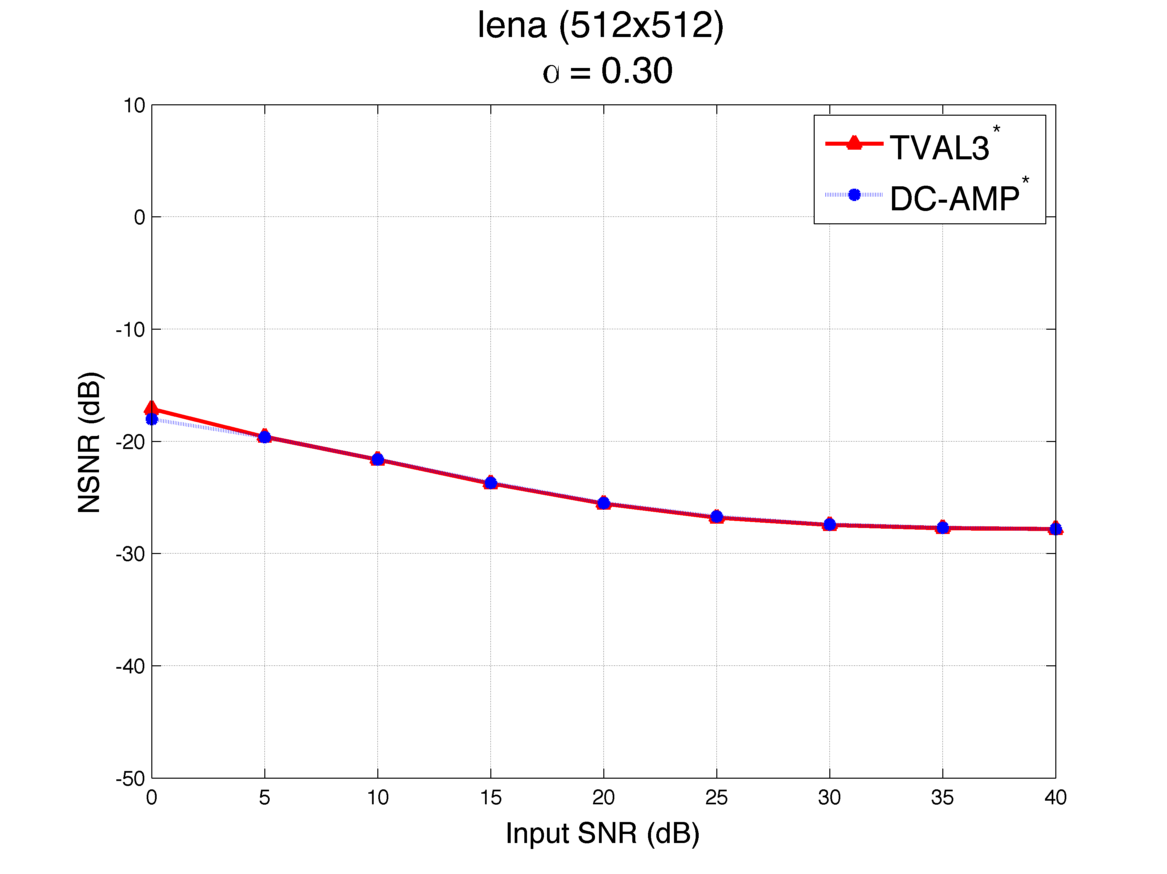}\\
		\includegraphics[width=0.45\textwidth, trim=35 10 45 25, clip=true]{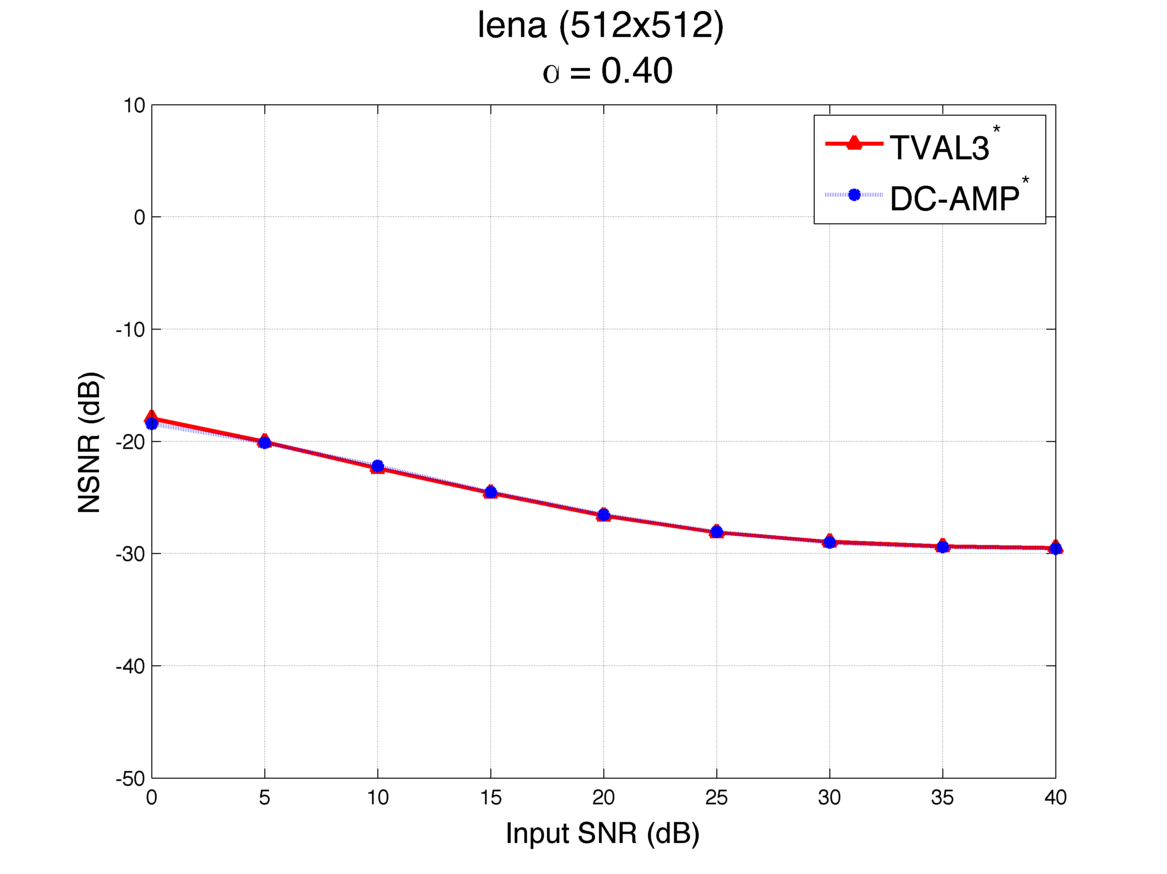}\quad
		\includegraphics[width=0.45\textwidth, trim=35 10 45 25, clip=true]{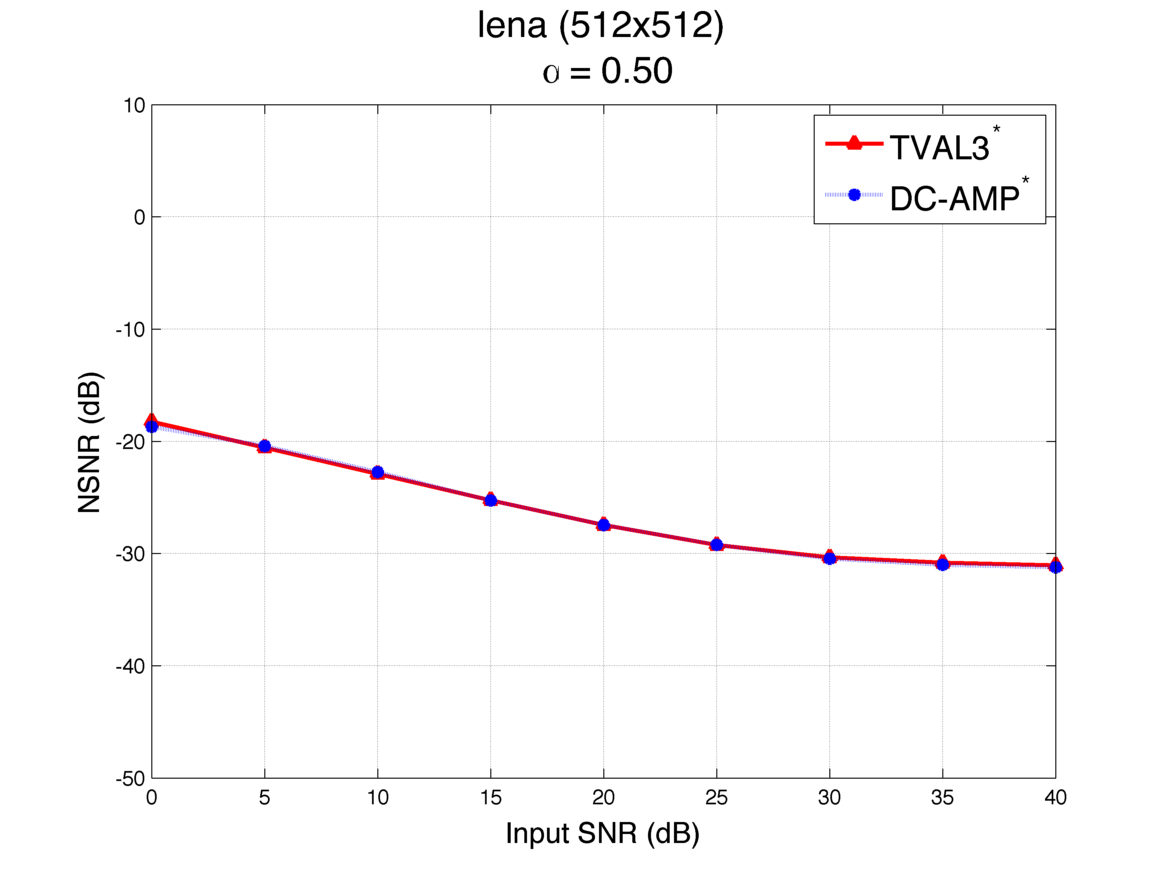}
	\caption[Comparison of the final reconstruction results for Lena]{Comparison of the final reconstruction results in NSNR as a function of the noise level in dB between TV-AL3 and DC-AMP for different measurement rates $\alpha$ for the Lena picture.\label{fig:lena_im}}
\end{figure}
\begin{table}
\centering
	\begin{tabular}{ | l || c | c | c | c | c | c |}
	\cline{2-7}
	\multicolumn{1}{l|}{~} & $\alpha = 0.05$ & $\alpha = 0.10$ & $\alpha = 0.20$ & $\alpha = 0.30$ & $\alpha = 0.40$ & $\alpha = 0.50$\\	
	\cline{2-7}		
	\multicolumn{1}{l|}{~} & \multicolumn{6}{c|}{ISNR = $\infty$~dB}\\
	\hline
		TVAL3 Optimal & \textbf{-19.74} & -22.51 & \textbf{-25.71} & -27.80 & -29.56 & -31.13 \\ 
DC-AMP Optimal & -19.33 & \textbf{-22.65} & -25.59 & \textbf{-27.87} & \textbf{-29.64} & \textbf{-31.32} \\ 

	\hline
	\multicolumn{1}{l|}{~} & \multicolumn{6}{c|}{ISNR = 40~dB}\\
	\hline
		TVAL3 Optimal & \textbf{-19.88} & -22.48 & \textbf{-25.68} & -27.82 & -29.51 & -31.06 \\ 
DC-AMP Optimal & -19.33 & \textbf{-22.65} & -25.57 & \textbf{-27.83} & \textbf{-29.58} & \textbf{-31.23} \\ 

	\hline
	\multicolumn{1}{l|}{~} & \multicolumn{6}{c|}{ISNR = 30~dB}\\
	\hline
		TVAL3 Optimal & \textbf{-19.66} & -22.40 & \textbf{-25.46} & \textbf{-27.46} & -28.97 & -30.34 \\ 
DC-AMP Optimal & -19.29 & \textbf{-22.58} & -25.34 & -27.41 & \textbf{-29.03} & \textbf{-30.46} \\ 

	\hline
	\multicolumn{1}{l|}{~} & \multicolumn{6}{c|}{ISNR = 25~dB}\\
	\hline
		TVAL3 Optimal & \textbf{-19.75} & \textbf{-22.31} & \textbf{-25.11} & \textbf{-26.81} & \textbf{-28.13} & \textbf{-29.25} \\ 
DC-AMP Optimal & -19.22 & -21.90 & -24.93 & -26.71 & -28.07 & -29.24 \\ 

	\hline
	\multicolumn{1}{l|}{~} & \multicolumn{6}{c|}{ISNR = 20~dB}\\
	\hline
		TVAL3 Optimal & \textbf{-19.53} & \textbf{-21.78} & \textbf{-24.15} & \textbf{-25.56} & \textbf{-26.63} & \textbf{-27.46} \\ 
DC-AMP Optimal & -19.05 & -21.36 & -24.01 & -25.50 & -26.54 & -27.46 \\ 

	\hline
	\multicolumn{1}{l|}{~} & \multicolumn{6}{c|}{ISNR = 15~dB}\\
	\hline
		TVAL3 Optimal & \textbf{-18.91} & \textbf{-20.71} & \textbf{-22.63} & \textbf{-23.77} & \textbf{-24.60} & -25.25 \\ 
DC-AMP Optimal & -18.43 & -20.40 & -22.48 & -23.70 & -24.53 & \textbf{-25.27} \\ 

	\hline
	\multicolumn{1}{l|}{~} & \multicolumn{6}{c|}{ISNR = 10~dB}\\
	\hline
		TVAL3 Optimal & \textbf{-17.64} & -19.08 & \textbf{-20.69} & \textbf{-21.64} & \textbf{-22.40} & \textbf{-22.91} \\ 
DC-AMP Optimal & -17.10 & \textbf{-19.09} & -20.61 & -21.62 & -22.20 & -22.75 \\ 

	\hline
	\multicolumn{1}{l|}{~} & \multicolumn{6}{c|}{ISNR = 5~dB}\\
	\hline
		TVAL3 Optimal & \textbf{-15.98} & -17.34 & -18.54 & -19.61 & -20.05 & \textbf{-20.55} \\ 
DC-AMP Optimal & -15.53 & \textbf{-17.65} & \textbf{-19.02} & \textbf{-19.63} & \textbf{-20.13} & -20.42 \\ 

	\hline
	\multicolumn{1}{l|}{~} & \multicolumn{6}{c|}{ISNR = 0~dB}\\
	\hline
		TVAL3 Optimal & -13.59 & -15.14 & -16.37 & -17.13 & -17.96 & -18.26 \\ 
DC-AMP Optimal & \textbf{-14.05} & \textbf{-15.90} & \textbf{-17.28} & \textbf{-18.03} & \textbf{-18.45} & \textbf{-18.72} \\ 

	\hline
	\end{tabular}
	\caption[Table of the final reconstruction results for Lena]{Table of the final reconstruction results in NSNR as a function of the noise level in dB between TV-AL3 and DC-AMP for different measurement rates $\alpha$ for the Lena picture.}
\end{table}
\clearpage
\subsubsection{\emph{barbara} Image}
\begin{figure}[H]
	\centering
		\includegraphics[width=0.35\textwidth, trim=0 17 0 0, clip=true]{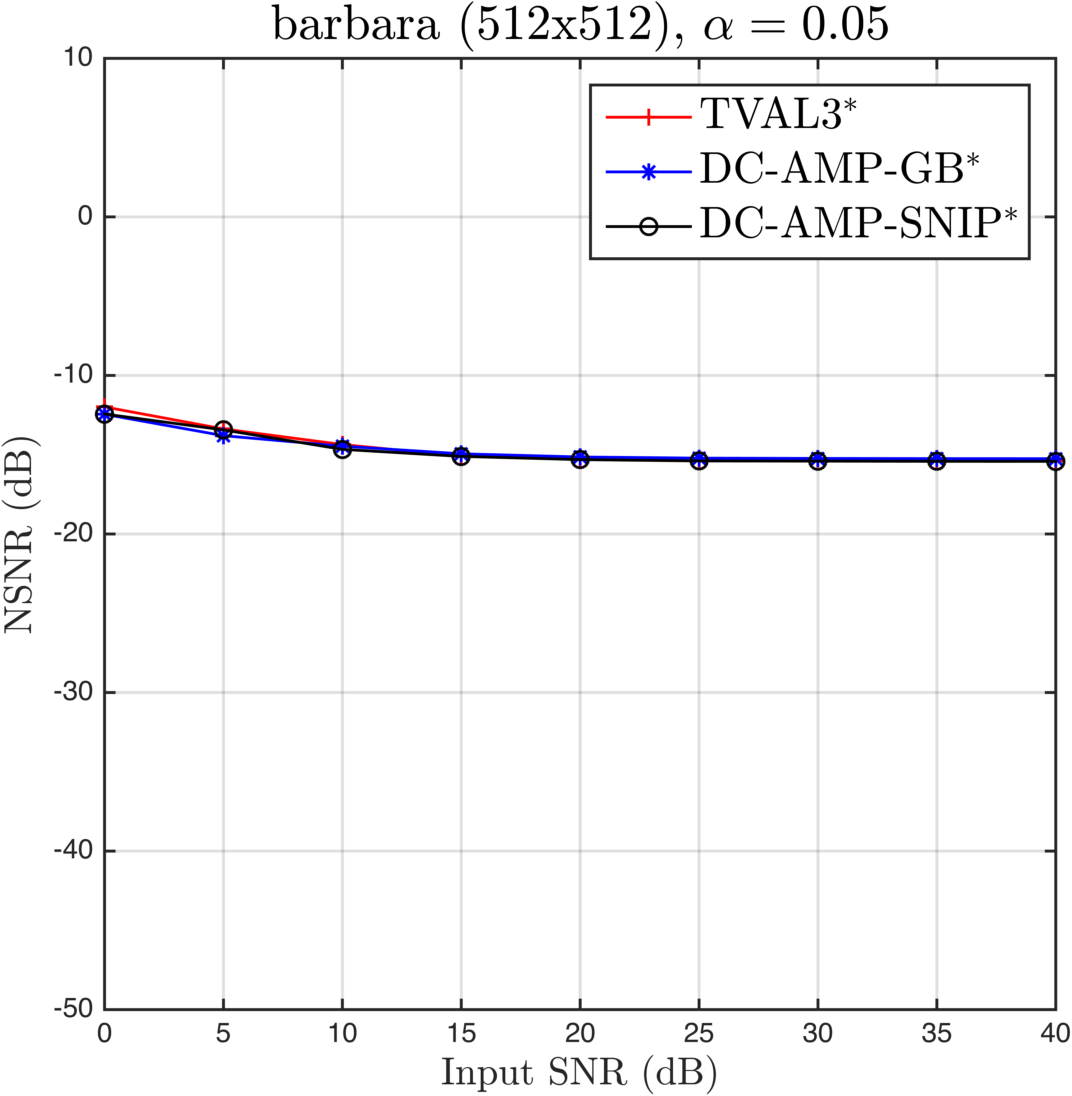}\quad
		\hspace{1.4cm}\includegraphics[width=0.45\textwidth, trim=35 30 45 0, clip=true]{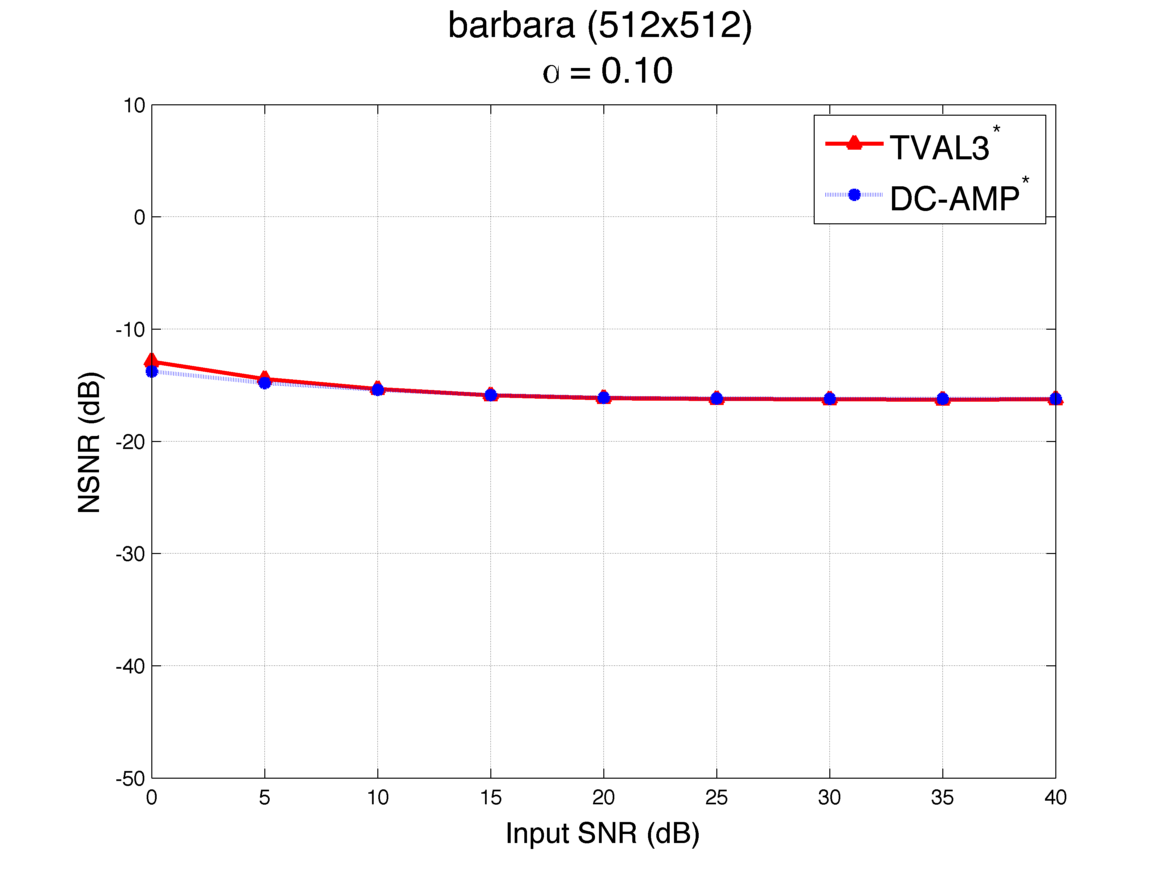}\\
		\includegraphics[width=0.45\textwidth, trim=35 30 45 25, clip=true]{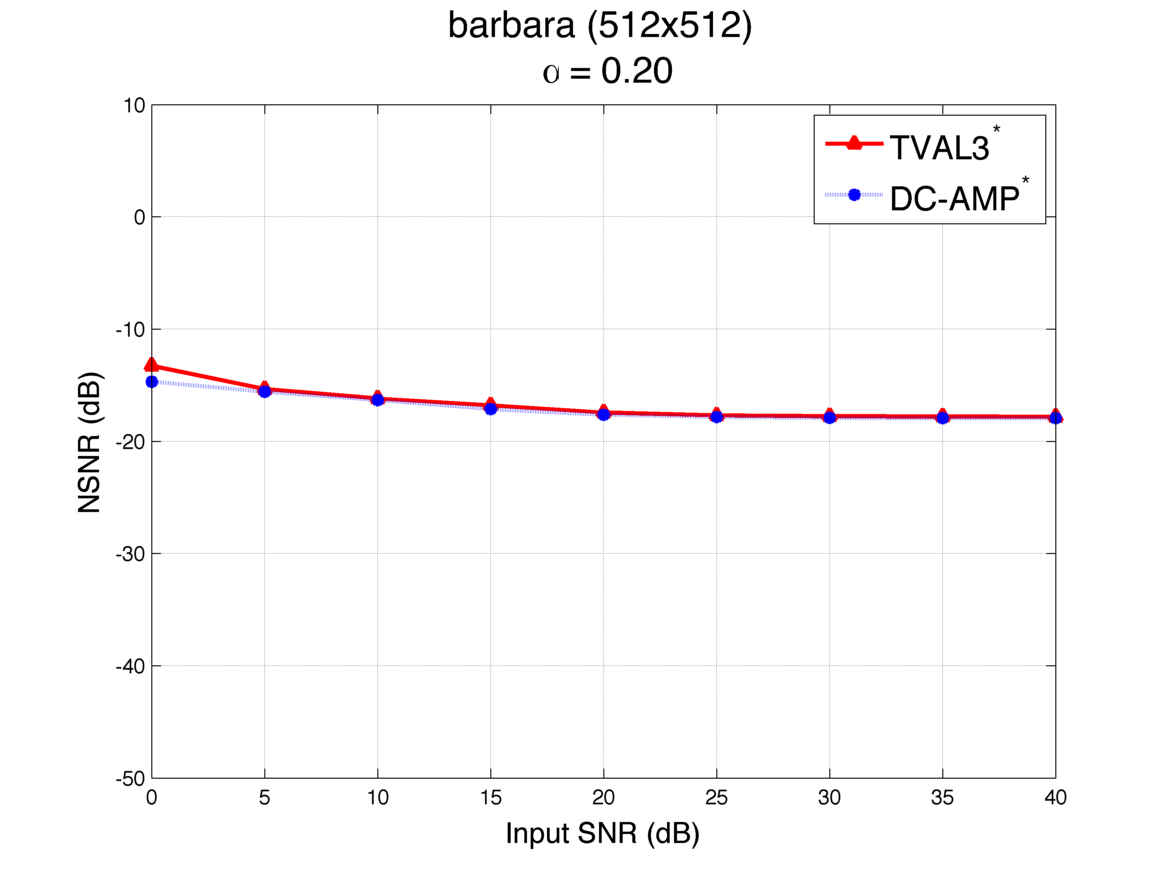}\quad
		\includegraphics[width=0.45\textwidth, trim=35 30 45 25, clip=true]{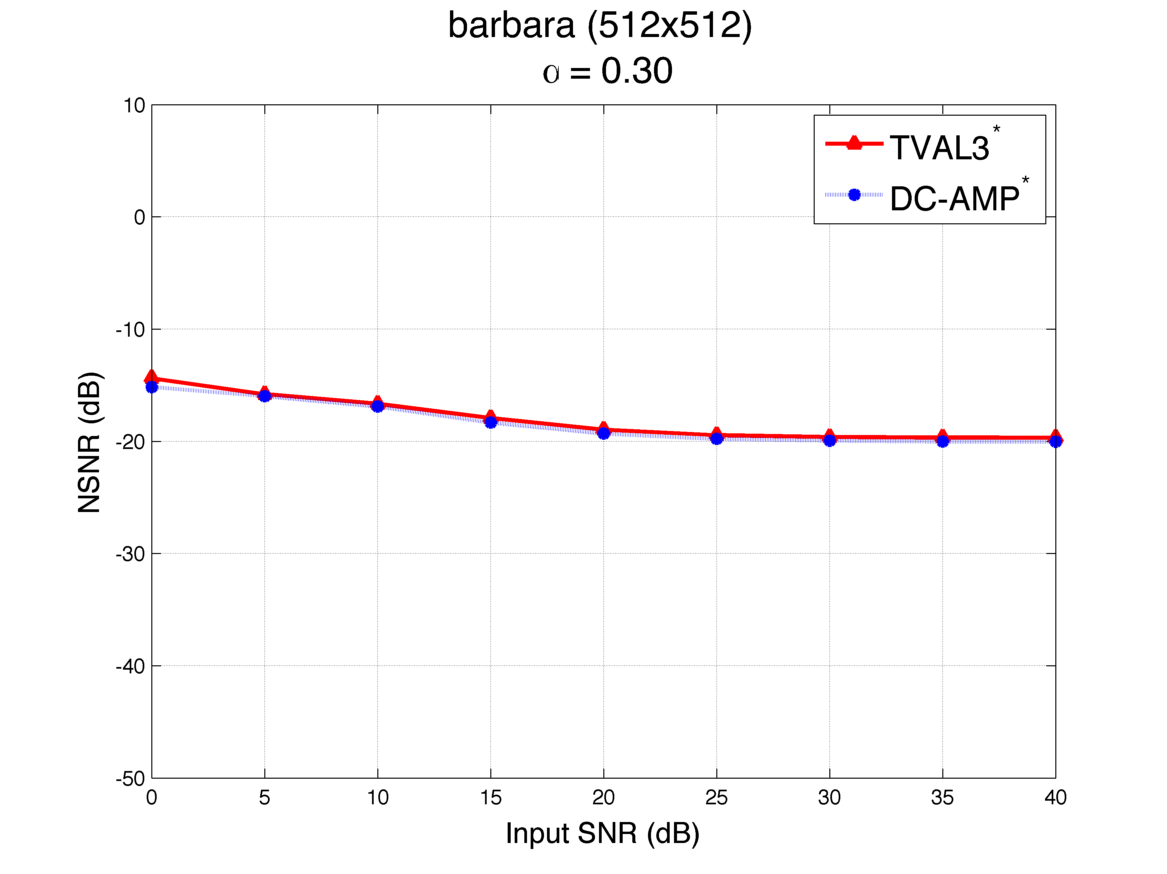}\\
		\includegraphics[width=0.45\textwidth, trim=35 10 45 25, clip=true]{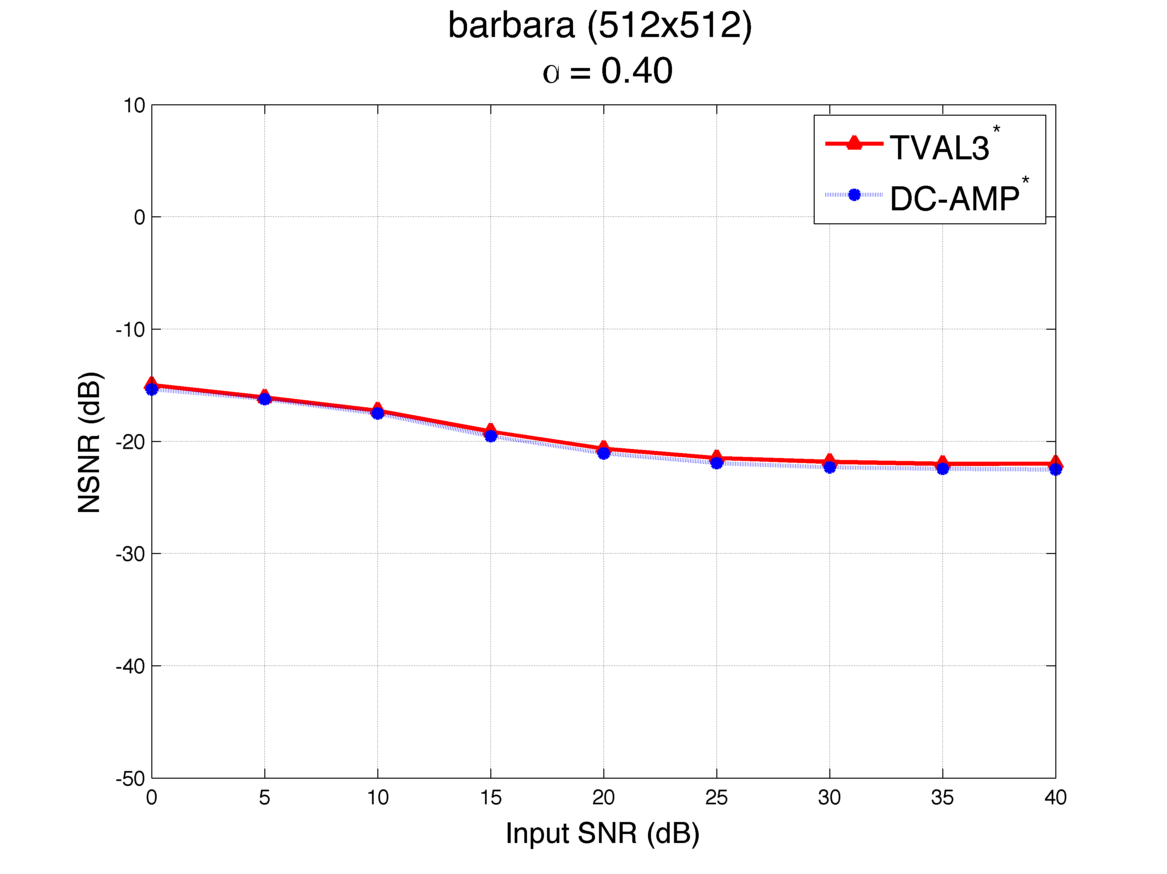}\quad
		\includegraphics[width=0.45\textwidth, trim=35 10 45 25, clip=true]{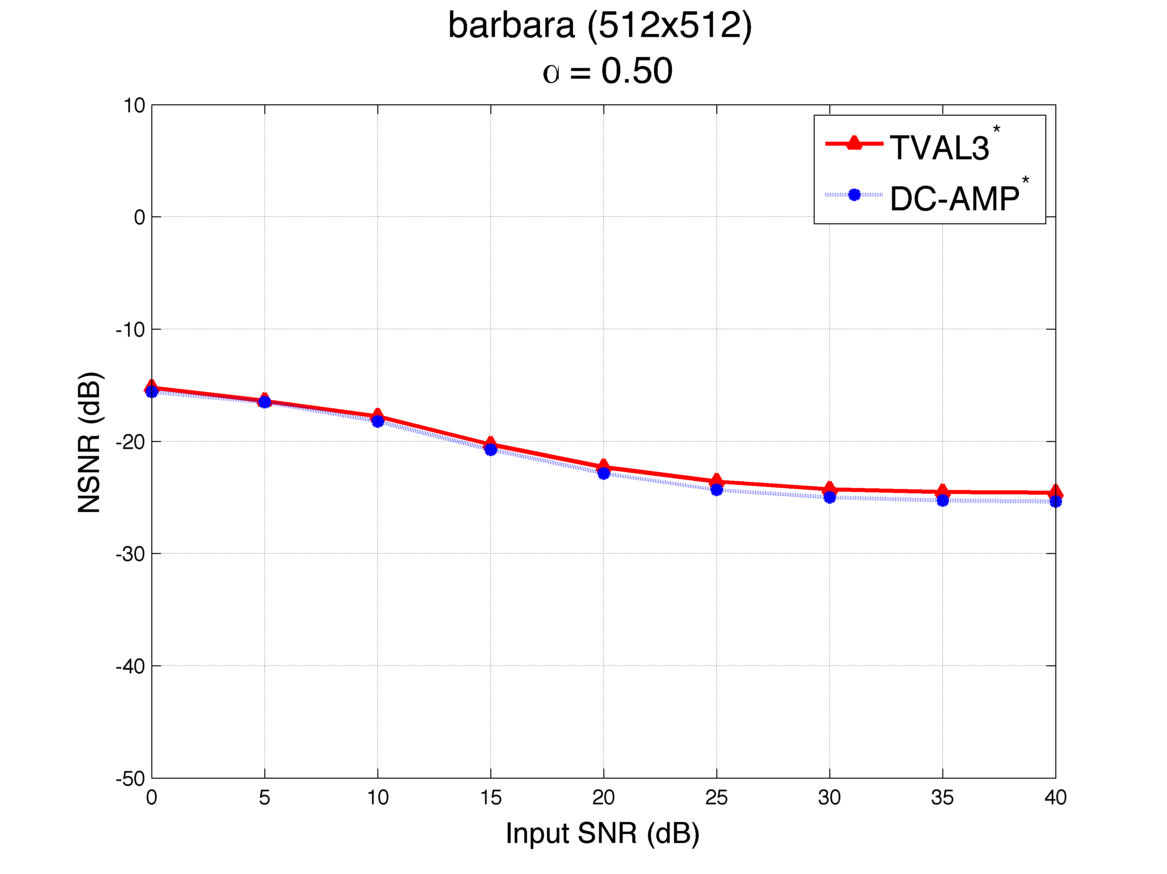}
	\caption[Comparison of the final reconstruction results for Barbara]{Comparison of the final reconstruction results in NSNR as a function of the noise level in dB between TV-AL3 and DC-AMP for different measurement rates $\alpha$ for the Barbara picture.}
\end{figure}
\begin{table}
\centering
	\begin{tabular}{ | l || c | c | c | c | c | c |}
	\cline{2-7}
	\multicolumn{1}{l|}{~} & $\alpha = 0.05$ & $\alpha = 0.10$ & $\alpha = 0.20$ & $\alpha = 0.30$ & $\alpha = 0.40$ & $\alpha = 0.50$\\	
	\cline{2-7}		
	\multicolumn{1}{l|}{~} & \multicolumn{6}{c|}{ISNR = $\infty$~dB}\\
	\hline
		TVAL3 Optimal & \textbf{-15.31} & \textbf{-16.27} & -17.80 & -19.70 & -22.00 & -24.63 \\ 
DC-AMP Optimal & -15.25 & -16.22 & \textbf{-17.94} & \textbf{-20.05} & \textbf{-22.54} & \textbf{-25.41} \\ 

	\hline
	\multicolumn{1}{l|}{~} & \multicolumn{6}{c|}{ISNR = 40~dB}\\
	\hline
		TVAL3 Optimal & \textbf{-15.38} & \textbf{-16.26} & -17.81 & -19.69 & -21.99 & -24.59 \\ 
DC-AMP Optimal & -15.25 & -16.22 & \textbf{-17.94} & \textbf{-20.03} & \textbf{-22.51} & \textbf{-25.37} \\ 

	\hline
	\multicolumn{1}{l|}{~} & \multicolumn{6}{c|}{ISNR = 30~dB}\\
	\hline
		TVAL3 Optimal & \textbf{-15.28} & \textbf{-16.26} & -17.77 & -19.62 & -21.83 & -24.28 \\ 
DC-AMP Optimal & -15.23 & -16.21 & \textbf{-17.91} & \textbf{-19.95} & \textbf{-22.33} & \textbf{-25.00} \\ 

	\hline
	\multicolumn{1}{l|}{~} & \multicolumn{6}{c|}{ISNR = 25~dB}\\
	\hline
		TVAL3 Optimal & \textbf{-15.35} & \textbf{-16.24} & -17.69 & -19.45 & -21.49 & -23.58 \\ 
DC-AMP Optimal & -15.22 & -16.19 & \textbf{-17.84} & \textbf{-19.78} & \textbf{-21.95} & \textbf{-24.32} \\ 

	\hline
	\multicolumn{1}{l|}{~} & \multicolumn{6}{c|}{ISNR = 20~dB}\\
	\hline
		TVAL3 Optimal & \textbf{-15.23} & \textbf{-16.16} & -17.44 & -18.97 & -20.67 & -22.31 \\ 
DC-AMP Optimal & -15.14 & -16.11 & \textbf{-17.63} & \textbf{-19.31} & \textbf{-21.07} & \textbf{-22.86} \\ 

	\hline
	\multicolumn{1}{l|}{~} & \multicolumn{6}{c|}{ISNR = 15~dB}\\
	\hline
		TVAL3 Optimal & \textbf{-15.09} & \textbf{-15.91} & -16.79 & -17.93 & -19.13 & -20.29 \\ 
DC-AMP Optimal & -14.94 & -15.88 & \textbf{-17.13} & \textbf{-18.33} & \textbf{-19.55} & \textbf{-20.75} \\ 

	\hline
	\multicolumn{1}{l|}{~} & \multicolumn{6}{c|}{ISNR = 10~dB}\\
	\hline
		TVAL3 Optimal & -14.36 & -15.34 & -16.20 & -16.66 & -17.27 & -17.79 \\ 
DC-AMP Optimal & \textbf{-14.46} & \textbf{-15.42} & \textbf{-16.32} & \textbf{-16.91} & \textbf{-17.53} & \textbf{-18.23} \\ 

	\hline
	\multicolumn{1}{l|}{~} & \multicolumn{6}{c|}{ISNR = 5~dB}\\
	\hline
		TVAL3 Optimal & -13.38 & -14.45 & -15.34 & -15.81 & -16.10 & -16.40 \\ 
DC-AMP Optimal & \textbf{-13.80} & \textbf{-14.80} & \textbf{-15.58} & \textbf{-15.98} & \textbf{-16.25} & \textbf{-16.52} \\ 

	\hline
	\multicolumn{1}{l|}{~} & \multicolumn{6}{c|}{ISNR = 0~dB}\\
	\hline
		TVAL3 Optimal & -11.98 & -12.90 & -13.28 & -14.39 & -14.98 & -15.23 \\ 
DC-AMP Optimal & \textbf{-12.45} & \textbf{-13.77} & \textbf{-14.69} & \textbf{-15.16} & \textbf{-15.37} & \textbf{-15.59} \\ 

	\hline
	\end{tabular}
	\caption[Table of the final reconstruction results for Barbara]{Table of the final reconstruction results in NSNR as a function of the noise level in dB between TV-AL3 and DC-AMP for different measurement rates $\alpha$ for the Barbara picture.}
\end{table}
\clearpage
\subsubsection{\emph{baboon} Image}
\begin{figure}[ht!]
	\centering
		\includegraphics[width=0.35\textwidth, trim=0 17 0 0, clip=true]{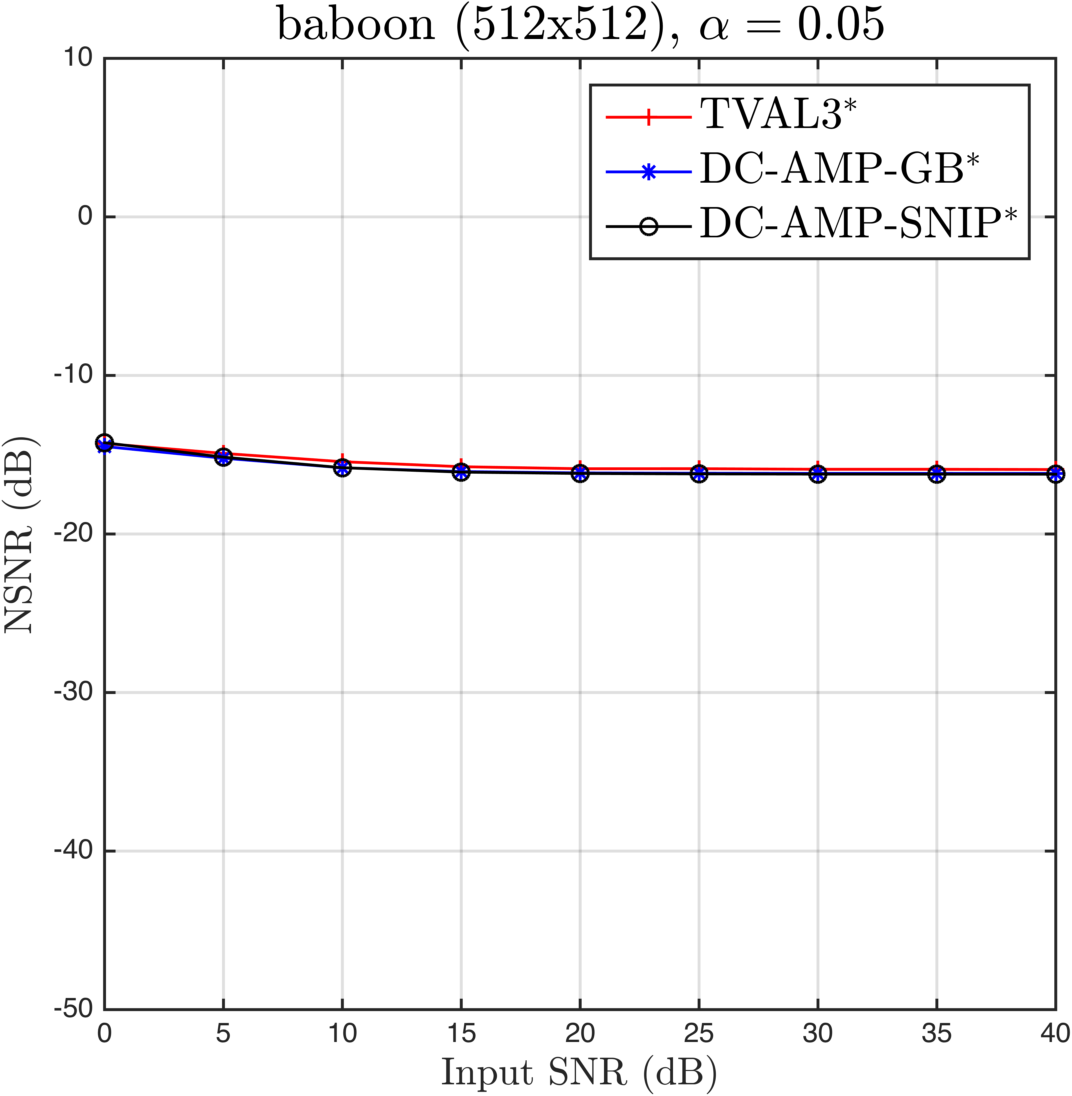}\quad
		\hspace{1.4cm}\includegraphics[width=0.45\textwidth, trim=35 30 45 0, clip=true]{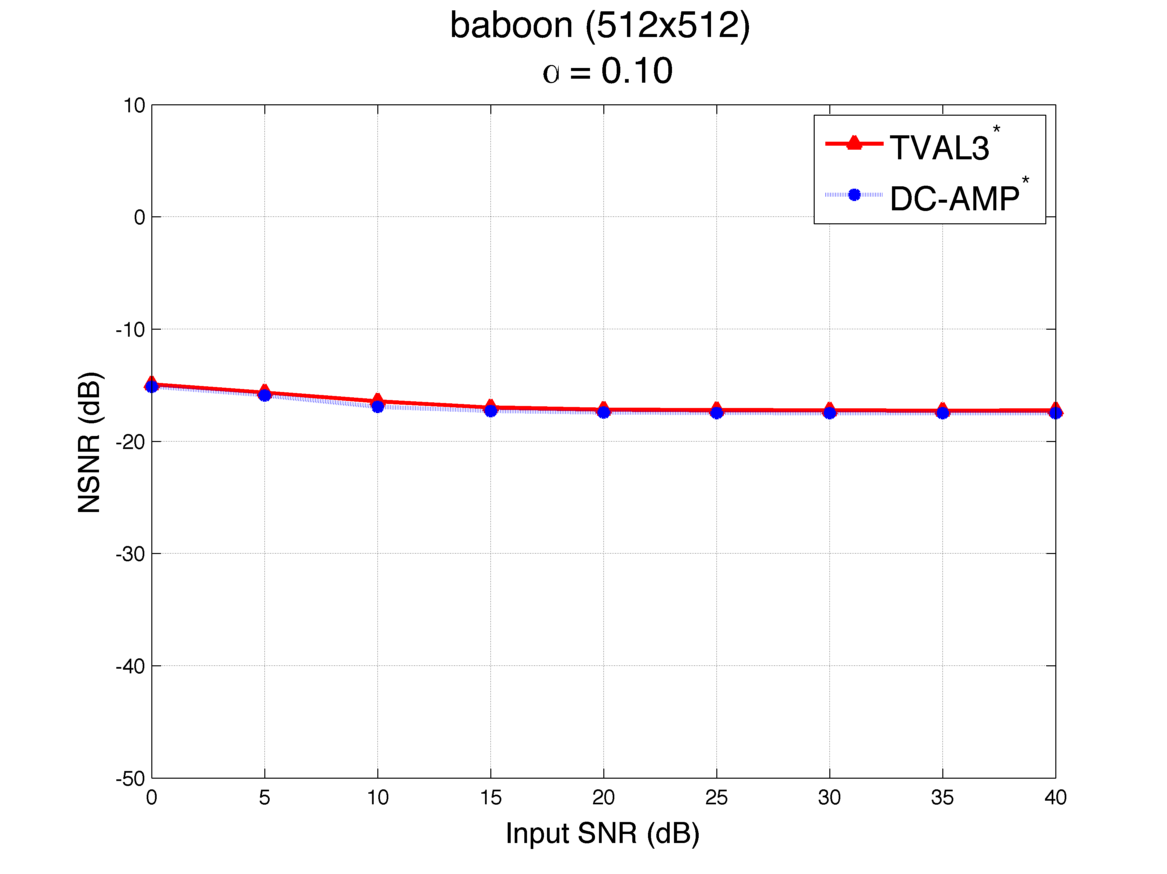}\\
		\includegraphics[width=0.45\textwidth, trim=35 30 45 25, clip=true]{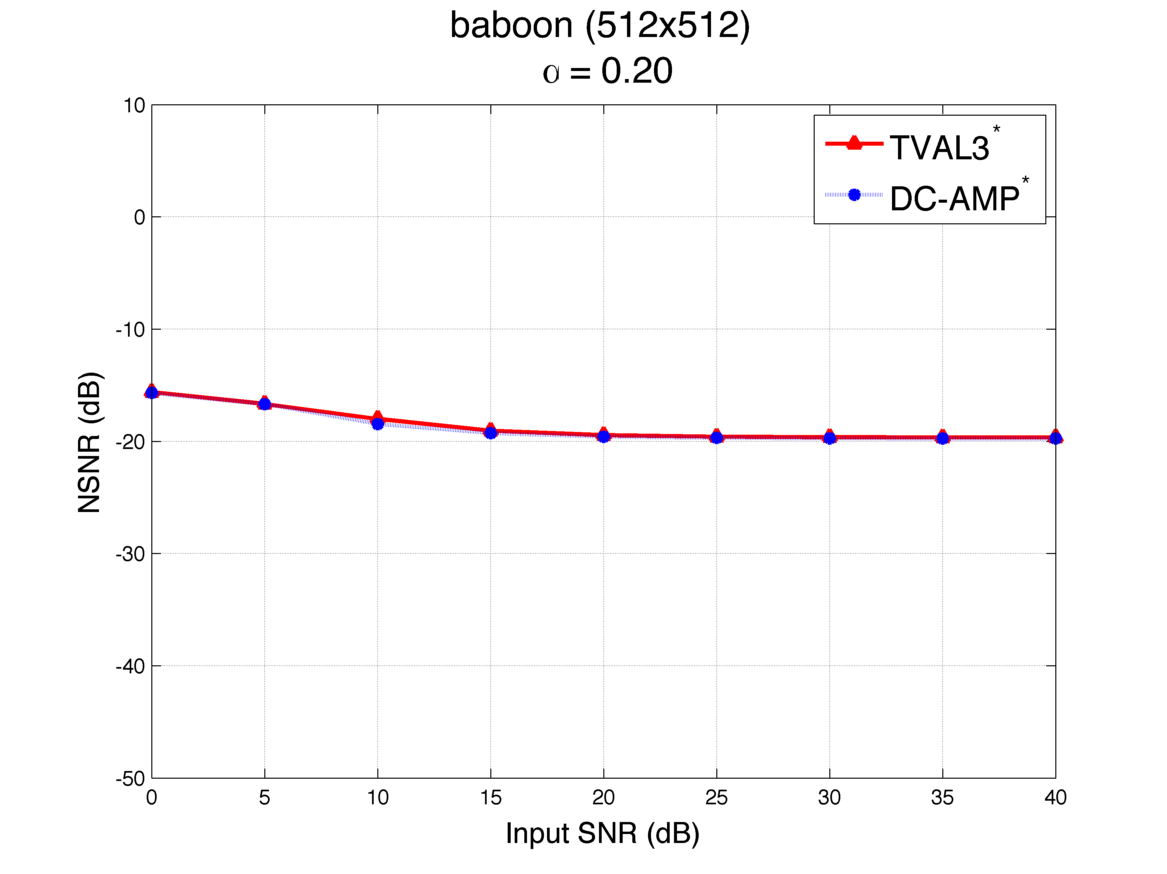}\quad
		\includegraphics[width=0.45\textwidth, trim=35 30 45 25, clip=true]{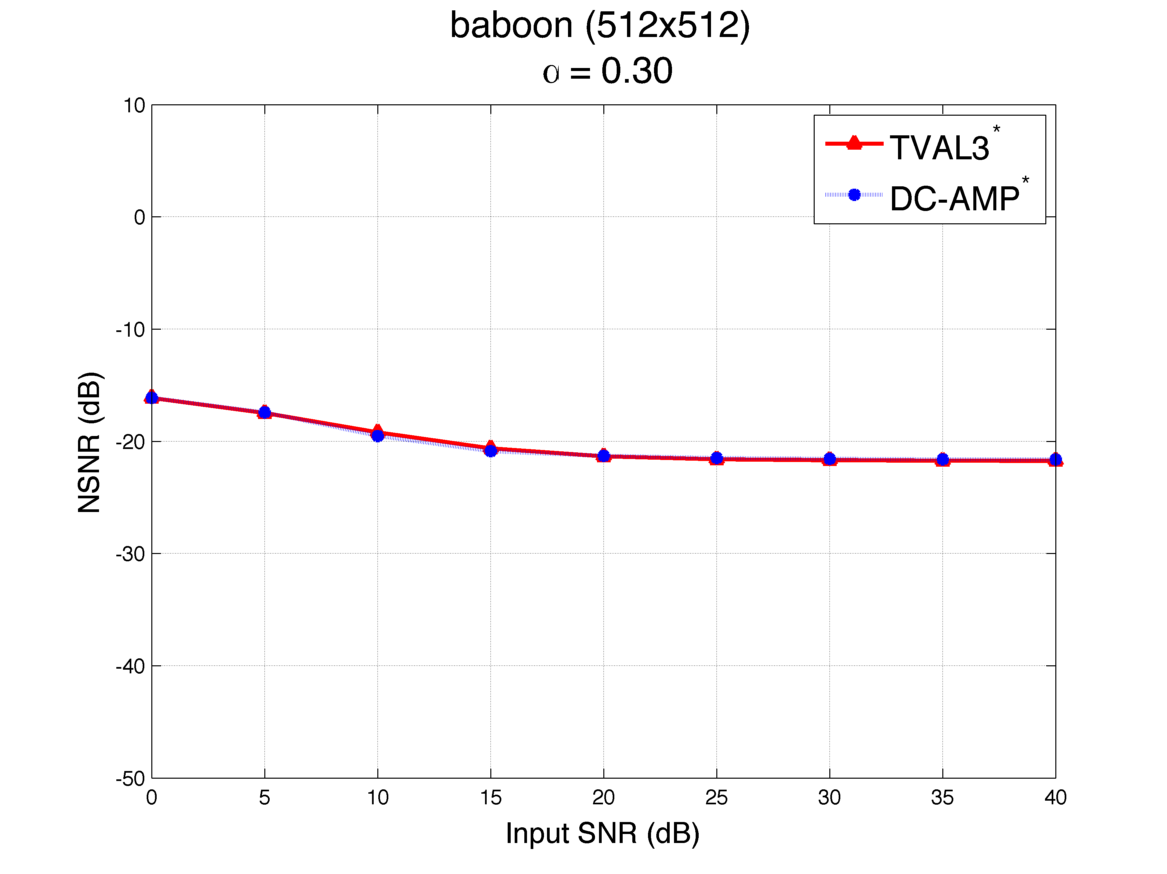}\\
		\includegraphics[width=0.45\textwidth, trim=35 10 45 25, clip=true]{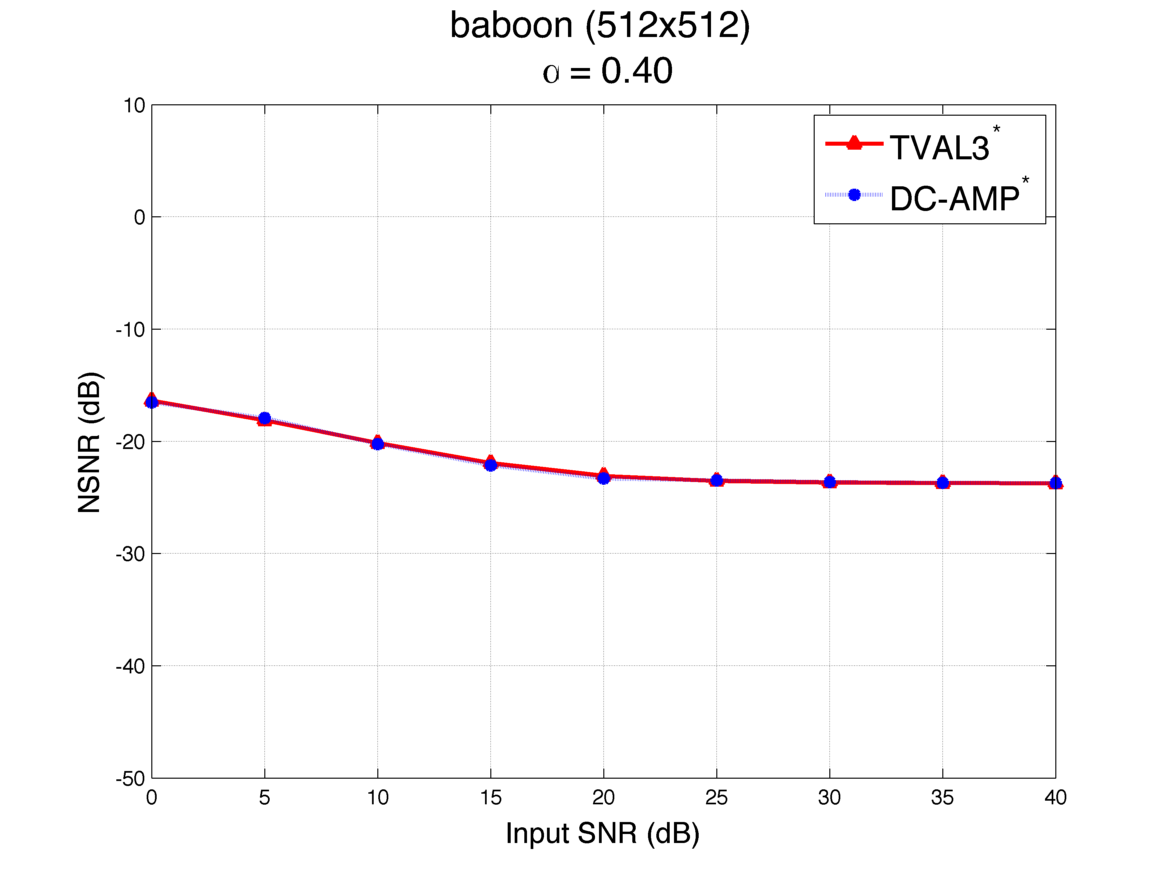}\quad
		\includegraphics[width=0.45\textwidth, trim=35 10 45 25, clip=true]{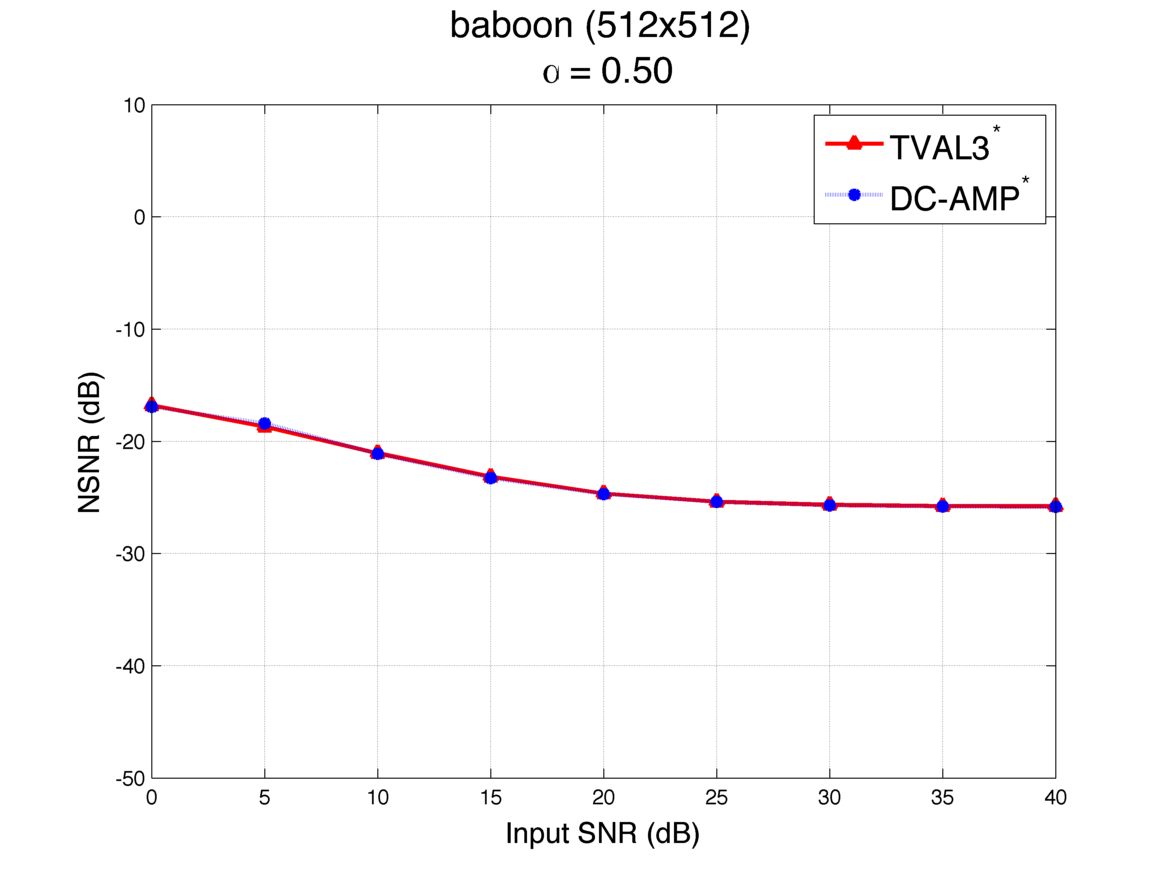}
		\caption[Comparison of the final reconstruction results for Baboon]{Comparison of the final reconstruction results in NSNR as a function of the noise level in dB between TV-AL3 and DC-AMP for different measurement rates $\alpha$ for the Baboon picture.}
\end{figure}
\begin{table}
\centering
	\begin{tabular}{ | l || c | c | c | c | c | c |}
	\cline{2-7}
	\multicolumn{1}{l|}{~} & $\alpha = 0.05$ & $\alpha = 0.10$ & $\alpha = 0.20$ & $\alpha = 0.30$ & $\alpha = 0.40$ & $\alpha = 0.50$\\	
	\cline{2-7}		
	\multicolumn{1}{l|}{~} & \multicolumn{6}{c|}{ISNR = $\infty$~dB}\\
	\hline
		TVAL3 Optimal & -15.94 & -17.28 & -19.65 & \textbf{-21.74} & \textbf{-23.75} & -25.79 \\ 
DC-AMP Optimal & \textbf{-16.18} & \textbf{-17.50} & \textbf{-19.76} & -21.63 & -23.74 & \textbf{-25.85} \\ 

	\hline
	\multicolumn{1}{l|}{~} & \multicolumn{6}{c|}{ISNR = 40~dB}\\
	\hline
		TVAL3 Optimal & -15.93 & -17.26 & -19.65 & \textbf{-21.75} & \textbf{-23.74} & -25.78 \\ 
DC-AMP Optimal & \textbf{-16.18} & \textbf{-17.50} & \textbf{-19.75} & -21.63 & -23.71 & \textbf{-25.84} \\ 

	\hline
	\multicolumn{1}{l|}{~} & \multicolumn{6}{c|}{ISNR = 30~dB}\\
	\hline
		TVAL3 Optimal & -15.92 & -17.24 & -19.64 & \textbf{-21.69} & \textbf{-23.67} & -25.67 \\ 
DC-AMP Optimal & \textbf{-16.18} & \textbf{-17.49} & \textbf{-19.74} & -21.57 & -23.64 & \textbf{-25.71} \\ 

	\hline
	\multicolumn{1}{l|}{~} & \multicolumn{6}{c|}{ISNR = 25~dB}\\
	\hline
		TVAL3 Optimal & -15.88 & -17.24 & -19.58 & \textbf{-21.60} & \textbf{-23.53} & -25.38 \\ 
DC-AMP Optimal & \textbf{-16.17} & \textbf{-17.48} & \textbf{-19.70} & -21.49 & -23.48 & \textbf{-25.39} \\ 

	\hline
	\multicolumn{1}{l|}{~} & \multicolumn{6}{c|}{ISNR = 20~dB}\\
	\hline
		TVAL3 Optimal & -15.88 & -17.18 & -19.45 & \textbf{-21.34} & -23.08 & -24.65 \\ 
DC-AMP Optimal & \textbf{-16.15} & \textbf{-17.44} & \textbf{-19.60} & -21.30 & \textbf{-23.32} & \textbf{-24.70} \\ 

	\hline
	\multicolumn{1}{l|}{~} & \multicolumn{6}{c|}{ISNR = 15~dB}\\
	\hline
		TVAL3 Optimal & -15.76 & -16.99 & -19.05 & -20.63 & -21.95 & -23.15 \\ 
DC-AMP Optimal & \textbf{-16.06} & \textbf{-17.30} & \textbf{-19.28} & \textbf{-20.89} & \textbf{-22.15} & \textbf{-23.27} \\ 

	\hline
	\multicolumn{1}{l|}{~} & \multicolumn{6}{c|}{ISNR = 10~dB}\\
	\hline
		TVAL3 Optimal & -15.44 & -16.44 & -18.00 & -19.21 & -20.15 & -21.06 \\ 
DC-AMP Optimal & \textbf{-15.82} & \textbf{-16.93} & \textbf{-18.46} & \textbf{-19.53} & \textbf{-20.24} & \textbf{-21.13} \\ 

	\hline
	\multicolumn{1}{l|}{~} & \multicolumn{6}{c|}{ISNR = 5~dB}\\
	\hline
		TVAL3 Optimal & -14.92 & -15.66 & -16.66 & \textbf{-17.49} & \textbf{-18.14} & \textbf{-18.70} \\ 
DC-AMP Optimal & \textbf{-15.22} & \textbf{-15.93} & \textbf{-16.70} & -17.42 & -17.93 & -18.40 \\ 

	\hline
	\multicolumn{1}{l|}{~} & \multicolumn{6}{c|}{ISNR = 0~dB}\\
	\hline
		TVAL3 Optimal & -14.29 & -14.93 & -15.61 & -16.13 & -16.39 & -16.80 \\ 
DC-AMP Optimal & \textbf{-14.48} & \textbf{-15.14} & \textbf{-15.69} & \textbf{-16.13} & \textbf{-16.53} & \textbf{-16.93} \\ 

	\hline
	\end{tabular}
	\caption[Table of the final reconstruction results for Baboon]{Comparison of the final reconstruction results in NSNR as a function of the noise level in dB between TV-AL3 and DC-AMP for different measurement rates $\alpha$ for the Baboon picture.}
\end{table}
\clearpage
\subsubsection{\emph{cameraman} Image}
\begin{figure}[ht!]
	\centering
		\includegraphics[width=0.35\textwidth, trim=0 17 0 0, clip=true]{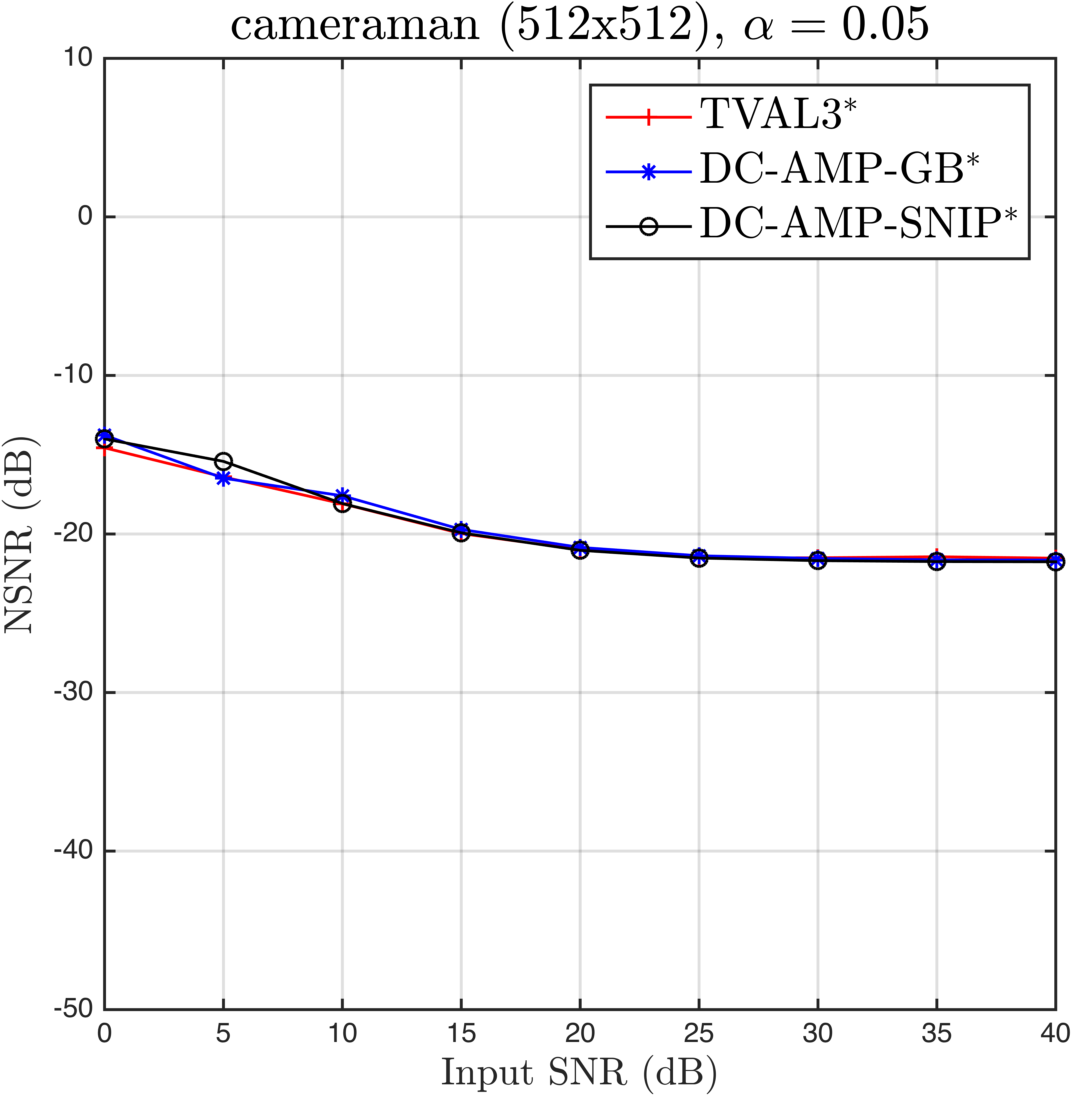}\quad
		\hspace{1.4cm}\includegraphics[width=0.45\textwidth, trim=35 30 45 0, clip=true]{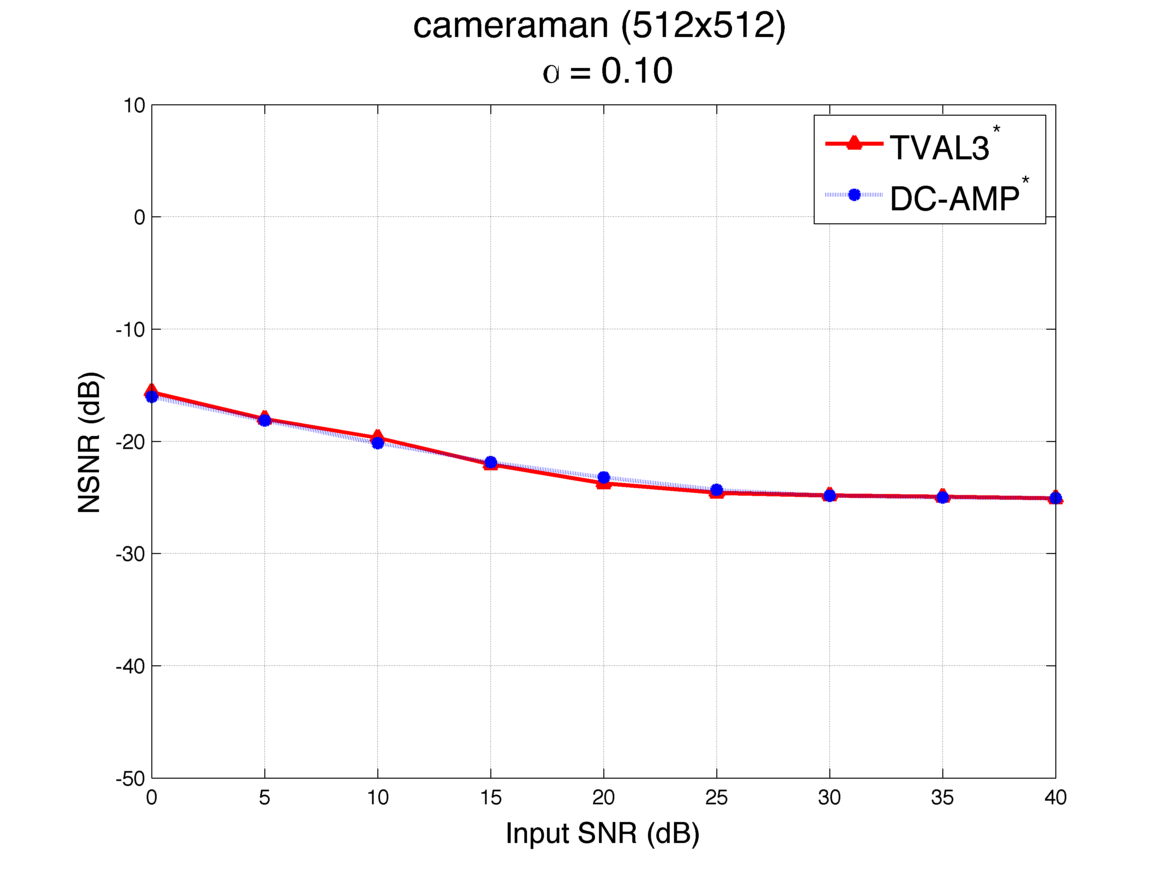}\\
		\includegraphics[width=0.45\textwidth, trim=35 30 45 25, clip=true]{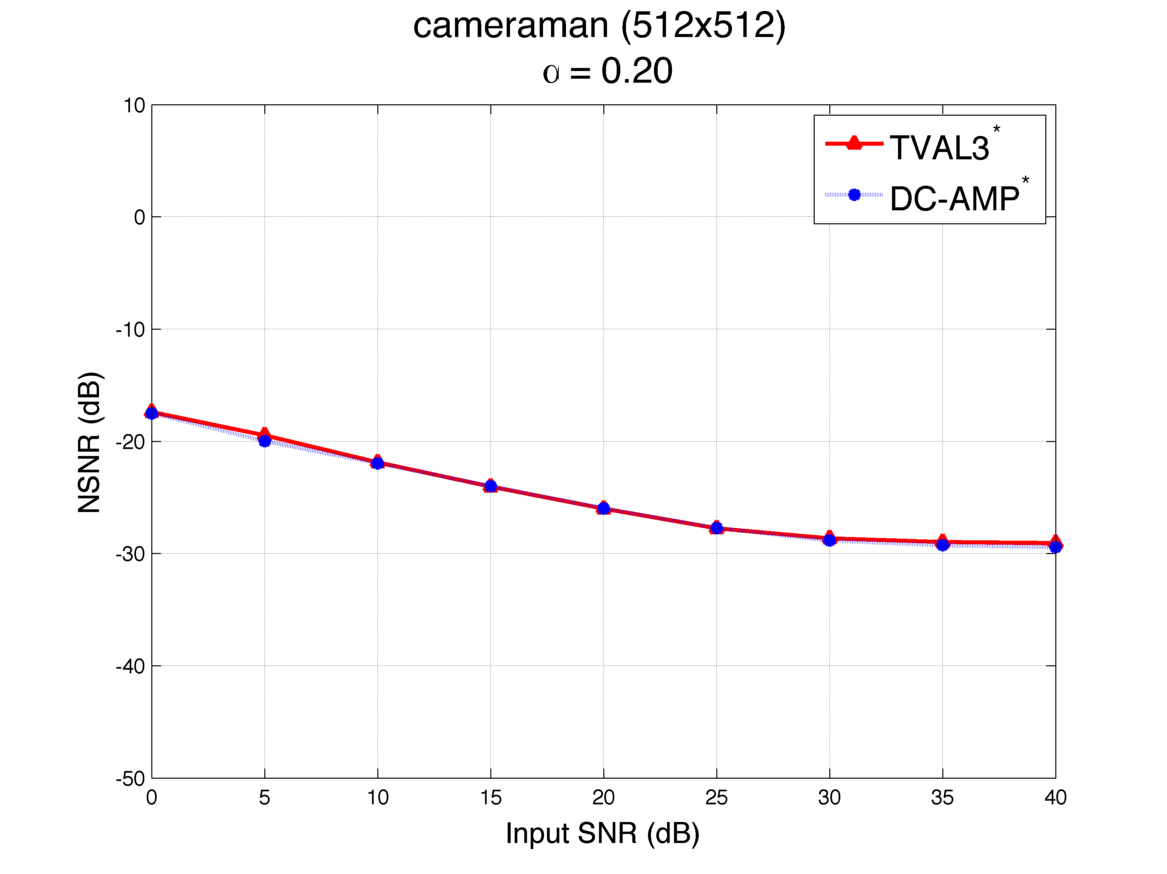}\quad
		\includegraphics[width=0.45\textwidth, trim=35 30 45 25, clip=true]{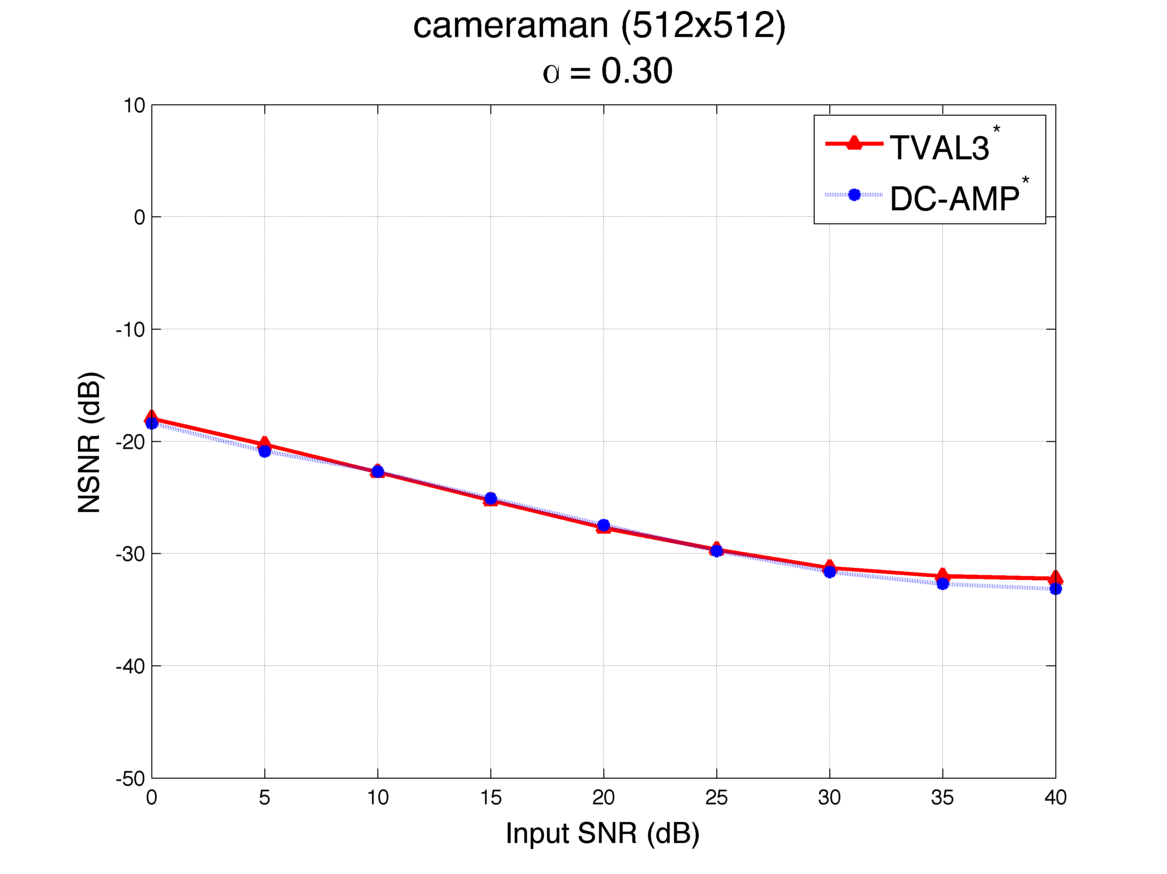}\\
		\includegraphics[width=0.45\textwidth, trim=35 10 45 25, clip=true]{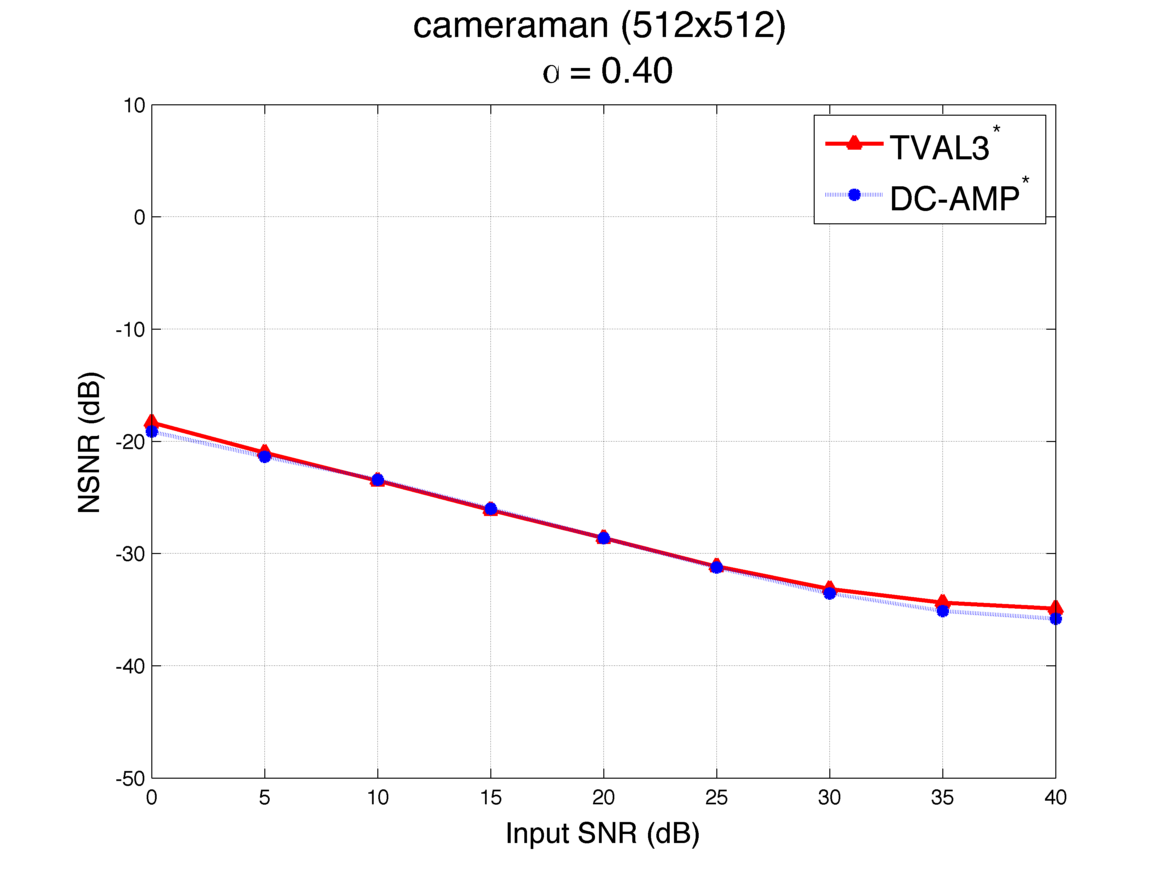}\quad
		\includegraphics[width=0.45\textwidth, trim=35 10 45 25, clip=true]{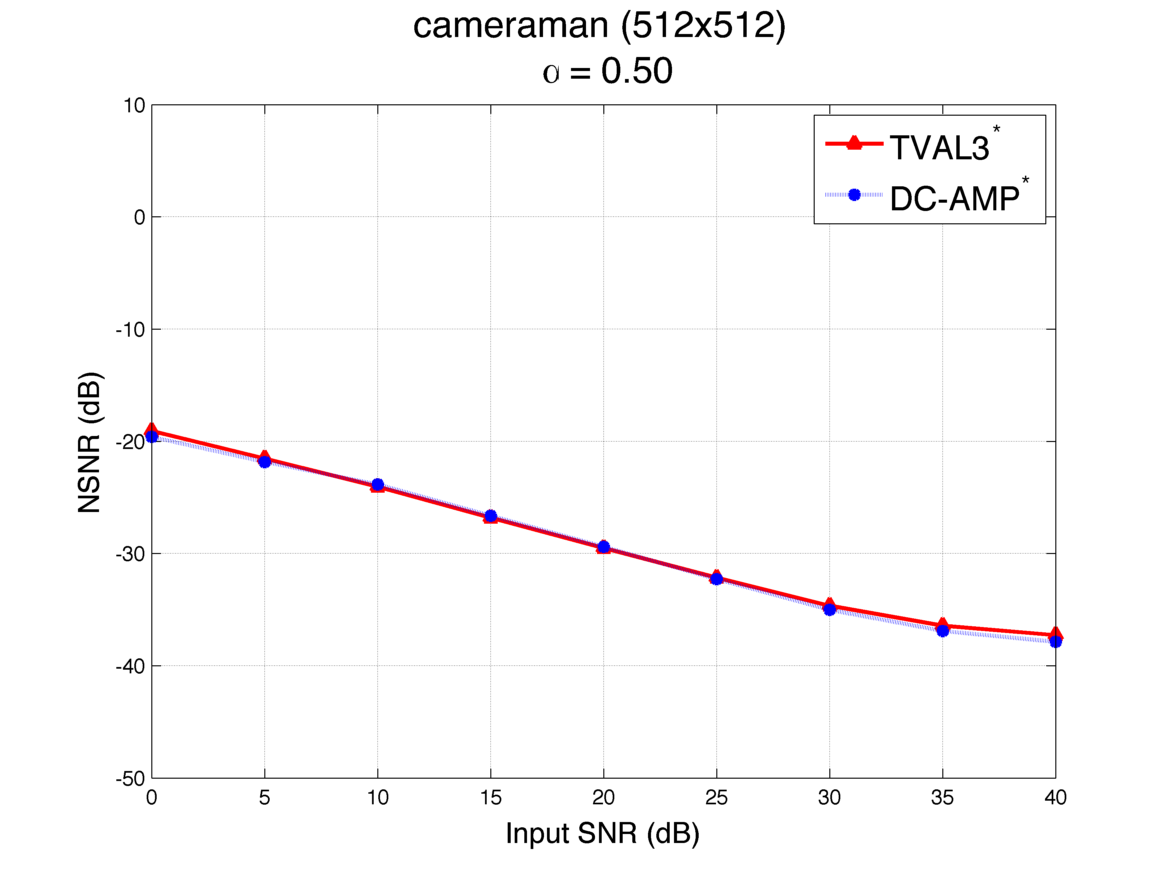}
		\caption[Comparison of the final reconstruction results for Cameraman]{Comparison of the final reconstruction results in NSNR as a function of the noise level in dB between TV-AL3 and DC-AMP for different measurement rates $\alpha$ for the Cameraman picture.}
\end{figure}
%
\begin{table}
\centering
	\begin{tabular}{ | l || c | c | c | c | c | c |}
	\cline{2-7}
	\multicolumn{1}{l|}{~} & $\alpha = 0.05$ & $\alpha = 0.10$ & $\alpha = 0.20$ & $\alpha = 0.30$ & $\alpha = 0.40$ & $\alpha = 0.50$\\	
	\cline{2-7}		
	\multicolumn{1}{l|}{~} & \multicolumn{6}{c|}{ISNR = $\infty$~dB}\\
	\hline
		TVAL3 Optimal & -21.43 & -25.02 & -29.11 & -32.42 & -35.12 & -37.67 \\ 
DC-AMP Optimal & \textbf{-21.65} & \textbf{-25.10} & \textbf{-29.50} & \textbf{-33.35} & \textbf{-36.18} & \textbf{-38.50} \\ 

	\hline
	\multicolumn{1}{l|}{~} & \multicolumn{6}{c|}{ISNR = 40~dB}\\
	\hline
		TVAL3 Optimal & -21.52 & \textbf{-25.09} & -29.09 & -32.24 & -34.91 & -37.27 \\ 
DC-AMP Optimal & \textbf{-21.64} & -25.07 & \textbf{-29.44} & \textbf{-33.16} & \textbf{-35.81} & \textbf{-37.88} \\ 

	\hline
	\multicolumn{1}{l|}{~} & \multicolumn{6}{c|}{ISNR = 30~dB}\\
	\hline
		TVAL3 Optimal & -21.50 & -24.83 & -28.66 & -31.28 & -33.17 & -34.67 \\ 
DC-AMP Optimal & \textbf{-21.54} & \textbf{-24.85} & \textbf{-28.84} & \textbf{-31.65} & \textbf{-33.57} & \textbf{-35.03} \\ 

	\hline
	\multicolumn{1}{l|}{~} & \multicolumn{6}{c|}{ISNR = 25~dB}\\
	\hline
		TVAL3 Optimal & \textbf{-21.51} & \textbf{-24.60} & \textbf{-27.77} & -29.66 & -31.15 & -32.15 \\ 
DC-AMP Optimal & -21.37 & -24.33 & -27.73 & \textbf{-29.78} & \textbf{-31.25} & \textbf{-32.28} \\ 

	\hline
	\multicolumn{1}{l|}{~} & \multicolumn{6}{c|}{ISNR = 20~dB}\\
	\hline
		TVAL3 Optimal & \textbf{-20.99} & \textbf{-23.75} & \textbf{-26.02} & \textbf{-27.72} & -28.62 & \textbf{-29.52} \\ 
DC-AMP Optimal & -20.85 & -23.21 & -25.98 & -27.47 & \textbf{-28.62} & -29.41 \\ 

	\hline
	\multicolumn{1}{l|}{~} & \multicolumn{6}{c|}{ISNR = 15~dB}\\
	\hline
		TVAL3 Optimal & \textbf{-19.97} & \textbf{-22.06} & \textbf{-24.05} & \textbf{-25.28} & \textbf{-26.14} & \textbf{-26.82} \\ 
DC-AMP Optimal & -19.71 & -21.87 & -24.00 & -25.08 & -26.01 & -26.64 \\ 

	\hline
	\multicolumn{1}{l|}{~} & \multicolumn{6}{c|}{ISNR = 10~dB}\\
	\hline
		TVAL3 Optimal & \textbf{-18.09} & -19.69 & -21.87 & \textbf{-22.75} & \textbf{-23.53} & \textbf{-24.04} \\ 
DC-AMP Optimal & -17.59 & \textbf{-20.17} & \textbf{-21.96} & -22.69 & -23.43 & -23.84 \\ 

	\hline
	\multicolumn{1}{l|}{~} & \multicolumn{6}{c|}{ISNR = 5~dB}\\
	\hline
		TVAL3 Optimal & -16.40 & -18.00 & -19.47 & -20.29 & -21.01 & -21.54 \\ 
DC-AMP Optimal & \textbf{-16.48} & \textbf{-18.15} & \textbf{-19.99} & \textbf{-20.90} & \textbf{-21.38} & \textbf{-21.85} \\ 

	\hline
	\multicolumn{1}{l|}{~} & \multicolumn{6}{c|}{ISNR = 0~dB}\\
	\hline
		TVAL3 Optimal & \textbf{-14.56} & -15.62 & -17.41 & -17.98 & -18.34 & -19.08 \\ 
DC-AMP Optimal & -13.77 & \textbf{-16.03} & \textbf{-17.51} & \textbf{-18.40} & \textbf{-19.13} & \textbf{-19.61} \\ 

	\hline
	\end{tabular}
	\caption[Table of the final reconstruction results for Cameraman]{Comparison of the final reconstruction results in NSNR as a function of the noise level in dB between TV-AL3 and DC-AMP for different measurement rates $\alpha$ for the Cameraman picture.}
\end{table}
\clearpage
\subsubsection{\emph{peppers} Image}
\begin{figure}[ht!]
	\centering
		\includegraphics[width=0.35\textwidth, trim=0 17 0 0, clip=true]{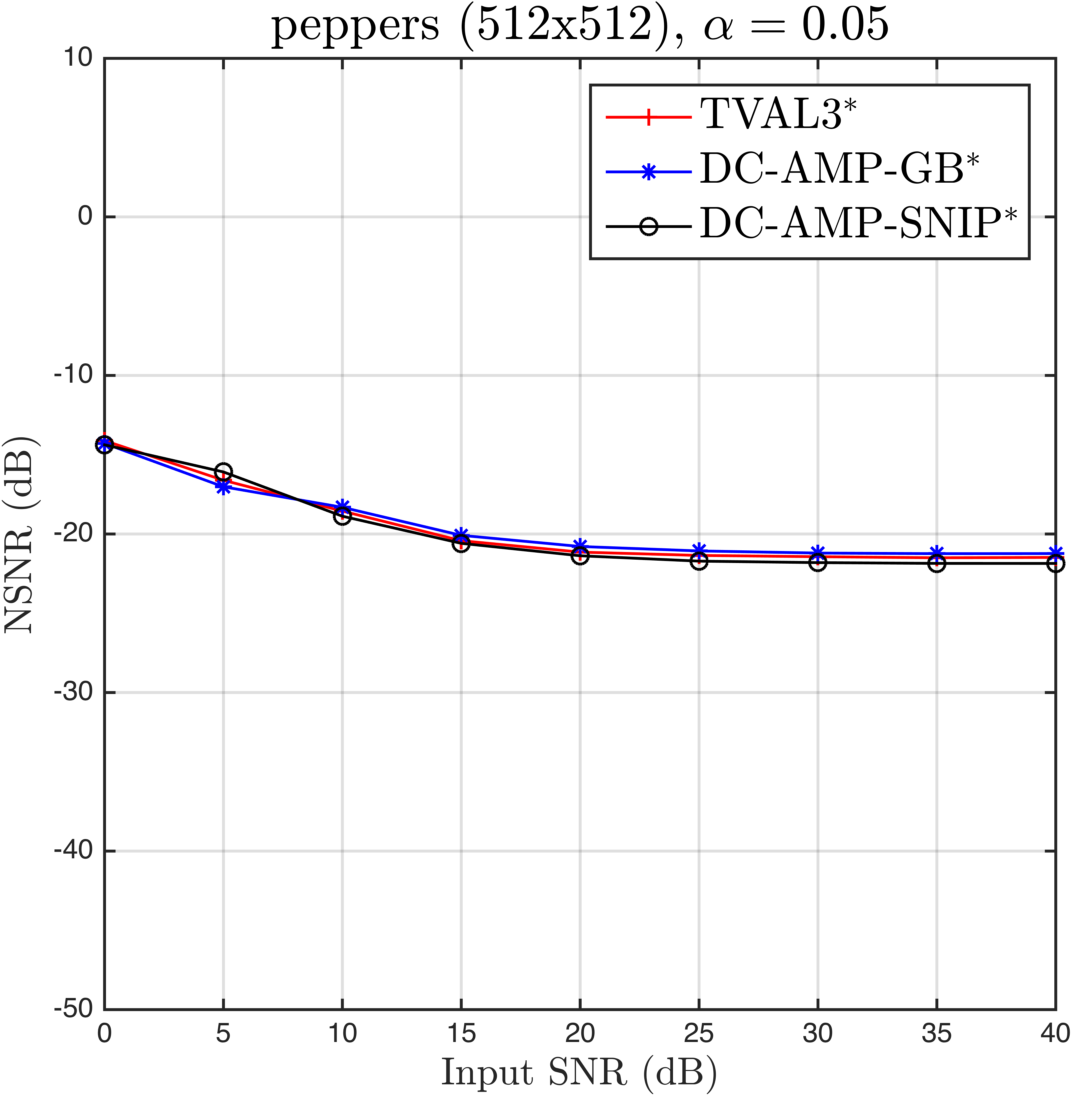}\quad
		\hspace{1.4cm}\includegraphics[width=0.45\textwidth, trim=35 30 45 0, clip=true]{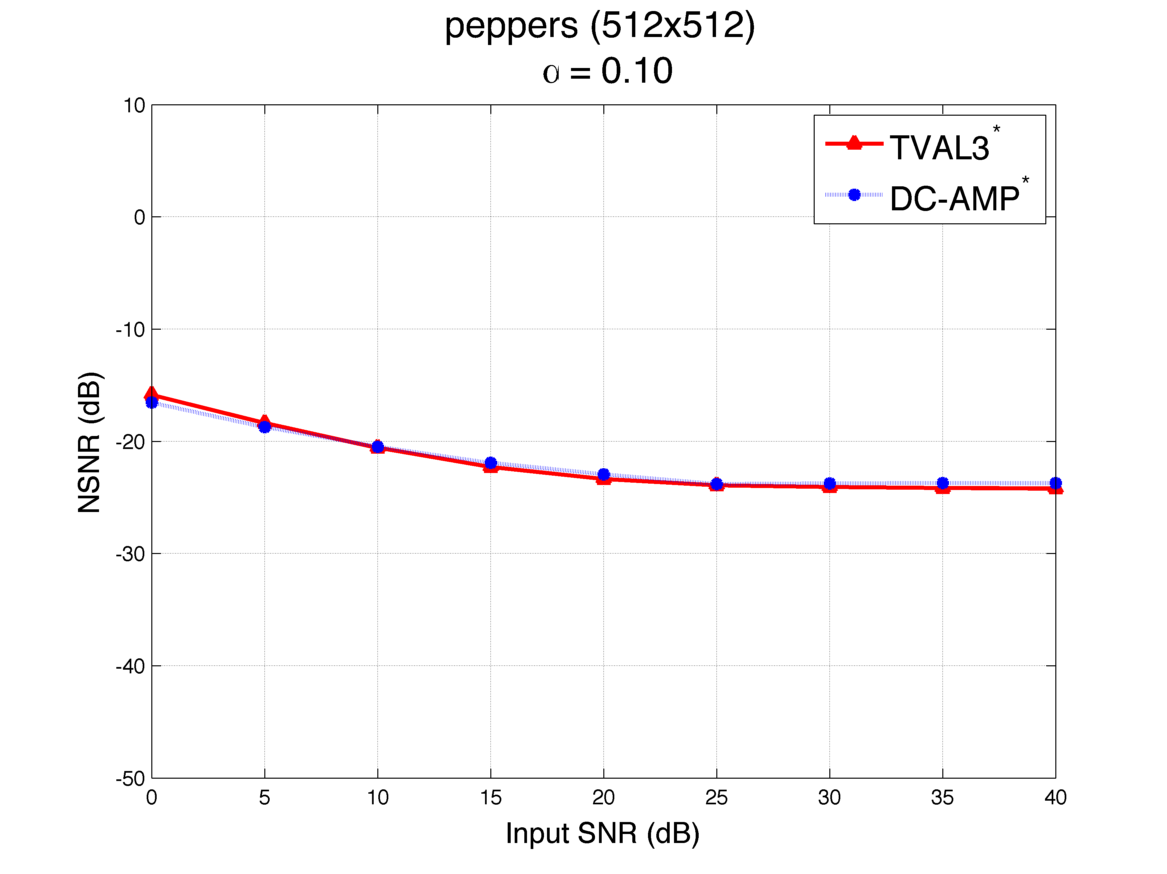}\\
		\includegraphics[width=0.45\textwidth, trim=35 30 45 25, clip=true]{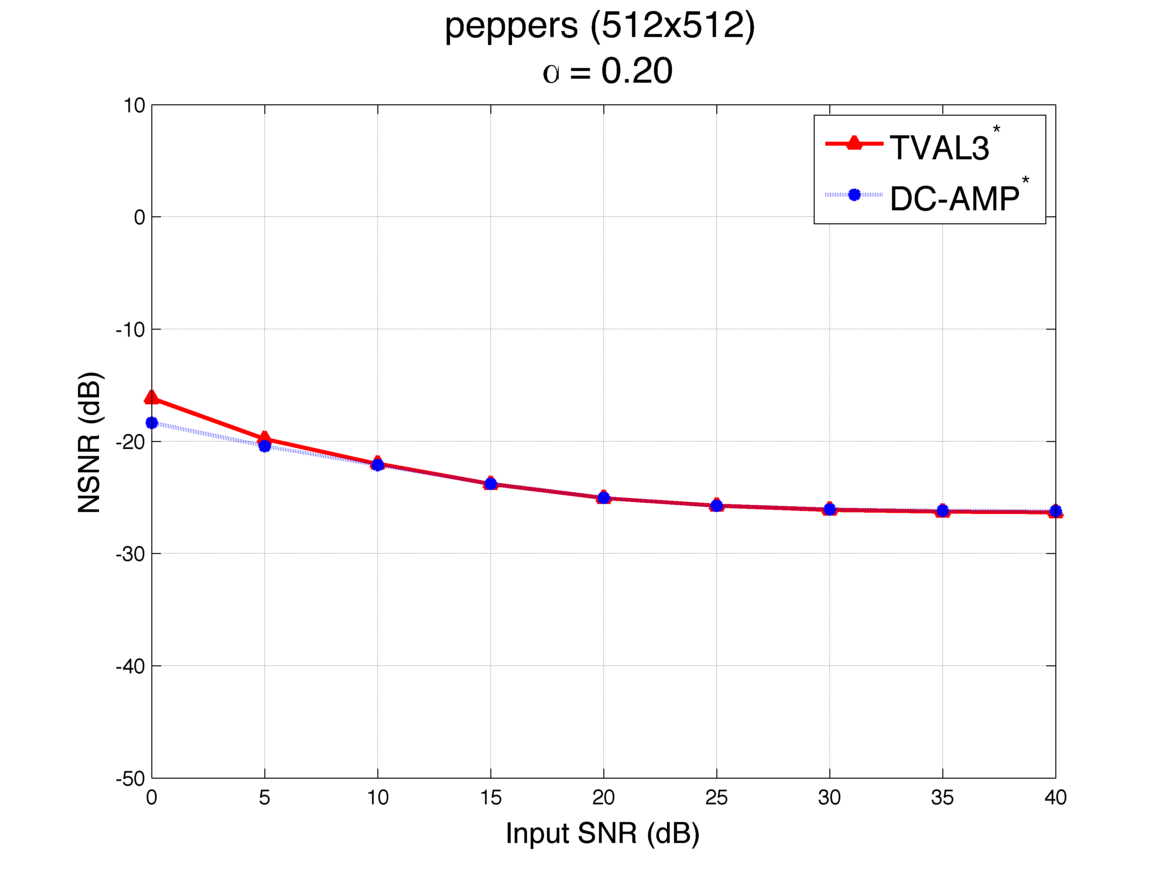}\quad
		\includegraphics[width=0.45\textwidth, trim=35 30 45 25, clip=true]{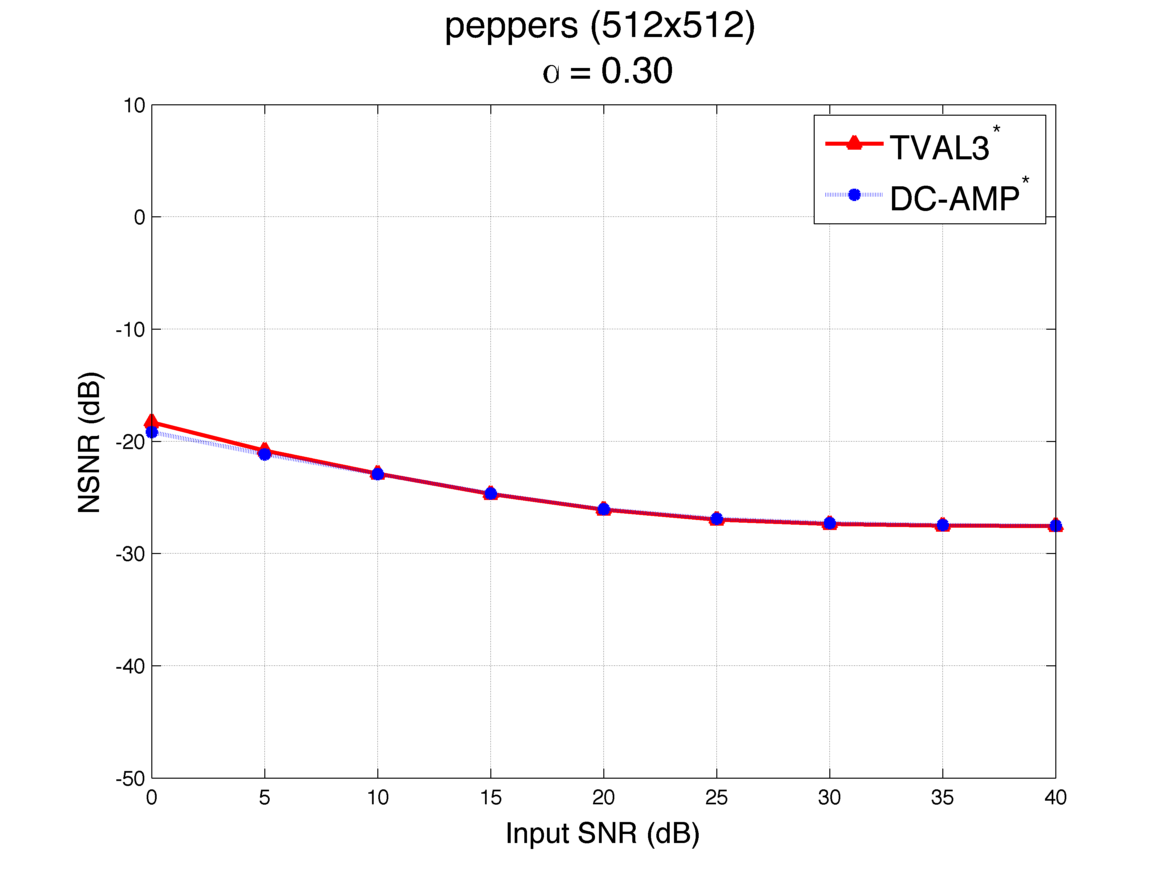}\\
		\includegraphics[width=0.45\textwidth, trim=35 10 45 25, clip=true]{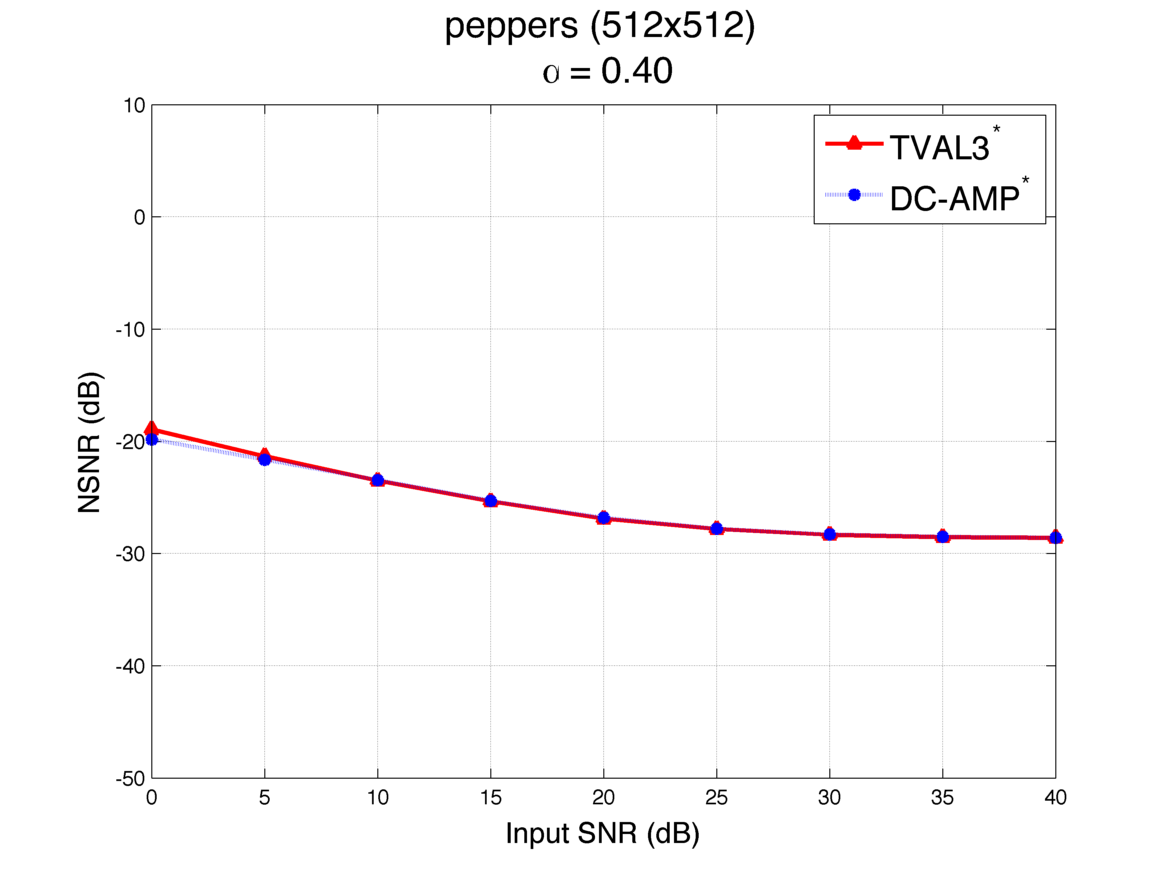}\quad
		\includegraphics[width=0.45\textwidth, trim=35 10 45 25, clip=true]{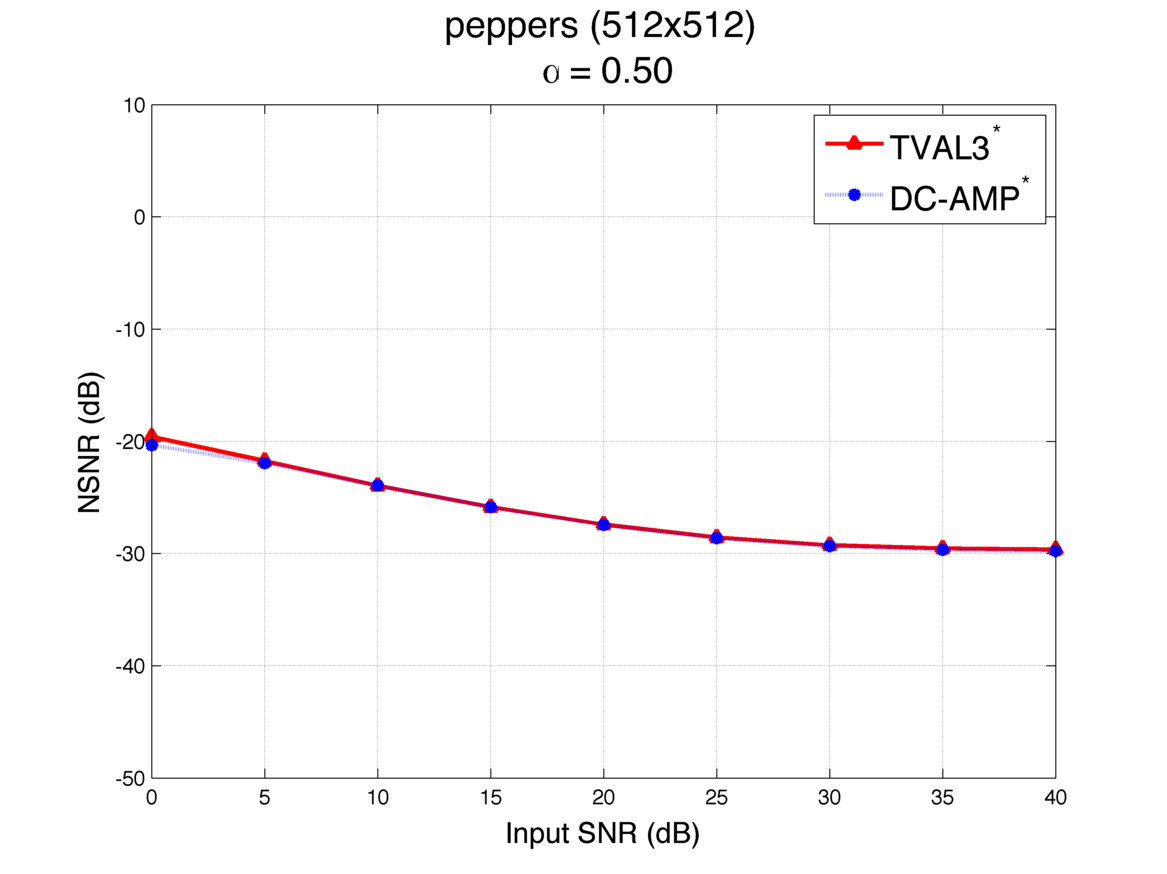}
		\caption[Comparison of the final reconstruction results for Peppers]{Comparison of the final reconstruction results in NSNR as a function of the noise level in dB between TV-AL3 and DC-AMP for different measurement rates $\alpha$ for the Peppers picture.}
\end{figure}
%
%
\begin{table}
\centering
	\begin{tabular}{ | l || c | c | c | c | c | c |}
	\cline{2-7}
	\multicolumn{1}{l|}{~} & $\alpha = 0.05$ & $\alpha = 0.10$ & $\alpha = 0.20$ & $\alpha = 0.30$ & $\alpha = 0.40$ & $\alpha = 0.50$\\	
	\cline{2-7}		
	\multicolumn{1}{l|}{~} & \multicolumn{6}{c|}{ISNR = $\infty$~dB}\\
	\hline
		TVAL3 Optimal & \textbf{-21.45} & \textbf{-24.22} & \textbf{-26.32} & \textbf{-27.59} & \textbf{-28.65} & -29.68 \\ 
DC-AMP Optimal & -21.25 & -23.74 & -26.24 & -27.55 & -28.65 & \textbf{-29.84} \\ 

	\hline
	\multicolumn{1}{l|}{~} & \multicolumn{6}{c|}{ISNR = 40~dB}\\
	\hline
		TVAL3 Optimal & \textbf{-21.48} & \textbf{-24.21} & \textbf{-26.34} & \textbf{-27.56} & -28.61 & -29.65 \\ 
DC-AMP Optimal & -21.23 & -23.73 & -26.22 & -27.52 & \textbf{-28.61} & \textbf{-29.79} \\ 

	\hline
	\multicolumn{1}{l|}{~} & \multicolumn{6}{c|}{ISNR = 30~dB}\\
	\hline
		TVAL3 Optimal & \textbf{-21.43} & \textbf{-24.08} & \textbf{-26.12} & \textbf{-27.38} & \textbf{-28.34} & -29.26 \\ 
DC-AMP Optimal & -21.20 & -23.76 & -26.07 & -27.30 & -28.29 & \textbf{-29.34} \\ 

	\hline
	\multicolumn{1}{l|}{~} & \multicolumn{6}{c|}{ISNR = 25~dB}\\
	\hline
		TVAL3 Optimal & \textbf{-21.35} & \textbf{-23.92} & -25.74 & \textbf{-26.99} & \textbf{-27.82} & -28.56 \\ 
DC-AMP Optimal & -21.07 & -23.81 & \textbf{-25.76} & -26.90 & -27.79 & \textbf{-28.64} \\ 

	\hline
	\multicolumn{1}{l|}{~} & \multicolumn{6}{c|}{ISNR = 20~dB}\\
	\hline
		TVAL3 Optimal & \textbf{-21.14} & \textbf{-23.37} & \textbf{-25.06} & \textbf{-26.10} & \textbf{-26.90} & -27.42 \\ 
DC-AMP Optimal & -20.79 & -22.94 & -25.05 & -26.05 & -26.80 & \textbf{-27.44} \\ 

	\hline
	\multicolumn{1}{l|}{~} & \multicolumn{6}{c|}{ISNR = 15~dB}\\
	\hline
		TVAL3 Optimal & \textbf{-20.42} & \textbf{-22.30} & \textbf{-23.81} & \textbf{-24.70} & \textbf{-25.36} & -25.86 \\ 
DC-AMP Optimal & -20.09 & -21.93 & -23.79 & -24.66 & -25.30 & \textbf{-25.88} \\ 

	\hline
	\multicolumn{1}{l|}{~} & \multicolumn{6}{c|}{ISNR = 10~dB}\\
	\hline
		TVAL3 Optimal & \textbf{-18.56} & \textbf{-20.59} & -22.01 & -22.89 & \textbf{-23.51} & \textbf{-23.96} \\ 
DC-AMP Optimal & -18.32 & -20.48 & \textbf{-22.11} & \textbf{-22.92} & -23.46 & -23.95 \\ 

	\hline
	\multicolumn{1}{l|}{~} & \multicolumn{6}{c|}{ISNR = 5~dB}\\
	\hline
		TVAL3 Optimal & -16.61 & -18.37 & -19.80 & -20.84 & -21.33 & -21.74 \\ 
DC-AMP Optimal & \textbf{-17.02} & \textbf{-18.74} & \textbf{-20.43} & \textbf{-21.17} & \textbf{-21.65} & \textbf{-21.94} \\ 

	\hline
	\multicolumn{1}{l|}{~} & \multicolumn{6}{c|}{ISNR = 0~dB}\\
	\hline
		TVAL3 Optimal & -14.11 & -15.86 & -16.19 & -18.32 & -18.94 & -19.61 \\ 
DC-AMP Optimal & \textbf{-14.29} & \textbf{-16.55} & \textbf{-18.35} & \textbf{-19.18} & \textbf{-19.84} & \textbf{-20.35} \\ 

	\hline
	\end{tabular}
	\caption[Table of the final reconstruction results for Peppers]{Comparison of the final reconstruction results in NSNR as a function of the noise level in dB between TV-AL3 and DC-AMP for different measurement rates $\alpha$ for the Peppers picture.\label{fig:peppers_im_table}}
\end{table}
\newpage
\section{Image reconstruction in compressive fluorescence microscopy}
\label{sec:microscopySec}
\begin{figure}[t]
\centering
\includegraphics[width=1\textwidth, trim=0 285 0 0, clip=true]{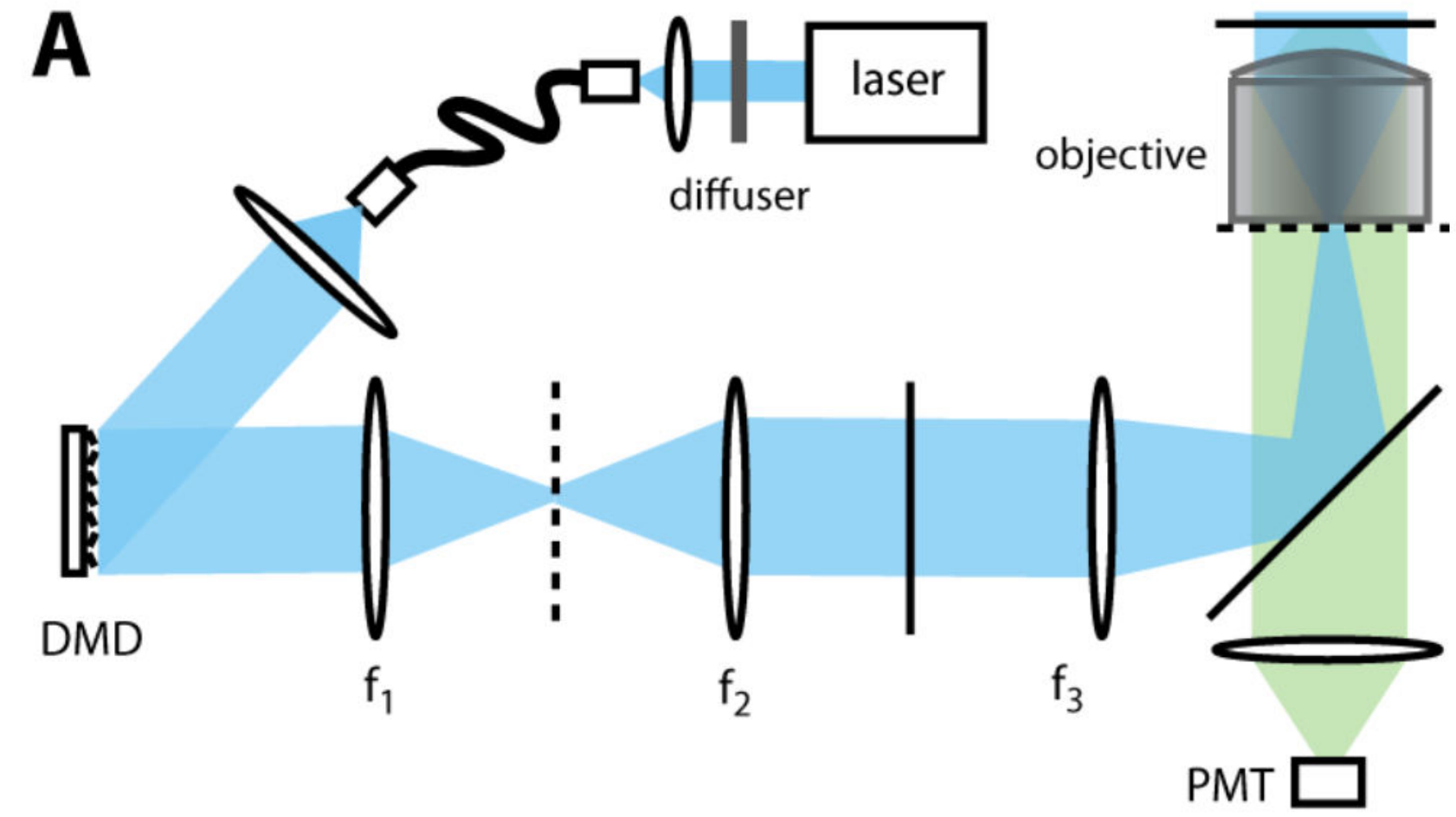}
\caption[Experimental setup for compressive fluorescence microscopy measurements]{Image taken from \cite{Studer26062012}. The experimental setup for compressive measurements of the fluorescent beads. A random selection of $2$-d binary $\{0,1\}$ (illumination or not) Hadamard patterns generated by a laser beam and a digital micromirror device DMD are successively projected onto the sample of interest. For each pattern, the beads that are illuminated are excited. These then emit light by fluorescence. Each measure (one per Hadamard bi-dimensional pattern) corresponds to perform the sum of all the resulting photons by converging all this emitted light on a single point, which intensity is measured by the objective that outputs a single scalar value $y_\mu$ proportional to the number of photons received.}
\label{figCh1:expSetup}
\end{figure}
We now present an application of approximate message-passing inference to image reconstruction in fluoresence microscopy.
The present work is part of an ongoing collaboration with Vincent Studer, Makhlad Chahid and Maxime Dahan who performed the experimental part of \cite{Studer26062012}. All the data analysed in sec.~\ref{sec:recResultsBeads} has been generated by them. 
\subsection{Introduction}
For the next, we call a {\it measure} (or measurement) the process of measuring the overall light intensity emitted by the beads after the excitation by one single $2$-d Hadamard pattern thanks to the setup Fig.~\ref{figCh1:expSetup}, and an {\it acquisition} the full process of getting $M$ different measures: one can try to reconstruct the image of the beads from one vector of $M$ measurements $\by$ obtained thanks to one acquisition. 

In the present problem, the aim is to locate fluorescent point-like beads on a plane (thus the signal is directly sparse in the pixel domain) thanks to compressive measurements i.e. from $M<N$ measurements, where $N$ is the number of pixels of the image to reconstruct.

Fluorescence microscopy has a great potential, especially in biological applications. Unfortunatly measurements may be costly in time making a single acquisition very long and thus compressed sensing is highly relevant here. As the image of the beads is sparse as seen from Fig.~\ref{figCh1:8192}, compressed sensing theoretically allows to reconstruct it from far fewer measurements than usual methods and thus to drastically speed up the acquisition. If acquisitions were fast enough thanks to compressed sensing, one could think of observing the dynamical behavior of small objects such as proteins and cells. This would require that their typical evolution time scale would be smaller than the acquisition time. Indeed, in order to reconstruct an image from an acquisition, it is essential that the measured system does not evolve from one measurement to the next or they would be incoherent and reconstruction impossible.

For example, even at very low temperatures (not even speaking of biologically relevant temperatures), a protein tertiary structure may evolve due to the thermal noise, and if one aim at oberving it in a particular configuration, fast acquisitions processes are essential or it will have time enough to relax to a new configuration during the acquisition. 

Compressed sensing could also allow to increase the spatial resolution of the images from a computational point of view. As the number of pixels $N$ fixes the maximum spatial resolution for the location of the beads and the classical sensing procedures require a number of measures that scale with $N$, an increase in resolution cost many more measures whereas with compressed sensing, only $O(\rho N)$ measurements are required, and thus the resolution for very sparse images with $\rho \ll 1$ can be improved at low cost.
\subsection{Experimental setup and algorithmic setting}
%
\begin{figure}[t]
\centering
\includegraphics[width=1\textwidth]{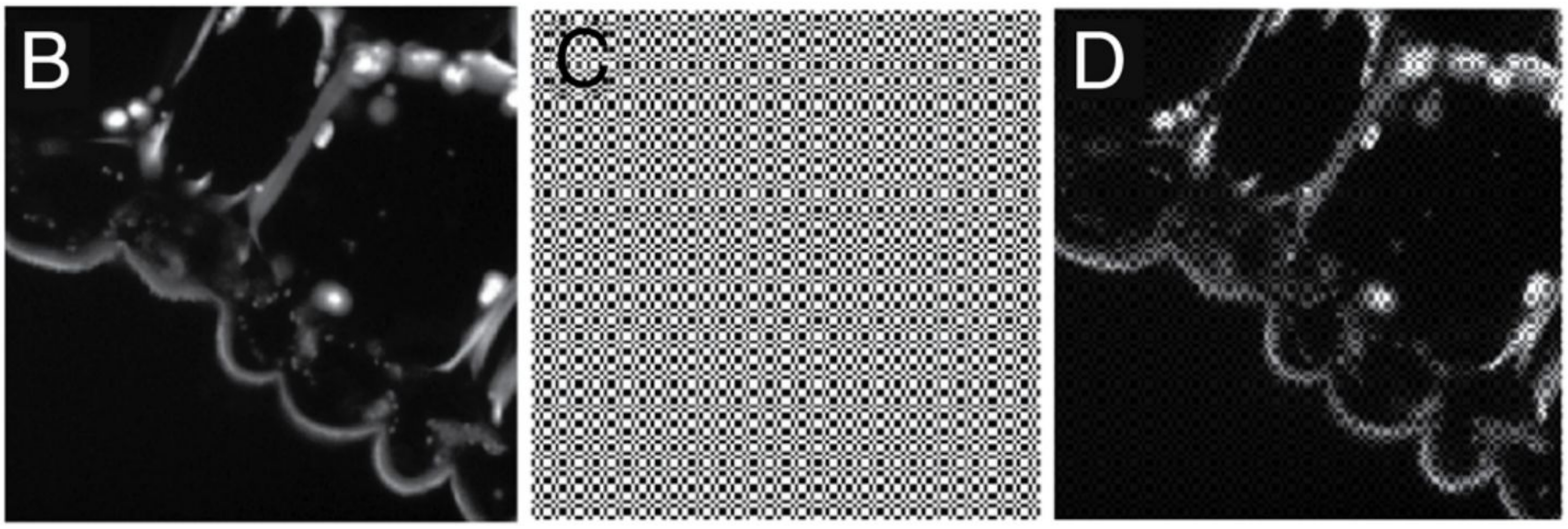}
\caption[Typical bi-dimensional Hadamard pattern used for the compressive measurements of the fluorescent beads]{Image taken from \cite{Studer26062012}. The left part is a biological sample, the center one is a typical randomly selected $2$-d Hadamard pattern (or mode) used for the compressive measurements of the fluorescent beads and the right one is the same pattern projected onto the biological sample: the lighter points are the ones that are excited.}
\label{figCh1:hadamardPattern}
\end{figure}
%
The signal processing problem treated in this chapter is the reconstruction of fluorescent beads randomly placed on a plane. The experimental setup of Fig.~\ref{figCh1:expSetup} is used for obtaining the compressive measurements of the beads: a random selection of $M$ $2$-d binary $\{0,1\}$ (illuminated or not) Hadamard patterns successively generated by a laser beam and a digital micromirror device are projected onto the plane where the beads are. For each pattern, the beads that are illuminated absorbe this light and then emit it back by fluorescence. The sum of all the resulting photons is performed by converging all this emitted light on a single point measured by an accurate intensity detector that output a scalar $y_\mu$ for each measurement, $\mu \in \{1,\ldots,M\}$. We refer to \cite{Studer26062012} for a detailed description of the experimental setup. A random $2$-d Hadamard pattern is represented at the center of Fig.~\ref{figCh1:hadamardPattern} and its projection on a biological sample is on the right part.

In the present setting, we assume that the linear model describing this experimental setup is given by (\ref{eqIntro:AWGNCS}). Indeed the $2$-d Hadamard transform $H_{2d}$ of a matrix $\bx$ of size $\sqrt{N}\times\sqrt{N}$ is defined in function of the Hadamard matrix $H$ as:
\begin{align}
	H_{2d}(\bx)_{ij} \defeq \(H(H\bx)^\intercal\)_{ij} = \sum_{k,u}^{\sqrt{N},\sqrt{N}} H_{ik}H_{ju} x_{uk} \label{eq:opFastBeads}
\end{align}
This operator can be easily written as a single matrix by rasterizing each $2$-d mode made of $\sqrt{N}$ lines of size $\sqrt{N}$ (see Fig.~\ref{figCh1:hadamardPattern}) into one line of size $N$ of $\bF$. Doing this for all the $M$ modes and considering the properly rasterized signal vector $\bx$ which is now of size $N$, we get back the system (\ref{eqIntro:AWGNCS}). 

But for the algorithm implementation, we want to use the fast Hadamard transform and thus the form (\ref{eq:opFastBeads}) of the operator $H_{2d}$ instead of its matrix representation as we are working with large data sets. Thus the operator $H_{2d}(\bx)$ which outputs the same vector as the direct matrix product $\bF \bx$ (which thus can be directly used in Fig.~\ref{algoCh1:AMP_op} without having to change anything) but that is constructed using the fast Hadamard transform is implemented by the following pseudo-code:
\begin{align}
	H_{2d}\bigg(&\txt{\tbf{input :} $\bx$ of size $N$, sets of $1$-d Hadamard modes indices $(I_1,I_2)$}\bigg)\nonumber \\
	&~~~~~1) ~\txt{"Derasterize" $\bx$ to make it of size $\sqrt{N}\times \sqrt{N}$}\nonumber\\
	&~~~~~2)~\txt{Apply the $2$-d fast Hadamard transform : } \tbf{p} = FT(FT(\bx)^{\intercal})\nonumber\\
	&~~~~~3) ~\txt{Rasterize $\tbf p$ to make it of size $N$}\nonumber\\
	&~~~~~4) ~\txt{Rescale the result to take into account the $\{0,1\}$ nature of the measurement}\nonumber\\ 
	&~~~~~~\txt{operator while the $FT$ operator is $\{-1,1\}$ : } \tbf z = \frac{1}{2} \(\tbf p + \sum_i^N x_i\)\nonumber\\
	&~~~~~5) ~\txt{Select the proper $M$ firsts modes : } \tbf u = \[\tbf z\(\sqrt{N}(I_1(i) - 1) + I_2(i) \)\]_i^M\nonumber\\	
	&\txt{\tbf{output :} $\tbf u$ of size $M$}
\end{align}
where $FT$ is the usual fast Hadamard transform. The required backward operator construction is related to what has been presented in sec.~\ref{sec:randomizationStructOp} and is implemented here as:
\begin{align}
	H_{2d}^{\intercal}\bigg(&\txt{\tbf{input :} $\by$ of size $M$, sets of $1$-d Hadamard mode indices $(I_1,I_2)$}\bigg)\nonumber \\
	&~~~~~1) ~\txt{Create a vector with the $M$ firsts $1$-d mode indices : } \tbf v=\[\sqrt{N}(I_1(i) - 1) + I_2(i)\]_i^M\nonumber\\
	&~~~~~2)~\txt{Define}~ \tbf f ~\txt{of size $N$ such that : } f_{v_i} = y_i~\forall i\in\{1,\ldots,M\}, f_{k} = 0 ~\forall k \not\in \tbf v \nonumber\\
	&~~~~~3) ~\txt{Derasterize $\tbf f$ to make it of size $\sqrt{N}\times\sqrt{N}$}\nonumber\\
	&~~~~~4) ~\txt{Apply the $2$-d fast Hadamard transform : } \tbf p = FT(FT(\tbf f)^{\intercal})\nonumber\\ 	
	&~~~~~5) ~\txt{Rasterize $\tbf p$ to make it of size $N$}\nonumber\\
	&~~~~~6)\txt{Rescale the result to take into account the $\{0,1\}$ nature of the measurement}\nonumber\\ 
	&~~~~~~\txt{operator while the $FT$ operator is $\{-1,1\}$ : } \tbf u = \frac{1}{2} \(\tbf p + \sum_\mu^M y_\mu\)\nonumber\\
	&\txt{\tbf{output :} $\tbf u$ of size $N$}
\end{align}
The modes-indices sets $(I_1,I_2)$ are selected by the experimentalists and define the scrambled sub-sampled $2$-d Hadamard transform used for acquisition (they contain respectively the $M$ $\{i\}$ and $\{j\}$ indices selected when using (\ref{eq:opFastBeads}), with $i,j\in \{1,\dots,\sqrt{N}\}$). $8192$ modes were selected for the data used in sec.~\ref{sec:recResultsBeads}. Only a sub-set of them can be selected for the reconstruction tests, changing $M$ in the previous operators. The fast implementation of the homogeneous approximate message-passing algorithm Fig.~\ref{algoCh1:AMP_op} (with $L=N, L_c=L_r=1, B=1$) is thus defined replacing the forward operator $O_\mu$ (\ref{eq_fastOpDefs1}) by the fast operator $H_{2d}$ and the backward one $O_i$ (\ref{eq_fastOpDefs22}) by $H_{2d}^\intercal$. The two others operators (\ref{eq_fastOpDefs11}), (\ref{eq_fastOpDefs2}) are also defined repectively by $H_{2d}$ and $H_{2d}^\intercal$ as $0$ or $1$ are invariant by the square operation. Let us now discuss a proposal for the denoisers that appear to have very good performances in the present problem.
\subsection{A proposal of denoisers for the reconstruction of point-like objects measured by compressive fluorescence microscopy}
In order to perform inference of the beads locations thanks to the approximate message-passing algorithm Fig.~\ref{algoCh1:AMP}, we propose here an exponential prior with approximate Gaussian sparsity for designing the denoisers following the procedure given in sec.~\ref{sec:cookAMP}. The exponential part approximates the discrete Poisson distribution associated with the number of informative photons emitted by the actual beads. This law is typical of sources which light emission is due to desexcitation processes, such as in fluorescence. The full prior that we assume factorizable over the pixels is thus given by:
\begin{align}
	P_0(\bx|\bsy \theta) = \prod_i^N P_0(x_i|\bsy \theta) = \prod_i^N \[\rho \lambda e^{-\lambda x_i}\mathbb{I}(x_i>0) + \mathcal{N}\(x_i|m,\sigma^2\)\] \label{priorBeads}
\end{align}
where $\bsy \theta \defeq [\lambda, \rho, m, \sigma^2]$. $\rho$ is the density of pixels of the picture on which beads are standing and $\mathbb{I}(x_i>0)$ enforces the pixels to have positive values. Of course this model is an approximation of the true signal generating process, but it appears empirically to reach very good performances. More complex prior models including correlations between pixels could be considered but this cannot be done with AMP, at least in its canonical form Fig.~\ref{algoCh1:AMP} which requires that the prior is factorizable over some subsets of signal components. Nevertheless, we will see in sec.~\ref{sec:firstNeighInt} how to include "a posteriori" some effective interaction between closeby pixels.

In the present context, we consider the measurement noise variance $\Delta \to 0$ as the background photon noise is already included into the Gaussian part of the prior. We could think also of using a strictly sparse prior replacing the Gaussian by a Dirac distribution and letting the noise variance have a finite value instead, but it appears empirically that it gives worst results than this approximate sparsity prior. This may come from the fact that the learning rules of the Gaussian parameters $(m,\sigma^2)$ are different that of the noise variance $\Delta$ and there are two instead of one. This remark is actually quite general, and we observed empirically in many situations that approximate sparsity without measurement noise gets better results than a sparse prior with noise. 
\subsection{Optimal Bayesian decision for the beads locations}
\label{sec:optimalBayesDecisionBeads}
A great advantage of the Bayesian framework with respect to convex optimization procedures is that it allows to directly estimate the probability that a pixel supports a bead or not, i.e. if this pixel is informative or belongs to the background noise. The posterior "noise" probability $P(x_i\in \mathcal{N})$ for a pixel $x_i$ to be pure noise (denoted by $\mathcal{N}$ in reference to the Gaussian part for the noise in the prior) is proportional to the prior probability of belonging to the background noise re-weighted by the AMP Gaussian field:
\begin{align}
	P(x_i \in \mathcal{N}|\bsy \theta, \Sigma^2_i,R_i) &= \frac{1}{z(\Sigma_i^2, R_i|\bsy\theta)}\int dx_i \mathcal{N}\(x_i|m,\sigma^2\) \mathcal{N}\(x_i|R_i,\Sigma_i^2\) \label{eq:noiseproba}\\
	z(\Sigma_i^2, R_i|\bsy \theta) &= \int dx_i \mathcal{N}\(x_i|R_i,\Sigma^2\) P_0(x_i|\bsy\theta)
\end{align}
where $z(\Sigma_i^2, R_i|\bsy\theta)$ is the posterior partition function of pixel $i$, $P_0$ is given by (\ref{priorBeads}) and $(\Sigma_i^2, R_i)$ are the usual moments controlling the Gaussian AMP field summarizing the likelihood constraints on $x_i$. The $(\Sigma_i^2, R_i)$ values are iteratively computed by AMP Fig.~\ref{algoCh1:AMP}. This expression is analytical and using (\ref{priorBeads}), (\ref{eq:noiseproba}) it becomes:
\begin{align}
	P(x_i \in \mathcal{N}|\bsy \theta, \Sigma^2_i,R_i) &= \[1+\rho\lambda\sqrt{\frac{\pi(\Sigma_i^2+\sigma^2)}{2}} e^{\frac{\lambda}{2} (\lambda\Sigma_i^2 - 2R_i) + \frac{(m - R_i)^2}{2(\Sigma_i^2+\sigma^2)}} \txt{erfc}\(\frac{\lambda\Sigma_i^2-R_i}{\sqrt{2\Sigma^2_i}}\) \]^{-1}
\end{align}
After convergence of the algorithm, in order to obtain the final estimate $\hat x_i$ for each pixel, we thus cancel all the final AMP posterior pixels estimates $a_i$ which final posterior noise probability is more than $0.5$. Doing this we keep only the supposed informative pixels from the AMP point of view:
\begin{align}
	\hat x_i = a_i^t ~ \mathbb{I}\(P(x_i \in \mathcal{N}|\bsy \theta^t, (\Sigma^t_i)^2,R_i^t) < 0.5\)
\end{align}
where $t$ is the final time step. This kind of decisions cannot be taken with $\ell_1$-minimization based solvers as they are not probabilistic algorithms, and arbitrary thresholding functions {\it must} be applied at the end or the results are very poor in this kind of highly noisy problems. In the experiments of sec.~\ref{sec:recResultsBeads}, the thresholding function applied to the $\ell_1$-minimization based solvers final estimates is such that we keep only the pixels that have an amplitude approximately $4$ times higher than the mean of the recovered overall picture, we cancel the other ones. This value of $4$ has been selected {\it empirically} to obtain the best possible match between the reconstructed and original pictures in the high measurement rate regime ($M=8192$ measurements in the results of sec.~\ref{sec:recResultsBeads}). Despite being probably suboptimal for other measurement rates, this thresholding function appears to output results close to the best ones at any rate, i.e. that are obtained when the optimization of this thresholding function is performed for every measurement rate and $\ell_1$-minimization solver independently.
\subsection{The learning equations}
\label{sec:learningMicroscopy}
In order to find the optimal values of the free parameters of the prior (\ref{priorBeads}), we use the expectation maximization strategy discussed in sec.~\ref{sec:EMlearning}. In the present case, all the quantities can be simply obtained directly from the posterior estimates and noise probability. Appropriate recursions that are very stable numerically and have simple interpretations are given by:
\begin{align}
	\rho^{t+1} &= \frac{1}{N} \sum_i^N \mathbb{I}\(P(x_i \in \mathcal{N}|\bsy \theta^t, (\Sigma^t_i)^2,R_i^t) < 0.5\) \label{eq:rhoMicroLearn}\\
	\lambda^{t+1} &= \[ \frac{1}{|\ba_{supp.}^t|} \sum_i^{|\ba_{supp.}^t|} a^t_{i,supp.} \]^{-1} = \frac{1}{<\ba_{supp.}^t>}\\
	m^{t+1} &= \frac{1}{|\ba_{noise}^t|} \sum_i^{|\ba_{noise}^t|} a^t_{i,noise}=<\ba_{noise}^t>
\end{align}
where $\ba_{supp.}^t \defeq \[a_i^t : P(x_i \in \mathcal{N}|\bsy\theta^t,(\Sigma_i^t)^2,R_i^t) < 0.5\]$ are the posterior estimates of the estimated support pixels of the beads at time $t$, $\ba_{noise}^t \defeq \[a_i^t : P(x_i \in \mathcal{N}|\bsy\theta^t,(\Sigma_i^t)^2,R_i^t) > 0.5)\]$ are the posterior estimates of the noise pixels (that are not in the support) at time $t$. The variance of the Gaussian could be learned as well in the same way but it appears empirically that fixing its value is more efficient. All these equalities are just coming from the very definitions of the different quantities. For example, the parameter $\lambda$ in the exponential distribution must be equal to the inverse of the mean of this distribution, and we naturally take only into account the values of the pixels that are considered in the support as the exponential is here to model the beads.
\subsection{Improvement using small first neighbor mean field interactions between pixels}
\label{sec:firstNeighInt}
It appears empirically that using a small first neighbor mean field interaction between pixels improve the results and allows AMP to recover perfectly the beads locations at smaller measurement rates. The trick is empirically done by adding to the moment $R_i^{t+1}$ controling the AMP field felt by the pixel $x_i$ at time $t+1$ a perturbation $\epsilon h_i^{t+1}$ where $\epsilon$ is a small $\in O(10^{-1})$ auxiliary parameter and $h_i^{t+1}$ is the mean field that takes into account the (up to) $4$ $x_i$'s neighbors states. It is defined as an extension of the so-called bilateral denoiser in statistical image processing and is defined as:
\begin{align}
	h_i^{t+1} &= \frac{\sum_{j \in \partial i} a_j^{t} ~ w_{ij}^t ~\mathbb{I}\(P(x_j \in \mathcal{N}|\bsy\theta^t,(\Sigma_j^t)^2,R_j^t) < 0.5 \)}{\sum_{j \in \partial i} w_{ij}^t~\mathbb{I}\(P(x_j \in \mathcal{N}|\bsy\theta^t,(\Sigma_j^t)^2,R_j^t) < 0.5 \)}	
\end{align}
where the weight given to the neighbor pixel $x_j$ of pixel $x_i$ is a Gaussian proportional to their posterior estimates difference, i.e. it gives higher weight to similar pixels:
\begin{align}
	w_{ij}^t = \mathcal{N}\(a_i^t\big|a_j^t,\sigma^2_w\)	
\end{align}
This field thus weakly shifts the AMP field of the $i^{th}$ pixel to higher values when its neighbors are considered being part of the support: $R_i^{t+1}=R_i^{t+1} + \epsilon h_i^{t+1}$. This mean field thus mimics in some sense the behavior of the algorithm discussed in sec.~\ref{sec:TV} but without changing the prior that remains fully factorizable over the pixels. It appears empirically that the improvement thanks to this strategy is only very weakly dependent on the $\sigma_w^2$ parameter and we take a uniform weight $\sigma_w^2\to\infty$ for the results presented in the next section.
\subsection{Reconstruction results on experimental data}
\label{sec:recResultsBeads}
We now present some results which compare the reconstruction performances of AMP and of two state-of-the-art $\ell_1$-minimization based solvers (NESTA \cite{journals/siamis/BeckerBC11} and fast iterative hard thresholding \cite{DBLP:journals/corr/abs-0805-0510}) and where the Bayesian optimal and heuristic thresholding functions discussed in sec.~\ref{sec:optimalBayesDecisionBeads} are applied to the final reconstructions made with AMP and the $\ell_1$-minimization based solvers respectively. Furthermore, each reconstructed support pixel is "highlighted" a posteriori by giving a positive value to its closest neighbors as well. The data used here has been obtained by the authors of \cite{Studer26062012} using the experimental setup described in Fig.~\ref{figCh1:expSetup}.

A first remark is that independently of the algorithm used or the measurement rate $\alpha$, there are always artefacts appearing on the up-left corner (two beads are always missing in the reconstruction) and for the positions of some beads as well. This must come from the experimental data set used here rather than the algorithms, as even at the higher rate $\alpha$, there remain these errors. So it is considered that these systematic errors are actually not errors for the algorithmic reconstructions.

\begin{figure}[H]
\centering
\includegraphics[width=.7\textwidth, trim=85 30 70 30, clip=true]{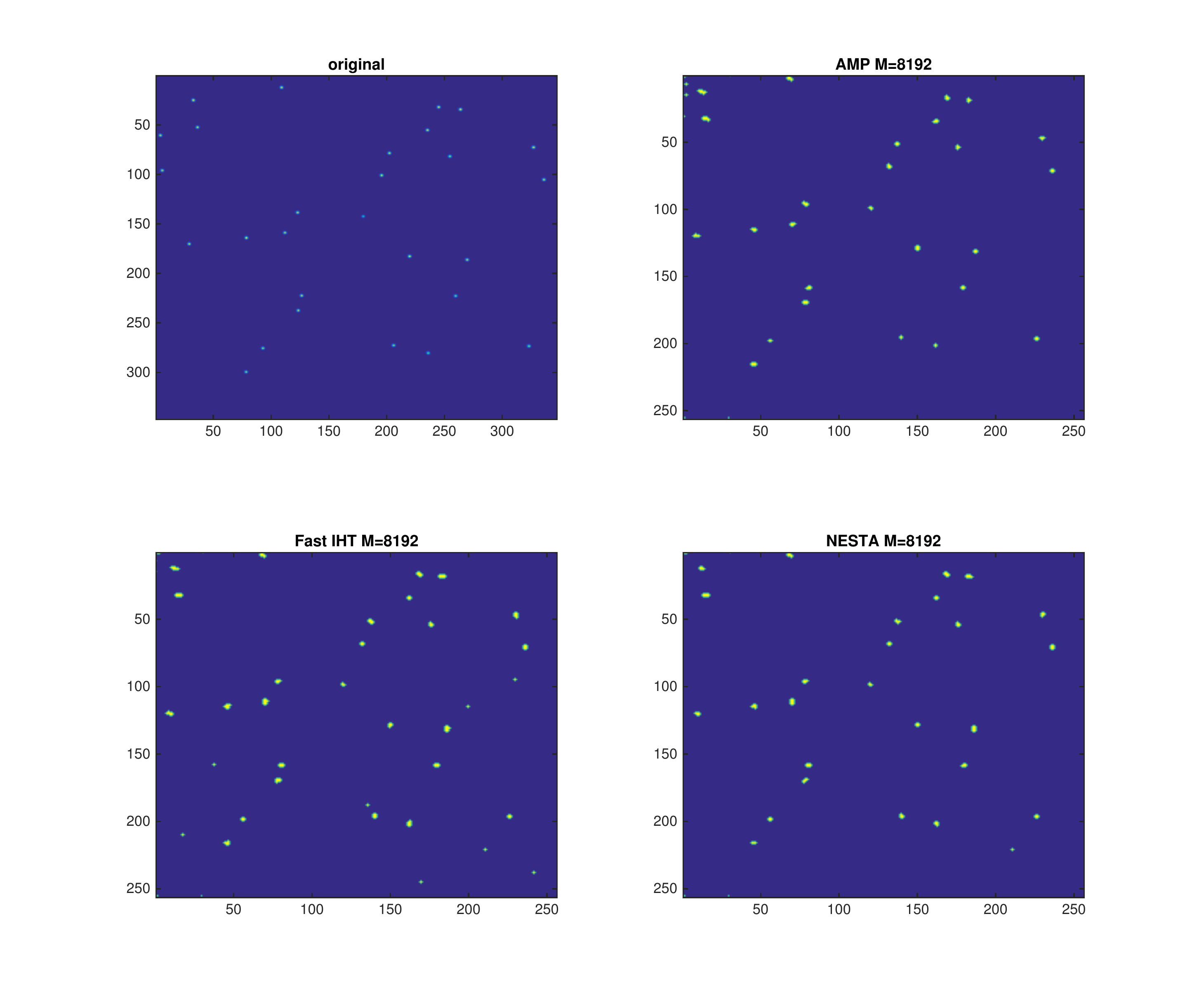}
\caption[Comparison of the beads location at $M=8192$]{Comparison of the reconstruction results of the $3$ algorithms used here: AMP, NESTA and Fast Iterative Hard Thresholding with the original picture. The number of measurements is here $M=8192$, $\alpha = 0.125$.}
\label{figCh1:8192}
\centering
\includegraphics[width=.7\textwidth, trim=85 30 70 30, clip=true]{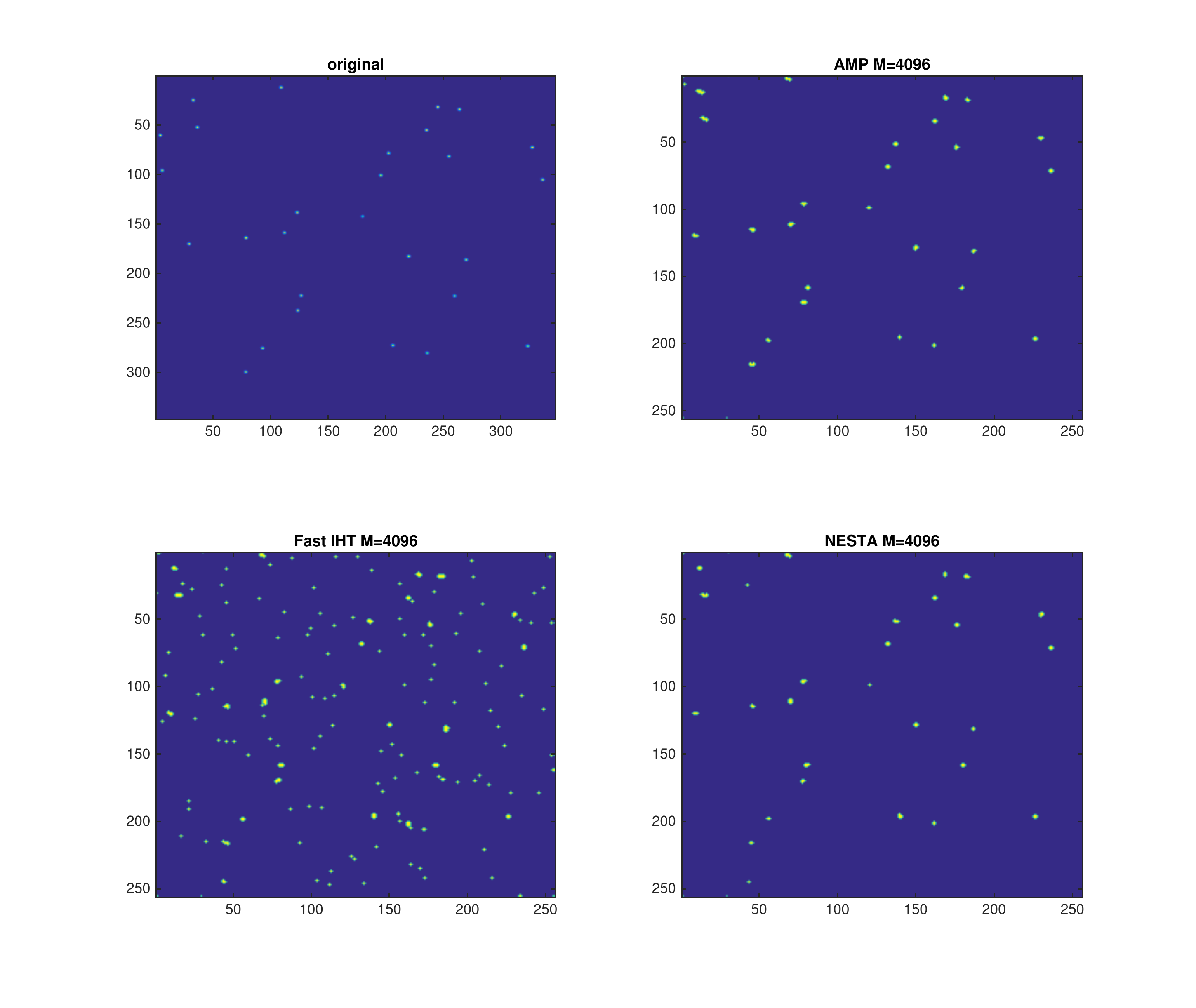}
\caption[Comparison of the beads location at $M=4096$]{Same as Fig.~\ref{figCh1:8192} with $M=4096$, $\alpha=0.0625$.}
\label{figCh1:4096}
\end{figure}

\begin{figure}[H]
\centering
\includegraphics[width=.7\textwidth, trim=85 30 70 30, clip=true]{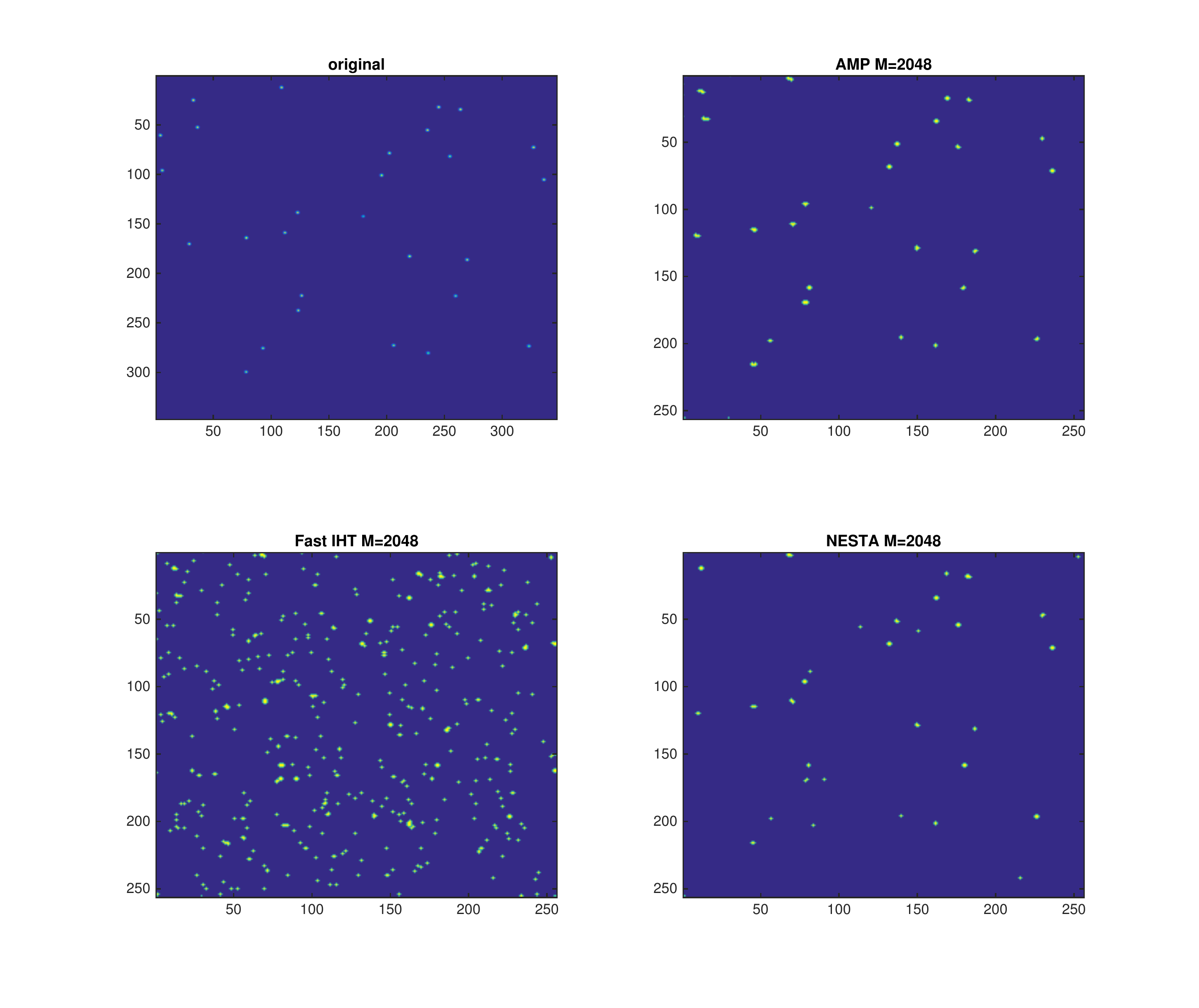}
\caption[Comparison of the beads location at $M=2048$]{Same as Fig.~\ref{figCh1:8192} with $M=2048$, $\alpha=0.031$.}
\label{figCh1:2048}
\centering
\includegraphics[width=.7\textwidth, trim=85 30 70 30, clip=true]{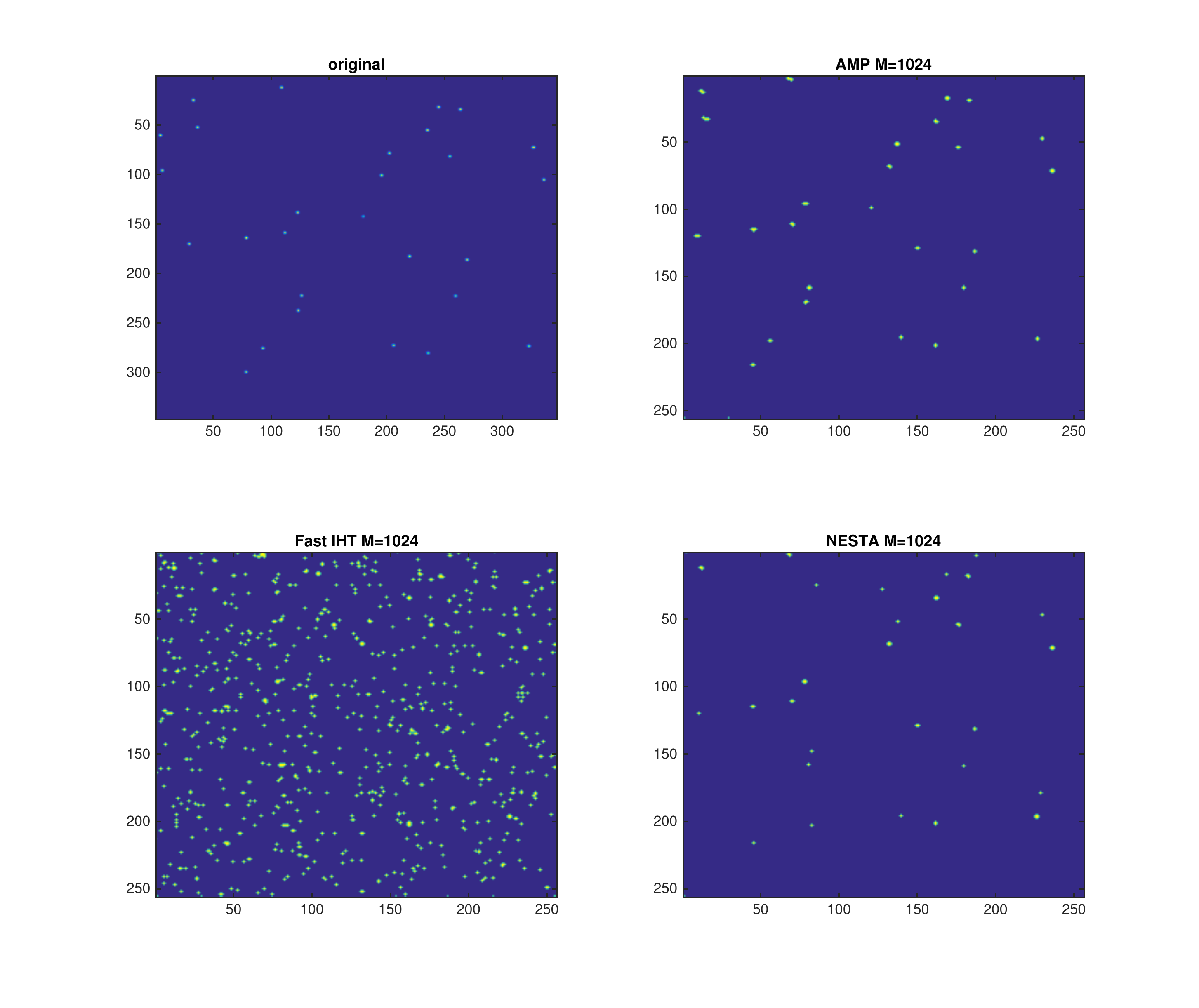}
\caption[Comparison of the beads location at $M=1024$]{Same as Fig.~\ref{figCh1:8192} with $M=1024$, $\alpha=0.016$.}
\label{figCh1:1024}
\end{figure}

\begin{figure}[H]
\centering
\includegraphics[width=.7\textwidth, trim=85 30 70 30, clip=true]{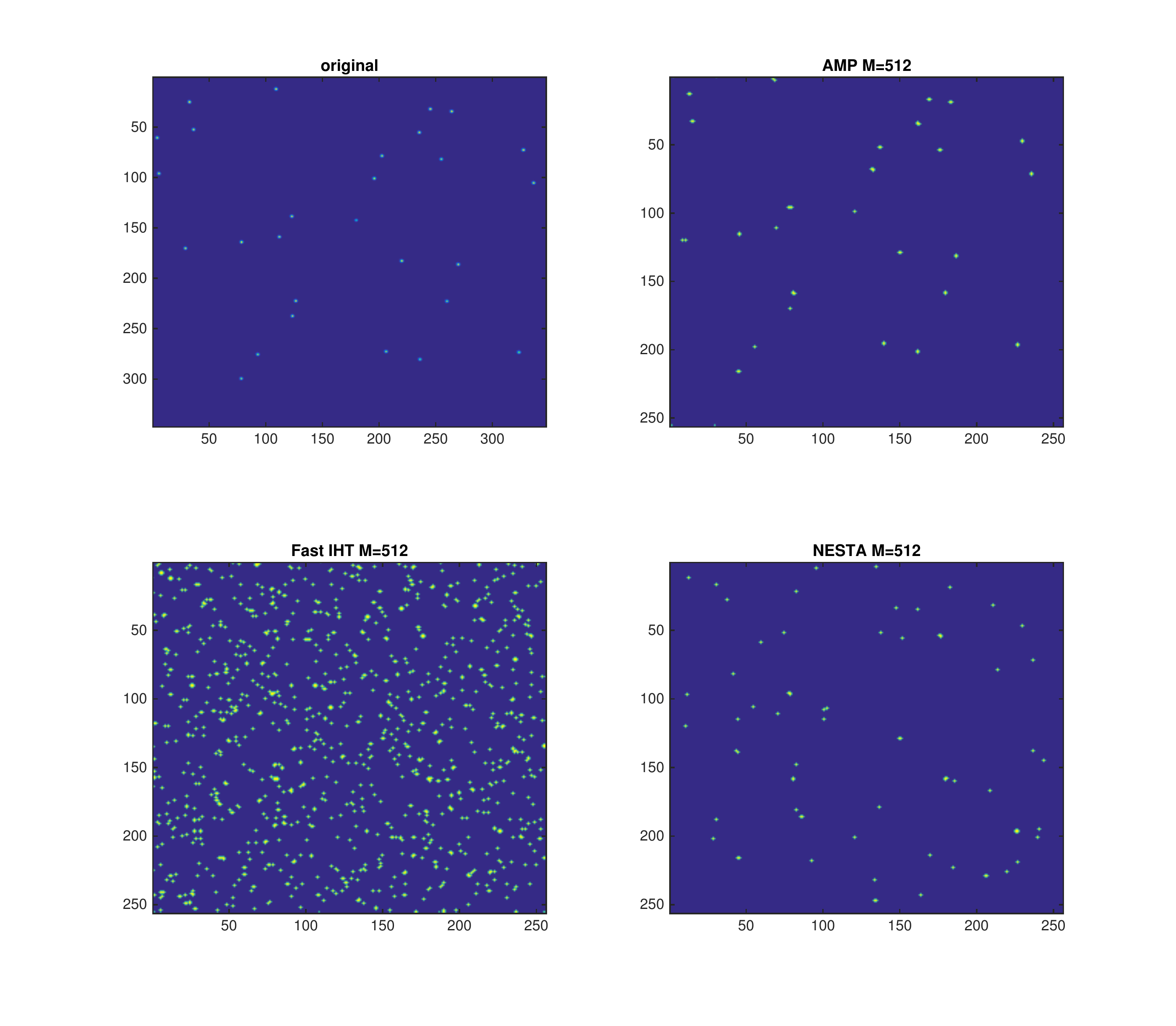}
\caption[Comparison of the beads location at $M=512$]{Same as Fig.~\ref{figCh1:8192} with $M=512$, $\alpha=0.0078$. This $\alpha$ is the limit of "perfect" beads location using the AMP algorithm.}
\label{figCh1:512}
%
%
\centering
\includegraphics[width=.7\textwidth, trim=85 30 70 30, clip=true]{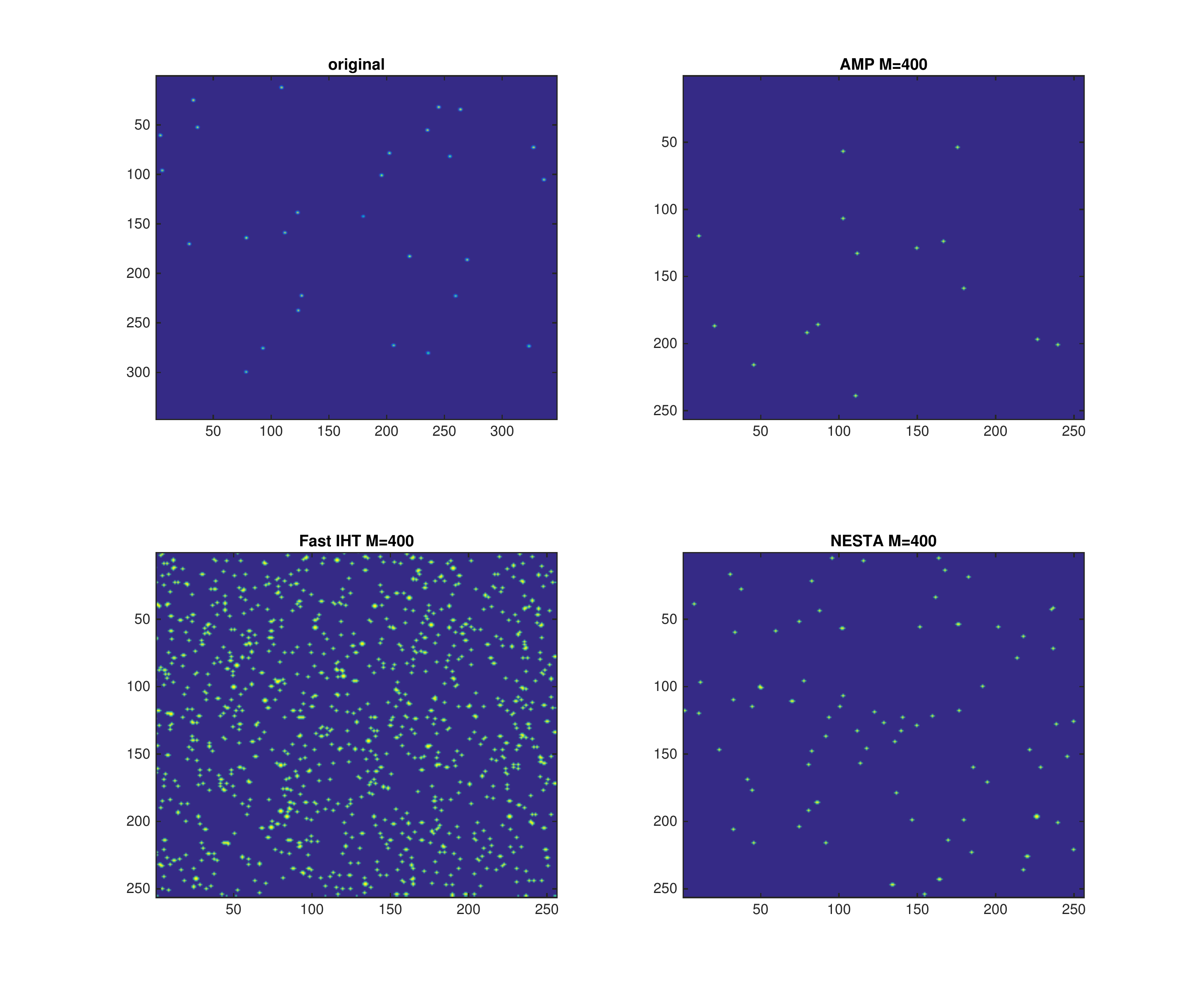}
\caption[Comparison of the beads location at $M=400$]{Same as Fig.~\ref{figCh1:8192} with $M=400$, $\alpha=0.0061$. We observe a continuous worsening of the AMP results.}
\label{figCh1:400}
\end{figure}

The results Fig.~\ref{figCh1:8192}-Fig.~\ref{figCh1:400} all show a clear advantage for AMP: its reconstruction and location of the beads is perfect (up to these systematic data-dependent errors) until $M=512$ whereas comparably good results (yet not perfect, whereas the AMP results are) are obtained with NESTA only for $M\ge4096$ or with FastIHT at $M\ge 8192$. So the gain with AMP is substantial, and the location is way more accurate. Furthermore, the speed of convergence of AMP is always $2$ to $10$ times faster than the convex optimization solvers used here. For $M<512$ approximatively, the AMP performances start to worsen continuously, and actual beads disappear while new "fake" ones start to appear as seen on the last figure Fig.~\ref{figCh1:400}.

An important remark is that when the previously discussed "TV-like" AMP algorithm of sec.~\ref{sec:TV} have been tried on the data, it appeared that the reconstruction performances were not comparable with the AMP implementation presented here. This is due to the point-like nature of the beads: the gradient-minimizing prior of sec.~\ref{sec:algoTV} tends to smoothen too much the background and makes totally disappear the beads for low measurement rates, whereas the present specifically designed prior (\ref{priorBeads}) does consider the precense of such ponctual objects.
\subsection{Concluding remarks and open questions}
We have studied how the AMP algorithm can be used for reconstruction of sparse images in the pixel domain measured by fluoresence microscopy in the compressive regime. This study is a proof-of-concept, which naturally extended the work of \cite{Studer26062012} where convex optimization is used for the reconstruction. A natural continuation is to try the algorithm on real biological samples, where the beads are put inside membranes of cells for example as in \cite{Studer26062012}. Furthemore, in the data set used for the present results, the beads were strongly excited and thus emitted a lot of photons. The question of wether the reconstruction by AMP is robust to a net lowering of the beads emission intensity is of great interest as stronger excitations means a longer exposition time of the sample to the light field, and thus an overall longer acquisition time. Another natural idea would be to try spatial coupling combined with Hadamard patterns as in chap.~\ref{chap:structuredOperators}. But as the noise is high (typically a relative intensity of $O(10^{-2}/10^{-3})$ with respect to the beads intensity in the present experiments), it is probable that there is no first order transition and thus that this strategy does not improve the performances as discussed in chap.~\ref{chap:appSparsity}. Nevertheless, the results are very encouraging and AMP seems to be a good option in this context.
%
\part{Coding theory}
\label{part:coding}
\chapter{Approximate message-passing decoder and capacity-achieving sparse superposition codes}
\label{sec:superCodes}
\vspace{.5cm}
We study the approximate message-passing decoder for sparse
superposition coding over the additive white Gaussian noise channel. While this coding scheme
asymptotically reaches the Shannon capacity, we show that the AMP iterative
decoder is limited by the BP phase transition, similar to what happens in low density parity check LDPC codes. We present and study two solutions to this problem, that both
allow to reach the Shannon capacity: $i)$ a non constant power
allocation and $ii)$ the use of spatially-coupled codes. We also
present extensive simulations that suggest that spatial coupling is
more robust and allows for better correction at finite code
lengths. Finally, we show empirically that the use of a fast
Hadamard-based operator allows for an efficient reconstruction, both
in terms of computational time and memory allocation, and the ability to deal
with very large signals.
\section{Introduction}
The error correction scheme called sparse superposition codes has
originally been introduced and studied in
\cite{barron2010sparse,barron2011analysis,joseph2012least} by Barron and Joseph who proved
the scheme to be capacity achieving over the additive white Gaussian
noise AWGN channel under maximum-a-posteriori $MAP$ decoding. In
\cite{barron2010sparse,barron2011analysis,joseph2012least}, an
iterative decoder called {\it adaptive successive decoder} was
presented, which was later improved in
\cite{barron2012high,choapproximate} by soft thresholding methods. The
idea is to decode a sparse vector with a special block structure over
the AWGN channel, represented in Fig.~\ref{figCh1:AWGN}. With these
decoders together
with the use of power allocation, the scheme was proved to be capacity achieving in a proper limit. However, finite blocklength
performances were far from ideal. In fact, it seemed that the
asymptotic results could be reproduced at any reasonable finite
lengths.

We propose instead an approximate message-passing decoder for sparse superposition codes. We will show that this decoder have much
better performances. In fact it allows better decoding than the iterative successive decoder at any reasonable finite length, and this even without power allocation. We present two
modifications of sparse superposition codes that allow AMP to be
asymptotically capacity achieving as well, while retaining good finite
block length properties. The first one is the addition of power
allocation to sparse superposition codes as done for the iterative successive decoder, and the second one, which is specific to the message-passing decoder, is the use of spatial coupling which appears to be even more promising.

We also present extensive numerical simulations and a study of a
practical scheme with Hadamard operators. The overall scheme is computationally efficient  and allows to practically reach near-to-capacity transmission rates with low error floors.
\subsection{Related works}
The phenomenology of these codes under AMP decoding, in particular the
sharp BP phase transition happening before the optimal threshold, has many
similarities with what appears in LDPC codes \cite{RichardsonUrbanke08}. It is not a priori trivial because LDPC are codes over finite fields, the
sparse superposition codes scheme works in the continuous framework
and LDPC codes are decoded by loopy belief propagation whereas sparse
superposition codes are decoded by AMP, a Gaussian
approximation of loopy BP, see sec.~\ref{sec:classicalDerivationAMP}. However, they arise due to a deep
connection to compressed sensing where these phenomena (phase
transition, spatial coupling, \ldots) are well known
\cite{KudekarPfister10,KrzakalaPRX2012,KrzakalaMezard12,DonohoJavanmard11} as discussed in sec.~\ref{sec:typicalPhaseTransitions}, and we shall make use of this connection extensively.

As we will see, the AMP algorithm is naturally applied
to sparse superposition codes as this scheme can be interpreted as a compressed sensing problem with structured sparsity. This scheme is actually the first example of error correction of a signal that is directly mapped to a compressed sensing problem. In the chap.~\ref{chap:robustErrorCorrection}, the approach is different as it is the noise that is reconstructed. The state evolution technique \cite{BayatiMontanari10} is unfortunatly not rigorous for the present AMP approach because of the structured sparsity of the signal, but in spite of that, we conjecture that it is exact.

Note that reconstruction of
structured signals is a new trend in compressed sensing theory that aims at going
beyond simple sparsity by introducing more complex structures in the
vector that is to be reconstructed. Other examples include group sparsity or tree
structure in the wavelet coefficients in image reconstruction \cite{modelBasedCSBaraniuk2008}.

A recent work of Rush, Greig and Venkataramanan
\cite{rush2015capacity} also studied AMP decoding in superposition
codes combined with power allocation. Using the same technics as in
\cite{BayatiMontanari10}, they proove rigorously that sparse superposition codes under AMP deconding is capacity-achieving if a proper power allocation is used. This
strengthen the claim that AMP is the tool of choice in the present
problem. We will see, however, that spatial coupling leads to even
better decoding results at finite size.
%
\subsection{Main contributions of the present study}
The main original results of the present study are listed below.

$\bullet$ A detailed derivation of the AMP decoder for sparse
  superposition codes for a generic power allocation. The derivation is
  self-contained and starts all the way from the canonical loopy BP
  equations, see sec.~\ref{sec:classicalDerivationAMP}.

$\bullet$ An analysis of the performance of the AMP decoder from the state
  evolution recursions. It is done in full
  generality with and without power allocation, and with and without
  spatial coupling. It is shown in
  particular that AMP, for simple sparse superposition codes, suffers from a
  phenomenon similar to those of BP with LDPC codes: there exist a sharp BP
  transition different from the optimum one of the code itself beyond which the decoder performance suddenly drops.

$\bullet$ An analysis of the optimum performance of sparse superposition
  codes using the non-rigorous replica method. This leads in
  particular to a single-letter formulation of the $MMSE$ estimate which we conjecture to be exact. The connection and
  consistency with the results coming from the state evolution
  approach is also underlined, see sec.~\ref{subsec:repIsSE}.

$\bullet$ The large section limit for the
  behavior of AMP is studied, and we compute its limit rate, the asymptotic BP threshold $R_{BP}^{\infty}<C$ where $C$ is the Shannon capacity of the channel. Studying as well the optimal threshold in this limit, we reconfirm using the replica method that these codes
  are Shannon capacity achieving.

$\bullet$ We also show that, with a proper power allocation, the BP threshold that was blocking the AMP decoder
  disappears so that AMP becomes asymptotically capacity achieving over the
  AWGN in the large section limit.

$\bullet$ Building on the connection with compressed sensing \cite{KrzakalaPRX2012,KrzakalaMezard12,DonohoJavanmard11} we also
  show that the use of spatial coupling \cite{KudekarRichardson12} for
  sparse superposition codes is an alternative way to obtain capacity
  achieving performances with AMP.

$\bullet$ We present an extensive numerical study at finite
  blocklength, showing that despite improvements of the scheme thanks
  to power allocation, a properly designed spatially-coupled coding
  matrix seems to allow better performances and robustness to noise for decoding over finite size signals.

$\bullet$ Furthermore we discuss a more practical scheme where the random coding operators are replaced by fast ones based on an Hadamard construction, see chap.~\ref{chap:structuredOperators}. We show that this allows
  a close to linear time algorithm able to deal with very large
  signals, yet performing very well at large rate for finite
  signals. We study the efficiency of these
  operators combined with sparse superposition codes with or without
  spatial coupling.

Finally, we note that this work differs from the mainstream of existing
literature. While a large part of the existing coding theory
literature provides theorems, part of this work, that using the
replica method, is based on statistical physics methods that are
conjectured to give exact results. While many results obtained with
these methods on a variety of problems have indeed been proven later
on, a general proof that these methods are rigorous is not known yet. Note, however, that the state evolution technique has been turned into a rigorous
tool under control in many similar cases \cite{BayatiMontanari10,javanmard2013state}. The present approach does not 
  verify the assumptions required for the proofs to be valid because of the structured sparsity of the signal, but nevertheless
  we conjecture that the analysis remains exact. We thus expect
that both the replica and state evolution analyzes are
exact and believe it is only a matter of time before they are fully proven.
%
%
%
\section{Sparse superposition codes}
\label{secCh1:Suc}
\begin{figure}[t]
\centering
\includegraphics[width=1\textwidth]{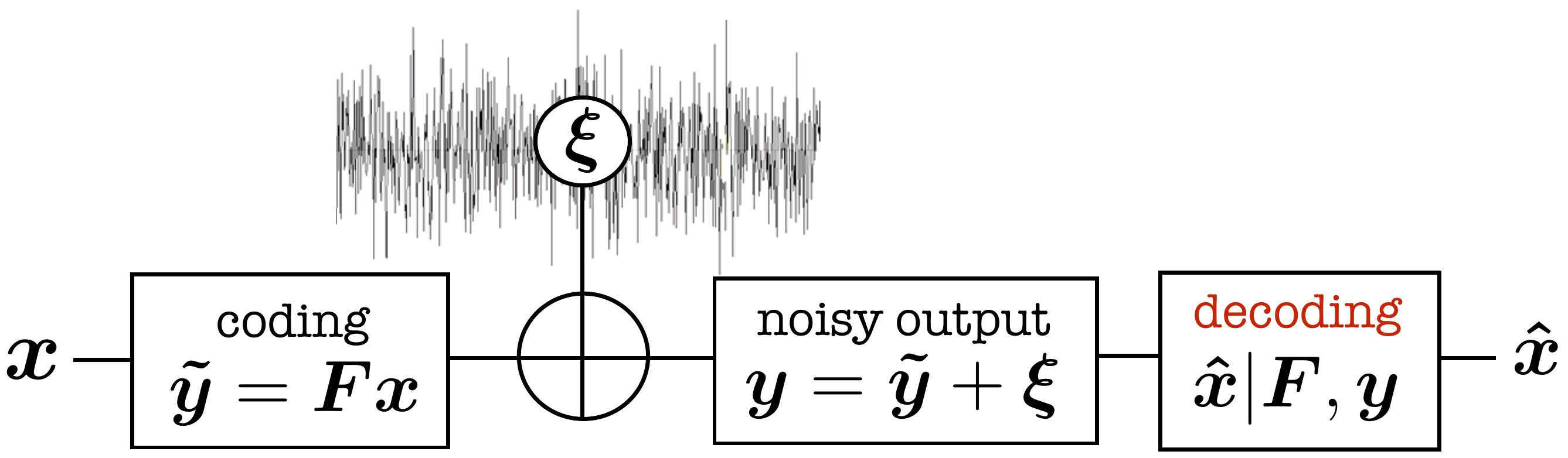}
\caption[Additive white Gaussian noise channel model]{Sending information through the AWGN channel with
superposition codes: the message $\bx$, created such that it has
only a single non zero element in each of its $L$ sections, is first
coded by a linear transform, $\tilde \by = \bF\bx$. The resulting codeword is then sent
through the AWGN channel that adds an i.i.d Gaussian noise $\bxi$
with zero mean and a given variance $\Delta$ to each components. The
receptor gets the corrupted codeword $\by$ and must estimate
$\hat \bx$ as close as possible from $\bx$ from the knowledge of
$\bF$ and $\by$. Perfect decoding happens if $\hat \bx = \bx$.}
\label{figCh1:AWGN}
\end{figure}
Suppose you want to send a generic message $\bs$ made of $L$ symbols through an AWGN channel, where each symbol belongs to an alphabet composed of $B$ letters : $\bs \defeq \[s_l : s_l \in \{1,\ldots,B\}\]_{l=1}^L$. Starting from a standard binary representation of $\bs$, it is of course trivial to encode it in this form.

An alternative and highly {\it sparse} representation is given by the sparse superposition codes scheme: the representation $\bx$ of this message ${\bs}$ is made of $L$
sections of size $B$, where only a {\it unique} value is $\not = 0$ in
each section at the location
corresponding to the original symbol. We will consider each non zero value to be positive as it can be interpreted as an input energy in the channel. Thus if the $i^{th}$ component of the
original message $\bs$ is the $k^{th}$ symbol of the alphabet,
the $i^{th}$ section of $\bx$ contains only zeros, except at the
position $k$ where there is a positive value (which amplitude depends on
the power allocation).

As an example, in the simplest setting where the {\it power
  allocation} is $c_l = 1 \ \forall \ l\in\{1,\ldots,L\}$ (where $c_l$ is
the positive constant appearing in the $l^{th}$ section), if
$\bs = [a,c,b,c]$, where the alphabet has only three symbols
$\{a,b,c\}$, then its sparse representation $\bx$ is made of four sections which are $\bx=[[100],[001],[010],[001]]$. The $l^{th}$ section of
$\bx$ will be denoted $\bx_l \defeq [x_i]_{i\in l}$ where $l$
is both the set of indices corresponding to the $1$-d components of $\bx$ in the
$l^{th}$ section or the index of the section depending on the context.

In sparse superposition codes, $\bx$ is then encoded through a linear
transform by application of an operator $\bF$ of dimension $M\times N$
to obtain a codeword $\tilde \by$ of dimension $M$, $\tilde \by = \bF \bx$ which is then sent through the Gaussian noisy
channel and the receiver gets a corrupted version of it. This is summarized in  Fig.~\ref{figCh1:AWGN}. 

The dimension of the coding operator $\bF$ is linked to the section size $B$ and the coding (or transmission) rate in bits per-channel use $R$. Defining $K\defeq\log_2(B^L)$ as the number of informative bits carried by the signal $\bx$ made of $L$ sections of size $B$ (i.e. its entropy (\ref{eqChIntro:entropy}) considering that all the messages are equiprobable, see sec.~\ref{sec:basicsInfoTheory}), we have:
\begin{align}
R&\defeq K/M\\
&=L\log_2(B)/(\alpha N) \\
&=\log_2(B)/(B\alpha)\label{eq1:alpha_0} \\
\Leftrightarrow \alpha &\defeq M/N= \log_2(B)/(RB)\label{eq1:alpha} 
\end{align}
In what follows, we will concentrate on i.i.d Gaussian
$\bF$ elements with $0$ mean and variance $v_\bF$ in order to be able to obtain analytical results. We always fix to $1$ the total
power sent through the channel
$P \defeq ||\tilde \by||_2^2 = <\tilde \by^2> =1$ by a proper rescaling of the variance of the elements of
$\bF$. The only relevant parameter is thus the signal-to-noise ratio:
\begin{equation} 
{\rm{snr}}\defeq P/{\Delta}=1/{\Delta}
\end{equation}
where $\Delta$ is the variance of the Gaussian noise of the
AGWN channel. It allows to define the Shannon capacity $C=\log_2(1+{\rm snr})/2$ of the channel (\ref{eq:AWGN_C}), see sec.~\ref{sec:capacityAWGN} for its derivation.

\begin{figure}[th!]
\centering
\includegraphics[width=.7\textwidth]{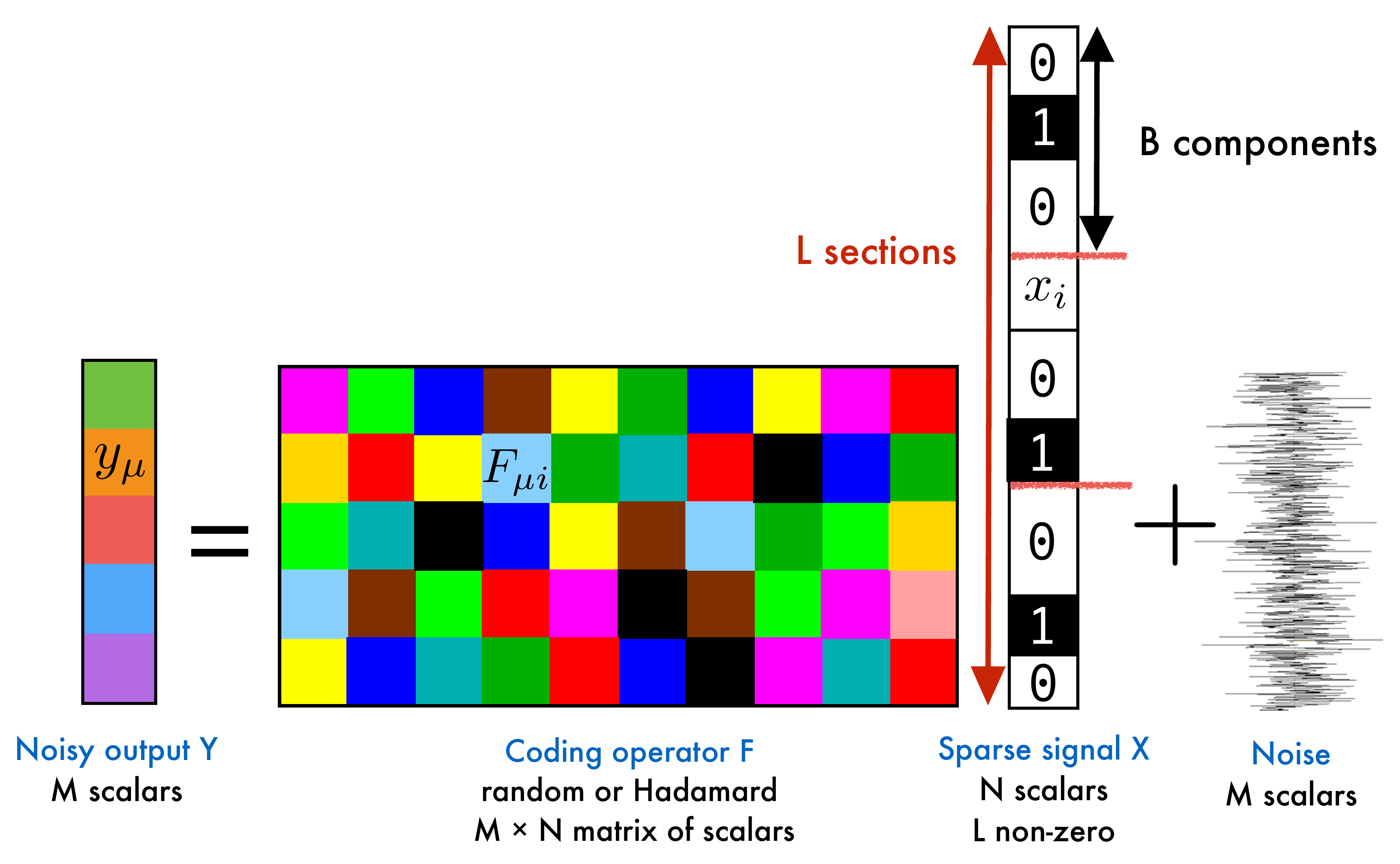}
\includegraphics[width=.7\textwidth]{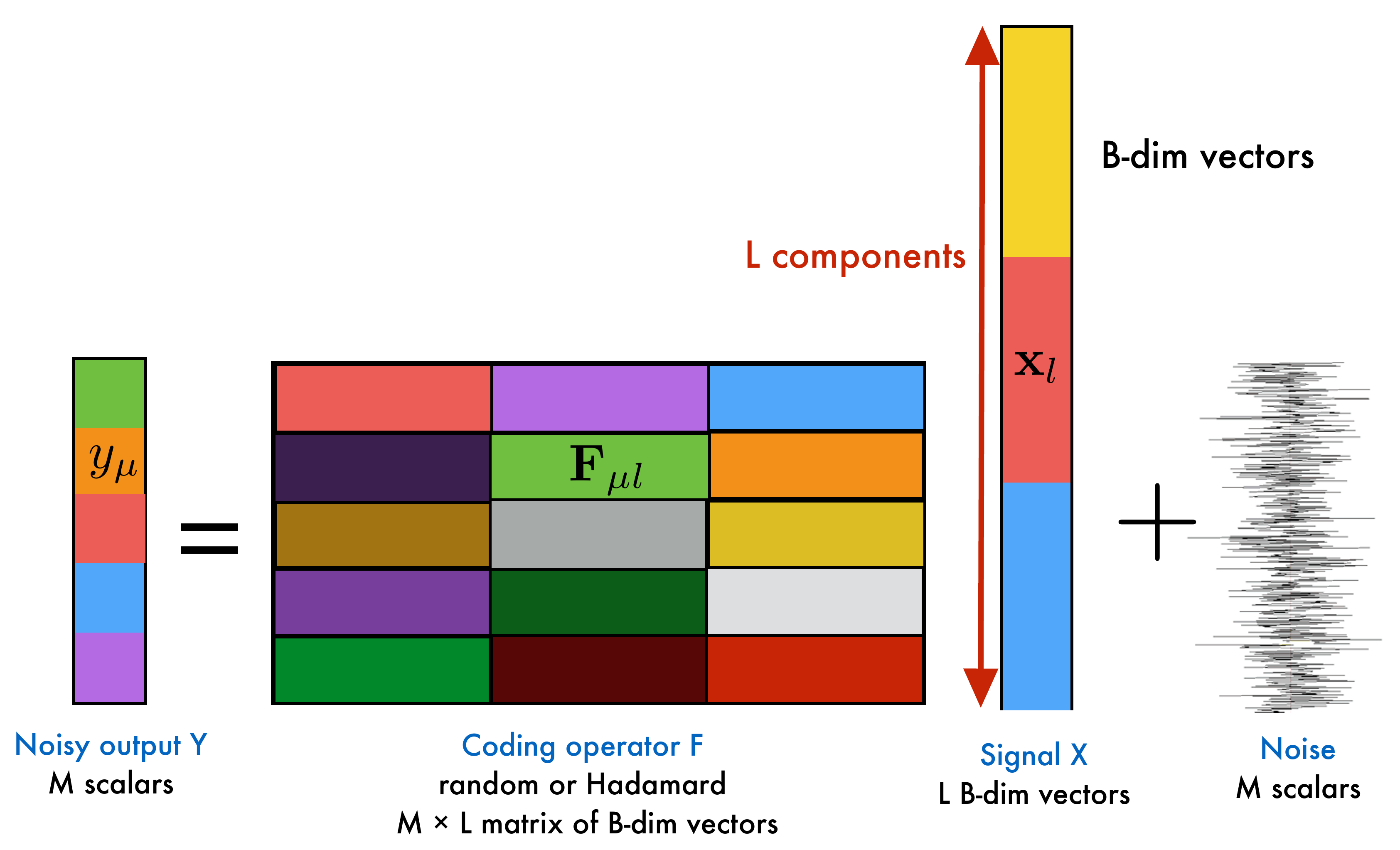}
\caption[Estimation problem associated to the decoding of the sparse signal over the AWGN channel and equivalence between the scalar and vectorial interpretations of the signal components]{\textbf{Up} : Representation of the estimation problem associated to the decoding of the sparse signal over the AWGN channel. All $1$-d variables in the same section $\bx_l=[x_i]_{i \in l}$ are strongly correlated due to the hard constraint that only one of them can be positive (or $1$ in this example). \textbf{Down} : Reinterpreting the same problem in terms of $B$-d variables. Now, the matrix elements of the previous figure are concatenated to form $B$-d vectors $\{\bF_{\mu l}\defeq[F_{\mu i}]_{i \in l}\}$ that are applied (using the usual scalar product for vectors) on the associated $B$-d vectors representing the new components of the signal, the sections. In this new setting, all the sections are uncorrelated.}
\label{figSuperCodes:op}
\end{figure}

The transmitted codeword $\tilde \by$ is corrupted through the model (\ref{eqIntro:AWGNCS}) by the AWGN channel. Let us now turn our attention to the decoding task. It is essentially a sparse linear estimation problem where we know $\by$ and need to estimate a sparse solution of $\by = \bF \bx + \bxi$. However the problem is different from the canonical compressed sensing problem \cite{Donoho:06} in that the elements of $\bx$ are strongly correlated by the constraint that only a single element in each section is non-zero, see Fig.~\ref{figSuperCodes:op}.
We thus prefer to think about the problem as a multidimensional one: each section $l \in \{1,2,\ldots,L\}$ made of $B$ $1$-d variables in $\bx$ is interpreted
as {\it a single} $B$-d variable for which we have a
strong prior information: it is zero in all dimensions {\it but} one
where there is a fixed positive known value. Given its length, we thus know
the vector must point in only one direction of the hypercube of
dimension $B$. In this new setting, instead of dealing with a $N$-d
vector $\bx$ with elements $\{x_i\}_i^N$, we deal with a $L$-d vector
$\bx$ which elements $\{\bx_l\}_l^L$ are $B$-d sections.

In this framework, the decoding problem becomes exactly of the kind considered in the Bayesian approach to compressed sensing, see sec.~\ref{sec:BayesianInferenceForCS} and for example \cite{Rangan10b,montanari2012graphical,KrzakalaPRX2012,KrzakalaMezard12,BarbierKrzakalaAllerton2012}. We can thus directly apply these techniques to the present problem. From now on, we always consider that the true ${\rm{snr}}$ is accessible to the channel user, and thus can be used in the algorithm.

We will be interested in two error estimators, the $MSE$ $E$ and the section error rate ${{SER}}$. They are defined respectively as the $MSE$ of the $1$-d variables and the fraction of wrongly decoded sections:
\begin{align}
E &\defeq \frac{1}{N} \sum_{i}^N (x_i - \hat x_i)^2 \label{eq1:biasedMSE}\\
{{SER}} &\defeq \frac{1}{L} \sum_{l}^L \mathbb{I}(\bx_l \neq \hat\bx_l) \label{eq1:SER}
\end{align}
where $\mathbb{I}\left(A\right)$ is the indicator function of the
event $A$ which is one if $A$ happens to be true, zero else and
$\hat \bx\defeq [\hat \bx_l]_l^L=[\hat x_i]_i^N$ is the final estimate of the signal by the decoder.
\section{Approximate message-passing decoder for superposition codes}
\begin{figure}
\centering
\includegraphics[width=.5\textwidth]{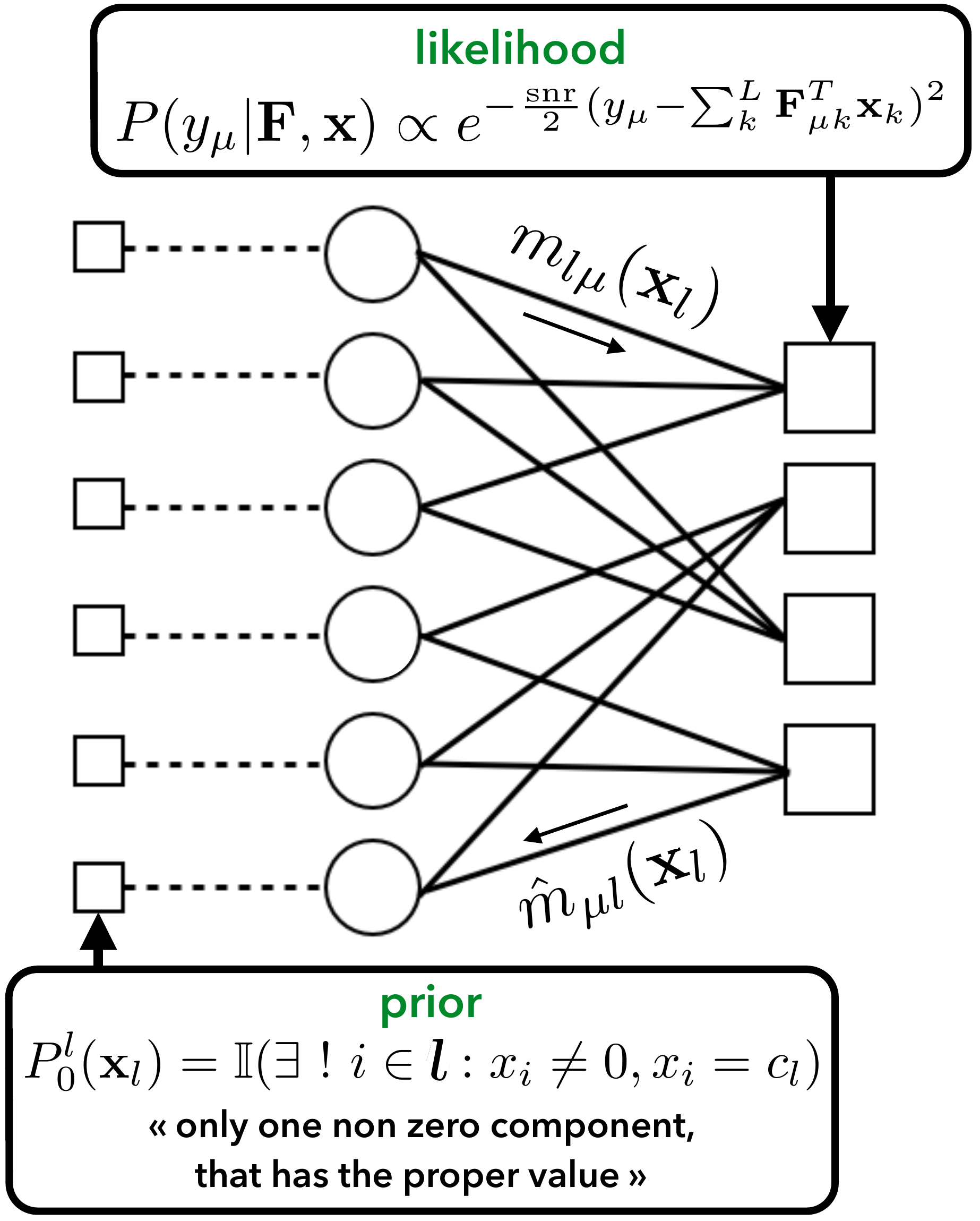}
\caption[Factor graph associated to the sparse
  superposition codes]{Factor graph associated to the sparse
  superposition codes. It is a bipartite graph where the variable estimates $\{\bx_l\}_l^L$ are
  represented by circles, the constraints (or factors) by squares. The
  variables are constrained by the $M$ likelihood factors that enforce the system $\by = \bF \bx$ to be fulfilled up to Gaussian fluctuations due to the Gaussian noise of the AWGN channel. The prior constraints enforce each section to have only one non-zero component, that must be the value fixed by the power allocation. In the homogeneous operator case, the
  variables are connected to all the likelihood factors
  and vice versa and in the spatially-coupled case, only to a finite
  fraction that depends on the spatial coupling ensemble, see
  Fig.~\ref{fig_opSpCoupling}. The factor-to-node $\hat m_{\mu l}(\bx_l)$ and node-to-factor $m_{l \mu}(\bx_l)$ cavity messages are represented. They should stand on the same edge as they depend on the same variable but we put them on distinct edges for readibility purpose.}
\label{fig_factorSC}
\end{figure}
The only problem-dependent objects in the AMP Fig.~\ref{algoCh1:AMP_op} are the denoising functions $\{f_{a_i}, f_{c_i}\}$. Let us derive them for any power allocation $\{c_l > 0\}_l^L$. Here, the prior that matches the signal distribution by enforcing the constraint of having only one known value $c_l>0$ per section is:
\begin{align}
P_0^l(\bx_l) &\defeq \frac{1}{B} \sum_{i \in l}^B \delta(x_i-c_l)\prod_{j \in l: j \neq i}^{B-1}\delta(x_j) \label{eq1:prior}
\end{align}
The denoisers which generic expressions are given by (\ref{eq1:fai}) and (\ref{eq1:fci}) are easily derived. We obtain the posterior average $a_i^{t}$ and variance $v_i^{t}$ of $x_i$ at step t of the algorithm:
\begin{align}
a_i^{t} &\defeq f_{a_i}((\bsy \Sigma_{l_i}^{t})^2,\bR_{l_i}^{t}) = c_{l_i}\frac{\exp\(-\frac{c_{l_i}(c_{l_i}-2R_i^{t})}{2(\Sigma_i^{t})^2}\)}{\sum_{j \in l_i}^B \exp\({-\frac{c_{l_i}(c_{l_i}-2R_j^{t})}{2(\Sigma_j^{t})^2}}\)} \label{eq1:meani}\\
v_i^{t} &\defeq f_{c_i}((\bsy \Sigma_{l_i}^{t})^2,\bR_{l_i}^{t}) =  a_i^{t} (c_{l_i} - a_i^{t} )
\end{align}
where $\bsy \Sigma_{l_i},\bR_{l_i}$ are the AMP fields of the section $l_i$ to which the $i^{th}$ $1$-d component of the signal belongs to. Combined with Fig.~\ref{algoCh1:AMP_op}, we thus get the full AMP algorithm for sparse superposition codes with associated graphical model given by Fig.~\ref{fig_factorSC}.
\subsection{The fast Hadamard-based coding operator}
\label{susec:fastHad}
In the present study, we use spatially-coupled operators constructed as in Fig.~\ref{fig_seededHadamard}.
\begin{figure}[!t]
\centering
\includegraphics[width=0.9\textwidth]{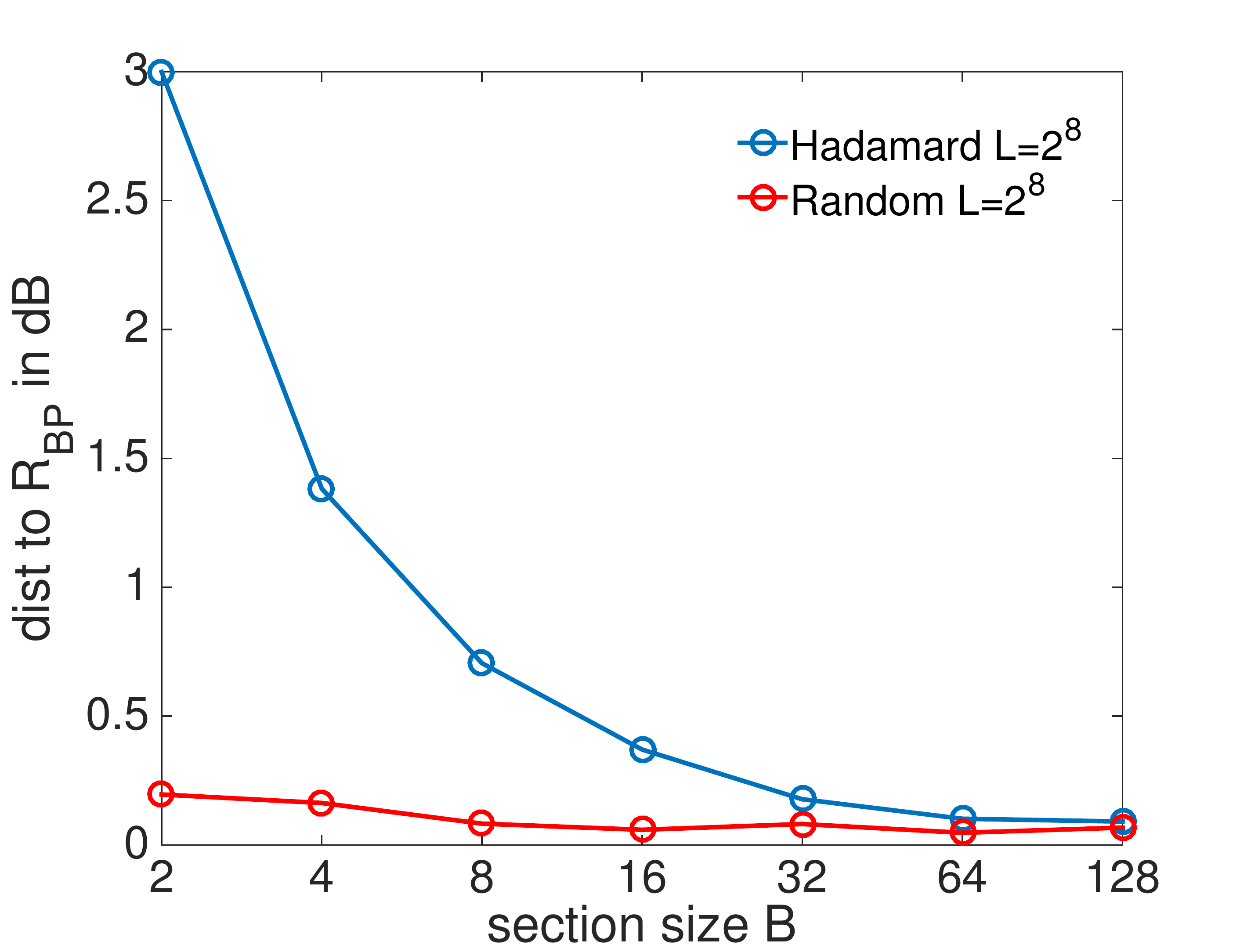}
\caption[Convergence of structured operators performances to the random operator ones]{Comparison between the distance in dB to the asymptotic BP threshold $R_{BP}({\rm{snr}}=100,B)$ at which the AMP decoder with homogeneous Hadamard-based coding operators (blue) or random i.i.d Gaussian ones (red) starts to reach an ${{SER}} <10^{-5}$, which essentially means perfect decoding for most of the instances. This is done for a fixed number of sections $L=2^8$ and ${\rm{snr}}=100$. Each point have been averaged over $100$ random instances. The BP threshold $R_{BP}({\rm{snr}}=100,B)$ is obtained by state evolution analysis for each $B$. Decoding with Hadamard-based operators works poorly when the signal density increases (i.e. when $B$ decreases), but matches quickly the random matrix performances as it decreases. Decoding with random Gaussian i.i.d matrices has a performance that is close to constant as a function of $B$ at fixed $L$ as it should (the relevant signal size is $L$).}\label{figCh1:distToRbp}
\end{figure}
Fig.~\ref{figCh1:distToRbp} shows that when the signal sparsity increases, i.e. when the section size $B$ increases, using Hadamard-based operators becomes quickly equivalent to using random i.i.d Gaussian ones in terms of performances (as observed in chap.~\ref{chap:structuredOperators}). For the figure, we have fixed the ${\rm snr}=100$ and then plotted the distance in dB to the BP threshold $R_{BP}({\rm snr}=100,B)$ at which the decoder starts to decode perfectly with Hadamard-based or random i.i.d Gaussian operators. We remind that $R_{BP}$ is defined as the highest rate until which AMP decoding is optimal without the need of non constant power allocation or spatial coupling. It appears that at low section size, it is advantageous to use random operators but as $B$ increases, structured operators quickly match their performances. The BP threshold is predicted by the state evolution analysis presented in Sec.~\ref{sec:SE}.
\newpage
\section{State evolution analysis for random i.i.d homogeneous operators with constant power allocation}
\label{sec:SE}
As usual, for the state evolution analysis we consider the case of an i.i.d Gausian matrix $\bF$, such that the recursions obtained in sec.~\ref{sec:stateEvolutionGeneric} are valid. Here we consider the matrix to be homogeneous and a constant power allocation. We will use the state evolution to predict the results with the Hadamard operator as well, as we have shown in chap.~\ref{chap:structuredOperators} that the analysis derived in the random i.i.d case is a good predictive tool of the behavior of the AMP decoder with structured operators, despite not perfect nor rigorous.
\newpage
\begin{figure}[H]
\centering
\includegraphics[width=0.7\linewidth, trim=0 25 0 0, clip=true]{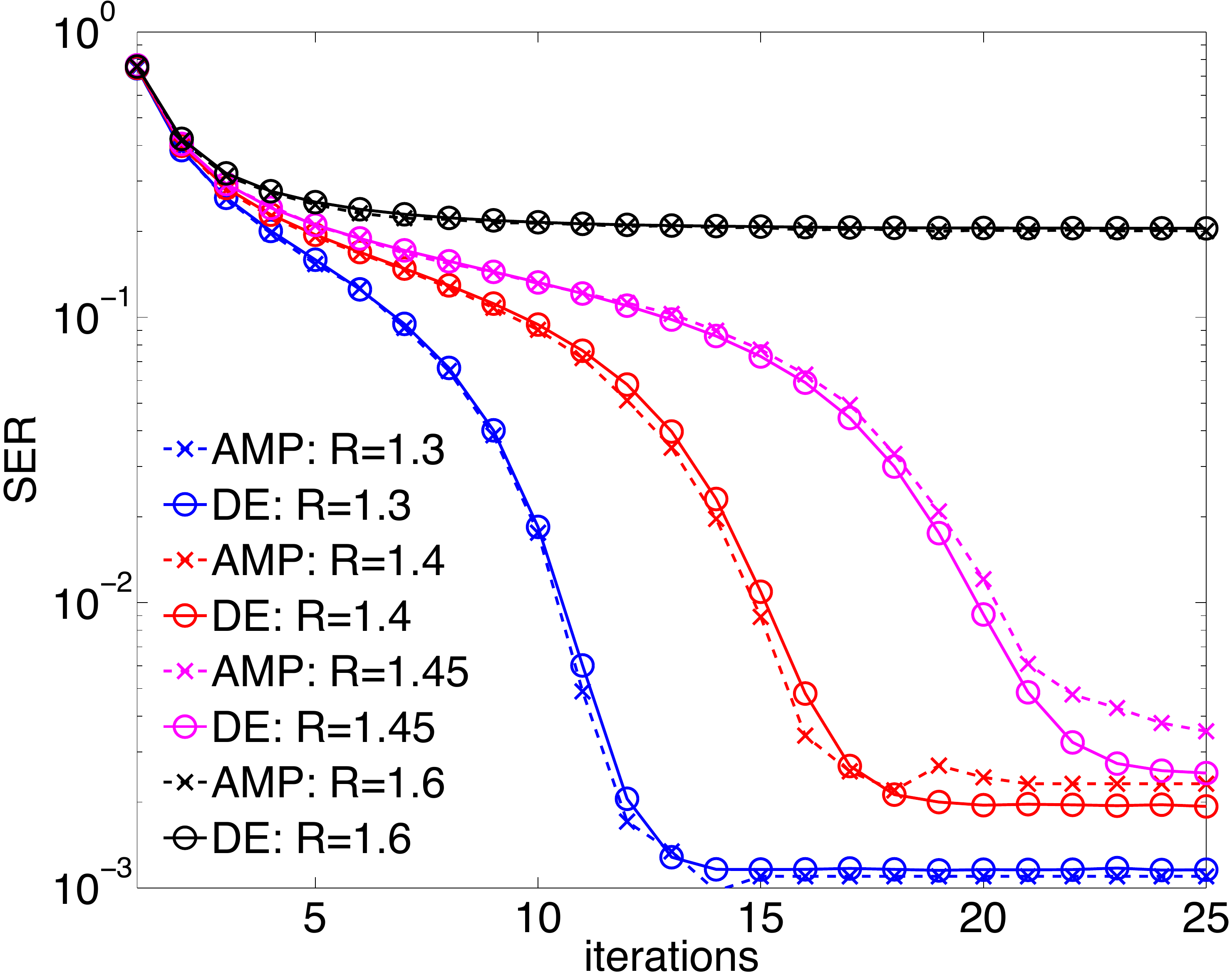}
\includegraphics[width=0.71\textwidth, trim=20 30 70 5, clip=true]{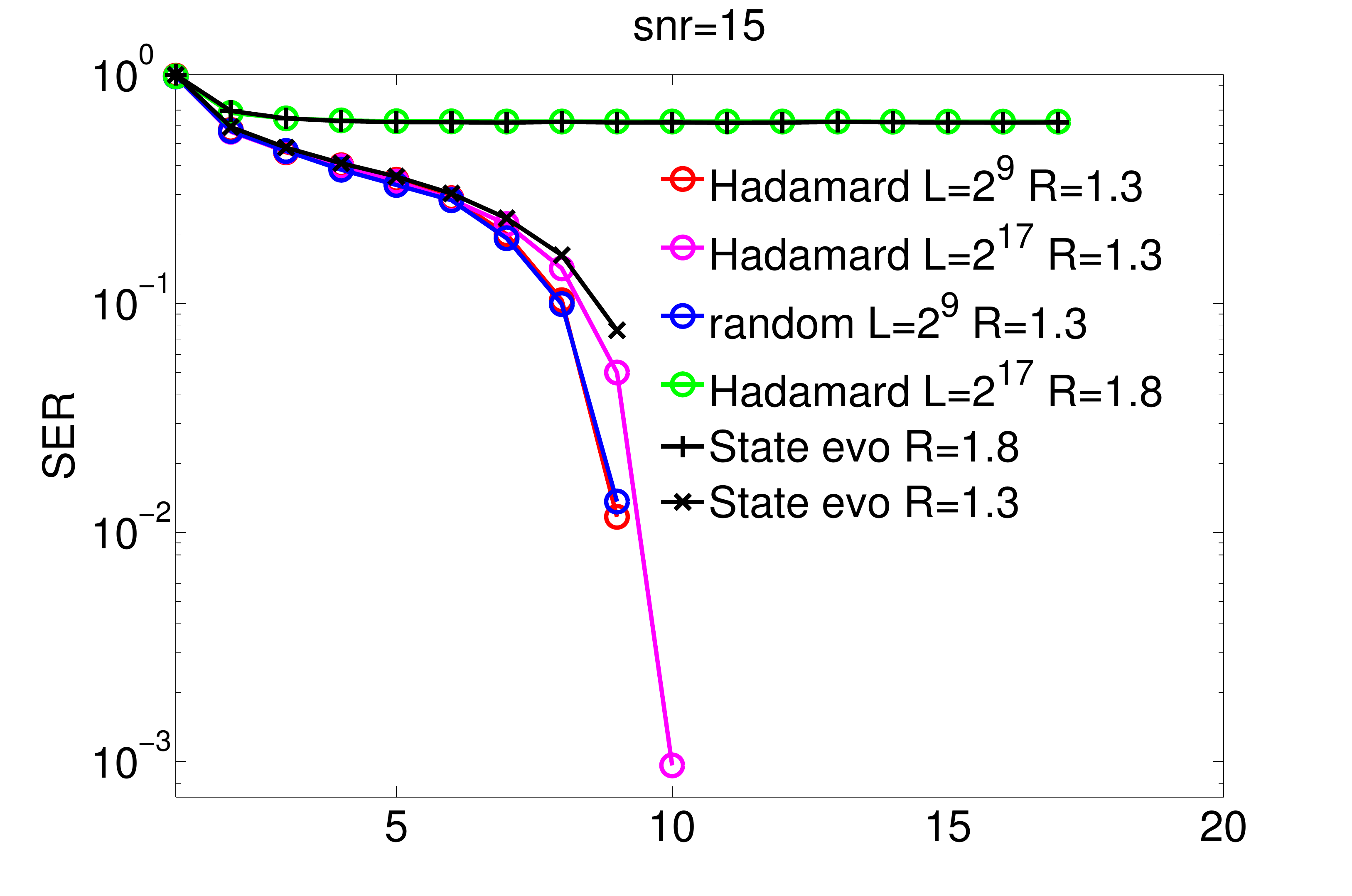}
\caption[State evolution and decoder performances with homogeneous matrices at ${\rm{snr}}=15$]{\tbf{Up} : The state evolution prediction (solid lines) of the ${{SER}}^t$ compared to the actual one of the algorithm on single instances for ${\rm{snr}}=15$, different rates and a section size $B=4$ (to be compared to the BP threshold at $R_{BP}({\rm{snr}}=15,B=4)=1.55$). The matrix is i.i.d Gaussian and we consider a constant power allocation $\{c_l=1\}_l^L$. The state evolution is computed by monte carlo with a sample size of $10^7$ and the signal
size for AMP is $L=2^{13}$. \tbf{Down} : The same as the upper plot with ${\rm{snr}}=15$, different rates $R$ (one above and one below $R_{BP}$) but with a larger section size $B=64$. The BP threshold is here $R_{BP}({\rm{snr}}=15,B=64)=1.47$. We consider different signal sizes $L$ and homogeneous Hadamard-based or purely random i.i.d Gaussian operators. The state evolution is computed by monte carlo with a sample size of $10^6$ as $B$ is larger and thus the monte carlo computation requires more time. The finite size curves stop without reaching a noise floor because the recovery is actually perfect and the final ${{SER}}=0$ which is due to the finite size effects. The same happens for the theoretical curves that reach $0$ due to finite numerical precision. We observe that when the signal size $L$ is big (and thus that we must use the Hadamard operator), the algorithm behavior follows the theoretical predictions closely.}
\label{illustrDE_15}
\end{figure}
\begin{figure}[!h]
\centering
\includegraphics[width=0.71\textwidth, trim=20 30 70 5, clip=true]{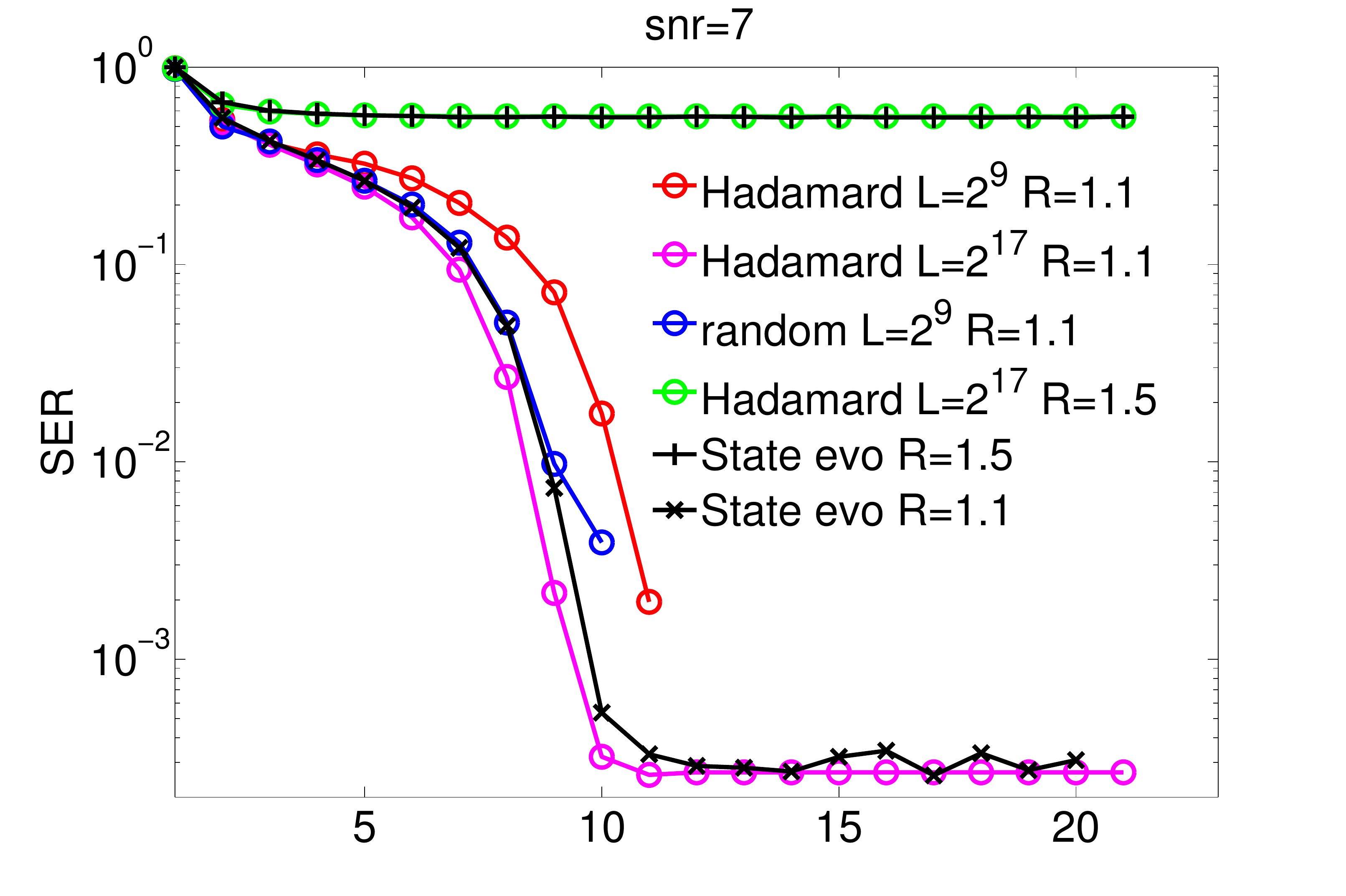}
\includegraphics[width=0.71\textwidth, trim=20 30 70 5, clip=true]{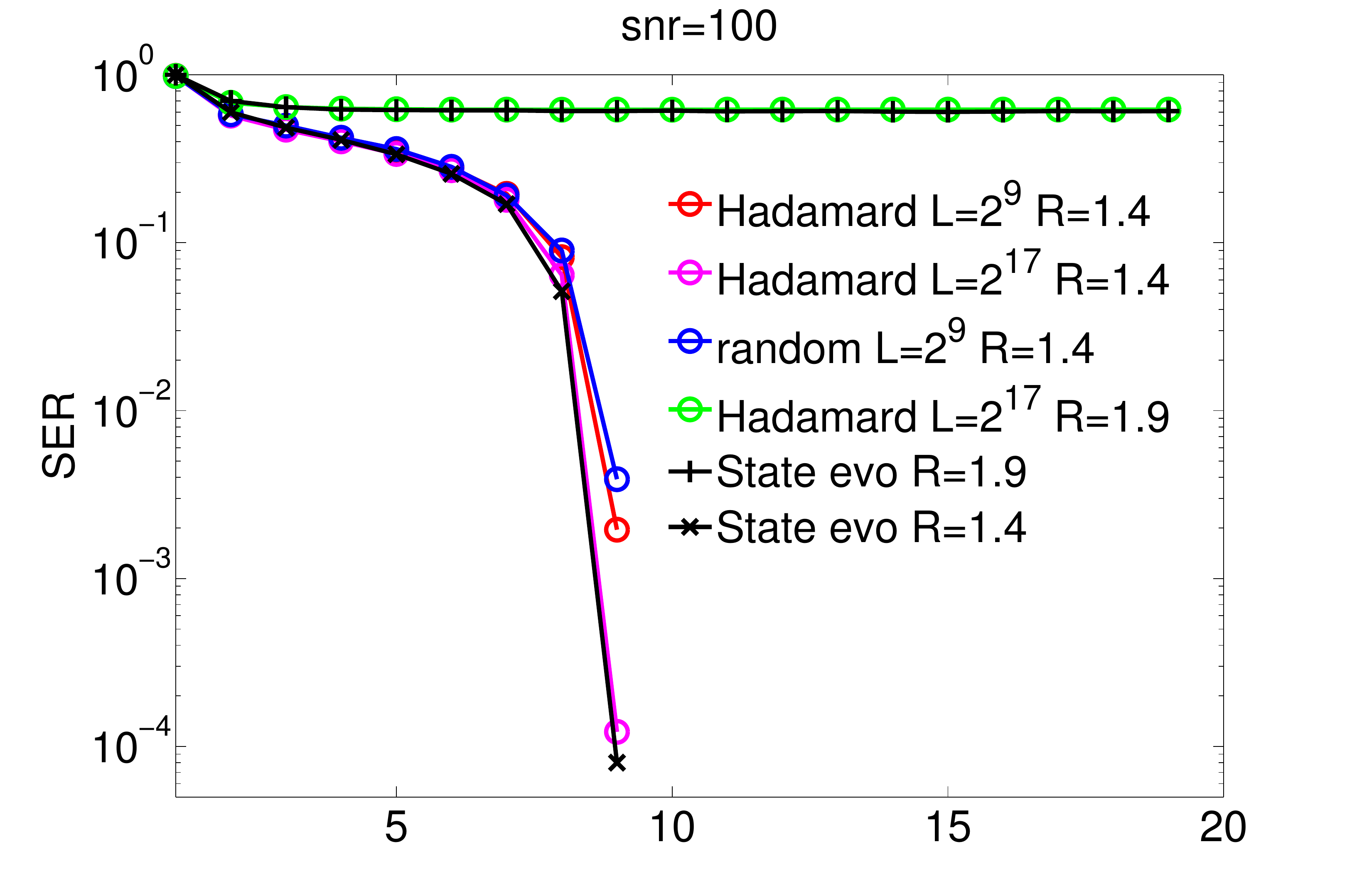}
\caption[State evolution and decoder performances with homogeneous matrices at ${\rm{snr}}=7/100$]{The state evolution prediction of the section error rate ${{SER}}^{t}$ (black curves), compared to the actual one of the AMP decoder for ${\rm{snr}}=7/100$, a section size $B=64$, different rates $R$ (one above and one below the BP threshold $R_{BP}({\rm{snr}}=7,B=64)=1.275, R_{BP}({\rm{snr}}=100,B=64)=1.625$), different signal sizes $L$ in the homogeneous Hadamard-based or purely random i.i.d Gaussian operator case with constant power allocation $\{c_l=1\}_l^L$. The state evolution is computed by monte carlo with a sample size of $10^6$.}
\label{figCh1:SEfull}
\end{figure}
We define $\tilde \Sigma^{t+1}(E^{t}) \defeq \Sigma^{t+1}(E^{t}) \sqrt{\log(B)}$ as this expression will be more convenient. $\Sigma^{t+1}(E^{t})$ is given by (\ref{eqChIntro:defSigma2}) where we use (\ref{eq1:alpha}) to express $\alpha$ in function of the rate. Starting from (\ref{eq_E1_1}), using (\ref{eqChIntro:defR}) and the prior (\ref{eq1:prior}) for sparse superposition codes, we get:
\begin{equation}
E^{t+1} = \frac{1}{B}\int \mathcal{D} \bz \left([f_{a_{1|1}}(\tilde\Sigma^{t+1})^2,\bz )-1]^{2}+(B-1)f_{a_{2|1}}((\tilde\Sigma^{t+1})^2,\bz )^{2}\right) \label{eq1:SE_MSE}
\end{equation}
where we define:
\begin{align}
f_{a_{i|i}}(\tilde\Sigma^2,\bz ) &\defeq  \left[1 + e^{-\frac{\log(B)}{\tilde\Sigma^2}} \sum_{1\le j\le B : j\ne i}^{B-1}  e^{\frac{\sqrt{\log(B)}(z_j-z_i)}{\tilde\Sigma}} \right]^{-1} \label{eq1:fa1fun} \\
f_{a_{j|i}}(\tilde\Sigma^2,\bz )  &\defeq  \left[1 + e^{\frac{\log(B)}{\tilde\Sigma^2} + \frac{\sqrt{\log(B)}(z_j - z_i)}{\tilde\Sigma}}  + \sum_{1\le k\le B : k \ne i,j }^{B-2} e^{\frac{\sqrt{\log(B)}(z_k-z_i)}{\tilde\Sigma}} \right]^{-1} \label{eq1:fa2fun}
\end{align}
Under the constant power allocation assumption, the quantity $f_{a_{i|i}}$ ($f_{a_{j|i}}$) can be interpreted as the asymptotic AMP posterior probability estimate of the $i^{th}$ component ($j^{th}$ component) to be the $1$ given that it is indeed the $1$ in the signal (given that it is actually the $i^{th}$ component that is the $1$ in the signal). In this approach, there is a one to one correspondance from the value of the $MSE$ to the ${{SER}}$ thanks to the mapping:
\begin{equation}
{{SER}}^{t+1} = \int  \mathcal{D}\bz \ \mathbb{I}\left(\exists \ j \in \{2,\ldots,B\} : f_{a_{j|1}}((\tilde \Sigma^{t+1})^2,\bz) > f_{a_{1|1}}((\tilde\Sigma^{t+1})^2,\bz) \right)
\label{eq1:SE_SER}
\end{equation}
From this equation, we can exactly predict the asymptotic $L\to\infty$ evolution in time of the algorithm, such as in Fig.~\ref{illustrDE_15} and Fig.~\ref{figCh1:SEfull}. The state evolution on these plots represent the iteration of (\ref{eq1:SE_SER}), (combined with (\ref{eq1:SE_MSE}), (\ref{eqChIntro:defSigma2})) for different experimental settings $({\rm{snr}},R,B)$ using homogeneous Hadamard-based or random i.i.d Gaussian matrices. (\ref{eq1:SE_SER}) and (\ref{eq1:SE_MSE}) are computed at each time step by monte carlo technique. We observe that for the ${\rm{snr}}=15/100, B=64$ cases, the experimental and theoretical curves stop at some iteration without reaching a noise floor. For the experimental curves, this is due to the fact that in order to observe an ${{SER}} \in O(\epsilon)$, there must be at least
$L\approx 1/\epsilon$ sections which is not the case for signals of reasonnable sizes, when the asymptotic ${{SER}}$ is very small. This finite size effect is actually in favor of the reconstruction performances. 
In fact, when the rate is below the BP threshold, the
decoding is usually perfect and is found to reach with high
probability ${{SER}}=0$. The black asymptotic curves should anyway reach a finite error floor but they do not because it is so low that the sample size used in the monte carlo computation sould be way too large to deal with by the same argument. But on Fig.~\ref{illustrDE_15}, for smaller $B=4$ the error floor is higher than for $B=64$ and thus we can see it numerically.

Another observation, natural from the definition of the state evolution as an asymptotic analysis, is that the theoretical and experimental results match better for larger signals. At rate $R>R_{BP}$ (green or black experimental curves and associated theoretical ones), we see that the AMP decoder does not reconstruct the signal and converges to an high ${{SER}}$ solution well predicted by the state evolution. On the contrary, below the threshold, the reconstruction works fine up to an error floor dependent on the parameters $(B,{\rm{snr}},R)$. We also observe as in chap.~\ref{chap:structuredOperators} that the state evolution predicts well the final performances of AMP with Hadamard-based operators.
\section{State evolution analysis for spatially-coupled i.i.d
  operators or with power allocation}
\label{sec:SEseeded}
Thanks to the spatial-coupling we can asymptotically reach the optimal rate. The total rate is a function of the rate of the blocks. From (\ref{eq_alphaRest}) combined with (\ref{eq1:alpha}) we deduce it:
\begin{equation}
R = \frac{L_c R_{rest} R_{seed} }{(L_r-1)R_{seed} + R_{rest} } \underset{L_c, L_r \to \infty}{\longrightarrow} R_{rest}<R_{opt}({\rm snr}, B)
\end{equation}
where $R_{rest}$ can be asymptotically as large as the Bayes optimal rate $R_{opt}({\rm snr}, B)$ defined as the highest rate until which the superposition codes scheme allows to decode, see sec.~\ref{sec:typicalPhaseTransitions}.
\begin{figure}[h!]
\centering
\includegraphics[width=0.71\textwidth, trim=10 30 65 5, clip=true]{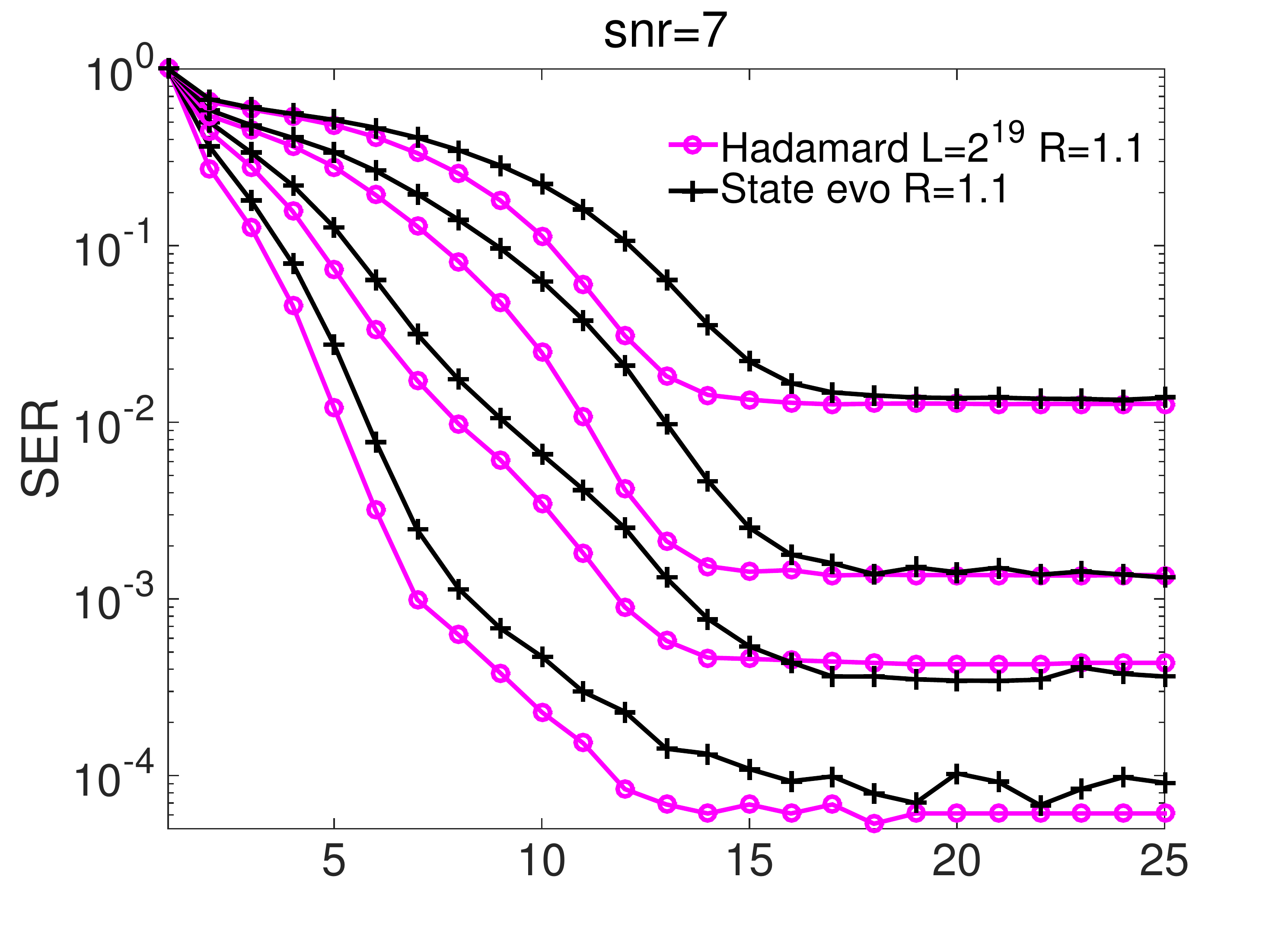}
\includegraphics[width=0.71\textwidth, trim=10 30 65 5, clip=true]{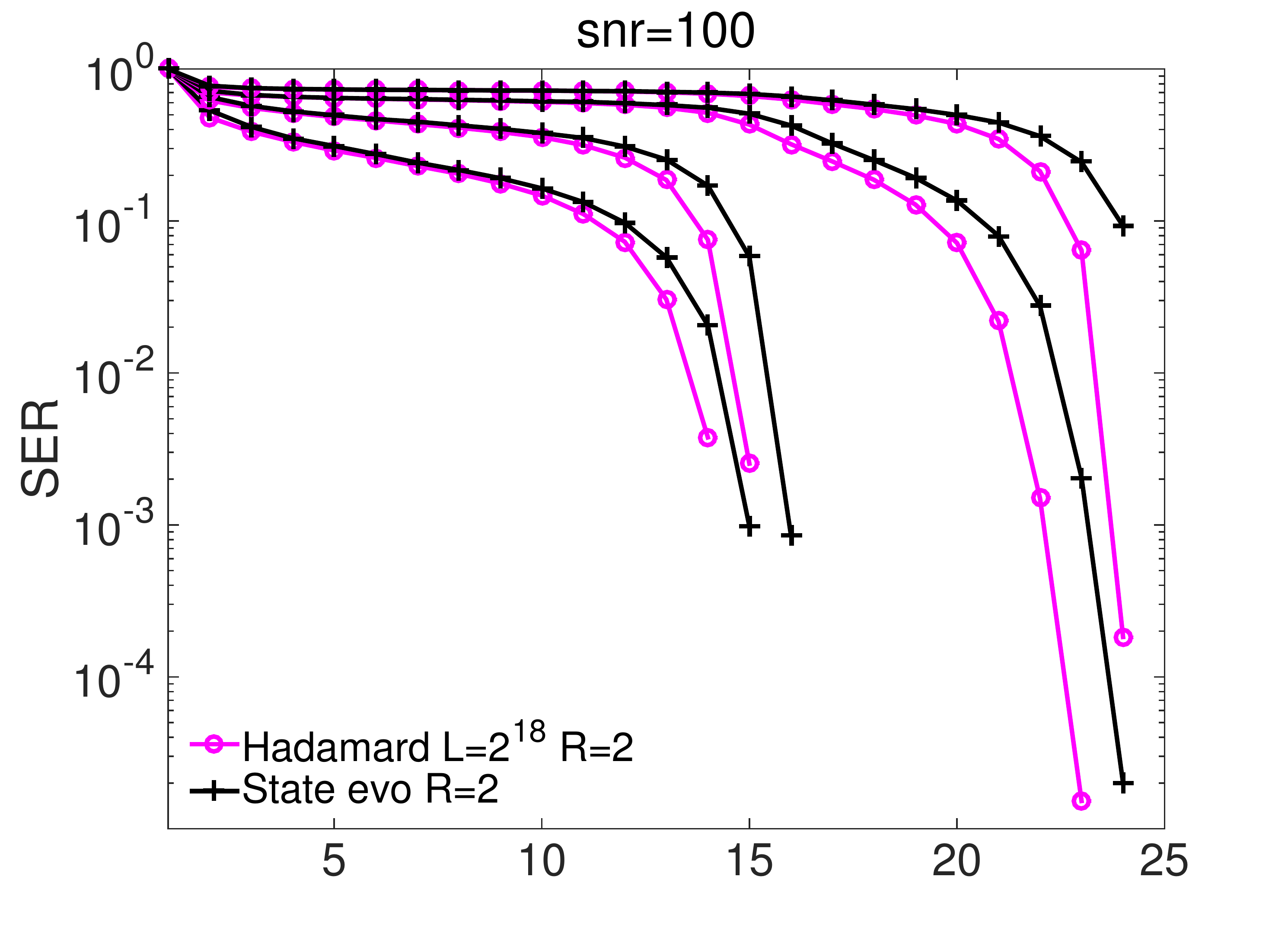}
\caption[State evolution for spatially-coupled superposition codes]{The state evolution prediction of the section error rate $\{{{SER}}^{t}_c\}_c^{L_c=4}$ for each of the four induced block of the signal (see Fig.~\ref{fig_seededHadamard}) as a function of time (black curves), compared to the actual error of the algorithm for ${\rm{snr}}=7/100$, different rates $R$, a section size $B=32$, different signal size $L$ values with a spatially-coupled Hadamard-based operator. The operator is drawn from the ensemble $(L_c=4,L_r=5,w=2,\sqrt{J}=0.6,R, \beta_{{\rm{seed}}}=1.5)$. The power allocation is constant. The state evolution is computed by monte carlo with a sample size of $10^6$. The finite size curves at high ${\rm{snr}}$ stop without reaching a noise floor because the recovery is actually perfect due to the finite size effects, the same happens for the theoretical curves due to finite numerical precision. In the low ${\rm{snr}}=7$ case, the error floor (different in each block) is well predicted by state evolution.}\label{figCh1:SEspc}
\end{figure}
Let us now derive the state evolution recursions in the spatially-coupled operator case.
As in the homogeneous case, defining the rescaled $\tilde\Sigma_c^{t+1} \defeq \Sigma_c^{t+1} \sqrt{\log(B)}$, from (\ref{eq_E1}), 
(\ref{eq_SEsigmaSeeded}) and (\ref{eqChIntro:defRseeded}), we obtain the following state evolution for the $MSE$ $E_c$ inside the block $c$ in the $L\to\infty$ limit:
\begin{align}
&E_c^{t+1} = \frac{1}{B}\int \mathcal{D} \bz \left([f_{a_{1|1}}((\tilde\Sigma_c^{t+1})^2,\bz )-1]^{2}+(B-1)f_{a_{2|1}}((\tilde\Sigma_c^{t+1})^2,\bz )^{2}\right) \\
&\tilde\Sigma_c^{t+1}\(\{E_{c'}^{t}\}_{c'}^{L_c}\) = \sqrt{\log(B)} \left[B\sum_{r}^{L_r} \frac{\alpha_{r} J_{rc}}{{L_c/{\rm{snr}}} + B\sum_{c'}^{L_c} J_{rc'}E_{c'}^{t}}\right]^{-1/2} \label{eq1:SEsigmaSeeded}
\end{align}
where the $f_a$ functions (\ref{eq1:fa1fun}), (\ref{eq1:fa2fun}) are defined in the previous section and where the mapping to the ${{SER}}^{t+1}_c$ per block is given by:
\begin{equation}
{{SER}}^{t+1}_c = \int  \mathcal{D}\bz \ \mathbb{I}\left(\exists \ j \in \{2,\ldots,B\} : f_{a_{j|1}}((\tilde\Sigma_c^{t+1})^2,\bz) > f_{a_{1|1}}((\tilde\Sigma_c^{t+1})^2,\bz) \right)  \ \forall\ c\in\{1,\ldots,L_c\}
\label{eq1:DE_E}
\end{equation}
Thanks to this analysis, we can now predict the asymptotic ${{SER}}$ per block in the signal estimate by the AMP decoder. Fig.~\ref{figCh1:SEspc} shows a comparison of the ${{SER}}$ per block $\{{{SER}}_c^{t}\}_c^{L_c}$ predicted by state evolution (black curves) with the actual ${{SER}}$ per block of the superposition codes with the AMP decoder combined with an Hadamard-based spatially-coupled operator on a single instance. The discrepancies between the theoretical and experimental curves come from the fact that state evolution is derived for random i.i.d Gaussian matrices, but the final error using these Hadamard operators is the same as predicted by state evolution as observed in the ${\rm snr} = 7$ case. In the high ${\rm{snr}}$ regime, the curves stop for the same reasons as the Fig.~\ref{figCh1:SEfull} of the previous section and it means that the decoding was perfect. As noted in chap.~\ref{chap:structuredOperators}, structured operators converge faster to the predicted final error than purely i.i.d matrices as predicted by the state evolution.
\subsection{State evolution for power allocated signals}
\label{sec:powA_SE}
\begin{figure}[!t]
\centering
\includegraphics[width=.9\textwidth]{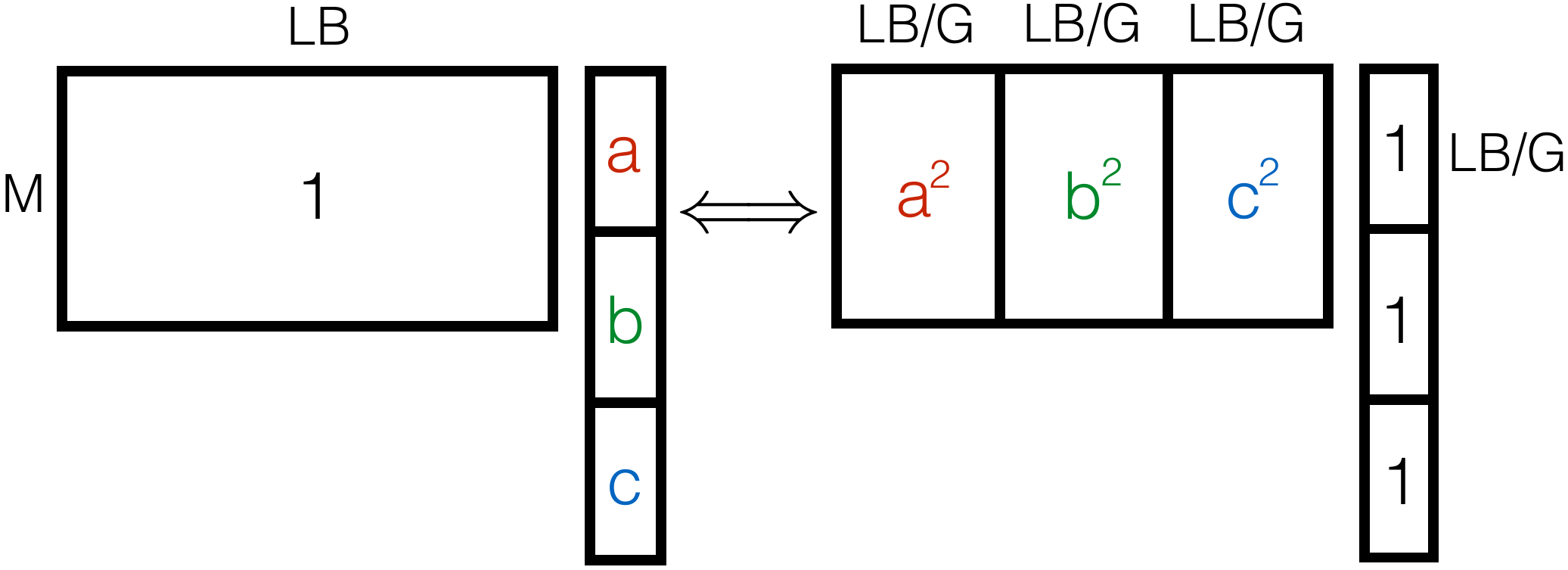}
\caption[Equivalence between non constant power allocation and a particular strutured operator]{The figure shows how to convert a system with any power allocated signal encoded through an homogeneous operator with elements of variance $1/L$ into an equivalent system with constant power allocation encoded by a structured operator. The values on the matrix represent the variance of the elements of the matrix (we drop here the rescaling by $1/L$), the values on the signal represent the non zero values of the sections that belong to a given group: here the signal is decomposed into $G=3$ groups, and all the sections inside the first group have a non zero value equal to a, and so on. The transformation is done by structuring the operator into block-columns, with as many blocks as different values in the power allocation, or groups: if a column of the original matrix acts on a component of a section where the non zero value is $u$, then this column variance is multiplied by $u^2$ in the new structured operator (such that the elements of this column are multiplied by $u$). The different sizes of the matrix blocks and signal groups are represented.}
\label{figCh1:equivPowaSpc}
\end{figure}

We now observe that we can trivially obtain the state evolution for any power allocation of the signal encoded with a random i.i.d Gaussian matrix from the previous analysis, thanks to the transformation of Fig.~\ref{figCh1:equivPowaSpc}: starting from an homogeneous matrix and a given power allocated signal, we convert the system into a structured matrix with a constant power allocated signal. 

Suppose the signal is decomposed into $G$ groups, where inside the group $g$, the power allocation is the same for all the sections belonging to this group and equals $c_g$. Now one must create a structured operator starting from the original homogeneous one, decomposing it into blocks with $LB/G$ columns and multiply all the elements of the block-column $g$ by $c_g$, as shown in Fig.~\ref{figCh1:equivPowaSpc}. The system with this new operator acting on a constant power allocated signal is totally equivalent to the original system and we have the state evolution of this new system from the previous analysis. Using (\ref{eq1:SEsigmaSeeded}) in the present setting, one has to be careful with the value of $\alpha_r$ defined as the number of lines over the number of columns of the block-line $r$ (which is unique). Here there is a unique value that equals $M/(N/G)=G\alpha$ where $\alpha$ is defined as the original measurement rate (\ref{eq1:alpha}). Given that, $L_c=G$ we finally obtain:
\begin{align}
&E_g^{t+1} = \frac{1}{B}\int \mathcal{D} \bz \left([f_{a_{1|1}}((\tilde \Sigma_g^{t+1})^2,\bz
  )-1]^{2}+(B-1)f_{a_{2|1}}((\tilde\Sigma_g^{t+1})^2,\bz )^{2}\right) 
\label{eq:SE_POWA_E}\\
&\tilde\Sigma_g^{t+1}\(\{E_{g'}^{t}\}_{g'}^G\) = \sqrt{\log(B)}
  \left[B\frac{\alpha c_g^2}{{1/{\rm{snr}}} + B/G\sum_{g'}^{G}
  c_{g'}^2E_{g'}^{t}}\right]^{-1/2}
\label{eq:SE_POWA_S}
\end{align}
The square $c_g^2$ appears because multiplying the matrix elements by $c_g$ multiply their variance by its square.
\section{Replica analysis and phase diagram}
\label{sec:replica}
\begin{figure}[!t]
\centering
\includegraphics[width=1\textwidth]{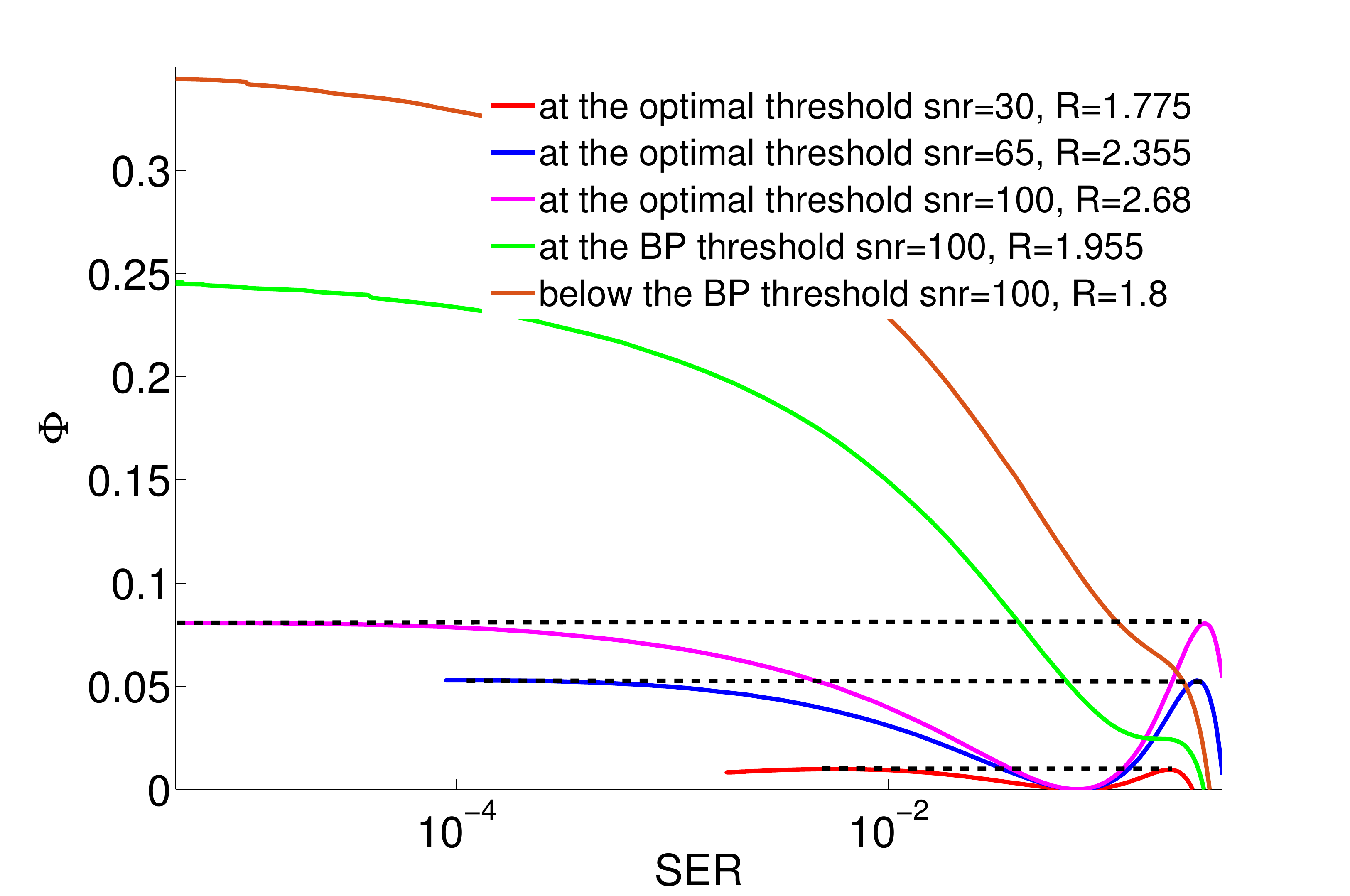}
\caption[Bethe free entropy for sparse superposition codes at the different transitions]{The Bethe free entropy (or potential) $\Phi({{SER}})$ for $B=2$, different rates and ${\rm{snr}}$. The
maxima of the curves correspond to the typical ${{SER}}$ which are fixed points of the state evolution equations (\ref{eq1:SE_SER}) for a given set of parameters $(R,B,{\rm{snr}})$. The global maximum is the (exponentially) most probable ${{SER}}$ solution, i.e. the equilibrium state refered also as the optimal ${{SER}}$. The curves are obtained by numerical integration of (\ref{eq1:freeEnt2_2}). The optimal threshold $R_{opt}(B,{\rm{snr}})$ is defined as the rate where the high and low error maxima have same height (i.e. same probability), see pink, blue and red curves. The BP threshold $R_{BP}(B,{\rm{snr}})$ is the rate at which the metastable state a high error that blocks the convergence of AMP appears, see green curve. The plot illustrates how the gap at the optimal threshold between the two maxima increases with the ${\rm{snr}}$. 
 $\textbf{snr}\boldsymbol{= 100}$ : Here for rates larger than $R>2.68$, the
optimal ${{SER}}$ jumps from a low value to a large $O(1)$ one (pink
curve). This defines the maximum possible rate (to compare here to
$C=3.3291$) below which acceptable performance can be obtained with AMP combined with spatial coupling or non constant power allocation. For
$R<2.68$, the ${{SER}}$ is much lower (and decay with $R$). The AMP
decoder Fig.~\ref{algoCh1:AMP_op} allows to perform an ascent
of this function. As long as the maximum is unique (i.e. for
$R<1.955$, see green curve), it will be able to achieve the predicted optimal 
performance with no need of spatial coupling or non constant power allocation in the large size limit, as in the case of the brown curve.}
\label{fig_freeEnt}
\end{figure}
Let us now compute the asymptotic $L\to\infty$ free entropy by the replica method. This potential of the ${{SER}}$ is derived in the constant power allocation case. We are more interested in this case as we will show later with Fig.~\ref{figCh1:compPowaVSspC} that anyway, the most efficient reconstruction scheme is with constant power allocation combined with spatial coupling. Pluging the prior (\ref{eq1:prior}) in the general free entropy expression under the matching prior condition (\ref{eq1:freeEnt2}), we directly obtain:
\begin{align}
\Phi_B(E) &= -\frac{\log_2(B)}{2R} \left(\log({1/{\rm{snr}}} + BE) + \frac{1 - BE}{{1/{\rm{snr}}} + BE} \right) \nonumber\\
&+ \int \mathcal{D}\bz \log\left( e^{\frac{\log(B)}{2\tilde\Sigma(E)^2} + \frac{\sqrt{\log(B)}z_1}{\tilde\Sigma(E)} } + \sum_{i = 2}^B e^{-\frac{\log(B)}{2\tilde\Sigma(E)^2} + \frac{\sqrt{\log(B)}z_i}{\tilde\Sigma(E)} }\right) \label{eq1:freeEnt2_2}
\end{align}
where $E$ is the $MSE$ and $\tilde\Sigma(E)^2\defeq\log(B)\Sigma(E)^2$ together with (\ref{eqChIntro:SIGMA2def}). Going from this expression to $\Phi_B({{SER}})$ is possible thanks to (\ref{eq1:SE_SER}) at the fixed point.

As discussed in sec.~\ref{sec:typicalPhaseTransitions}, the ${{SER}}$ values associated with the maxima of this potential correspond to fixed points of the state evolution equations (\ref{eq1:SE_SER}). The information brought by this analysis that is not explicitly included in the state evolution analysis is the identification of the phase in which the system is (easy/hard/impossible inference) for a given set of parameters $(R,B,{\rm{snr}})$. In particular, the hard phase can only be identified knowing where is situated the global maxima. This infomation cannot be extracted from the state evolution: in the hard phase, (\ref{eq1:SE_SER}) will converge to a local maxima that depends on its initialization, but without telling which of the two is the global one. The state evolution can thus identify the appearance of the hard phase at the BP transition, when the second fixed point appear but it cannot identify the optimal transition as it requires to know the relative height of the maxima.

An example of this potential (\ref{eq1:freeEnt2_2}) in the $(B=2,{\rm{snr}}=100)$ case for various $R$ is shown on Fig.~\ref{fig_freeEnt}. As discussed in sec.~\ref{sec:typicalPhaseTransitions}, the AMP algorithm follows a dynamic that can be {\it interpreted} as a gradient ascent of this free entropy, which starts the ascent from an high error state, i.e. a random guess for the signal estimate. The brown curve thus corresponds to an easy case as the global maximum is unique and corresponds to a low error state. The green curve corresponds to the BP threshold, which marks the appearance of the hard inference phase. For higher rate, the problem is to reach the global maximum despite the high error metastable state: it is achieved using spatial coupling. The pink curve is the optimal transition which marks the entrance in the impossible inference phase. Below this rate, the AMP algorithm combined with spatial coupling or well designed power allocation is theoretically able to decode. 

The blue and red curves also correspond to optimal transitions at higher noise levels and we notice that as the ${\rm{snr}}$ decreases the relative height between the maxima decreases and the basin of attraction of the maxima tends to be more flat. This explains why it is easier to decode finite size signals closer to the optimal threshold with spatial coupling for larger ${\rm{snr}}$: as the basin of attraction of the equilibrium has a more pronounced gradient and the global maximum is higher, the dynamic climbs more easily to the maximum. The solutions associated to the maxima of this potential are exponentially more probable than the other ones (the probability of any state is proportional to the exponential of its free entropy times the system size) but at finite size, the factors gained thanks to an higher maximum can help a lot the algorithm convergence. For example let us assume that we transmit information at a rate $R_{BP}<R<R_{opt}$ which is such that the difference in the free entropy of the equilibrium and metastable states is $\Delta\Phi >0$. Furthemore the system size is $L$. It implies that the ratio of the probability $P_{eq.}$ of the equilibrium state over the metastable state's one $P_{meta.}$ is $P_{eq.}/P_{meta.}\propto \exp\(L\Delta\Phi\)$. If the difference $\Delta\Phi\to\Delta\Phi/\tau$ is divided by $\tau>0$ because we increase the rate to get closer to $R_{opt}$ or because the ${\rm{snr}}$ decreases, the system size must be multiplied by $\tau$ as well to keep this ratio constant, thus this $\Delta\Phi$ does matter at small finite size.

Furthermore, the $SER$ gap separating the high and low error states decreases as well with the ${\rm{snr}}$ which implies that the error floor of the decoding increases. At some value of ${\rm{snr}}$, there are no more two maxima for any rate and the transition becomes continuous as we observed in chap.~\ref{chap:appSparsity}.
\subsection{Large section limit of the superposition codes by analogy with the random energy model}
\label{subsec_largeBrep}
In order to get the asymptotic behavior in the section size of this
potential, we need to compute the asymptotic value
$I\defeq \lim_{B\to\infty}I_B$ of the integral $I_B$ that appears in
(\ref{eq1:freeEnt2_2}). We shall drop the dependency of $\tilde\Sigma$ in $E$ to avoid confusions and compute:
\begin{align}
I_B &\defeq \int \mathcal{D}\bz \log\underbrace{\Bigg( e^{\frac{\log(B)}{2\tilde\Sigma^2} + \frac{\sqrt{\log(B)}z_1}{\tilde\Sigma} } + \sum_{i = 2}^B e^{-\frac{\log(B)}{2\tilde\Sigma^2} + \frac{\sqrt{\log(B)}z_i}{\tilde\Sigma} }\Bigg)}_{\defeq K_B(\bz)} \\
&= \mathbb{E}_{\bz} \{\log(K_B(\bz))\} \label{eq1:logKB}
\end{align}
We shall adopt here the vocabulary of statistical mechanics
\cite{mezard2009information}: this is formally a problem of computing the
average of the logarithm of a partition function $K_B(\bz)$ of a system with $B$
(disordered) states. Indeed, one can re-write (\ref{eq1:logKB}) as:
\begin{align}
I_B &= -\frac{\log(B)}{2\tilde\Sigma^2} + \int \mathcal{D}\bz \log\left(
e^{\frac{\log(B)}{\tilde\Sigma^2} + \frac{\sqrt{\log(B)}z_1}{\tilde\Sigma} } +
  \sum_{i = 2}^B e^{\frac{\sqrt{\log(B)}z_i}{\tilde\Sigma} }\right) \\
&= -\frac{\log(B)}{2\tilde\Sigma^2} +\int \mathcal{D}\bz \log\bigg( {\cal  Z}_1(z_1) + {\cal Z}_{2}\(\{z_i\}_{i=2}^B\)\bigg)  \label{eq:lastIB}
\end{align}
where:
\begin{align}
{\cal  Z}_1(z_1)&\defeq \exp\left( \frac{\log(B)}{\tilde\Sigma^2} +\frac{\sqrt{\log(B)} z_1}{\tilde\Sigma}\right) \\
{\cal Z}_{2}\(\{z_i\}_{i=2}^B\)&\defeq\sum_{i=2}^B \exp\left(\frac{\sqrt{\log(B)} z_i}{\tilde\Sigma}\right) \label{eq:ZREM}
\end{align}
In fact ${\cal Z}_2$ is formally known as a random energy model
in the statistical physics literature
\cite{derrida1980random,mezard2009information}, a statistical physics
model where i.i.d energy levels are drawn from some given
distribution. This analogy can be further refined by writing the energy
as $u_i=-\sqrt{\log(B)}z_i$ and by denoting $\tilde\Sigma$ as the
temperature. In this case, a standard result
\cite{derrida1980random,mezard2009information,arous2005limit} is:

$\bullet$ The asymptotic limit for large $B$ of ${\cal J}\defeq\log_B({\cal Z}_2)$ exists, and is concentrated (i.e. it does not
  depend on the disorder $\bz$ realization).

$\bullet$ It is equal to ${\cal J} = \sqrt{2}/\tilde\Sigma \ \mathbb{I}\left(\tilde\Sigma<1/\sqrt{2}\right)+ \(1/(2\tilde\Sigma^2) + 1\) \ \mathbb{I}\left(\tilde\Sigma>1/\sqrt{2}\right)$.

We can thus now obtain the value of the integral by comparing
${\cal Z}_{1}$ and ${\cal Z}_{2}$ and keeping only the dominant
term. First let us consider the case where $\tilde\Sigma<1/\sqrt{2}$ and $B$ is large:
\begin{align}
\log_B\left( {\cal  Z}_1 + {\cal Z}_{2}\right)&= \log_B ( {\cal Z}_{1}) + \log_B \( 1 +\frac{{\cal{Z}}_2}{{\cal Z}_{1}}\) \\
&\approx\log_B ( {\cal Z}_{1})+\frac{1}{\log(B)}\exp\(-\frac{\log(B)}{\tilde\Sigma^2}-\frac{\sqrt{\log(B)}z_1}{\tilde\Sigma} +\frac{\log(B)\sqrt{2}}{\tilde\Sigma}\)\\
&\approx \log_B ( {\cal Z}_{1})\\
\Rightarrow\lim_{B\to \infty}\frac{I_B}{\log(B)}&=-\frac{1}{2\tilde\Sigma^2}+ \frac{1}{\log(B)} \int\mathcal{D}\bz\(\frac{\log(B)}{\tilde\Sigma^2}+\frac{\sqrt{\log(B)}z_1}{\tilde\Sigma}\)\\
&=\frac{1}{2\tilde\Sigma^2}
\end{align}
using (\ref{eq:lastIB}) and where $\log_B$ is the base $B$ logarithm. If, however, $\tilde\Sigma>1/\sqrt{2}$ and $B$ is large, then:
\begin{align}
\log_B\left( {\cal  Z}_1 + {\cal Z}_{2}\right)&= \log_B ( {\cal Z}_{2}) + \log_B \( 1 +\frac{{\cal{Z}}_1}{{\cal Z}_{2}}\) \\
&\approx\log_B ( {\cal Z}_{2})+\frac{1}{\log(B)}\exp\(\frac{\log(B)}{\tilde\Sigma^2}+\frac{\sqrt{\log(B)}z_1}{\tilde\Sigma} -\frac{\log(B)}{2\tilde\Sigma^2} - \log(B)\)\\
&\approx \log_B ( {\cal Z}_{2})\\
\Rightarrow\lim_{B\to \infty}\frac{I_B}{\log(B)}&=-\frac{1}{2\tilde\Sigma^2} + \frac{1}{2\tilde\Sigma^2} + 1=1
\end{align}
This leads to:
\begin{align}
\lim_{B\to \infty} \frac {I_B}{\log(B)} &=\frac 1{2\tilde\Sigma^2} \ \mathbb{I}\left(\tilde\Sigma< 1/\sqrt{2}\right) +\mathbb{I}\left(\tilde\Sigma > 1/\sqrt{2}\right)
\end{align}
From these results combined with (\ref{eq1:freeEnt2_2}), we now can give the asymptotic value of the potential:
\begin{align}
\phi(E) &\defeq \lim_{B\to \infty} \(\frac{\Phi_B(E)}{\log(B)} \)\\
&= -\frac{1}{2R\log(2)}
\left(\log({1/{\rm{snr}}} + BE) + \frac{1 - BE}{{1/{\rm{snr}}} + BE}
\right) + \max\left(1,\frac 1{2\tilde\Sigma^2(E)}\right)
\label{Phi_largeB}
\end{align}
where (\ref{eqChIntro:SIGMA2def}) with (\ref{eq1:alpha}) implies:
\begin{align}
	\tilde\Sigma^2(E) = R\log(2) (1/{\rm{snr}} + B E) \label{eq:defSIGMATILDESQ}
\end{align}
We define $\tilde E\defeq BE$. Let us now look at the extrema of this potential. We see that we have
to distinghish between the high error case
($\tilde\Sigma>1/\sqrt{2}$ so that $\tilde E>1/(2R\log(2)) - 1/{\rm{snr}}$) and the low
error one ($\tilde\Sigma<1/\sqrt{2}$, so that $\tilde E<1/(2R\log(2)) - 1/{\rm{snr}}$).

In the high error case, the derivative $\partial_{\tilde E}\phi(\tilde E)$ of the potential is zero when:
\begin{align}
\frac 1{2R\log(2)} \( \frac 1{1/{\rm{snr}}+\tilde E} - \frac{
1/{\rm{snr}}+1}{(1/{\rm{snr}}+\tilde E)^2}\)=0
\end{align}
which happens when $\tilde E=1$. Therefore, if both the condition $\tilde E=1$ and
$\tilde E>1/(2R\log(2)) - 1/{\rm{snr}}$ are met, there is a stable extremum of the replica potential at $\tilde E=1$. The existence of this high-error
extremum thus requires $1/(2R\log(2)) - 1/{\rm{snr}}<1$, and we thus define
the critical rate beyond which the state at $\tilde E=1$ is stable:
\begin{equation}
R_{BP}^{\infty}({\rm{snr}})\defeq [({1/{\rm{snr}}} + 1)2\log(2)]^{-1} \label{BPcrit}
\end{equation}
Since we initilize the recursion at $\tilde E=1$ when we attempt to
reconstruct the signal with AMP, we see that $R_{BP}^{\infty}({\rm{snr}})$ is a crucial limit for the reconstruction ability by message-passing.

In the low error case, the derivative of the potential is zero when:
\begin{equation}
  \frac 1{2R\log(2)} \( \frac 1{1/{\rm{snr}}+\tilde E} - \frac{
    1/{\rm{snr}}+1}{(1/{\rm{snr}}+\tilde E)^2}\)=-\frac {1}{2R\log(2)}
  \frac{1}{(1/{\rm{snr}} +\tilde E)^2}
\end{equation}
which happens when $\tilde E=0$. Hence, there is another extremum with
zero error. Let us determine which of these two is dominant. We
have:
\begin{align}
\phi(\tilde E=0) &=-  \frac 1{2R\log(2)} \( \log(1/{\rm{snr}}) + {\rm{snr}} \) +
  \frac {\rm snr}{2R\log(2)} \\
  &= \frac{\log_2({\rm{snr}})}{2R}\\
\phi(\tilde E=1) &=-  \frac {\log_2(1/{\rm{snr}} + 1)}{2R} + 1
\end{align}
The perfect reconstruction extremum is dominant as long as:
\begin{equation}
\log_2({\rm{snr}}) > 2R - \log_2(1+1/{\rm{snr}})
\end{equation}
or equivalently when:
\begin{equation}
R < \frac 12 \log_2{(1+{\rm{snr}})}=C
\end{equation}
where we recognize the expression of the Shannon capacity (\ref{eq:AWGN_C}). As discussed in sec.~\ref{sec:typicalPhaseTransitions}, the optimal transition of the code is defined as the rate where the two maxima have same height, i.e. at $R=C$: these
results are thus confirming that, at large value of $B$, the correct Bayes optimal value
of the section error rate tends to zero and to a perfect
reconstruction, at least as long as the rate remains below the
Shannon capacity after which, of course, this could not be true
anymore. This confirms, using the replica method, the results by
\cite{barron2010sparse,barron2011analysis} that these codes are
capacity achieving.
\subsection{Alternative derivation of the large section limit via the replica method}
We now re-derive the results of the previous section, that the superposition codes are capacity achieving, using the replica method to compute $I\defeq \lim_{B\to\infty}I_B$ that appears in (\ref{eq1:freeEnt2_2}). The computation is performed at fixed $\tilde\Sigma$ which plays again the role of a temperature. The replica method is appropriate because we have to average the logarithm of the partition function (\ref{eq1:logKB}) over the disorder $\mathbb{E}_{\bz}\{\log\(K_B(\bz)\)\}$, here the Gaussian i.i.d vector $\bz$. Starting from (\ref{eq1:logKB}), we can re-write $K_B$ as:
\begin{align}
K_B(\bz) &\defeq \exp\(\frac{\log(B)}{2\tilde\Sigma^2} + \frac{\sqrt{\log(B)}z_1}{\tilde\Sigma} \) + \sum_{i = 2}^B \exp\(-\frac{\log(B)}{2\tilde\Sigma^2} + \frac{\sqrt{\log(B)}z_i}{\tilde\Sigma} \)\\
&= \sum_{i }^B \exp\(-\frac{1}{\tilde\Sigma}\left(\frac{\log(B)}{2\tilde\Sigma}(1 - 2\delta_{i,1}) - \sqrt{\log(B)}z_i \right)\) \\
&\eqdef \sum_{i}^B \exp\(-\frac{h_i(z_i)}{\tilde\Sigma}\) \label{eq_KB}
\end{align}
Meanwhile $Z$ given by (\ref{eq1:fullZ}) is the full (random) partition function of the overall signal, $K_B$ can be interpreted as the partition function of one single section of size $B$.
An important difference with the random energy model (\ref{eq:ZREM}) of the previous section is that here there is a favored section state distinct from the other ones (noted state $1$), corresponding to the actual state of the section in the original signal. It has been treated apart in the previous section but we kept it here in the "energy states" $\{h_i\}_i^B$. From the statistics of $z_i$ we get the one of $h_i$:
\begin{align}
z_i &\sim \mathcal{N}(z_i|0,1) \\
\Rightarrow h_i &\sim \mathcal{N}\(h_i\bigg|\frac{(1 - 2\delta_{i,1})\log(B)}{2\tilde\Sigma}, \log(B)\)
\end{align}
The average of $K_B$ with respect to $\bz$ can thus be replaced by the average over $\bh$, the vector of independent energy states (independent because the $\{z_i\}_{i}^B$ are). We use again the replica trick for computing $I_B=\mathbb{E}_{\bh}\{\log\(K_B(\bh)\)\}$ as $B$ diverges. We thus need the average replicated partition function as in the section sec.~\ref{sec:replicaAnalyisGeneric}:
\begin{align}
I&\defeq\lim_{B\to \infty}\mathbb{E}_{\bh}\{\log K_B(\bh)\} \\
&= \lim_{B\to \infty}\lim_{n\to 0}\frac{\mathbb{E}_{\bh}\{K_B^n\}-1}{n} \label{eq_repTrickK}\\
%
\mathbb{E}_{\bh}\{K_B^n\} &= \mathbb{E}_{\bh} \bigg\{\sum_{i_1,..,i_n }^{B,..,B} \exp\(-\frac{1}{\tilde\Sigma}(h_{i_1} + \ldots + h_{i_n})\)\bigg\} \\
&= \mathbb{E}_{\bh} \bigg\{\sum_{i_1,..,i_n }^{B,..,B} \prod_{j}^B \exp\(-\frac{h_j}{\tilde\Sigma} \sum_{a}^n\delta_{j,i_a}\)\bigg\}
\end{align}
\begin{align}
&= \sum_{i_1,..,i_n }^{B,..,B} \prod_{j}^B \mathbb{E}_{h_j} \bigg\{\exp\(-\frac{h_j}{\tilde\Sigma} \sum_{a}^n\delta_{j,i_a}\)\bigg\}\\
&= \sum_{i_1,..,i_n }^{B,..,B} \exp\(\frac{\log(B)}{2\tilde\Sigma^2}\sum_{j}^B\left(\sum_{a,b}^{n,n}\delta_{j,i_a}\delta_{j,i_b} - (1 - 2\delta_{j,1})\sum_{a}^n\delta_{j,i_a} \right)\)\\
&= \sum_{i_1,..,i_n }^{B,..,B}  \exp\(\frac{\log(B)}{2\tilde\Sigma^2}\left(\sum_{a,b}^{n,n}\delta_{i_a,i_b} - \sum_{j}^B\sum_{a}^n\delta_{j,i_a} (1 - 2 \delta_{j,1}) \right)\) \\
&= \sum_{i_1,..,i_n }^{B,..,B} \exp\(\frac{\log(B)}{2\tilde\Sigma^2}\left(\sum_{a,b}^{n,n}\delta_{i_a,i_b} - n + 2\sum_{a}^n\delta_{1,i_a} \right)\) \label{eq_KB3}
\end{align}
We now define new macroscopic order parameters: 
\begin{align}
q_{ab} &\defeq \delta_{i_a,i_b} \ \forall\ (a,b)\label{eq_repOrderParamBbig_q}\\ 
m_a &\defeq \delta_{i_a,1} \ \forall\ a \label{eq_repOrderParamBbig}
\end{align}
The first one indicates if two replicas are in the same state or not, the second one if a given replica is in the favored state $1$. As in sec.~\ref{sec:replicaAnalyisGeneric}, we replace the microscopic sums over the single replica states by sums over the macroscopic replica order parameters (\ref{eq_repOrderParamBbig_q}), (\ref{eq_repOrderParamBbig}) which become the new variables. The definitions of these must be fulfilled so the sums are restricted over the subspace matching the order parameters definitions (\ref{eq_repOrderParamBbig_q}), (\ref{eq_repOrderParamBbig}). In the sec.~\ref{sec:replicaAnalyisGeneric}, this condition was enforced by the introduction of Dirac delta functions in the integral through (\ref{eq1:1Fourier}), here it is simpler because we are in a discrete case. We deduce from (\ref{eq_KB3}):
\begin{equation}
\mathbb{E}_{\bh} \{K_B^n\} = \sum_{\bq,\bm} \exp\(\frac{\log(B)}{2\tilde\Sigma^2} \left(\sum_{a,b}^{n,n} q_{ab} + 2\sum_{a}^n m_a - n + 2\tilde\Sigma^2 s_{\bq,\bm}\right)\)
\end{equation}
where we have introduced the entropy associated to these new order parameters: $s_{\bq,\bm} \defeq S_{\bq,\bm}/\log(B)$ where $S_{\bq,\bm}$ is the logarithm of the number of microscopic configurations (or states) of the replicas compatible with $\bq$ and $\bm$ definitions at the same time, where $\bq \defeq \[q_{ab}\]_{a,b}^{n,n}$ and $\bm \defeq \[m_a\]_a^n$. We use as before the replica symmetric ansatz where each replica is considered equivalent:
\begin{align}
q_{ab} &= q + (1-q)\delta_{a,b} \ \forall\ (a,b)\\ 
m_a &= m \ \forall \ a
\end{align}
It allows to simplify the average replicated partition function:
\begin{align}
\mathbb{E}_{\bh} \{K^n_B\}&= \sum_{q,m} \exp\Bigg(n\log(B) \underbrace{\bigg[\frac{(n-1)q + 2m + \frac{2\tilde\Sigma^2}{n} s_{q,m}}{2\tilde\Sigma^2}\bigg]}_{\defeq \tilde I(q,m)}\Bigg) \\
&\eqdef \sum_{q,m} \exp\(n\log(B)\tilde I(q,m)\) \label{eq_saddleK}
\end{align}
Looking at (\ref{eq_repOrderParamBbig_q}), (\ref{eq_repOrderParamBbig}), there are a priori four different possible ansatz, corresponding to four different macroscopic states of the section: $(q=m=0), (q=m=1), (q=0, m=1)$ and $(q=1, m=0)$ but actually, only three possibilities remain as the state $(q=0, m=1)$ has no meaning: the replicas cannot be all in different states ($q=0$) and all in the favored one ($m=1$) at the same time. Thus it remains:

$\bullet ~(q=m=0)$ : all the replicas are in different states but none of them are in the favored one 1. \label{eq_mq0}\\
$\bullet ~(q=m=1)$ : all the replicas are in the favored state 1.\label{eq_mq1}\\
$\bullet ~(q=1, m=0)$ : all the replicas are in the same state, which is not the favored one. \label{eq_thirdAnsatz}

The last ansatz can be forgotten as the computation shows that it always leads to lower free entropy than the two other ansatz. This is understandable as there should be a symmetry among all the "wrong" states (different from $1$) as none of them is special with respect to the other ones, so the replicated system should not choose a particular one spontaneously. It leaves two ansatz. The previous sum $\sum_{q,m}$ is performed by the saddle point method as $B \to \infty$, assuming as previously (see sec.~\ref{sec:replicaAnalyisGeneric}) the commutativity of the limits in (\ref{eq_repTrickK}). From (\ref{eq_repTrickK}), the "section free entropy density" $I/\log(B)$ associated to each state is thus $\tilde I(q^*,m^*)$ where $(q^*,m^*)$ is choosen among the two possible ansatz. The integral $I$ is thus:
\begin{equation}
I=\log(B)\max_{(q^*,m^*)}\[\tilde I(q^*,m^*)\] \label{eq_maxTildeI}
\end{equation}
Let's compute the value of $\tilde I$ for the two remaining ansatz as $n\to 0$ in order to find the maximum:
\begin{align}
(q^*=m^*=0) &\Rightarrow s_{0,0} = \log ((B-1)^n)/\log(B) \approx n \\
&\Rightarrow \tilde I(E|q^*=m^*=0) \approx 1\\
(q^*=m^*=1) &\Rightarrow s_{1,1} = \log (1)/\log(B) = 0 \\
&\Rightarrow \tilde I(E|q^*=m^*=1)= (2\tilde\Sigma(E)^2)^{-1}
\end{align}
where $\tilde\Sigma(E)^2$ is given by (\ref{eq:defSIGMATILDESQ}). From these results combined with (\ref{eq1:freeEnt2_2}), defining $\tilde E=BE$, the rescaled potential $\phi(\tilde E)$ and the function $g(\tilde E)$ as:
\begin{align}
g(\tilde E)&\defeq -\frac{1}{2R\log(2)} \left(\log({1/{\rm{snr}}} + \tilde E) + \frac{1 - \tilde E}{{1/{\rm{snr}}} + \tilde E} \right) \label{eq_g}\\
\phi(\tilde E|q^*,m^*)&\defeq \lim_{B\to\infty} \frac{\Phi_B(\tilde E|q^*,m^*)}{\log(B) }\\
&= g(\tilde E) + \tilde I(\tilde E|q^*,m^*)\label{eq_phiResc}
\end{align}
we get the final potential of the sparse superposition codes in the large section size limit:
\begin{align}
\phi_0(\tilde E) &\defeq \phi(\tilde E|q^*=m^*=0) = g(\tilde E) + 1 \label{eq_phi0}\\
\phi_1(\tilde E) &\defeq \phi(\tilde E|q^*=m^*=1) = g(\tilde E) + \big(2\log(2) R({1/{\rm{snr}}} + \tilde E)\big)^{-1}\label{eq_phi1}\\
\phi(\tilde E) &= \txt{max}\(\phi_0(\tilde E), \phi_1(\tilde E)\)
\end{align}
where the actual potential $\phi(\tilde E)$ for a given error $\tilde E$, rate $R$ and ${\rm{snr}}$ is the maximum of the two ansatz-dependent potentials. These two potentials give the statistical weight of two different regimes
(or pure states) \cite{mezard2009information,krzakalaGibbsStates06} that
have respectively a probability $\propto \exp(L\log(B)\phi_0(\tilde E))$ and
$\propto \exp(L\log(B)\phi_1(\tilde E))$.
\subsubsection{The belief propagation transition}
The BP transition $R_{BP}$ is defined as the rate until which AMP
without spatial-coupling or power allocation is Bayes optimal, see the green curve on
Fig.~\ref{fig_freeEnt}. It corresponds to the lowest rate for which there exist two maxima of the Bethe free entropy. So to find it we equate the free entropies of the two pure states (\ref{eq_phi0}), (\ref{eq_phi1}):
\begin{align}
\phi_0(\tilde E) &= \phi_1(\tilde E) \\
\Rightarrow R_c(\tilde E) &= [({1/{\rm{snr}}} + \tilde E)2\log(2)]^{-1} \label{eq:BPtransInf2}
\end{align}
$R_c(\tilde E)$ is the critical line until which it exists only one maximum and thus one state, or equivalently where appears the second maximum. Above it $R>R_c$ (but before the static transition) there are two distincts maxima. As in the previous section, a particular role is played by the value $\tilde E= 1$ in which AMP is initialized on real reconstructions. So $R_c(\tilde E=1)$ gives the asymptotic $B\to\infty$ BP transition:
\begin{align}
R_{BP}^{\infty}({\rm{snr}})&\defeq R_c(1)= [({1/{\rm{snr}}} + 1)2\log(2)]^{-1}
\end{align}
we find back (\ref{BPcrit}). Above $R>R_{BP}^{\infty}({\rm{snr}})$, we are in the hard or impossible phase, below it is the easy regime. The formula (\ref{eq:BPtransInf2}) can be interpreted the other way around: we consider a practical situation where the rate is fixed above the critical one $R>R_{BP}^\infty$ (at fixed ${\rm{snr}}$) and we are in the hard phase such that there are two maxima. From (\ref{eq:BPtransInf2}) we can define the critical $\tilde E_c(R)=\txt{max}\([2R\log(2)]^{-1}-1/{\rm{snr}},0\)$, where the free entropy expression changes for a given rate:
\begin{align}
\tilde E<\tilde E_c &\Rightarrow \phi_1(\tilde E) > \phi_0(\tilde E) \\
\tilde E>\tilde E_c &\Rightarrow \phi_1(\tilde E) < \phi_0(\tilde E)
\end{align}
Thus we can write the potential at fixed rate as:
\begin{align}
	\phi(\tilde E) = \phi_0(\tilde E) \mathbb{I}\(\tilde E>\tilde E_c\)+\phi_1(\tilde E) \mathbb{I}\(\tilde E<\tilde E_c\)
\end{align}
From the fixed point equations of the potential:
\begin{align}
\tilde E<\tilde E_c &\Rightarrow \frac{\partial\phi_1(\tilde E)}{\partial \tilde E} = 0\Rightarrow \tilde E=0\\
\tilde E>\tilde E_c &\Rightarrow \frac{\partial\phi_0(\tilde E)}{\partial \tilde E} = 0\Rightarrow \tilde E=1
\end{align}
we see that in the hard phase, it coexists asymptotically two maxima such that the first corresponds to a perfect reconstruction, the other to the metastable failure state: we can identify $\phi_1(\tilde E)$ as the free entropy corresponding to the perfect reconstruction state, $\phi_0(\tilde E)$ to the failure one. But when does the hard phase stop and the impossible one starts, i.e. where is the optimal transition?
\subsubsection{The optimal transition}
The Bayes optimal rate is defined as the rate where the two distinct maxima have same height, which means they have the same statistical weight:
\begin{align}
\phi(\tilde E=0) &= \phi(\tilde E=1)\\
\Rightarrow\phi_1(\tilde E=0) &= \phi_0(\tilde E=1) \\
\Rightarrow \log ({1/{\rm{snr}}}) &=\log({1/{\rm{snr}}} + 1) - 2R_{opt}\log(2) \\
\Rightarrow R_{opt} &= \frac{1}{2}\log_2(1 + {\rm{snr}})= C
\end{align}
where we recognize the Shannon capacity (\ref{eq:AWGN_C}). The optimal transition of the superposition codes scheme is thus asymptotically given by the capacity and between $R_{BP}^{\infty}$ and $C$ is the hard phase.
\subsection{Results from the replica analysis}
\begin{figure}[h!]
\centering
\includegraphics[width=0.5\textwidth]{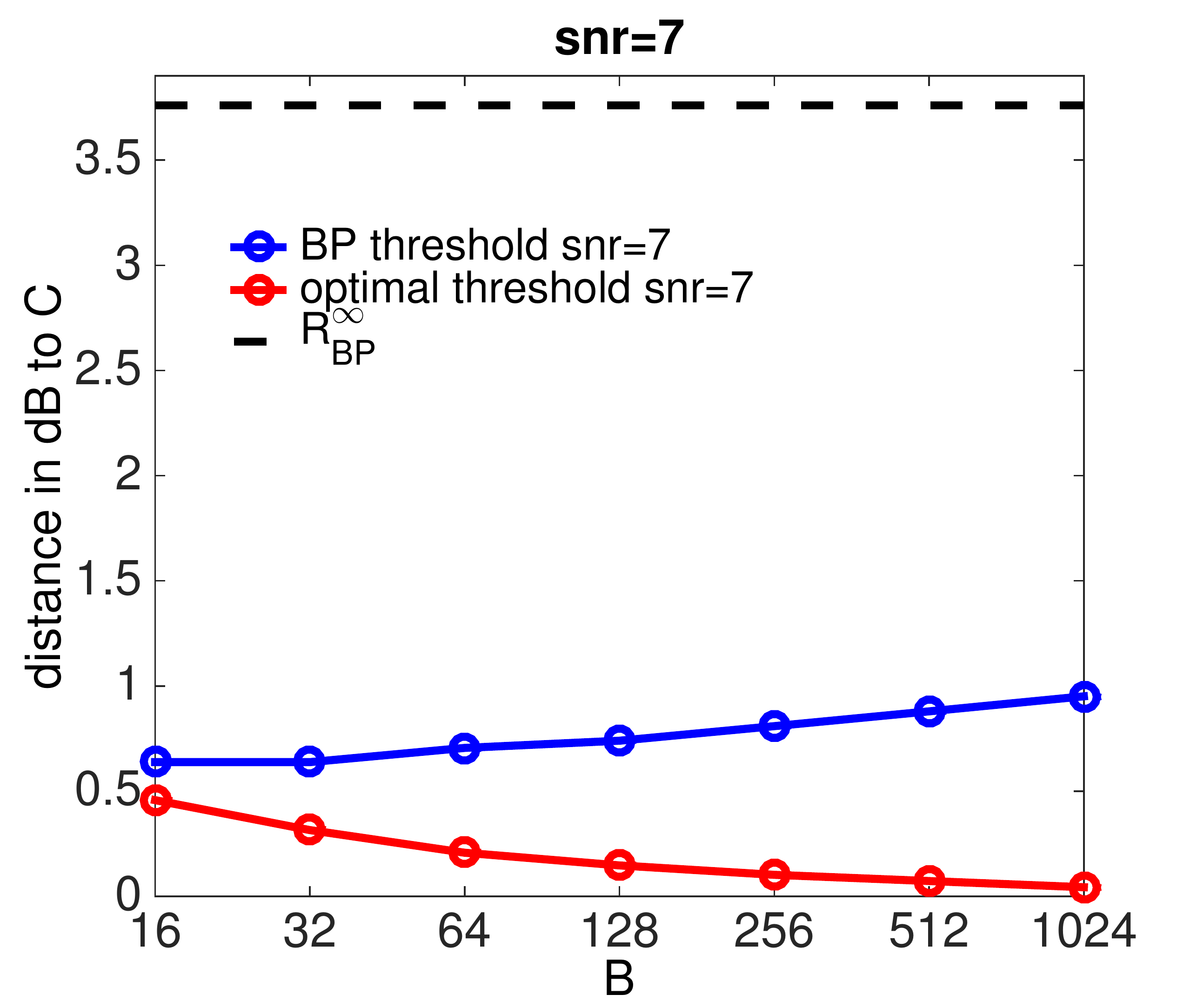}
\includegraphics[width=0.5\textwidth]{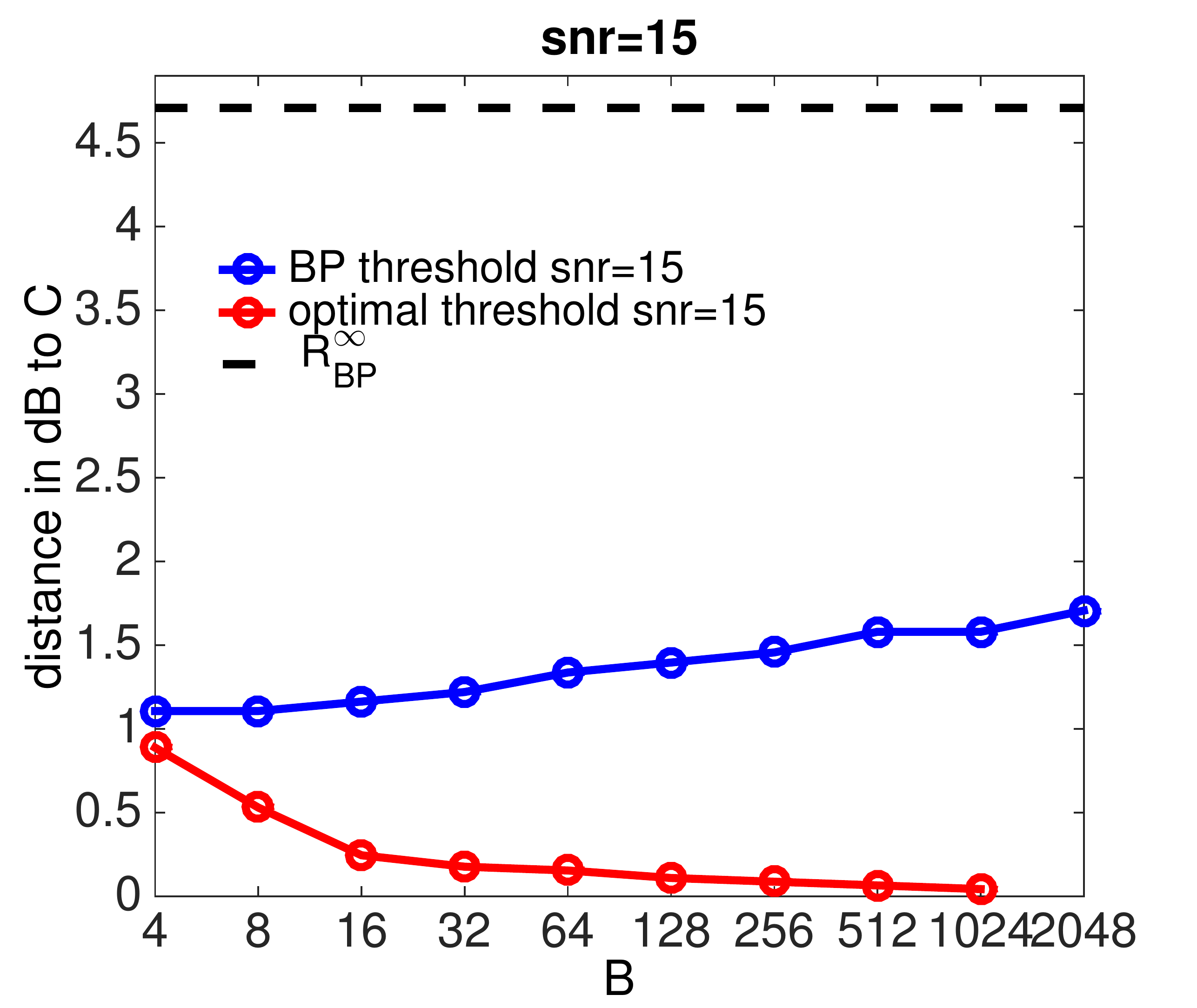}
\includegraphics[width=0.5\textwidth]{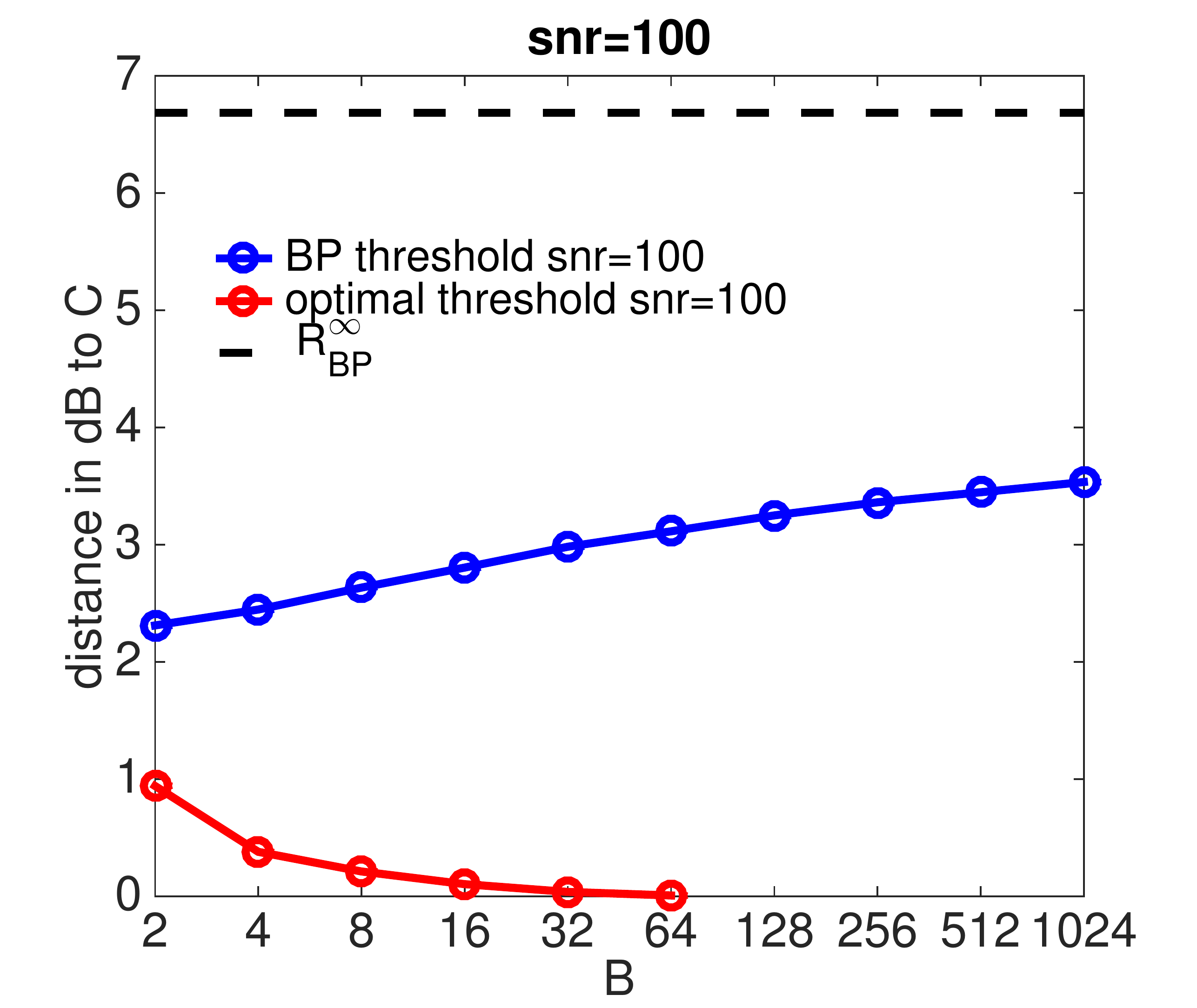}
\caption[Phase diagrams of superposition codes at various ${\rm snr}$]{All the points are computed from the Bethe free entropy (\ref{eq1:freeEnt2_2}) where the integral is computed by monte carlo. These are the phase diagrams of the superposition codes for different ${\rm snr}$, where the $x$ axis is the section size $B$, the $y$ axis is the distance to the capacity $C$ in dB. The blue and red curves are respectively the BP and optimal transitions. The black dashed line is the asymptotic value $B\to\infty$ of the BP threshold $R_{BP}^{\infty}({\rm snr})$ (\ref{BPcrit}).}
\label{figCh1:diagsDist}
\end{figure}
\begin{figure}[!t]
\centering
\includegraphics[width=0.34\textwidth]{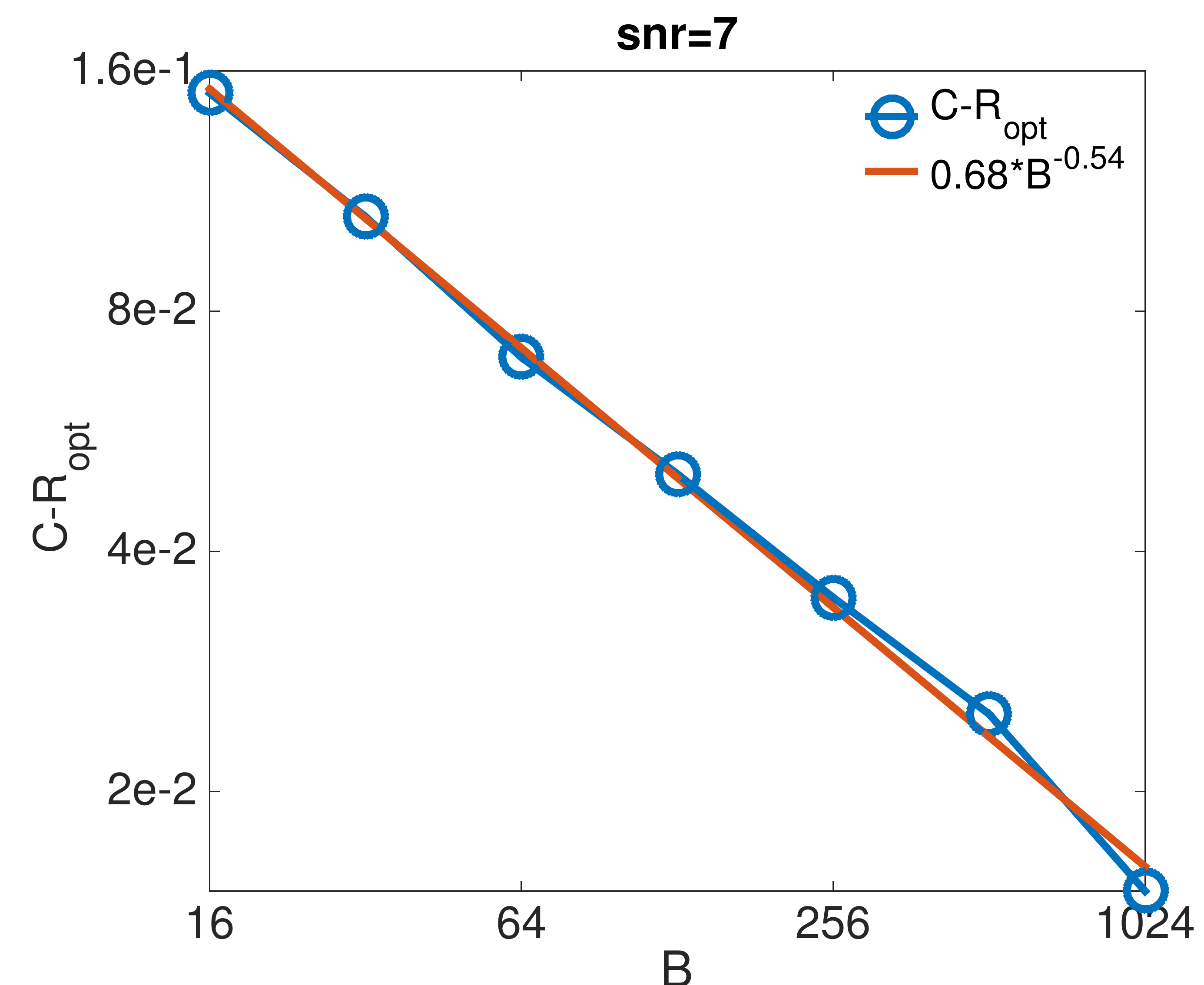}
\includegraphics[width=0.32\textwidth, trim=50 0 0 0, clip=true]{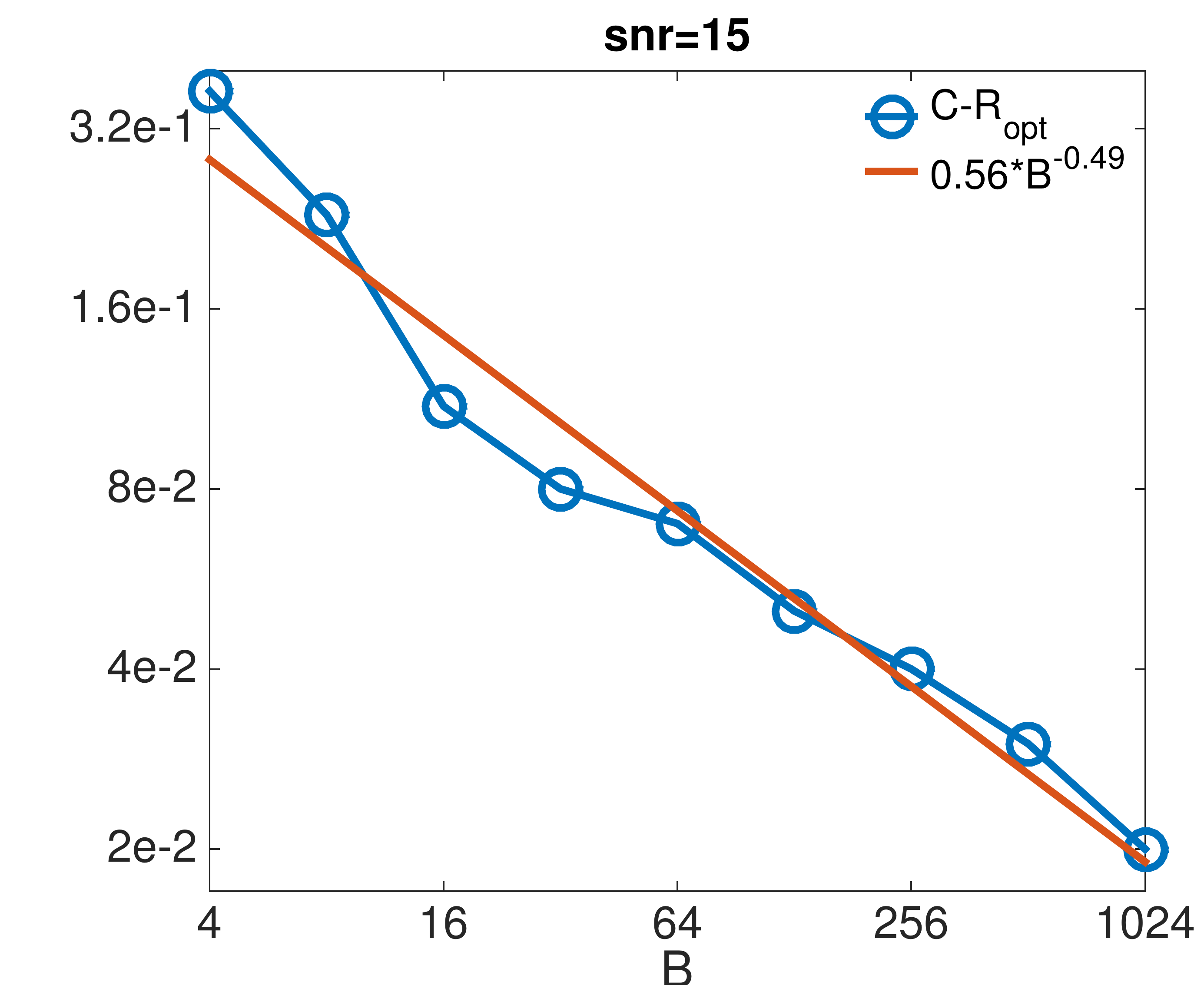}
\includegraphics[width=0.32\textwidth, trim=50 0 0 0, clip=true]{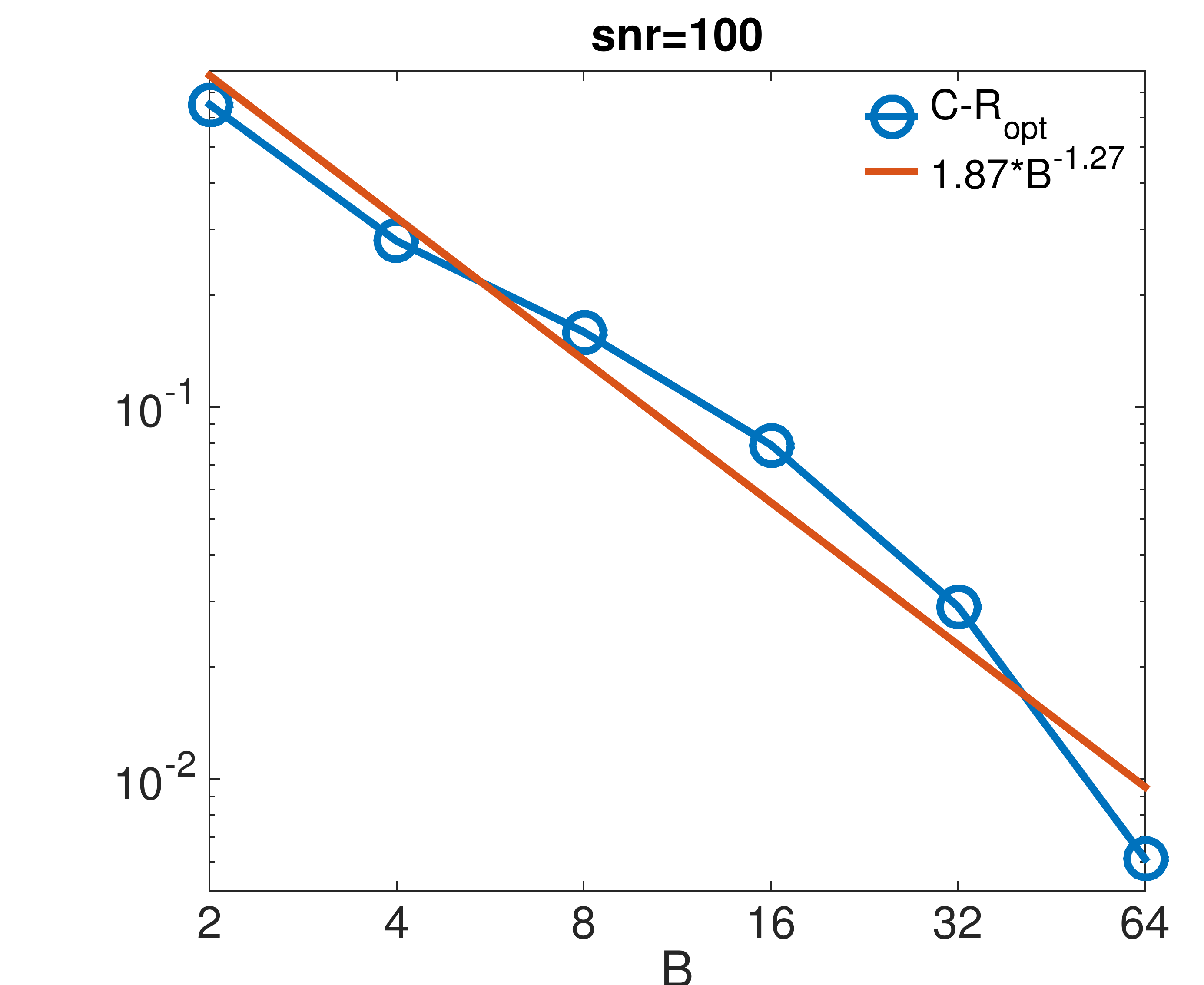}
\includegraphics[width=0.34\textwidth]{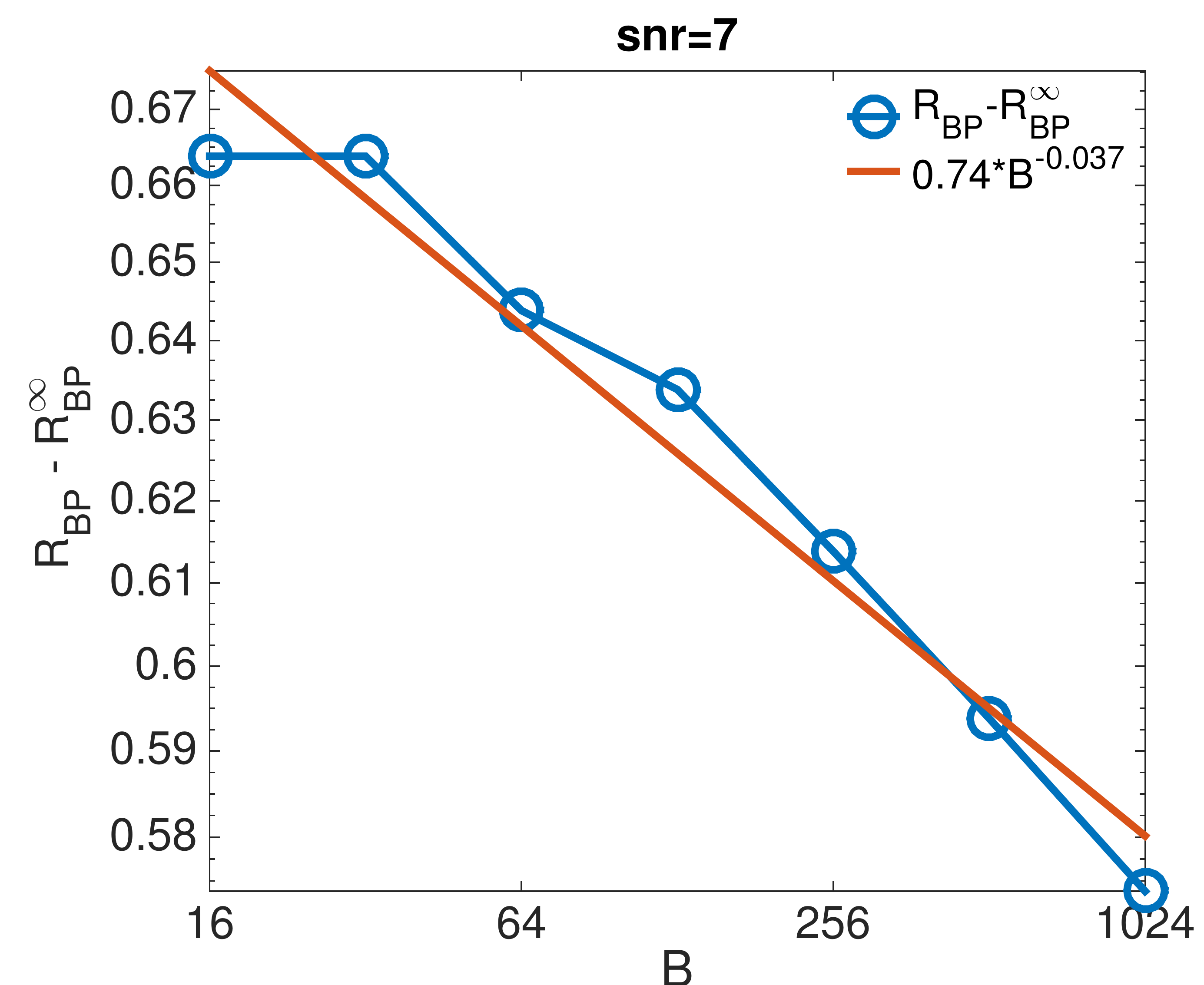} 
\includegraphics[width=0.32\textwidth, trim=45 0 0 0, clip=true]{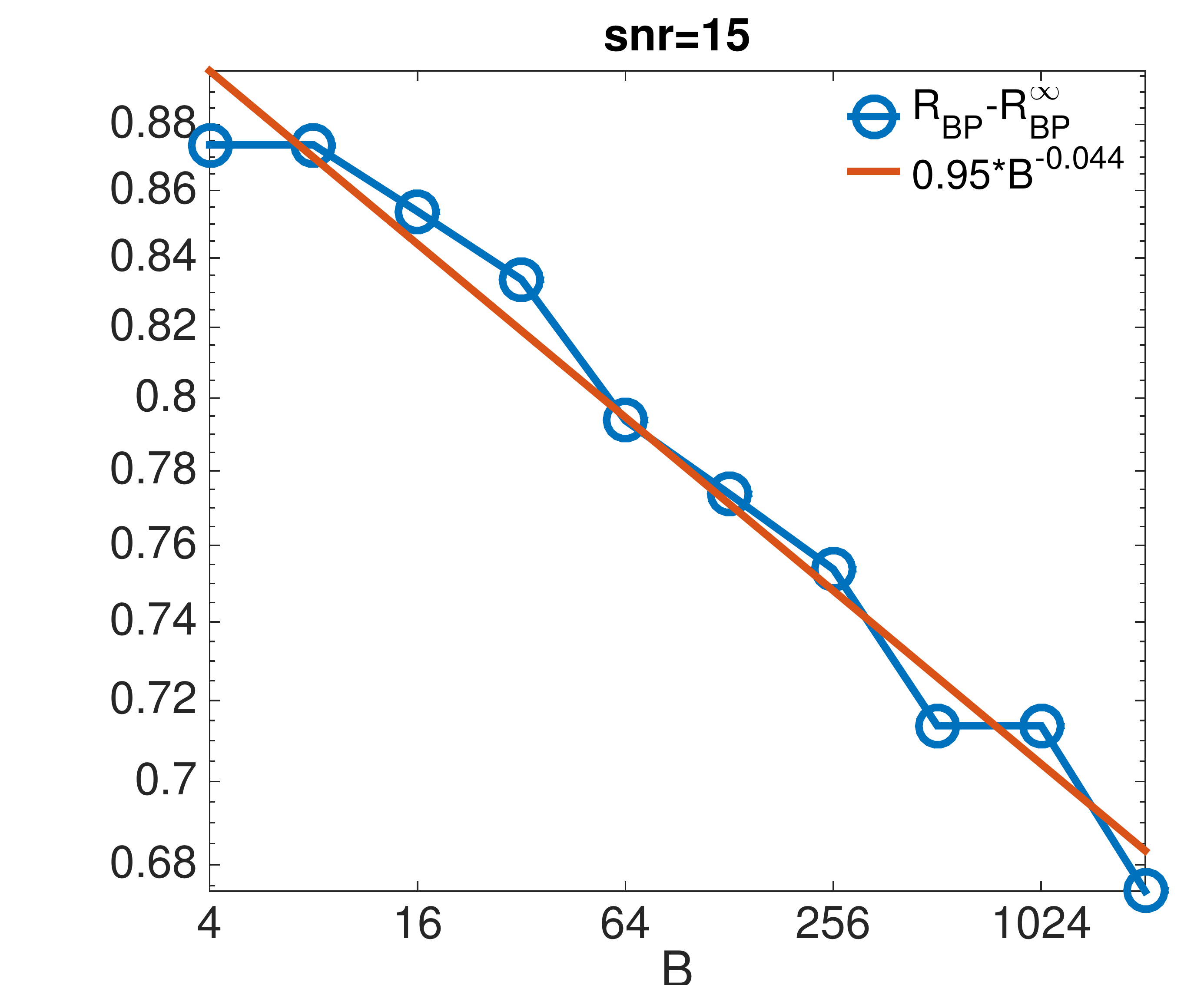}
\includegraphics[width=0.32\textwidth, trim=45 0 0 0, clip=true]{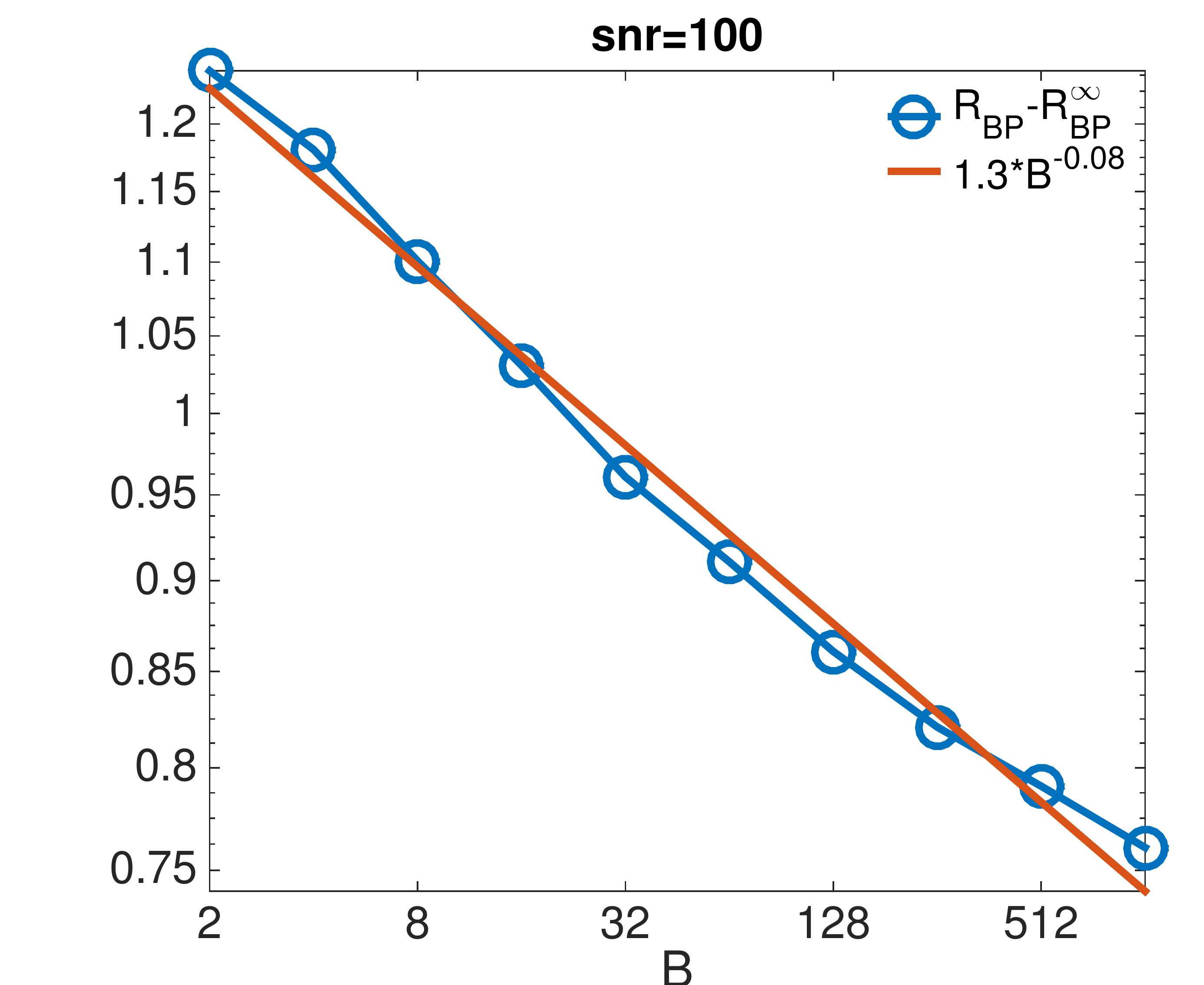}
\caption[Convergence rate of the optimal transition of superposition codes to the capacity and of the BP transition to its asymptotic value]{All the blue points
  are computed by replica method from the potential
  (\ref{eq1:freeEnt2_2}) where the integral is computed by monte carlo. \textbf{Upper plots :} These plots show how
  fast the optimal transition $R_{opt}(B|{\rm{snr}})$ is approaching the capacity. We plot
  the difference $C({\rm{snr}})-R_{opt}(B|{\rm{snr}})$ as a function of $B$ in double
  logarithmic scale. The lines are guides for the eyes, and should not
  be taken as serious fits. They strongly suggest, however, a power
  law behavior. \textbf{Lower plots :} We did the same for
  $R_{BP}(B|{\rm{snr}})$. We plot the difference $R_{BP}(B|{\rm{snr}})-R_{BP}^{\infty}({\rm{snr}})$ as a
  function of $B$ in double logarithmic scale, where $R_{BP}^{\infty}({\rm{snr}})$
  is the asymptotic BP transition computed by replica method, see (\ref{BPcrit}). In every cases, we observe a
  behavior quite well predicted by a power law as well. The lines are again
  guides for the eyes, and the very low values of the exponents suggest
  a very slow logarithmic behavior.}\label{figCh1:CminRscaling}
\end{figure}
From this analysis, we can extract the phase diagram of the superposition codes scheme. Fig.~\ref{figCh1:diagsDist} presents phase diagrams for different ${\rm snr}$ values. The blue curve is the BP transition extracted from the potential (\ref{eq1:freeEnt2_2}) which marks the end of optimality of the AMP decoder without spatial coupling or proper power allocation, while the red curve is the optimal transition: the highest rate until decoding is theoretically possible. The black dashed curve is the asymptotic BP transition (\ref{BPcrit}).

\begin{figure}[!t]
\centering
\includegraphics[width=0.6\textwidth]{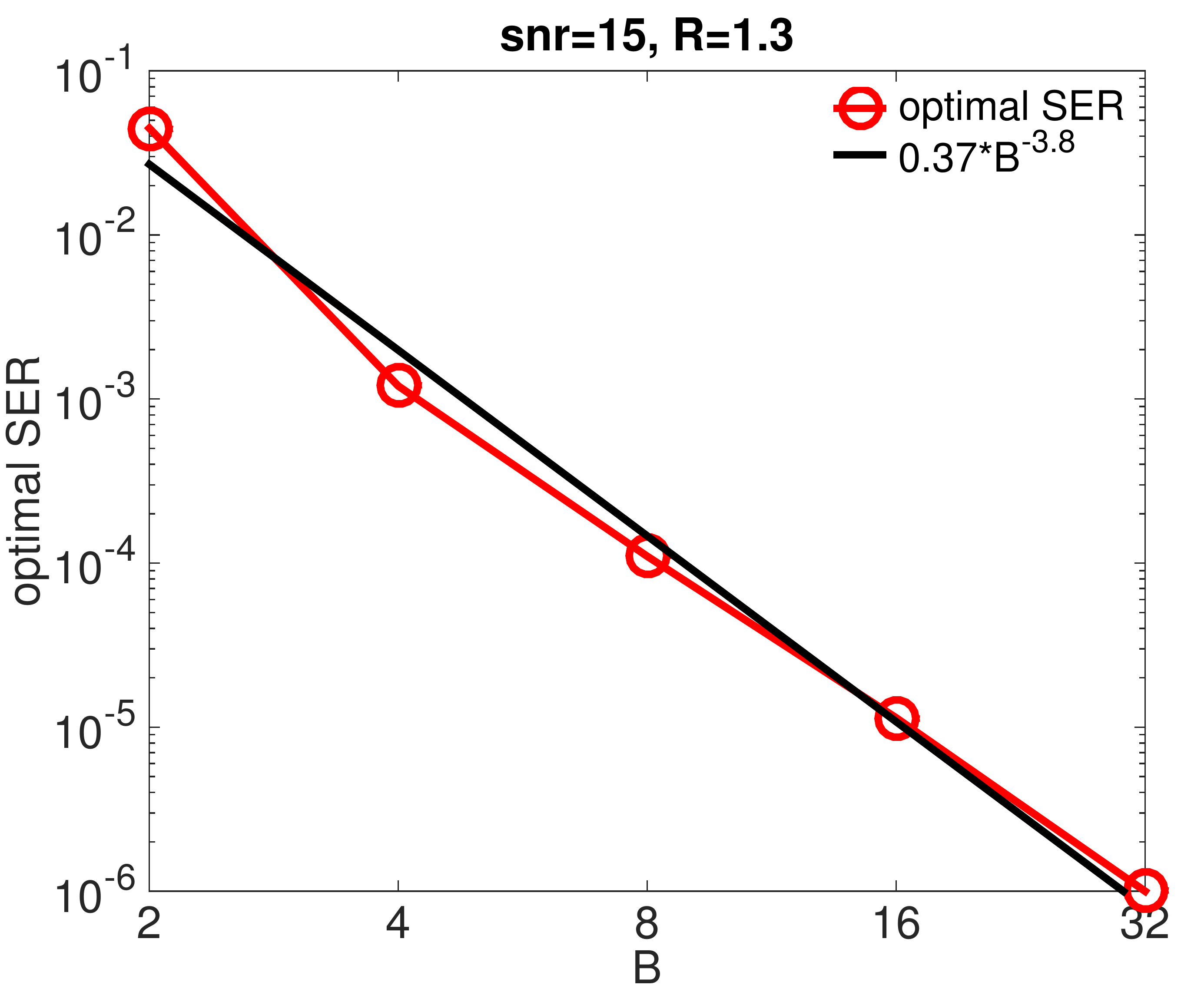}
\includegraphics[width=0.6\textwidth]{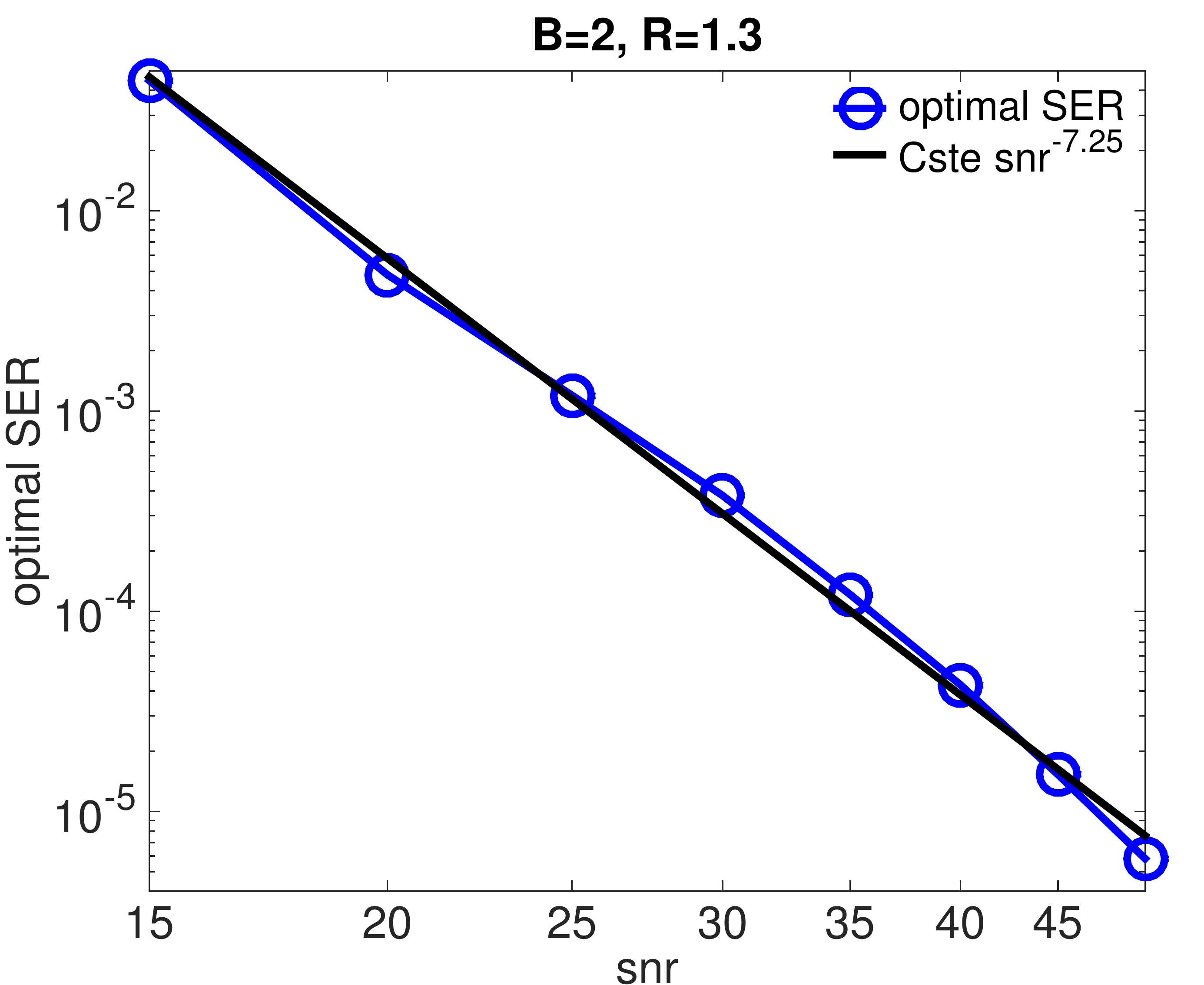}
\caption[Optimal section error rate as a function of the section size and the ${\rm{snr}}$]{On these plots we show how the optimal $SER$ changes when $B$ or the ${\rm{snr}}$ increase according to the replica theory (\ref{eq1:freeEnt2_2}). Both curves are plots
at fixed rate $R=1.3$ and are in double logarithmic scale. The red curve is function of $B$ at fixed ${\rm{snr}}=15$, the blue one at fixed $B=2$ as a function of the ${\rm{snr}}$. The best linear fit is on top of the curves (Cste is a constant).\label{figCh1:optSER}}
\end{figure}

\begin{figure}[!t]
\centering
\includegraphics[width=0.47\textwidth, trim=0 0 35 5, clip=true]{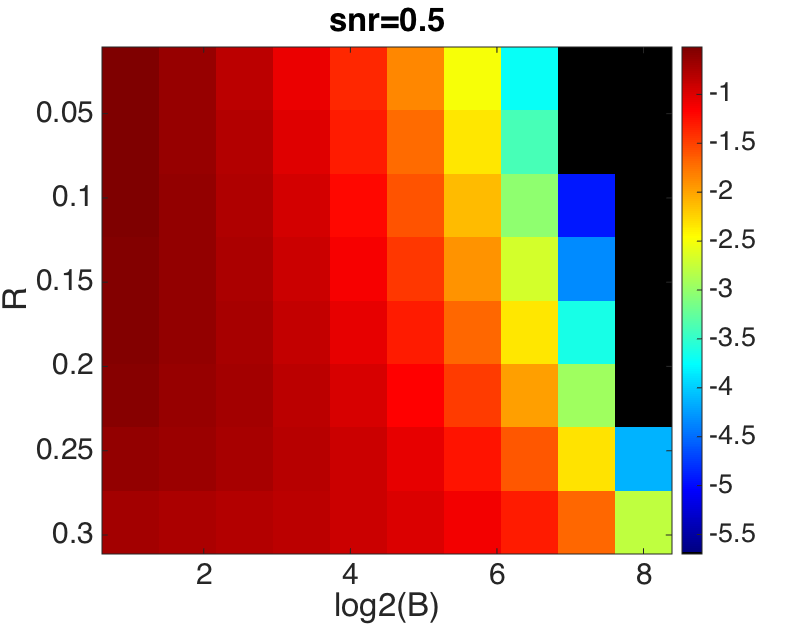}
\includegraphics[width=0.47\textwidth, trim=0 0 35 5, clip=true]{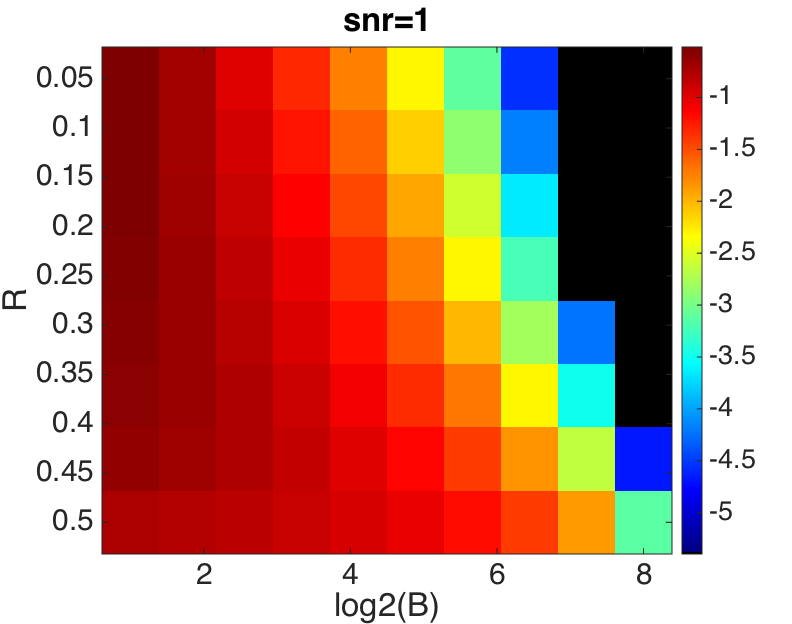}
\includegraphics[width=0.47\textwidth, trim=0 0 35 5, clip=true]{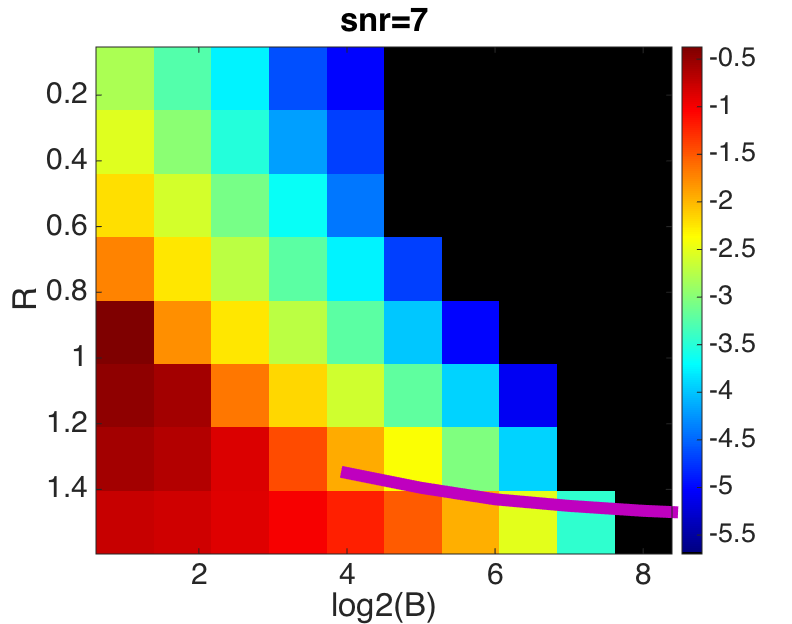}
\includegraphics[width=0.47\textwidth, trim=0 0 35 5, clip=true]{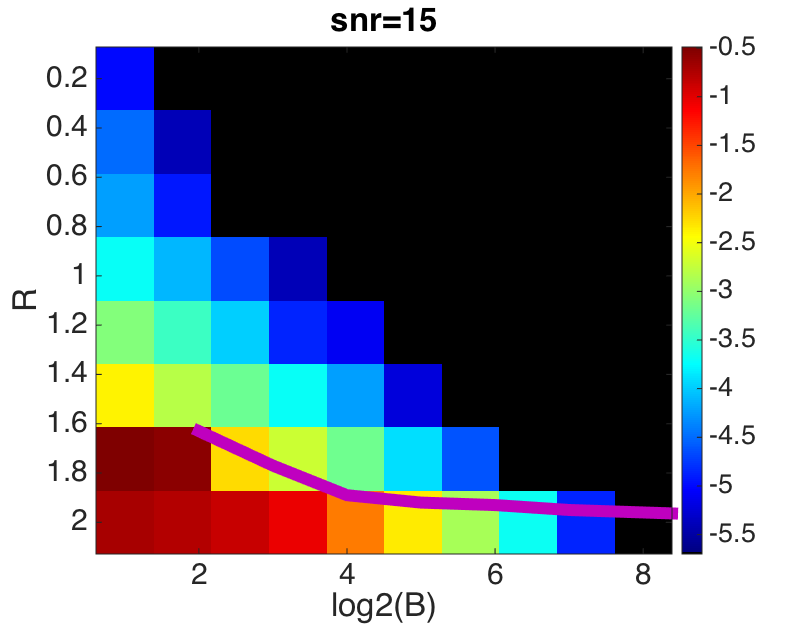} 
\caption[Optimal section error rate for superposition codes in the ($R,B$) plane]{On this plot, we show the logarithm in base $10$ of the section error rate corresponding to the lowest ${SER}$ maximum of the replica potential (\ref{eq1:freeEnt2_2}) in the $(R,B)$ plane for different ${\rm snr}$ values. The values are obtained from the state evolution recursion (\ref{eq1:SE_SER}) starting from the solution (i.e. with an initial error equal to $0$). The recursions (\ref{eq1:SE_MSE}), (\ref{eq1:SE_SER}) are computed by monte carlo with a sample size of $5B\times10^5$. The black squares correspond to points where the computed value is ${SER}=0$ which actually means a value that is lower to $(5\times10^5)^{-1}$ with high probability. The solid pink curve on the two lower plots correspond to the optimal rates $R_{opt}(B,{{\rm snr}})$ as in Fig.~\ref{figCh1:diagsDist}. In the two upper plots that correspond to high noise regimes, there is no transition at all (the AMP decoder is thus always Bayes optimal, at least for these manageable section sizes $B$) and the optimal ${SER}$ is a smooth increasing function of the rate $R$ at fixed $B$ and decreasing function of $B$ at fixed $R$. The ${SER}$ values in the two lower plots corresponding to low noise regimes matche the optimal ${SER}$ as long as it is for a rate $R < R_{opt}(B,{{\rm snr}})$ lower than the optimal one. For higher rates, the maximum of the potential corresponding to the plotted ${SER}$ values is not the global maximum and thus cannot be reached, even with spatial coupling that works asymptotically until the optimal rate. For $B$ smaller than $4$ (resp. $2$) on the ${\rm snr}=7$ (resp. ${\rm snr}=15$) plot, there is no sharp transition and the represented ${SER}$ value is the optimal one and can be reached by AMP without spatial coupling, as in the high noise regime.}\label{figCh1:phaseDiagsFinalSER}
\end{figure}

A first observation is that the BP transition is converging quite slowly to its asymptotic value compared to the optimal one that converges faster to the capacity, which is good. We also note that the section size where start the transitions (and thus marks the appearance of the hard regime where the AMP decoder without spatial coupling is not Bayes optimal anymore) gets larger as the ${\rm snr}$ decreases. When the ${\rm snr}$ is not too large, we see that the optimal and BP transitions almost coincide at small $B$ values, such as for $B=16$ at ${\rm snr}=7$ and $B=4$ for ${\rm snr}=15$. Below this section size value, there are no more sharp phase transitions as only one maximum exists in the potential (\ref{eq1:freeEnt2_2}) and the dedoder, even without spatial coupling, is optimal at any rate: the $SER$ increases continuously with the rate. As the ${\rm snr}$ increases, the curves split sooner until they remain different $\forall \ B$ such as in the ${\rm snr}=100$ case. See Fig.~\ref{figCh1:optSER} and Fig.~\ref{figCh1:phaseDiagsFinalSER} for more details on the achievable values of the $SER$.

Fig.~\ref{figCh1:CminRscaling} gives details on the scalings of the convergence of the first order transitions to their asymptotic values. It seems that the rate of convergence of both transitions can be well approximated by power laws. On Fig.~\ref{figCh1:CminRscaling} we present the differences between the points of Fig.~\ref{figCh1:diagsDist} and their asymptotic $B\to\infty$ values which are the capacity $C$ for the optimal transition (as shown in the two previous sections) and $B_{BP}^{\infty}$ (\ref{BPcrit}) for the BP transition. It appears that the scaling exponents amplitude tend to increase with the ${\rm snr}$. 

Fig.~\ref{figCh1:optSER} represents how the optimal $SER$ evolves at fixed rate $R=1.3$ and ${\rm snr}=15$ as a function of the section size $B$ (upper plot) and at fixed rate $R=1.3$ and $B=2$ as a function of the ${\rm snr}$ (lower plot). The observations are similar: in both cases, the curves seems to be well approximated by power laws with exponents given on the plots. The points are extracted from the replica potential (\ref{eq1:freeEnt2_2}).

Fig.~\ref{figCh1:phaseDiagsFinalSER} quantifies the optimal performances asymptotically attainable by the decoder, obtained from the state evolution analysis by initializing the recursion (\ref{eq1:SE_MSE}) at $E^{t=0}=0$ and using (\ref{eq1:SE_SER}). We plot the base $10$ logarithm of the ${SER}$ corresponding to the lower ${SER}$ maximum of the potential (\ref{eq1:freeEnt2_2}) as a function of the rate and the section size $B$ (again the state evolution and replica analyzes are equivalent to determine the fixed points as shown in \ref{subsec:repIsSE}). In high noise regimes, the depicted ${SER}$ is always reachable by AMP without the need of spatial coupling as there are no sharp transitions. For lower noise regimes, the plotted ${SER}$ matches the optimal one as long as $R<R_{opt}(B,{\rm snr})$ where the optimal transitions correspond to the solid pink curves. When there is no optimal transition (for a $B$ before that the pink line starts), the ${SER}$ is the optimal one and AMP can always reach it. The upper plot of Fig.~\ref{figCh1:optSER} is a cut in the ${\rm snr}=15$ plot.
\section{Optimality of the approximate message-passing decoder with a proper power allocation}
\label{secCh1:AMP-PowA}
In this section, we shall discuss a particular power allocation that
allows AMP to be capacity achieving in the large size $L\gg 1$ limit, without
the need for spatial coupling. We shall work again in the large
section size $B\gg 1$ limit as well.

We first divide the system into $G$ groups, see Fig.~\ref{figCh1:equivPowaSpc}. For our analysis, each of
these groups has to be large enough and must contain many sections, each of these sections being itself large so that $1\ll B\ll L_G$, $1\ll G\ll L$ where $L_G\defeq L/G$ is the number of sections per group. Now, in
each of these groups, we use a different power allocation: the non zero values of the sections inside the group $g$ are all equal to $c_g$. This is
precisly the case which we have studied in Sec.~\ref{sec:powA_SE}, so
we can apply the corresponding state evolution in a straightforward manner.

Our claim is that we can use the following power allocation: 
\begin{align}
c_g=\frac{2^{-\frac{Cg}{G}}}{Z} \ \forall \ g\in\{1,\ldots,G\}
\end{align}
where $C=\frac 12 \log_2{\(1+{\rm snr}\)}$ is the Shannon capacity. We choose $Z$ such that the power of the signal equals one:
\begin{align}
	\frac{1}{G}\sum_{g}^G c_g^2=1 \label{eq:normalizationPoweAlloc}
\end{align}
With this definition, we have:
\begin{equation}
Z^2 =  \frac{2^{-\frac{2C}G} \(1-2^{-2C}\)}{G(1-2^{-\frac{2C}G})} \label{eq:Z2}
\end{equation}
It will be useful to know the following identity:
\begin{equation}
\frac 1G \sum_{g}^{\tilde g} c_g^2= \frac {1-2^{-\frac{2C\tilde g}G}}{1-2^{-2C}}
\label{mysum}
\end{equation}
Now, we want to show that, if we have decoded all the sections before the
section $\tilde g$ at time $t$, then we will be able to decode section $\tilde g$
as well. If we can show this, then starting from $\tilde g=1$ we will
have a succession of decoding until all is decoded, and we would have
shown that this power allocation works. In this situation, using
(\ref{eq:SE_POWA_S}) and the expression of the rate $R$ (\ref{eq1:alpha_0}), we have
for the section $\tilde g$:
\begin{equation} (\tilde\Sigma_{\tilde g}^{t+1})^2 =
R \log(2) \left[\frac{{1/{\rm{snr}}} +    {\cal E}_{\tilde g -1}}{c_{\tilde g}^2}\right]
\end{equation}
with:
\begin{equation}  
 {\cal E}_{\tilde g} \defeq 1 - \frac 1G \sum_{g}^{\tilde g} c_g^2  \label{eq1:calE}
\end{equation}
where we have used (\ref{eq:normalizationPoweAlloc}) with our assumption of having already decoded until $\tilde g-1$ included at time $t$: $\tilde E_g^t = B E_g^t= \mathbb{I}(g \ge \tilde g)$. (\ref{eq1:calE}) is the average (rescaled by $B$) $MSE$ if all has been decoded until
$\tilde g$ included: it is given by the initial total rescaled $MSE$ $\tilde E^{t=0}=1$ from which we have
to remove what has been already decoded. We now ask if the group
${\tilde g}$ can be decoded as well. The evolution of the error in
this group is given by (\ref{eq:SE_POWA_E}) and we have seen in
sec.~\ref{subsec_largeBrep}, that the condition for a perfect decoding
in the large $B$ limit is simply that $\tilde\Sigma_{\tilde g}^2<1/2$ which remains true per group as the only coupling with the other groups in the state evolution (\ref{eq:SE_POWA_E}) is through the "temperature" $\tilde\Sigma_{\tilde g}^2$. We thus need the
following to be satisfied (as long as $R<C$):
\begin{equation} 
R \log(2) \left[\frac{ 1/{\rm snr} + {\cal E}_{\tilde g-1}}{c_{\tilde g}^2}\right] <\frac 12
\label{condition}
\end{equation} 
If this condition is satisfied, there is no local BP transition to block
the AMP reconstruction in the group ${\tilde g}$, then the decoder
will move to the next group, etc. We thus need this condition to
be correct $\forall \ {\tilde g} \in \{1,\ldots,G\}$. Let us perform the large $G$ limit (remembering that $g/G$ stays however finite). Using (\ref{eq:Z2}) we have:
\begin{align}
c_g^2 &= \frac{2^{-\frac{2Cg}{G} } }{Z^2}\\
	&=\frac{ G (1-2^{-\frac{2C}G})}{2^{-\frac{2C}G}
  \(1-2^{-2C}\)}2^{-\frac {2Cg}G}\\
  &= \frac{ G (1-2^{-\frac{2C}G})}{  \(1-2^{-2C}\)}2^{-\frac
  {2C(g-1)}G}\\
  &\approx G\frac{ 2^{-\frac
  {2C(g-1)}G} }{  \(1-2^{-2C}\)} \(\log(2) 2C/G + O(1/G^2)\)
  \\&\approx \frac {2C \log(2)~2^{-\frac {2C(g-1)}G}}{1-2^{-2C}} + O(1/G) \label{eq1:cgsq}
\end{align}
Now, we note from (\ref{eq:AWGN_C}) that the ${\rm snr}$ can be written as
$\rm snr=2^{2C}-1=\frac {1-2^{-2C}}{2^{-2C}}$ so plugging (\ref{mysum}) inside (\ref{eq1:calE}) we have:
\begin{align}
1/{\rm snr}+{\cal E}_{\tilde g-1} &= \frac {2^{-2C}}{1-2^{-2C}} + 1 - \frac {1-2^{-\frac{2C(\tilde   g-1)}G}}{1-2^{-2C}}\\
&=\frac {2^{-\frac{2C(\tilde   g-1)}G}}{1-2^{-2C}}
\end{align}
Therefore to leading order, we have using (\ref{eq1:cgsq}) that:
\begin{equation}
\frac{1/{\rm snr} + {\cal E}_{\tilde g-1}}{c_{\tilde g}^2} \approx \frac 1{2C \log(2)}
\end{equation}
so that the condition (\ref{condition}) becomes for large $G$:
\begin{equation}
\frac {R \log(2)}{2C \log(2)} = \frac R{2C}<\frac 12
\end{equation}
or equivalently, $R<C$. This shows that, with proper power allocation
and as long as $R<C$, a local minimum asymptotically cannot exist in the potential, or equivalently, that the AMP decoder cannot be
stuck in such a spurious minimum: it will reach the solution with perfect reconstruction ${SER}=0$.
\section{Numerical experiments for finite size signals}
\label{sec:numerics}
\begin{figure}[!t]
\centering
\includegraphics[width=0.45\textwidth, trim=42 35 74 8, clip=true]{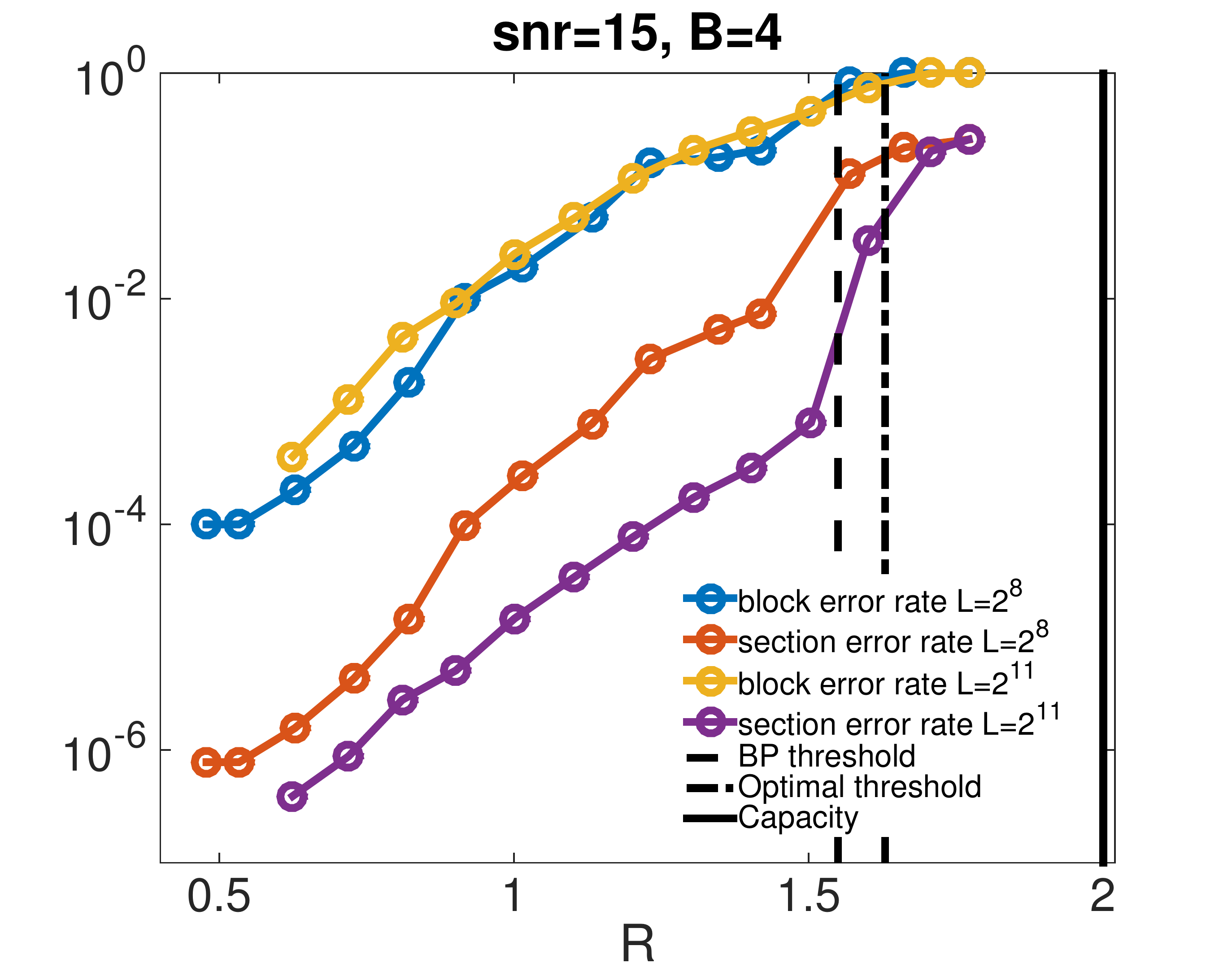}
\includegraphics[width=0.45\textwidth, trim=42 35 74 8, clip=true]{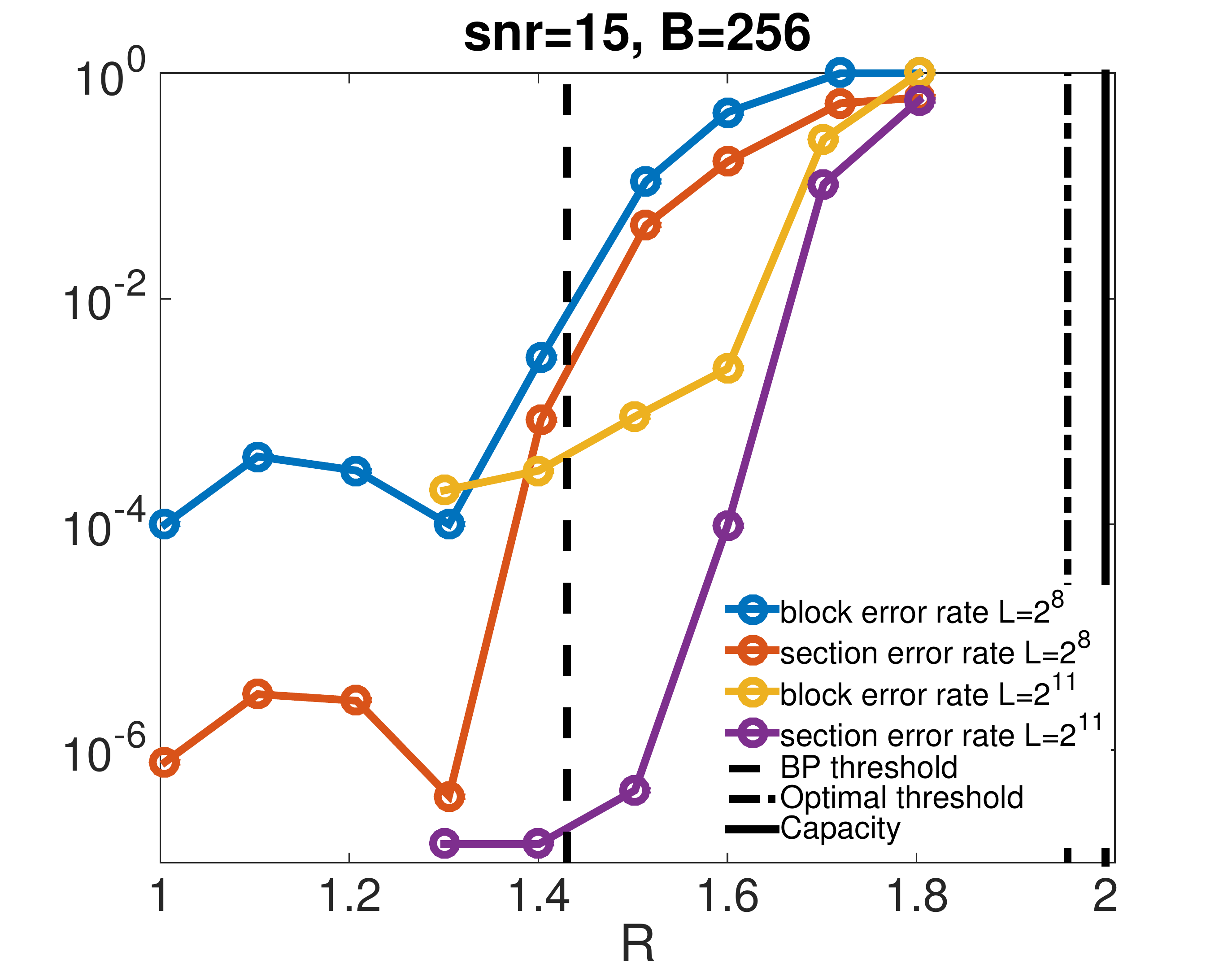}
\includegraphics[width=0.45\textwidth, trim=42 2 74 8, clip=true]{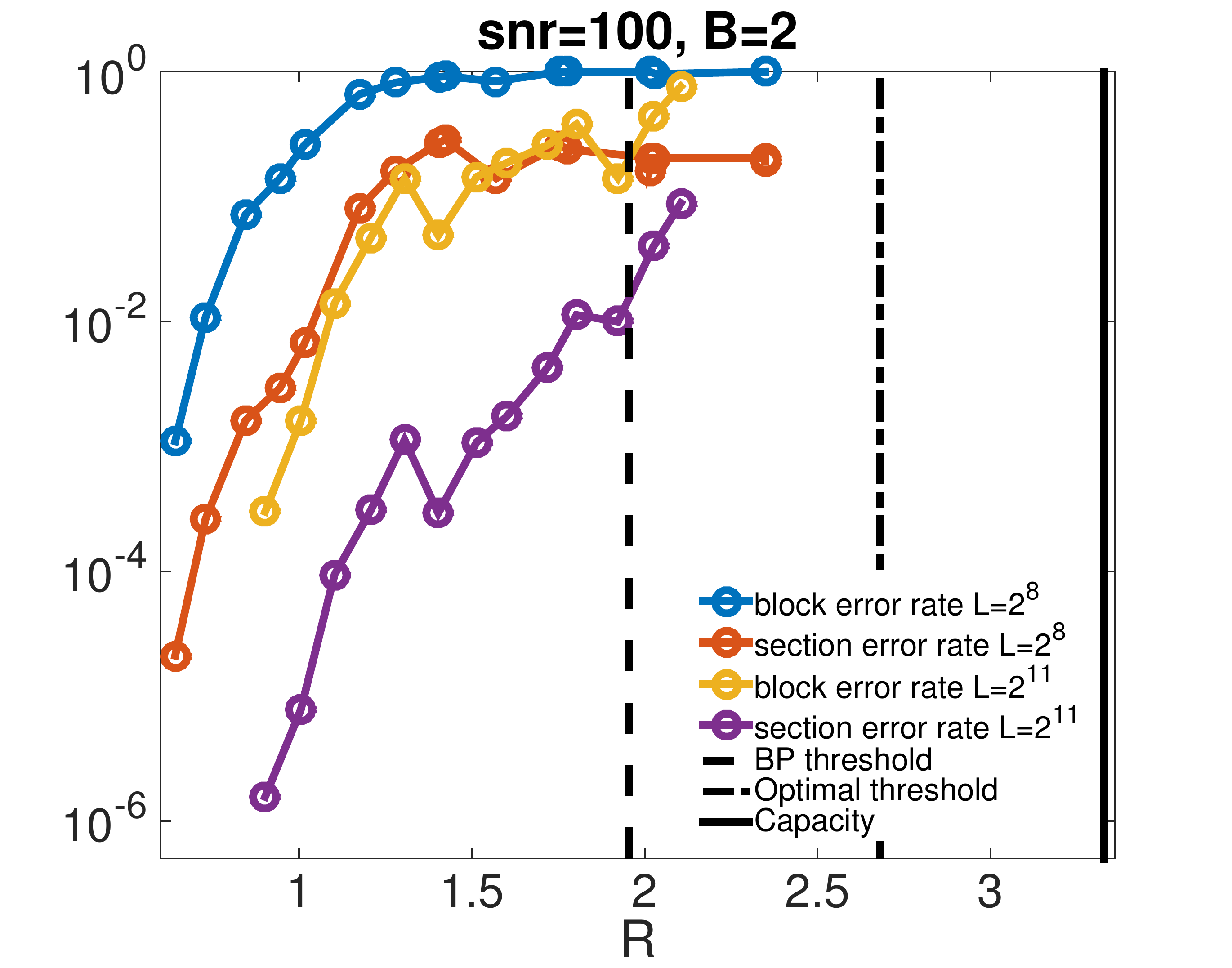}
\includegraphics[width=0.45\textwidth, trim=42 2 74 8, clip=true]{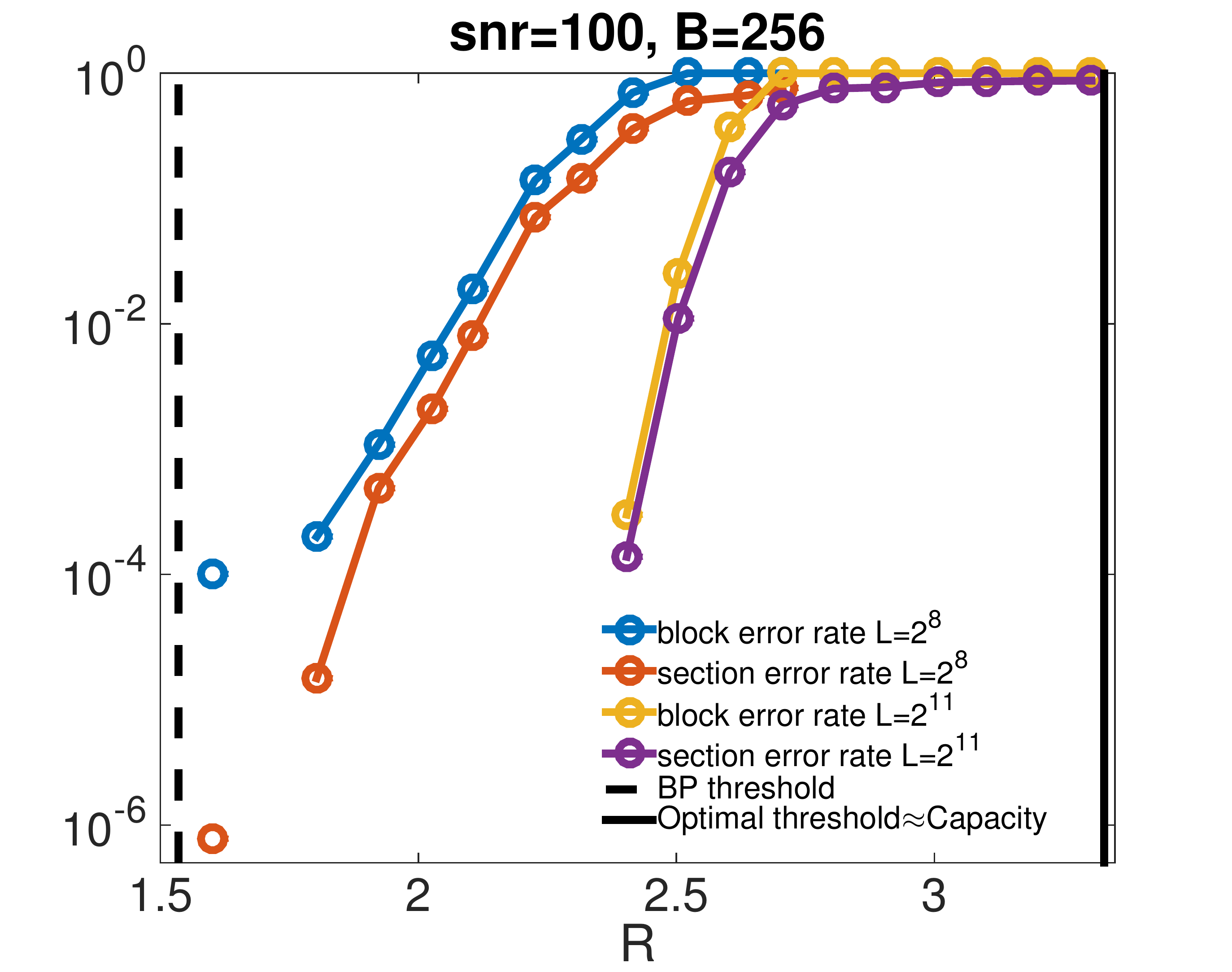} 
\caption[Block and section error rates and finite size effects for superposition codes]{On this plot, we show the empirical block error rate and average section error rate of the
  superposition codes using the AMP decoder combined with spatially-coupled
  Hadamard-based operators for two different ${\rm{snr}}$, signal sizes $L$
  and section sizes $B$. The block error rate is the fraction of the
  $10^4$ random instances we ran for each point that have not been
  perfectly reconstructed, i.e. in these instances at least one section
  has not been well recontructed and the final ${{SER}} > 0$. The ${{SER}}$ has been averaged over the $10^4$ random instances.
  The convergence
  criterion is that the mean change in the variables estimates between
  two consecutive iterations $\delta < 10^{-8}$ and the
  maximum number of iterations is $t_{max}=3000$. The upper plots
  are for ${\rm{snr}}=100$, the lower for ${\rm{snr}}=15$ (notice the
  different x axes). The first dashed black line is the BP transition
  obtained by state evolution analysis, the second one is the optimal
  transition obtained by the replica method from the Bethe free entropy (\ref{eq1:freeEnt2_2}) and the solid black
  line is the capacity. In the (${\rm{snr}}=100, B=256$) case, the
  optimal transition is so close to the capacity that we plot a single
  line. For such sizes, the block error rate is $0$ for rates lower
  than the lowest represented one. The
  spatially-coupled operators used for the experiments are drawn from the ensemble
  $(L_c=16,L_r=17,w=2,\sqrt{J}=0.4,R, \beta_{{{seed}}}=1.8)$.}\label{figCh1:finiteSizeSeeded}
\end{figure}
We now present a number of numerical experiments testing the performance and behavior of the AMP decoder in different practical scenarios with finite size signals. The first experiment Fig.~\ref{figCh1:finiteSizeSeeded} quantifies the influence of the finite size effects over the superposition codes scheme with spatially-coupled Hadamard-based operators, decoded by AMP. For each plot, we fix the ${\rm snr}$ and the alphabet size $B$ and repeat $10^4$ decoding experiments per point with each time a different signal with constant power allocation and operator drawn from the ensemble $(L_c=16,L_r=17,w=2,\sqrt{J}=0.4,R, \beta_{{{seed}}}=1.8)$. The curves present the empirical block error rate (blue and yellow curves) which is the fraction of instances that have not been perfectly decoded, i.e. such that the final ${SER} >0$, and the average ${SER}$ (red and purple curves). This is done for two different sizes $L=2^8$ and $L=2^{11}$. When the curves stop, it means that the empirical block error rate (and thus the section error rate as well) is actually $0$. In reality it should reach a noise floor $<10^{-4}$ but does not because of the same reasons explained in sec.~\ref{sec:SE}. The dashed lines are the BP transition $R_{BP}({\rm snr}, B)$ and optimal transition $R_{opt}({\rm snr}, B)$ extracted respectively from the state evolution analysis and potential (\ref{eq1:freeEnt2_2}) and the solid black line is the capacity $C({\rm snr})$. Thanks to the fact that at large enough section size $B$, the gap between the BP transition and capacity is consequent, it leaves room for the spatially-coupled AMP decoder to beat the transition, allowing to decode at $R>R_{BP}$ as in LDPC codes. For small section size $B$, the gap is too small to get real improvement over the full operators. We also note the previsible fact that as the signal size $L$ increases, the results are improving: one can decode closer to the asymptotic transitions and reach a lower error floor. For $B=256$, the sharp phase transition between the phases where decoding is possible and impossible by AMP with spatial coupling is clear and gets sharper as $L$ increases.

\begin{figure}[t!]
\centering
  \includegraphics[width=1\textwidth]{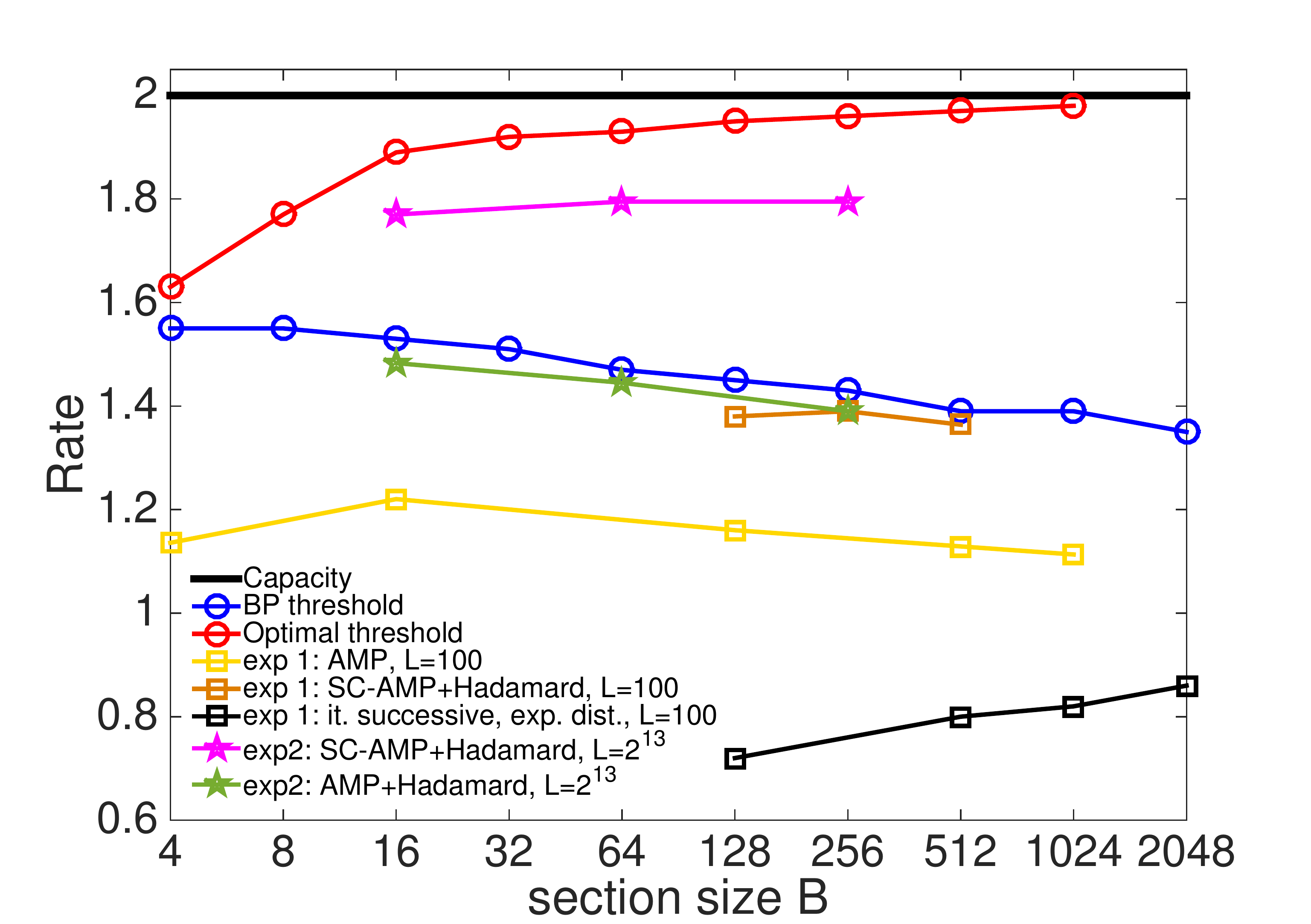}
  \caption[Phase diagram of superposition codes with finite size results]{Phase diagram and experimental results for superposition
    codes at finite size $L$ for ${\rm{snr}}=15$
    compared to the asymptotic results. The solid black line is the capacity which
    bounds the performance of any reconstruction algorithm for this
    ${\rm{snr}}$, the blue line is the BP transition $R_{BP}({\rm{snr}}=15,B)$ obtained by state
    evolution analysis and the red line
    is the Bayesian optimal transition $R_{opt}({\rm{snr}}=15,B)$ obtained by
    from the potential (\ref{eq1:freeEnt2_2}). The
    yellow, black and brown curves are results of the following
    experiment (exp 1): decode $10^4$ random instances and identify
    the empirical transition curve between a phase where the empirical probability
    $P\({{SER}}>10^{-1}\)<10^{-3}$ (below the line) from a phase where
    $P\({{SER}}>10^{-1}\)\ge10^{-3}$ (more than 9 instances have failed over the
    $10^4$ ones). The green and pink curves are the result of the
    second protocol (exp 2) which is a relaxed version of exp 1 with
    $10^2$ random instances and $P\({{SER}}>10^{-1}\)<10^{-1}$ below the line,
    $P\({{SER}}>10^{-1}\)\ge10^{-1}$ above. Note that in our experiments
    ${{SER}}<10^{-1}$ essentially means ${{SER}}=0$ at these
    sizes. The yellow curve compares our results with the iterative
    successive decoder (black curve) of
    \cite{barron2010sparse,barron2011analysis} where the number of
    sections $L=100$. Note that these data, taken from
    \cite{barron2010sparse,barron2011analysis}, have been generated
    with an exponential power allocation rather than the constant one
    we used. Compared with the yellow curve (AMP with the same
    value of $L$) the better quality of AMP reconstruction is
    clear. The green and pink curves are here to show the efficiency
    of the Hadamard-based operators with AMP with (pink curve) or without
    (green curve) spatial coupling. For the experimental results, the
    maximum number of iterations of the algorithm is arbitrarily fixed
    to $t_{max}=500$. The parameters used for the spatially-coupled
    operators are
    $(L_r=16, L_c=17, w=2,
    \sqrt{J}=0.3,R, \beta_{{{seed}}}=1.2)$.}\label{figCh1:phaseDiagSC}
\end{figure}

The next experiment Fig.~\ref{figCh1:phaseDiagSC} is the phase diagram 
for superposition codes at fixed ${\rm{snr}}=15$ like on Fig.~\ref{figCh1:diagsDist} but where we added on top finite size results. The asymptotic rates that can be reached
are shown as a function of $B$ (blue line for the BP transition, red one for the optimal rate). The solid black line is the capacity. Comparing the black and yellow curves, it is clear that even without spatial coupling or power allocation, AMP outperforms the iterative successive decoder of \cite{barron2010sparse} for practical
$B$ values. With the Hadamard-based spatially-coupled AMP algorithm, this is true for any $B$ and is even more
pronounced (brown curve). The green (pink) curve shows that the homogeneous (spatially-coupled) Hadamard-based operator has very good performances for reasonably large signals, corresponding here to a
blocklength $M<64000$ (the blocklength is the size of the transmitted codeword $\tilde \by$).

\begin{figure}[!t]
\centering
\includegraphics[width=.95\textwidth, trim=45 5 45 27, clip=true]{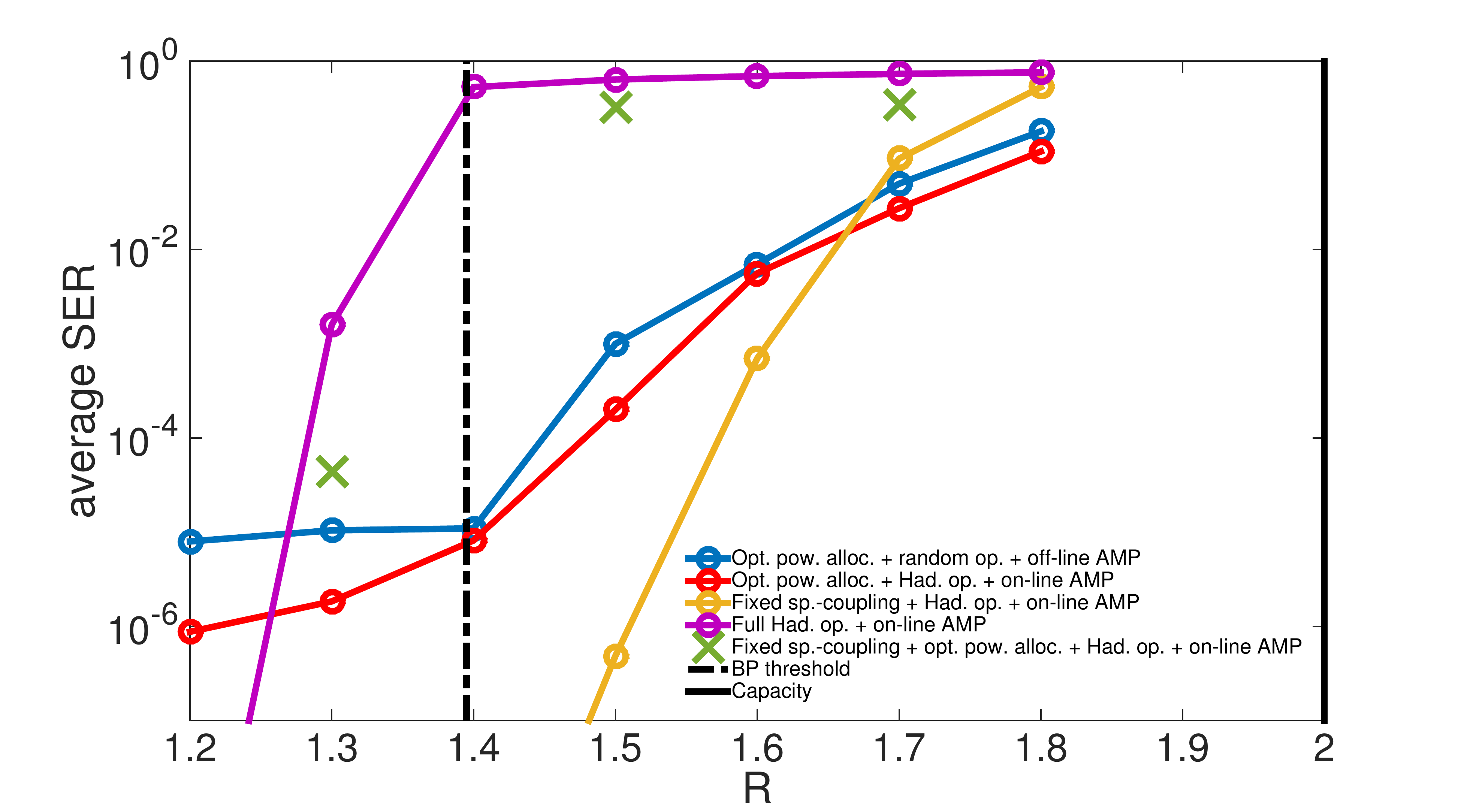}
\caption[Comparison between power allocation and spatial coupling]{The average section error rate ${{SER}}$ in logarithmic scale as a function of the rate $R$ for different settings, all at fixed (${\rm{snr}}=15, B=512, L=1024$). The black dashed curve identifies the BP transition, the highest rate until which AMP can asymptotically perform well {\it{without}} spatial coupling or non constant power allocation, the black solid line is the Shannon capacity. \textbf{Blue curve :} It corresponds to the results of Fig.~3 of \cite{rush2015capacity}: the points are averaged over $10^3$ instances of random experiments using i.i.d Gaussian matrices and an optimized power allocation scheme where the parameters defining the power allocation are optimized for each rates. The values of the parameters and the associated power allocation scheme can be found in \cite{rush2015capacity}. The denomination off-line AMP refers to the AMP decoder update rules of \cite{rush2015capacity} that are different than ours and require an off-line pre-processing part as opposed to our procedure where all the quantities are computed on-line without any need of pre-processing. \textbf{Red curve :} We reproduced exactly the same experiment (with same power allocation scheme and parameters) as for the blue curve with two important differences: $i)$ we used our on-line AMP decoder instead of their off-line implementation and $ii)$ we used an Hadamard-based homogeneous operator instead of a random i.i.d Gaussian one. In addition, we runned $10^4$ instances instead of $10^3$ as we obtained an average ${{SER}}$ equals to $0$ for the two first points. \textbf{Purple curve :} This experiment with $10^4$ instances per point is with an Hadamard-based homogeneous operator with on-line AMP decoding of constant power allocated signals. As it should, the decoder does not work anymore for $R>R_{BP}$. \textbf{Yellow curve :} The points of this experiment have been averaged over $10^4$ instances. In this setting, we used our on-line AMP decoder and generated the signals with constant power allocation. We replaced the homogeneous operator by a spatially-coupled Hadamard-based operator, described in Fig.~\ref{fig_seededHadamard}. The parameters defining the ensemble from which the operator is randomly generated are fixed once for all for the all experimental curve, as oposed to the power allocation curves where parameters have been optimized for each point. The ensemble is here given by $(L_c=16,L_r=17,w=2,\sqrt{J}=0.4,R, \beta_{{{seed}}}=1.4)$. \textbf{Green crosses :} These points have been averaged over $10^4$ instances. We used the same spatially-coupled Hadamard-based operator ensemble as for the yellow curve for decoding power allocated signals with same power allocation scheme as the blue and red curves. When the purple and yellow curves fall, it means that the points values are $0$. The codeword size for all these curves is between $5\times10^3$ to $9\times10^3$.\label{figCh1:compPowaVSspC}}
\end{figure}

Finally, the last experiment Fig.~\ref{figCh1:compPowaVSspC} is a comparison of the efficiency of the AMP decoder combined with spatial coupling or an optimized power allocation coming from \cite{rush2015capacity}. We repeated their experiments and also compared the results to a spatial coupling strategy. Comparing the results with Hadamard-based operators, given by the red and yellow curves for power allocation and spatial coupling respectively, it is clear that spatial coupling (despite being not optimized at each rate as it is done for the power allocation) greatly outperforms an optimized power allocation scheme. 

In addition, we see that our red curve corresponding to the optimized power allocation homogeneously outperforms the results of \cite{rush2015capacity} with exactly the same parameters, given by the blue curve. As we have numerically shown that Hadamard-based operators gets same final performances as random ones as used in \cite{rush2015capacity} (see chap.~\ref{chap:structuredOperators} and Fig.~\ref{figCh1:distToRbp}), the difference in performances must come from the AMP implementation: in our decoder that we denote by on-line decoder, there is no need of pre-processing computations but in the decoder of \cite{rush2015capacity} denoted by off-line, quantities need to be computed in advance. 

The advantage of spatial coupling over power allocation is independent of the decoder and the fact that we use Hadamard-based operators, as it outperforms the red curve as well which is also obtained with our on-line decoder and Hadamard-based operators. This is true at any rate except at very high values where spatial coupling does not decode at all, meanwhile the very first components of the signal are found by power allocation strategy as their power is very large. But it is not a really useful regime as only a small part of the signal is decoded anyway. The green points show that a mixed strategy of spatial coupling with optimized power allocation does not perform well compared to individual strategies. This is easily understood from the Fig.~\ref{figCh1:equivPowaSpc}: a power allocation modifies the spatial coupling and worsen its original design. In addition we notice that at low rates, a power allocation strategy performs worst than constant power allocation as the very last components with very low power are rarely decoded.
\section{Concluding remarks}
We have fully derived and studied the approximate message-passing decoder, combined with spatial coupling or power allocation, for the sparse superposition error correction scheme over the additive white Gaussian noise channel. Links have been established between the present problem and compressed sensing with structured sparsity.

On the theoretical side, we have computed the potential of the scheme thanks to the heuristic replica method and have shown that the scheme is capacity achieving in a proper limit. The analysis shows that there exist a sharp phase transition blocking the decoding by message-passing before the capacity but that the optimal Bayesian decoder obtained by combining message-passing to spatial coupling or power allocation can reach the capacity as the section size of the signal increases. We have also derived the state evolution recursions associated to the message-passing decoder, with or without spatial coupling and power allocation. The optimal performances have been studied and it appeared that the error decrease and the rates of convergence of the various transitions to their asymptotic values follow power laws.

On the more practical and experimental side, we have presented an efficient and capacity achieving solver based on spatially-coupled fast Hadamard-based operators. It allows to deal with very large instances and performs as well as random coding operators. Intensive numerical experiments have shown that a well designed spatial coupling performs way better than an optimized power allocation of the signal, both in terms of reconstruction error and robustness to noise. Finite size performances of the decoder under spatial coupling have been studied and it appeared that even for small signals, spatial coupling allows to obtain very good perfomances. In addition, we have shown that the message-passing decoder without spatial coupling beats the iterative successive decoder of Barron and Joseph for any manageable size and that its performances with spatial coupling are way better for any section size.

The scheme should be now compared in a more systematic way to other state-of-the-art error correction schemes over the additive white Gaussian noise channel. On the application side, from the structure of the reconstructed signal itself in superpostion codes, we can also interpret the problem as a structured group testing problem where one is looking for the only individual that has some property (for example infected) in each group, the sections of the signal.
\chapter{Robust error correction for real-valued signals via message-passing decoding and spatial coupling}
\label{chap:robustErrorCorrection}
\vspace{1cm}
In the previous chapter we studied error correction over the additive white Gaussian noise channel. Some sparse discrete signal was encoded using a linear transform to get a real codeword sent through the channel. But let us imagine now that the noisy channel is even less reliable than the AWGN one as it adds gross Gaussian distributed errors in addition of the background AWGN. Is it possible anyway to send information reliably through such channel?
The real transmitted signal in this new model can be interpreted as the codeword of another coding scheme for the AWGN. So if we manage to correct these gross errors, we come back to the original AWGN channel error correction problem and sparse superposition codes (or any other scheme for the AWGN) can be used afterwards.

As for the sparse superposition codes scheme, we will use the approximate message-passing algorithm as an efficient decoder. We will show that the error correction and its robustness towards noise can be enhanced considerably thanks to spatially-coupled coding matrices. We discuss the performances in the large signal limit using previous results on state evolution, as well as for finite size signals through numerical simulations. Even for relatively small sizes, the approach proposed here outperforms convex-relaxation-based decoders. 
%
%
%
%
%
%
%
%
%
%
%
%
%
%
%
%
%
%
%
\section{Introduction}
Although information is discrete in the classical coding theory, there are
situations of interest where one should consider real-valued signals,
such as scrambling of discrete time analog signals for privacy
\cite{1056050}, network \cite{feizi2011power,shintre2008real} or
jointed source and channel coding \cite{grangetto2005joint}, or in the
impulse noise cancellation in orthogonal frequency division
multiplexing systems \cite{4595196}. We consider here a real channel model which adds gross errors on a fraction of elements and a small noise on all of them. The real signal could also be interpreted as the codeword of a previous error correction scheme, and the full error correction becomes the concatenation of two distinct schemes: the first for the background noise, on top of which we use the present scheme to correct the gross errors.

To perform error correction for such real signals, a compressed sensing based scheme has been proposed by Donoho and Huo \cite{959265} and Candes and Tao \cite{CandesTao:05}. Here we reconsider this problem taking full advantage of the approximate message-passing decoder and spatial coupling coding design.

The problem is easily stated. One is given a real-valued signal $\bs$,
and a channel that adds gross errors to a fraction of elements. Is there a way to encode the signal such that these errors can be corrected? Can this approach still be used when the channel is in addition adding a small AWGN to all elements (a situation arguably much closer to some real
channels \cite{959265,Candes:2008:HRE:2263476.2273354,lampe2011bursty})?
The method proposed in \cite{959265,CandesTao:05,Candes:2008:HRE:2263476.2273354} is to first multiply the signal $\bs$ by a random matrix in order to create a
codeword of larger dimension, and then to use the classical compressed sensing approach, based on convex-relaxation decoding, in order to correct the
errors of transmission.

In the present chapter, we replace
the convex-relaxation decoding by the AMP decoder that uses the available prior
information about the error statistical properties \cite{DonohoMaleki10,KrzakalaPRX2012} as opposed to convex optimization based solvers, see sec.~\ref{sec:advantagesDisadvantagesConvex}. This provides a significant
improvement in performances. Then we consider an approximately sparse channel
where, in addition to the gross errors on a fraction of elements, there is a small AWGN over all components. We will show that the performances of the AMP 
decoder are stable under this additional noise, as already discussed in chap.~\ref{chap:appSparsity}. Finally we will use spatially-coupled measurement matrices in the decoding, which allow to further enhance the possibility for error correction (and up to its
information theoretical limit in the case of strictly sparse
noise).
\section{Compressed sensing based error correction}
Consider a real-valued vector of information $\bs \in
\mathbb{R}^N$, encode this vector by a full-rank real $M \times N$
matrix $\tbf A$, with $\gamma \defeq M/N > 1$ being the encoding rate (the redundancy or "over-sampling" introduced in the code), so that the codeword is
$\by=\tbf A \bs \in \mathbb{R}^M$. The aim is to recover $\bs$ lowering as much as possible the encoding rate. Since $\tbf A$ is full rank, one can recover
the original signal $\bs$ from $\by$ multiplying it by the
pseudo-inverse of $\bA$. The codeword is then sent through a noisy channel and gives rise to the corrupted codeword $\bytilde=\by+\bre$ where $\bre$ is i.i.d with a distribution:
\be 
	P(\bre) = \prod_i^M \[\rho {\cal N}(e_i|0,1+\epsilon)+ (1-\rho) {\cal N}(e_i|0,\epsilon) \] \label{noise}
\ee
where $0<\rho<1$. So the noise distribution is of the approximate sparsity form studied in chap.~\ref{chap:appSparsity}. We thus have a fraction $\rho$ of elements with
gross (variance $1+\epsilon$) errors, the rest having small (variance
$\epsilon$) amplitudes. One then considers a full rank
"parity-check"-like matrix $\bF$ such that $\bF \tbf A=\bsy 0$. We construct such a
pair of matrices by first choosing a $R \times M$ matrix $\bF$ with
i.i.d Gaussian distributed elements of zero mean and variance
$1/M$ (or variance specified by the seeding matrix, see Fig.~\ref{fig_opSpCoupling}), the kernel of $\bF$ is then the range of the encoding matrix $\tbf A$. Note that
  \cite{CandesTao:05,Candes:2008:HRE:2263476.2273354} take $\tbf A$ as the
  random matrix, but here we choose the opposite in order to be able to implement the
  spatially-coupled decoding. One must have $R \le M-N$ for the couple $(\bF,\bA)$ to exist, and in order to minimize the encoding rate $\gamma$, we take from now on $R\defeq M-N$. The application of $\bF$ to the corrupted codeword $\bytilde$ results in the real-valued vector $\bh$ given by:
\begin{align} 
	\bh&= \bF (\by+\bre) \\
	&= \bF (\bA\bs+\bre)\\
	&=\bF \bre 
	\label{CS-eq}
\end{align}
where $\bh$ has dimension $R$ and $\bre$ is an approximately sparse vector of dimension $M$. This is a compressed sensing problem (\ref{eqIntro:AWGNCS}) similar to what have been studied in chap.~\ref{chap:appSparsity}: reconstruct the
approximately sparse $M$-d error vector $\bre$ given $R$ of its linear projections (measurements) $\bh$. In the context of compressed sensing $\bF$ is the measurement matrix. 

\begin{figure}
\centering
\includegraphics{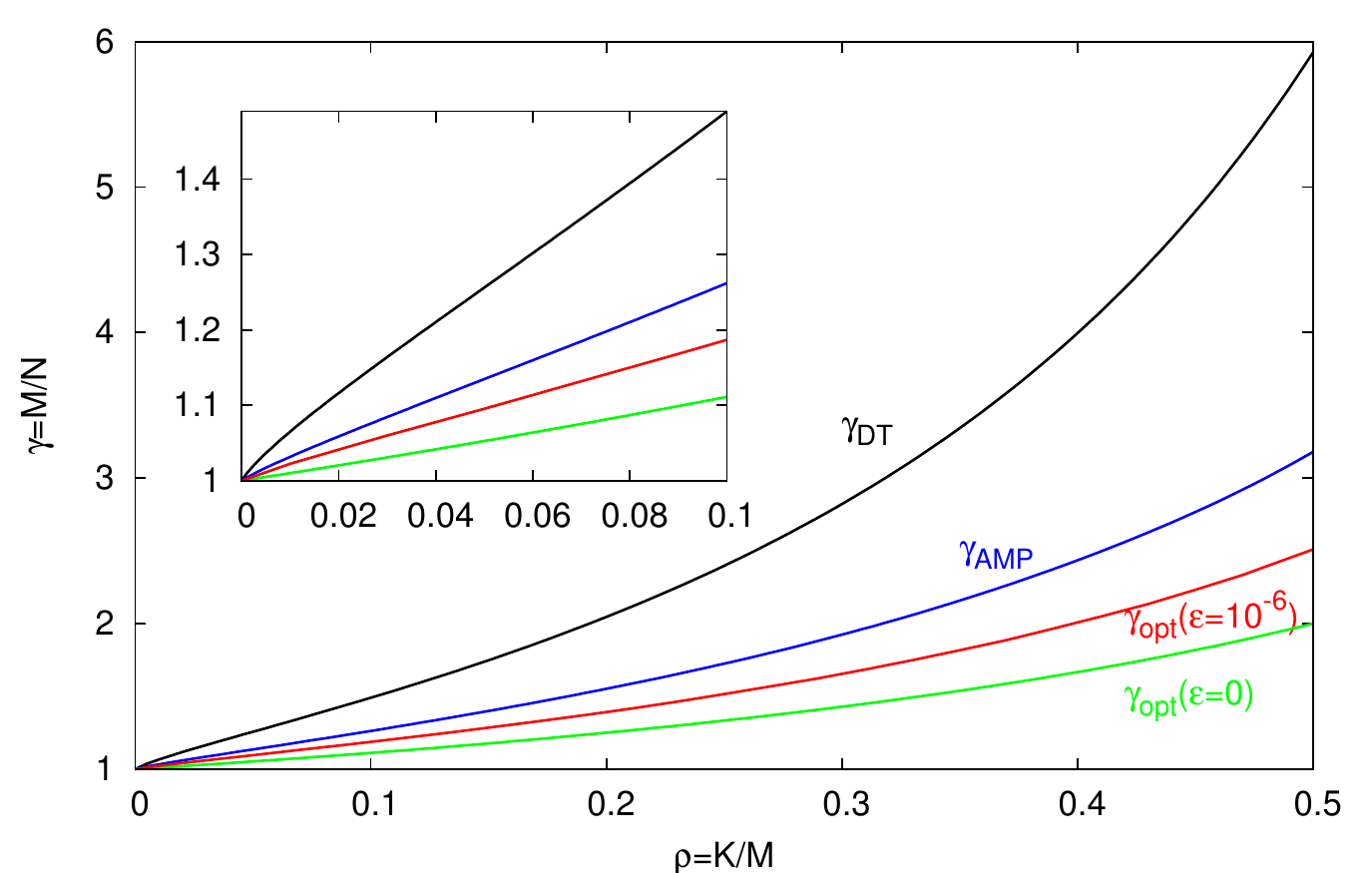}
\caption[Phase diagram for error correction of real-valued signals corrupted by approximately sparse Gaussian noise]{\label{fig:PhaseDiagram} Phase diagram showing the encoding
  rate $\gamma=M/N$ necessary to perform error correction over a
  channel with noise described by (\ref{noise}), plotted as a
  function of the noise sparsity $\rho$, zoom in the inset. The black (top) curve
  $\gamma_{DT}$ depicts the limit of performance of the $\ell_1$-minimization
  approach for $\epsilon=0$, i.e. when the noise is strictly sparse. The blue (2nd from top) curve shows the limit of performance of the Bayesian approximate message-passing approach for $\epsilon=0$. Note that up to about $\epsilon\lesssim 10^{-5}$
  this curve does not change visibly, see Fig.~\ref{fig_phase_diagram}. The green (bottom) curve,
  given by $\gamma_{opt}=1/(1-\rho)$, depicts the lowest possible encoding rate for
  which exact decoding is possible for $\epsilon=0$. Above the red (3rd
  from top) line, error correction with $MSE$ comparable to
  $\epsilon=10^{-6}$ is possible with the Bayes optimal estimation of
  the error vector. These two rates $\gamma_{opt}$ can be reached in the limit of large signal size 
  using the spatially-coupled Bayesian AMP approach.}
\end{figure}

Let us first review the possibility of this error-correction scheme when the error $\bre$ is exactly sparse, i.e. $\epsilon=0$.  Using an
intractable $\ell_0$ minimization, the gross error $\bre$ in
(\ref{CS-eq}) can be found exactly as long as $R=M-N>M\rho$. So error
correction in real-valued signals corrupted by strictly sparse gross noise is
possible (but hard, see sec.~\ref{sec:typicalPhaseTransitions}) for encoding rates $\gamma>\gamma_{opt}=1/(1-\rho)$. Popular tractable $\ell_1$-minimization, as used
in \cite{CandesTao:05,Candes:2008:HRE:2263476.2273354}, recovers the error $\bre$ exactly when $M-N \ge \alpha_{DT} M$, where $\alpha_{DT} $ is the Donoho-Tanner measurement rate
\cite{Donoho05072005}, see sec.~\ref{sec:typicalPhaseTransitions}, or equivalently when the encoding rate is larger
than $\gamma \ge \gamma_{DT} = 1/(1-\alpha_{DT} )$. These two
transitions are depicted in Fig.~\ref{fig:PhaseDiagram} and one can
see that $\gamma_{DT}$ is considerably larger than $\gamma_{opt}$. A first step to improvement is to decode with an approximate message-passing approach. 
\subsection{Performance of the approximate message-passing decoder}
As we fall exactly under the setting studied in the chap.~\ref{chap:appSparsity}, the posterior estimates of the noise components are directly given by the AMP decoder Fig.~\ref{algoCh1:AMP_op} with the operators definitions (\ref{eq_fastOpDefs11}), (\ref{eq_fastOpDefs1}), (\ref{eq_fastOpDefs2}), (\ref{eq_fastOpDefs22}) together with the denoisers (\ref{eq:fa_appSparsity}), (\ref{eq:fc_appSparsity}) where their parameters are given in the present case by:
\begin{align}
w_1&=\rho\\
\sigma_1^2&=1 +\epsilon\\
w_2&=1-\rho \\
\sigma_2^2&=\epsilon 
\end{align}
When parameters $\rho, \epsilon,\gamma$ are fixed whereas $M\to\infty$ and $\bF$ is an homogeneous random i.i.d matrix, the evolution of the AMP algorithm can be described exactly using the state evolution given by (\ref{Et}) with initialization $E^{t=0}=\txt{Var}_{P_0}(e_i)=\rho+\epsilon$. 
\subsubsection{State evolution for homogeneous matrices}
The state evolution analysis of the Bayesian AMP for Gauss-Bernoulli noise $\bre$ (that is,
with $\epsilon=0$) has been considered in great details in
\cite{KrzakalaPRX2012,KrzakalaMezard12}. In that case AMP reconstructs
{\it perfectly} the solution in a region larger than the
$\ell_1$-minimization and up to the spinodal transition
$\alpha_{BP}$, see sec.~\ref{sec:typicalPhaseTransitions}. In the notation of the present problem, this leads exact decoding for considerably lower encoding rates: the
resulting $\gamma_{BP}=1/(1-\alpha_{BP})$ is shown in blue in
Fig.~\ref{fig:PhaseDiagram} (where it is denoted $\gamma_{AMP}$). The advantage with respect to the $\ell_1$-minimization decoding is clear. For a fraction of $\rho=0.1$ of gross elements, for instance, the improvement goes from a necessary coding rate $\gamma_{DT} \approx 1.490$ for $\ell_1$-minimization based decoders to $\gamma_{BP} \approx
1.262$ with AMP decoding with homogeneous matrices.

As already discussed in sec.~\ref{sec:advantagesDisadvantagesConvex}, nevertheless, the $\ell_1$
performance is independent of the distribution of the gross error,
whereas the Bayesian AMP uses it. The properties of the channel are, however, often well
known, in which case the improvement depicted if
Fig.~\ref{fig:PhaseDiagram} is indeed achievable.

To assess how robust are these results towards approximately sparse noisy channels (nonzero value of $\epsilon$ in (\ref{noise})) we
use the state evolution analysis that was performed in chap.~\ref{chap:appSparsity}. It was shown that for about
$\epsilon\lesssim 10^{-5}$ and $\alpha>\alpha_{BP}$ the AMP
algorithm leads to reconstruction with $MSE$ comparable to
$\epsilon$, see Fig.~\ref{fig_MSE2}. This shows that the AMP approach is actually very robust to such noise.
\subsubsection{State evolution for spatially-coupled matrices}
Despite the advantage of the AMP-based decoding over the $\ell_1$-minimization, it is
still not asymptotically optimal since $\gamma_{BP}> \gamma_{opt}$, and one ideally aims
to perform error correction with smallest possible encoding rates. In order to do so, we use a spatially-coupled measurement matrix $\bF$. In the present framework of error correction of real-valued signals, the spatial coupling can be implemented by first constructing the matrix $\bF$ of the form Fig.~\ref{fig_opSpCoupling}, then 
determining the encoding matrix $\tbf A$ as the null space of the matrix $\bF$. 

The state evolution for these spatially-coupled matrices is given by (\ref{Et_seeded_appSparsity}), (\ref{eq_SEsigmaSeeded_appSparsity}). The results are
shown again in Fig.~\ref{fig:PhaseDiagram} in green (lower-most) curve
for $\epsilon=0$ and in red (3rd from top) curve for
$\epsilon=10^{-6}$. The conclusion is that with a properly spatially-coupled matrix $\bF$, one can perform error correction using AMP down to these very low encoding rates.
\begin{figure}[!ht]
\centering
\includegraphics[width=1\textwidth]{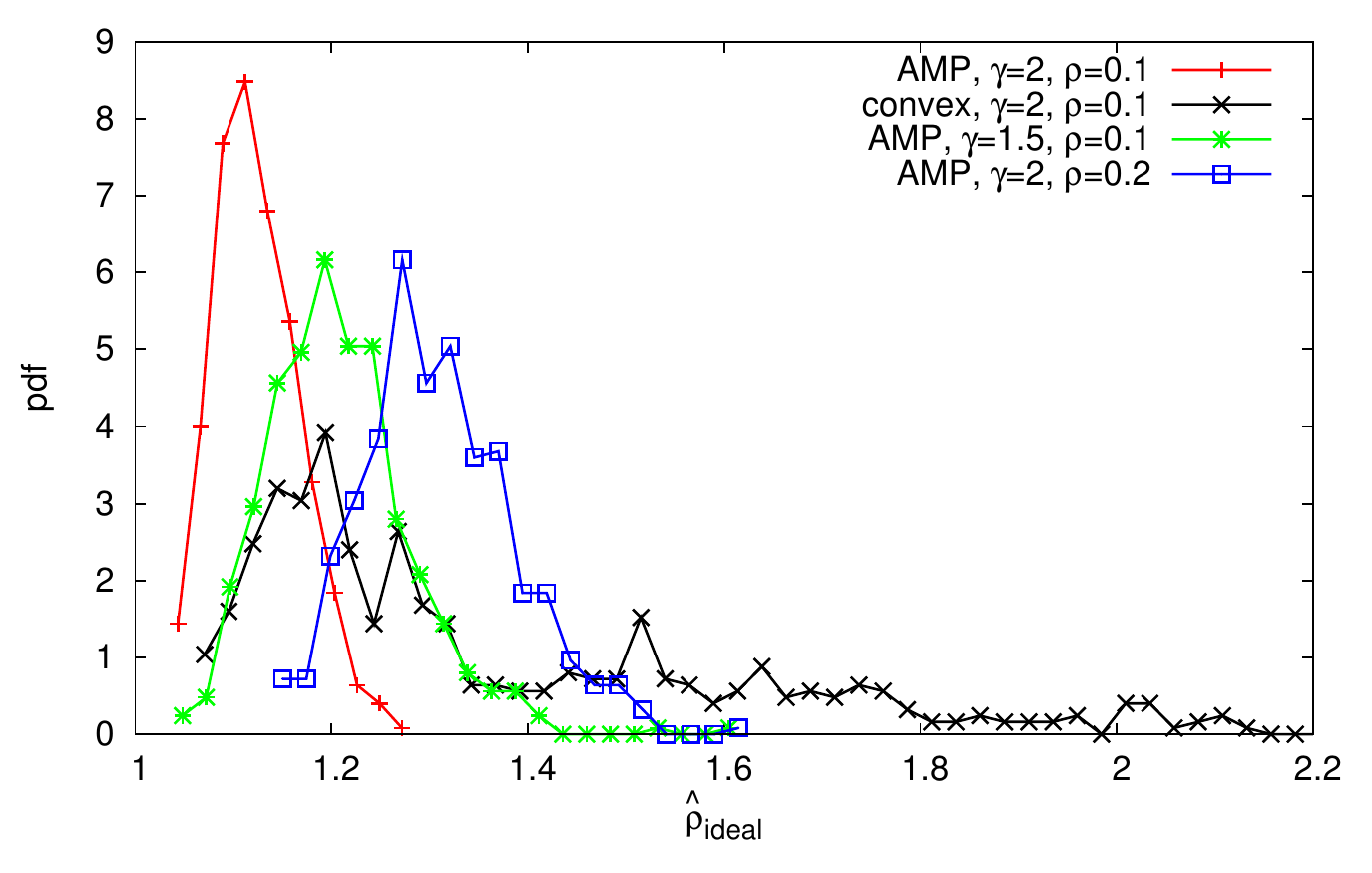}
\caption[Robustness to noise of the error correction of real-valued signals corrupted by an approximately sparse Gaussian channel]{\label{fig:dist:rho} Robustness to noise of the error
  correction of real signals of size
  $N=256$. We compare the performances of the AMP-based and $\ell_1$ decoders,
  using $\epsilon=10^{-6}$ and homogeneous Gaussian i.i.d matrices $\bF$. The figure
  shows the probability density (estimated over $500$ instances) of the
  robustness ratio (\ref{rhoideal}), called ${\hat \rho}_{ideal}$ in
  \cite{Candes:2008:HRE:2263476.2273354} at different values of
  the encoding rate $\gamma$ and gross noise sparsity $\rho$. For
  coding rate $\gamma=2$ and gross noise sparsity
  $\rho=0.1$, both methods (AMP in red and $\ell_1$ in black)
  are giving values close to one. However AMP is better: on
  average it gives $1.12$ versus $1.36$ for $\ell_1$, which empirical distribution has larger tails than the one of AMP. Furthermore, AMP still
  performs very well when the fraction of gross errors is doubled
  (blue curve, with $\rho=0.2$) or when the coding rate $\gamma$ is lower
  (green curve, with $\gamma=1.5$). In both cases, the
  $\ell_1$-based reconstruction gives poor results, an
  average value  ${\hat \rho}_{ideal} =37.1$ for
  ($\rho=0.2,\gamma=2$), versus $1.30$ for AMP, and ${\hat \rho}_{ideal} = 18.3$ for
  ($\rho=0.1,\gamma=1.5$) versus $1.20$ for AMP.}
\end{figure}
\section{Numerical tests and finite size study}
The asymptotic guarantees given in the last sections are encouraging,
but evaluating analytically finite $M$ corrections is intrinsically
more difficult and hence we withdraw to numerical verifications of the
achievable encoding rates for sizes relevant for practical
applications.

For numerical verifications, we use a randomly generated $N$-d Gaussian signal
$\bs$ with zero mean and unit variance (the algorithm is {\it not} using this information) and a channel noise distributed according to
(\ref{noise}), information that we know and use in the algorithm. We use the Bayesian AMP algorithm to
estimate the error $\hat \bre$. As already discussed, for exactly sparse channel $\epsilon=0$, the exact reconstruction of $\bre$ is possible and hence $\bs$
can be recovered exactly. For an approximately sparse channel with $\epsilon>0$, we use the AMP estimate of the error $\hat \bre$ to compute the estimate of $\tbf A \hat \bx$ and finally use the pseudoinverse of $\tbf A$ to estimate 
the signal $\hat \bx$. We compare to the $\ell_1$ decoding approach (including the performances-improving reprojection step) as developed in \cite{CandesTao:05,Candes:2008:HRE:2263476.2273354}.

\begin{figure}[t]
\centering
\includegraphics[width=1\textwidth]{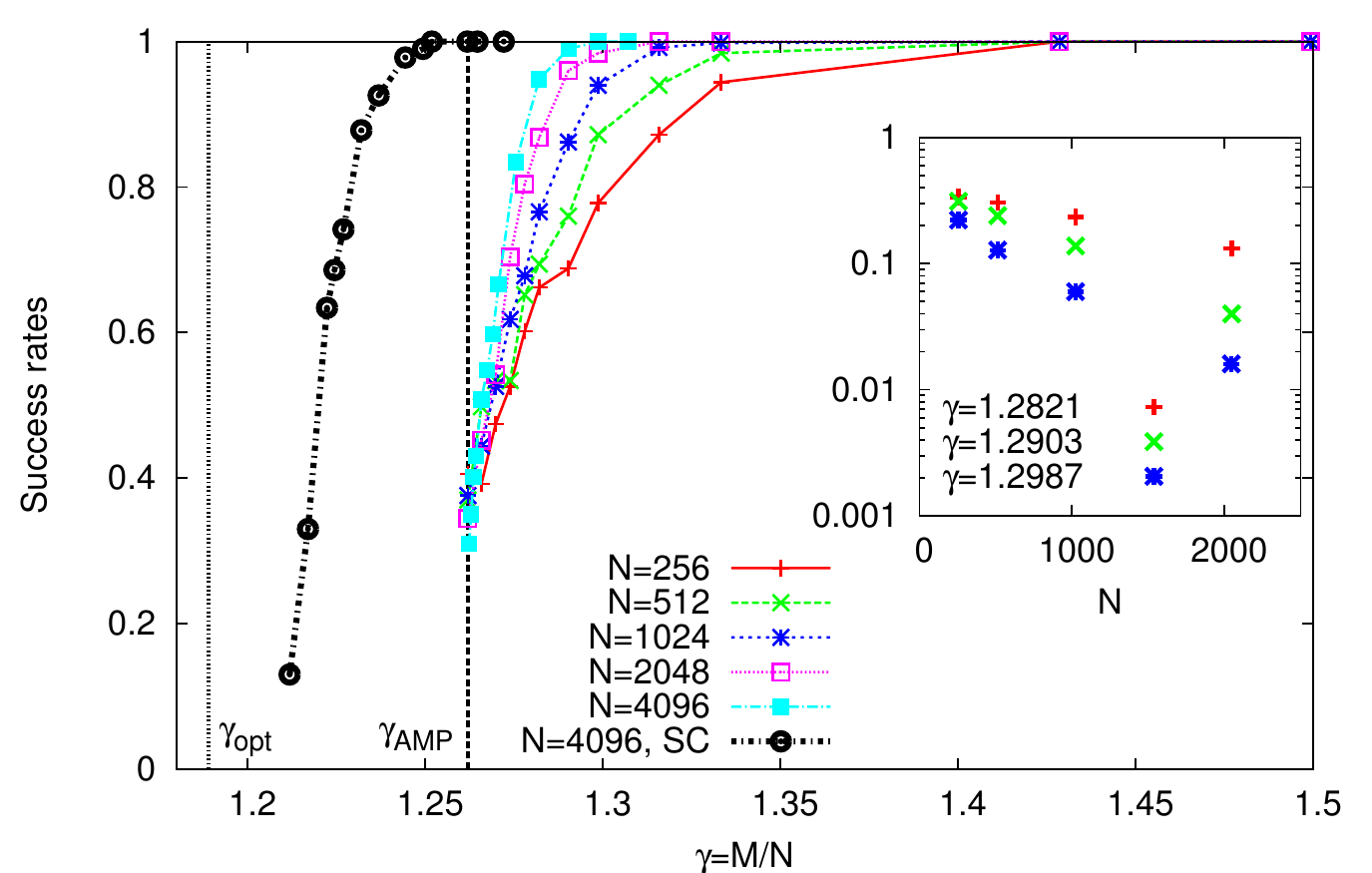}
\caption[Success rates of decoding with AMP over the approximately sparse Gaussian channel]{\label{fig:frac} Success rates of the decoding over $500$
  instances for different signal sizes $N$ with noise parameters
  $\rho=0.1$, $\epsilon=10^{-6}$, both with spatially-coupled matrices $\bF$ (SC)
  and homogeneous random Gaussian i.i.d ones, as a function of the coding rate
  $\gamma$. The vertical lines represent the limiting asymptotic
  coding rate for AMP with homogeneous ($\gamma_{AMP}$) and seeding matrices ($\gamma_{opt}$)
  respectively. The maximum number of iterations in these simulations
  is set to $1000$. An instance is considered successful if the final
  mean square error of the reconstructed signal is less than
  $10^{-5}$. The inner plot shows the empirical probability of failure over these instances for three different
  values of encoding rate $\gamma$. It decays exponentially with the signal size and the decay exponent-amplitude increases with the coding rate.}
\end{figure}

%
The data for $N=256$ are shown in Fig.~\ref{fig:dist:rho} where the
performance of the AMP decoding is compared to the
$\ell_1$-based decoding of \cite{Candes:2008:HRE:2263476.2273354}.
Following \cite{Candes:2008:HRE:2263476.2273354} we introduce an
estimator of the robustness to noise called ${\hat \rho}_{ideal}$ as 
the ratio of the $MSE$ of the
reconstructed signal $\hat \bx$ with the $MSE$ of the "ideal"
reconstruction $\hat \bx_{ideal}$,
where the pseudoinverse of $\tbf A$ is applied to $\by$ that was corrupted
only by the small AWGN without gross errors at all:
\be
{\hat \rho}_{ideal} \defeq \frac{||\hat \bx -  \bs||_2}{||\hat \bx_{ideal} -
  \bs||_2}
\label{rhoideal}
\ee 
Fig.~\ref{fig:dist:rho} depicts the histogram of ${\hat \rho}_{ideal}$ over 500 random instances of the problem. 
We find that in all the cases we have tried with AMP (which were all in
the favorable region of the asymptotic phase diagram, the easy phase above the BP transition), the robustness
estimator ${\hat \rho}_{ideal}$ is very close to unity, even at
these relatively small sizes.  Moreover the robustness estimator of
AMP was always on average closer to unity than the one based on
$\ell_1$ estimation and the distribution more peaked, thus
demonstrating the advantage of the Bayesian AMP reconstruction in
terms of performance, and noise robustness. Another important point is
that the probability of an unsuccessful reconstruction is decaying
exponentially fast when the system size increases. It is also decaying
faster as $\gamma$ increases (see Fig.~\ref{fig:frac}, inset).

\begin{figure*}[!ht]
\centering
\includegraphics[width=0.97\textwidth]{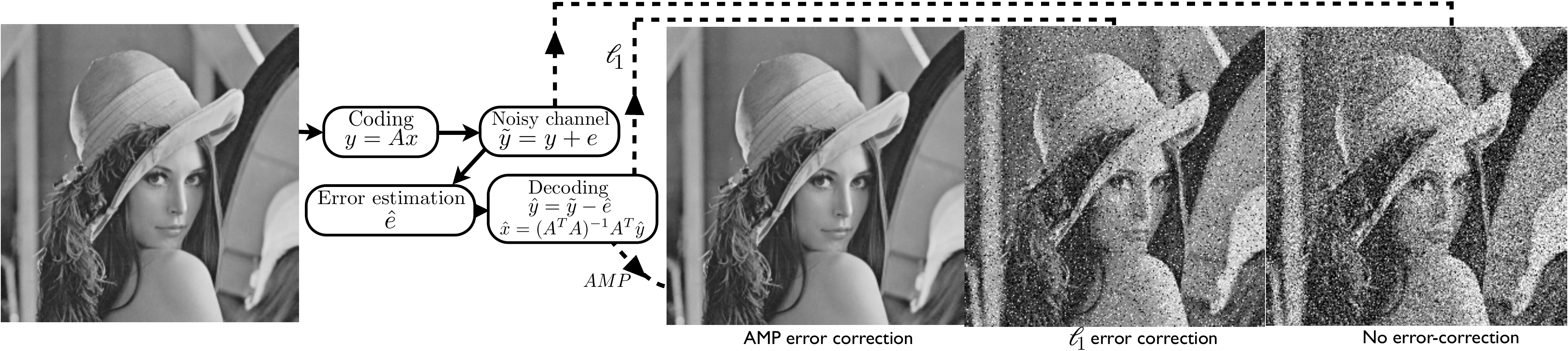}
\caption[Decoding Lena corrupted by the approximately sparse Gaussian noise channel]{\label{fig:Lena} Illustration of the error correction scheme
  with the spatially-coupled AMP approach and the $\ell_1$-based
  method of \cite{Candes:2008:HRE:2263476.2273354}, applied to the benchmark Lena picture. The original $256\times256$
  image is decomposed in patches of size $N=64^2$. The noisy channel
  is given by (\ref{noise}) using $\rho=0.1$, $\epsilon=10^{-6}$
  and the coding rate is $\gamma=1.256$ (to be compared with $\gamma_{opt}(\rho=0.1,\epsilon=10^{-6})=1.184$) is used together with a spatially-coupled matrix $\bF$ with parameters ($L_c=L_r=10$, $w=3$, $\sqrt{J}=\sqrt{0.2}$,
  $\alpha_{seed}=0.22$, $\alpha_{rest}=0.1830$). While error correction is close to perfect with AMP, the results are as poor as no error correction at all with an $\ell_1$ convex optimization solver.}
\end{figure*}

We also applied spatial coupling by choosing the matrix $\bF$
as described in Fig.~\ref{fig_opSpCoupling} drawn from the ensemble ($L_c=L_r=10$, $w=3$, $\sqrt{J}=\sqrt{0.2}$, $\alpha_{seed}=0.22$) and varying $\alpha_{rest}$. While such
parameters are far from the limit $L\to \infty$, $w\to \infty$, $w/L\to 0$ in which the optimal performance is guaranteed, we still obtain
a considerable improvement in the achievable encoding rate, as shown in Fig.~\ref{fig:frac}. In Fig.~\ref{fig:Lena} we give a more visual illustration of the performance of the spatially-coupled AMP
decoder for $N=4096$. For gross error sparsity $\rho=0.1$ and small
error variance $\epsilon=10^{-6}$ we were able to perform reliable
transmission at coding rate $\gamma= 1.256$. This has to be
compared with the original approach of
\cite{Candes:2008:HRE:2263476.2273354} which only allows, even for an infinite size system and in absence of small noise, an asymptotic $\gamma_{DT}(\rho=0.1) \approx 1.5$.
\section{Discussion}
We have studied an error-correction scheme based on AMP for real channels that corrupt a fraction of the codeword by gross errors. It appeared that this scheme is highly robust to an additional AWGN. Spatial coupling allows to reach close to optimal results. Nevertheless, an important question remains: how to properly optimize the spatial coupling on finite size systems? Furthermore, the present scheme can be combined with the structured operators developed in chap.~\ref{chap:structuredOperators} in order to work with very large signals. It would be also interesting to study the combined use of the present strategy with sparse superposition codes for joint source-channel coding problems.
%
%
%
%
%
%
%
%
%
%
%
%
%
%
%
\backmatter
\cleardoublepage
\bibliographystyle{plainnat}
\bibliography{PhD_jeanBarbier_AllInOne.bbl}

\begin{thebibliography}{100}
\providecommand{\url}[1]{#1}
\csname url@samestyle\endcsname
\providecommand{\newblock}{\relax}
\providecommand{\bibinfo}[2]{#2}
\providecommand{\BIBentrySTDinterwordspacing}{\spaceskip=0pt\relax}
\providecommand{\BIBentryALTinterwordstretchfactor}{4}
\providecommand{\BIBentryALTinterwordspacing}{\spaceskip=\fontdimen2\font plus
\BIBentryALTinterwordstretchfactor\fontdimen3\font minus
  \fontdimen4\font\relax}
\providecommand{\BIBforeignlanguage}[2]{{%
\expandafter\ifx\csname l@#1\endcsname\relax
\typeout{** WARNING: IEEEtran.bst: No hyphenation pattern has been}%
\typeout{** loaded for the language `#1'. Using the pattern for}%
\typeout{** the default language instead.}%
\else
\language=\csname l@#1\endcsname
\fi
#2}}
\providecommand{\BIBdecl}{\relax}
\BIBdecl

\bibitem{barbier2013hard}
J.~Barbier, F.~Krzakala, L.~Zdeborov{\'a}, and P.~Zhang, ``The hard-core model
  on random graphs revisited,'' in \emph{Journal of Physics: Conference
  Series}, vol. 473.\hskip 1em plus 0.5em minus 0.4em\relax IOP Publishing,
  2013, p. 012021.

\bibitem{BarbierKrzakalaAllerton2012}
J.~Barbier, F.~Krzakala, M.~M{\'e}zard, and L.~Zdeborov{\'a}, ``Compressed
  sensing of approximately-sparse signals: Phase transitions and optimal
  reconstruction,'' in \emph{50th Annual Allerton Conference on Communication,
  Control, and Computing}, 2012.

\bibitem{barbierSchulkeKrzakala}
\BIBentryALTinterwordspacing
J.~Barbier, C.~Schülke, and F.~Krzakala, ``Approximate message-passing with
  spatially coupled structured operators, with applications to compressed
  sensing and sparse superposition codes,'' \emph{Journal of Statistical
  Mechanics: Theory and Experiment}, vol. 2015, no.~5, p. P05013, 2015.
  [Online]. Available: \url{http://stacks.iop.org/1742-5468/2015/i=5/a=P05013}
\BIBentrySTDinterwordspacing

\bibitem{barbierRobustErrorCorrection13}
J.~Barbier, F.~Krzakala, L.~Zdeborova, and P.~Zhang, ``Robust error correction
  for real-valued signals via message-passing decoding and spatial coupling,''
  in \emph{Information Theory Workshop (ITW), 2013 IEEE}, Sept 2013, pp. 1--5.

\bibitem{barbier2014replica}
J.~Barbier and F.~Krzakala, ``Replica analysis and approximate message passing
  decoder for superposition codes,'' in \emph{2014 IEEE International Symposium
  on Information Theory}, 2014.

\bibitem{DBLP:journals/corr/BarbierK15}
\BIBentryALTinterwordspacing
------, ``Approximate message-passing decoder and capacity-achieving sparse
  superposition codes,'' \emph{CoRR}, vol. abs/1503.08040, 2015. [Online].
  Available: \url{http://arxiv.org/abs/1503.08040}
\BIBentrySTDinterwordspacing

\bibitem{mackay2003information}
D.~J. MacKay, \emph{Information theory, inference, and learning
  algorithms}.\hskip 1em plus 0.5em minus 0.4em\relax Citeseer, 2003, vol.~7.

\bibitem{james2013introduction}
\BIBentryALTinterwordspacing
G.~James, D.~Witten, T.~Hastie, and R.~Tibshirani, \emph{An Introduction to
  Statistical Learning: with Applications in R}, ser. Springer Texts in
  Statistics.\hskip 1em plus 0.5em minus 0.4em\relax Springer New York, 2013.
  [Online]. Available: \url{http://books.google.fr/books?id=qcI\_AAAAQBAJ}
\BIBentrySTDinterwordspacing

\bibitem{DBLP:journals/corr/TramelKGM14}
\BIBentryALTinterwordspacing
E.~W. Tramel, S.~Kumar, A.~Giurgiu, and A.~Montanari, ``Statistical estimation:
  From denoising to sparse regression and hidden cliques,'' \emph{CoRR}, vol.
  abs/1409.5557, 2014. [Online]. Available:
  \url{http://arxiv.org/abs/1409.5557}
\BIBentrySTDinterwordspacing

\bibitem{wasserman2010statistics}
\BIBentryALTinterwordspacing
L.~Wasserman, \emph{All of statistics : a concise course in statistical
  inference}.\hskip 1em plus 0.5em minus 0.4em\relax New York: Springer, 2010.
  [Online]. Available:
  \url{http://www.amazon.de/All-Statistics-Statistical-Inference-Springer/dp/1441923225/ref=sr_1_2?ie=UTF8&qid=1356099149&sr=8-2}
\BIBentrySTDinterwordspacing

\bibitem{DBLP:journals/corr/SchulkeCZ14}
\BIBentryALTinterwordspacing
C.~Sch{\"{u}}lke, F.~Caltagirone, and L.~Zdeborov{\'{a}}, ``Blind sensor
  calibration using approximate message passing,'' \emph{CoRR}, vol.
  abs/1406.5903, 2014. [Online]. Available:
  \url{http://arxiv.org/abs/1406.5903}
\BIBentrySTDinterwordspacing

\bibitem{sakellariou2013inverse}
J.~Sakellariou, ``Inverse inference in the asymmetric ising model,'' Ph.D.
  dissertation, Universit{\'e} Paris-Sud 11, 2013.

\bibitem{guggiola2015minimal}
A.~Guggiola and G.~Semerjian, ``Minimal contagious sets in random regular
  graphs,'' \emph{Journal of Statistical Physics}, vol. 158, no.~2, pp.
  300--358, 2015.

\bibitem{DBLP:journals/corr/KrzakalaMMNSZZ13}
\BIBentryALTinterwordspacing
F.~Krzakala, C.~Moore, E.~Mossel, J.~Neeman, A.~Sly, L.~Zdeborov{\'{a}}, and
  P.~Zhang, ``Spectral redemption: clustering sparse networks,'' \emph{CoRR},
  vol. abs/1306.5550, 2013. [Online]. Available:
  \url{http://arxiv.org/abs/1306.5550}
\BIBentrySTDinterwordspacing

\bibitem{2014arXiv1406.1880S}
A.~{Saade}, F.~{Krzakala}, and L.~{Zdeborov{\'a}}, ``{Spectral Clustering of
  Graphs with the Bethe Hessian},'' \emph{ArXiv e-prints}, Jun. 2014.

\bibitem{Fortunato201075}
\BIBentryALTinterwordspacing
S.~Fortunato, ``Community detection in graphs,'' \emph{Physics Reports}, vol.
  486, no. 3–5, pp. 75 -- 174, 2010. [Online]. Available:
  \url{http://www.sciencedirect.com/science/article/pii/S0370157309002841}
\BIBentrySTDinterwordspacing

\bibitem{2014arXiv1409.2290D}
A.~{Decelle}, J.~{H{\"u}ttel}, A.~{Saade}, and C.~{Moore}, ``{Computational
  Complexity, Phase Transitions, and Message-Passing for Community
  Detection},'' \emph{ArXiv e-prints}, Sep. 2014.

\bibitem{moore2011nature}
C.~Moore and S.~Mertens, \emph{The nature of computation}.\hskip 1em plus 0.5em
  minus 0.4em\relax Oxford University Press, 2011.

\bibitem{Wasserman:2006:NS:1202956}
L.~Wasserman, \emph{All of Nonparametric Statistics (Springer Texts in
  Statistics)}.\hskip 1em plus 0.5em minus 0.4em\relax Secaucus, NJ, USA:
  Springer-Verlag New York, Inc., 2006.

\bibitem{2015arXiv150503941M}
Y.~{Ma}, D.~{Baron}, and A.~{Beirami}, ``{Mismatched Estimation in Large Linear
  Systems},'' \emph{ArXiv e-prints}, May 2015.

\bibitem{DBLP:journals/corr/TramelDK15}
\BIBentryALTinterwordspacing
E.~W. Tramel, A.~Dr{\'{e}}meau, and F.~Krzakala, ``Approximate message passing
  with restricted boltzmann machine priors,'' \emph{CoRR}, vol. abs/1502.06470,
  2015. [Online]. Available: \url{http://arxiv.org/abs/1502.06470}
\BIBentrySTDinterwordspacing

\bibitem{2011PhRvL.107f5701D}
A.~{Decelle}, F.~{Krzakala}, C.~{Moore}, and L.~{Zdeborov{\'a}}, ``{Inference
  and Phase Transitions in the Detection of Modules in Sparse Networks},''
  \emph{Physical Review Letters}, vol. 107, no.~6, p. 065701, Aug. 2011.

\bibitem{richardson2008modern}
T.~Richardson and R.~Urbanke, \emph{Modern coding theory}.\hskip 1em plus 0.5em
  minus 0.4em\relax Cambridge University Press, 2008.

\bibitem{mezard2009information}
M.~M{\'e}zard and A.~Montanari, \emph{Information, physics, and
  computation}.\hskip 1em plus 0.5em minus 0.4em\relax Oxford University Press,
  2009.

\bibitem{2010PhDT474S}
V.~{Sessak}, ``{Inverse problems in spin models},'' Ph.D. dissertation, PhD
  Thesis, 2010, 2010.

\bibitem{2012JSMTE..08..015R}
F.~{Ricci-Tersenghi}, ``{The Bethe approximation for solving the inverse Ising
  problem: a comparison with other inference methods},'' \emph{Journal of
  Statistical Mechanics: Theory and Experiment}, vol.~8, p.~15, Aug. 2012.

\bibitem{KrzakalaZdeborova09}
F.~Krzakala and L.~Zdeborov\'a, ``Hiding quiet solutions in random constraint
  satisfaction problems,'' \emph{Phys. Rev. Lett.}, vol. 102, p. 238701, 2009.

\bibitem{modelBasedCSBaraniuk2008}
\BIBentryALTinterwordspacing
R.~G. Baraniuk, V.~Cevher, M.~F. Duarte, and C.~Hegde, ``Model-based
  compressive sensing,'' \emph{CoRR}, vol. abs/0808.3572, 2008. [Online].
  Available: \url{http://arxiv.org/abs/0808.3572}
\BIBentrySTDinterwordspacing

\bibitem{Candes:2005um}
E.~Candes, J.~Romberg, and T.~Tao, ``{Stable Signal Recovery from Incomplete
  and Inaccurate Measurements},'' \emph{arXiv.org}, Mar. 2005.

\bibitem{CandesTao:06}
E.~J. Cand\`es and T.~Tao, ``{Near-Optimal Signal Recovery From Random
  Projections: Universal Encoding Strategies?}'' \emph{IEEE Trans. Inform.
  Theory}, vol.~52, p. 5406, 2006.

\bibitem{CandesRombergTao06}
E.~Cand\`es, J.~Romberg, and T.~Tao, ``Robust uncertainty principles: Exact
  signal reconstruction from highly incomplete frequency information,''
  \emph{IEEE Trans. Inform. Theory}, vol.~52, pp. 489--509, 2006.

\bibitem{Donoho:06}
D.~L. Donoho, ``{Compressed sensing},'' \emph{IEEE Trans. Inform. Theory},
  vol.~52, p. 1289, 2006.

\bibitem{4472240}
E.~Candes and M.~Wakin, ``An introduction to compressive sampling,''
  \emph{Signal Processing Magazine, IEEE}, vol.~25, no.~2, pp. 21--30, March
  2008.

\bibitem{KrzakalaPRX2012}
F.~Krzakala, M.~M{\'e}zard, F.~Sausset, Y.~Sun, and L.~Zdeborov{\'a},
  ``Statistical physics-based reconstruction in compressed sensing,''
  \emph{Phys. Rev. X}, vol.~2, p. 021005, 2012.

\bibitem{KrzakalaMezard12}
------, ``Probabilistic reconstruction in compressed sensing: Algorithms, phase
  diagrams, and threshold achieving matrices,'' \emph{Journal of Statistical
  Mechanics: Theory and Experiment}, vol. 2012, no.~8, p. P08009, August 2012.

\bibitem{Lustig07compressedsensing}
M.~Lustig, D.~L. Donoho, J.~M. Santos, and J.~M. Pauly, ``Compressed sensing
  mri,'' in \emph{IEEE SIGNAL PROCESSING MAGAZINE}, 2007.

\bibitem{Wiaux21052009}
\BIBentryALTinterwordspacing
Y.~Wiaux, L.~Jacques, G.~Puy, A.~M.~M. Scaife, and P.~Vandergheynst,
  ``Compressed sensing imaging techniques for radio interferometry,''
  \emph{Monthly Notices of the Royal Astronomical Society}, vol. 395, no.~3,
  pp. 1733--1742, 2009. [Online]. Available:
  \url{http://mnras.oxfordjournals.org/content/395/3/1733.abstract}
\BIBentrySTDinterwordspacing

\bibitem{duarte2008single}
M.~F. Duarte, M.~A. Davenport, D.~Takhar, J.~N. Laska, T.~Sun, K.~F. Kelly, and
  R.~G. Baraniuk, ``Single-pixel imaging via compressive sampling,''
  \emph{Signal Processing Magazine, IEEE}, vol.~25, no.~2, pp. 83--91, 2008.

\bibitem{2014NatSR...4E5552L}
A.~{Liutkus}, D.~{Martina}, S.~{Popoff}, G.~{Chardon}, O.~{Katz}, G.~{Lerosey},
  S.~{Gigan}, L.~{Daudet}, and I.~{Carron}, ``{Imaging With Nature: Compressive
  Imaging Using a Multiply Scattering Medium},'' \emph{Scientific Reports},
  vol.~4, p. 5552, Jul. 2014.

\bibitem{2015arXiv150203324D}
A.~{Dremeau}, A.~{Liutkus}, D.~{Martina}, O.~{Katz}, C.~{Schulke},
  F.~{Krzakala}, S.~{Gigan}, and L.~{Daudet}, ``{Reference-less measurement of
  the transmission matrix of a highly scattering material using a DMD and phase
  retrieval techniques},'' \emph{ArXiv e-prints}, Feb. 2015.

\bibitem{2012ASAJ..132.1521C}
G.~{Chardon}, L.~{Daudet}, A.~{Peillot}, F.~{Ollivier}, N.~{Bertin}, and
  R.~{Gribonval}, ``{Near-field acoustic holography using sparse regularization
  and compressive sampling principles},'' \emph{Acoustical Society of America
  Journal}, vol. 132, p. 1521, 2012.

\bibitem{MignotDaudet11}
R.~Mignot, L.~Daudet, and F.~Ollivier, ``Compressed sensing for acoustic
  response reconstruction: Interpolation of the early part,'' in \emph{IEEE
  Workshop on Applications of Signal Processing to Audio and Acoustics
  (WASPAA)}, 2011, pp. 225--228.

\bibitem{DBLP:journals/corr/abs-1302-0189}
\BIBentryALTinterwordspacing
P.~Zhang, F.~Krzakala, M.~M{\'{e}}zard, and L.~Zdeborov{\'{a}}, ``Non-adaptive
  pooling strategies for detection of rare faulty items,'' \emph{CoRR}, vol.
  abs/1302.0189, 2013. [Online]. Available:
  \url{http://arxiv.org/abs/1302.0189}
\BIBentrySTDinterwordspacing

\bibitem{doi:10.1021/oc5000404}
\BIBentryALTinterwordspacing
J.~N. Sanders, X.~Andrade, and A.~Aspuru-Guzik, ``Compressed sensing for the
  fast computation of matrices: Application to molecular vibrations,''
  \emph{ACS Central Science}, vol.~1, no.~1, pp. 24--32, 2015. [Online].
  Available: \url{http://dx.doi.org/10.1021/oc5000404}
\BIBentrySTDinterwordspacing

\bibitem{papadimitriou1994computational}
\BIBentryALTinterwordspacing
C.~Papadimitriou, \emph{Computational Complexity}, ser. Theoretical computer
  science.\hskip 1em plus 0.5em minus 0.4em\relax Addison-Wesley, 1994.
  [Online]. Available: \url{http://books.google.fr/books?id=JogZAQAAIAAJ}
\BIBentrySTDinterwordspacing

\bibitem{Boyd:2004:CO:993483}
S.~Boyd and L.~Vandenberghe, \emph{Convex Optimization}.\hskip 1em plus 0.5em
  minus 0.4em\relax New York, NY, USA: Cambridge University Press, 2004.

\bibitem{van2009convex}
E.~van~den Berg, ``Convex optimization for generalized sparse recovery,'' Ph.D.
  dissertation, The University of British Columbia (Vancouver, 2009.

\bibitem{journals/corr/abs-1007-3753}
\BIBentryALTinterwordspacing
A.~Y. Yang, A.~Ganesh, Z.~Zhou, S.~Sastry, and Y.~Ma, ``A review of fast
  l1-minimization algorithms for robust face recognition,'' \emph{CoRR}, vol.
  abs/1007.3753, 2010. [Online]. Available:
  \url{http://dblp.uni-trier.de/db/journals/corr/corr1007.html#abs-1007-3753}
\BIBentrySTDinterwordspacing

\bibitem{2009RSPTA.367.4273D}
D.~{Donoho} and J.~{Tanner}, ``{Observed universality of phase transitions in
  high-dimensional geometry, with implications for modern data analysis and
  signal processing},'' \emph{Royal Society of London Philosophical
  Transactions Series A}, vol. 367, pp. 4273--4293, Nov. 2009.

\bibitem{decodingByLinearProgCandesTao}
E.~Candes and T.~Tao, ``Decoding by linear programming,'' \emph{Information
  Theory, IEEE Transactions on}, vol.~51, no.~12, pp. 4203--4215, Dec 2005.

\bibitem{Candes06robustuncertainty}
E.~J. Candès, J.~Romberg, and T.~Tao, ``Robust uncertainty principles: Exact
  signal reconstruction from highly incomplete frequency information,'' 2006.

\bibitem{journals/siamis/BeckerBC11}
\BIBentryALTinterwordspacing
S.~Becker, J.~Bobin, and E.~J. Candès, ``Nesta: A fast and accurate
  first-order method for sparse recovery.'' \emph{SIAM J. Imaging Sciences},
  vol.~4, no.~1, pp. 1--39, 2011. [Online]. Available:
  \url{http://dblp.uni-trier.de/db/journals/siamis/siamis4.html#BeckerBC11}
\BIBentrySTDinterwordspacing

\bibitem{NishimoriBook}
H.~Nishimori, \emph{Statistical Physics of Spin Glasses and Information
  Processing}.\hskip 1em plus 0.5em minus 0.4em\relax Oxford: Oxford University
  Press, 2001.

\bibitem{shannon48}
C.~Shannon, ``A mathematical theory of communication,'' \emph{Bell System
  Technical Journal}, vol.~27, pp. 379--423, 623--656, 1948.

\bibitem{GribonvalChardon11}
R.~Gribonval, G.~Chardon, and L.~Daudet, ``Blind calibration for compressed
  sensing by convex optimization,'' in \emph{IEEE International Conference on
  Acoustics, Speech and Signal Processing (ICASSP)}, 2012, pp. 2713 --2716.

\bibitem{KrzakalaMezard13b}
F.~Krzakala, M.~M\'ezard, and L.~Zdeborov\'a, ``Compressed sensing under matrix
  uncertainty: Optimum thresholds and robust approximate message passing,''
  2012, arXiv:1301.0901 [cs.IT], ICASSP 2013.

\bibitem{KabashimaKMSZ14}
\BIBentryALTinterwordspacing
Y.~Kabashima, F.~Krzakala, M.~M{\'{e}}zard, A.~Sakata., and L.~Zdeborov{\'{a}},
  ``Phase transitions and sample complexity in bayes-optimal matrix
  factorization,'' \emph{CoRR}, vol. abs/1402.1298, 2014. [Online]. Available:
  \url{http://arxiv.org/abs/1402.1298}
\BIBentrySTDinterwordspacing

\bibitem{DBLP:journals/corr/SchulkeCKZ13}
\BIBentryALTinterwordspacing
C.~Sch{\"{u}}lke, F.~Caltagirone, F.~Krzakala, and L.~Zdeborov{\'{a}}, ``Blind
  calibration in compressed sensing using message passing algorithms,''
  \emph{CoRR}, vol. abs/1306.4355, 2013. [Online]. Available:
  \url{http://arxiv.org/abs/1306.4355}
\BIBentrySTDinterwordspacing

\bibitem{DBLP:journals/corr/abs-1301-5898}
\BIBentryALTinterwordspacing
F.~Krzakala, M.~M{\'{e}}zard, and L.~Zdeborov{\'{a}}, ``Phase diagram and
  approximate message passing for blind calibration and dictionary learning,''
  \emph{CoRR}, vol. abs/1301.5898, 2013. [Online]. Available:
  \url{http://arxiv.org/abs/1301.5898}
\BIBentrySTDinterwordspacing

\bibitem{DBLP:journals/corr/BilenPGD13}
\BIBentryALTinterwordspacing
C.~Bilen, G.~Puy, R.~Gribonval, and L.~Daudet, ``Convex optimization approaches
  for blind sensor calibration using sparsity,'' \emph{CoRR}, vol.
  abs/1308.5354, 2013. [Online]. Available:
  \url{http://arxiv.org/abs/1308.5354}
\BIBentrySTDinterwordspacing

\bibitem{FigueiredoCOurse}
\BIBentryALTinterwordspacing
M.~Figueiredo, ``Lecture notes on bayesian estimation and classification.''
  [Online]. Available:
  \url{http://www.lx.it.pt/~mtf/learning/Bayes_lecture_notes.pdf}
\BIBentrySTDinterwordspacing

\bibitem{KrzakalaCOurse}
\BIBentryALTinterwordspacing
F.~Krzakala, L.~Zdeborova, M.~C. Angelini, and F.~Caltagirone, ``Statistical
  physics of inference and bayesian estimation.'' [Online]. Available:
  \url{http://indico.ictp.it/event/a14244/material/10/0.pdf}
\BIBentrySTDinterwordspacing

\bibitem{DBLP:journals/corr/abs-0704-2857}
\BIBentryALTinterwordspacing
A.~Montanari and R.~L. Urbanke, ``Modern coding theory: The statistical
  mechanics and computer science point of view,'' \emph{CoRR}, vol.
  abs/0704.2857, 2007. [Online]. Available:
  \url{http://arxiv.org/abs/0704.2857}
\BIBentrySTDinterwordspacing

\bibitem{zecchinaKsatScience}
M.~M. Zecchina~R. and P.~G., ``Analytic and algorithmic solution of random
  satisfiability problems,'' \emph{Science}, vol. 297, no. 812, 2002.

\bibitem{2001AIPC..553...89S}
D.~{Saad}, Y.~{Kabashima}, and T.~{Murayama}, ``{Public key cryptography and
  error correcting codes as Ising models},'' in \emph{American Institute of
  Physics Conference Series}, ser. American Institute of Physics Conference
  Series, vol. 553, Feb. 2001, pp. 89--94.

\bibitem{2002PhRvE..66d6120F}
S.~{Franz}, M.~{Leone}, A.~{Montanari}, and F.~{Ricci-Tersenghi}, ``{Dynamic
  phase transition for decoding algorithms},'' vol.~66, no.~4, p. 046120, Oct.
  2002.

\bibitem{MezardParisi87b}
M.~M{\'e}zard, G.~Parisi, and M.~A. Virasoro, \emph{Spin-Glass Theory and
  Beyond}.\hskip 1em plus 0.5em minus 0.4em\relax Singapore: World Scientific,
  1987, vol.~9.

\bibitem{2001PhyA..302...14S}
N.~{Sourlas}, ``{Statistical mechanics and capacity-approaching
  error-correcting codes},'' \emph{Physica A Statistical Mechanics and its
  Applications}, vol. 302, pp. 14--21, Dec. 2001.

\bibitem{SOURLAS_errorCorrection_book}
\BIBentryALTinterwordspacing
N.~Sourlas, ``\BIBforeignlanguage{English}{Statistical mechanics and
  error-correcting codes},'' in \emph{\BIBforeignlanguage{English}{From
  Statistical Physics to Statistical Inference and Back}}, ser. NATO ASI
  Series, P.~Grassberger and J.-P. Nadal, Eds.\hskip 1em plus 0.5em minus
  0.4em\relax Springer Netherlands, 1994, vol. 428, pp. 195--204. [Online].
  Available: \url{http://dx.doi.org/10.1007/978-94-011-1068-6_12}
\BIBentrySTDinterwordspacing

\bibitem{0295-5075-45-1-097}
\BIBentryALTinterwordspacing
Y.~Kabashima and D.~Saad, ``Statistical mechanics of error-correcting codes,''
  \emph{EPL (Europhysics Letters)}, vol.~45, no.~1, p.~97, 1999. [Online].
  Available: \url{http://stacks.iop.org/0295-5075/45/i=1/a=097}
\BIBentrySTDinterwordspacing

\bibitem{PhysRevE.60.132}
\BIBentryALTinterwordspacing
H.~Nishimori and K.~Y.~M. Wong, ``Statistical mechanics of image restoration
  and error-correcting codes,'' \emph{Phys. Rev. E}, vol.~60, pp. 132--144, Jul
  1999. [Online]. Available:
  \url{http://link.aps.org/doi/10.1103/PhysRevE.60.132}
\BIBentrySTDinterwordspacing

\bibitem{2006PhRvE..73b6122M}
G.~{Migliorini} and D.~{Saad}, ``{Finite-connectivity spin-glass phase diagrams
  and low-density parity check codes},'' vol.~73, no.~2, p. 026122, Feb. 2006.

\bibitem{1999PhRvE..60.5352V}
R.~{Vicente}, D.~{Saad}, and Y.~{Kabashima}, ``{Finite-connectivity systems as
  error-correcting codes},'' vol.~60, pp. 5352--5366, Nov. 1999.

\bibitem{2007JPhA...4012259A}
R.~C. {Alamino} and D.~{Saad}, ``{Statistical mechanics analysis of LDPC coding
  in MIMO Gaussian channels},'' \emph{Journal of Physics A Mathematical
  General}, vol.~40, pp. 12\,259--12\,279, Oct. 2007.

\bibitem{2000JPhA...33.1675K}
I.~{Kanter} and D.~{Saad}, ``{Finite-size effects and error-free communication
  in Gaussian channels},'' \emph{Journal of Physics A Mathematical General},
  vol.~33, pp. 1675--1681, Mar. 2000.

\bibitem{DBLP:journals/corr/ManoelKTZ14}
\BIBentryALTinterwordspacing
A.~Manoel, F.~Krzakala, E.~W. Tramel, and L.~Zdeborov{\'{a}}, ``Sparse
  estimation with the swept approximated message-passing algorithm,''
  \emph{CoRR}, vol. abs/1406.4311, 2014. [Online]. Available:
  \url{http://arxiv.org/abs/1406.4311}
\BIBentrySTDinterwordspacing

\bibitem{DBLP:journals/corr/CaltagironeKZ14}
\BIBentryALTinterwordspacing
F.~Caltagirone, F.~Krzakala, and L.~Zdeborov{\'{a}}, ``On convergence of
  approximate message passing,'' \emph{CoRR}, vol. abs/1401.6384, 2014.
  [Online]. Available: \url{http://arxiv.org/abs/1401.6384}
\BIBentrySTDinterwordspacing

\bibitem{DBLP:journals/corr/KrzakalaMTZ14}
\BIBentryALTinterwordspacing
F.~Krzakala, A.~Manoel, E.~W. Tramel, and L.~Zdeborov{\'{a}}, ``Variational
  free energies for compressed sensing,'' \emph{CoRR}, vol. abs/1402.1384,
  2014. [Online]. Available: \url{http://arxiv.org/abs/1402.1384}
\BIBentrySTDinterwordspacing

\bibitem{Opper_and_Saad:2001}
M.~Opper and D.~Saad, \emph{{Advanced Mean Field Methods : Theory and
  Practice}}.\hskip 1em plus 0.5em minus 0.4em\relax MIT press, 2001.

\bibitem{DBLP:journals/corr/abs-0805-0510}
\BIBentryALTinterwordspacing
T.~Blumensath and M.~E. Davies, ``Iterative hard thresholding for compressed
  sensing,'' \emph{CoRR}, vol. abs/0805.0510, 2008. [Online]. Available:
  \url{http://arxiv.org/abs/0805.0510}
\BIBentrySTDinterwordspacing

\bibitem{krzakalaGibbsStates06}
\BIBentryALTinterwordspacing
F.~Krzakala, A.~Montanari, F.~Ricci-Tersenghi, G.~Semerjian, and L.~Zdeborová,
  ``Gibbs states and the set of solutions of random constraint satisfaction
  problems,'' \emph{CoRR}, vol. abs/cond-mat/0612365, 2006. [Online].
  Available:
  \url{http://dblp.uni-trier.de/db/journals/corr/corr0612.html#abs-cond-mat-0612365}
\BIBentrySTDinterwordspacing

\bibitem{journals/corr/abs-0806-4112}
\BIBentryALTinterwordspacing
L.~Zdeborová, ``Statistical physics of hard optimization problems,''
  \emph{CoRR}, vol. abs/0806.4112, 2008. [Online]. Available:
  \url{http://dblp.uni-trier.de/db/journals/corr/corr0806.html#abs-0806-4112}
\BIBentrySTDinterwordspacing

\bibitem{Yedidia:2005:CFA:2263425.2271720}
\BIBentryALTinterwordspacing
J.~S. Yedidia, W.~T. Freeman, and Y.~Weiss, ``Constructing free-energy
  approximations and generalized belief propagation algorithms,'' \emph{IEEE
  Trans. Inf. Theor.}, vol.~51, no.~7, pp. 2282--2312, Jul. 2005. [Online].
  Available: \url{http://dx.doi.org/10.1109/TIT.2005.850085}
\BIBentrySTDinterwordspacing

\bibitem{2015arXiv150503504Z}
H.-J. {Zhou} and W.-M. {Zheng}, ``{Loop-corrected belief propagation for
  lattice spin models},'' \emph{ArXiv e-prints}, May 2015.

\bibitem{MezardParisiZecchina02}
M.~M{\'e}zard, G.~Parisi, and R.~Zecchina, ``Analytic and algorithmic solution
  of random satisfiability problems,'' \emph{Science}, vol. 297, pp. 812--815,
  2002.

\bibitem{DBLP:journals/corr/cs-CC-0212002}
\BIBentryALTinterwordspacing
A.~Braunstein, M.~M{\'{e}}zard, and R.~Zecchina, ``Survey propagation: an
  algorithm for satisfiability,'' \emph{CoRR}, vol. cs.CC/0212002, 2002.
  [Online]. Available: \url{http://arxiv.org/abs/cs.CC/0212002}
\BIBentrySTDinterwordspacing

\bibitem{2002cond.mat.12451B}
A.~{Braunstein}, M.~{Mezard}, M.~{Weigt}, and R.~{Zecchina}, ``{Constraint
  Satisfaction by Survey Propagation},'' \emph{eprint arXiv:cond-mat/0212451},
  Dec. 2002.

\bibitem{yedidia2003understanding}
J.~S. Yedidia, W.~T. Freeman, and Y.~Weiss, ``Understanding belief propagation
  and its generalizations,'' \emph{Exploring artificial intelligence in the new
  millennium}, vol.~8, pp. 236--239, 2003.

\bibitem{Rangan10b}
S.~Rangan, ``Generalized approximate message passing for estimation with random
  linear mixing,'' in \emph{Proc. of the IEEE Int. Symp. on Inform. Theory
  (ISIT)}, 2011, pp. 2168 --2172.

\bibitem{DonohoMaleki10}
D.~L. Donoho, A.~Maleki, and A.~Montanari, ``Message passing algorithms for
  compressed sensing: I. motivation and construction,'' in \emph{IEEE
  Information Theory Workshop (ITW)}, 2010, pp. 1 --5.

\bibitem{BaronSarvotham10}
D.~Baron, S.~Sarvotham, and R.~Baraniuk, ``Bayesian compressive sensing via
  belief propagation,'' \emph{IEEE Transactions on Signal Processing}, vol.~58,
  no.~1, pp. 269 -- 280, 2010.

\bibitem{ThoulessAnderson77}
D.~J. Thouless, P.~W. Anderson, and R.~G. Palmer, ``Solution of `solvable model
  of a spin-glass','' \emph{Phil. Mag.}, vol.~35, pp. 593--601, 1977.

\bibitem{DBLP:journals/corr/abs-cs-0503070}
\BIBentryALTinterwordspacing
J.~P. Neirotti and D.~Saad, ``Improved message passing for inference in densely
  connected systems,'' \emph{CoRR}, vol. abs/cs/0503070, 2005. [Online].
  Available: \url{http://arxiv.org/abs/cs/0503070}
\BIBentrySTDinterwordspacing

\bibitem{GuoWang06}
D.~Guo and C.-C. Wang, ``Asymptotic mean-square optimality of belief
  propagation for sparse linear systems,'' \emph{Information Theory Workshop,
  2006. ITW '06 Chengdu.}, pp. 194--198, 2006.

\bibitem{Rangan10}
S.~Rangan, ``Estimation with random linear mixing, belief propagation and
  compressed sensing,'' in \emph{Information Sciences and Systems (CISS), 2010
  44th Annual Conference on}, 2010, pp. 1 --6.

\bibitem{rangan2010estimation}
------, ``Estimation with random linear mixing, belief propagation and
  compressed sensing,'' in \emph{Information Sciences and Systems (CISS), 2010
  44th Annual Conference on}.\hskip 1em plus 0.5em minus 0.4em\relax IEEE,
  2010, pp. 1--6.

\bibitem{DonohoMaleki09}
D.~L. Donoho, A.~Maleki, and A.~Montanari, ``Message-passing algorithms for
  compressed sensing,'' \emph{Proc. Natl. Acad. Sci.}, vol. 106, no.~45, pp.
  18\,914--18\,919, 2009.

\bibitem{sudderth2010nonparametric}
E.~B. Sudderth, A.~T. Ihler, M.~Isard, W.~T. Freeman, and A.~S. Willsky,
  ``Nonparametric belief propagation,'' \emph{Communications of the ACM},
  vol.~53, no.~10, pp. 95--103, 2010.

\bibitem{DBLP:journals/corr/VilaSRKZ14}
\BIBentryALTinterwordspacing
J.~P. Vila, P.~Schniter, S.~Rangan, F.~Krzakala, and L.~Zdeborov{\'{a}},
  ``Adaptive damping and mean removal for the generalized approximate message
  passing algorithm,'' \emph{CoRR}, vol. abs/1412.2005, 2014. [Online].
  Available: \url{http://arxiv.org/abs/1412.2005}
\BIBentrySTDinterwordspacing

\bibitem{DBLP:journals/corr/GuoX15}
\BIBentryALTinterwordspacing
Q.~Guo and J.~Xi, ``Approximate message passing with unitary transformation,''
  \emph{CoRR}, vol. abs/1504.04799, 2015. [Online]. Available:
  \url{http://arxiv.org/abs/1504.04799}
\BIBentrySTDinterwordspacing

\bibitem{WuVerdu11}
Y.~Wu and S.~Verdu, ``Optimal phase transitions in compressed sensing,'' 2011,
  arXiv:1111.6822v1 [cs.IT].

\bibitem{RanganFletcherGoyal09}
S.~Rangan, A.~Fletcher, and V.~Goyal, ``Asymptotic analysis of map estimation
  via the replica method and applications to compressed sensing,''
  \emph{arXiv:0906.3234v2}, 2009.

\bibitem{GuoBaron09}
D.~Guo, D.~Baron, and S.~Shamai, ``A single-letter characterization of optimal
  noisy compressed sensing,'' in \emph{47th Annual Allerton Conference on
  Communication, Control, and Computing}, 2009, pp. 52 -- 59.

\bibitem{BayatiMontanari10}
M.~Bayati and A.~Montanari, ``The dynamics of message passing on dense graphs,
  with applications to compressed sensing,'' \emph{IEEE Transactions on
  Information Theory}, vol.~57, no.~2, pp. 764 --785, 2011.

\bibitem{GuoWang07}
D.~Guo and C.-C. Wang, ``Random sparse linear system observed via arbitrary
  channels: A decoupling principle,'' \emph{Proc. IEEE Int. Symp. Inform. Th.,
  Nice, France}, pp. 946--950, 2007.

\bibitem{WenW14}
\BIBentryALTinterwordspacing
C.~Wen and K.~Wong, ``Analysis of compressed sensing with spatially-coupled
  orthogonal matrices,'' \emph{CoRR}, vol. abs/1402.3215, 2014. [Online].
  Available: \url{http://arxiv.org/abs/1402.3215}
\BIBentrySTDinterwordspacing

\bibitem{FelstromZigangirov99}
A.~Jimenez~Felstrom and K.~Zigangirov, ``Time-varying periodic convolutional
  codes with low-density parity-check matrix,'' \emph{Information Theory, IEEE
  Transactions on}, vol.~45, no.~6, pp. 2181 --2191, 1999.

\bibitem{KudekarRichardson10}
S.~Kudekar, T.~Richardson, and R.~Urbanke, ``Threshold saturation via spatial
  coupling: Why convolutional ldpc ensembles perform so well over the bec,'' in
  \emph{Proc. of the IEEE Int. Symposium on Information Theory (ISIT),}, 2010,
  pp. 684--688.

\bibitem{KudekarRichardson12}
------, ``Spatially coupled ensembles universally achieve capacity under belief
  propagation,'' 2012, arXiv:1201.2999v1 [cs.IT].

\bibitem{DonohoJavanmard11}
D.~Donoho, A.~Javanmard, and A.~Montanari, ``Information-theoretically optimal
  compressed sensing via spatial coupling and approximate message passing,'' in
  \emph{Proc. of the IEEE Int. Symposium on Information Theory (ISIT)}, 2012,
  pp. 1231--1235.

\bibitem{DBLP:journals/corr/abs-1301-5676}
\BIBentryALTinterwordspacing
A.~Giurgiu, N.~Macris, and R.~L. Urbanke, ``Spatial coupling as a proof
  technique,'' \emph{CoRR}, vol. abs/1301.5676, 2013. [Online]. Available:
  \url{http://arxiv.org/abs/1301.5676}
\BIBentrySTDinterwordspacing

\bibitem{DBLP:journals/corr/abs-1105-0785}
\BIBentryALTinterwordspacing
S.~H. Hassani, N.~Macris, and R.~L. Urbanke, ``Coupled graphical models and
  their thresholds,'' \emph{CoRR}, vol. abs/1105.0785, 2011. [Online].
  Available: \url{http://arxiv.org/abs/1105.0785}
\BIBentrySTDinterwordspacing

\bibitem{EPFL-REPORT-195693}
D.~Achlioptas, S.~Hamed~Hassani, N.~Macris, and R.~Urbanke, ``New {B}ounds for
  {R}andom {C}onstraint {S}atisfaction {P}roblems via {S}patial {C}oupling,''
  Tech. Rep., 2013.

\bibitem{DBLP:journals/corr/abs-1207-2853}
\BIBentryALTinterwordspacing
M.~C. Angelini, F.~Ricci{-}Tersenghi, and Y.~Kabashima, ``Compressed sensing
  with sparse, structured matrices,'' \emph{CoRR}, vol. abs/1207.2853, 2012.
  [Online]. Available: \url{http://arxiv.org/abs/1207.2853}
\BIBentrySTDinterwordspacing

\bibitem{caltagirone2014properties}
F.~Caltagirone and L.~Zdeborov{\'a}, ``Properties of spatial coupling in
  compressed sensing,'' \emph{arXiv preprint arXiv:1401.6380}, 2014.

\bibitem{journals/corr/CaltagironeFMZ13}
\BIBentryALTinterwordspacing
F.~Caltagirone, S.~Franz, R.~Morris, and L.~Zdeborová, ``Dynamics and
  termination cost of spatially coupled mean-field models.'' \emph{CoRR}, vol.
  abs/1310.2121, 2013. [Online]. Available:
  \url{http://dblp.uni-trier.de/db/journals/corr/corr1310.html#CaltagironeFMZ13}
\BIBentrySTDinterwordspacing

\bibitem{montanari2012graphical}
A.~Montanari, ``Graphical models concepts in compressed sensing,''
  \emph{Compressed Sensing: Theory and Applications}, pp. 394--438, 2012.

\bibitem{VilaSchniter11}
J.~P. Vila and P.~Schniter, ``Expectation-maximization bernoulli-gaussian
  approximate message passing,'' in \emph{Proc. Asilomar Conf. on Signals,
  Systems, and Computers (Pacific Grove, CA)}, 2011.

\bibitem{KudekarPfister10}
S.~Kudekar and H.~Pfister, ``The effect of spatial coupling on compressive
  sensing,'' in \emph{Communication, Control, and Computing (Allerton)}, 2010,
  pp. 347--353.

\bibitem{YedlaJian12}
A.~Yedla, Y.~Jian, P.~Nguyen, and H.~Pfister, ``A simple proof of threshold
  saturation for coupled scalar recursions,'' 2012, arXiv:1204.5703v1 [cs.IT].

\bibitem{AmraouiMontanari2007}
\BIBentryALTinterwordspacing
A.~Amraoui, A.~Montanari, and R.~Urbanke, ``How to find good finite-length
  codes: from art towards science,'' \emph{European Transactions on
  Telecommunications}, vol.~18, no.~5, pp. 491--508, 2007. [Online]. Available:
  \url{http://dx.doi.org/10.1002/ett.1182}
\BIBentrySTDinterwordspacing

\bibitem{DBLP:journals/corr/abs-1108-2632}
\BIBentryALTinterwordspacing
S.~Som and P.~Schniter, ``Compressive imaging using approximate message passing
  and a markov-tree prior,'' \emph{CoRR}, vol. abs/1108.2632, 2011. [Online].
  Available: \url{http://arxiv.org/abs/1108.2632}
\BIBentrySTDinterwordspacing

\bibitem{DBLP:journals/corr/VilaSM15}
\BIBentryALTinterwordspacing
J.~P. Vila, P.~Schniter, and J.~Meola, ``Hyperspectral unmixing via turbo
  bilinear approximate message passing,'' \emph{CoRR}, vol. abs/1502.06435,
  2015. [Online]. Available: \url{http://arxiv.org/abs/1502.06435}
\BIBentrySTDinterwordspacing

\bibitem{DBLP:journals/tip/DongSLMH14}
\BIBentryALTinterwordspacing
W.~Dong, G.~Shi, X.~Li, Y.~Ma, and F.~Huang, ``Compressive sensing via nonlocal
  low-rank regularization,'' \emph{{IEEE} Transactions on Image Processing},
  vol.~23, no.~8, pp. 3618--3632, 2014. [Online]. Available:
  \url{http://dx.doi.org/10.1109/TIP.2014.2329449}
\BIBentrySTDinterwordspacing

\bibitem{DBLP:journals/tsp/TanMB15}
\BIBentryALTinterwordspacing
J.~Tan, Y.~Ma, and D.~Baron, ``Compressive imaging via approximate message
  passing with image denoising,'' \emph{{IEEE} Transactions on Signal
  Processing}, vol.~63, no.~8, pp. 2085--2092, 2015. [Online]. Available:
  \url{http://dx.doi.org/10.1109/TSP.2015.2408558}
\BIBentrySTDinterwordspacing

\bibitem{DBLP:journals/corr/MetzlerMB14}
\BIBentryALTinterwordspacing
C.~A. Metzler, A.~Maleki, and R.~G. Baraniuk, ``From denoising to compressed
  sensing,'' \emph{CoRR}, vol. abs/1406.4175, 2014. [Online]. Available:
  \url{http://arxiv.org/abs/1406.4175}
\BIBentrySTDinterwordspacing

\bibitem{do2008fast}
T.~T. Do, T.~D. Tran, and L.~Gan, ``Fast compressive sampling with structurally
  random matrices,'' in \emph{Acoustics, Speech and Signal Processing, 2008.
  ICASSP 2008. IEEE International Conference on}.\hskip 1em plus 0.5em minus
  0.4em\relax IEEE, 2008, pp. 3369--3372.

\bibitem{JavanmardMontanari12}
A.~Javanmard and A.~Montanari, ``Subsampling at information theoretically
  optimal rates,'' 2012, arXiv:1202.2525v1 [cs.IT].

\bibitem{kamilovsparse}
U.~S. Kamilov, A.~Bourquard, and M.~Unser, ``Sparse image deconvolution with
  message passing,'' sPARSE 2013.

\bibitem{vehkapera2013analysis}
M.~Vehkapera, Y.~Kabashima, and S.~Chatterjee, ``Analysis of regularized ls
  reconstruction and random matrix ensembles in compressed sensing,''
  \emph{arXiv preprint arXiv:1312.0256}, 2013.

\bibitem{wen2014analysis}
C.~Wen and K.~Wong, ``Analysis of compressed sensing with spatially-coupled
  orthogonal matrices,'' \emph{arXiv preprint arXiv:1402.3215}, 2014.

\bibitem{maleki2012asymptotic}
A.~Maleki, L.~Anitori, Z.~Yang, and R.~Baraniuk, ``Asymptotic analysis of
  complex lasso via complex approximate message passing (camp),'' 2012.

\bibitem{Schniter2012compressive}
P.~Schniter and S.~Rangan, ``Compressive phase retrieval via generalized
  approximate message passing,'' in \emph{Communication, Control, and Computing
  (Allerton), 2012 50th Annual Allerton Conference on}.\hskip 1em plus 0.5em
  minus 0.4em\relax IEEE, 2012, pp. 815--822.

\bibitem{DBLP:journals/corr/abs-1108-0477}
\BIBentryALTinterwordspacing
A.~Maleki, L.~Anitori, Z.~Yang, and R.~G. Baraniuk, ``Asymptotic analysis of
  complex {LASSO} via complex approximate message passing {(CAMP)},''
  \emph{CoRR}, vol. abs/1108.0477, 2011. [Online]. Available:
  \url{http://arxiv.org/abs/1108.0477}
\BIBentrySTDinterwordspacing

\bibitem{DBLP:journals/corr/SchniterR14}
\BIBentryALTinterwordspacing
P.~Schniter and S.~Rangan, ``Compressive phase retrieval via generalized
  approximate message passing,'' \emph{CoRR}, vol. abs/1405.5618, 2014.
  [Online]. Available: \url{http://arxiv.org/abs/1405.5618}
\BIBentrySTDinterwordspacing

\bibitem{bayati2012universality}
M.~Bayati, M.~Lelarge, and A.~Montanari, ``Universality in polytope phase
  transitions and message passing algorithms,'' \emph{arXiv preprint
  arXiv:1207.7321}, 2012.

\bibitem{Li2009}
C.~Li, ``An efficient algorithm for total variation regularization with
  applications to the single pixel camera and compressive sensing,'' Master's
  thesis, Rice University, Sep. 2009.

\bibitem{DJM2013}
D.~Donoho, I.~Johnstone, and A.~Montanari, ``Accurate prediction of phase
  transitions in compressed sensing via a connection to minimax denoising,''
  \emph{arXiv Preprint}, no. 1111.1041, January 2013.

\bibitem{BP2013}
M.~A. Borgerding and P.~Schniter, ``Generalized approximate message passing for
  the cosparse analysis model,'' \emph{arXiv Preprint}, no. 1312.3968, December
  2013.

\bibitem{KJL2014}
J.~Kang, H.~Jung, H.-N. Lee, and K.~Kim, ``Spike-and-slab approximate
  message-passing for high-dimensional picewise-constant recovery,''
  \emph{arXiv Preprint}, no. 1408.3930, August 2014.

\bibitem{Studer26062012}
\BIBentryALTinterwordspacing
V.~Studer, J.~Bobin, M.~Chahid, H.~S. Mousavi, E.~Candes, and M.~Dahan,
  ``Compressive fluorescence microscopy for biological and hyperspectral
  imaging,'' \emph{Proceedings of the National Academy of Sciences}, vol. 109,
  no.~26, pp. E1679--E1687, 2012. [Online]. Available:
  \url{http://www.pnas.org/content/109/26/E1679.abstract}
\BIBentrySTDinterwordspacing

\bibitem{barron2010sparse}
A.~R. Barron and A.~Joseph, ``Sparse superposition codes: Fast and reliable at
  rates approaching capacity with gaussian noise,'' \emph{Manuscript. Available
  at ‚Äúhttp://www. stat. yale. edu/~ arb4}, 2010.

\bibitem{barron2011analysis}
------, ``Analysis of fast sparse superposition codes,'' in \emph{Information
  Theory Proceedings (ISIT), 2011 IEEE International Symposium on}.\hskip 1em
  plus 0.5em minus 0.4em\relax IEEE, 2011, pp. 1772--1776.

\bibitem{joseph2012least}
A.~Joseph and A.~R. Barron, ``Least squares superposition codes of moderate
  dictionary size are reliable at rates up to capacity,'' \emph{Information
  Theory, IEEE Transactions on}, vol.~58, no.~5, pp. 2541--2557, 2012.

\bibitem{barron2012high}
A.~R. Barron and S.~Cho, ``High-rate sparse superposition codes with
  iteratively optimal estimates,'' in \emph{Information Theory Proceedings
  (ISIT), 2012 IEEE International Symposium on}.\hskip 1em plus 0.5em minus
  0.4em\relax IEEE, 2012, pp. 120--124.

\bibitem{choapproximate}
S.~Cho and A.~Barron, ``Approximate iterative bayes optimal estimates for
  high-rate sparse superposition codes.''

\bibitem{RichardsonUrbanke08}
T.~Richardson and R.~Urbanke, \emph{{Modern Coding Theory}}.\hskip 1em plus
  0.5em minus 0.4em\relax Cambridge University Press, 2008.

\bibitem{rush2015capacity}
C.~Rush, A.~Greig, and R.~Venkataramanan, ``Capacity-achieving sparse
  superposition codes via approximate message passing decoding,'' \emph{arXiv
  preprint arXiv:1501.05892}, 2015.

\bibitem{javanmard2013state}
A.~Javanmard and A.~Montanari, ``State evolution for general approximate
  message passing algorithms, with applications to spatial coupling,''
  \emph{Information and Inference}, p. iat004, 2013.

\bibitem{derrida1980random}
B.~Derrida, ``Random-energy model: Limit of a family of disordered models,''
  \emph{Physical Review Letters}, vol.~45, no.~2, pp. 79--82, 1980.

\bibitem{arous2005limit}
G.~B. Arous, L.~V. Bogachev, and S.~A. Molchanov, ``Limit theorems for sums of
  random exponentials,'' \emph{Probability theory and related fields}, vol.
  132, no.~4, pp. 579--612, 2005.

\bibitem{1056050}
A.~Wyner, ``An analog scrambling scheme which does not expand bandwidth, part
  i: Discrete time,'' \emph{Information Theory, IEEE Transactions on}, vol.~25,
  no.~3, pp. 261--274, 1979.

\bibitem{feizi2011power}
S.~Feizi and M.~Medard, ``A power efficient sensing/communication scheme: Joint
  source-channel-network coding by using compressive sensing,'' in
  \emph{Communication, Control, and Computing (Allerton), 2011 49th Annual
  Allerton Conference on}.\hskip 1em plus 0.5em minus 0.4em\relax IEEE, 2011,
  pp. 1048--1054.

\bibitem{shintre2008real}
S.~Shintre, S.~Katti, S.~Jaggi, B.~Dey, D.~Katabi, and M.~Medard, ``Real and
  complex network codes: Promises and challenges,'' 2008.

\bibitem{grangetto2005joint}
M.~Grangetto, P.~Cosman, and G.~Olmo, ``Joint source/channel coding and map
  decoding of arithmetic codes,'' \emph{Communications, IEEE Transactions on},
  vol.~53, no.~6, pp. 1007--1016, 2005.

\bibitem{4595196}
G.~Caire, T.~Al-Naffouri, and A.~Narayanan, ``Impulse noise cancellation in
  ofdm: an application of compressed sensing,'' in \emph{Information Theory,
  2008. ISIT 2008. IEEE International Symposium on}, 2008, pp. 1293--1297.

\bibitem{959265}
D.~Donoho and X.~Huo, ``Uncertainty principles and ideal atomic
  decomposition,'' \emph{Information Theory, IEEE Transactions on}, vol.~47,
  no.~7, pp. 2845--2862, 2001.

\bibitem{CandesTao:05}
E.~J. Cand{\`e}s and T.~Tao, ``{Decoding by linear programming},'' \emph{IEEE
  Trans. Inform. Theory}, vol.~51, p. 4203, 2005.

\bibitem{Candes:2008:HRE:2263476.2273354}
\BIBentryALTinterwordspacing
E.~J. Candes and P.~A. Randall, ``Highly robust error correction byconvex
  programming,'' \emph{IEEE Trans. Inf. Theor.}, vol.~54, no.~7, pp.
  2829--2840, Jul. 2008. [Online]. Available:
  \url{http://dx.doi.org/10.1109/TIT.2008.924688}
\BIBentrySTDinterwordspacing

\bibitem{lampe2011bursty}
L.~Lampe, ``Bursty impulse noise detection by compressed sensing,'' in
  \emph{Power Line Communications and Its Applications (ISPLC), 2011 IEEE
  International Symposium on}.\hskip 1em plus 0.5em minus 0.4em\relax IEEE,
  2011, pp. 29--34.

\bibitem{Donoho05072005}
D.~L. Donoho and J.~Tanner, ``Sparse nonnegative solution of underdetermined
  linear equations by linear programming,'' \emph{Proc. Natl. Acad. Sci.}, vol.
  102, no.~27, pp. 9446--9451, 2005.

\end{thebibliography}
\addcontentsline{toc}{chapter}{Bibliography}
\end{document}